\documentclass[preprint,authoryear,12pt]{elsarticle}

\usepackage{graphicx}

\usepackage{amssymb}

\usepackage{lineno}

\journal{J. Theor. Biol.}

\usepackage{color}
\usepackage{times}
\usepackage{txfonts}

\newcommand{\Detail}[1]{}

\newcommand{\script}[1]{{\mbox{\scriptsize #1}}}
\newcommand{\CITE}[1]{ \citep{#1}}

\newcommand{\Eqno}[1]{{(#1)}}
\newcommand{\Eq}[1]{Eq. \Eqno{#1}}
\newcommand{\Eqs}[1]{Eqs. {#1}}
\newcommand{\Ref}[1]{\Eqno{\ref{#1}}}

\newcommand{\EqPunc}[1]{}
\newcommand{\EqPeriod}[1]{}
\newcommand{\Fig}[1]{Fig. {#1}}

\newcommand{\Figs}[1]{Figs. {#1}}

\newcommand{\FigureInText}[1]{}
\newcommand{\FigureLegends}[1]{#1}
\newcommand{\FigureInLegends}[1]{}
\newcommand{\TableInText}[1]{}
\newcommand{\TableInLegends}[1]{#1}

\newcommand{\EQUATION}[1]{Equation \Eqno{#1}}
\newcommand{\EQUATIONS}[1]{Equations {#1}}

\renewcommand{\FigureInText}[1]{#1}
\renewcommand{\FigureInLegends}[1]{#1}
\newcommand{\NoFigureInLegends}[1]{#1}
\FigureInLegends{
\renewcommand{\NoFigureInLegends}[1]{}
}
\renewcommand{\FigureLegends}[1]{#1}
\newcommand{\NoFigureInText}[1]{#1}
\FigureInText{
\renewcommand{\NoFigureInText}[1]{}
}

\renewcommand{\TableInText}[1]{#1}
\renewcommand{\TableInLegends}[1]{#1}
\newcommand{\NoTableInLegends}[1]{#1}
\TableInLegends{
\renewcommand{\NoTableInLegends}[1]{}
}

\newcommand{\ColorRed}{\color{red}}
\newcommand{\DefColor}{\color{black}}
\newcommand{\BF}[1]{\textbf{#1}}

\newcommand{\RED}[1]{\textcolor{red}{#1}}

\newcommand{\Red}[1]{\textcolor{red}{#1}}
\renewcommand{\ColorRed}{}
\renewcommand{\DefColor}{}
\renewcommand{\Red}[1]{#1}
\renewcommand{\RED}[1]{#1}

\newcommand{\TextFig}[1]{#1}
\newcommand{\SupFig}[1]{#1}

\newcommand{\SupplementaryMaterial}[1]{#1}

\begin{document}


\begin{frontmatter}



\title{{Selection maintaining protein stability at equilibrium}}


\author{Sanzo Miyazawa}
\ead{sanzo.miyazawa@gmail.com}
\address{6-5-607 Miyanodai, Sakura, Chiba 285-0857, Japan}

\begin{abstract}

The common understanding of protein evolution
has been that neutral mutations 
are fixed by random drift, and
a proportion of neutral mutations
depending on the strength of structural and functional constraints
primarily determines evolutionary rate.
Recently it was 
indicated 
that fitness costs due to misfolded proteins 
are a determinant of evolutionary rate
and selection originating in protein stability 
is a driving force of protein evolution.
Here we examine protein evolution under the selection 
maintaining protein stability.

Protein fitness is 
a generic form of fitness costs due to  
misfolded proteins;
$s = \kappa \exp(\Delta G / kT) ( 1 - \exp(\Delta\Delta G / kT))$, where 
$s$ and $\Delta\Delta G$ are selective advantage and 
stability change 
of a mutant protein, 
$\Delta G$ is the folding free energy of the wild-type protein,
and $\kappa$ is a parameter representing protein abundance and indispensability. 
The distribution of $\Delta\Delta G$ is approximated 
to be a bi-Gaussian distribution,
which represents structurally slightly- or highly-constrained sites.
Also, the mean of the distribution is negatively proportional
to $\Delta G$.

The evolution of this gene has an equilibrium point ($\Delta G_e$) of
protein stability, the range of which is consistent with observed values
in the ProTherm database.
The probability distribution
of $K_a/K_s$, the ratio of 
nonsynonymous to synonymous substitution rate per site,
over fixed mutants in the vicinity of the equilibrium 
shows that 
nearly neutral selection is predominant only in low-abundant, non-essential proteins
of $\Delta G_e > -2.5$ kcal/mol. 
In the other proteins, positive selection on stabilizing mutations
is significant to maintain protein stability at equilibrium
as well as random drift on slightly negative mutations,
although the average $\langle K_a/K_s \rangle$ is less than 1.
Slow evolutionary rates can be caused by
both high protein abundance/indispensability and large effective population size, 
which
produces positive shifts of $\Delta\Delta G$ through decreasing $\Delta G_e$,
and strong structural constraints, which 
directly make $\Delta\Delta G$ more positive.
Protein abundance/indispensability more 
affect 
evolutionary rate for 
less constrained proteins, and
structural constraint for less abundant, less essential proteins.
The effect of protein indispensability on evolutionary rate
may be hidden by the variation of protein abundance and detected only
in low-abundant proteins.
\Red{
Also,
\RED{
protein stability ($-\Delta G_e/kT$)
} 
and $\langle K_a/K_s \rangle$ are predicted to decrease as growth temperature increases.
} 

\vspace*{1em}
\noindent
Highlights
\begin{itemize}
\item Protein stability is kept at equilibrium by random drift and positive selection.
\item Neutral selection is predominant only for low-abundant, non-essential proteins.
\item Protein abundance more decreases evolutionary rate for less-constrained proteins.
\item Structural constraint more decreases 
evolutionary
rate for less-abundant, less-essential proteins.
\Red{
\item 
\RED{
Protein stability ($-\Delta G_e/kT$)
} 
and $\langle K_a/K_s \rangle$ are predicted to decrease as growth temperature increases.
} 
\end{itemize}

\end{abstract}


\end{frontmatter}



\section{Introduction}

The common understanding of protein evolution
has been that amino acid substitutions observed in homologous proteins are
neutral \CITE{K:68,K:69,KO:71,KO:74} or slightly deleterious
\CITE{O:73,O:92}, and random drift is a primary force to fix amino acid
substitutions in population.
The rate of protein evolution has been understood
to be determined primarily by the proportion of neutral mutations, which 
may be measured by the ratio of nonsynonymous to synonymous substitution rate
per site ($K_a/K_s$) \CITE{MY:80} and determined by
functional density\CITE{Z:76} 
weighted by the relative variability at specific-function sites of a protein
\CITE{GM:80}.
Since then, these theories have
been widely accepted, however,
recently a question has been raised on
whether the diversity of protein
evolutionary rate among genes
can be explained only by
the proportion and the variability of 
specific-function sites,
and molecular and population-genetic
constraints on protein evolutionary rate have been
explored.

Recent works have revealed
that protein evolutionary rate is correlated with 
gene expression level;
highly expressed genes evolve slowly, accounting for
as much as 34\% of rate variation in yeast\CITE{PPH:01}.
Of course,
there are many reports that support a principle of lower evolution rate 
for stronger functional density.
Broadly expressed proteins in many tissues
tend to evolve slower than tissue-specific ones\CITE{KIM:95,DM:00}.
The connectivity of well-conserved proteins in a network is shown
\CITE{FHSSF:02} to be negatively correlated
with their rate of evolution, because a greater
proportion of the protein sequence is directly involved in its function.
A fitness cost due to protein--protein misinteraction 
affects the evolutionary rate of surface residues\CITE{YLZZ:12}.
Protein dispensability in yeast is correlated with the rate of evolution\CITE{HF:01,HF:03},
although there is a report insisting on no correlation between them\CITE{PPH:03}.
Other reports indicate that
the correlation between gene dispensability and evolutionary
rate, although low, is significant\CITE{ZH:05,WHFKGEF:05,JRWK:02}.

It was proposed\CITE{DBAWA:05,DW:08,GDBWHD:11} that
low substitution rates of highly expressed genes
could be explained by fitness costs due to functional loss and toxicity\CITE{SDD:08,GDBWHD:11}
of misfolded proteins.
Misfolding reduces the concentration of functional proteins,
and wastes cellular time and energy on production of useless proteins.
Also misfolded proteins form insoluble aggregates\CITE{GDBWHD:11}.
Fitness cost due to misfolded proteins is larger for highly expressed genes 
than for less expressed ones. 

Fitness cost due to misfolded proteins 
was formulated\CITE{DW:08,GDBWHD:11} to be related to the proportion of misfolded
proteins.  Knowledge of protein folding indicates that
protein folding primarily occurs in two-state transition\CITE{MJ:82,MJ:82B}, 
which means
that the ensemble of protein conformations are a mixture of completely folded and unfolded conformations.
Free energy ($\Delta G$) of protein stability, which is equal to the free energy of 
the denatured state subtracted from that of the native state, 
and stability change ($\Delta\Delta G$) due to 
amino acid substitutions 
are 
collected in the ProTherm database\CITE{KBGPKUS:06}, although the data are not sufficient. 
Prediction methods, however, for $\Delta\Delta G$ are improved enough to 
reproduce real distributions of $\Delta\Delta G$\CITE{SBSNRS:05,YDD:07}. 
Therefore, on the biophysical basis, the distribution of fitness
can be estimated and protein evolution can be studied. 
Shakhnovich group 
studied protein evolution on the basis of knowledge of protein folding
\CITE{SS:14,DSKS:14} and showed\CITE{SRS:12} that 
the negative correlation between
protein abundance and $K_a/K_s$
was caused by 
the distribution of $\Delta\Delta G$
that negatively correlates with the $\Delta G$ of a wild type.
Also, it was shown\CITE{SLS:13} that highly abundant proteins had to be more stable
than low abundant ones.
Relationship between evolutionary rate and protein stability is studied
from various points of view \CITE{EJW:15,FK:15}.

Here we study relationship between evolutionary rate and
selection on protein stability in a monoclonal approximation.
A fitness assumed here for a protein is a generic form to which all formulations 
\CITE{DW:08,GDBWHD:11,SRS:12,SLS:13,SS:14,DSKS:14}
previously employed 
for protein fitness are reduced 
in the condition of $\exp (\beta \Delta G) \ll 1$, 
which is satisfied
in the typical range of folding free energies
shown in \Fig{\ref{fig: observed_distribution_of_protein_stabilities}};
$\beta = 1 / (kT)$, $k$ is the Boltzmann constant and $T$ is absolute temperature. 
The generic form of Malthusian fitness of a protein-coding gene
is $m \equiv - \kappa \exp(\beta\Delta G)$,
where 
$\kappa$ is a parameter, which may be a function of protein abundance and dispensability; 
see Methods for details.
The distribution of stability change $\Delta\Delta G$ due to single amino acid substitutions
is approximated as a weighted sum of two Gaussian functions
that was shown\CITE{TSSST:07} 
to well reproduce actual distributions of $\Delta\Delta G$. 
One of the two Gaussian functions describes substitutions at structurally less-constrained surface sites, 
and the other at more-constrained core sites of proteins. 
The proportion of less-constrained surface sites is a parameter ($\theta$).

The fixation probability of a mutant with $\Delta\Delta G$ can be calculated 
for a duploid population with effective population size $N_e$\CITE{CK:70}.  
In the population of genes with such a fitness 
protein stability is evolutionarily maintained at equilibrium,
and equilibrium stability ($\Delta G_e$) negatively correlates with 
protein abundance/dispensability ($\kappa$). 
\Red{
The range of $\Delta G_e$ is consistent with the observed range of folding free energies
shown in \Fig{\ref{fig: observed_distribution_of_protein_stabilities}}. 
} 

The probability density functions (PDF) of 
$K_a/K_s$,
the ratio of nonsynonymous 
to synonymous substitution rate per site\CITE{MY:80}, 
at equilibrium and also in the vicinity of
equilibrium are numerically examined over a whole
domain of the parameters, $0 \leq \log 4N_e\kappa \leq 20$ and $0 \leq \theta \leq 1$.
The dependences of evolutionary rate on protein abundance/dispensability and on 
structural constraint are quantitatively described, and it is shown that both factors
cannot be ignored on protein evolutionary rate, although
protein abundance/indispensability more affect evolutionary rate for
less constrained proteins, and
structural constraint for less abundant, less essential proteins.
Like protein abundance, protein indispensability must correlate with evolutionary rate,
but a correlation between them
may be hidden by the variation of protein abundance
as well as effective population size,
and detected only in low-abundant proteins.
It has also become clear that 
nearly neutral selection is predominant only 
in low-abundant, non-essential proteins with 
$\log 4N_e\kappa< 2$ or $\Delta G_e > -2.5$ kcal/mol, and
in the other proteins
positive selection is significant to more stabilize a less-stable wild type.
Also, a significant amount of slightly negative mutants are fixed in population by random drift.
This view of protein evolution is contrary to the previous understanding.
\Red{
The present model based on 
a biophysical knowledge of protein stability
also indicates that 
\RED{
protein stability ($-\beta\Delta G_e$)
} 
and the average of $K_a/K_s$ decrease
as growth temperature increases.
} 

\FigureInText{

\TextFig{

\begin{figure*}[ht]
\FigureInLegends{
\centerline{
\includegraphics*[width=90mm,angle=0]{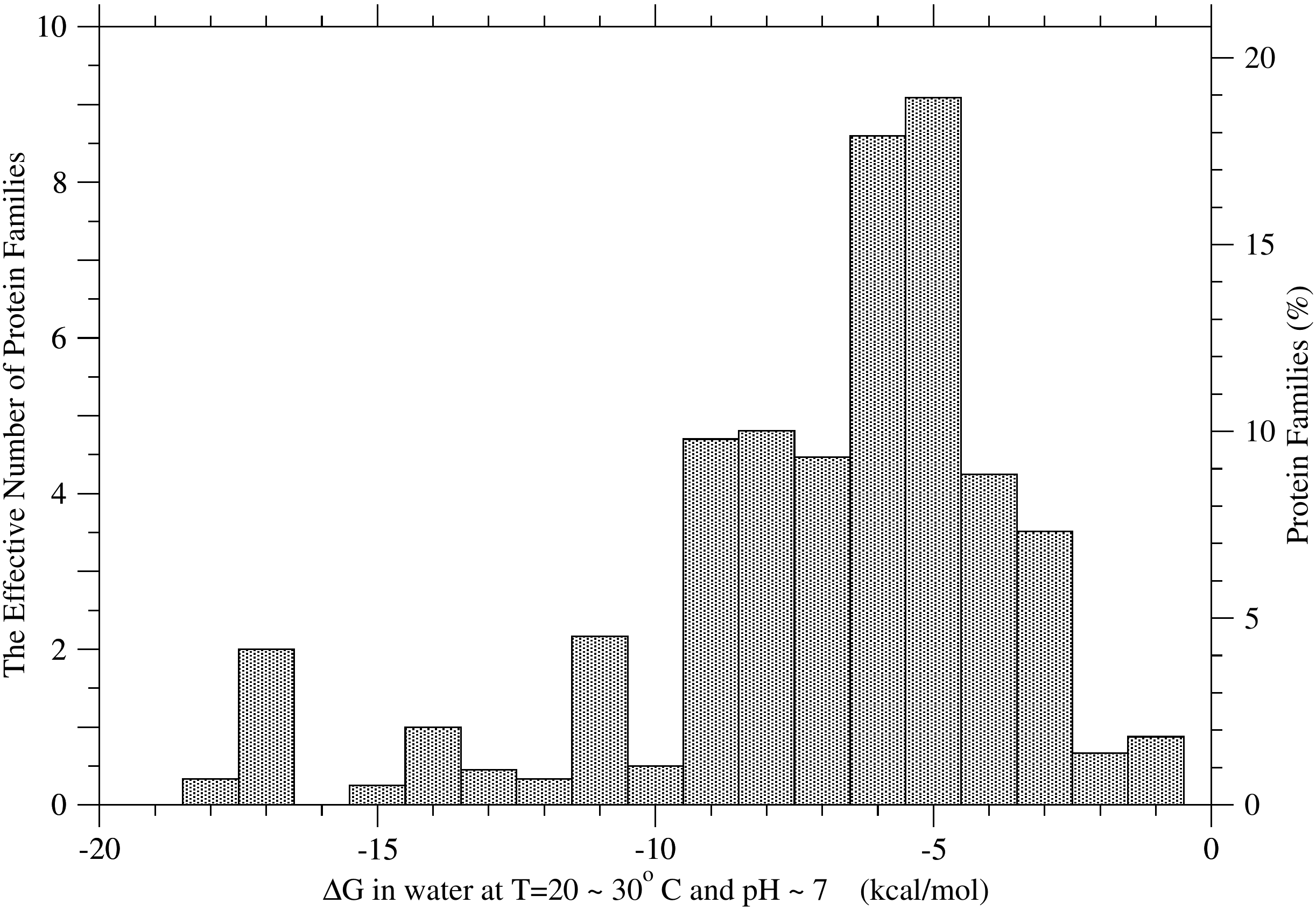}
}
} 
\FigureLegends{
\vspace*{1em}
\caption{
\label{fig: observed_distribution_of_protein_stabilities}
\Red{
\BF{Distribution of folding free energies of monomeric protein families.}
} 
Stability data of monomeric proteins for which the item of dG\_H2O or dG
was obtained in the experimental condition of $6.7 \leq \textrm{pH} \leq 7.3$
and $20^{\circ}C \leq T \leq 30^{\circ}C$ and their folding-unfolding transition is two state and reversible
are extracted from the ProTherm\CITE{KBGPKUS:06};
in the case of dG only thermal transition data are used.
Thermophilic proteins, and proteins observed with salts or additives are also removed.
An equal sampling weight is assigned to each species of homologous protein,
and the total sampling weight of each protein family is normalized to
one.  In the case in which multiple data exist for the same species of protein,
its sampling weight is divided to each of the data.
However, proteins whose stabilities are known
may be samples biased from the protein universe.
The value, $\Delta G_e = -5.24$ kcal/mol, of equilibrium stability at the representative parameter values, $\log 4N_e\kappa = 7.55$
and $\theta = 0.53$, agrees with the most probable value of $\Delta G$ in the distribution above.
Also, the range of $\Delta G$ shown above is consistent with
that range, $-2$ to $-12.5$ kcal/mol, expected from the present model.
The kcal/mol unit is used for $\Delta G$.
A similar distribution was also compiled \CITE{ZCS:07}.
}
} 
\end{figure*}

\clearpage

} 

} 

\section{Methods}

\subsection{Fitness costs due to misfolded proteins}

Misfolding can impose costs in three distinct ways\CITE{GDBWHD:11}; loss of function,
diversion of protein synthesis resources away from essential
proteins, and toxicity of the misfolded molecules.
Fitness cost due to functional loss 
was formulated\CITE{DW:08} by taking account of protein dispensability.
Assuming that fitness cost of each gene is additive in the Malthusian fitness scale,
the total Malthusian fitness of a genome was estimated as
\begin{eqnarray}
m_{\script{dispensability}} &\equiv& - \sum_i \gamma_i (1 - f^{\script{native}}_i )
			\label{eq: def_m_dispensability}
\end{eqnarray}
where $-\gamma_i$ is defined as $-\gamma_i \equiv \log ($deletion-strain growth rate / max growth rate$)$, and $f^{\script{native}}_i$ is
the fraction of the native conformation for gene $i$.

Protein folding primarily occurs in the two-state transition, which means
that protein conformations are a mixture of completely folded and unfolded conformations
\CITE{MJ:82,MJ:82B}.
Therefore, if the completely folded (native) state is more stable by
a free energy difference $\Delta G$ than the unfolded (denatured) state,
then the native fraction in the conformational ensemble will be equal to
\begin{eqnarray}
	f^{\script{native}} &=& \frac{e^{- \beta \Delta G}}{1 + e^{- \beta \Delta G}}
	\label{eq: fraction_of_native_state}
\end{eqnarray}
where $\beta = 1 / kT$; $k$ is the Boltzmann constant and $T$ is absolute temperature.

Thus, \Eq{\ref{eq: def_m_dispensability}} for the Malthusian fitness of a genome 
can be transformed as follows in terms of the folding free energy $\Delta G$ 
of the native conformation:
\begin{eqnarray}
m_{\script{dispensability}}
		&=&  - \sum_i \gamma_i \frac{e^{\beta\Delta G_i}}{e^{\beta\Delta G_i} + 1}
			\label{eq: m_dispensability}
\end{eqnarray}
\Red{
Because of 
$\exp (\beta \Delta G) \ll 1$ in the typical range of folding free energies
shown in \Fig{\ref{fig: observed_distribution_of_protein_stabilities}}, 
} 
the above definition of fitness is approximated by 
\begin{eqnarray}
m_{\script{dispensability}} &=& -  \sum_i \gamma_i [ e^{\beta\Delta G_i} - O(e^{2 \beta\Delta G_i}) ] 
\end{eqnarray}

Drummond and Wilke\CITE{DW:08} took notice of toxicity of misfolded proteins as well as 
diversion of protein synthesis resources, and
formulated the Malthusian fitness ($m_{\script{misfolds}}$) of 
a genome to be negatively proportional to the total amount of misfolded proteins, which
must be produced to obtain the necessary amount of folded proteins\CITE{SRS:12}.
\begin{eqnarray}
m_{\script{misfolds}} 
	&=& - c \sum_i A_i \frac{1 - f^{\script{native}}_i } {f^{\script{native}}_i } 
	\label{eq: def_m_toxicity}
	\\
	&=& - c \sum_i A_i e^{\beta \Delta G_i}
			\label{eq: m_misfold}
\end{eqnarray}
where $c$ is a positive constant and assumed to be $c=0.0001$,
and $A_i$ is the abundance of protein $i$.

\subsection{Fitness of a linear metabolic pathway}

Serohijos and Shakhnovich \CITE{SS:14} examined the evolution of a linear metabolic pathway whose Wrightian 
fitness was defined as
\begin{eqnarray}
w_{\script{linear pathway}} &\equiv& w_{\script{flux}} + w_{\script{misfolds}} 
		\\
 w_{\script{flux}} &\equiv&
		\frac{\sum_i \varepsilon_i A_i^{-1}}
			{\sum_i \varepsilon_i (A_i f^{\script{native}}_i)^{-1}}
				\label{ eq: w_flux}
		\\
w_{\script{misfolds}} &\equiv& - c \sum_i A_i (1 - f^{\script{native}}_i ) 
\end{eqnarray}
where $\varepsilon_i$ was defined as enzyme efficiency and assumed to be $\varepsilon_i = 1$.
The $w_{\script{flux}}$ is a fitness originating from the enzymatic flux of a linear metabolic pathway,
and $w_{\script{misfolds}}$ represents the effect of toxicity of misfolded proteins, and
is the same functional form as \Eq{\ref{eq: def_m_dispensability}}, 
although \Eq{\ref{eq: def_m_dispensability}} is a definition for Malthusian fitness.
Then, the Malthusian fitness corresponding to the Wrightian fitness above can be represented as
\begin{eqnarray}
\lefteqn{
m_{\script{linear\_pathway}} 
} \\
	&=& \log \, [ \, 
		\{ 1 + \sum_i \frac{\varepsilon_i A_i^{-1}} { \sum_i \varepsilon_i A_i^{-1} } e^{\beta \Delta G_i} \}^{-1}
		-c \sum_i A_i e^{\beta \Delta G_i} (1 + e^{\beta \Delta G_i})^{-1}
		\, ]		
		\nonumber
		\\
	&=& - \sum_i \{ \frac { \varepsilon_i A_i^{-1} }  { (\sum_i \varepsilon_i A_i^{-1}) } 
		+ c A_i \} e^{\beta \Delta G_i}
		\nonumber
		\\
	& &	+ O (   ( 
		\sum_i \{ \frac { \varepsilon_i A_i^{-1} }  { (\sum_i \varepsilon_i A_i^{-1}) } 
		+ c A_i \} e^{\beta \Delta G_i}  
		)^2	
		   )
\end{eqnarray}
Because $c A_i \leq 0.459$\CITE{SS:14}, $\Delta G < -3$ and 
\linebreak
$\sum_{i=1}^{10}c A_i \exp(\beta \Delta G_i) < 0.03$, 
the higher order terms can be neglected in this case.
However,
the fitness costs due to the flux and misfolded proteins
may be formulated to be additive in the Malthusian scale rather than 
in the Wrightian scale, employing \Eq{\ref{eq: m_misfold}} for
the fitness cost due to misfolded proteins;
\begin{eqnarray} 
m_{\script{linear\_pathway}} &\equiv&
	 - \sum_i \, \{ \, \frac{\varepsilon_i A_i^{-1}} { \sum_i \varepsilon_i A_i^{-1} } 
		+ c A_i \, \} \, e^{\beta\Delta G_i}
\end{eqnarray} 

\subsection{Other formulations of protein fitness}

Also, the following simple definition for fitness 
to maintain protein stability
was used\CITE{DSKS:14}:
\begin{eqnarray}
	w &\propto& f_{\script{native}}   
		\\
	m &=& - e^{\beta \Delta G} + O( e^{2 \beta \Delta G} ) + \textrm{constant}
\end{eqnarray}

In addition, \Eq{\ref{eq: m_dispensability}} for functional loss was employed 
with $\gamma_i \Rightarrow c A_i$ 
to represent toxicity of misfolded proteins in \CITE{SRS:12,SLS:13}.

\subsection{A generic form of protein fitness}

Thus, all expressions above for Malthusian fitness of protein can be well approximated
by the following expression, 
\Red{
because of $\exp (\beta \Delta G) \ll 1$ 
in the typical range of folding free energies
shown in \Fig{\ref{fig: observed_distribution_of_protein_stabilities}}. 
} 
\begin{eqnarray}
	m &\equiv& - \sum_i \kappa_i e^{\beta \Delta G_i}
	\hspace*{1em}
	\textrm{ with } \kappa_i \geq 0
	\label{eq: def_Malthusian_fitness}
\end{eqnarray}
where $\kappa_i$ is a parameter.
If the fitness costs of
functional loss and toxicity due to misfolded proteins are taken into account, 
$\kappa_i$ will be defined as
\begin{eqnarray}
	\kappa_i &=& c A_i + \gamma_i \geq 0
\end{eqnarray}
assuming their additivity in the Malthusian fitness scale.

The selective advantage of a mutant, in which 
each protein is destabilized by $\Delta \Delta G_i$, to the wild type can be
represented by
\begin{eqnarray}
 s	&\equiv& m^{\script{mutant}} - m^{\script{wildtype}} 
	= \sum_i s_i
	\\
 s_i	
	&=& \kappa_i e^{\beta \Delta G_i} ( 1 - e^{\beta \Delta \Delta G_i} )
	\hspace*{1em}
	\textrm{ with } \kappa_i \geq 0
		\label{eq: selective_advantage}
\end{eqnarray}

\section{Results}

\subsection{Protein stability and fitness}

Here, we consider the evolution of a single protein-coding gene in which 
the selective advantage of mutant proteins in Malthusian parameters is 
assumed to be 
\begin{eqnarray}
	s
	&=& \kappa e^{\beta \Delta G} (1 - e^{\beta \Delta \Delta G} )
	\hspace*{1em}
	\textrm{ with } \kappa \geq 0
	\label{eq: def_s}
	\label{eq: 4sNe}
	\label{eq: def_of_s}
\end{eqnarray}
and therefore
$s$ is upper-bounded by
\begin{eqnarray}
        s &\leq& \kappa e^{\beta \Delta G}
                \label{eq: s_max}
\end{eqnarray}
where $\Delta G$ is the stability of a wild-type protein,
$\Delta\Delta G$ is a stability change of a mutant protein,
$\beta = 1 / kT$; unless specified, $\beta = 1/ 0.593$ kcal$^{-1}$mol
corresponding to $T = 298 ^{\circ}$K.
$\kappa$ is a parameter 
whose meaning may depend on the situation; 
refer to Method for details.
If the fitness costs of
functional loss and toxicity due to misfolded proteins are both taken into account 
and assumed to be additive in the Malthusian fitness scale,
$\kappa$ will be defined as
\begin{eqnarray}
        \kappa &=& c A + \gamma
        \label{eq: def_kappa}
\end{eqnarray}
where 
$c$ is fitness cost per misfolded protein\CITE{DW:08,GDBWHD:11},
$A$ is the cellular abundance of the protein\CITE{DW:08,GDBWHD:11}, 
and $\gamma$ is indispensability\CITE{DW:08} and
defined to be $\gamma = - \log ($deletion-strain growth rate / max growth rate$)$. 
\EQUATION{\ref{eq: def_s}} indicates that
the selective advantage $s$ is upper-bounded by $\kappa \exp(\beta \Delta G)$.
The parameter $\kappa$ is assumed in the present analysis to take 
values in the range of 
$0 \leq \log 4 N_e \kappa \leq 20$ with effective population size $N_e$,
taking account of the values of the parameters, $c \sim 10^{-4}$\CITE{DW:08}, $10<A<10^{6}$\CITE{GHBHBDOW:03},
$\gamma = 10$ for essential genes\CITE{DW:08}, and 
$N_e \sim 10^4$ to $10^5$ for vertebrates, $\sim 10^5$ to $10^6$ for invertebrates,
$\sim 10^7$ to $10^8$ for unicellular eukaryotes, and $> 10^8$ for prokaryotes \CITE{LC:03}.
The above ranges of the parameters indicate that
the effect of protein indispensability ($\gamma$) may be hidden 
by the variation of protein abundance ($c A$) as well as effective population size ($N_e$),
and may be detected only in low-abundant proteins.

Based on measurements of stability changes due to single amino acid substitutions in
proteins, which are collected in the ProTherm database\CITE{KBGPKUS:06},
Serohijos et al.\CITE{SRS:12} 
reported that the distribution of $\Delta \Delta G$ is approximately a Gaussian distribution 
with mean = 1 kcal/mol and standard deviation = 1.7 kcal/mol.
In addition, it was shown\CITE{SRS:12} that the mean of $\Delta\Delta G$ is negatively proportional
to $\Delta G$, and that this dependence of the mean of $\Delta\Delta G$ on $\Delta G$
is not large but still important to cause the observed negative correlation between
protein abundance and evolutionary rate.
On the other hand,
Tokuriki et al. \CITE{TSSST:07} computationally predicted $\Delta\Delta G$ for
all possible single amino acid substitutions
in 21 different globular, single domain proteins,
and showed that 
the predicted distributions of $\Delta \Delta G$ 
were strikingly similar despite a range of protein sizes and folds and
largely follow a bi-Gaussian distribution:
one of the two Gaussian distributions 
results from substitutions on protein surfaces and
is a narrow distribution with a mildly destabilizing mean $\Delta \Delta G$,
whereas the other due to substitutions in protein cores is a wider 
distribution with a stronger destabilizing mean\CITE{TSSST:07}.

\FigureInText{

\TextFig{

\begin{figure*}[ht]
\FigureInLegends{
\centerline{
\includegraphics*[width=90mm,angle=0]{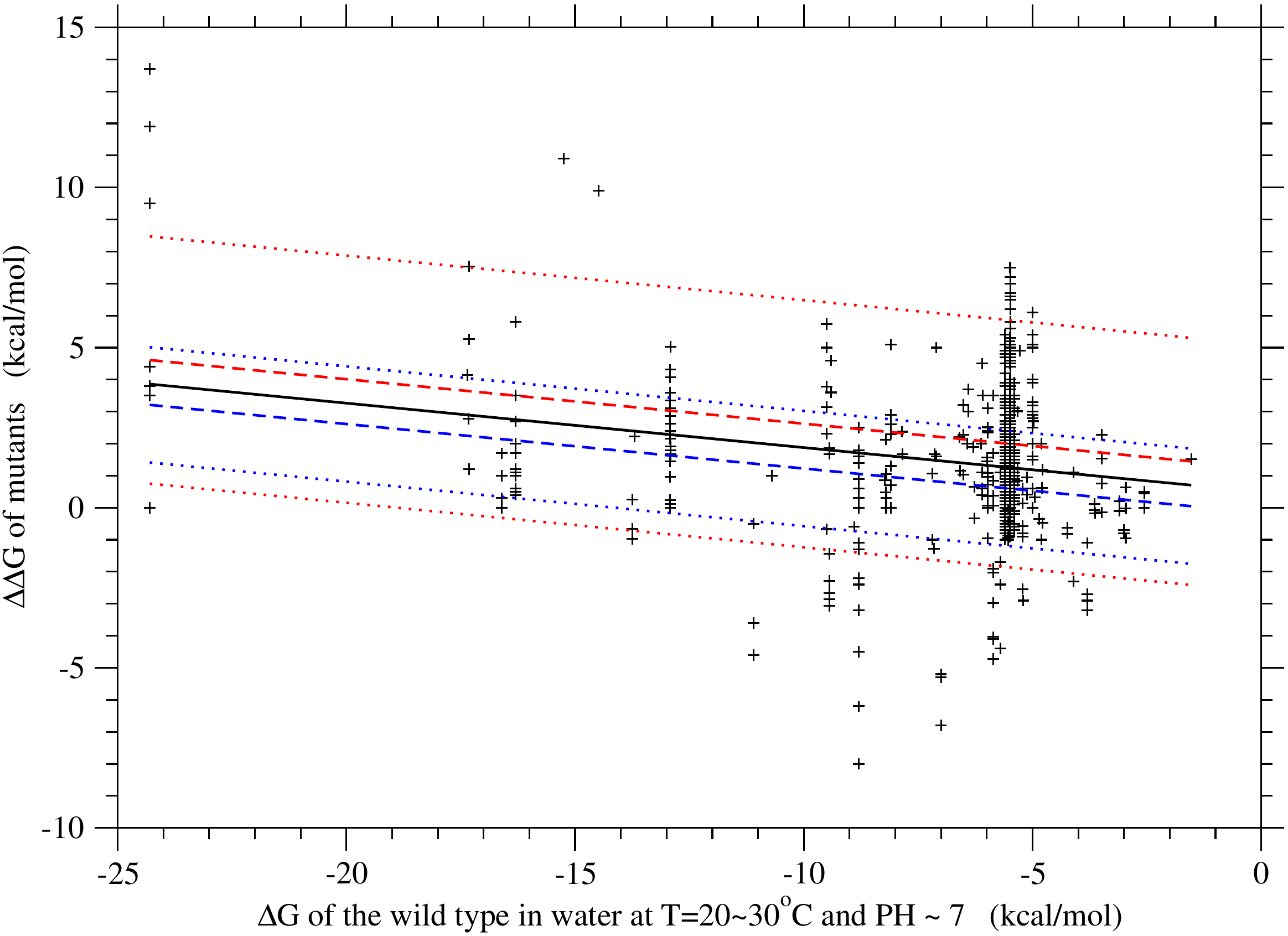}
}
} 
\FigureLegends{
\vspace*{1em}
\caption{
\label{fig: ddG_vs_dG}
\Red{
\BF{Dependence of stability changes, $\Delta\Delta G$, due to single amino acid substitutions
on the protein stability, $\Delta G$, of the wild type.}
} 
A solid line shows the regression line,
$\Delta\Delta G = -0.14 \Delta G + 0.49$;
the correlation coefficient and p-value are equal to $-0.20$ and $< 10^{-7}$, respectively.
Broken lines show two means of bi-Gaussian distributions, $\mu_s$ in blue and $\mu_c$ in red.
Blue dotted lines show $\mu_s \pm 2 \sigma_s$ and red dotted lines $\mu_c \pm 2 \sigma_c$.
See \Eqs{\Ref{eq: def_bi-Gaussian}, \Ref{eq: def_ms} and \Ref{eq: def_mc}} for the bi-Gaussian distribution.
Stability data of single amino acid mutants for which the items dG\_H2O and ddG\_H2O or dG and ddG
were obtained in the experimental condition of 
$6.7 \leq \textrm{pH} \leq 7.3$ and $20^{\circ}C \leq T \leq 30^{\circ}C$
and their folding-unfolding transitions are two state and reversible
are extracted from the ProTherm\CITE{KBGPKUS:06}.
In the case of dG only thermal transition data are used.
In the case in which multiple data exist for the same protein, only one of them
is used.
The kcal/mol unit is used for $\Delta\Delta G$ and $\Delta G$.
A similar distribution was also compiled \CITE{SRS:12}.
}
} 
\end{figure*}

\clearpage

} 

} 

Here, according to \CITE{TSSST:07}, 
the distribution of $\Delta\Delta G$ due to single amino acid substitutions is 
approximated as a bi-Gaussian function with
the dependence of mean $\Delta\Delta G$ on $\Delta G$, 
in order to examine the effects of structural constraint on evolutionary rate.
The probability density function (PDF) of $\Delta \Delta G$, $p(\Delta \Delta G)$, 
for nonsynonymous 
substitutions is assumed to be
\begin{eqnarray}
p(\Delta \Delta G) &=& \theta \mathcal{N}(\mu_s, \sigma_s)
		+ (1 - \theta) \mathcal{N}(\mu_c, \sigma_c)
	\label{eq: def_bi-Gaussian}
\end{eqnarray}
where $0 \leq \theta \leq 1$, and $\mathcal{N}(\mu, \sigma)$
is a normal distribution with mean $\mu$ and standard deviation $\sigma$.
Since the majority of substitutions appear to be single nucleotide substitutions,
the values of the standard deviations 
($\sigma_s$ and $\sigma_c$)
estimated in \CITE{TSSST:07} for single nucleotide substitutions
are employed here; in kcal/mol units,
\begin{eqnarray}
\mu_s &=& -0.14 \, \Delta G - 0.17	
\hspace*{1em}, \hspace*{1em}
\sigma_s = 0.90 
	\label{eq: def_ms}
	\\
\mu_c &=& -0.14 \, \Delta G + 1.23	
\hspace*{1em}, \hspace*{1em}	
\sigma_c = 1.93 
	\label{eq: def_mc}
\end{eqnarray}
\Red{
To analyze the dependences of the means, $\mu_s$ and $\mu_c$, on $\Delta G$,
we plotted the observed values of $\Delta\Delta G$ of single amino acid mutants
against $\Delta G$ of the wild type, which are collected in the ProTherm
database\CITE{KBGPKUS:06}; the same analysis was done by \CITE{SRS:12}.
\Fig{\ref{fig: ddG_vs_dG}} shows a significant dependence 
of $\Delta\Delta G$ on $\Delta G$;
the regression line is $\mu = -0.14 \Delta G + 0.49$.
} 
The linear slopes 
of $\mu_s$ and $\mu_c$ are taken to be equal to 
the slope
($-0.14$) 
of the regression line.  
The intercepts have been estimated
to satisfy the following two conditions.
\begin{enumerate}
\item \EQUATIONS{\Ref{eq: def_ms} and \Ref{eq: def_mc}} satisfy
$\mu_s(\Delta G_0) = 0.56$ and $\mu_c(\Delta G_0) = 1.96$,
which were estimated for single nucleotide substitutions in \CITE{TSSST:07},
at a certain value ($\Delta G_0$) of $\Delta G$.
\item The total mean of the two Gaussian functions agrees with
the regression line, $\mu = -0.14 \Delta G + 0.49$.
The value of $\theta$ is taken to be $0.53$, which is
equal to
the average of $\theta$ over proteins used in \CITE{TSSST:07}. 
\end{enumerate}
A representative value, 
$7.550$,
of $\log 4N_e \kappa$ is determined
in such way that the equilibrium value of $\Delta G$ 
is equal to $\Delta G_{0} = -5.24$ introduced above;
$\Delta G_e$ is explicitly defined later.
\Red{
It is interesting that this value $\Delta G_e = -5.24$ kcal/mol
agrees with the most probable value of $\Delta G$ in the observed distribution of protein stabilities
shown in \Fig{\ref{fig: observed_distribution_of_protein_stabilities}}.
} 
The fraction $\theta$ of less-constrained residues such as most residues on protein surface
is correlated with protein length for globular, monomeric proteins; 
$
\theta = 1.27 - 0.33 \cdot \log_{10}(\textrm{protein length}) 
			\textrm{ for }50 \leq \textrm{length} \leq 330 
$
\CITE{TSSST:07}.  However, residues taking part in protein--protein interactions 
may be regarded as core residues rather than surface residues.

The dependence of the PDF, $p(\Delta\Delta G)$, of $\Delta\Delta G$ on $\theta$ is shown 
in \Fig{\ref{fig: pdf_of_ddG_at_dGe}}. 
Also, the PDF of selective advantage, 
$p(4N_e s) = - p(\Delta\Delta G) d \Delta\Delta G / d 4N_e s$, 
is shown in 
\Fig{\ref{sfig: pdf_of_4sNe}}
to have a peak at a small, positive value of selective advantage, 
which moves toward more positive values as 
$\theta$ and/or $4N_e \kappa$ increase.

\FigureInText{

\TextFig{

\begin{figure*}[ht]
\FigureInLegends{
\centerline{
\includegraphics*[width=90mm,angle=0]{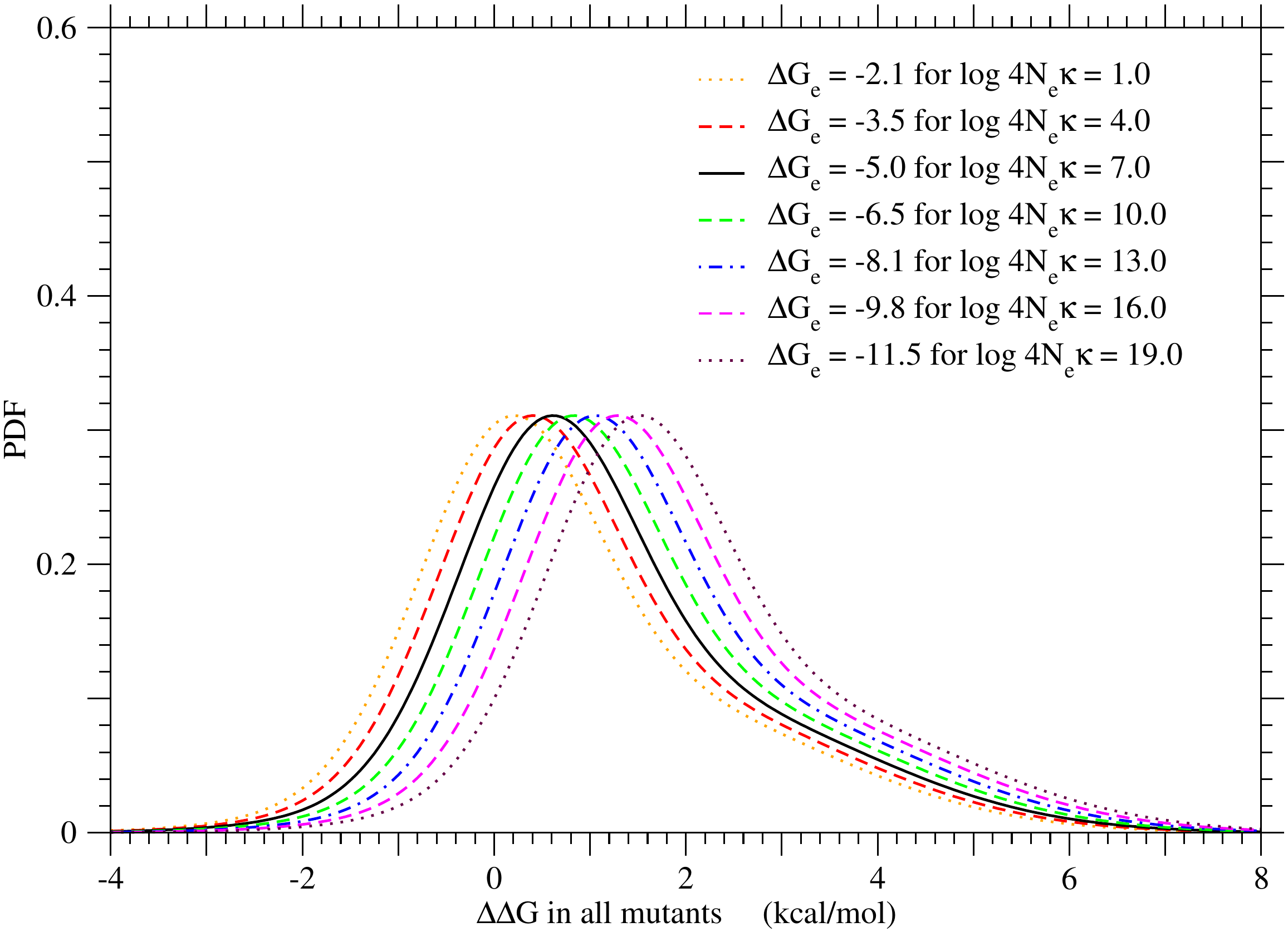}
\includegraphics*[width=90mm,angle=0]{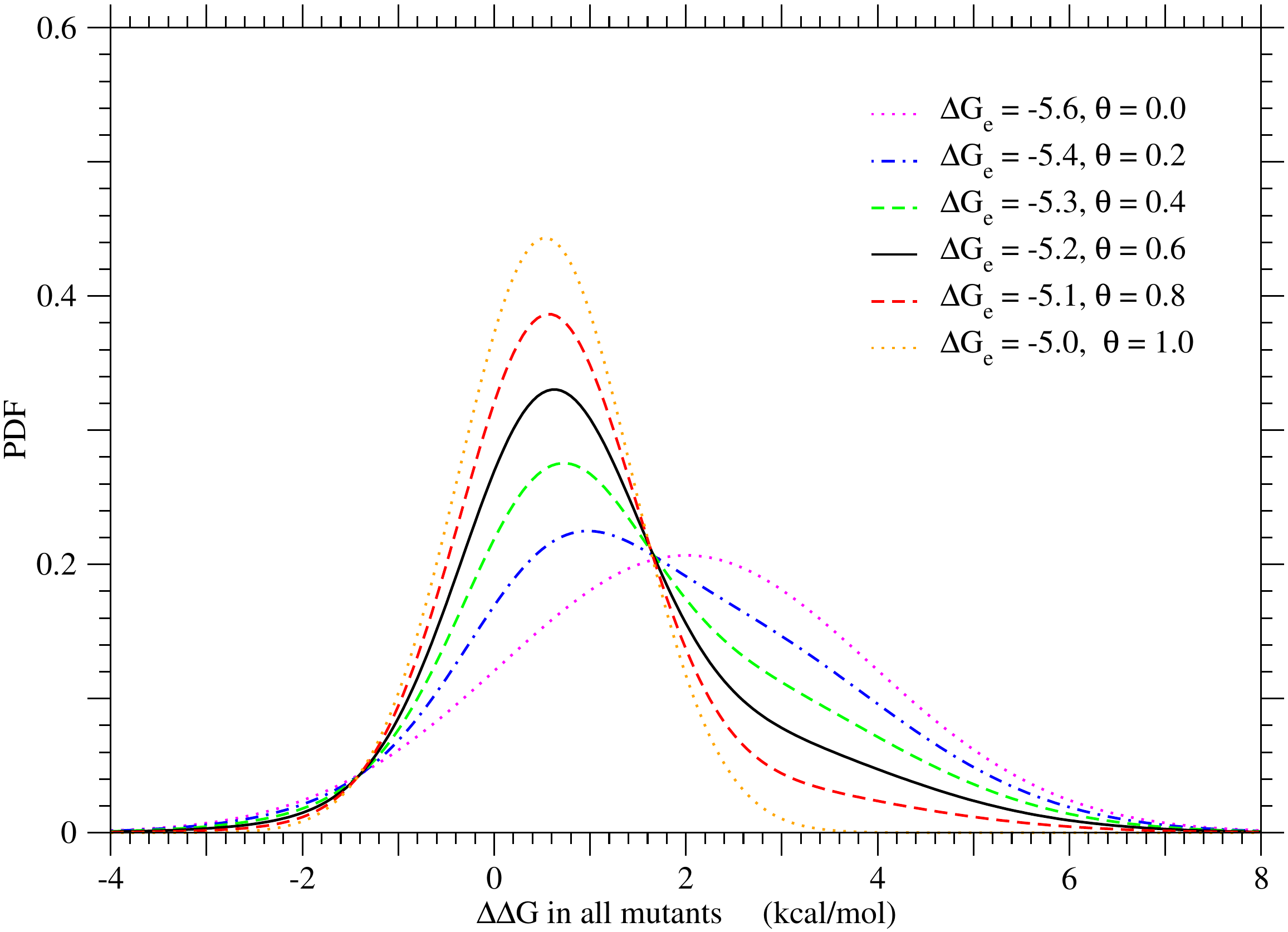}
}
\centerline{
\includegraphics*[width=90mm,angle=0]{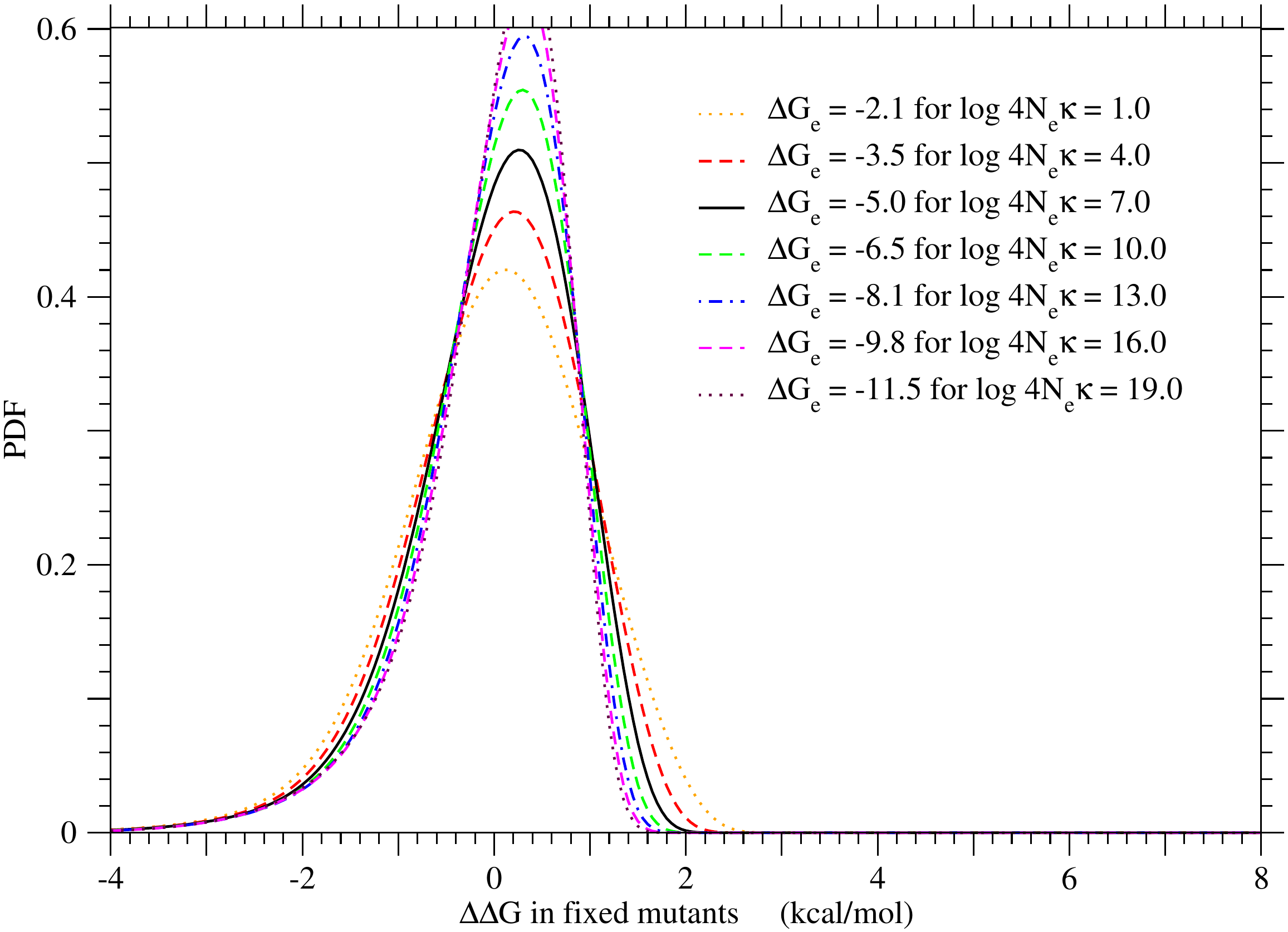}
\includegraphics*[width=90mm,angle=0]{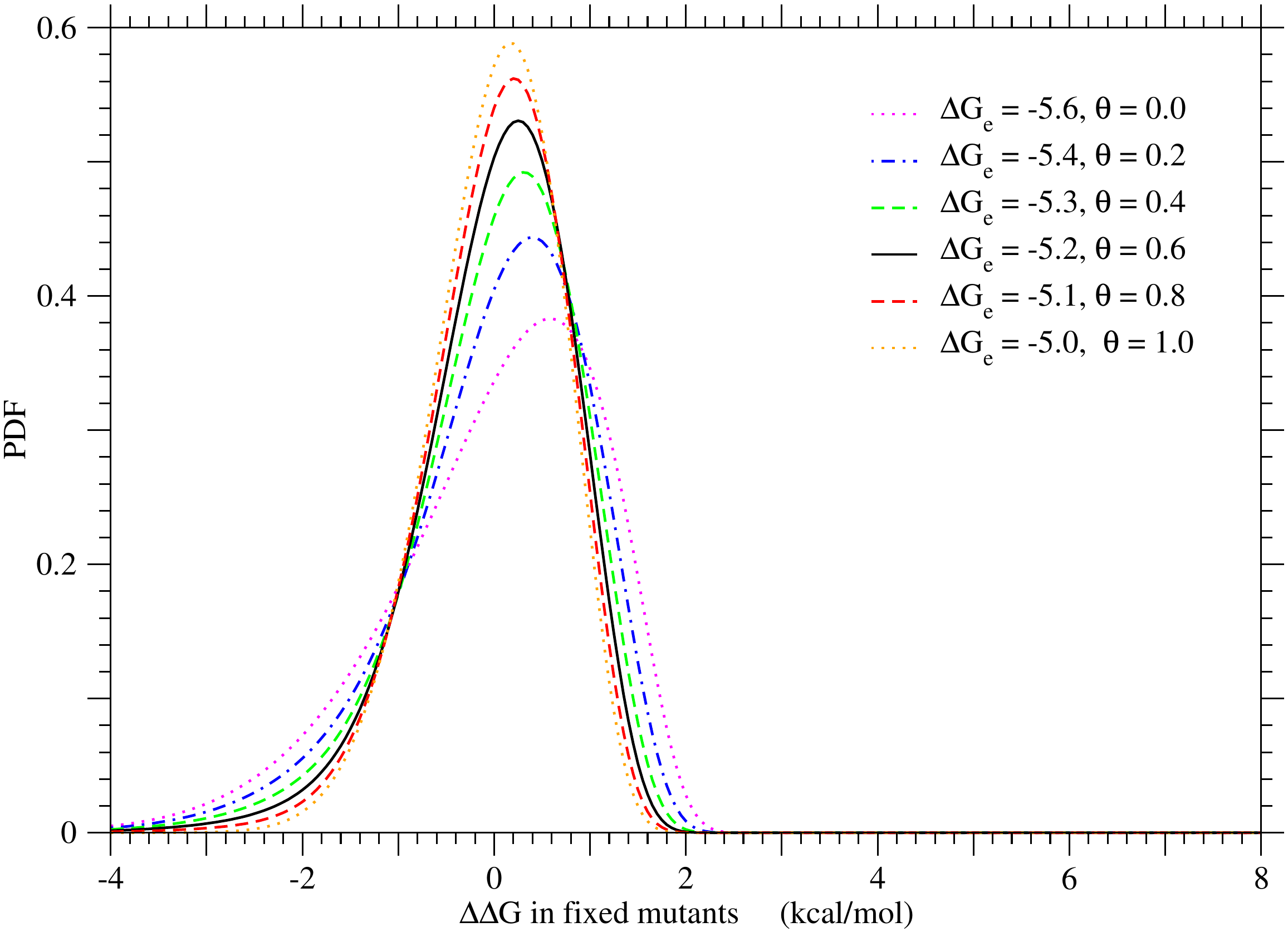}
}
} 
\FigureLegends{
\vspace*{1em}
\caption{
\label{fig: pdf_of_ddG_at_dGe}
\label{fig: pdf_of_ddG_fixed_at_dGe}
\label{fig: pdf_of_ddG}
\BF{PDFs of stability changes, $\Delta\Delta G$, due to single amino acid substitutions 
in all mutants and in fixed mutants
at equilibrium of protein stability, $\Delta G = \Delta G_e$.}
The PDF of $\Delta\Delta G$ due to single amino acid substitutions
in all arising mutants is assumed to be bi-Gaussian; see \Eq{\ref{eq: def_bi-Gaussian}}.
Unless specified, 
$\log 4N_e\kappa = 7.55$ and $\theta = 0.53$
are employed.
The kcal/mol unit is used for $\Delta\Delta G$ and $\Delta G_e$.
}
} 
\end{figure*}

\clearpage

} 

} 

\vspace*{1em}

\subsection{Equilibrium state of protein stability in protein evolution}

The fixation probability $u$ for a mutant gene with selective advantage $s$ 
and gene frequency $q$
in a duploid system of effective population size $N_e$ was given 
as a function of $4N_e s$ and $q$ by\CITE{CK:70} 
\begin{eqnarray}
	u(4N_e s)
	&=& \frac{1 - e^{-4N_e s q }} {1 - e^{-4N_e s} } 
		\label{eq: Fixation_prob}
\end{eqnarray}
where $q = 1/ (2N)$ for a single mutant gene in a population of size $N$.
Population size is taken to be $N = 10^6$.
The ratio of the substitution rate per nonsynonymous site ($K_a$) for 
nonsynonymous substitutions with selective advantage $s$
to the substitution rate per synonymous site ($K_s$) for nonsynonymous substitutions with $s = 0$ 
is 
\begin{eqnarray}
	\frac{K_a}{K_s} &=& \frac{u(4N_e s)}{u(0)} =  \frac{u(4N_e s)}{q}	\hspace*{1em} \textrm{ with } q = \frac{1}{2N}
	\label{eq: def_Ka_over_Ks}
		\\
		&\simeq& \frac{4N_e s}{1 - e^{-4N_e s}}	 \hspace*{1em} \textrm{ for } \frac{| 4N_e s q |}{2} \ll 1
		\label{eq: approx_Ka_over_Ks}
\end{eqnarray}
assuming that synonymous substitutions are completely neutral
and mutation rates at both types of sites are the same.
\Eqs{\Ref{eq: 4sNe} and \Ref{eq: Fixation_prob}}  indicate that 
$4 N_e \kappa$ can be regarded as
a single parameter for $K_a/K_s$.  
Furthermore, if the dependence of the mean $\Delta \Delta G$ on $\Delta G$
could be neglected, $4 N_e \kappa \exp(\beta \Delta G)$ could be regarded as a single parameter.
In the range of $| 4N_e s q | / 2 \ll 1$, both $K_a/K_s$ and 
the PDF of $K_a/K_s$ do not depend on $q = 1/(2N)$; see \Eqs{\Ref{eq: approx_Ka_over_Ks}
and \Ref{seq: approx_pdf_of_Ka_over_Ks}}.

The PDF of $\Delta\Delta G$ of fixed mutant genes, $p(\Delta\Delta G_{\script{fixed}})$, is
\begin{eqnarray}
 p(\Delta\Delta G_{\script{fixed}}) &\equiv& p(\Delta\Delta G) \frac{u(4N_e s)}{\langle u \rangle}
	\label{eq: PDF_of_ddG_fixed}
	\\
 \langle u \rangle &\equiv& \int_{-\infty}^{\infty} u(4N_e s) p(\Delta\Delta G) d\Delta\Delta G 
\end{eqnarray}
where $\langle u \rangle$ is the average fixation rate.
\Fig{\ref{fig: pdf_of_ddG_fixed_at_dGe}} shows the PDF of $\Delta\Delta G$ of fixed mutant genes.
The PDF of $4N_e s$ in fixed mutants is also shown in \Fig{\ref{sfig: pdf_of_4sNe_fixed}};
$p(4N_e s_{\script{fixed}}) = - p(\Delta\Delta G_{\script{fixed}}) d \Delta\Delta G / d 4N_e s$.
Then, the average of $\Delta\Delta G$ in fixed mutant genes can be calculated; 
$
\langle \Delta\Delta G \rangle_{\script{fixed}} \equiv
	\int_{-\infty}^{\infty} \Delta\Delta G \, p(\Delta\Delta G_{\script{fixed}}) \, d\Delta\Delta G 
$.

\FigureInText{

\TextFig{

\begin{figure*}[ht]
\FigureInLegends{
\centerline{
\includegraphics*[width=90mm,angle=0]{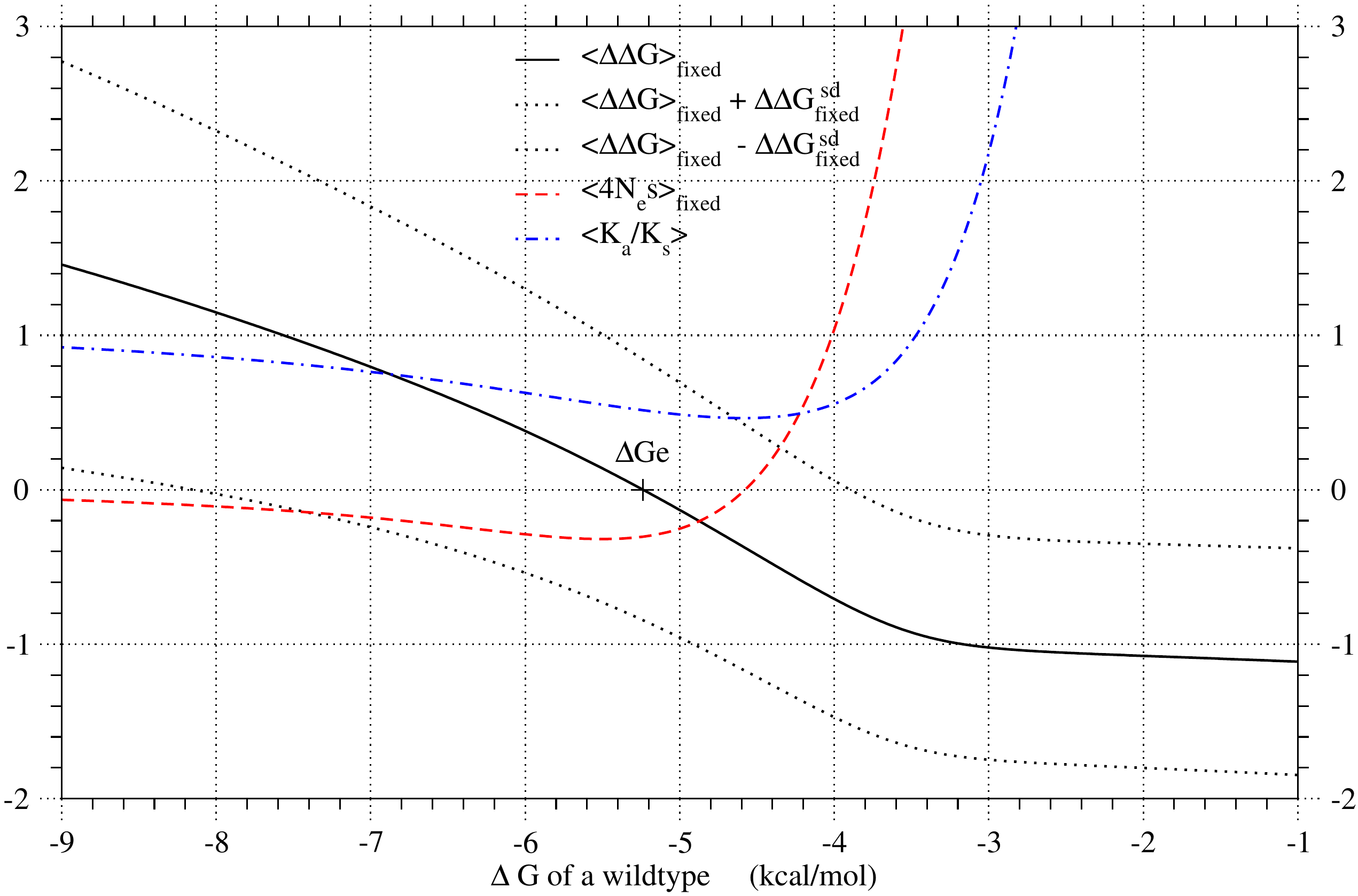}
}
} 
\FigureLegends{
\vspace*{1em}
\caption{
\label{fig: ave_ddG_vs_dG}
\label{fig: ave_ddG_vs_dG_fixed}
\BF{The average, $\langle \Delta\Delta G \rangle_{\script{fixed}}$, of stability changes over fixed mutants versus protein stability, $\Delta G$, of the wild type.}
$\Delta G_e$, where $\langle \Delta\Delta G \rangle = 0$, is 
the stable equilibrium value of folding free energy, $\Delta G$,
in protein evolution. 
The averages of $\Delta\Delta G$, $4N_e s$, and $K_a/Ks$ over fixed mutants are
plotted against protein stability, $\Delta G$, of the wild type by solid, broken, and dash-dot lines,
respectively.
Thick dotted lines show the values of $\langle \Delta\Delta G \rangle_{\script{fixed}} \pm \Delta\Delta G^{\script{sd}}_{\script{fixed}}$,
where $\Delta\Delta G^{\script{sd}}_{\script{fixed}}$ is the standard deviation of $\Delta\Delta G$ over fixed mutants.
$\log 4N_e\kappa = 7.55$ and $\theta = 0.53$
are employed.
The kcal/mol unit is used for $\Delta\Delta G$ and $\Delta G$.
}
} 
\end{figure*}

\clearpage

} 

} 

\Fig{\ref{fig: ave_ddG_vs_dG}} shows the average of the $\Delta\Delta G$ over fixed mutant genes,
$\langle \Delta\Delta G \rangle_{\script{fixed}}$,
to monotonically decrease with $\Delta G$, indicating that 
the temporal process of $\Delta G$ is stable 
at $\langle \Delta\Delta G \rangle_{\script{fixed}}(\Delta G_e) = 0$
due to the balance between random drift on 
destabilizing mutations
and positive selection on stabilizing mutations; 
$\Delta G_e$ is the folding free energy at the equilibrium state.
If a wild-type protein 
becomes less stable than the equilibrium, $\Delta G > \Delta G_e$,
more stabilizing mutants will fix due to
primarily positive selection and secondarily random drift,
because stabilizing mutants will increase 
due to negative shifts of $\Delta\Delta G$
and also the effect of stability change on selective advantage
will be more amplified; see 
\Eqs{\Ref{eq: def_ms} and \Ref{eq: def_mc}} 
for the dependence of $\Delta\Delta G$ on $\Delta G$, 
and \Eq{\ref{eq: def_s}} for the fitness of stability change.
As shown in \Fig{\ref{sfig: pdf_of_Ka_over_Ks_around_dGe}},
the probability of $K_a/K_s > 1.0$, that is, positive selection, 
significantly increases as $\Delta G$ becomes 
more positive than the equilibrium stability $\Delta G_e$.
On the other hand, if
a wild-type protein
becomes more stable than the equilibrium, $\Delta G < \Delta G_e$,
more destabilizing mutants will fix due to random drift, 
because destabilizing mutants will increase 
due to positive shifts of $\Delta\Delta G$
and also 
more destabilizing mutants become nearly neutral
due to the less-amplified effect of stability change on selective advantage.
As shown later,
the PDF of $K_a/K_s$ in the vicinity of equilibrium
confirms this mechanism for maintaining protein stability at equilibrium.

It was claimed \CITE{SRS:12,SLS:13} that the equilibrium point would correspond to the minimum of 
the average fixation probability. 
However, in \Fig{\ref{fig: ave_ddG_vs_dG}} 
for $\log 4N_e\kappa = 7.550$ and $\theta = 0.53$,
the average $\langle s \rangle_{\script{fixed}}$ of selective advantage in fixed mutants
has a minimum at $\Delta G = -5.50$ kcal/mol and changes its sign at $\Delta G = -4.58$ kcal/mol,
where the average $\langle K_a / K_s \rangle = \langle u \rangle / q$ has a minimum and 
which is more positive than 
the equilibrium stability $\Delta G_e = -5.24$ kcal/mol.
In other words,
\Figs{\ref{fig: ave_ddG_vs_dG} and \ref{sfig: dependence_of_ave_Ka_over_Ks_on_dG}} show that
the values of $\Delta G$ at $\langle \Delta\Delta G \rangle_{\script{fixed}} = 0$
and at the minimum of $\langle K_a / K_s \rangle$ may be close but differ from each other, and
indicate that
the value of $\Delta G$ corresponding to the minimum of $\langle K_a / K_s \rangle$
is not a good approximation for the equilibrium stability, 
because 
$\langle K_a/K_s \rangle$ gently changes
in the vicinity of the equilibrium stability as shown in \Fig{\ref{sfig: dependence_of_ave_Ka_over_Ks_on_dG}}.

\FigureInText{

\TextFig{

\begin{figure*}[ht]
\FigureInLegends{
\centerline{
\includegraphics*[width=90mm,angle=0]{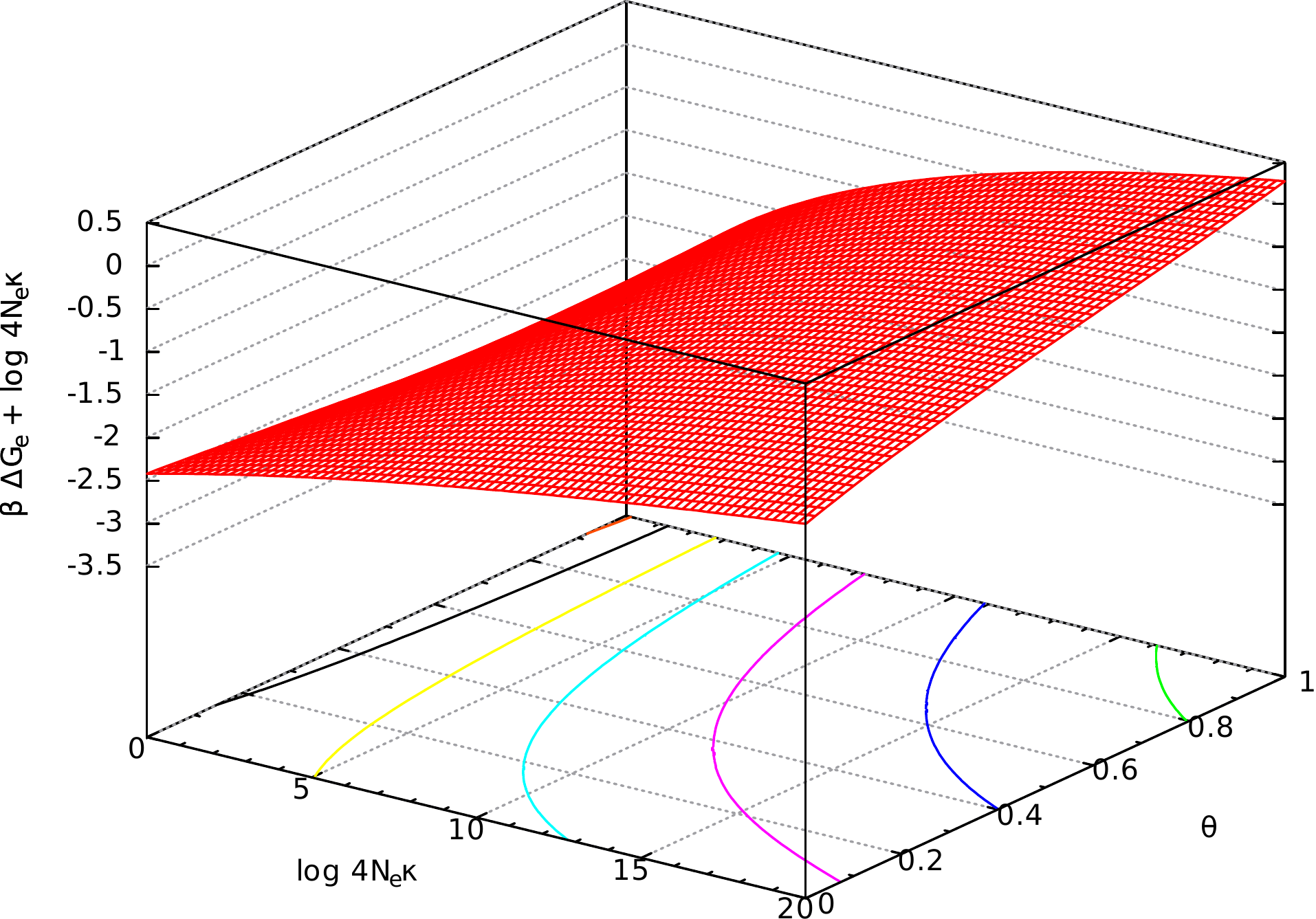}
\includegraphics*[width=90mm,angle=0]{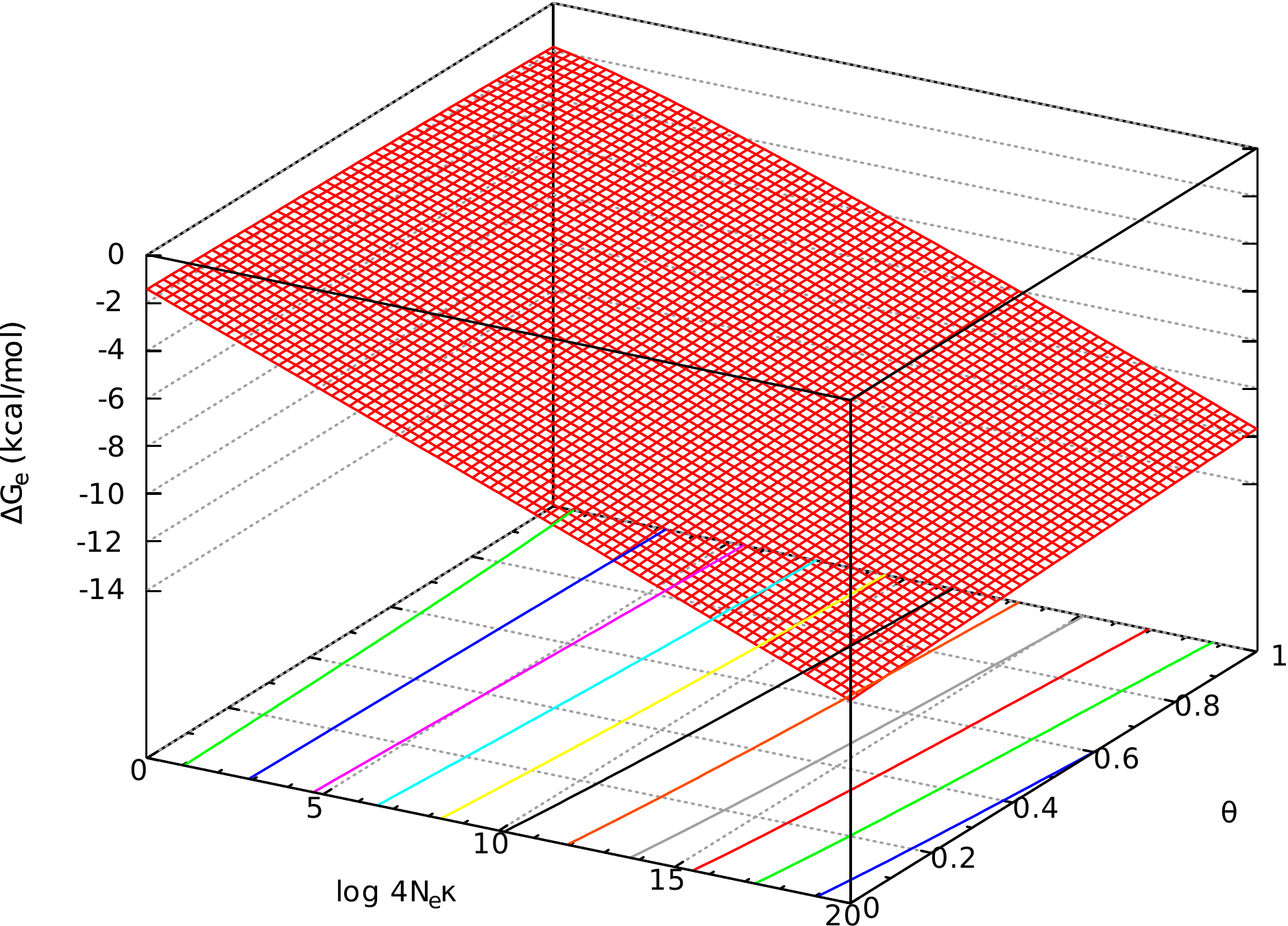}
}
} 
\FigureLegends{
\vspace*{1em}
\caption{
\label{fig: dependence_of_dGe_on_4Nekappa_and_theta}
\BF{Dependence of equilibrium stability, $\Delta G_e$, on parameters, $4N_e\kappa$ and $\theta$.}
$\Delta G_e$ is the equilibrium value of folding free energy, $\Delta G$,
in protein evolution.
The value of $\beta\Delta G_e + \log 4N_e\kappa$ is the upper bound of $\log 4N_e s$,
and would be constant if the mean of $\Delta\Delta G$ in all arising mutants did not depend on $\Delta G$;
see \Eq{\ref{eq: def_s}}.
The kcal/mol unit is used for $\Delta G_e$.
}
} 
\end{figure*}

\clearpage

} 

} 

\vspace*{1em}
\noindent
\subsection{Equilibrium stability, $\Delta G_e$}

The equilibrium value, $\Delta G_e$, of $\Delta G$ that satisfies 
$\langle \Delta\Delta G_{\script{fixed}} \rangle$ = 0 in fixed mutants
depends directly on $\theta$ and indirectly on $4N_e\kappa$ through fixation probability; see
\Eqs{\Ref{eq: def_s} and \Ref{eq: Fixation_prob}}.
As shown in \Fig{\ref{fig: dependence_of_dGe_on_4Nekappa_and_theta}},
$\Delta G_e$ depends weakly on $\theta$.
On the other hand, 
$\Delta G_e$ depends more strongly on and is almost negatively proportional to $\log 4N_e\kappa$,
as also shown in real proteins\CITE{SLS:13}.
If the dependence of the means, $\mu_s$ and $\mu_c$ in \Eqs{\Ref{eq: def_ms} and \Ref{eq: def_mc}}, 
of $\Delta \Delta G$ in all mutants on $\Delta G$
could be neglected, $4 N_e \kappa \exp(\beta \Delta G)$ could be regarded as a single parameter,
and so $4 N_e \kappa \exp(\beta \Delta G_e)$ would be constant, irrespective of $4 N_e \kappa$. 
Thus, the dependence of $\log 4 N_e \kappa \exp(\beta \Delta G_e)$ on $\log 4 N_e \kappa$ shown in
\Fig{\ref{fig: dependence_of_dGe_on_4Nekappa_and_theta}} is caused solely by 
the linear dependence of the means $\mu_s$ and $\mu_c$ of $\Delta\Delta G$ on $\Delta G$
\CITE{SRS:12}.
It is interesting to know that as $\log 4N_e \kappa $ varies from $0$ to $20$, $\Delta G_e$ 
changes from $-1.5$ to $-12.5$ kcal/mol, 
\Red{
the range of which is consistent with
experimental values of protein folding free energies
shown in \Fig{\ref{fig: observed_distribution_of_protein_stabilities}}.
} 

\vspace*{1em}
\noindent
\subsection{$K_a/K_s$ at equilibrium, $\Delta G = \Delta G_e$}

\EQUATIONS{\Ref{eq: def_ms} and \Ref{eq: def_mc}} indicate that
the distribution of $\Delta\Delta G$ shifts toward the positive direction
as $\Delta G$ becomes more negative. 
Hence, increasing $4N_e\kappa$ that makes $\Delta G_e$ more negative results in
positive shifts of the distribution of $\Delta\Delta G$, which
increase destabilizing mutations.
In addition, as indicated by \Eq{\ref{eq: def_s}}, the upper bound of $4N_e s$, $4 N_e \kappa \exp(\beta \Delta G_e)$,
scales the effect of $\Delta\Delta G$ on protein fitness.
The larger $4 N_e \kappa \exp(\beta \Delta G_e)$ is,  the larger the effect of $\Delta\Delta G$ on selective advantage becomes.
Thus, the increase of $4N_e \kappa \exp(\beta \Delta G_e)$ caused by the increase of $\kappa$ and/or $N_e$ 
increases both destabilizing mutations and their fitness costs,
and results in
slow evolutionary rates for proteins with large $\kappa$ and/or $N_e$.
In other words, highly expressed and indispensable genes, and genes with a large effective population size 
must evolve slowly.
On the other hand,
the decrease of $\theta$, that is, the increase of highly constrained residues
directly shifts the average of $\Delta\Delta G$ in all arising mutants 
toward the positive direction, and causes slow evolutionary rates.

\FigureInText{

\TextFig{

\begin{figure*}[ht]
\FigureInLegends{
\centerline{
\includegraphics*[width=90mm,angle=0]{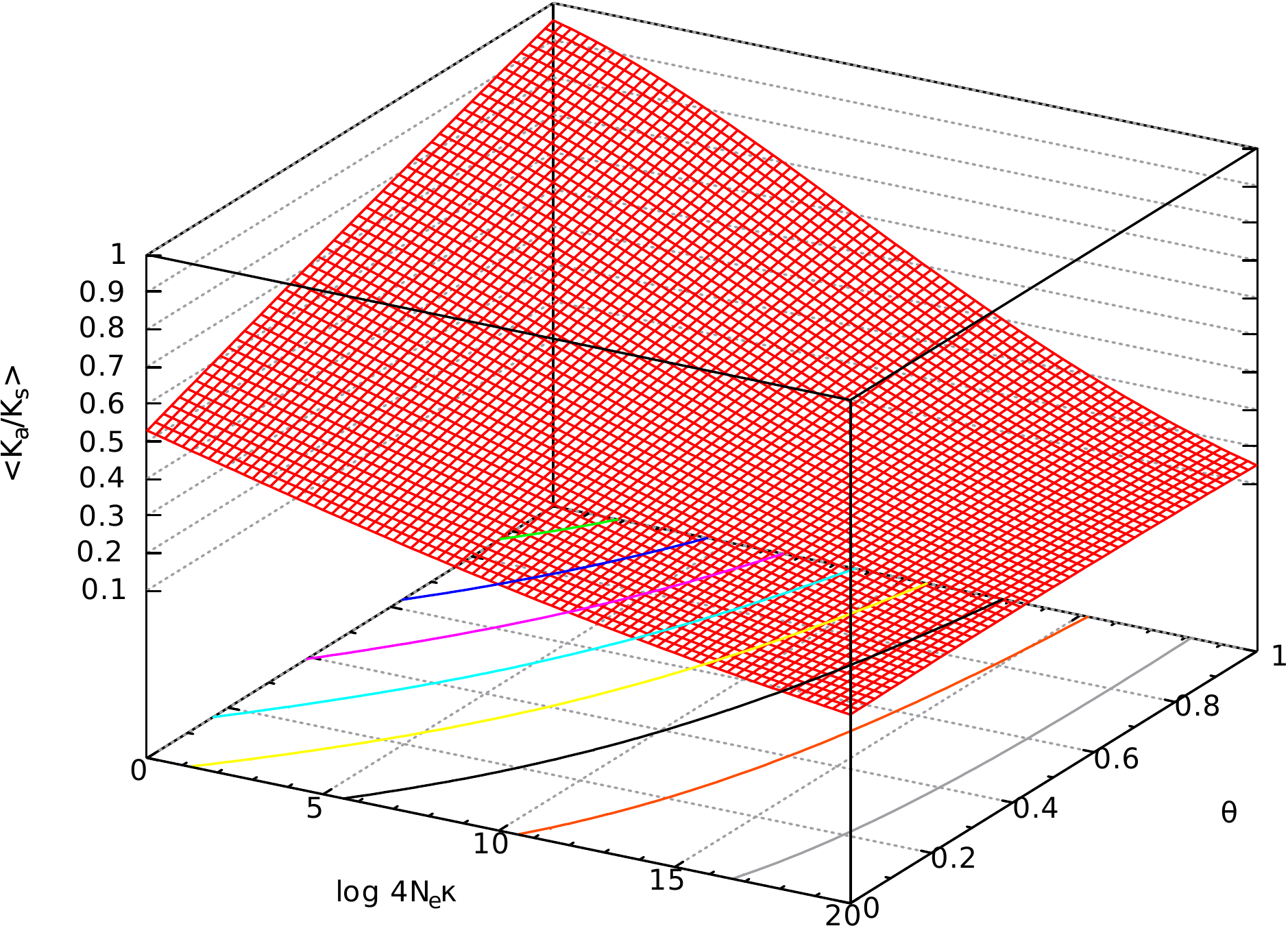}
\includegraphics*[width=90mm,angle=0]{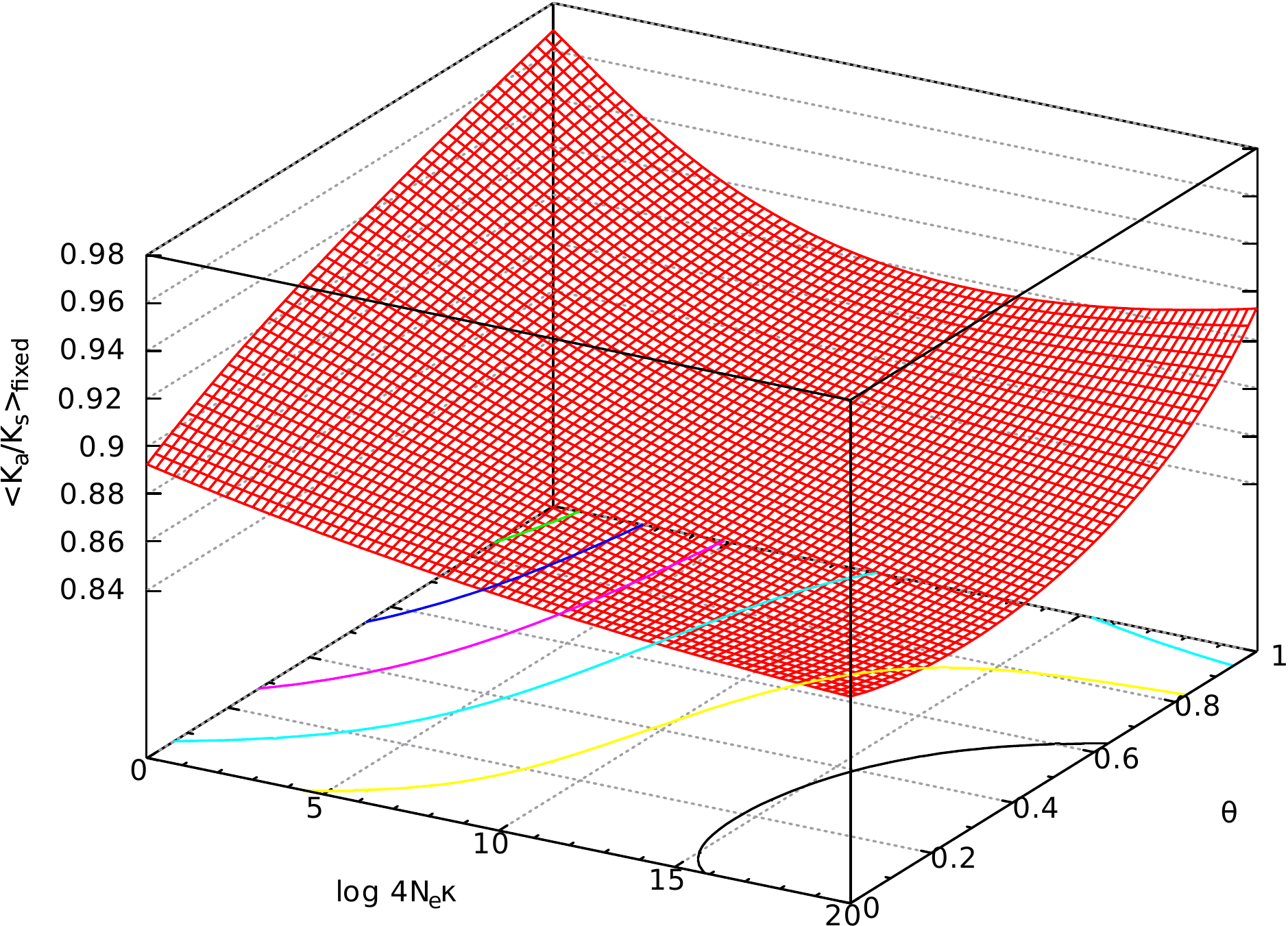}
}
} 
\FigureLegends{
\vspace*{1em}
\caption{
\label{fig: dependence_of_ave_Ka_over_Ks_on_4Nekappa_and_theta}
\label{fig: dependence_of_ave_Ka_over_Ks_fixed_on_4Nekappa_and_theta}
\BF{The average of $K_a / K_s$ over all mutants or over fixed mutants only at equilibrium of protein stability, $\Delta G = \Delta G_e$.}
}
} 
\end{figure*}

\clearpage

} 

} 

The average of $K_a/K_s$ over all mutants, which can be observed as the ratio of
average nonsynonymous substitution rate per nonsynonymous site to 
average synonymous substitution rate per synonymous site, 
and also that over fixed mutants only are shown in 
\Fig{\ref{fig: dependence_of_ave_Ka_over_Ks_on_4Nekappa_and_theta}}.
At any value of $\theta$, $\langle K_a/K_s \rangle$ decreases as  
$\log 4N_e\kappa$ increases, explaining the observed relationship 
that highly expressed and indispensable genes 
evolve slowly
\CITE{DBAWA:05,DW:08,SRS:12}.
Likewise, at any value of $\log 4N_e\kappa$,
$\langle K_a/K_s \rangle$ increases as
$\theta$ increases.
In other words,
the more structurally constrained a protein is,
the more slowly it evolves.
The effect of protein abundance/indispensability on evolutionary rate
is more remarkable for less constrained proteins 
and the effect of structural constraint 
is more remarkable for less abundant, less essential proteins.

The average of $K_a/K_s$ over all mutants, $\langle K_a/K_s \rangle$, is less than 1.0 
on the whole domain shown in the figure, indicating that
the average of $K_a/K_s$ over a long time interval and over many sites 
should not show any positive selection.
Even the average of $K_a/K_s$ over fixed mutants is less than 1.0, and
falls into a narrow range of 0.97--0.85, which is much narrower 
than a range of 0.96--0.15 for that over all mutants;
the average of $K_a/K_s$ over fixed mutants is equal to 
$\langle (K_a/K_s)^2 \rangle / \langle K_a/K_s \rangle$, and 
as a matter of course must be
equal to or larger than the averages of $K_a/K_s$ over all mutants.
However, the average of $K_a/K_s$ over a short time interval and over
a small number of sites may exhibit values larger than one.
In \Fig{\ref{fig: dependence_of_pdf_of_Ka_over_Ks_on_4Nekappa_and_theta}},
the PDFs of $K_a/K_s$ 
for all mutants and also for fixed mutants only
are shown;
$p(K_a/K_s) = p(4N_e s) d (4N_e s) / d (K_a/K_s)$.
A significant fraction of fixed mutants fix with $K_a/K_s > 1$.

\FigureInText{

\TextFig{

\begin{figure*}[ht]
\FigureInLegends{
\centerline{
\includegraphics*[width=90mm,angle=0]{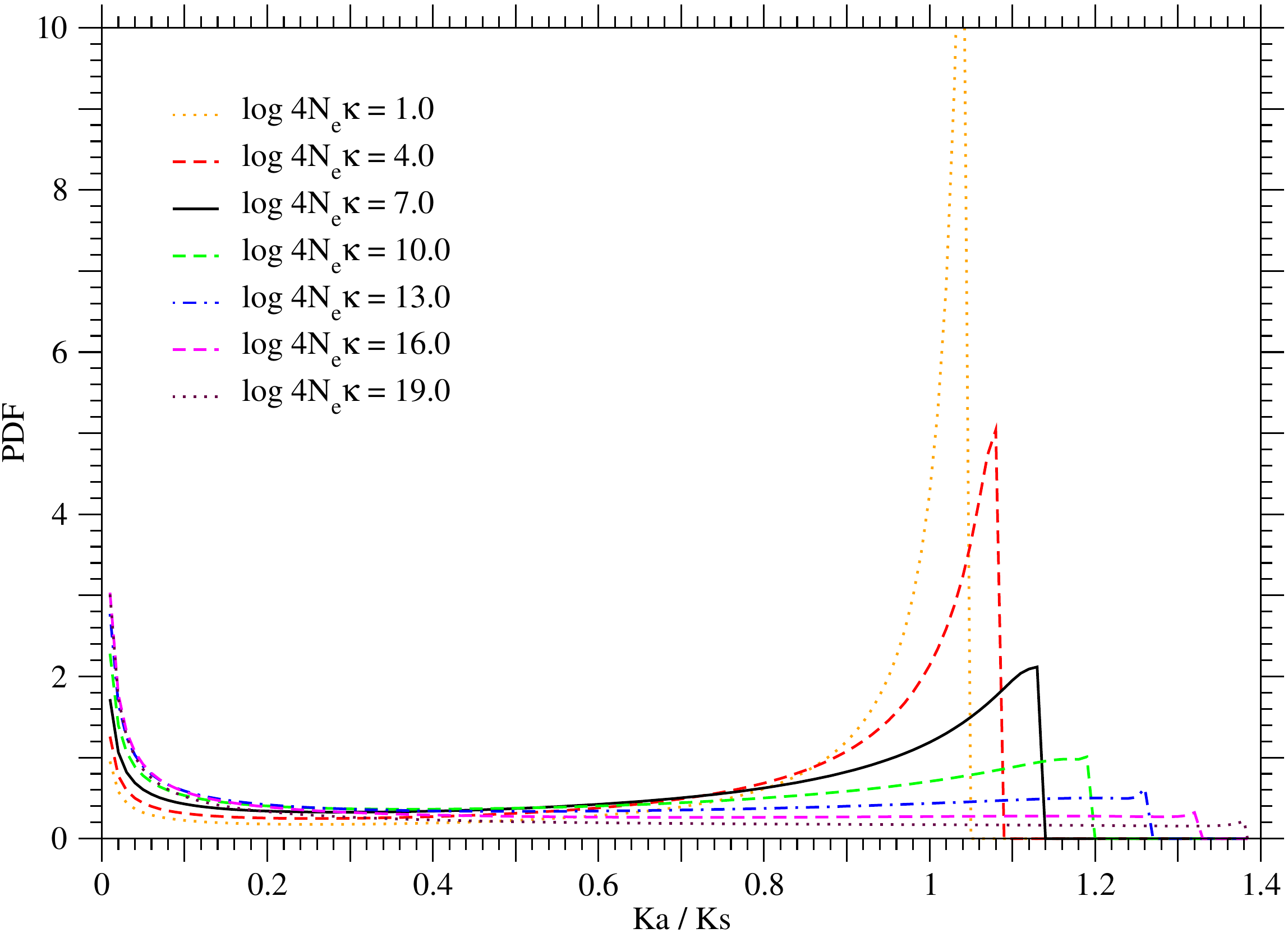}
\includegraphics*[width=90mm,angle=0]{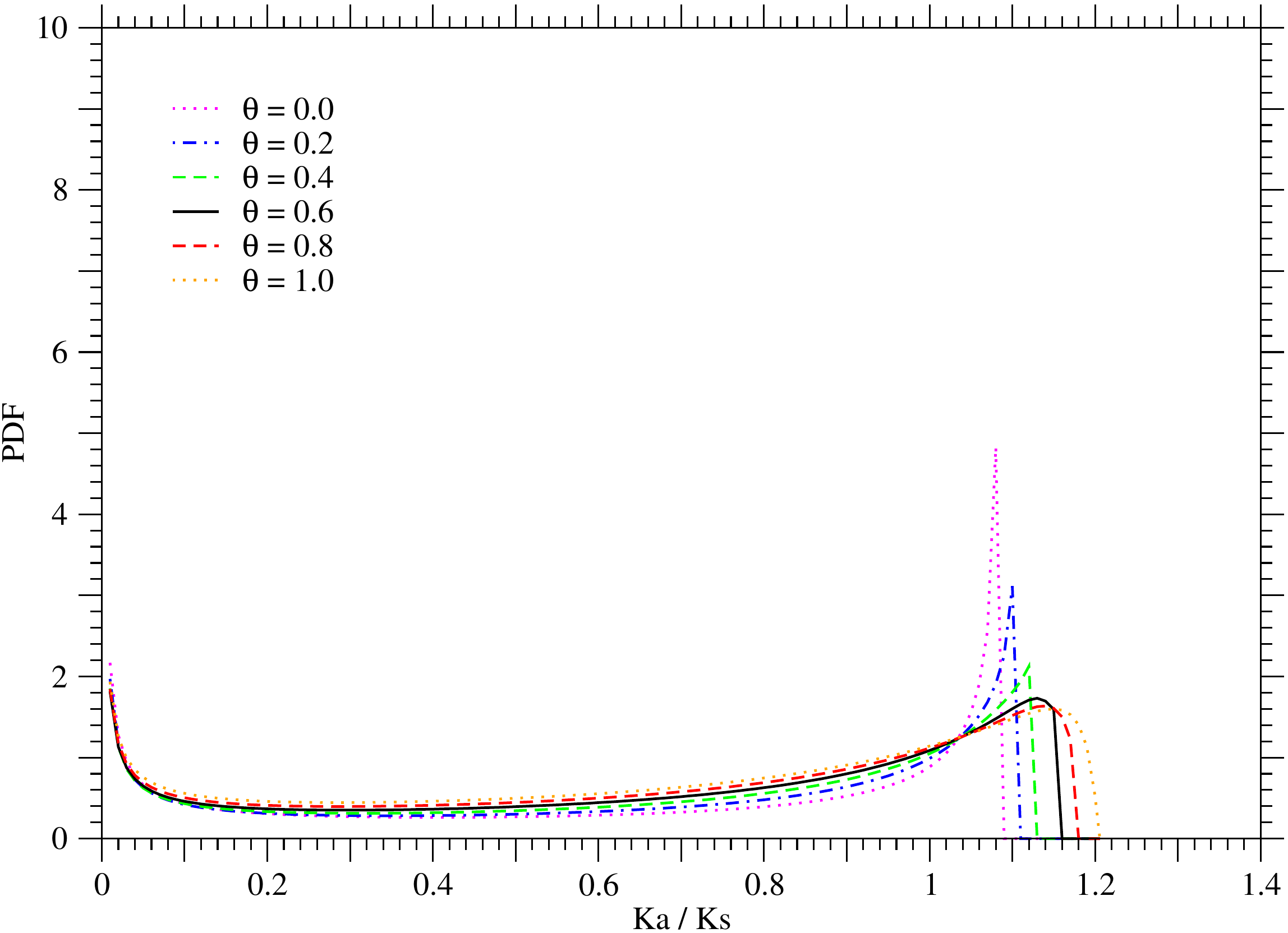}
}
\centerline{
\includegraphics*[width=90mm,angle=0]{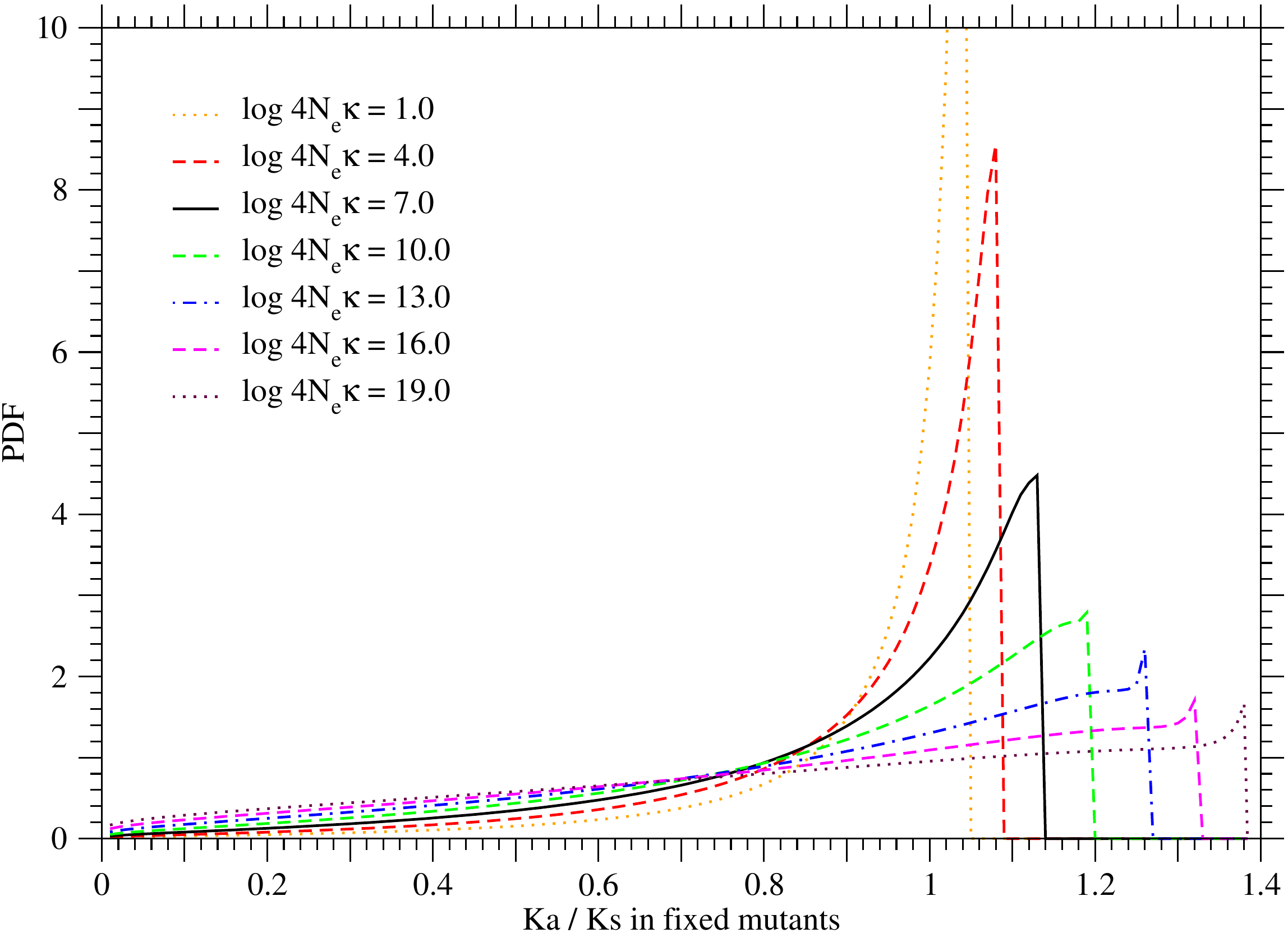}
\includegraphics*[width=90mm,angle=0]{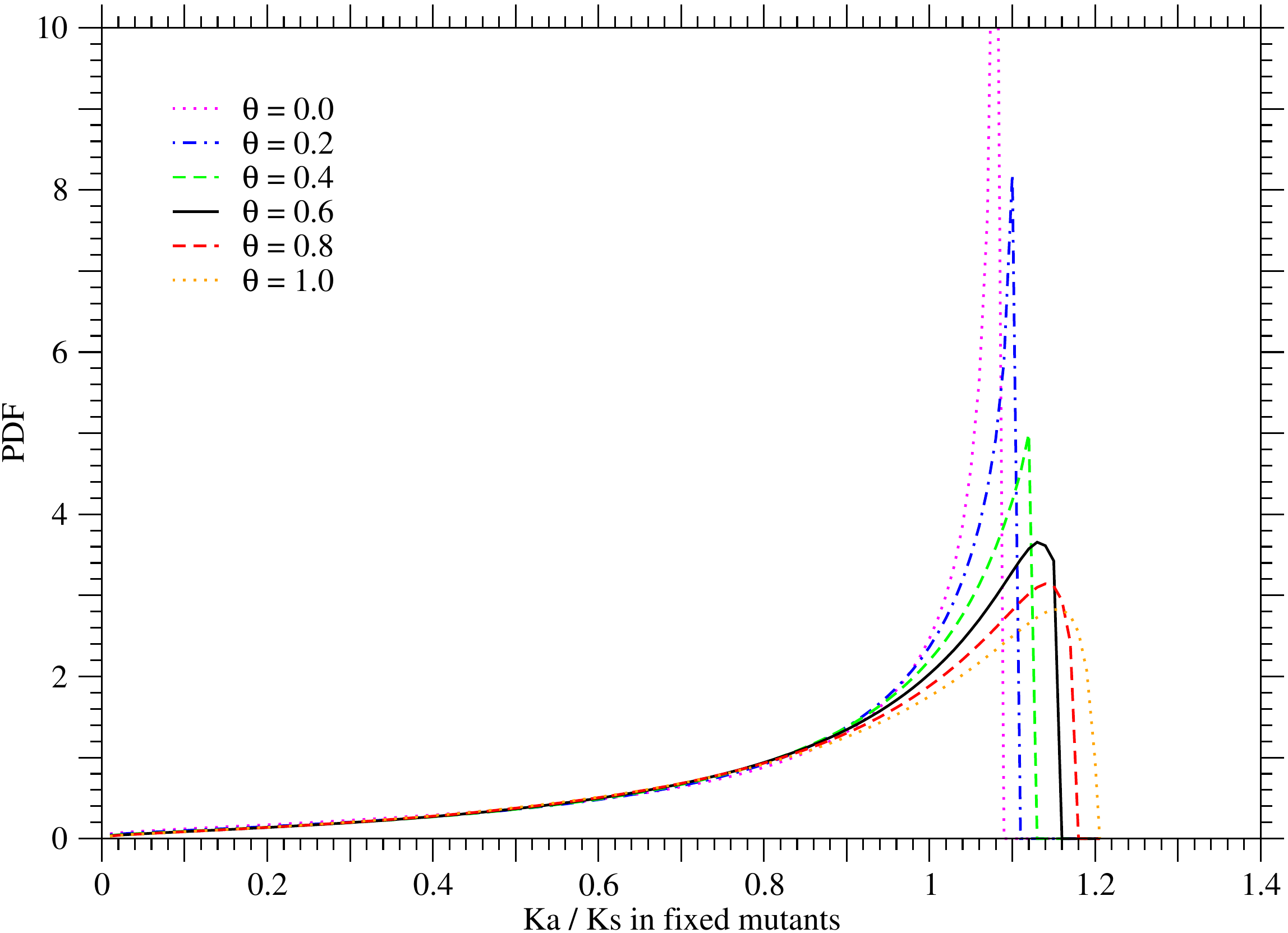}
}
} 
\FigureLegends{
\vspace*{1em}
\caption{
\label{fig: dependence_of_pdf_of_Ka_over_Ks_on_4Nekappa_and_theta}
\label{fig: dependence_of_pdf_of_Ka_over_Ks_fixed_on_4Nekappa_and_theta}
\BF{PDFs of $K_a / K_s$ in all mutants and in fixed mutants only
at equilibrium of protein stability, $\Delta G = \Delta G_e$.}
Unless specified, 
$\log 4N_e\kappa = 7.55$ and $\theta = 0.53$
are employed.
}
} 
\end{figure*}

\clearpage

} 

} 

Arbitrarily, the value of $K_a/K_s$ is categorized into four classes;
negative, slightly negative, nearly neutral, and positive selection categories whose 
$K_a/K_s$ are within the range of
$K_a/K_s \leq 0.5$, $0.5< K_a/K_s \leq 0.95$, $0.95< K_a/K_s \leq 1.05$, and $ 1.05 < K_a/K_s$,
respectively.
Then, the probabilities of each selection category in all mutants and in fixed mutants 
are calculated and shown in 
\Figs{\ref{sfig: prob_of_each_selection_category_in_all_mutants} 
and \ref{fig: prob_of_each_selection_category_in_fixed_mutants}}, respectively.
At the largest abundance ($\log 4N_e\kappa = 20$) 
most arising mutations are negative mutations whose $K_a/K_s$ are less than 0.5.
This is reasonable, because at this condition 
the wild-type protein is very stable with
low equilibrium values $\Delta G_e$ as shown in \Fig{\ref{fig: dependence_of_dGe_on_4Nekappa_and_theta}},
and therefore most mutations destabilize the wild type
and tend to be removed from population.
Most fixed mutants are 
positive mutants or slightly negative mutants fixed by random drift.
Nearly neutral mutants are less than 3\% of all mutants, and
less than 15\% of fixed mutants. 

\FigureInText{

\TextFig{

\begin{figure*}[ht]
\FigureInLegends{
\centerline{
\includegraphics*[width=90mm,angle=0]{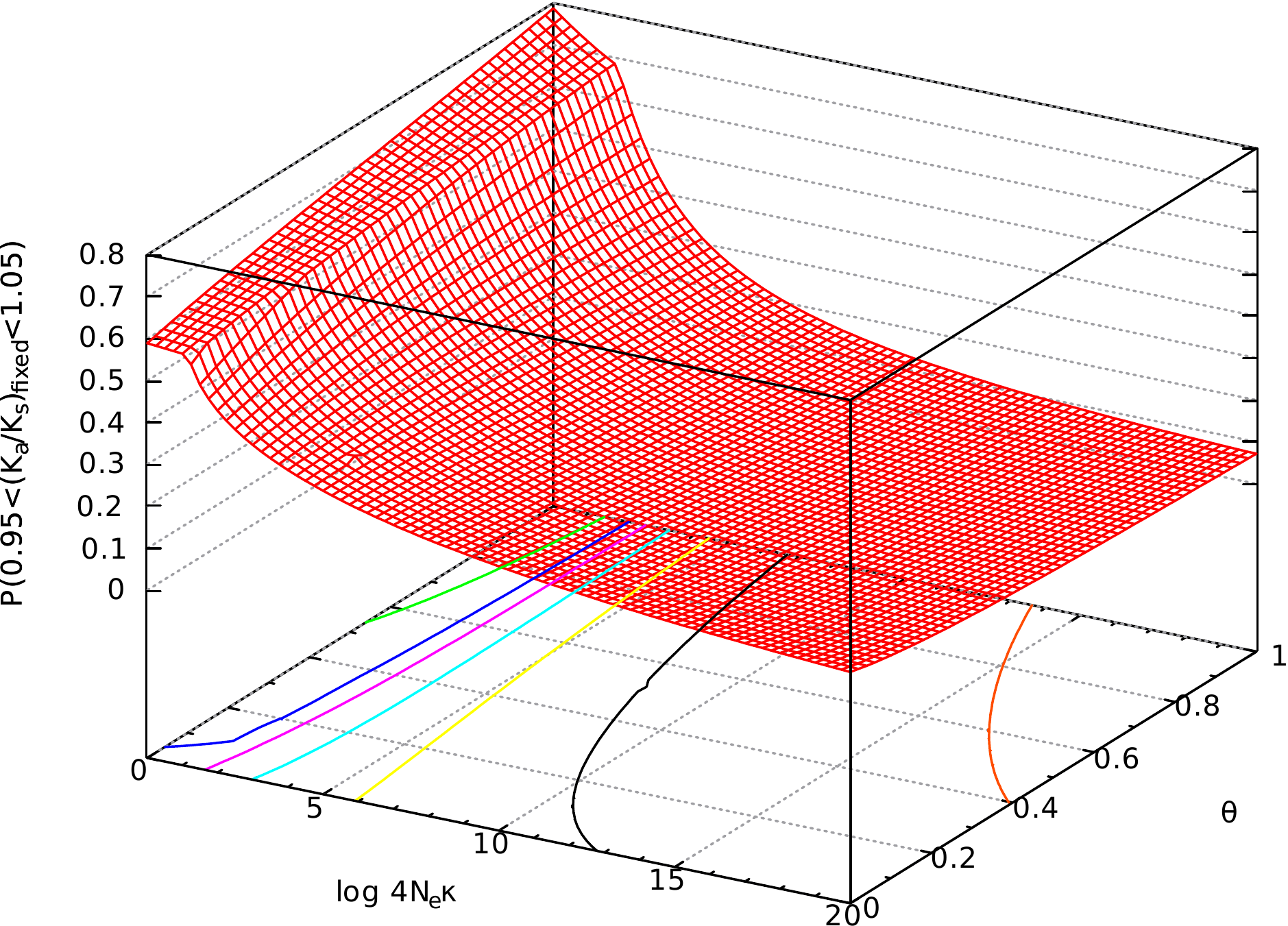}
\includegraphics*[width=90mm,angle=0]{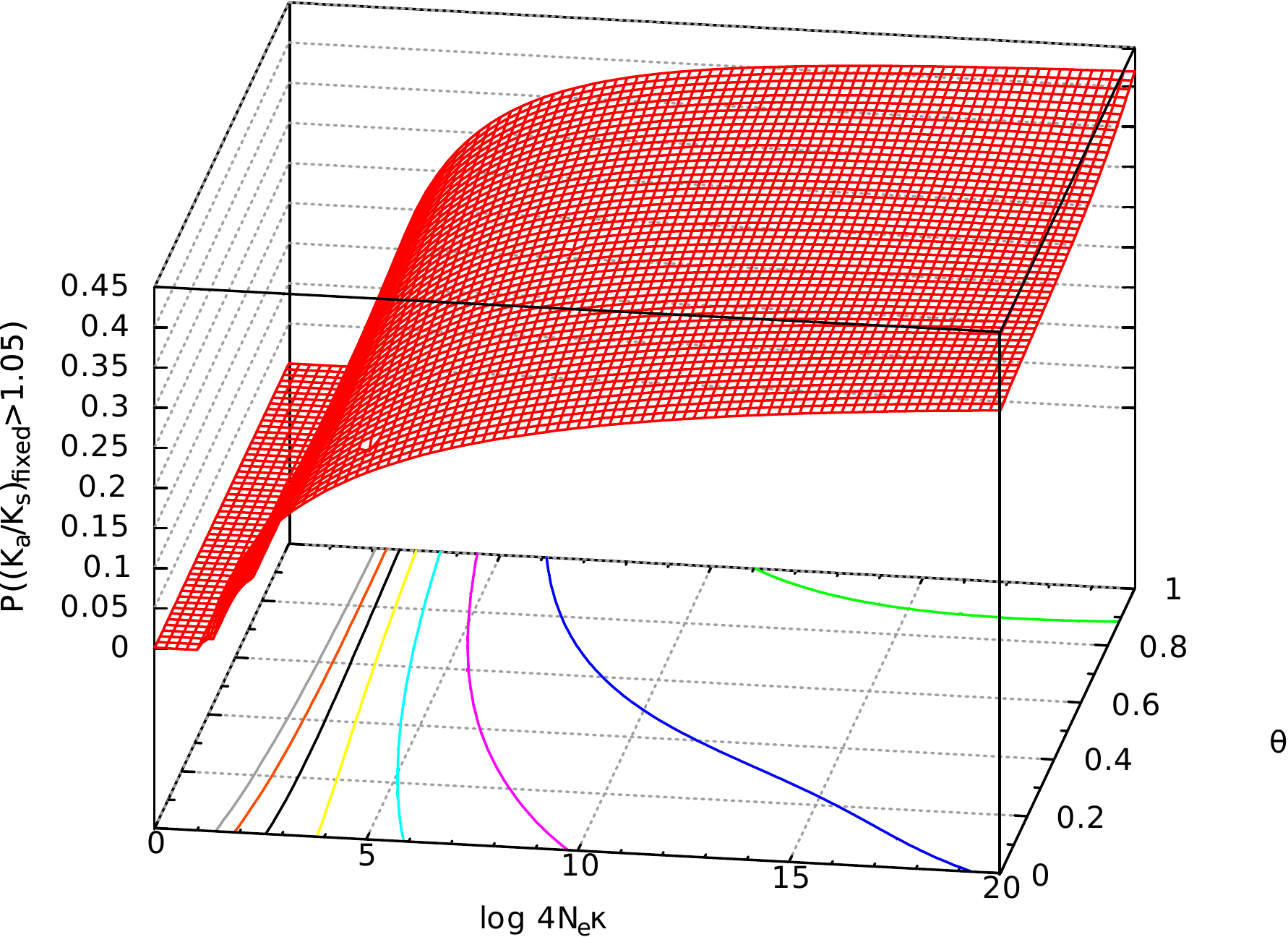}
}
\centerline{
\includegraphics*[width=90mm,angle=0]{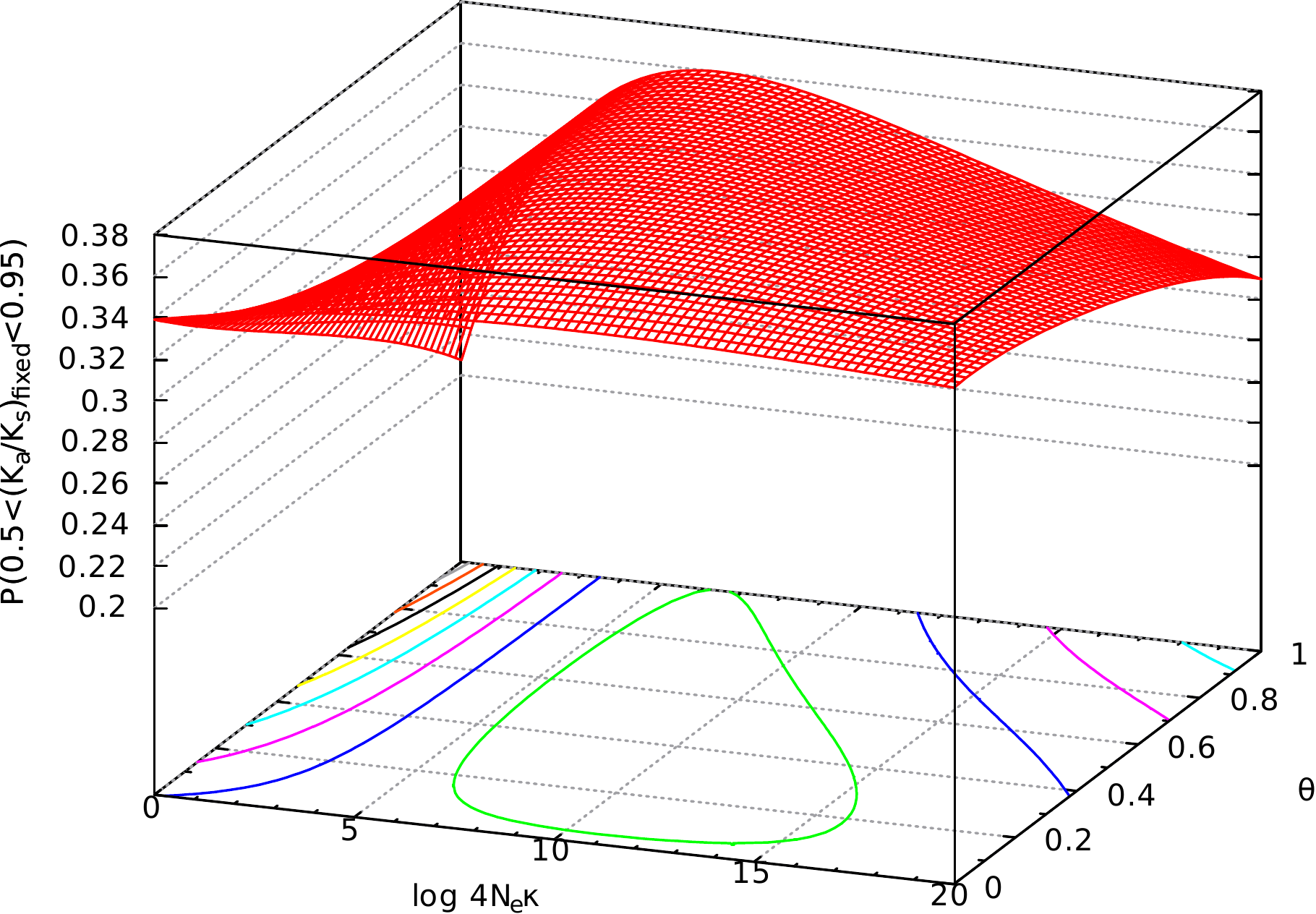}
\includegraphics*[width=90mm,angle=0]{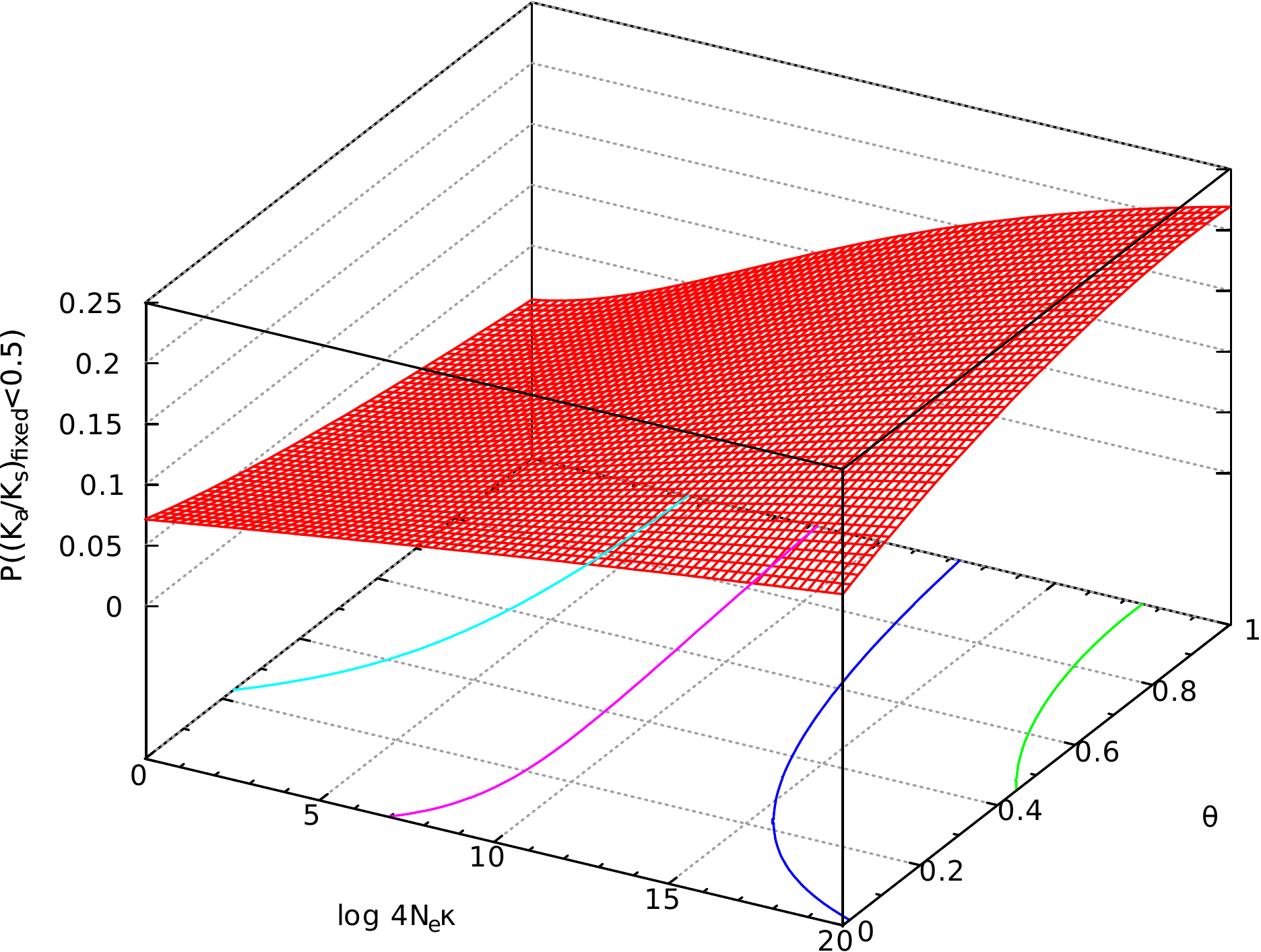}
}
} 
\FigureLegends{
\vspace*{1em}
\caption{
\label{fig: prob_of_each_selection_category_in_fixed_mutants}
\BF{Probability of each selection category in fixed mutants at equilibrium of protein stability, $\Delta G = \Delta G_e$.}
Arbitrarily, the value of $K_a/K_s$ is categorized into four classes;
negative, slightly negative, nearly neutral, and positive selection categories in which
$K_a/K_s$ is within the ranges of
$K_a/K_s \leq 0.5$, $0.5< K_a/K_s \leq 0.95$, $0.95< K_a/K_s \leq 1.05$, and $ 1.05 < K_a/K_s$,
respectively.
}
} 
\end{figure*}

\clearpage

} 

} 

On the other hand, 
at the other extreme of $\log 4N_e\kappa < 2$,
there are no mutations of the positive selection category, this is because
the upper bound of $K_a/K_s$, which corresponds to the upper bound 
($4N_e\kappa \exp(\beta\Delta G)$) of $4N_e s$, 
at the equilibrium stability $\Delta G_e$ becomes less than 1.05
that is the lower bound for the positive selection category; see \Eq{\ref{eq: def_s}}.
The significant amount of mutations become nearly neutral.
As $\theta$ changes from 1 to 0, that is, structural constraints increase,
the proportion of nearly neutral mutations changes 
from 0.75(0.56) to 0.31(0.22),
and instead negative mutations increase and most of them are removed from population.
Thus, the selection mechanism for fixing 
stabilizing mutants in little expressed, non-essential genes 
($\log 4N_e\kappa < 2$) 
is not positive selection but nearly neutral selection, that is,
random drift.

The probability of each selection category in fixed mutants depends strongly on $4N_e\kappa$,
but much less on $\theta$.
Current common understanding is that
amino acid substitutions in protein evolution are either neutral\CITE{K:68} or lethal, 
at most slightly deleterious\CITE{O:73} or lethal, unless functional selection operates and
functional changes occur.
On the contrary, nearly neutral fixations are predominant only in proteins with 
$\log 4N_e\kappa < 2$ or $\Delta G_e > -2.5$ kcal/mol,
and positive selection is significant in 
the other proteins.
On the other hand, slightly negative selection is always significant.   
An interesting result is that 
the effects of structural constraint 
on $K_a/K_s$ 
are the most remarkable in proteins with 
$\log 4N_e\kappa < 2$ or instead $\Delta G_e > -2.5$ kcal/mol
in which nearly neutral fixations are predominant.

\vspace*{1em}
\noindent
\subsection{$K_a/K_s$ in the vicinity of equilibrium}

In \Fig{\ref{fig: ave_ddG_vs_dG}}, the 
$\langle \Delta\Delta G \rangle_{\script{fixed}} \pm $ standard deviation of $\Delta\Delta G$ of
fixed mutants are also drawn. The standard deviation of $\Delta\Delta G$ of fixed mutants
is equal to 0.84 kcal/mol at the equilibrium, $\Delta G_e$, indicating that protein stability $\Delta G$
fluctuates more or less within $\Delta G_e \pm 0.84$ kcal/mol instantaneously. 
Such a deviation from the equilibrium 
must be canceled 
by compensatory substitutions that consecutively occur, 
otherwise the protein stability would far depart from its equilibrium point.

In \Figs{\ref{fig: prob_of_each_selection_category_in_fixed_mutants;_dependence_on_kappa_and_dG} and
\ref{fig: prob_of_each_selection_category_in_fixed_mutants;_dependence_on_theta_and_dG} }
and \Figs{\ref{sfig: prob_of_each_selection_category_in_all_mutants;_dependence_on_kappa_and_dG} and
\ref{sfig: prob_of_each_selection_category_in_all_mutants;_dependence_on_theta_and_dG}},
the probabilities of each selection category in fixed mutants and in all arising mutants 
are shown as a function of $\Delta G$ and $4N_e\kappa$ or $\theta$, respectively.
The range of $\Delta G$ around $\Delta G_e$ shown by a blue line on the surface grid
is within two times of the standard deviation of $\Delta\Delta G$ over fixed mutants 
at $\Delta G_e$.

\FigureInText{

\TextFig{

\begin{figure*}[ht]
\FigureInLegends{
\centerline{
\includegraphics*[width=87mm,angle=0]{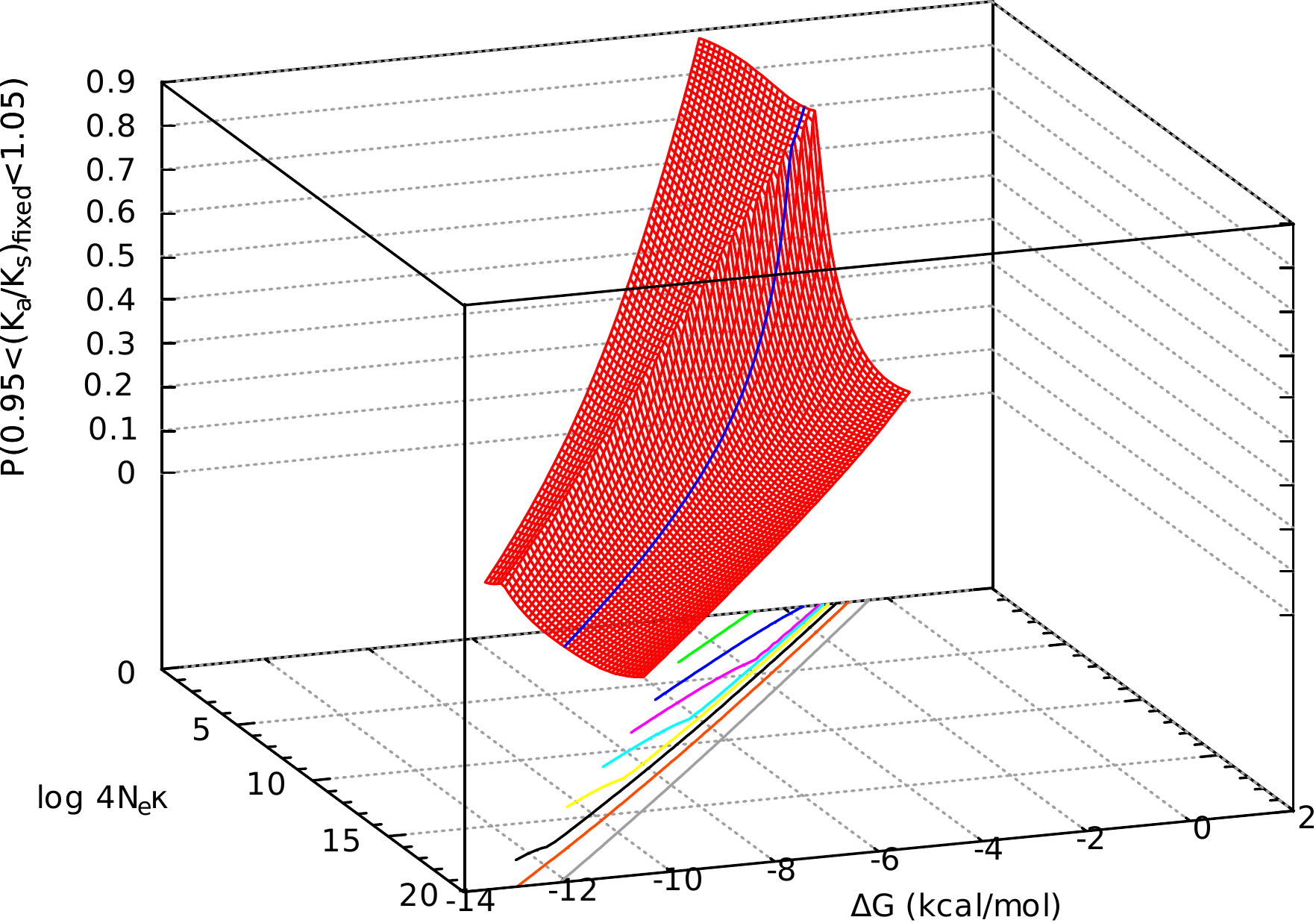}
\includegraphics*[width=90mm,angle=0]{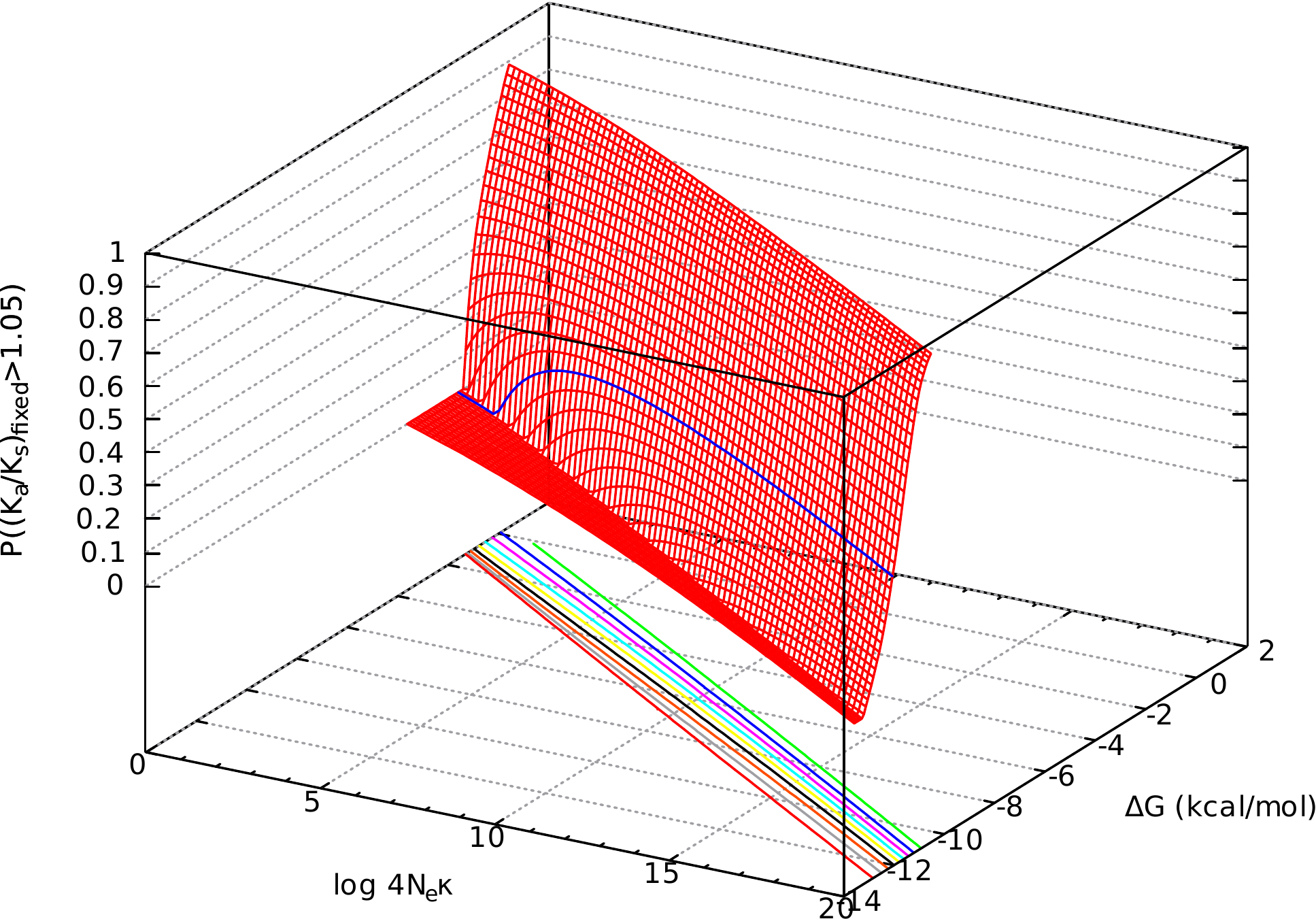}
}
\centerline{
\includegraphics*[width=90mm,angle=0]{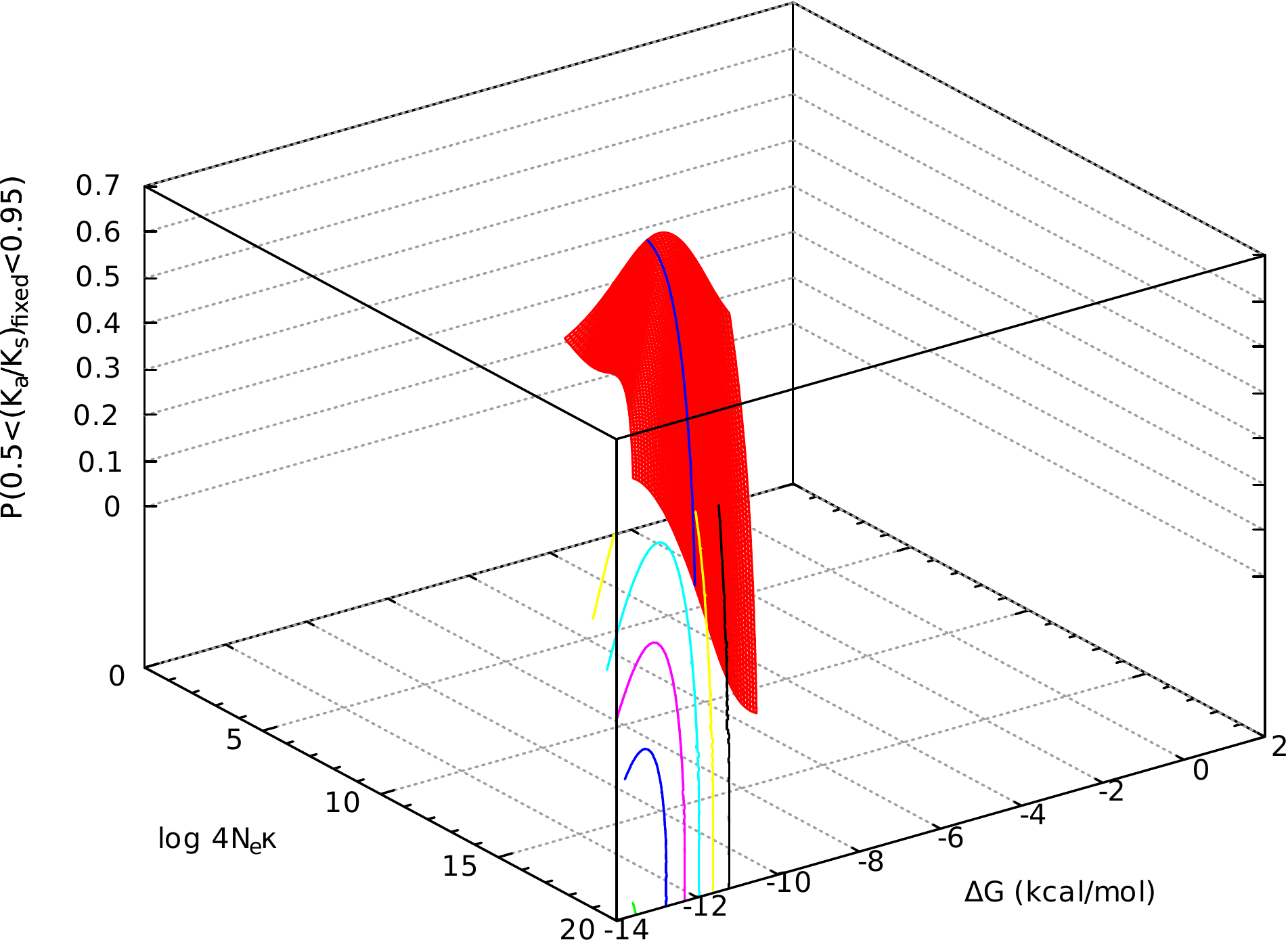}
\includegraphics*[width=90mm,angle=0]{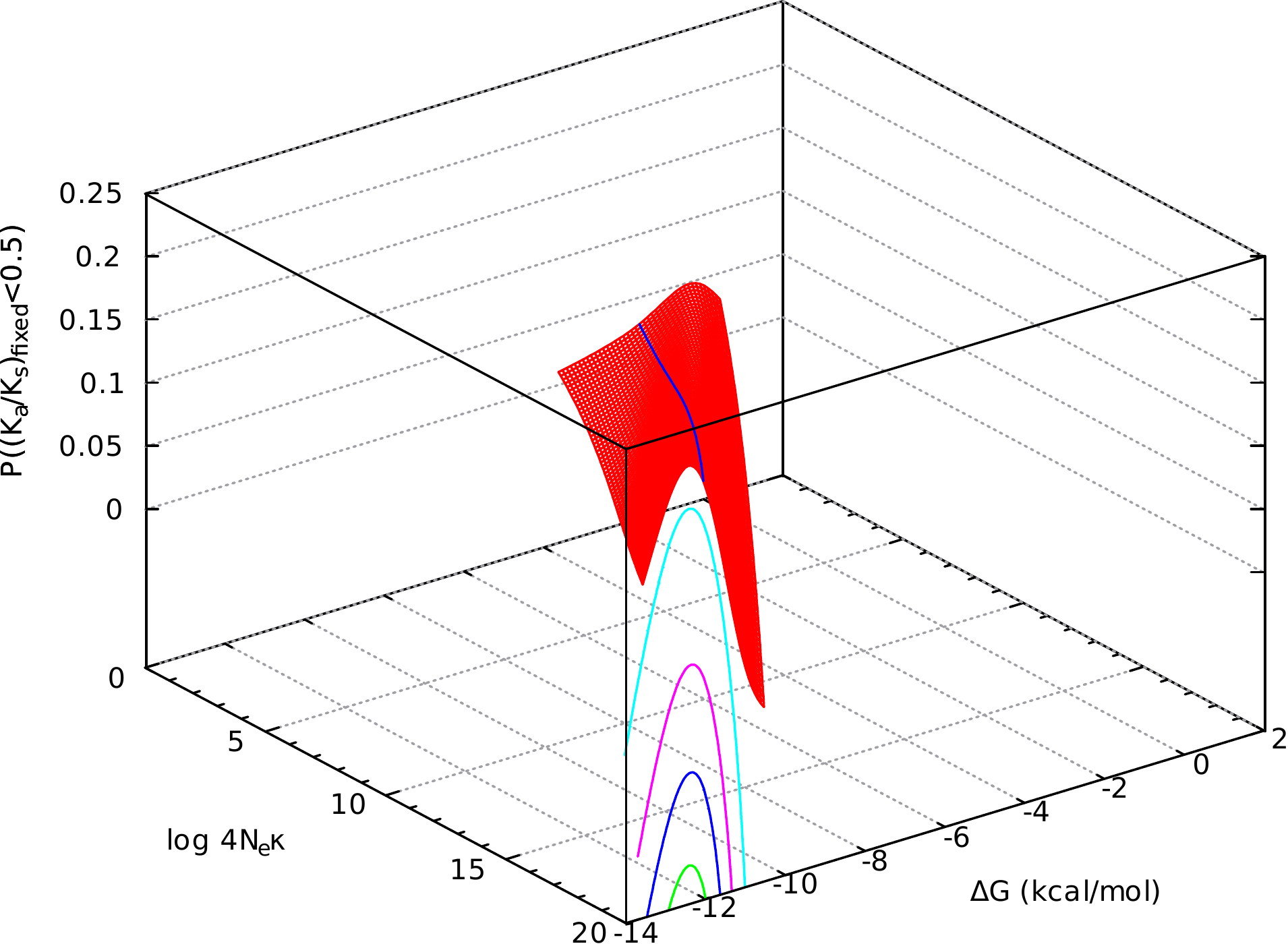}
}
} 
\FigureLegends{
\vspace*{1em}
\caption{
\label{fig: prob_of_each_selection_category_in_fixed_mutants;_dependence_on_kappa_and_dG}
\BF{Dependence of the probability of each selection category in fixed mutants on $4N_e\kappa$ and $\Delta G$.}
A blue line on the surface grid shows $\Delta G=\Delta G_e$, which is the equilibrium value of $\Delta G$
in protein evolution.
The range of $\Delta G$ shown in the figures is
$| \Delta G - \Delta G_e | < 2 \cdot \Delta\Delta G^{\script{sd}}_{\script{fixed}}$,
where $\Delta\Delta G^{\script{sd}}_{\script{fixed}}$ is the standard deviation of $\Delta\Delta G$ over fixed mutants at $ \Delta G = \Delta G_e$.
Arbitrarily, the value of $K_a/K_s$ is categorized into four classes;
negative, slightly negative, nearly neutral, and positive selection categories in which
$K_a/K_s$ is within the ranges of
$K_a/K_s \leq 0.5$, $0.5< K_a/K_s \leq 0.95$, $0.95< K_a/K_s \leq 1.05$, and $ 1.05 < K_a/K_s$,
respectively.
$\theta = 0.53$
is employed.
The kcal/mol unit is used for $\Delta G$.
}
} 
\end{figure*}

\clearpage

} 

\TextFig{

\begin{figure*}[ht]
\FigureInLegends{
\centerline{
\includegraphics*[width=83mm,angle=0]{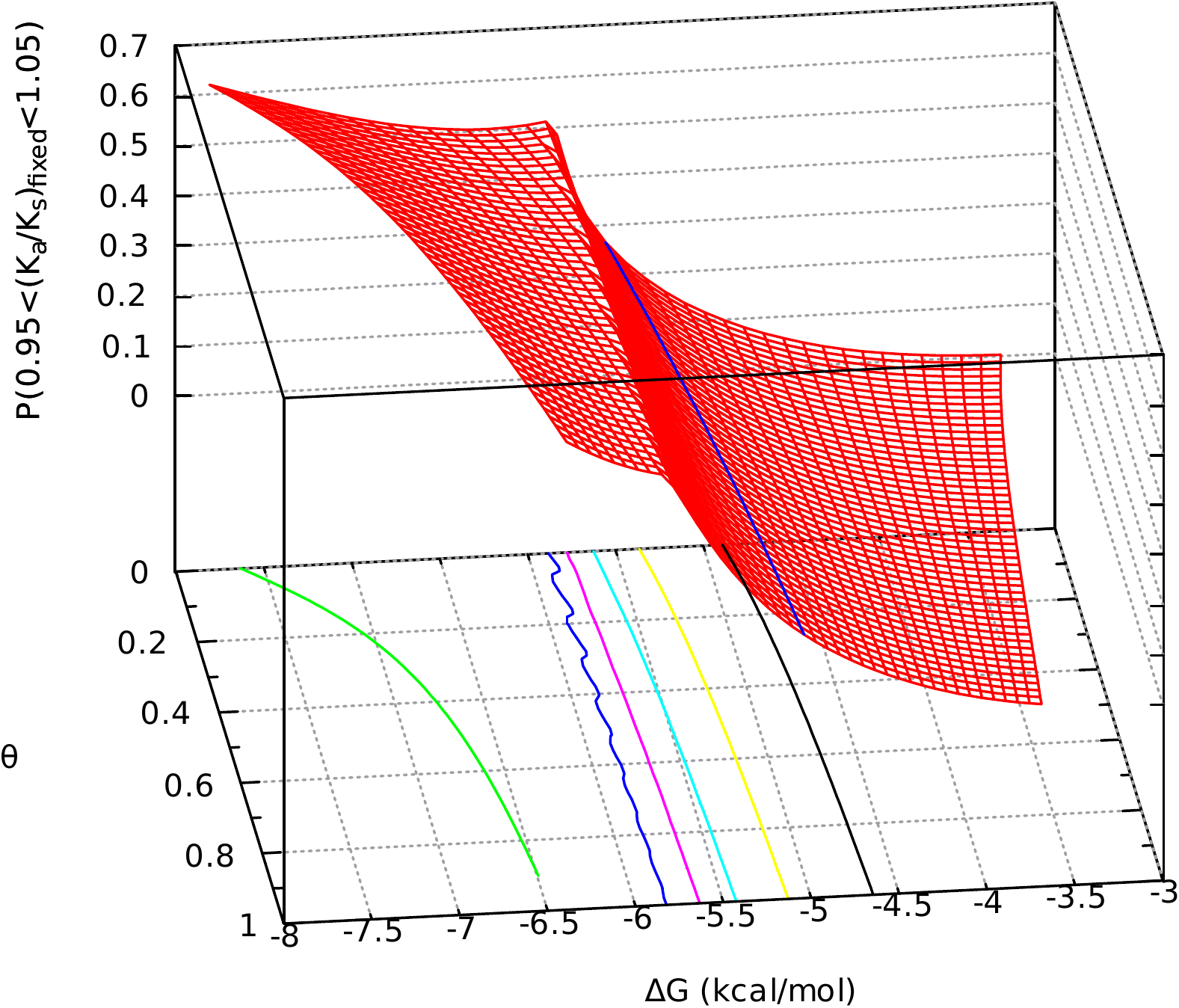}
\includegraphics*[width=90mm,angle=0]{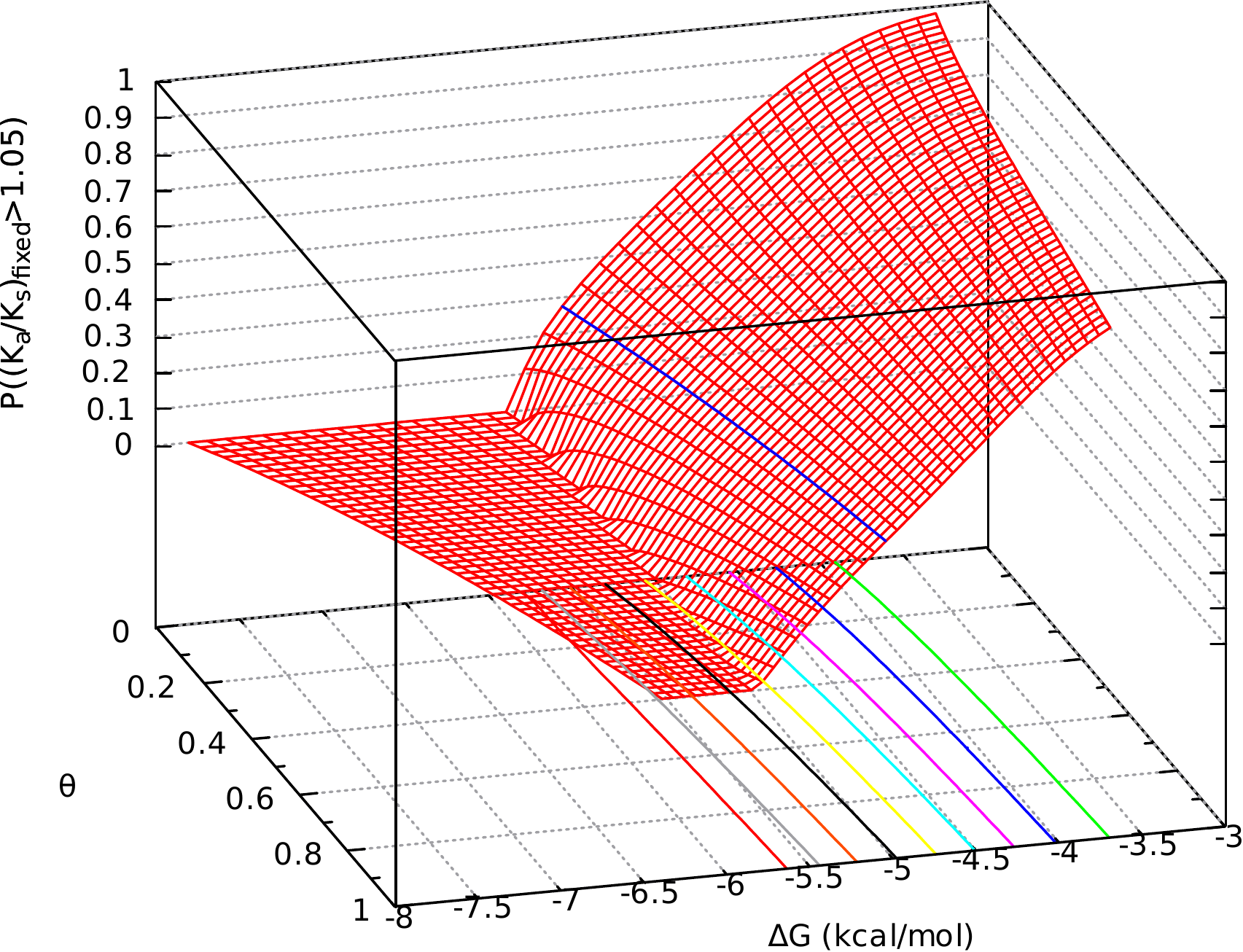}
}
\centerline{
\includegraphics*[width=89mm,angle=0]{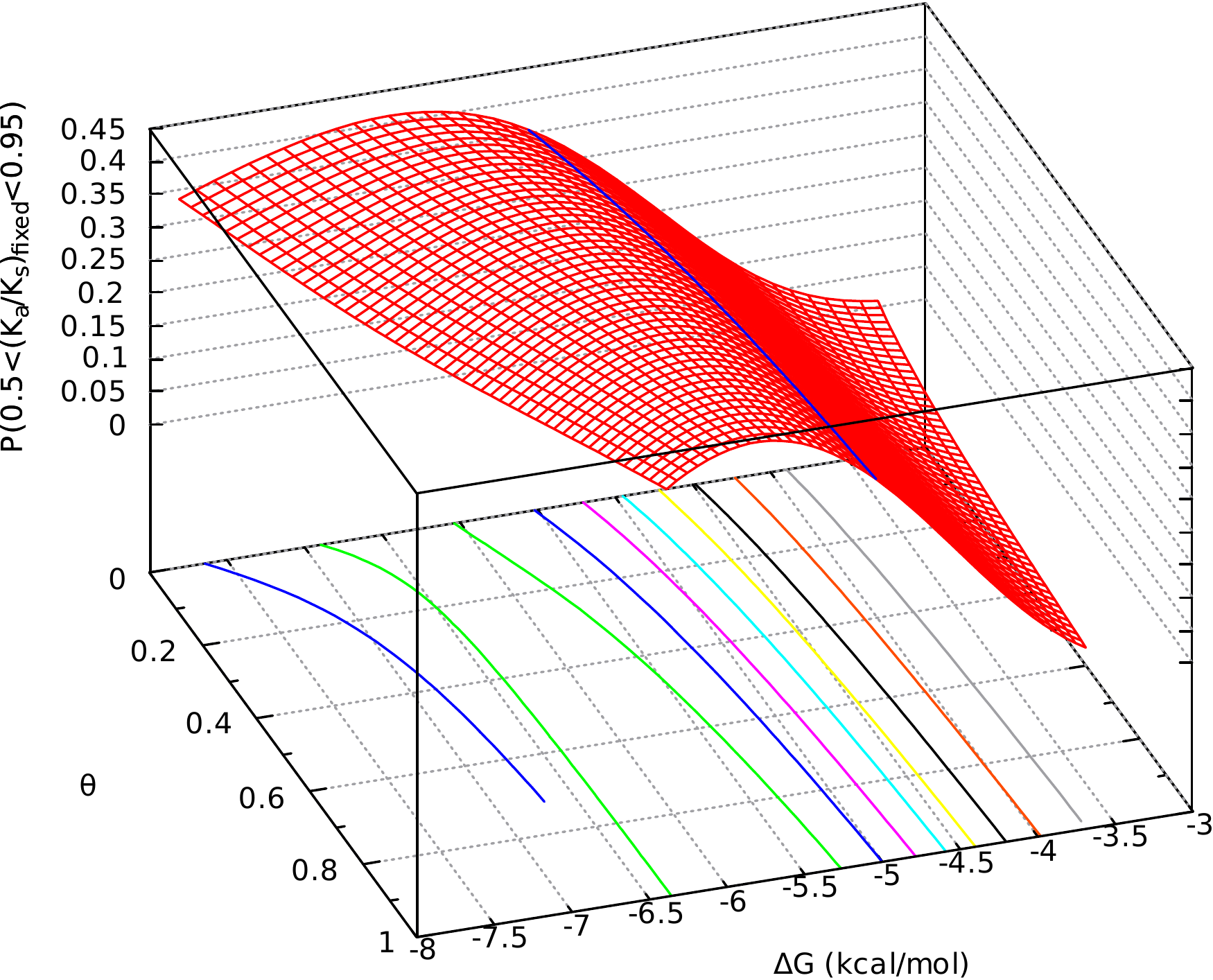}
\includegraphics*[width=90mm,angle=0]{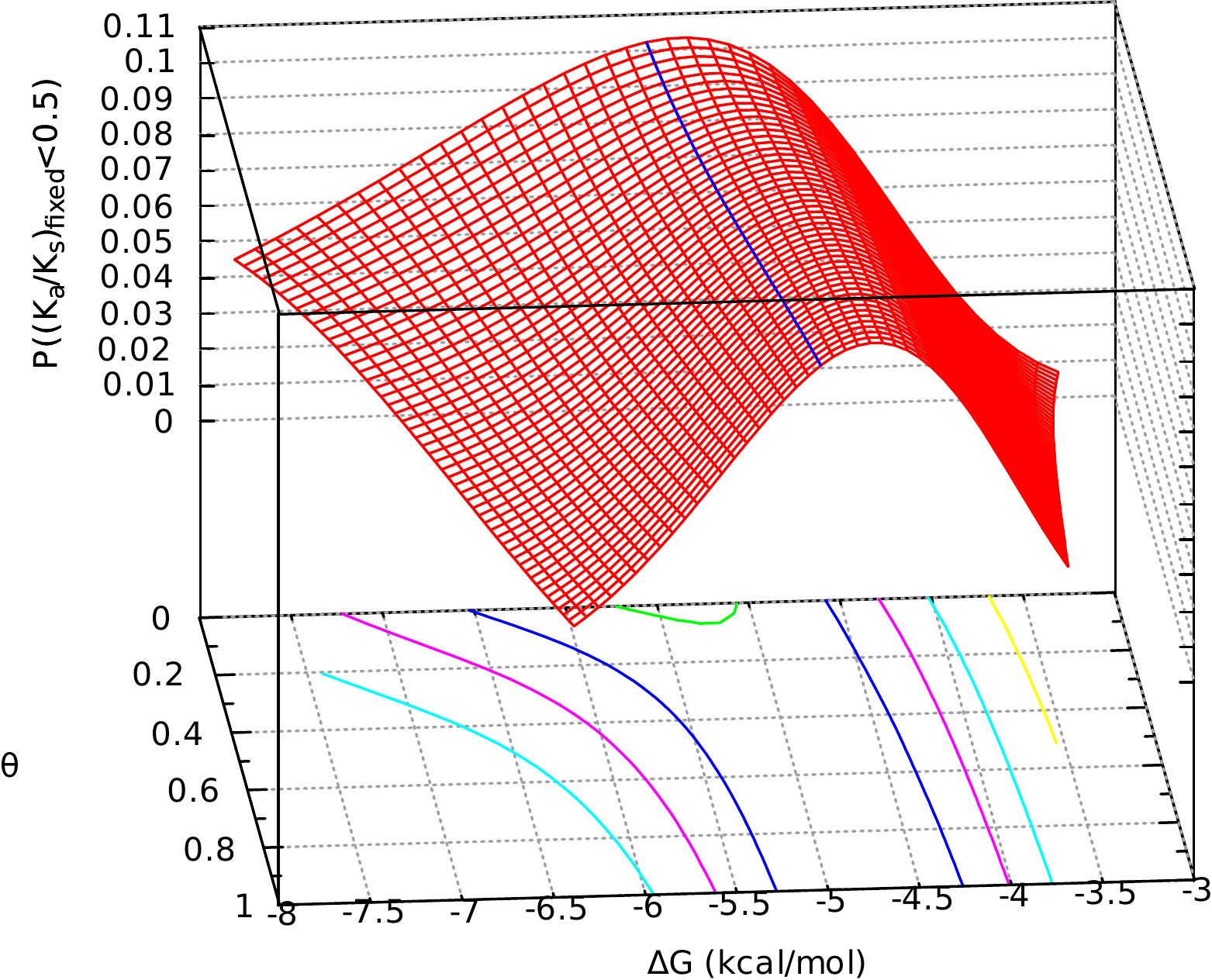}
}
} 
\FigureLegends{
\vspace*{1em}
\caption{
\label{fig: prob_of_each_selection_category_in_fixed_mutants;_dependence_on_theta_and_dG}
\BF{Dependence of the probability of each selection category in fixed mutants on $\theta$ and $\Delta G$.}
A blue line on the surface grid shows $\Delta G=\Delta G_e$, which is the equilibrium value of $\Delta G$
in protein evolution.
The range of $\Delta G$ shown in the figures is
$| \Delta G - \Delta G_e | < 2 \cdot \Delta\Delta G^{\script{sd}}_{\script{fixed}}$,
where $\Delta\Delta G^{\script{sd}}_{\script{fixed}}$ is the standard deviation of $\Delta\Delta G$ over fixed mutants at $ \Delta G = \Delta G_e$.
Arbitrarily, the value of $K_a/K_s$ is categorized into four classes;
negative, slightly negative, nearly neutral, and positive selection categories in which
$K_a/K_s$ is within the ranges of
$K_a/K_s \leq 0.5$, $0.5< K_a/K_s \leq 0.95$, $0.95< K_a/K_s \leq 1.05$, and $ 1.05 < K_a/K_s$,
respectively.
$\log 4N_e\kappa = 7.55$
is employed.
The kcal/mol unit is used for $\Delta G$.
}
} 
\end{figure*}

\clearpage

} 

} 

As indicated by \Eqs{\Ref{eq: def_ms} and \Ref{eq: def_mc}}, 
it is shown in
\Figs{\ref{sfig: prob_of_each_selection_category_in_all_mutants;_dependence_on_kappa_and_dG} and
\ref{sfig: prob_of_each_selection_category_in_all_mutants;_dependence_on_theta_and_dG}}
that stabilizing mutations increase due to negative shifts of $\Delta\Delta G$
as the wild type becomes less stable than the equilibrium, $\Delta G > \Delta G_e$,
and that destabilizing mutations increase due to positive shifts of $\Delta\Delta G$ 
as $\Delta G < \Delta G_e$.
In addition, as indicated by \Eq{\ref{eq: def_s}},
it is shown in 
\Figs{\ref{fig: prob_of_each_selection_category_in_fixed_mutants;_dependence_on_kappa_and_dG} and
\ref{fig: prob_of_each_selection_category_in_fixed_mutants;_dependence_on_theta_and_dG}}
that positive selection on stabilizing mutants is more amplified as $\Delta G > \Delta G_e$,
and that 
more destabilizing mutations become nearly neutral
due to the less-amplified effect of stability change on selective advantage as $\Delta G < \Delta G_e$. 
This is a mechanism of maintaining protein stability at equilibrium.

\FigureInText{

\TextFig{

\begin{figure*}[ht]
\FigureInLegends{
\centerline{
\includegraphics*[width=90mm,angle=0]{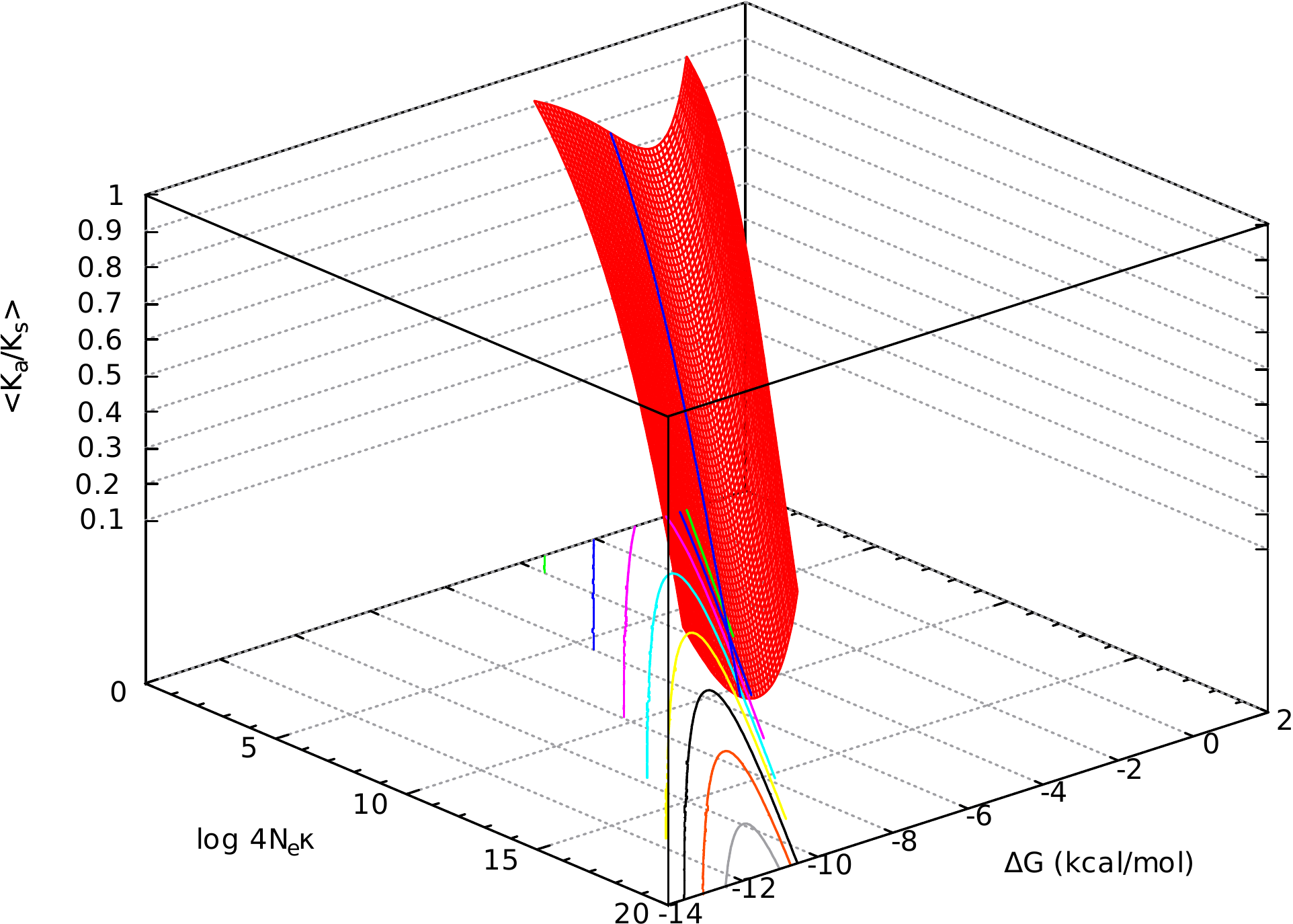}
\includegraphics*[width=85mm,angle=0]{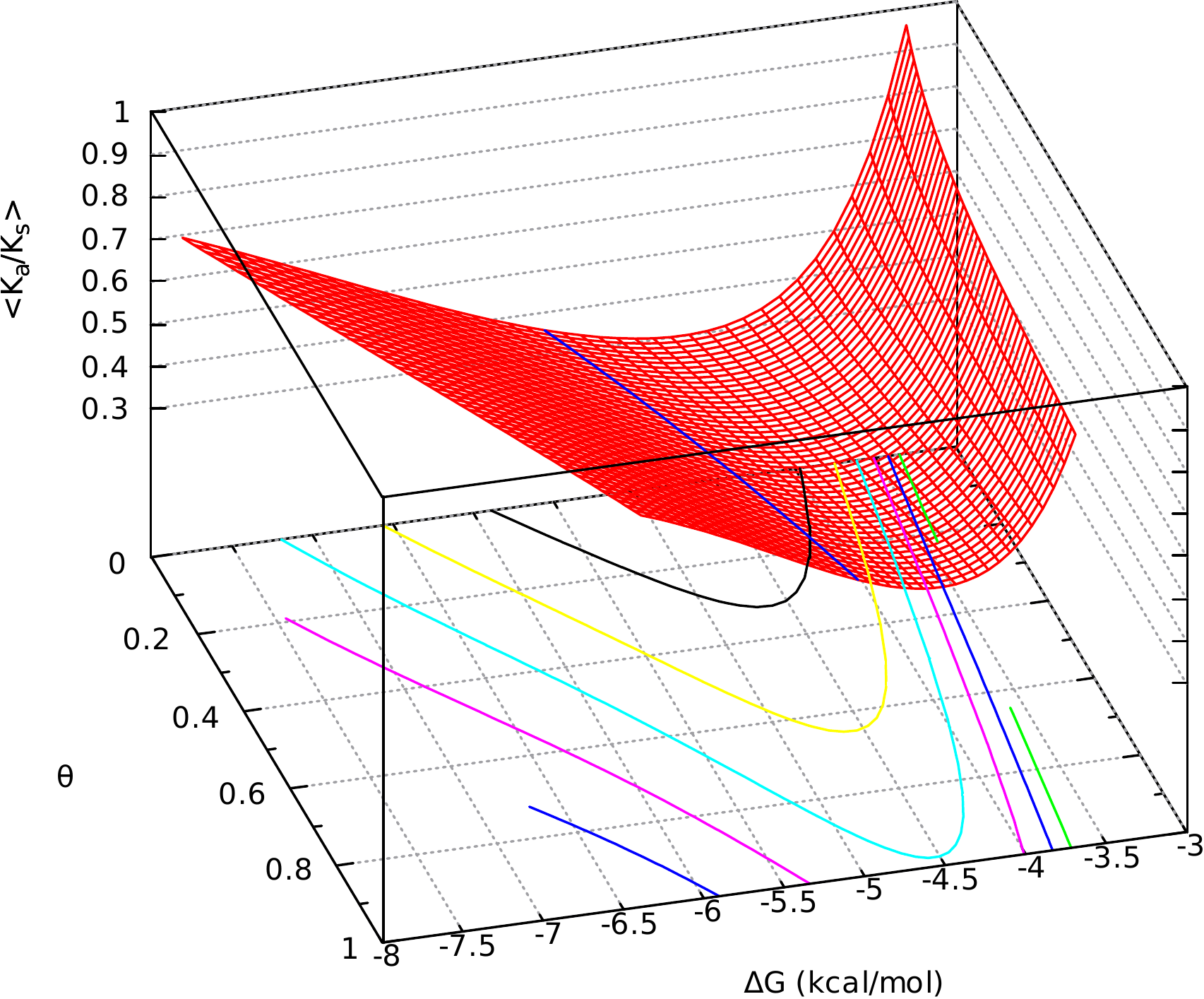}
}
\centerline{
\includegraphics*[width=88mm,angle=0]{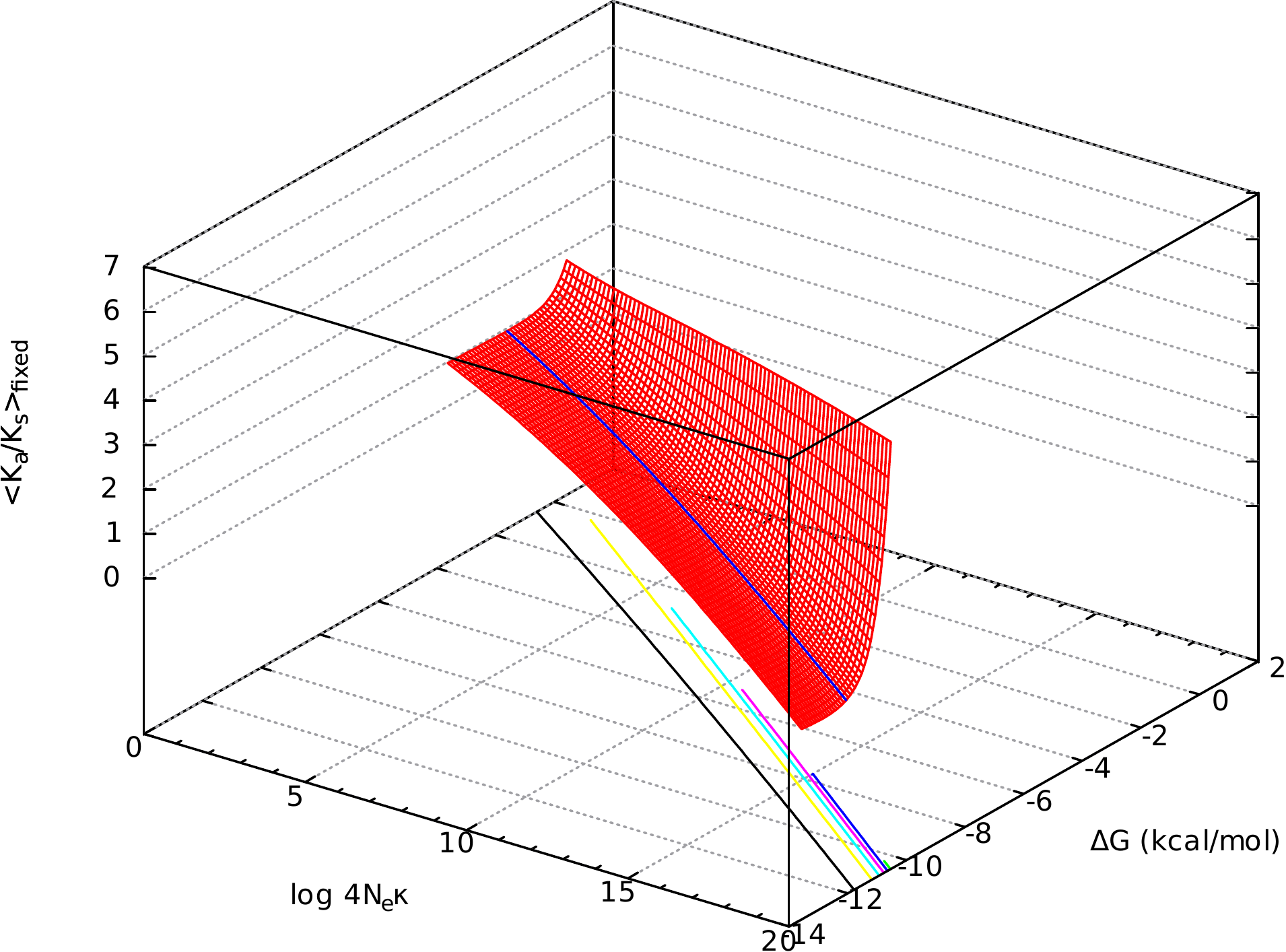}
\includegraphics*[width=90mm,angle=0]{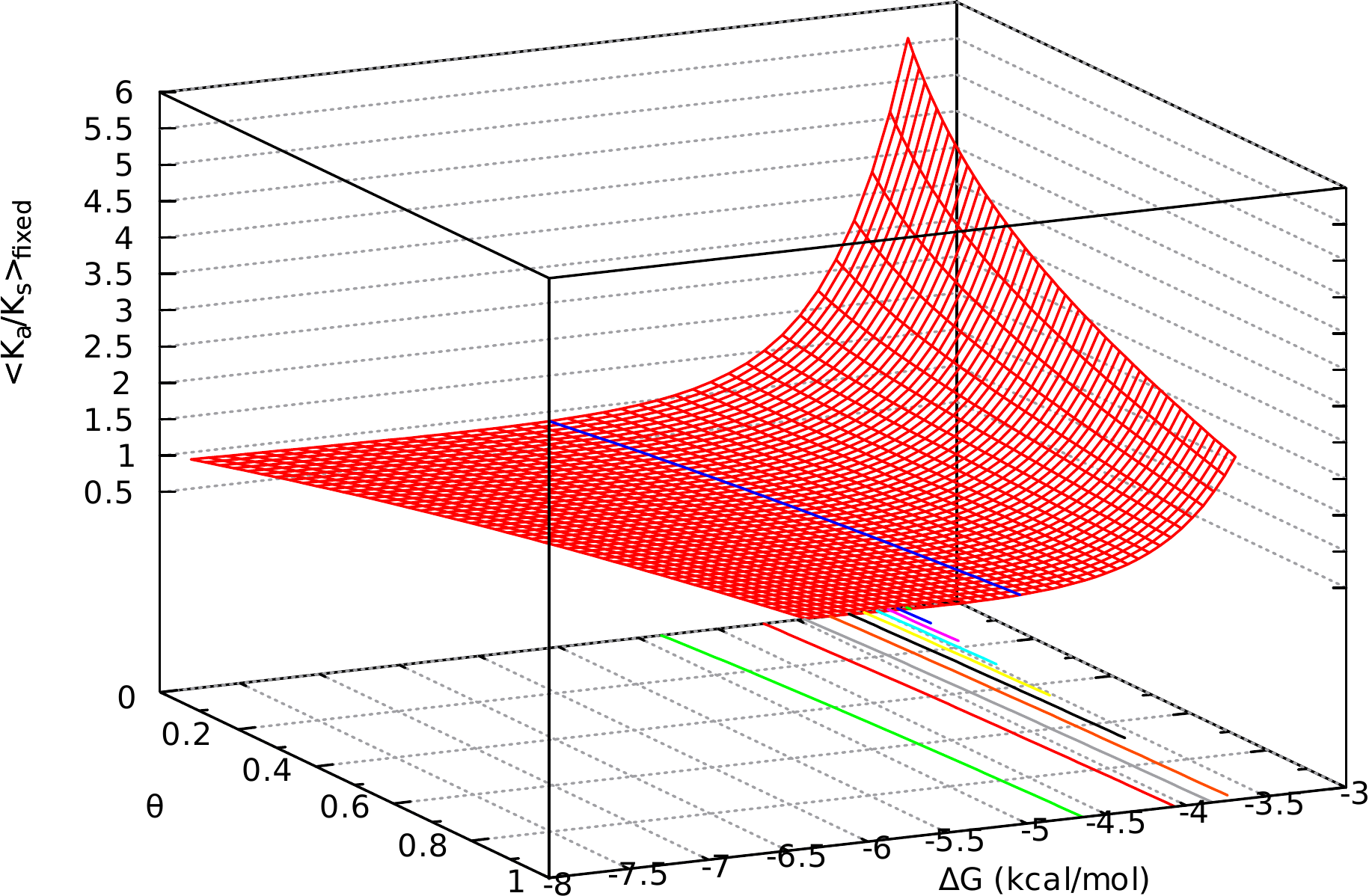}
}
} 
\FigureLegends{
\vspace*{1em}
\caption{
\label{fig: dependence_of_ave_Ka_over_Ks_on_dG}
\label{fig: dependence_of_ave_Ka_over_Ks_fixed_on_dG}
\BF{Dependence of the average of $K_a / K_s$ over all mutants or over fixed mutants only on 
protein stability, $\Delta G$, of the wild type.}
A blue line on the surface grid shows $\Delta G=\Delta G_e$, which is the equilibrium value of $\Delta G$
in protein evolution.
The range of $\Delta G$ shown in the figures is
$| \Delta G - \Delta G_e | < 2 \cdot \Delta\Delta G^{\script{sd}}_{\script{fixed}}$,
where $\Delta\Delta G^{\script{sd}}_{\script{fixed}}$ is the standard deviation of $\Delta\Delta G$ over fixed mutants at $\Delta G = \Delta G_e$.
Unless specified, 
$\log 4N_e\kappa = 7.55$ and $\theta = 0.53$
are employed.
The kcal/mol unit is used for $\Delta G$.
}
} 
\end{figure*}

\clearpage

} 

} 

\subsection{Lower bound of $K_a/K_s$ for adaptive substitutions on protein function}

The observed value of $K_a/K_s > 1$ is often used to 
indicate functional selection. 
The averages of $K_a/K_s$ over all mutants and even over 
fixed mutants are less than 1 as shown in
\Fig{\ref{fig: dependence_of_ave_Ka_over_Ks_on_4Nekappa_and_theta}}. 
Therefore, the average of $K_a/K_s$ over long time or many sites does not indicate
positive selection.
However, the probability of $K_a/K_s > 1$ is not negligible as shown in 
\Figs{\ref{fig: dependence_of_pdf_of_Ka_over_Ks_on_4Nekappa_and_theta} and 
\ref{fig: prob_of_each_selection_category_in_fixed_mutants}}.
Then, a question is how large $K_a/K_s$ due to selection on protein stability can be.

The distribution of $K_a/K_s$ significantly changes with $\Delta G$,
as shown in \Figs{\ref{sfig: pdf_of_Ka_over_Ks_around_dGe} and \ref{fig: dependence_of_ave_Ka_over_Ks_on_dG}}.
It may be appropriate to see the average of $K_a/K_s$, $\langle K_a/K_s \rangle_{\script{fixed}}$, 
in mutants fixed at $\Delta G > \Delta G_e$,
because the upper bound of $K_a/K_s$ becomes larger for $\Delta G > \Delta G_e$ than at the equilibrium,
and also positive mutants must fix to improve the protein stability of the wild type. 
\Fig{\ref{fig: dependence_of_ave_Ka_over_Ks_fixed_on_dG}}
shows that 
$\langle K_a/K_s \rangle_{\script{fixed}}$ can be very large
for proteins with low equilibrium stabilities (large $4N_e\kappa$ and small $\theta$),
although $\langle K_a/K_s \rangle_{\script{fixed}} \sim 1$ in $\Delta G < \Delta G_e$
in which nearly neutral and slightly negative selections are predominant;
1.7(1.2) at $\Delta G_e + \Delta\Delta G^{\script{sd}}_{\script{fixed}}$ and 6.1(5.6) at $\Delta G_e + 2 \cdot \Delta\Delta G^{\script{sd}}_{\script{fixed}}$ 
for $\log 4N_e\kappa = 20$ ($\theta = 0.0$),
where $\Delta\Delta G^{\script{sd}}_{\script{fixed}}$ means the standard deviation of $\Delta\Delta G$ in fixed mutants at $\Delta G_e$.
The 85 \% of fixed mutants have $\Delta\Delta G$ within the standard deviation.
Therefore, a lower bound for adaptive substitutions may be taken to be 
about 1.7, which is almost equal to the upper bound of $K_a/K_s$ at the equilibrium for $\log 4N_e\kappa = 20$ and $\theta=1$;
see \Fig{\ref{sfig: maximum_Ka_over_Ks_on_4Nekappa_and_theta}}.
However, as shown in \Fig{\ref{sfig: maximum_Ka_over_Ks_on_4Nekappa_and_theta}},
the more genes are expressed and/or the stronger structural constraints are, 
the larger the upper bound of $K_a/K_s$ at the equilibrium is.
Judging of adaptive changes 
may need not only $K_a/K_s > 1$ but also other supporting evidences;
such that substitutions are localized at specific sites.

\FigureInText{

\TextFig{

\FigureInLegends{\newpage}

\begin{figure*}[ht]
\FigureInLegends{
\centerline{
\includegraphics*[width=90mm,angle=0]{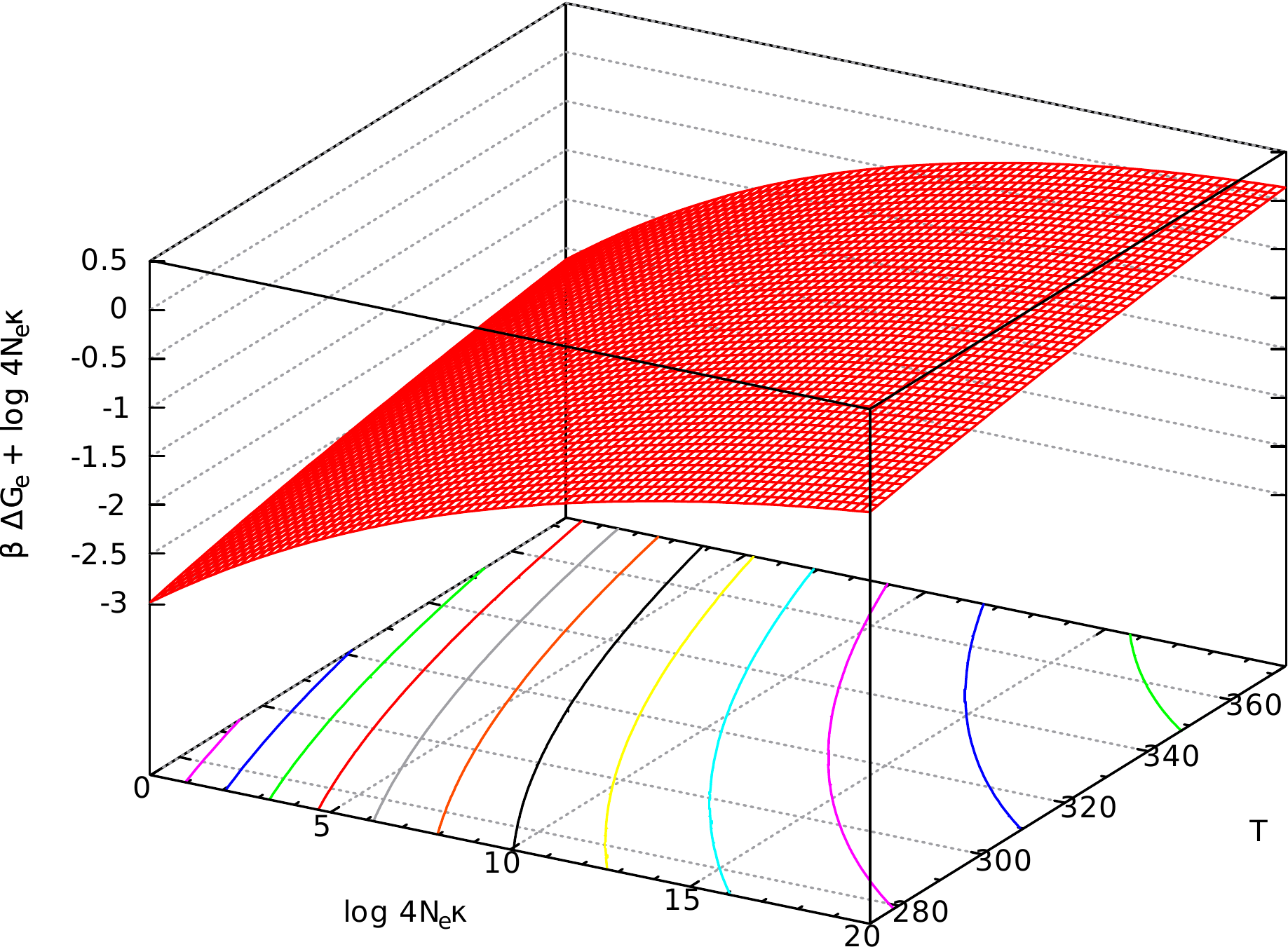}
\includegraphics*[width=90mm,angle=0]{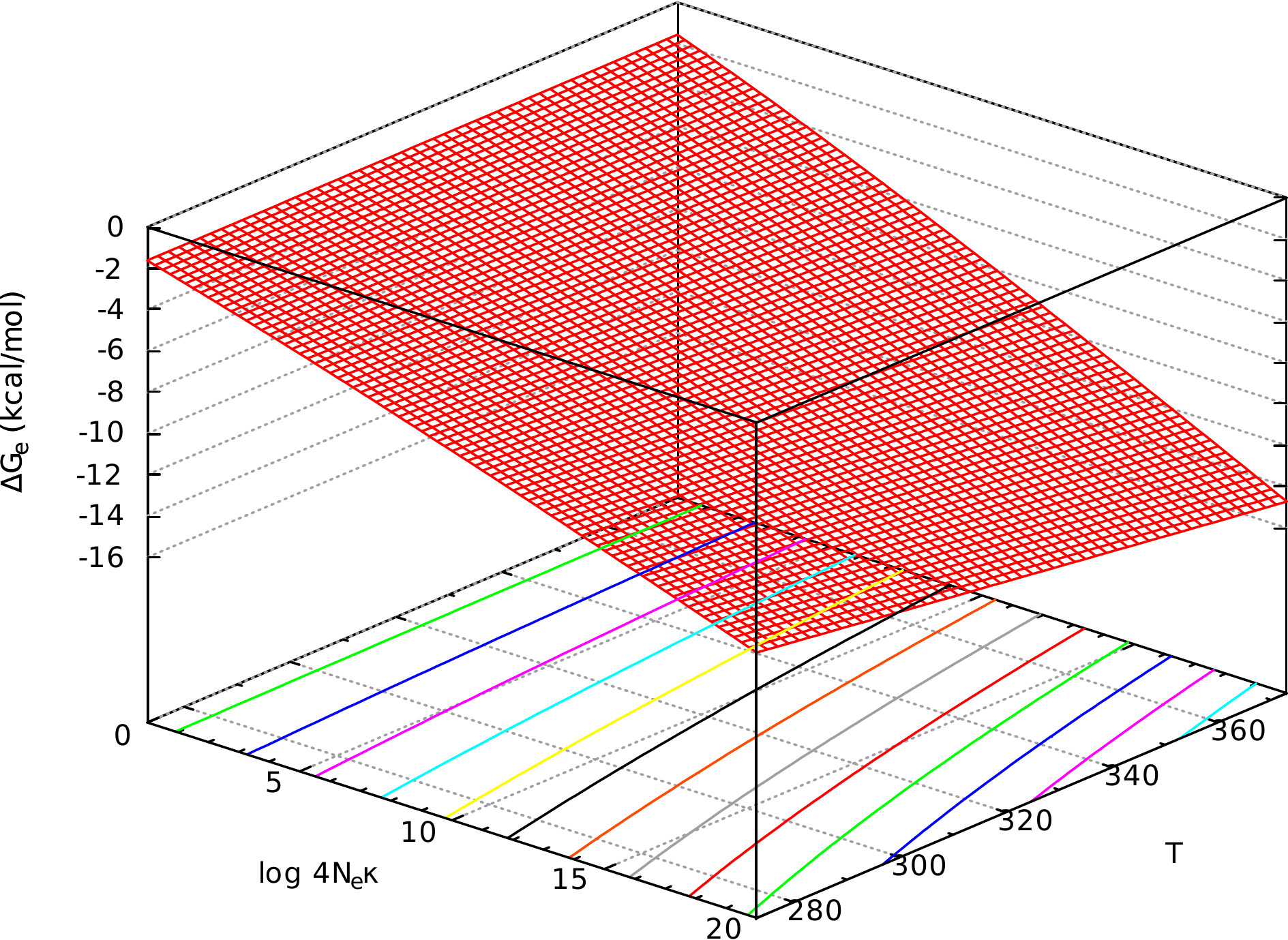}
}
} 
\FigureLegends{
\vspace*{1em}
\caption{
\label{fig: dependence_of_max4Nes_on_4Nekappa_and_T}
\label{fig: dependence_of_dGe_on_4Nekappa_and_T}
\Red{
\BF{Dependence of equilibrium stability, $\Delta G_e$, on parameters, $4N_e\kappa$ and $T$.}
} 
$\Delta G_e$ is the equilibrium value of folding free energy, $\Delta G$,
in protein evolution.
$T$ is absolute temperature; $\beta = 1 / kT$, where $k$ is the Boltzmann constant.
\EQUATIONS{\Ref{eq: def_bi-Gaussian}, \Ref{eq: def_ms} and \Ref{eq: def_mc}} are assumed for
the distribution of $\Delta\Delta G$ and its dependency on $\Delta G$; they are assumed to be independent of $T$.
$\theta = 0.53$
is employed.
The value of $\beta\Delta G_e + \log 4N_e\kappa$ is the upper bound of $\log 4N_e s$,
and would not depend on $\log 4N_e \kappa$ if the mean of $\Delta\Delta G$ in all arising mutants did not depend on $\Delta G$;
see \Eq{\ref{eq: def_s}}.
The kcal/mol unit is used for $\Delta G_e$.
}
} 
\end{figure*}

\clearpage

} 

} 

\ColorRed
\subsection{Dependences of protein stability ($\beta \Delta G_e$) and 
evolutionary rate ($\langle K_a/K_s \rangle$) on growth temperature}

It is natural that the folding free energies, $\Delta G_e$, of proteins
in organisms growing at higher temperatures must be lower than
those at the normal temperature, 
in order to attain the same stabilities and fitnesses as in the normal temperature.
\EQUATIONS{\Ref{eq: fraction_of_native_state} and \Ref{eq: def_Malthusian_fitness}} indicate that
the same stability and fitness will be attained if $\beta \Delta G_e$ is constant.
It means that it is sufficient for $\Delta G_e$ at the $100^{\circ}C$ 
to 
\RED{
decrease 
} 
$373/298 = 1.25$ times of that at the normal temperature ($25^{\circ}C$).
Is this a figure expected for folding free energies of thermophilic proteins at high growth temperature? 
It is not enough data of $\Delta G$ at high temperature in the ProTherm\CITE{KBGPKUS:06} to answer this question;
$\Delta G(T=75^{\circ}C) = 10.76$ kcal/mol for oxidized and $4.3$ for reduced CuA domain of cytochrome oxidase from Thermus thermophilus\CITE{WMSFWG:98} 
and $\Delta G(T=60^{\circ}C) = 13.01$ kcal/mol for pyrrolidone carboxyl peptidase from Pyrococcus\CITE{ONNTKYY:98}.
The present model indicates that $\beta \Delta G_e$ 
\RED{
slightly increases
} 
as growth temperature increases.

In \Figs{\ref{fig: dependence_of_max4Nes_on_4Nekappa_and_T} and \ref{sfig: dependence_of_max4Nes_on_theta_and_T}},
$\beta \Delta G_e + \log 4N_e \kappa$
is shown as a function of absolute temperature $T$ and $\log 4N_e \kappa$ or $\theta$,
assuming that the distribution of $\Delta\Delta G$ and its dependency on $\Delta G$ do not depend on $T$, 
that is, \Eqs{\Ref{eq: def_bi-Gaussian}, \Ref{eq: def_ms} and \Ref{eq: def_mc}}.
At fixed values of $\log 4N_e \kappa$ and $\theta$, $\beta \Delta G_e + \log 4N_e \kappa$ 
\RED{
increases
} 
as $T$ increases,
meaning that 
protein stability, $-\beta\Delta G$, 
\RED{
decreases 
} 
as growth temperature increases.
This tendency is 
\RED{
slightly
} 
larger at 
\RED{
smaller
} 
values of $\log 4N_e \kappa$, that is, for 
\RED{
less 
} 
abundant proteins.

The effects of growth temperature on $K_a/K_s$ are shown in 
\Fig{\ref{sfig: dependence_of_ave_Ka_over_Ks_on_4Nekappa_and_T}}.
The present model predicts that $\langle K_a/K_s \rangle$ decreases
as growth temperature increases unless any other parameter does not change.

\DefColor

\section{Discussion}

Recently, fitness costs due to misfolded proteins have been widely noticed,
particularly neurological disorder linked to misfolded protein toxicity\CITE{BGCBFZTRDS:02}.
Fitness costs that originate in functional loss\CITE{GDBWHD:11} and 
in diversion of protein synthesis and aggregation of proteins
have been evaluated\CITE{DW:08} to be related to the proportion of misfolded
proteins.  
Also, previous studies indicate that
factors that relate protein stability to protein fitness
are protein abundance, protein indispensability, and 
structural constraints of protein.
Current knowledge of protein folding can provide an exact
formulation for the proportion of misfolded proteins
as a function of folding free energy, 
and reasonable predictions\CITE{SBSNRS:05,YDD:07,TSSST:07} of stability changes 
due to single amino acid substitutions in protein native structures. 
Thus, on the basis of knowledge of protein biophysics 
it became possible to study 
the effects of amino acid substitutions on protein stability and 
then the evolution of protein\CITE{DW:08,SRS:12,SS:14,EJW:15,FK:15}.

Here, 
the effects of protein abundance and indispensability ($\kappa$) 
and of structural constraint ($\theta$) 
on protein evolutionary rate ($K_a/K_s$) have been examined in detail.
Both the effects are represented with different functional forms.
Structural constraints affect the distribution of 
stability change $\Delta\Delta G$ due to mutations.
On the other hand,
protein abundance/indispensability affects the effectiveness of 
stability change on protein fitness 
as well as the distribution of $\Delta\Delta G$.

The common understanding of protein evolution
has been that amino acid substitutions found in homologous proteins are
selectively neutral \CITE{K:68,K:69,KO:71,KO:74} or slightly deleterious
\CITE{O:73,O:92}, and random drift is a primary force to fix amino acid
substitutions in population.
However,
there is a selection maintaining protein stability at equilibrium 
\CITE{DW:08,SRS:12,SS:14}.
From the present analysis of the PDF of $K_a/K_s$, 
it has become clear how the equilibrium of stability is maintained;
see \Figs{\ref{fig: prob_of_each_selection_category_in_fixed_mutants;_dependence_on_kappa_and_dG} and
\ref{fig: prob_of_each_selection_category_in_fixed_mutants;_dependence_on_theta_and_dG}}.
In less-stable proteins of $\Delta G > \Delta G_e$,
more stabilizing mutations fix due to positive selection,
because negative shifts of $\Delta\Delta G$ increase stabilizing mutants
and also more amplify the effect of stability change on selective advantage; 
see \Eqs{\Ref{eq: def_ms}, \Ref{eq: def_mc} and \Ref{eq: def_s}}.
In more-stable proteins of $\Delta G < \Delta G_e$,
more destabilizing mutants 
are fixed by
random drift,
because positive shifts of $\Delta\Delta G$ increase destabilizing mutants
and also make more destabilizing mutants nearly neutral
with the less-amplified effect of stability change on selective advantage.
It has been revealed that
contrary to the neutral theory
nearly neutral selection is predominant only 
in low-abundant, non-essential proteins with $\log 4N_e\kappa < 2$ 
or with low equilibrium stability ($\Delta G_e > -2.5$ kcal/mol);
see \Fig{\ref{fig: prob_of_each_selection_category_in_fixed_mutants}}.

The average $\langle K_a/K_s \rangle$ 
and even $\langle K_a/K_s \rangle_{\script{fixed}}$
at equilibrium stability $\Delta G = \Delta G_e$
are less than one over the whole parameter range; see \Fig{\ref{fig: dependence_of_ave_Ka_over_Ks_on_4Nekappa_and_theta}}.
Hence, as far as selection is on protein stability,
the average of $K_a/K_s$ over a long time interval and over many sites
will be expected to be less than one, if all synonymous mutations are neutral\CITE{SW:15}.
However, because the probability of $K_a/K_s > 1$ is significant,
branches with $K_a/K_s > 1$ in phylogenetic trees may be observed,
as observed in a population dynamics simulation\CITE{SS:14}, 
even though synonymous mutations are neutral
and no adaptive selection operates on protein function.
According to the  present estimate, a lower bound of $K_a/K_s$ to indicate
adaptive substitutions
must be at least as large as 1.7.

Protein equilibrium stability ($\Delta G_e$) has been clearly 
described here as a function of $4N_e\kappa$ and $\theta$.
The more expressed a gene is (the larger $4N_e\kappa$ is), 
the stabler the wild-type protein at equilibrium is (the more negative $\Delta G_e$ becomes);
see \Fig{\ref{fig: dependence_of_dGe_on_4Nekappa_and_theta}}.
The decrease of $\Delta G_e$ 
shifts the distribution of $\Delta\Delta G$ toward the positive direction, 
generating more highly destabilizing mutants; see \Eqs{\Ref{eq: def_ms} and \Ref{eq: def_mc}}.
In addition,
as $4N_e\kappa$ increases,
the net effect, $4N_e\kappa \exp(\beta\Delta G_e)$, 
increases and more amplifies
the effects of stability changes ($\Delta\Delta G$) on selective advantage ($s$);
see \Fig{\ref{fig: dependence_of_dGe_on_4Nekappa_and_theta}} and \Eq{\ref{eq: def_s}}.
As a result, highly expressed and indispensable genes, and genes with a large effective population size 
evolve slowly; 
see \Fig{\ref{fig: dependence_of_ave_Ka_over_Ks_on_4Nekappa_and_theta}}.
However, if the distribution of $\Delta\Delta G$ did not depend on $\Delta G$,
$4N_e\kappa \exp(\beta\Delta G_e)$ would be constant, and
$K_a/K_s$ would not depend on $4N_e\kappa$, that is, protein abundance/indispensability 
and effective population size.

On the other hand, structural constraints on protein affect
protein evolutionary rate
by changing
the distribution of $\Delta\Delta G$ due to 
amino acid substitutions.
As shown in \Fig{\ref{fig: dependence_of_ave_Ka_over_Ks_on_4Nekappa_and_theta}},
at any value of $\log 4N_e\kappa$,
$\langle K_a/K_s \rangle$ decreases as
$\theta$ decreases.
In other word,
the more a protein is structurally constrained,
the more slowly it evolves, as claimed by Zuckerkandl\CITE{Z:76}.
\Fig{\ref{fig: dependence_of_ave_Ka_over_Ks_on_4Nekappa_and_theta}} shows that
the effect of protein abundance/indispensability on evolutionary rate
is more remarkable for less constrained proteins,
and the effect of structural constraint
is more remarkable for less abundant, less essential proteins.

In the result,
the average of $K_a/K_s$ over all arising mutants 
decreases roughly by $0.4--0.8$ as $\log 4N_e\kappa$ increases from 0 to 20;
see \Fig{\ref{fig: dependence_of_ave_Ka_over_Ks_on_4Nekappa_and_theta}}.
On the other hand, it decreases by $0.1--0.4$ as 
the proportion of the residues of the surface type, 
$\theta$, 
decreases from 1 to 0.
For monomeric, globular proteins, the proportion of protein surface may range from 0.7 to 0.45.
Thus, in typical globular proteins, 
protein abundance/indispensability may cause larger differences of evolutionary rate 
between proteins than structural constraint.
However, proteins that interact with other molecules 
on protein surface
effectively reduce residues of the protein-surface type \CITE{FX:09}.
Both protein abundance/indispensability and structural constraint must be taken into account
for protein evolutionary rate.

Protein abundance and indispensability
both affect evolutionary rate similarly through protein fitness.
It was shown in real proteins that
protein abundance correlates with evolutionary rate\CITE{PPH:01}. 
The present model of protein fitness (\Eq{\ref{eq: def_s}}) also indicates that
protein indispensability must correlate with evolutionary rate \CITE{HF:01,HF:03},
but a correlation between them 
may be hidden by the variation of protein abundance
and detected only in low-abundant proteins \CITE{PPH:03}; see \Eq{\ref{eq: def_kappa}}.
\Red{
In addition, effective population size must 
affect $\Delta G_e$ and $\langle K_a/K_s \rangle$
together with $\kappa$ as $4N_e\kappa$.
} 

In the present model,
protein equilibrium stability ($\Delta G_e$) and
evolutionary rate ($\langle K_a/K_s \rangle$) 
are predictable from $\theta$ and $4N_e \kappa$.  
The proportion of the surface type of residues may be estimated as 
those whose surface accessibility
values (ASA) are less than 0.25 \CITE{TSSST:07}, but 
experimental measurements of protein abundance, indispensability, 
and effective population size to determine $4N_e \kappa$ 
may be relatively hard.  
Instead the experimental value of
protein stability may be employed as equilibrium stability
to predict evolutionary rate and others, although it is not an independent variable.
\Fig{\ref{fig: dependence_of_ave_Ka_over_Ks_on_dGe_and_theta}}
shows evolutionary rate as a function of $\Delta G_e$ and $\theta$.
\Red{
Needless to say, mutational effects on $\Delta\Delta G$, 
such as $\theta$ and the distribution of $\Delta\Delta G$, 
must be well estimated for various categories of proteins\CITE{FK:15} 
to obtain successful predictions.
Also, accurate estimations of $\Delta G$ for various proteins 
are needed to examine the present predictions. 
It is interesting to examine if 
\RED{
protein stability ($-\beta\Delta G$)
} 
and $K_a/K_s$ 
\RED{
decrease 
} 
as growth temperature increases.
} 

\FigureInText{

\TextFig{

\begin{figure*}[ht]
\FigureInLegends{
\centerline{
\includegraphics*[width=90mm,angle=0]{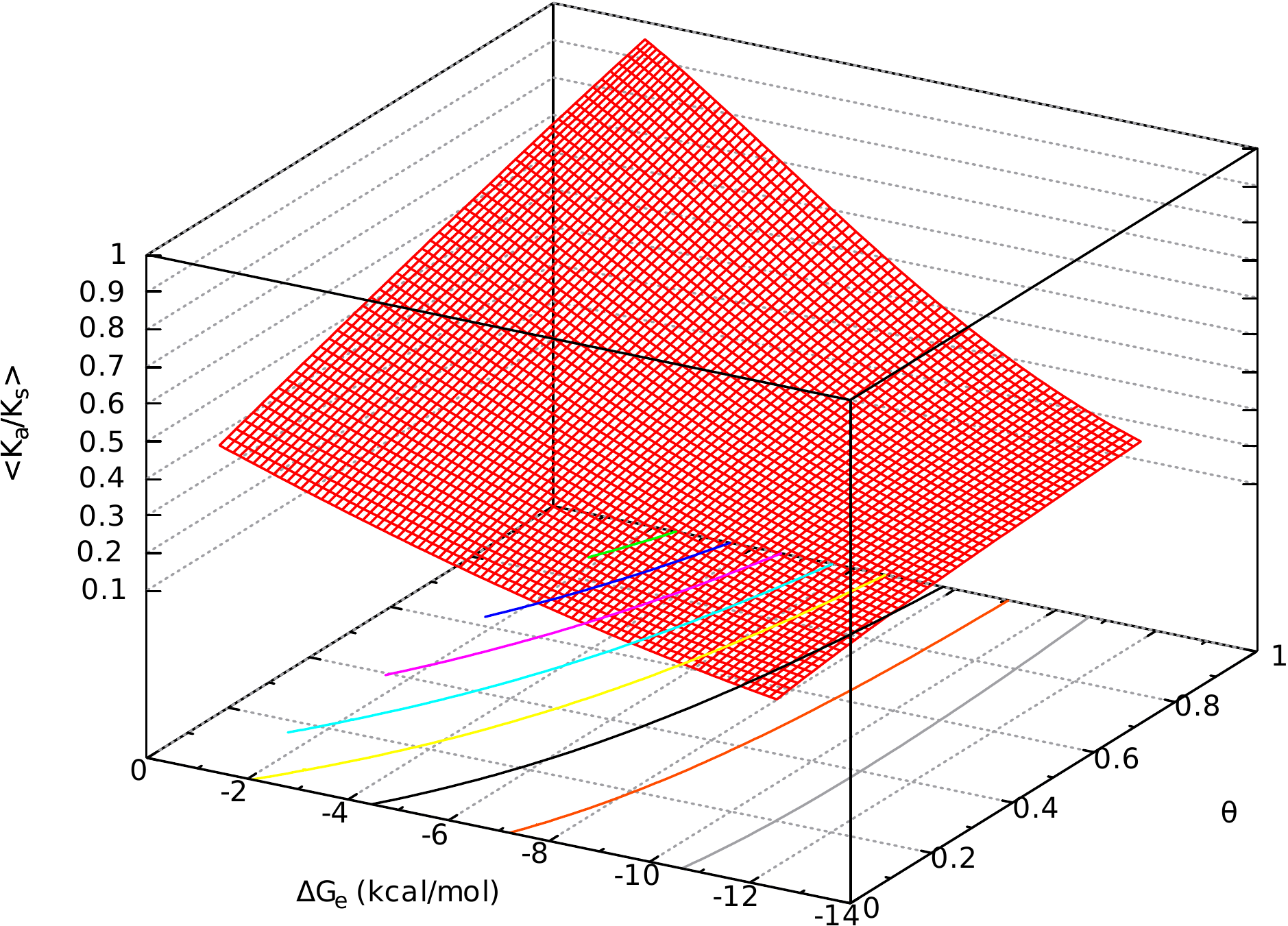}
}
} 
\FigureLegends{
\vspace*{1em}
\caption{
\label{fig: dependence_of_ave_Ka_over_Ks_on_dGe_and_theta}
\BF{The average of $K_a / K_s$ over all mutants as a function of $\Delta G_e$ and $\theta$.}
}
} 
\end{figure*}

\clearpage

} 

} 

\section{Conclusions}

\begin{itemize}
\item 
\Red{
The range, 
$-2$ to $-12.5$ kcal/mol,	
of equilibrium values, $\Delta G_e$, of protein stability
calculated with the present fitness model is consistent with 
the distribution of experimental values 
shown in \Fig{\ref{fig: observed_distribution_of_protein_stabilities}}.
} 

\item 
Contrary to the neutral theory,
nearly neutral selection is predominant only 
in low-abundant, non-essential proteins of $\log 4N_e\kappa < 2$ or $\Delta G_e > -2.5$ kcal/mol.
In the other proteins, positive selection on stabilizing mutations
\RED{
is
} 
significant to maintain protein stability at equilibrium 
as well as random drift on slightly negative mutations. 
However,
$\langle K_a/K_s \rangle$ and
even $\langle K_a/K_s \rangle_{\script{fixed}}$ at $\Delta G = \Delta G_e$ are less than 1.

\item
Protein abundance/indispensability ($\kappa$) and effective population size ($N_e$) 
more affect evolutionary rate for less constrained proteins, and
structural constraint ($1- \theta$) for less abundant, less essential proteins.

\item
Protein indispensability must negatively correlate with evolutionary rate 
like protein abundance, but the correlation between them
may be hidden by the variation of protein abundance
and detected only in low-abundant proteins.

\item
Evolutionary rates of proteins may be predicted from
equilibrium stability ($\Delta G_e$)
and structural constraints (PDF of $\Delta\Delta G$) 
of the protein.

\item
\Red{
The present model indicates that 
\RED{
protein stability ($-\beta\Delta G_e$) 
} 
and $\langle K_a/K_s \rangle$ decrease as growth temperature increases.
} 

\end{itemize}

\bibliographystyle{elsarticle-harv}
\bibliography{jnames_with_dots,MolEvol,Protein,Bioinfo,SM}

\begin{thebibliography}{45}
\expandafter\ifx\csname natexlab\endcsname\relax\def\natexlab#1{#1}\fi
\expandafter\ifx\csname url\endcsname\relax
  \def\url#1{\texttt{#1}}\fi
\expandafter\ifx\csname urlprefix\endcsname\relax\def\urlprefix{URL }\fi

\bibitem[{Bucciantini et~al.(2002)Bucciantini, Giannoni, Chiti, Baroni,
  Formigli, Zurdo, Taddei, Ramponi, Dobson, and Stefani}]{BGCBFZTRDS:02}
Bucciantini, M., Giannoni, E., Chiti, F., Baroni, F., Formigli, L., Zurdo, J.,
  Taddei, N., Ramponi, G., Dobson, C.~M., Stefani, M., 2002. Inherent toxicity
  of aggregates implies a common mechanism for protein misfolding diseases.
  Nature 416, 507--511.

\bibitem[{Crow and Kimura(1970)}]{CK:70}
Crow, J.~F., Kimura, M., 1970. An Introduction to population genetics theory.
  Harper \& Row publishers, New York.

\bibitem[{Dasmeh et~al.(2014)Dasmeh, Serohijos, Kepp, and
  Shakhnovich}]{DSKS:14}
Dasmeh, P., Serohijos, A.~W., Kepp, K.~P., Shakhnovich, E.~I., 2014. The
  influence of selection for protein stability on {dN/dS} estimations. Genome
  Biol. Evol. 6, 2956--2967.

\bibitem[{Drummond et~al.(2005)Drummond, Bloom, Adami, Wilke, and
  Arnold}]{DBAWA:05}
Drummond, D.~A., Bloom, J.~D., Adami, C., Wilke, C.~O., Arnold, F.~H., 2005.
  Why highly expressed proteins evolve slowly. Proc. Natl. Acad. Sci. USA
  102~(40), 14338--14343.

\bibitem[{Drummond and Wilke(2008)}]{DW:08}
Drummond, D.~A., Wilke, C.~O., 2008. Mistranslation-induced protein misfolding
  as a dominant constraint on coding-sequence evolution. Cell 134~(2), 341 --
  352.
\newline\urlprefix\url{http://dx.doi.org/10.1016/j.cell.2008.05.042}

\bibitem[{Duret and Mouchiro(2000)}]{DM:00}
Duret, L., Mouchiro, D., 2000. Determinants of substitution rates in mammalian
  genes: expression pattern affects selection intensity but not mutation rate.
  Mol. Biol. Evol. 17, 68--85.

\bibitem[{Echave et~al.(2015)Echave, Jackson, and Wilke}]{EJW:15}
Echave, J., Jackson, E.~L., Wilke, C.~O., 2015. Relationship between protein
  thermodynamic constraints and variation of evolutionary rates among sites.
  Phys. Biol. 12~(2), 025002.
\newline\urlprefix\url{http://stacks.iop.org/1478-3975/12/i=2/a=025002}

\bibitem[{Faure and Koonin(2015)}]{FK:15}
Faure, G., Koonin, E.~V., 2015. Universal distribution of mutational effects on
  protein stability, uncoupling of protein robustness from sequence evolution
  and distinct evolutionary modes of prokaryotic and eukaryotic proteins. Phys.
  Biol. 12~(3), 035001.
\newline\urlprefix\url{http://stacks.iop.org/1478-3975/12/i=3/a=035001}

\bibitem[{Franzosa and Xia(2009)}]{FX:09}
Franzosa, E.~A., Xia, Y., 2009. Structural determinants of protein evolution
  are context-sensitive at the residue level. Mol. Biol. Evol. 26, 2387--2395.

\bibitem[{Fraser et~al.(2002)Fraser, Hirsh, Steinmetz, Scharfe, and
  Feldman}]{FHSSF:02}
Fraser, H.~B., Hirsh, A.~E., Steinmetz, L.~M., Scharfe, C., Feldman, M.~W.,
  2002. Evolutionary rate in the protein interaction network. Science 296,
  750--752.

\bibitem[{Geiler-Samerotte et~al.(2011)Geiler-Samerotte, Dion, Budnik, Wang,
  Hartl, and Drummond}]{GDBWHD:11}
Geiler-Samerotte, K.~A., Dion, M.~F., Budnik, B.~A., Wang, S.~M., Hartl, D.~L.,
  Drummond, D.~A., 2011. Misfolded proteins impose a dosage-dependent fitness
  cost and trigger a cytosolic unfolded protein response in yeast. Proc. Natl.
  Acad. Sci. USA 108, 680--685.

\bibitem[{Ghaemmaghami et~al.(2003)Ghaemmaghami, Huh, Bower, Howson, Belle,
  Dephoure, \'{O}Shea, and Weissman}]{GHBHBDOW:03}
Ghaemmaghami, S., Huh, W.-K., Bower, K., Howson, R.~W., Belle, A., Dephoure,
  N., \'{O}Shea, E.~K., Weissman, J.~S., 2003. Global analysis of protein
  expression in yeast. Nature 425, 737--741.

\bibitem[{Go and Miyazawa(1980)}]{GM:80}
Go, M., Miyazawa, S., 1980. Relationship between mutability, polarity and
  exteriority of amino acid residues in protein evolution. Int. J. Peptide
  Protein Res. 15, 211--224.

\bibitem[{Hirsh and Fraser(2001)}]{HF:01}
Hirsh, A.~E., Fraser, H.~B., 2001. Protein dispensability and rate of
  evolution. Nature 411, 1047--1049.

\bibitem[{Hirsh and Fraser(2003)}]{HF:03}
Hirsh, A.~E., Fraser, H.~B., 2003. Genomic function (communication arising):
  Rate of evolution and gene dispensability. Nature 421, 497--498.

\bibitem[{Jordan et~al.(2002)Jordan, Rogozin, Wolf, and Koonin}]{JRWK:02}
Jordan, I.~K., Rogozin, I.~B., Wolf, Y.~I., Koonin, E.~V., 2002. Essential
  genes are more evolutionarily conserved than are nonessential genes in
  bacteria. Genome Res. 12, 962--968.

\bibitem[{Kimura(1968)}]{K:68}
Kimura, M., 1968. Evolutionary rate at the molecular level. Nature 217,
  624--626.

\bibitem[{Kimura(1969)}]{K:69}
Kimura, M., 1969. The rate of molecular evolution considered from the
  standpoint of population genetics. Proc. Natl. Acad. Sci. USA 63, 1181--1188.

\bibitem[{Kimura and Ohta(1971)}]{KO:71}
Kimura, M., Ohta, T., 1971. Protein polymorphism as a phase of molecular
  evolution. Nature 229, 467--469.

\bibitem[{Kimura and Ohta(1974)}]{KO:74}
Kimura, M., Ohta, T., 1974. On some principles governing molecular evolution.
  Proc. Natl. Acad. Sci. USA 71, 2848--2852.

\bibitem[{Kuma et~al.(1995)Kuma, Iwabe, and Miyata}]{KIM:95}
Kuma, K., Iwabe, N., Miyata, T., 1995. Functional constraints against
  variations on molecules from the tissue level: slowly evolving brain-specific
  genes demonstrated by protein kinase and immunoglobulin supergene families.
  Mol. Biol. Evol. 12, 123--130.

\bibitem[{Kumar et~al.(2006)Kumar, Bava, Gromiha, Prabakaran, Kitajima,
  Uedaira, and Sarai}]{KBGPKUS:06}
Kumar, M., Bava, K., Gromiha, M., Prabakaran, P., Kitajima, K., Uedaira, H.,
  Sarai, A., 2006. {ProTherm} and {ProNIT}: thermodynamic databases for
  proteins and protein-nucleic acid interactions. Nucl. Acid Res. 34,
  D204--D206.

\bibitem[{Lynch and Conery(2003)}]{LC:03}
Lynch, M., Conery, J.~S., 2003. The origins of genome complexity. Science 302,
  1401--1404.

\bibitem[{Miyata and Yasunaga(1980)}]{MY:80}
Miyata, T., Yasunaga, T., 1980. Molecular evolution of m{RNA}: a method for
  estimating evolutionary rates of synonymous and amino acid substitutions from
  homologous nucleotide sequences and its applications. J. Mol. Evol. 16,
  23--36.

\bibitem[{Miyazawa and Jernigan(1982{\natexlab{a}})}]{MJ:82}
Miyazawa, S., Jernigan, R.~L., 1982{\natexlab{a}}. Equilibrium folding and
  unfolding pathways for a model protein. Biopolymers 21, 1333--1363.

\bibitem[{Miyazawa and Jernigan(1982{\natexlab{b}})}]{MJ:82B}
Miyazawa, S., Jernigan, R.~L., 1982{\natexlab{b}}. Most probable intermediates
  in protein folding-unfolding with a non-interacting globule-coil model.
  Biochemistry 21, 5203--5213.

\bibitem[{Ogasahara et~al.(1998)Ogasahara, Nakamura, Nakura, Tsunasawa, Kato,
  Yoshimoto, and Yutani}]{ONNTKYY:98}
Ogasahara, K., Nakamura, M., Nakura, S., Tsunasawa, S., Kato, I., Yoshimoto,
  T., Yutani, K., 1998. The unusually slow unfolding rate causes the high
  stability of pyrrolidone carboxyl peptidase from a hyperthermophile,
  {Pyrococcus} furiosus: equilibrium and kinetic studies of guanidine
  hydrochloride-induced unfolding and refolding. Biochemistry 37, 17537--17544.

\bibitem[{Ohta(1973)}]{O:73}
Ohta, T., 1973. Slightly deleterious mutant substitutions in evolution. Nature
  246, 96--98.

\bibitem[{Ohta(1992)}]{O:92}
Ohta, T., 1992. The nearly neutral theory of molecular evolution. Annu. Rev.
  Ecol. Syst. 23, 263--286.

\bibitem[{P\'{a}l et~al.(2001)P\'{a}l, Papp, and Hurst}]{PPH:01}
P\'{a}l, C., Papp, B., Hurst, L.~D., 2001. Highly expressed genes in yeast
  evolve slowly. Genetics 158, 927--931.

\bibitem[{P\'{a}l et~al.(2003)P\'{a}l, Papp, and Hurst}]{PPH:03}
P\'{a}l, C., Papp, B., Hurst, L.~D., 2003. Genomic function (communication
  arising): Rate of evolution and gene dispensability. Nature 421, 496--497.

\bibitem[{Schymkowitz et~al.(2005)Schymkowitz, Borg, Stricher, Nys, Rousseau,
  and Serrano}]{SBSNRS:05}
Schymkowitz, J., Borg, J., Stricher, F., Nys, R., Rousseau, F., Serrano, L.,
  2005. The {FoldX} web server: an online force field. Nucl. Acid Res. 33,
  W382--W388.

\bibitem[{Serohijos et~al.(2012)Serohijos, Rimas, and Shakhnovich}]{SRS:12}
Serohijos, A., Rimas, Z., Shakhnovich, E., 2012. Protein biophysics explains
  why highly abundant proteins evolve slowly. Cell Reports 2~(2), 249 -- 256.
\newline\urlprefix\url{http://dx.doi.org/10.1016/j.celrep.2012.06.022}

\bibitem[{Serohijos et~al.(2013)Serohijos, S. Y. {Ryan Lee}, and
  Shakhnovich}]{SLS:13}
Serohijos, A., S. Y. {Ryan Lee}, Shakhnovich, E., 2013. Highly abundant
  proteins favor more stable 3{D} structures in yeast. Biophys. J.
  104~(3), L1 -- L3.
\newline\urlprefix\url{http://dx.doi.org/10.1016/j.bpj.2012.11.3838}

\bibitem[{Serohijos and Shakhnovich(2014)}]{SS:14}
Serohijos, A.~W., Shakhnovich, E.~I., 2014. Contribution of selection for
  protein folding stability in shaping the patterns of polymorphisms in coding
  regions. Mol. Biol. Evol. 31, 165--176.

\bibitem[{Spielman and Wilke(2015)}]{SW:15}
Spielman, S.~J., Wilke, C.~O., 2015. The relationship between {dN/dS} and
  scaled selection coefficients. Mol. Biol. Evol. 32, 1097--1108.

\bibitem[{Stoebel et~al.(2008)Stoebel, Dean, and Dykhuizen}]{SDD:08}
Stoebel, D.~M., Dean, A.~M., Dykhuizen, D.~E., 2008. The cost of expression of
  {Escherichia} coli lac operon proteins ss in the process, not in the product.
  Genetics 178, 1653--1660.

\bibitem[{Tokuriki et~al.(2007)Tokuriki, Stricher, Schymkowitz, Serrano, and
  Tawfik}]{TSSST:07}
Tokuriki, N., Stricher, F., Schymkowitz, J., Serrano, L., Tawfik, D.~S., 2007.
  The stability effects of protein mutations appear to be universally
  distributed. J. Mol. Biol. 369, 1318--1332.

\bibitem[{Wall et~al.(2005)Wall, Hirsh, Fraser, Kumm, Giaever, Eisen, and
  Feldman}]{WHFKGEF:05}
Wall, D.~P., Hirsh, A.~E., Fraser, H.~B., Kumm, J., Giaever, G., Eisen, M.~B.,
  Feldman, M.~W., 2005. Functional genomic analysis of the rates of protein
  evolution. Proc. Natl. Acad. Sci. USA 102, 5483--5488.

\bibitem[{Wittung-Stafshede et~al.(1998)Wittung-Stafshede, Malmstrom, Sanders,
  Fee, Winkler, and Gray}]{WMSFWG:98}
Wittung-Stafshede, P., Malmstrom, B.~G., Sanders, D., Fee, J.~A., Winkler,
  J.~R., Gray, H.~B., 1998. Effect of redox state on the folding free energy of
  a thermostable electron-transfer metalloprotein: the {CuA} domain of
  cytochrome oxidase from thermus thermophilus. Biochemistry 37, 3172--3177.

\bibitem[{Yang et~al.(2012)Yang, Liao, Zhuang, and Zhang}]{YLZZ:12}
Yang, J.-R., Liao, B.-Y., Zhuang, S.-M., Zhang, J., 2012. Protein
  misinteraction avoidance causes highly expressed proteins to evolve slowly.
  Proc. Natl. Acad. Sci. USA 109, E831--E840.

\bibitem[{Yin et~al.(2007)Yin, Ding, and Dokholyan}]{YDD:07}
Yin, S., Ding, F., Dokholyan, N.~V., 2007. Eris: an automated estimator of
  protein stability. Nature Methods 4, 466--467.

\bibitem[{Zeldovich et~al.(2007)Zeldovich, Chen, and Shakhnovich}]{ZCS:07}
Zeldovich, K.~B., Chen, P., Shakhnovich, E.~I., 2007. Protein stability imposes
  limits on organism complexity and speed of molecular evolution. Proc. Natl.
  Acad. Sci. USA 104, 16152--16157.

\bibitem[{Zhang and He(2005)}]{ZH:05}
Zhang, J., He, X., 2005. Significant impact of protein dispensability on the
  instantaneous rate of protein evolution. Mol. Biol. Evol. 22, 1147--1155.

\bibitem[{Zuckerkandl(1976)}]{Z:76}
Zuckerkandl, E., 1976. Evolutionary processes and evolutionary noise at the
  molecular level. J. Mol. Evol. 7, 167--183.

\end{thebibliography}







\clearpage

\NoFigureInText{

\newpage
\section*{Figure Legends}

\renewcommand{\TextFig}[1]{#1}
\renewcommand{\SupFig}[1]{}

\renewcommand{\FigureInLegends}[1]{}
\renewcommand{\FigureLegends}[1]{#1}
\setcounter{figure}{0}

\SupFig{

\FigureInLegends{\clearpage\newpage}

\begin{figure*}[ht]
\FigureInLegends{
\centerline{
\includegraphics*[width=90mm,angle=0]{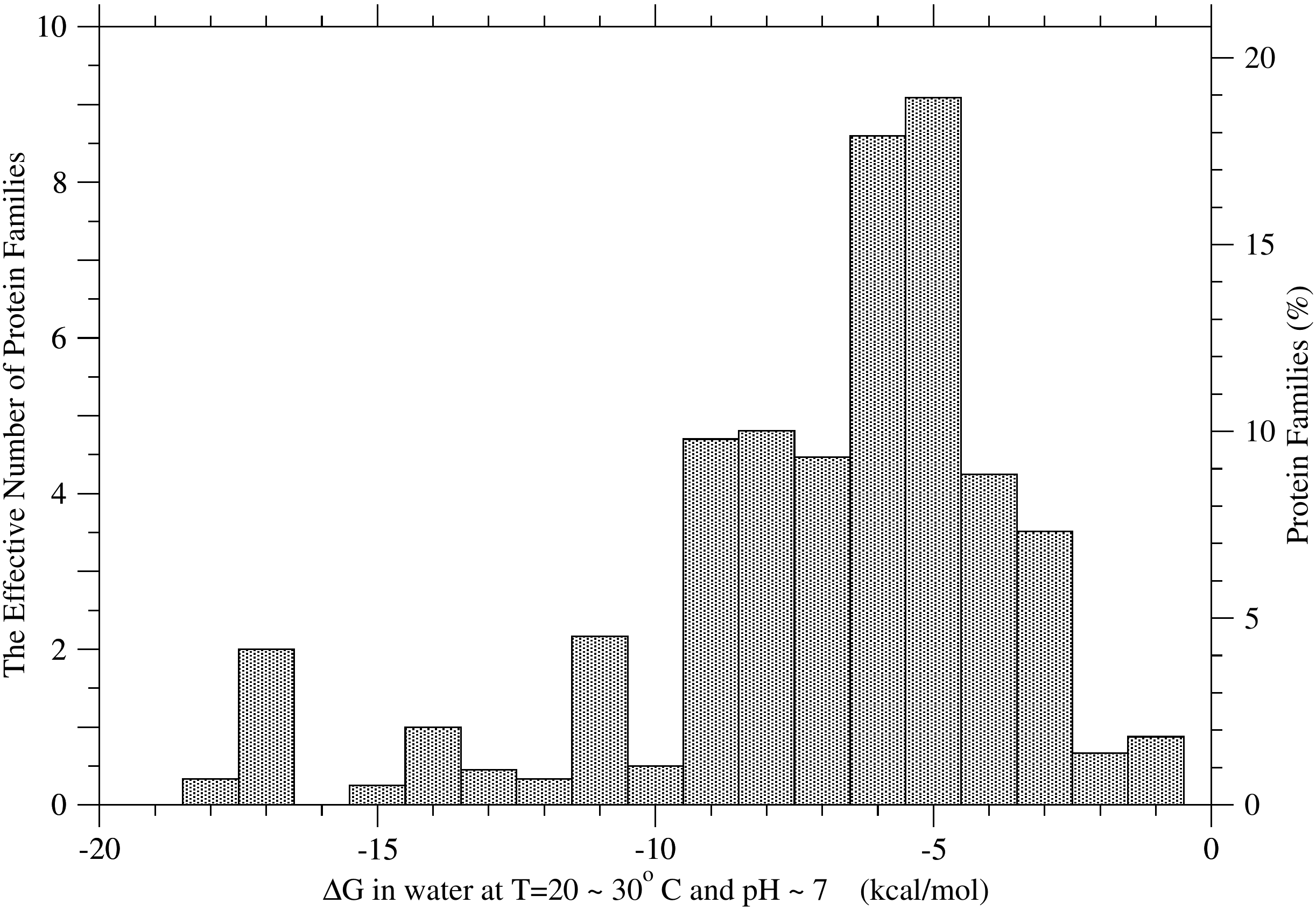}
}
} 
\vspace*{1em}
\caption{
\label{sfig: observed_distribution_of_protein_stabilities}
\FigureLegends{
\Red{
\BF{Distribution of folding free energies of monomeric protein families.}
} 
Stability data of monomeric proteins for which the item of dG\_H2O or dG
was obtained in the experimental condition of $6.7 \leq \textrm{pH} \leq 7.3$ 
and $20^{\circ}C \leq T \leq 30^{\circ}C$ and their folding-unfolding transition is two state and reversible
are extracted from the ProTherm\CITE{KBGPKUS:06};
in the case of dG only thermal transition data are used. 
Thermophilic proteins, and proteins observed with salts or additives are also removed.
An equal sampling weight is assigned to each species of homologous protein,
and the total sampling weight of each protein family is normalized to
one.  In the case in which multiple data exist for the same species of protein, 
its sampling weight is divided to each of the data. 
However, proteins whose stabilities are known 
may be samples biased from the protein universe.
The value, $\Delta G_e = -5.24$ kcal/mol, of equilibrium stability at the representative parameter values, $\log 4N_e\kappa = 7.550$
and $\theta = 0.53$, agrees with the most probable value of $\Delta G$ in the distribution above.
Also, the range of $\Delta G$ shown above is consistent with
that range, $-2$ to $-12.5$ kcal/mol, expected from the present model.
The kcal/mol unit is used for $\Delta G$.
A similar distribution was also compiled \CITE{ZCS:07}.
} 
}
\end{figure*}

\FigureInLegends{\clearpage\newpage}

\begin{figure*}[ht]
\FigureInLegends{
\centerline{
\includegraphics*[width=90mm,angle=0]{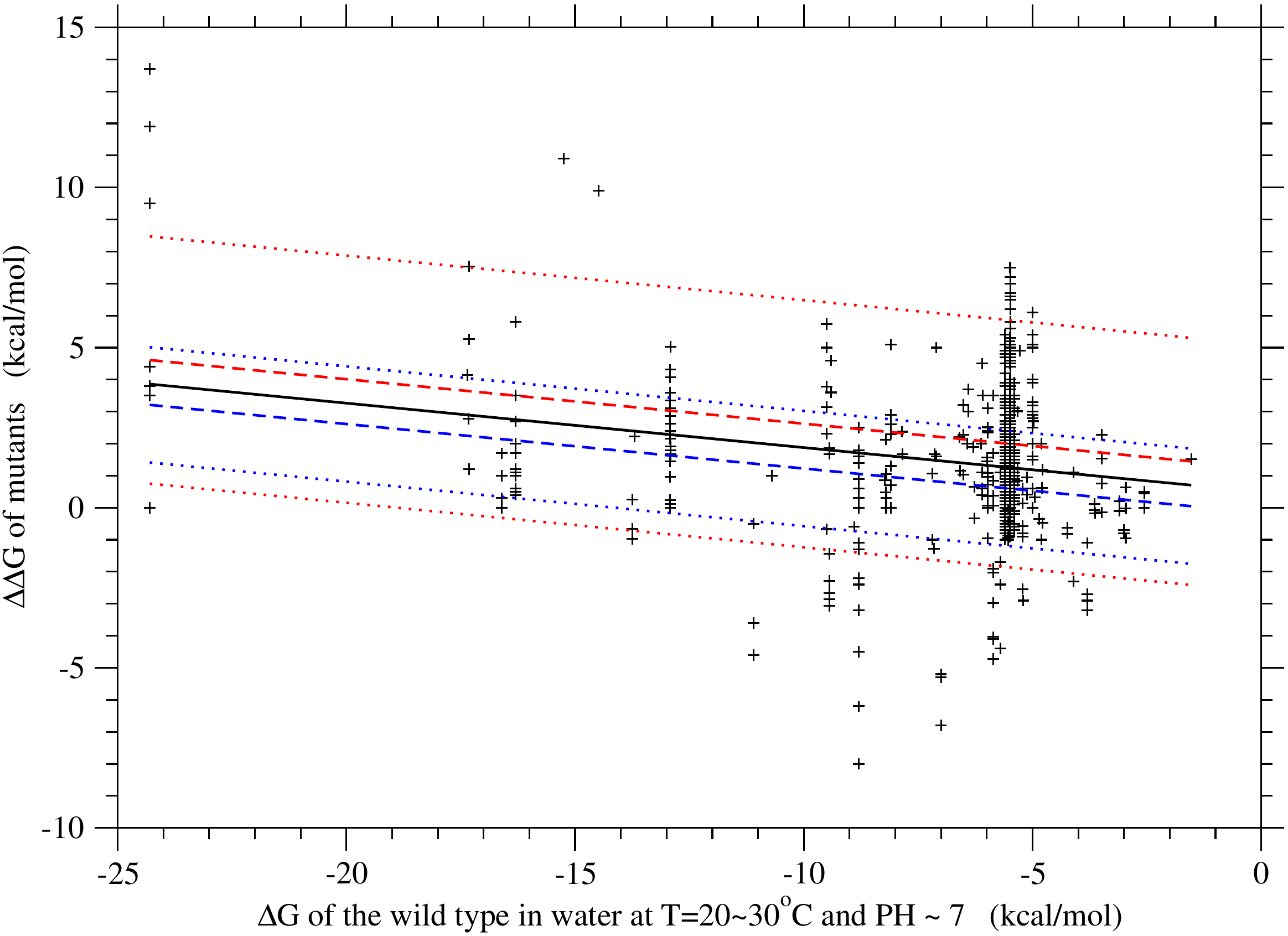}
}
} 
\vspace*{1em}
\caption{
\label{sfig: ddG_vs_dG}
\FigureLegends{
\Red{
\BF{Dependence of stability changes, $\Delta\Delta G$, due to single amino acid substitutions
on the protein stability, $\Delta G$, of the wild type.}
} 
A solid line shows the regression line, 
$\Delta\Delta G = -0.139 \Delta G + 0.490$; 
the correlation coefficient and p-value are equal to $-0.20$ and $< 10^{-7}$, respectively.
Broken lines show two means of bi-Gaussian distributions, $\mu_s$ in blue and $\mu_c$ in red.
Blue dotted lines show $\mu_s \pm 2 \sigma_s$ and red dotted lines $\mu_c \pm 2 \sigma_c$.
See \Eqs{\Ref{seq: def_bi-Gaussian}, \Ref{seq: def_ms} and \Ref{seq: def_mc}} for the bi-Gaussian distribution.
Stability data of single amino acid mutants for which the items dG\_H2O and ddG\_H2O or dG and ddG 
were obtained in the experimental condition of 
$6.7 \leq \textrm{pH} \leq 7.3$ and $20^{\circ}C \leq T \leq 30^{\circ}C$
and their folding-unfolding transitions are two state and reversible
are extracted from the ProTherm\CITE{KBGPKUS:06}.
In the case of dG only thermal transition data are used.
In the case in which multiple data exist for the same protein, only one of them
is used.
The kcal/mol unit is used for $\Delta\Delta G$ and $\Delta G$.
A similar distribution was also compiled \CITE{SRS:12}.
} 
}
\end{figure*}

\FigureInLegends{\newpage}

\begin{figure*}[ht]
\FigureInLegends{
\centerline{
\includegraphics*[width=90mm,angle=0]{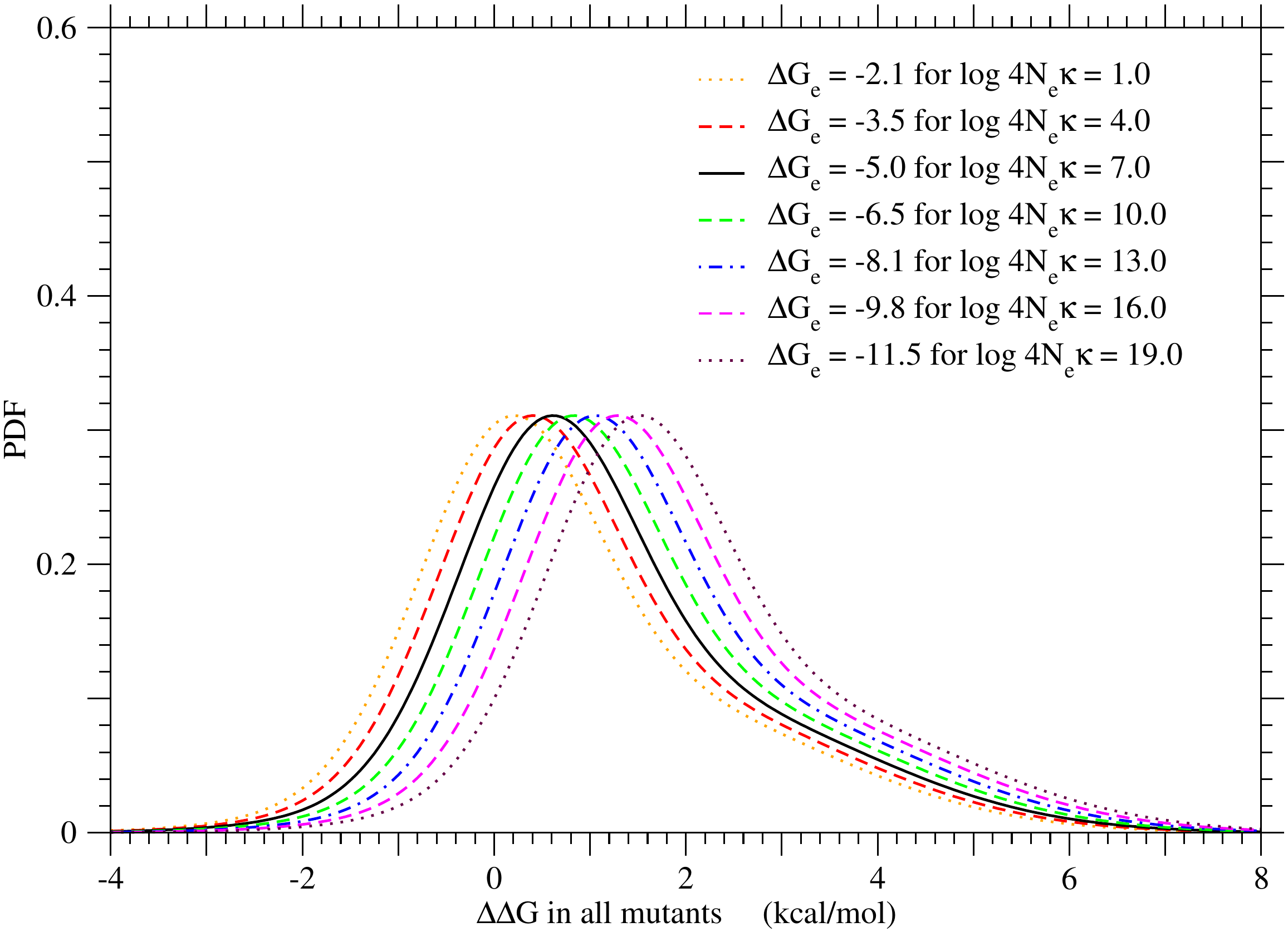}
\includegraphics*[width=90mm,angle=0]{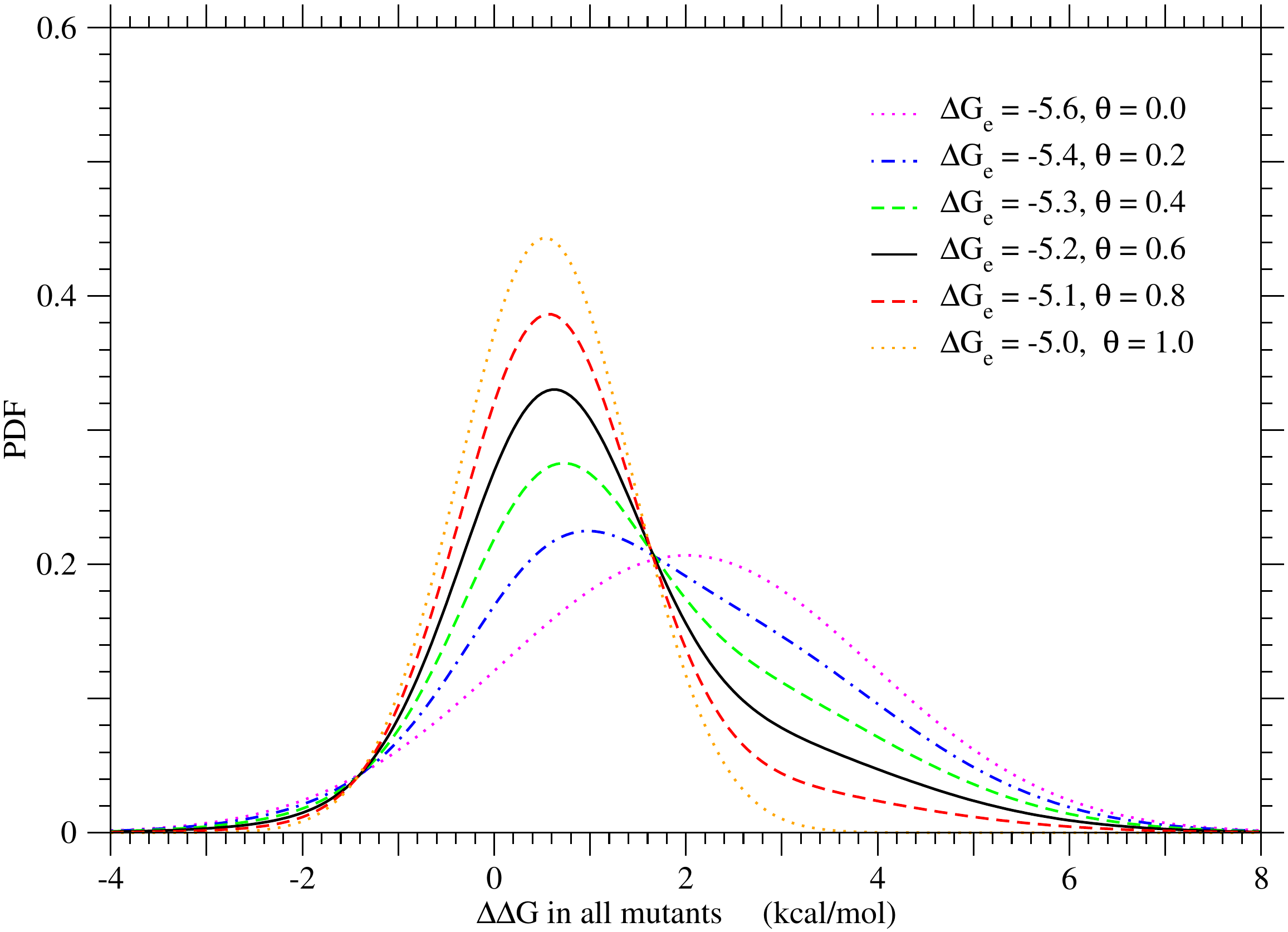}
}
\centerline{
\includegraphics*[width=90mm,angle=0]{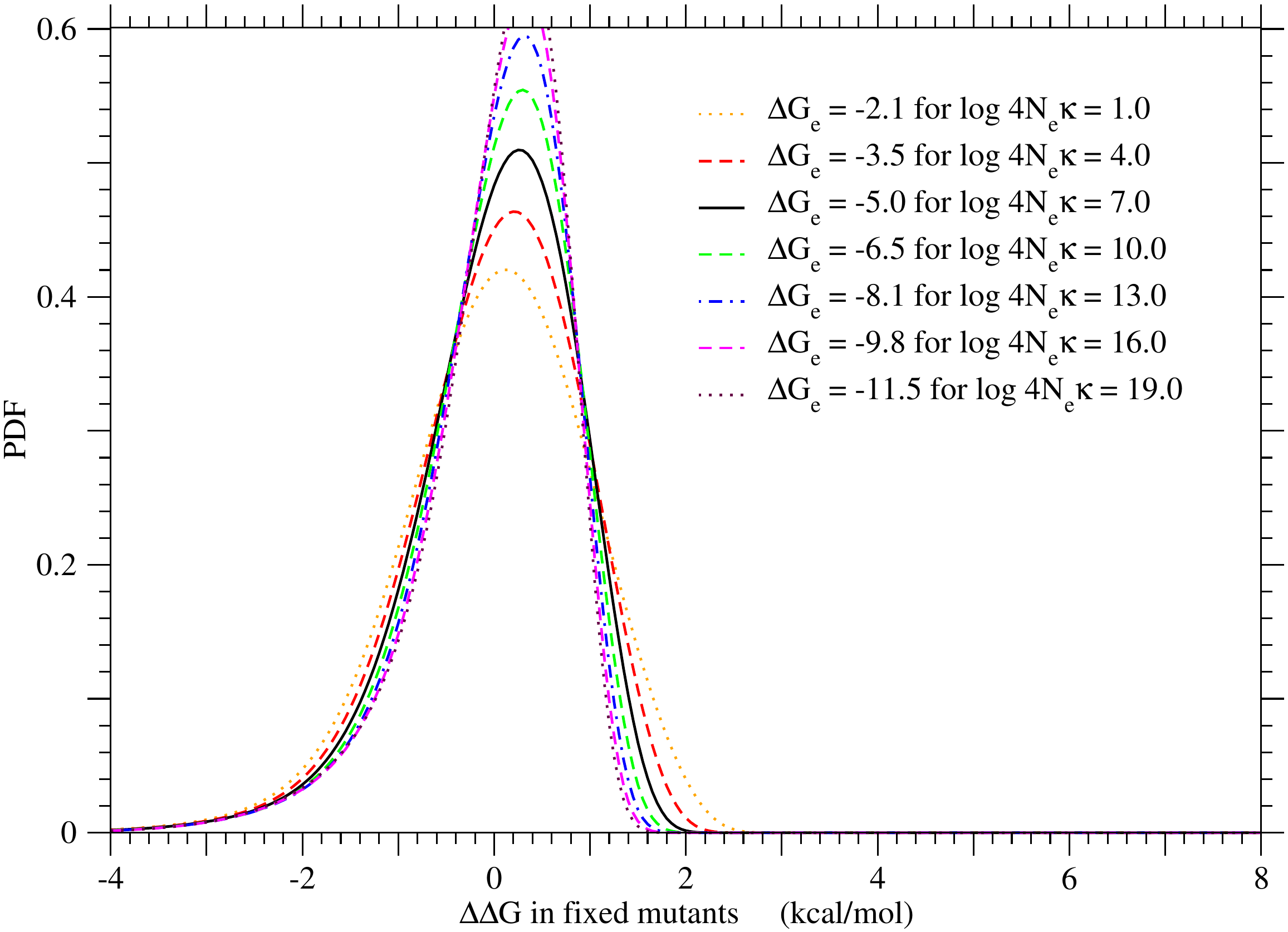}
\includegraphics*[width=90mm,angle=0]{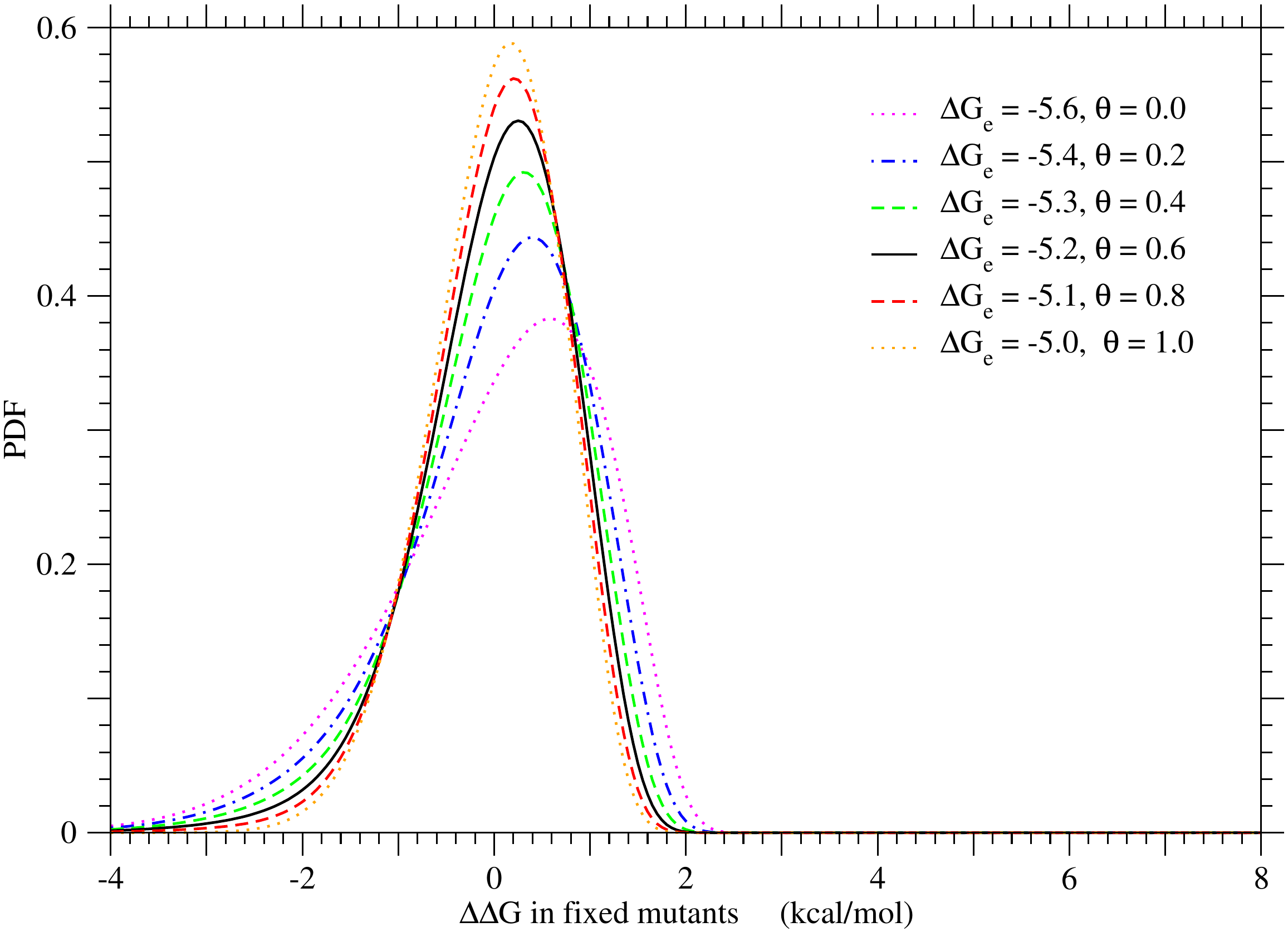}
}
} 
\vspace*{1em}
\caption{
\label{sfig: pdf_of_ddG_at_dGe}
\label{sfig: pdf_of_ddG_fixed_at_dGe}
\label{sfig: pdf_of_ddG}
\FigureLegends{
\BF{PDFs of stability changes, $\Delta\Delta G$, due to single amino acid substitutions
in all mutants and in fixed mutants
at equilibrium of protein stability, $\Delta G = \Delta G_e$.}
The PDF of $\Delta\Delta G$ due to single amino acid substitutions
in all arising mutants is assumed to be bi-Gaussian; see \Eq{\ref{seq: def_bi-Gaussian}}.
Unless specified, 
$\log 4N_e\kappa = 7.550$ and $\theta = 0.53$
are employed.
The kcal/mol unit is used for $\Delta\Delta G$ and $\Delta G_e$.
} 
}
\end{figure*}

\FigureInLegends{\newpage}

\begin{figure*}[ht]
\FigureInLegends{
\centerline{
\includegraphics*[width=90mm,angle=0]{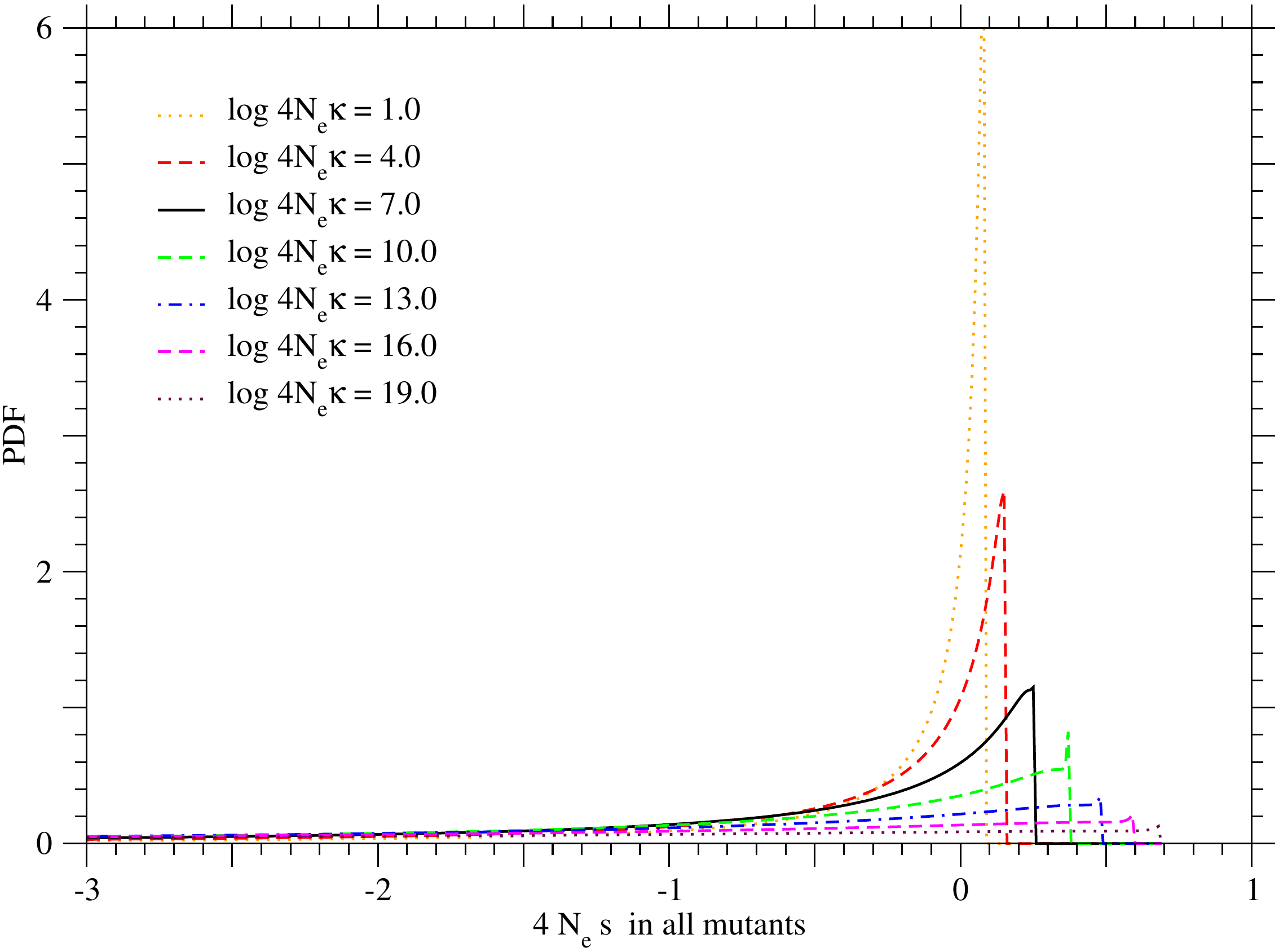}
\includegraphics*[width=90mm,angle=0]{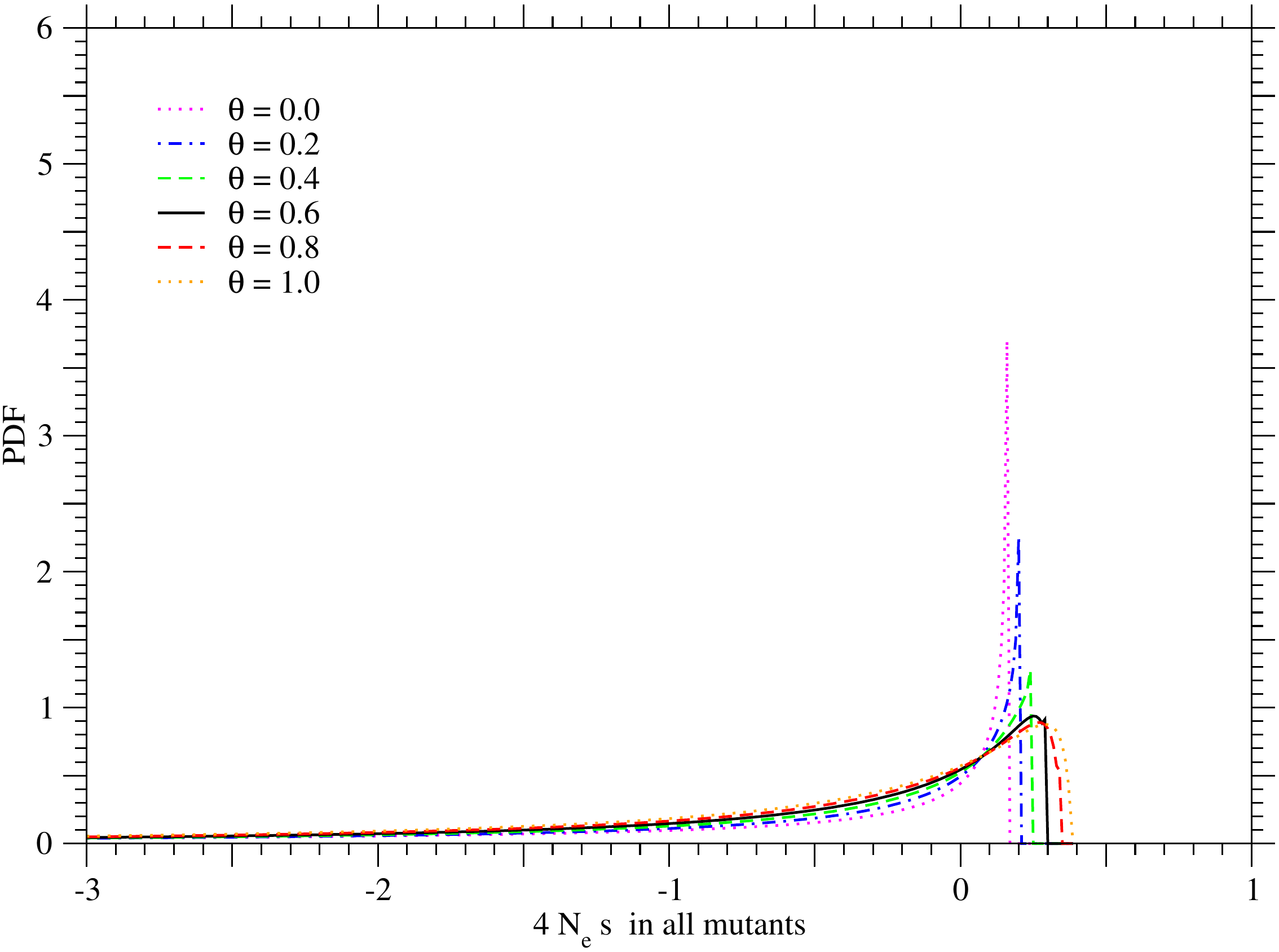}
}
\centerline{
\includegraphics*[width=90mm,angle=0]{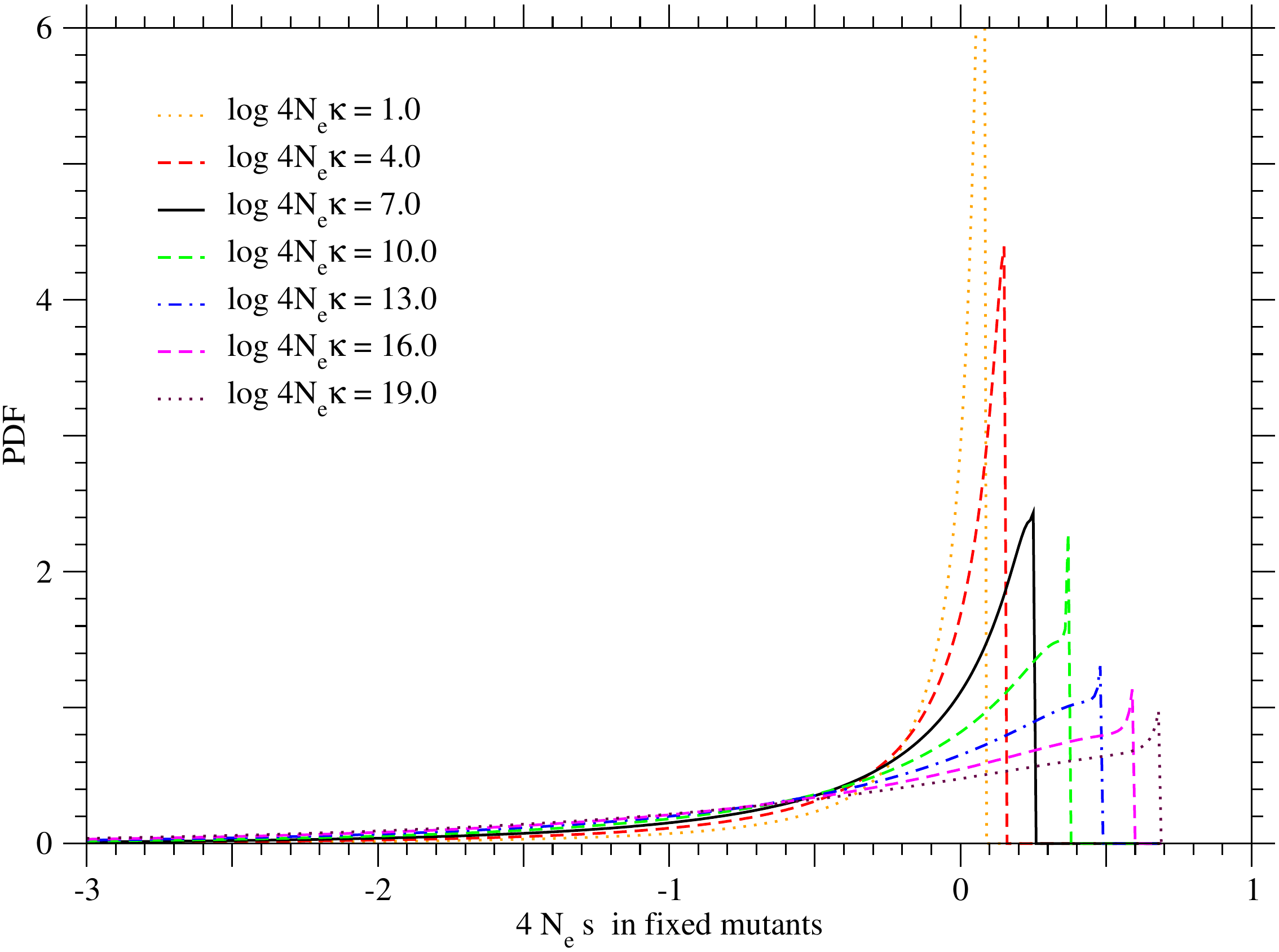}
\includegraphics*[width=90mm,angle=0]{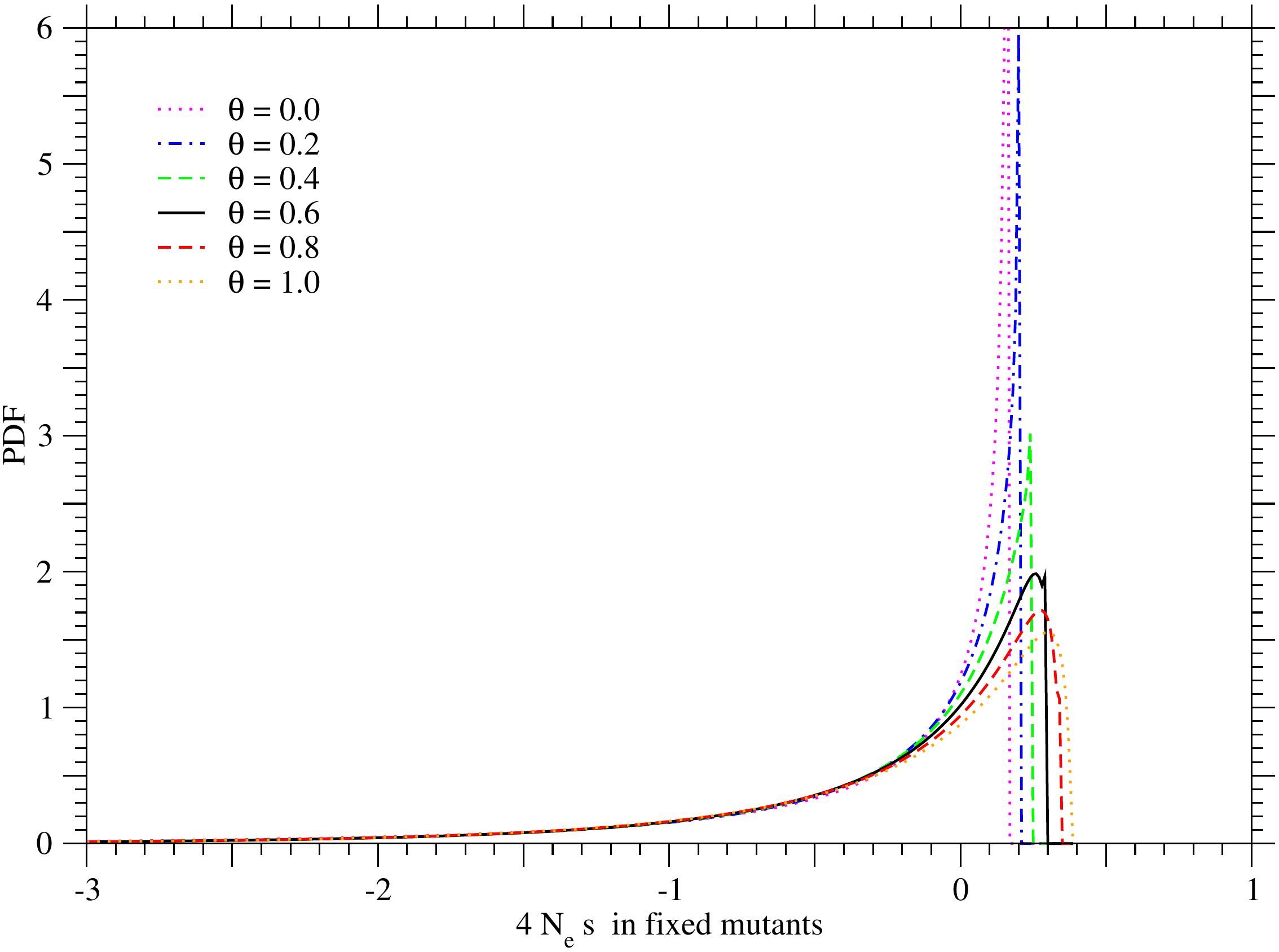}
}
} 
\vspace*{1em}
\caption{
\label{sfig: pdf_of_4sNe_at_dGe}
\label{sfig: pdf_of_4sNe}
\label{sfig: pdf_of_4sNe_fixed}
\label{sfig: pdf_of_4sNe_fixed_at_dGe}
\FigureLegends{
\BF{PDFs of $4N_e s$ in all mutants and in fixed mutants at equilibrium of protein stability, $\Delta G = \Delta G_e$.}
Unless specified, 
$\log 4N_e\kappa = 7.550$ and $\theta = 0.53$
are employed.
} 
}
\end{figure*}

\FigureInLegends{\newpage}

\begin{figure*}[ht]
\FigureInLegends{
\centerline{
\includegraphics*[width=90mm,angle=0]{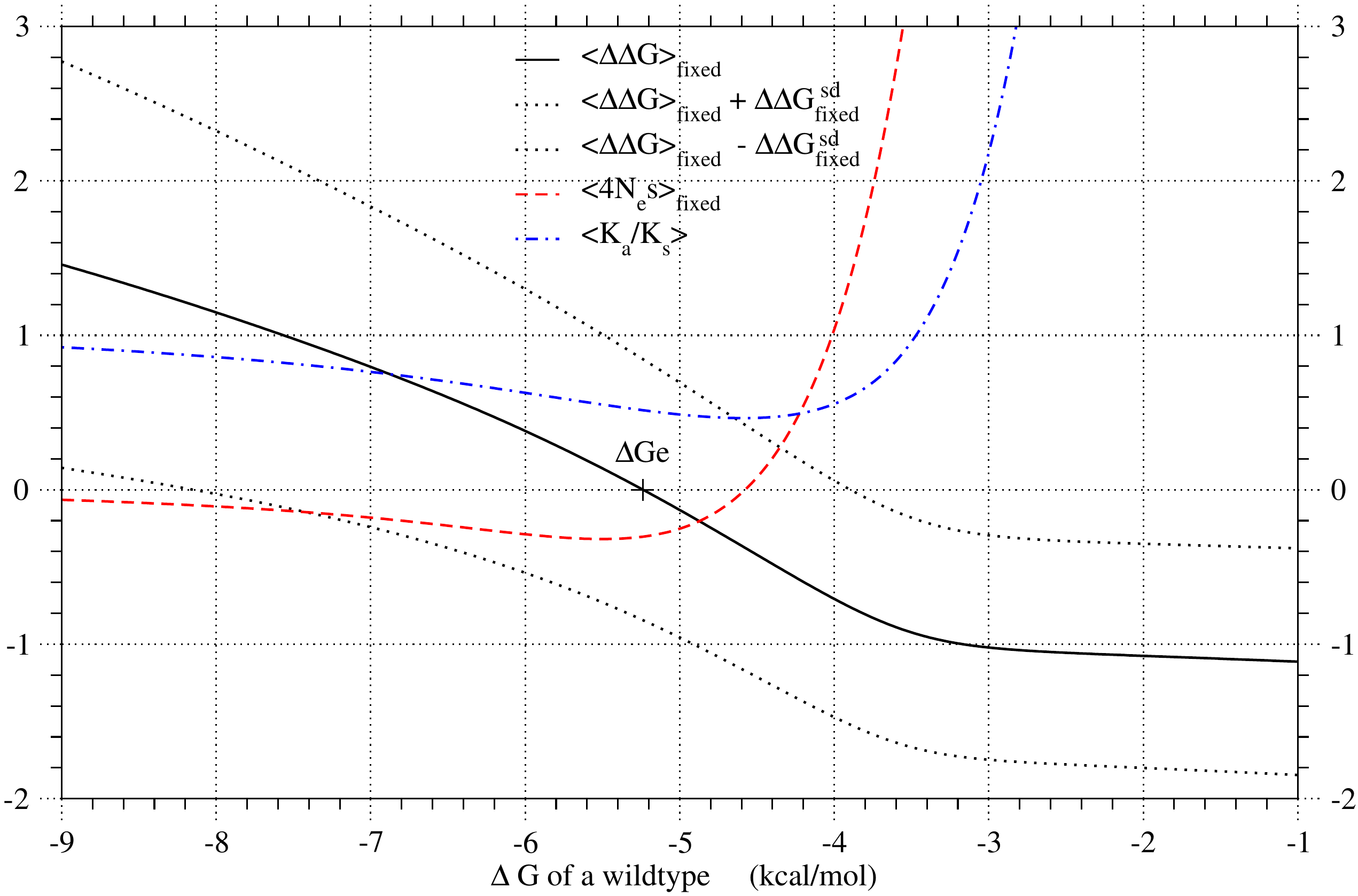}
}
} 
\vspace*{1em}
\caption{
\label{sfig: ave_ddG_vs_dG}
\label{sfig: ave_ddG_vs_dG_fixed}
\FigureLegends{
\BF{The average, $\langle \Delta\Delta G \rangle_{\script{fixed}}$, of stability changes over fixed mutants versus protein stability, $\Delta G$, of the wild type.}
$\Delta G_e$, where $\langle \Delta\Delta G \rangle = 0$, is the stable equilibrium value
of folding free energy, $\Delta G$, in protein evolution. 
The averages of $\Delta\Delta G$, $4N_e s$, and $K_a/Ks$ over fixed mutants are
plotted against protein stability, $\Delta G$, of the wild type by solid, broken, and dash-dot lines,
respectively.
Thick dotted lines show the values of $\langle \Delta\Delta G \rangle_{\script{fixed}} \pm \Delta\Delta G^{\script{sd}}_{\script{fixed}}$,
where $\Delta\Delta G^{\script{sd}}_{\script{fixed}}$ is the standard deviation of $\Delta\Delta G$ over fixed mutants.
$\log 4N_e\kappa = 7.550$ and $\theta = 0.53$
are employed.
The kcal/mol unit is used for $\Delta\Delta G$.
} 
}
\end{figure*}

\FigureInLegends{\newpage}

\begin{figure*}[ht]
\FigureInLegends{
\centerline{
\includegraphics*[width=90mm,angle=0]{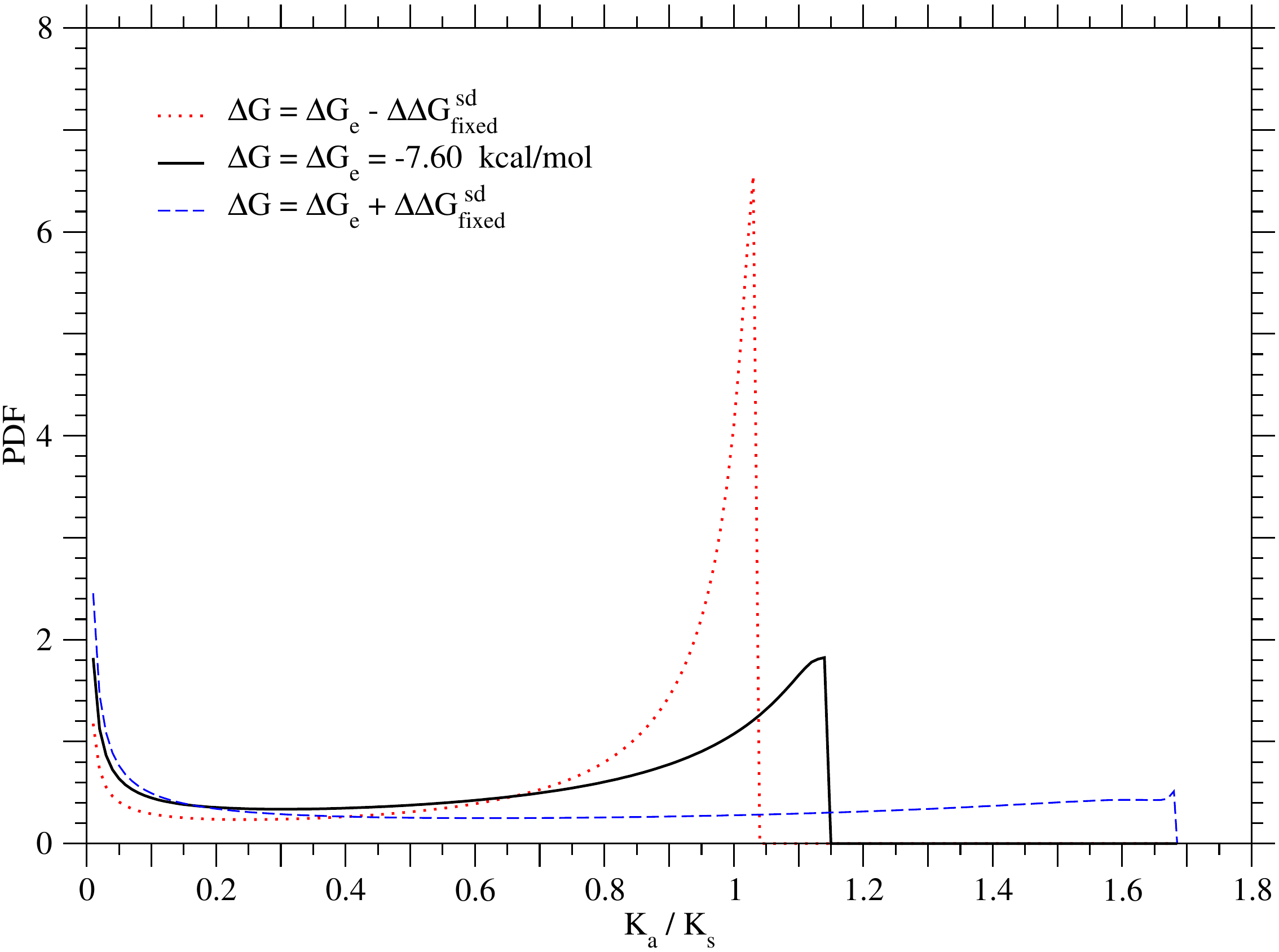}
}
\centerline{
\includegraphics*[width=90mm,angle=0]{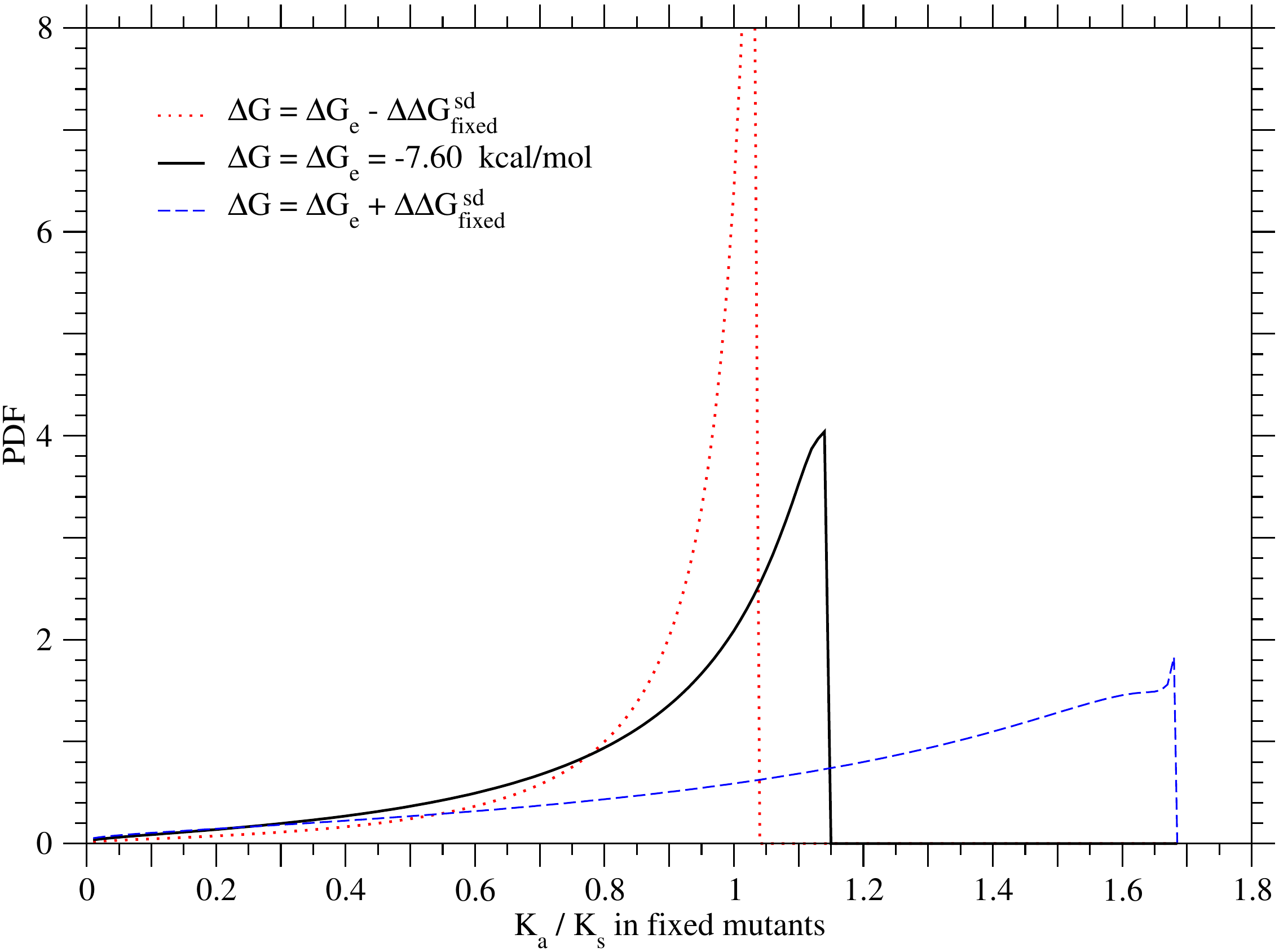}
}
} 
\vspace*{1em}
\caption{
\label{sfig: pdf_of_Ka_over_Ks_around_dGe}
\label{sfig: pdf_of_Ka_over_Ks_fixed_around_dGe}
\FigureLegends{
\BF{Dependence of the PDF of $K_a/K_s$ on protein stability, $\Delta G$, of the wild type in all mutants or in fixed mutants only.}
$\Delta\Delta G^{\script{sd}}_{\script{fixed}}$ is the standard deviation 
(0.84 kcal/mol)	
of $\Delta\Delta G$
over fixed mutants at $\Delta G = \Delta G_e$.
$\log 4N_e\kappa = 7.550$ and $\theta = 0.53$
are employed.
The kcal/mol unit is used for $\Delta G_e$.
} 
}
\end{figure*}

\FigureInLegends{\clearpage\newpage}

\begin{figure*}[ht]
\FigureInLegends{
\centerline{
\includegraphics*[width=90mm,angle=0]{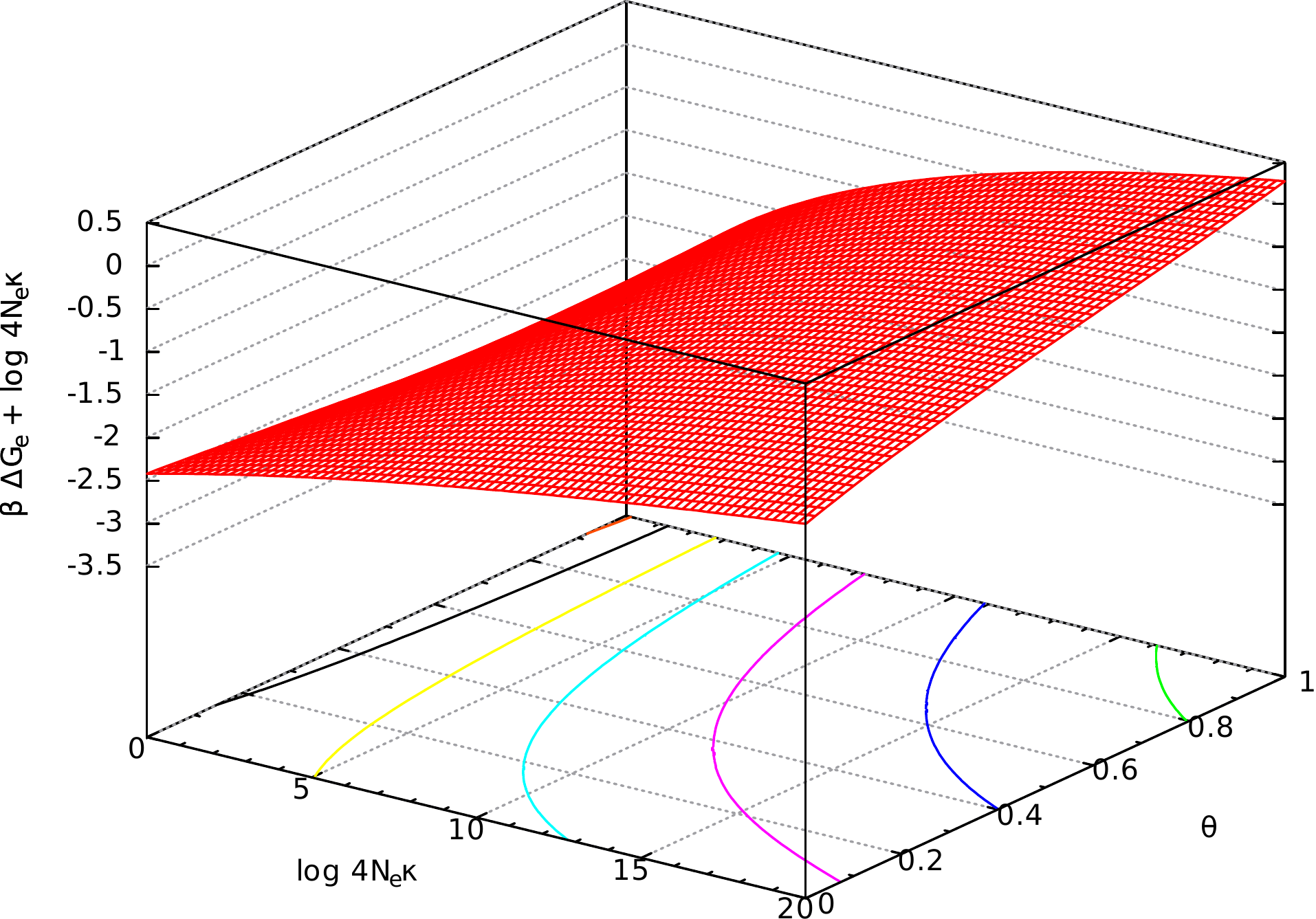}
\includegraphics*[width=90mm,angle=0]{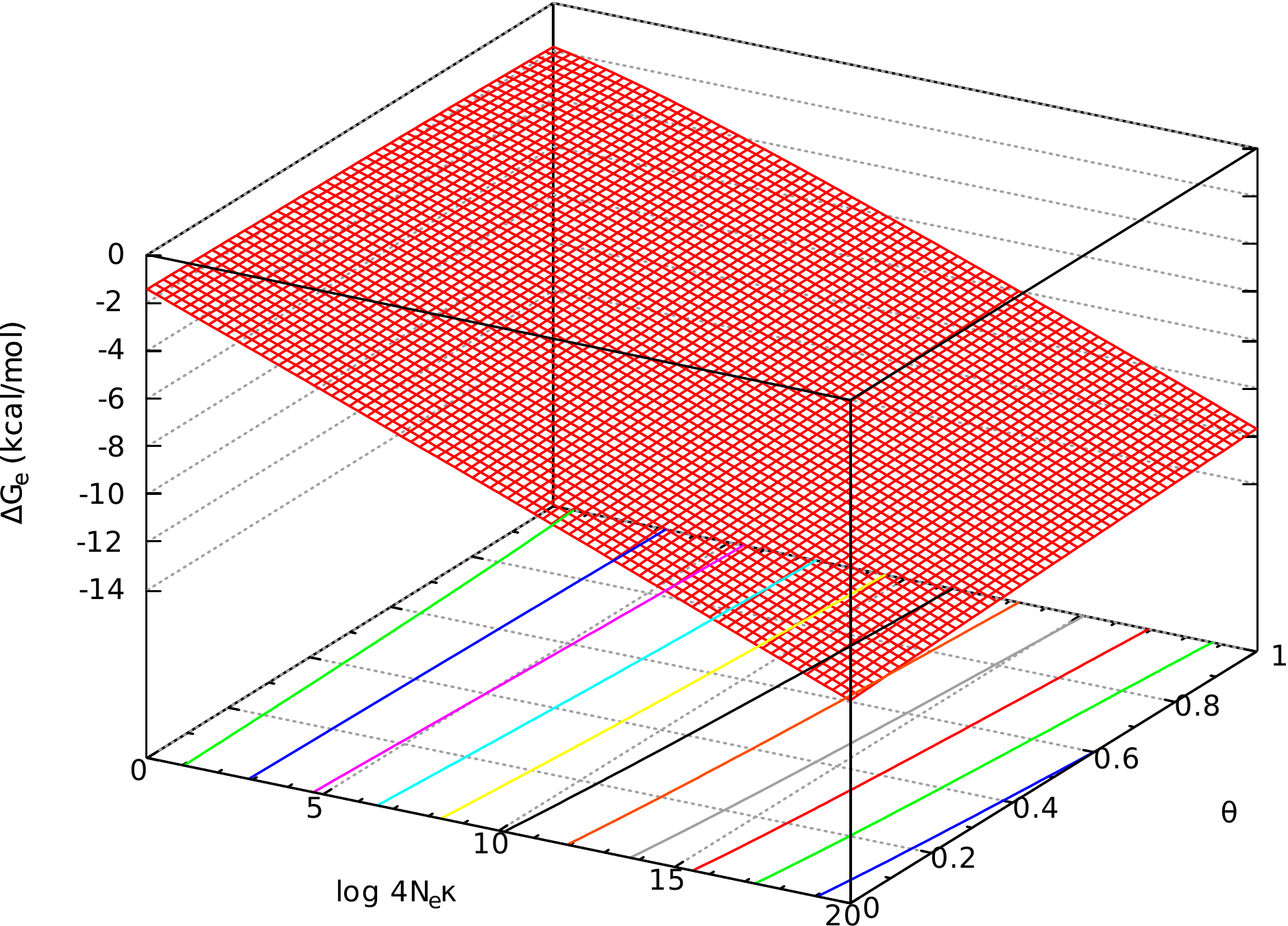}
}
} 
\vspace*{1em}
\caption{
\label{sfig: dependence_of_dGe_on_4Nekappa_and_theta}
\FigureLegends{
\BF{Dependence of equilibrium stability, $\Delta G_e$, on parameters, $4N_e\kappa$ and $\theta$.}
$\Delta G_e$ is the equilibrium value of folding free energy, $\Delta G$,
in protein evolution.
The value of $\beta\Delta G_e + \log 4N_e\kappa$ is the upper bound of $\log 4N_e s$,
and would be constant if the mean of $\Delta\Delta G$ in all arising mutants did not depend on $\Delta G$; 
see \Eq{\ref{seq: def_s}}.
The kcal/mol unit is used for $\Delta G_e$.
} 
}
\end{figure*}

\FigureInLegends{\clearpage\newpage}

\begin{figure*}[ht]
\FigureInLegends{
\centerline{
\includegraphics*[width=90mm,angle=0]{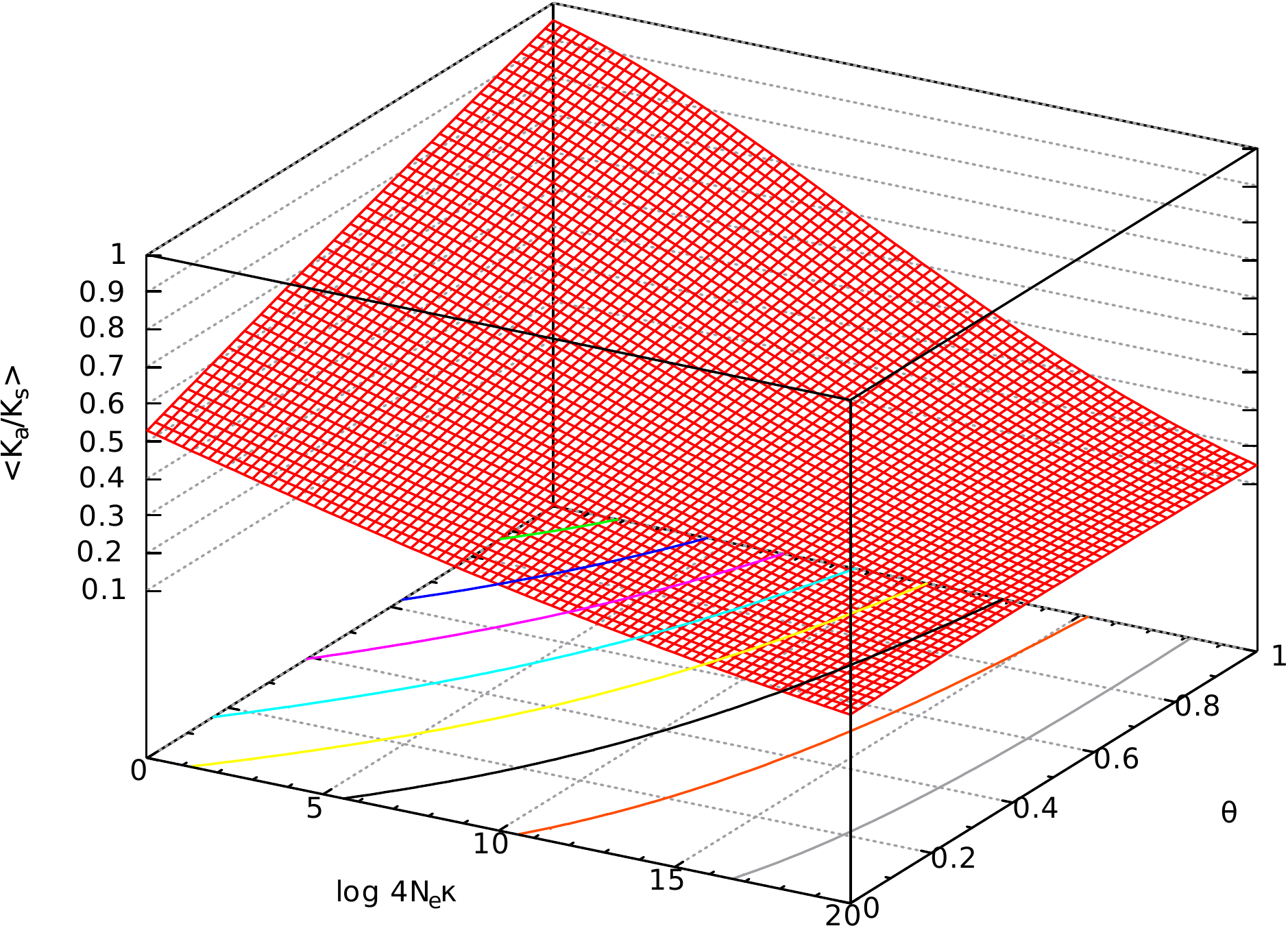}
\includegraphics*[width=90mm,angle=0]{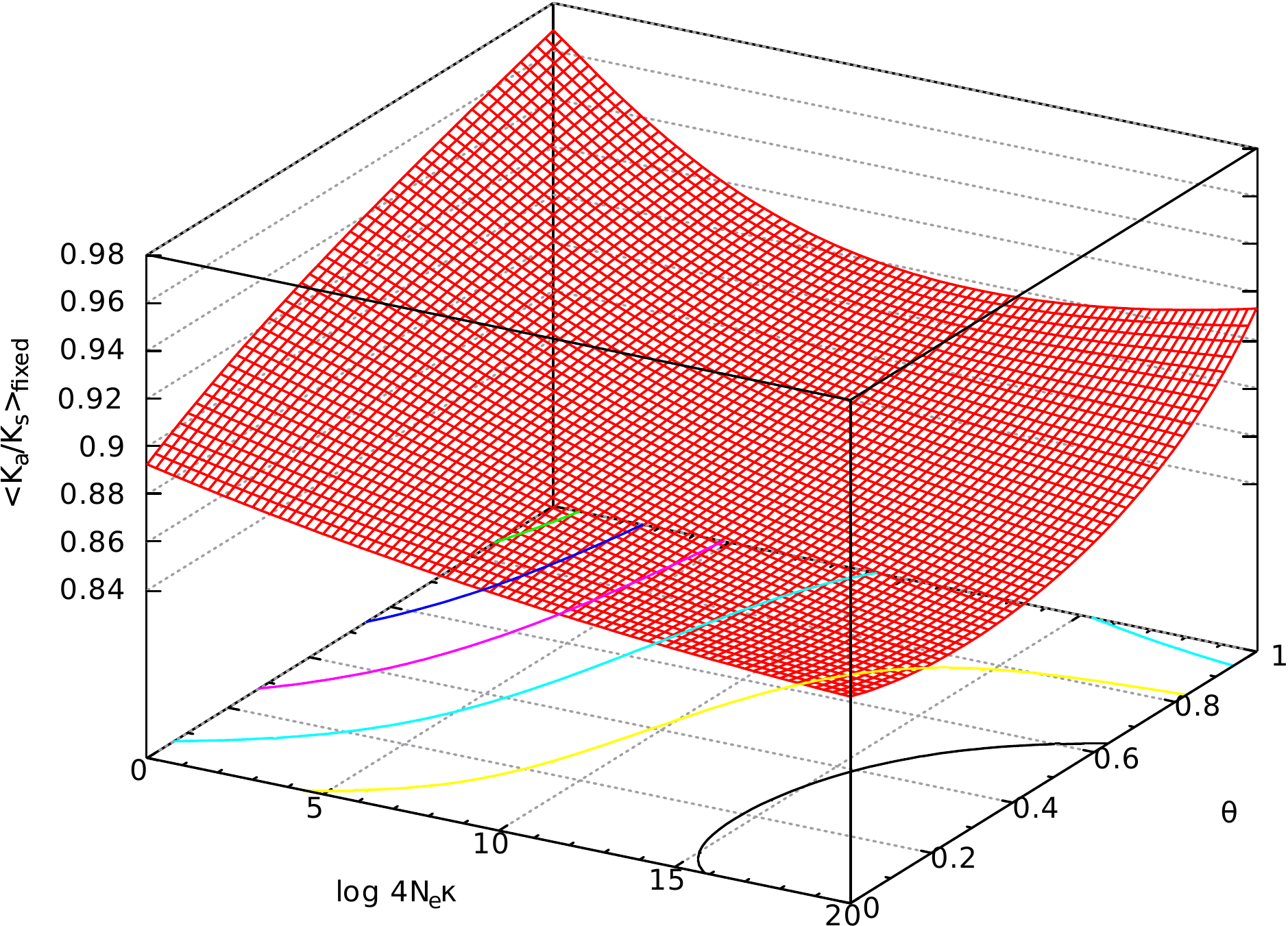}
}
} 
\vspace*{1em}
\caption{
\label{sfig: dependence_of_ave_Ka_over_Ks_on_4Nekappa_and_theta}
\label{sfig: dependence_of_ave_Ka_over_Ks_fixed_on_4Nekappa_and_theta}
\FigureLegends{
\BF{The average of $K_a / K_s$ over all mutants or over fixed mutants only at equilibrium of protein stability, $\Delta G = \Delta G_e$.}
} 
}
\end{figure*}

\FigureInLegends{\newpage}

\begin{figure*}[ht]
\FigureInLegends{
\centerline{
\includegraphics*[width=90mm,angle=0]{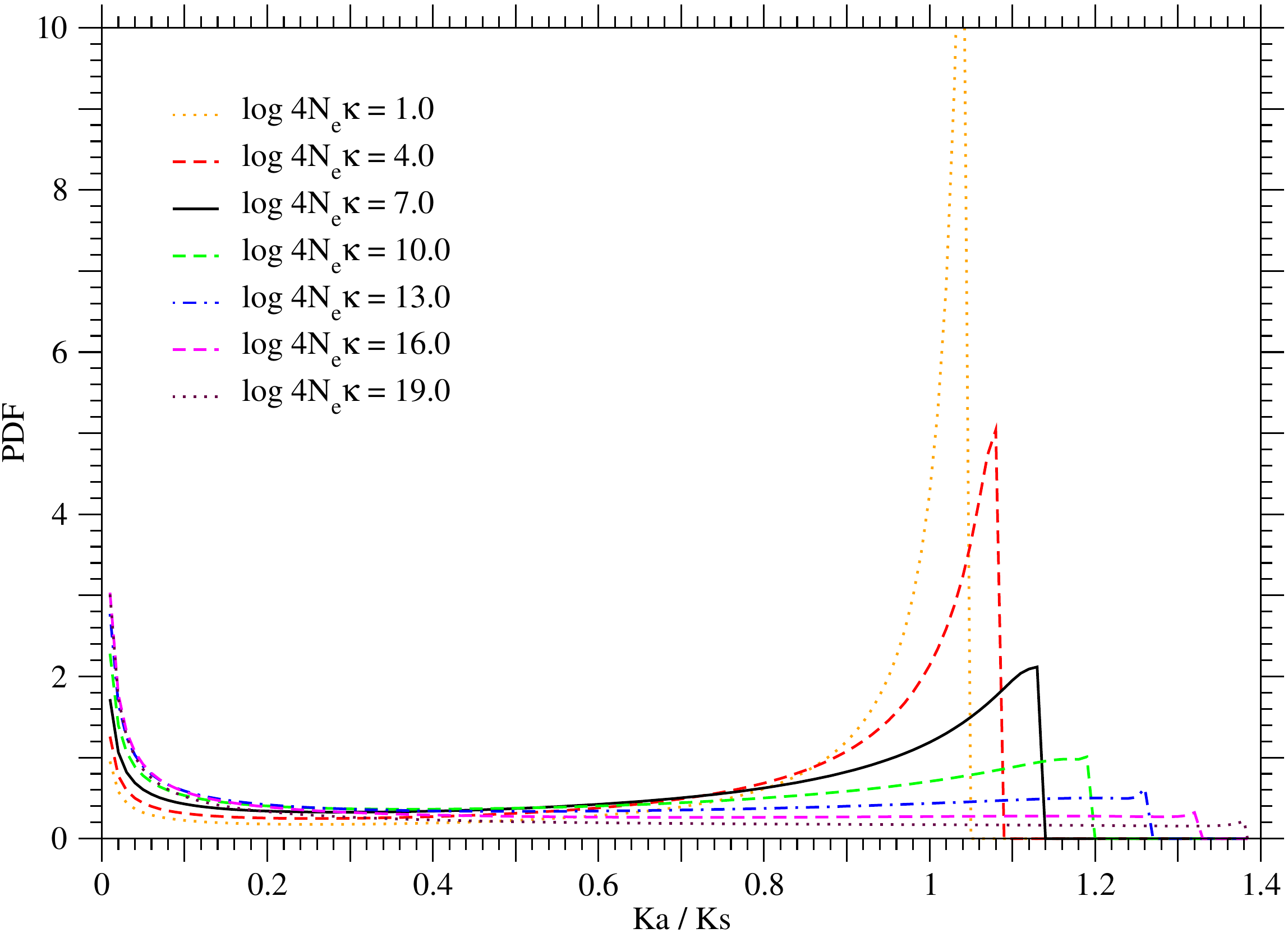}
\includegraphics*[width=90mm,angle=0]{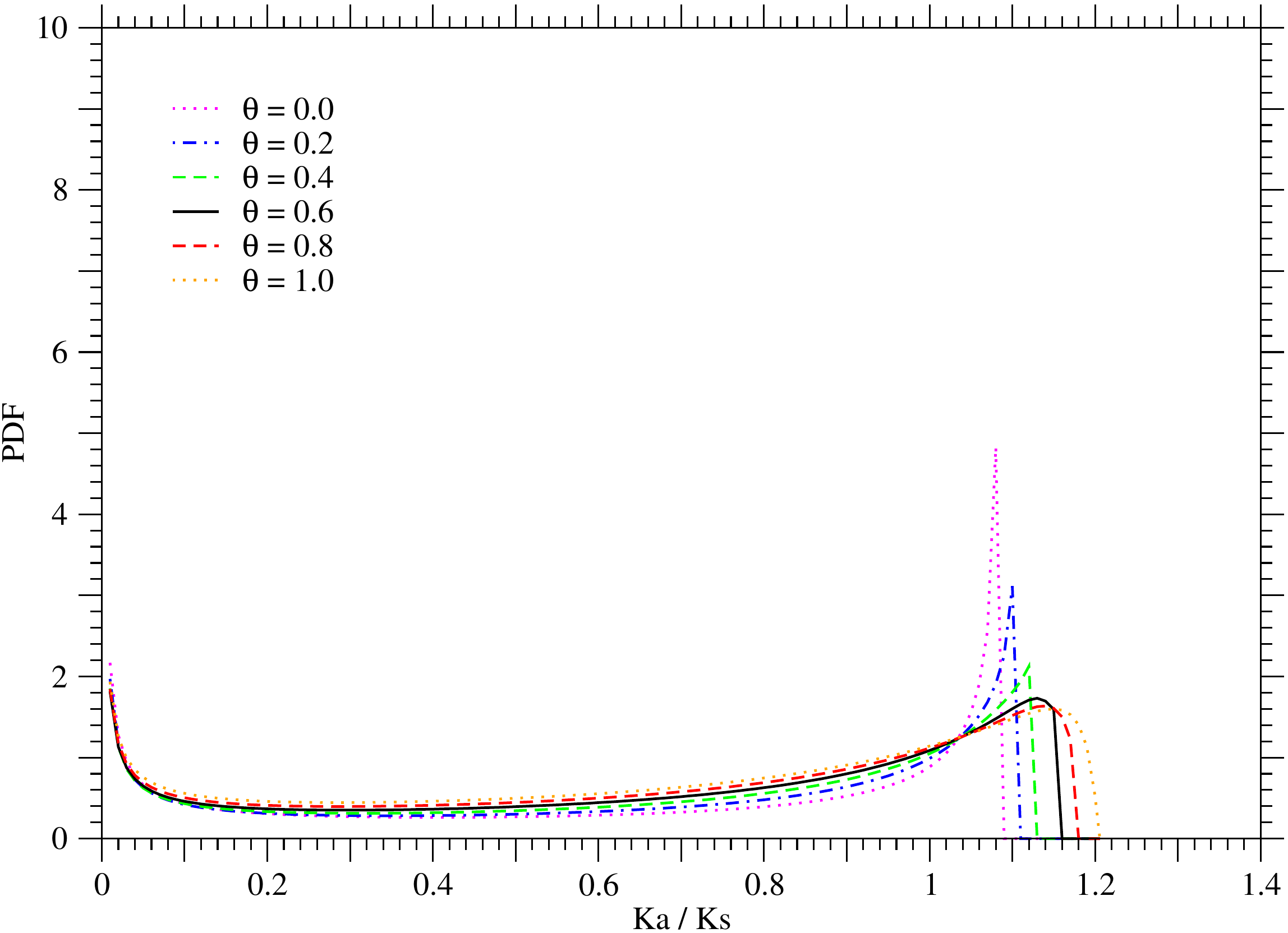}
}
\centerline{
\includegraphics*[width=90mm,angle=0]{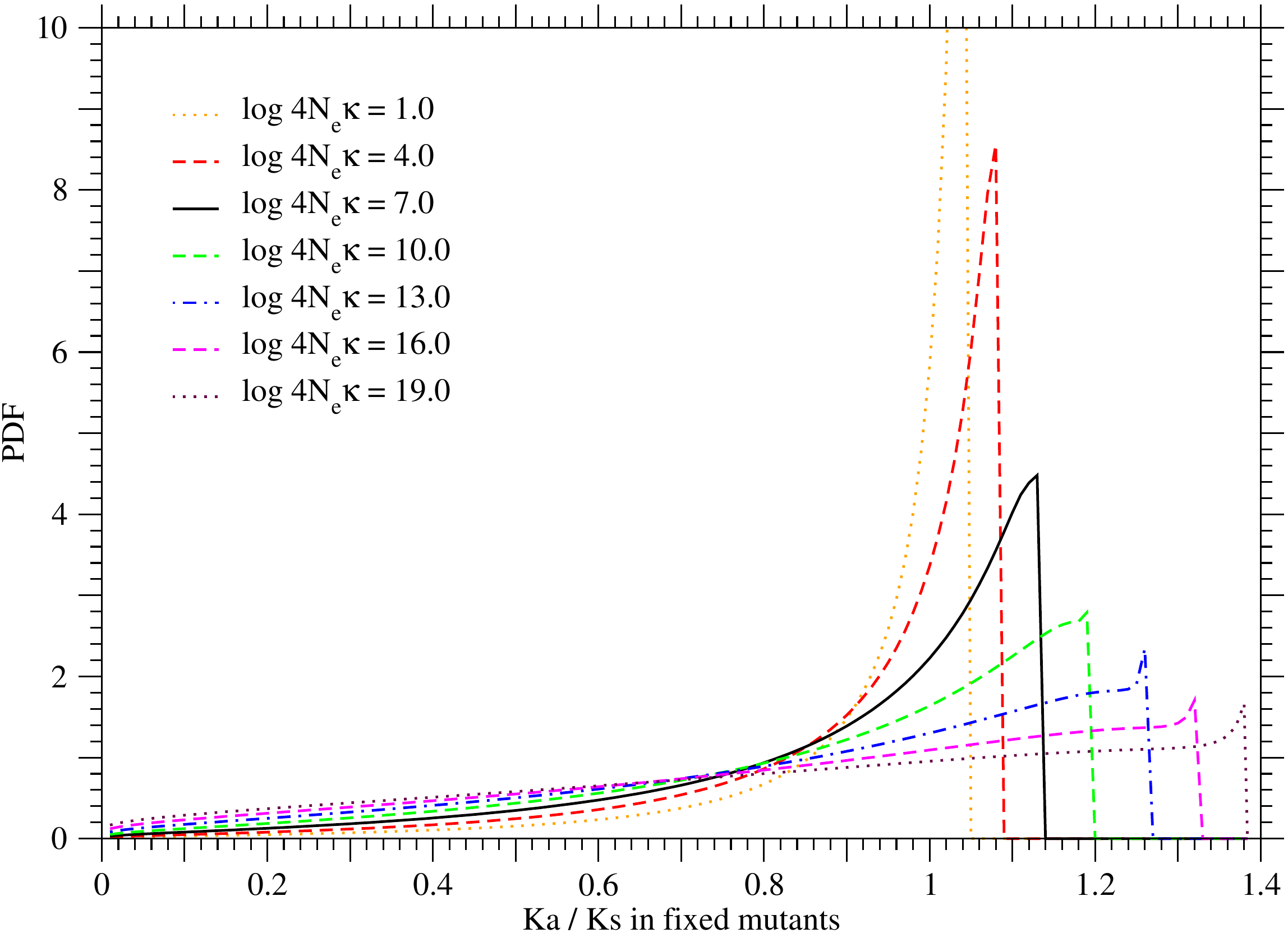}
\includegraphics*[width=90mm,angle=0]{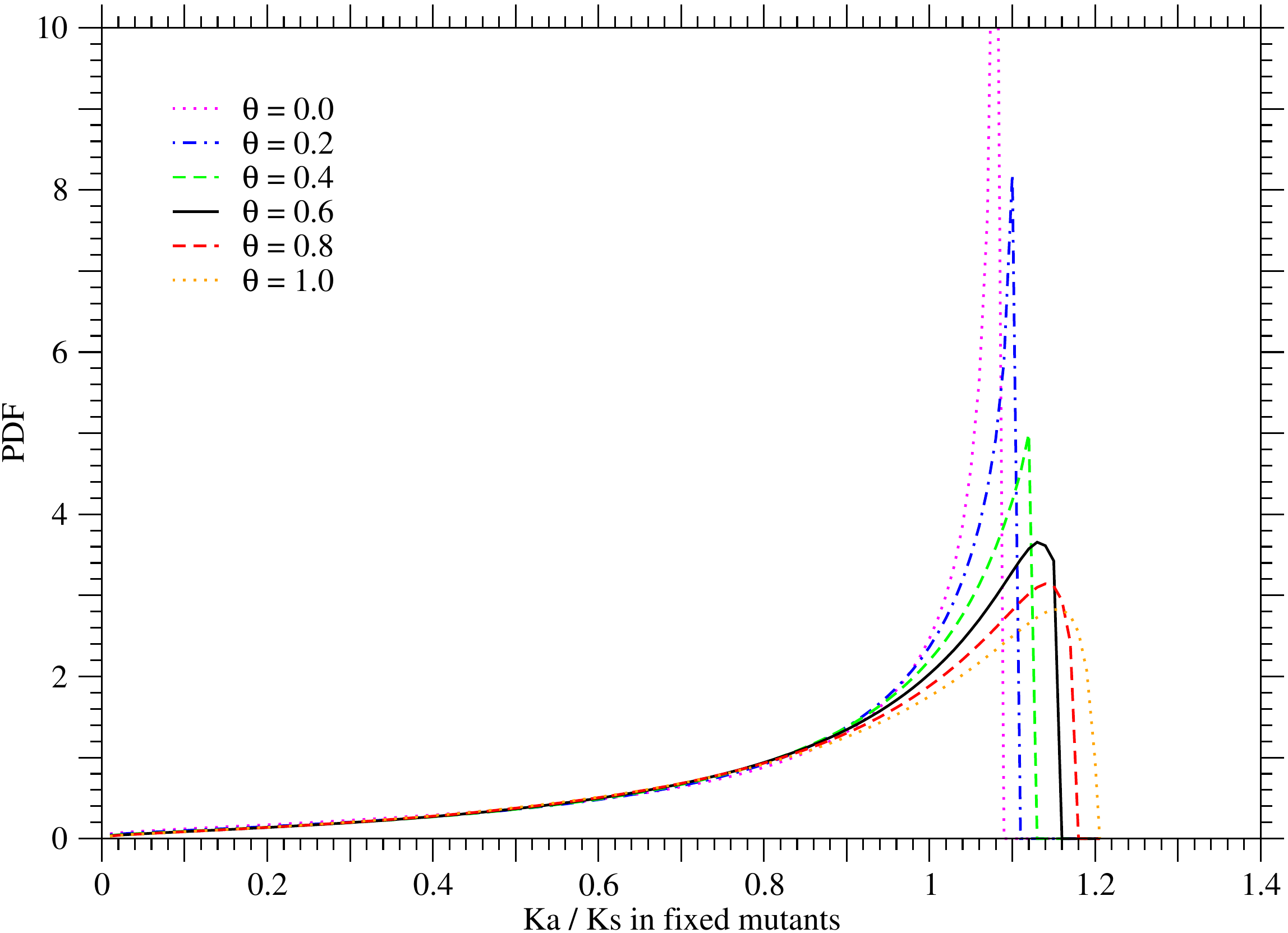}
}
} 
\vspace*{1em}
\caption{
\label{sfig: dependence_of_pdf_of_Ka_over_Ks_on_4Nekappa_and_theta}
\label{sfig: dependence_of_pdf_of_Ka_over_Ks_fixed_on_4Nekappa_and_theta}
\FigureLegends{
\BF{PDFs of $K_a / K_s$ in all mutants and in fixed mutants only
at equilibrium of protein stability, $\Delta G = \Delta G_e$.}
Unless specified, 
$\log 4N_e\kappa = 7.550$ and $\theta = 0.53$
are employed.
} 
}
\end{figure*}

\FigureInLegends{\newpage}

\begin{figure*}[ht]
\FigureInLegends{
\centerline{
\includegraphics*[width=87mm,angle=0]{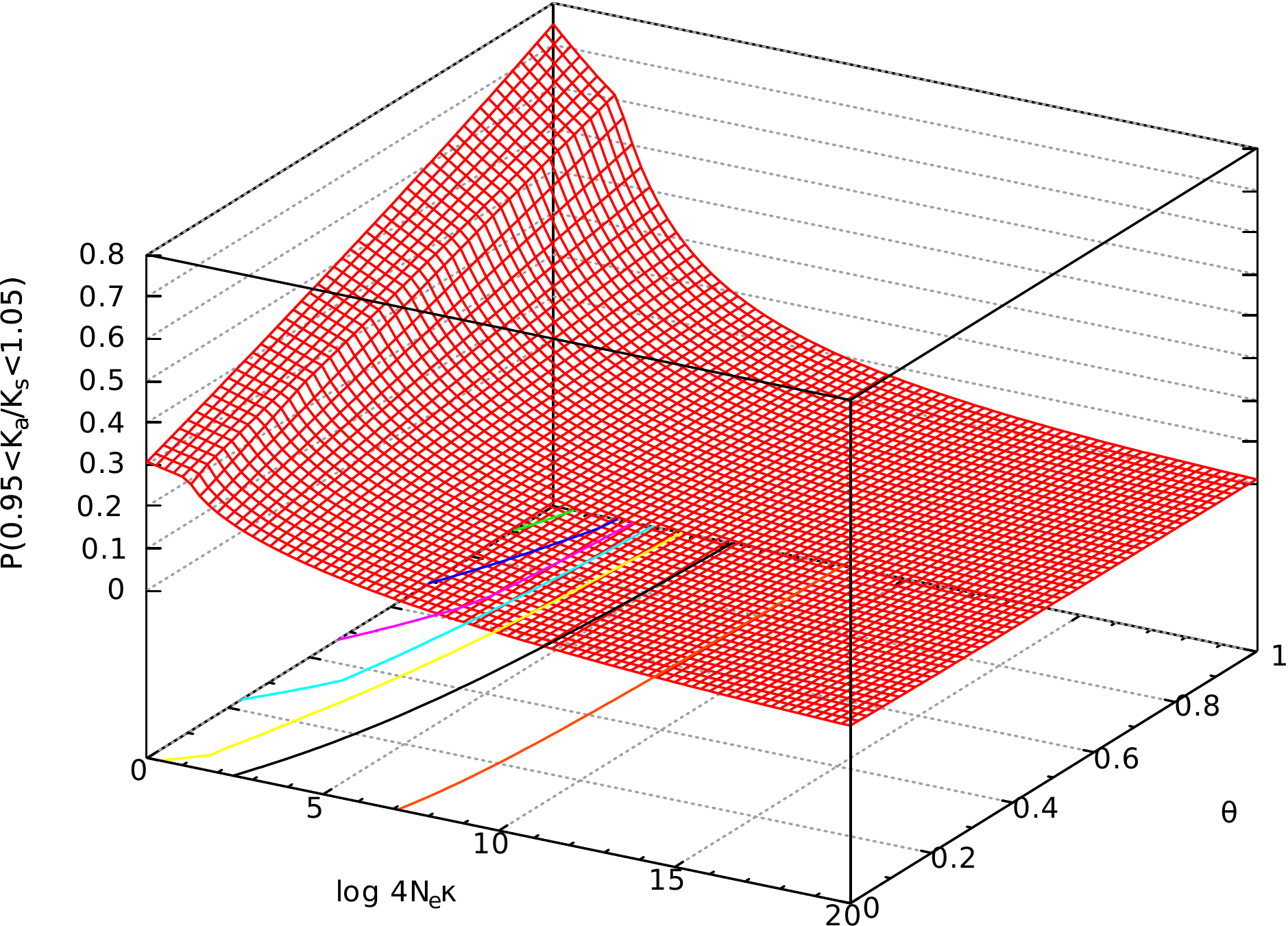}
\includegraphics*[width=90mm,angle=0]{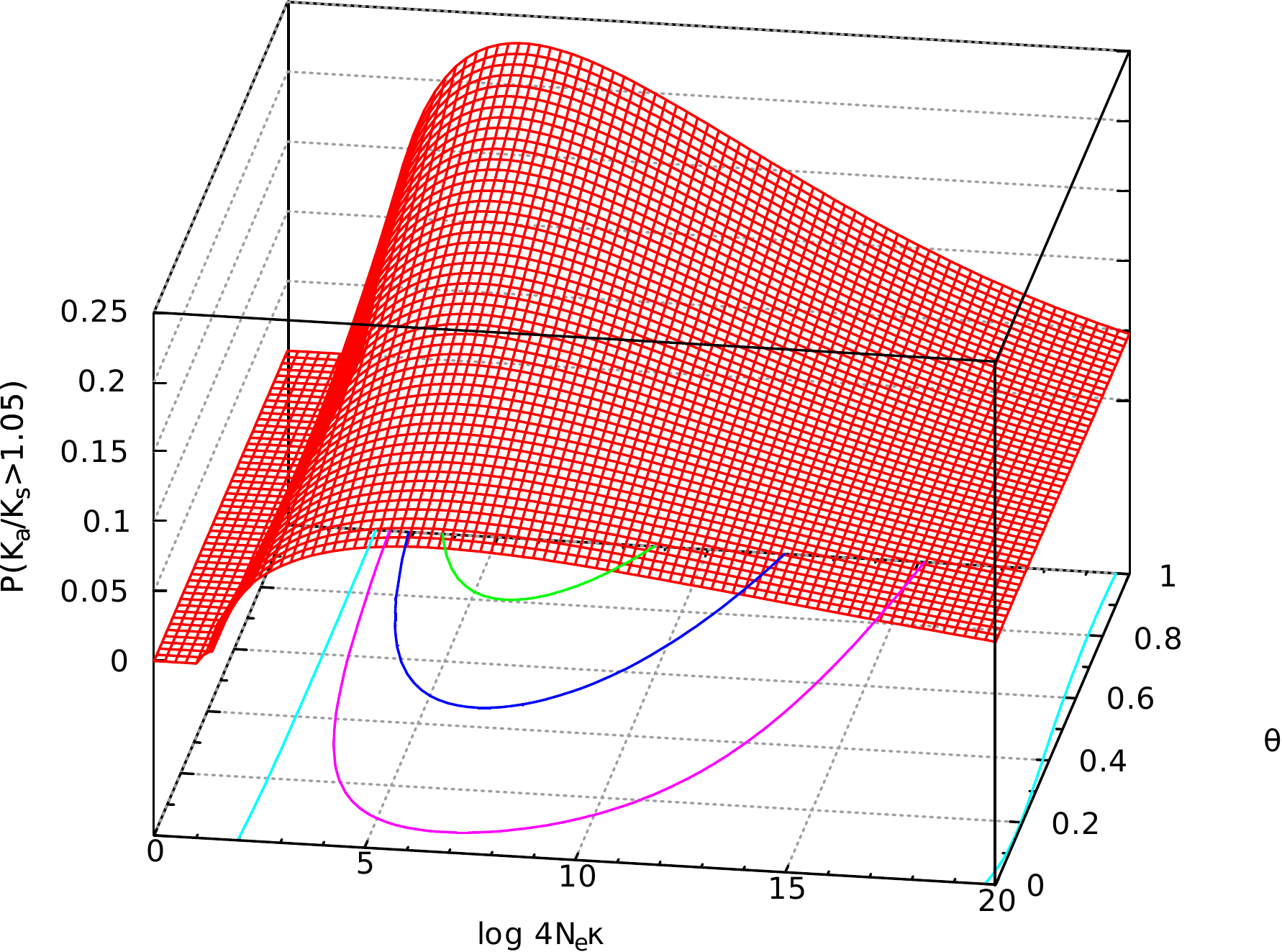}
}
\centerline{
\includegraphics*[width=88mm,angle=0]{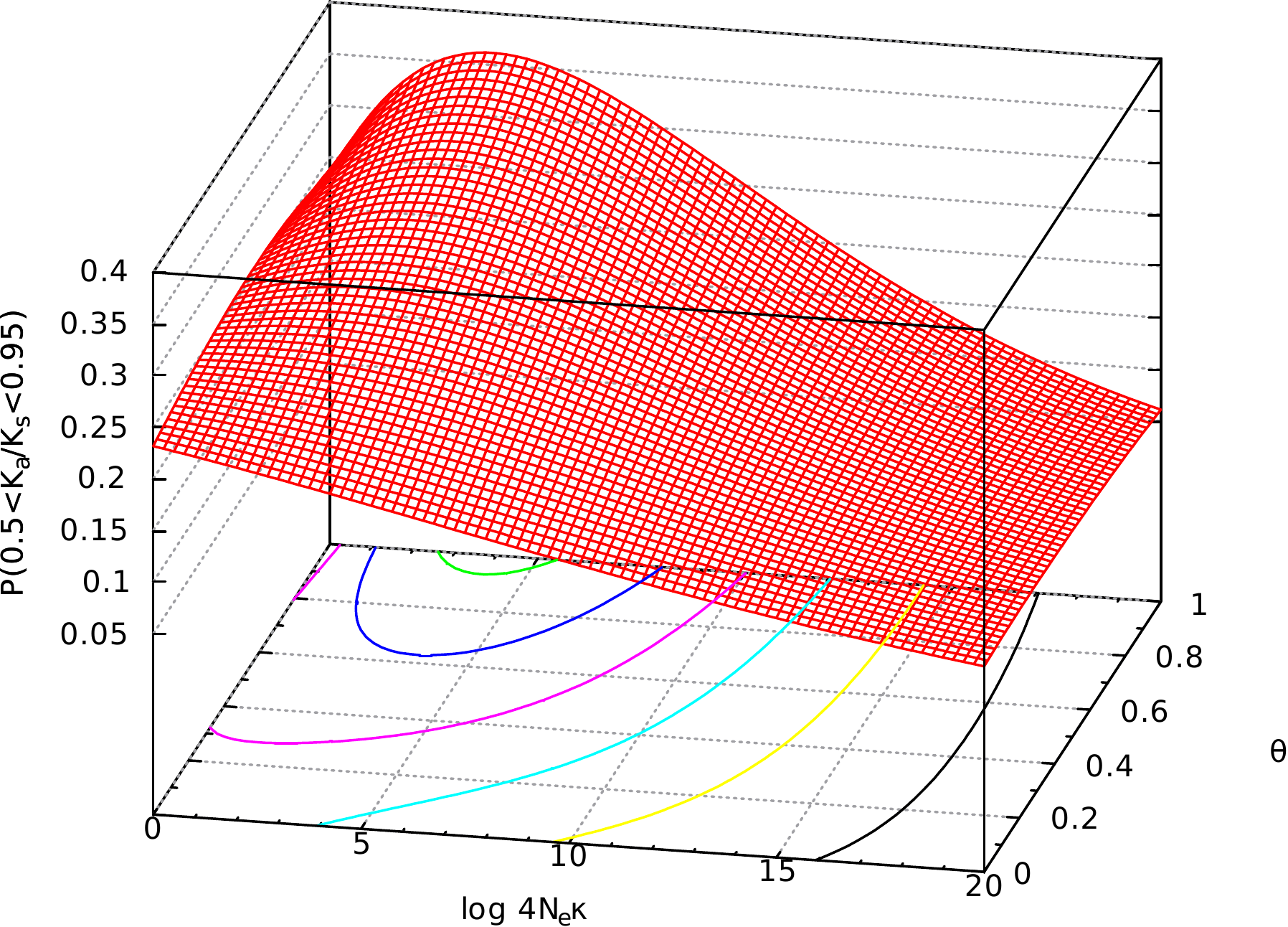}
\includegraphics*[width=90mm,angle=0]{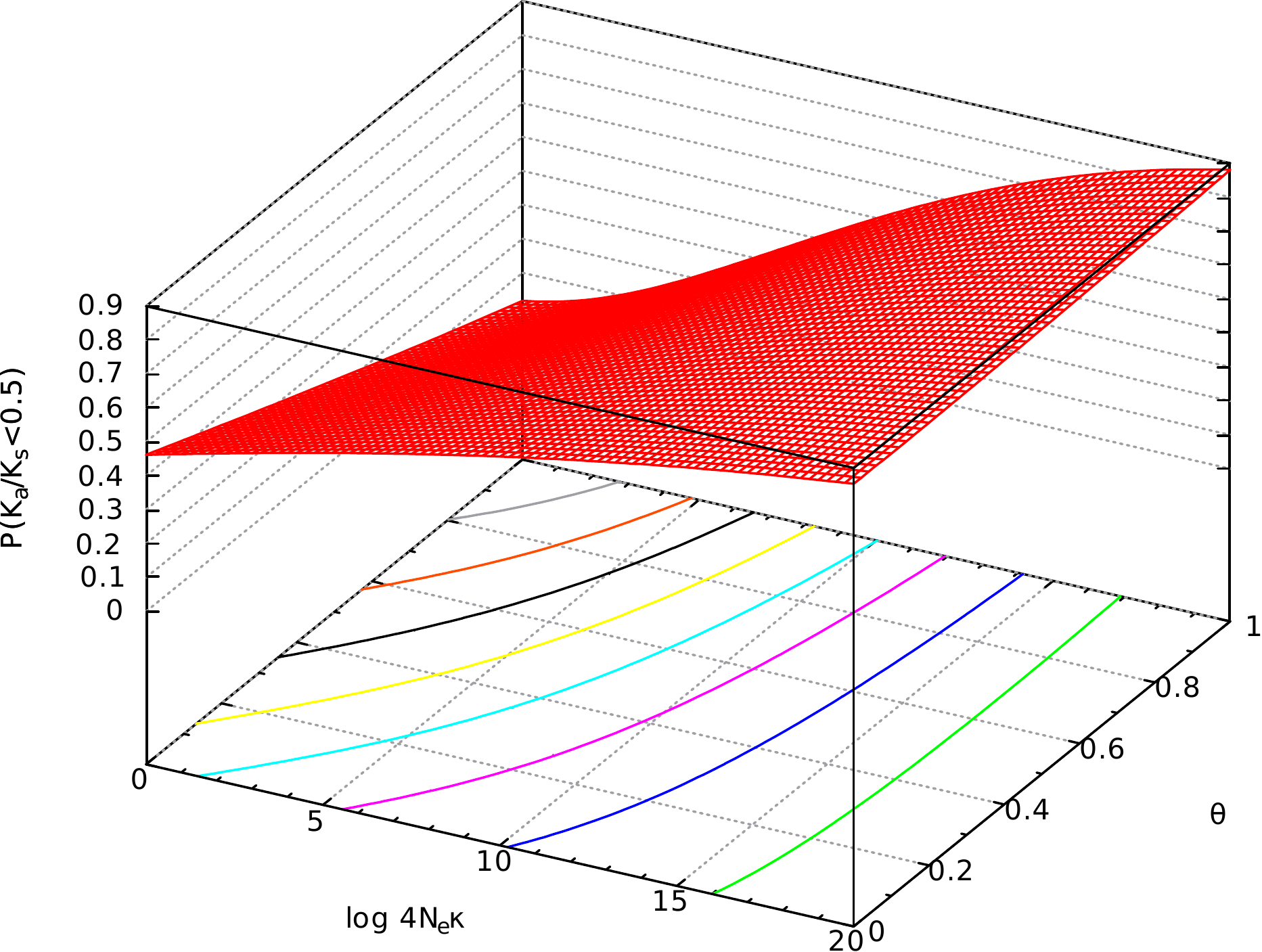}
}
} 
\vspace*{1em}
\caption{
\label{sfig: prob_of_each_selection_category_in_all_mutants}
\FigureLegends{
\BF{Probability of each selection category in all mutants at equilibrium of protein stability, $\Delta G = \Delta G_e$.}
Arbitrarily, the value of $K_a/K_s$ is categorized into four classes;
negative, slightly negative, nearly neutral, and positive selection categories in which
$K_a/K_s$ is within the ranges of
$K_a/K_s \leq 0.5$, $0.5< K_a/K_s \leq 0.95$, $0.95< K_a/K_s \leq 1.05$, and $ 1.05 < K_a/K_s$,
respectively.
} 
}
\end{figure*}

\FigureInLegends{\newpage}

\begin{figure*}[ht]
\FigureInLegends{
\centerline{
\includegraphics*[width=90mm,angle=0]{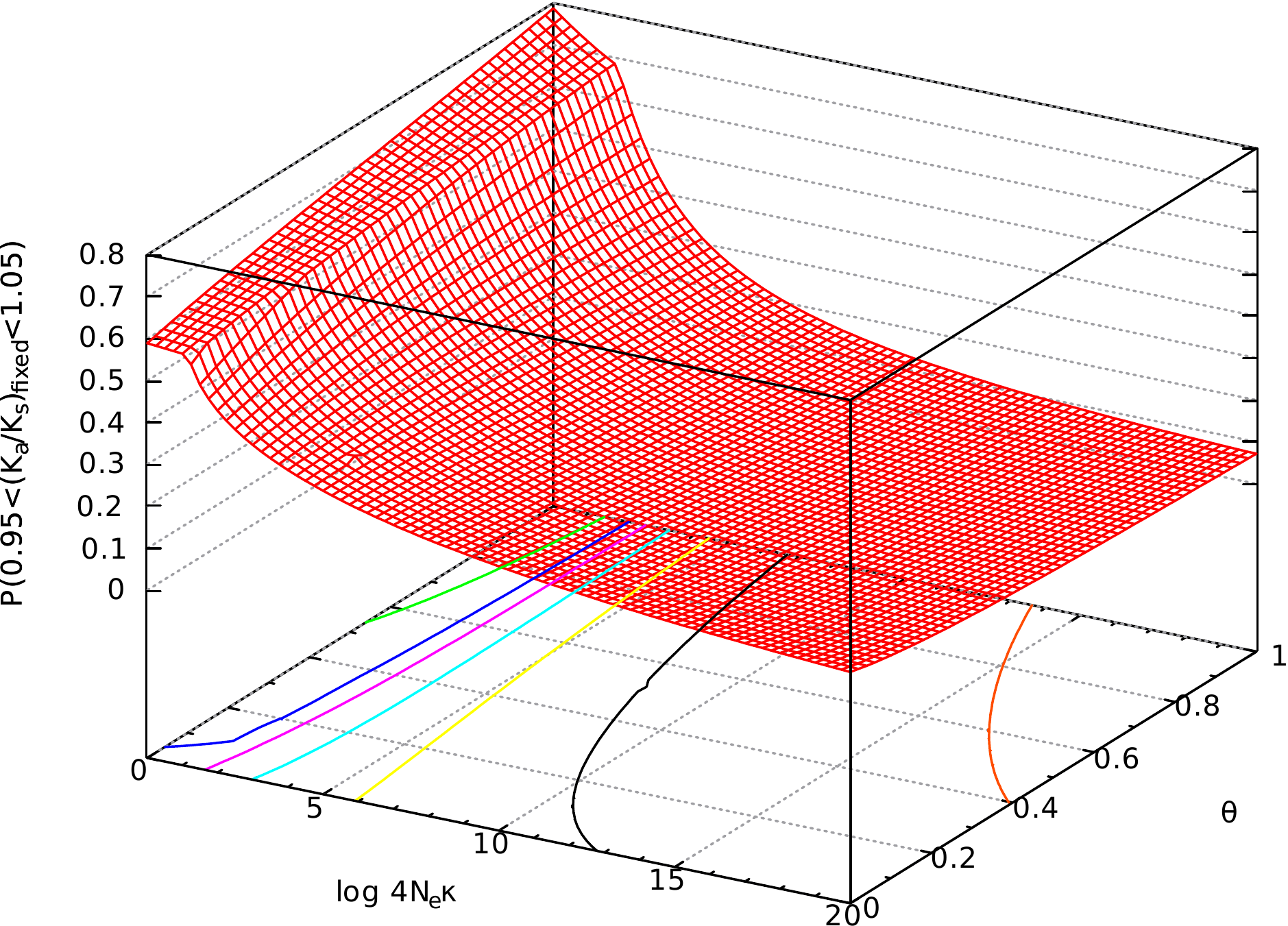}
\includegraphics*[width=90mm,angle=0]{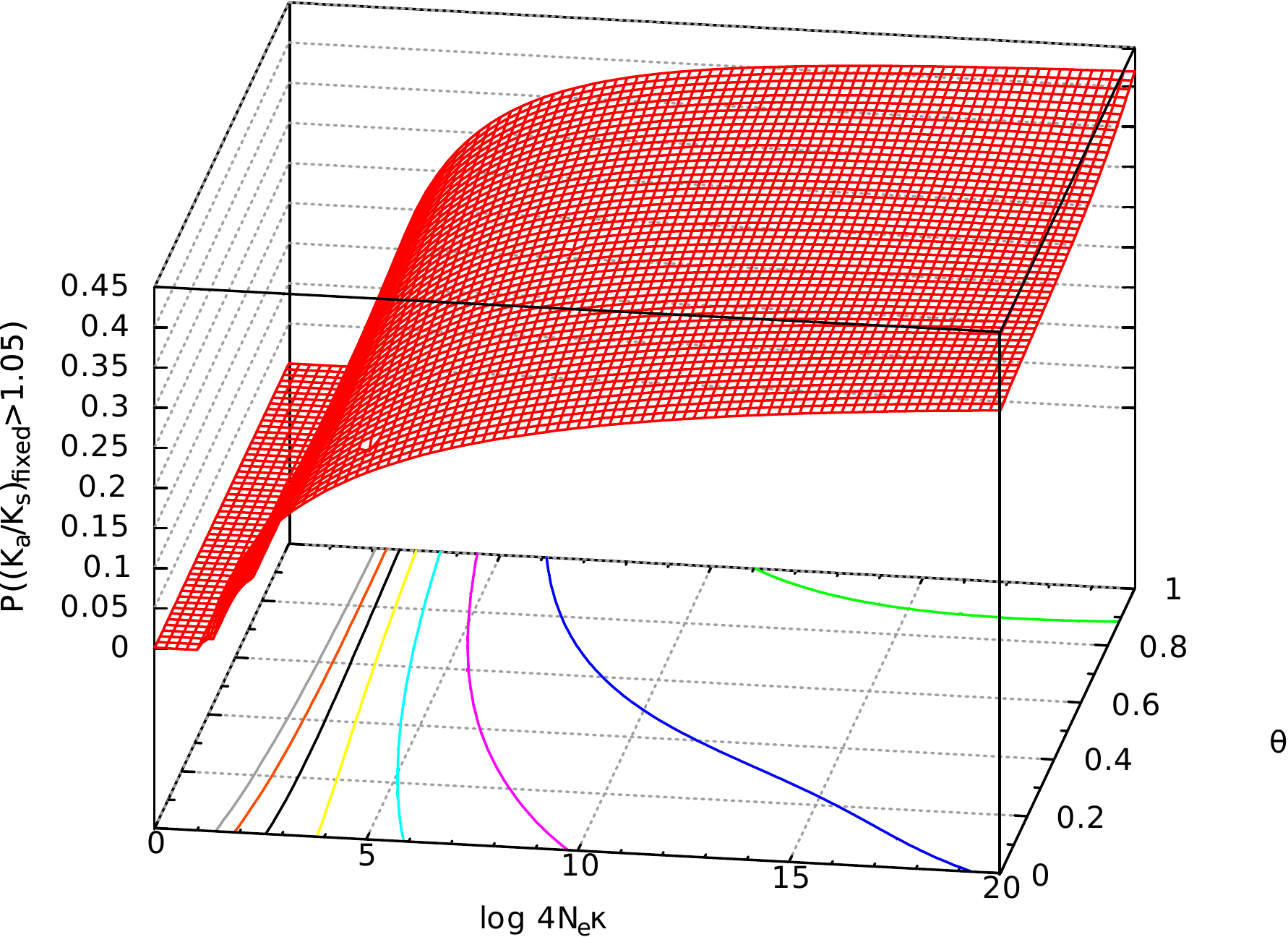}
}
\centerline{
\includegraphics*[width=90mm,angle=0]{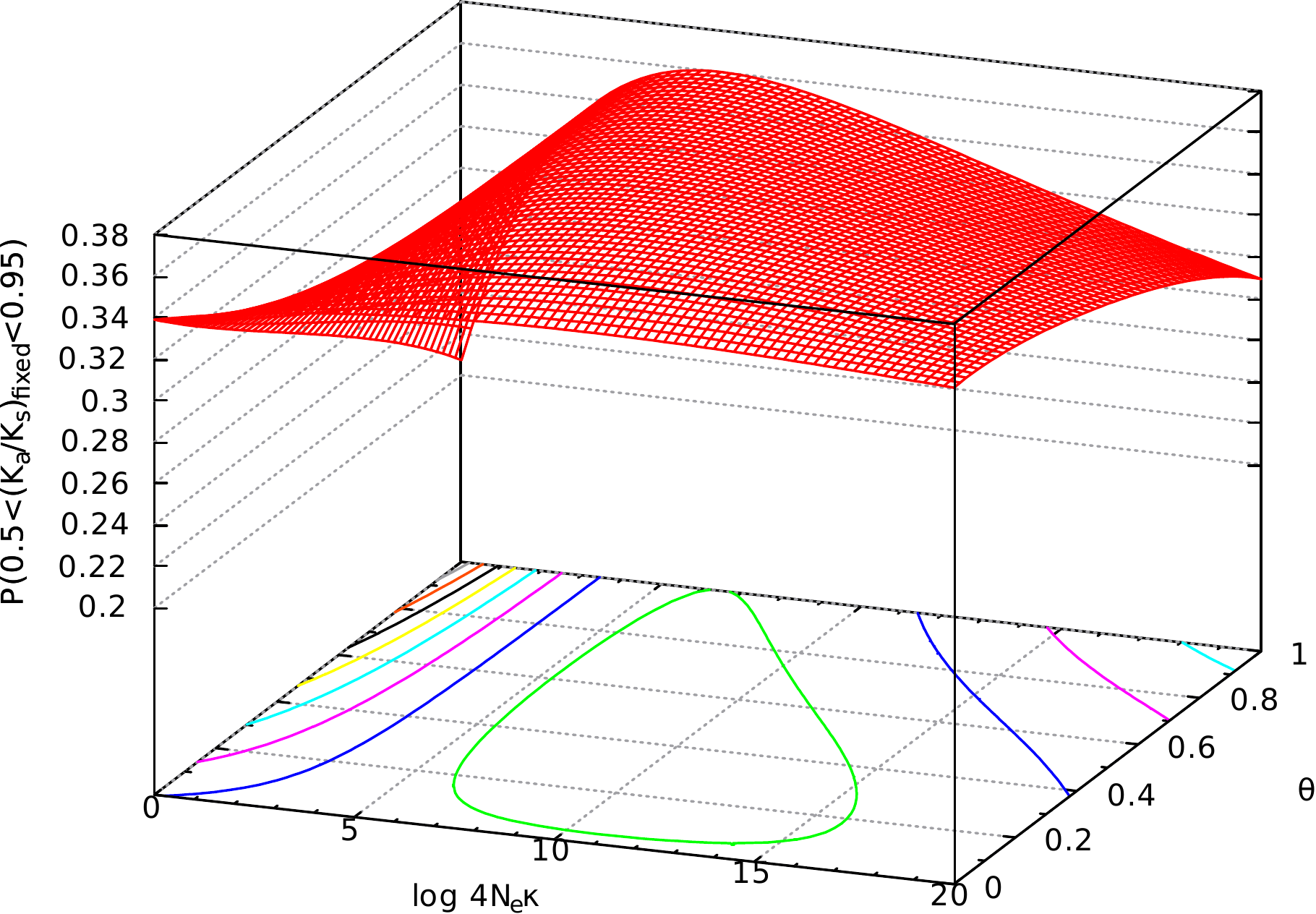}
\includegraphics*[width=90mm,angle=0]{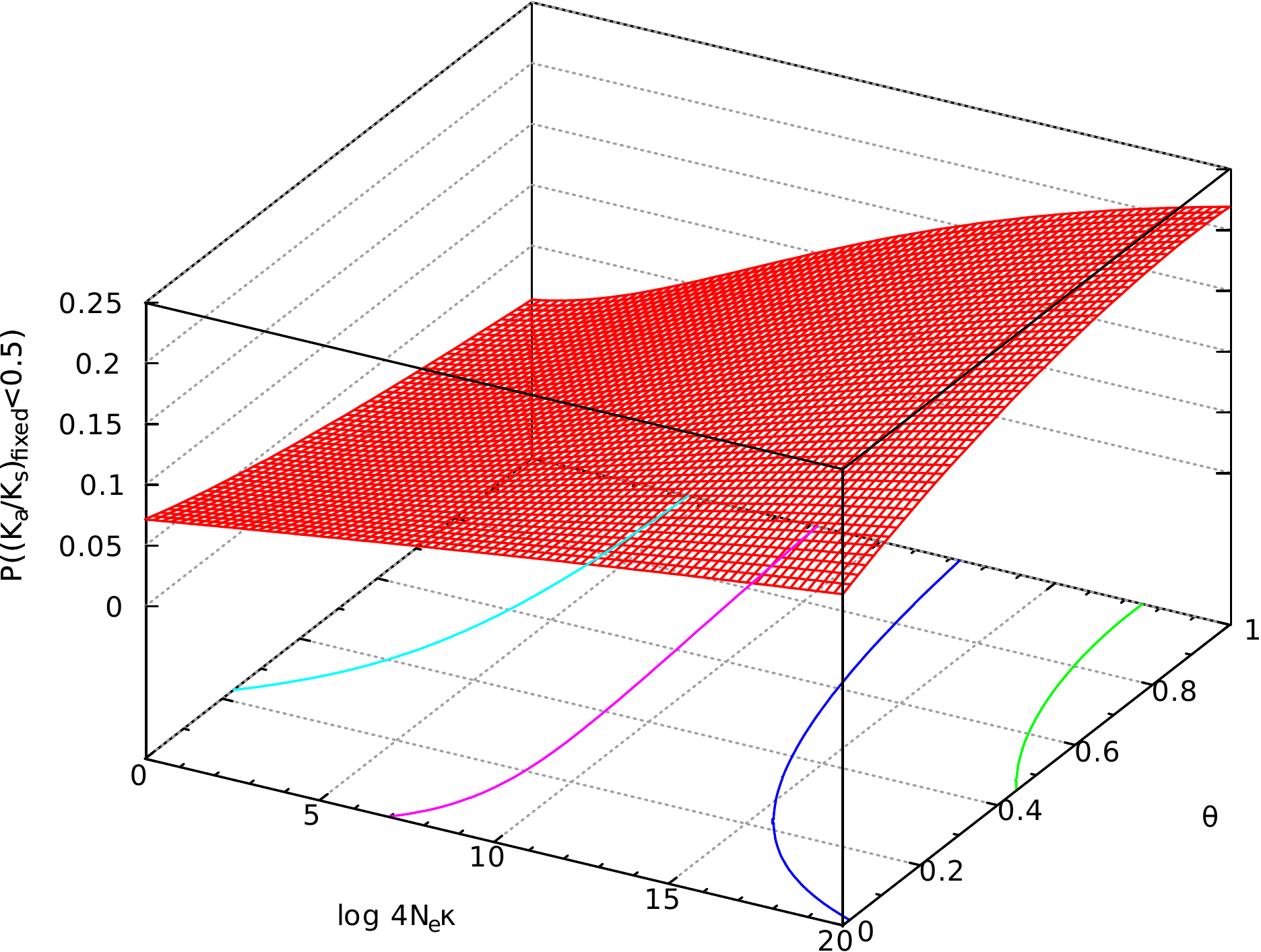}
}
} 
\vspace*{1em}
\caption{
\label{sfig: prob_of_each_selection_category_in_fixed_mutants}
\FigureLegends{
\BF{Probability of each selection category in fixed mutants at equilibrium of protein stability, $\Delta G = \Delta G_e$.}
Arbitrarily, the value of $K_a/K_s$ is categorized into four classes;
negative, slightly negative, nearly neutral, and positive selection categories in which
$K_a/K_s$ is within the ranges of
$K_a/K_s \leq 0.5$, $0.5< K_a/K_s \leq 0.95$, $0.95< K_a/K_s \leq 1.05$, and $ 1.05 < K_a/K_s$,
respectively.
} 
}
\end{figure*}

\FigureInLegends{\newpage}

\begin{figure*}[ht]
\FigureInLegends{
\centerline{
\includegraphics*[width=80mm,angle=0]{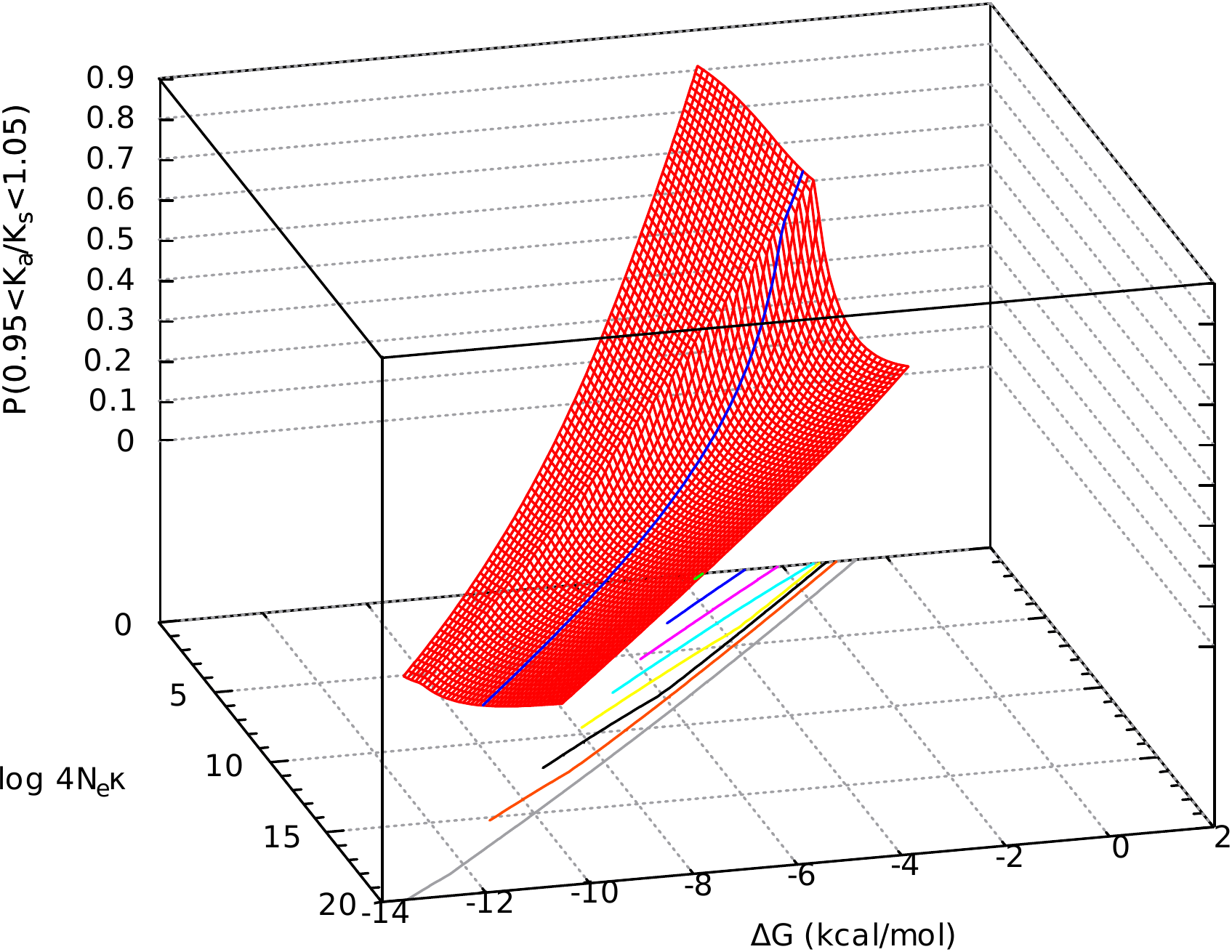}
\includegraphics*[width=90mm,angle=0]{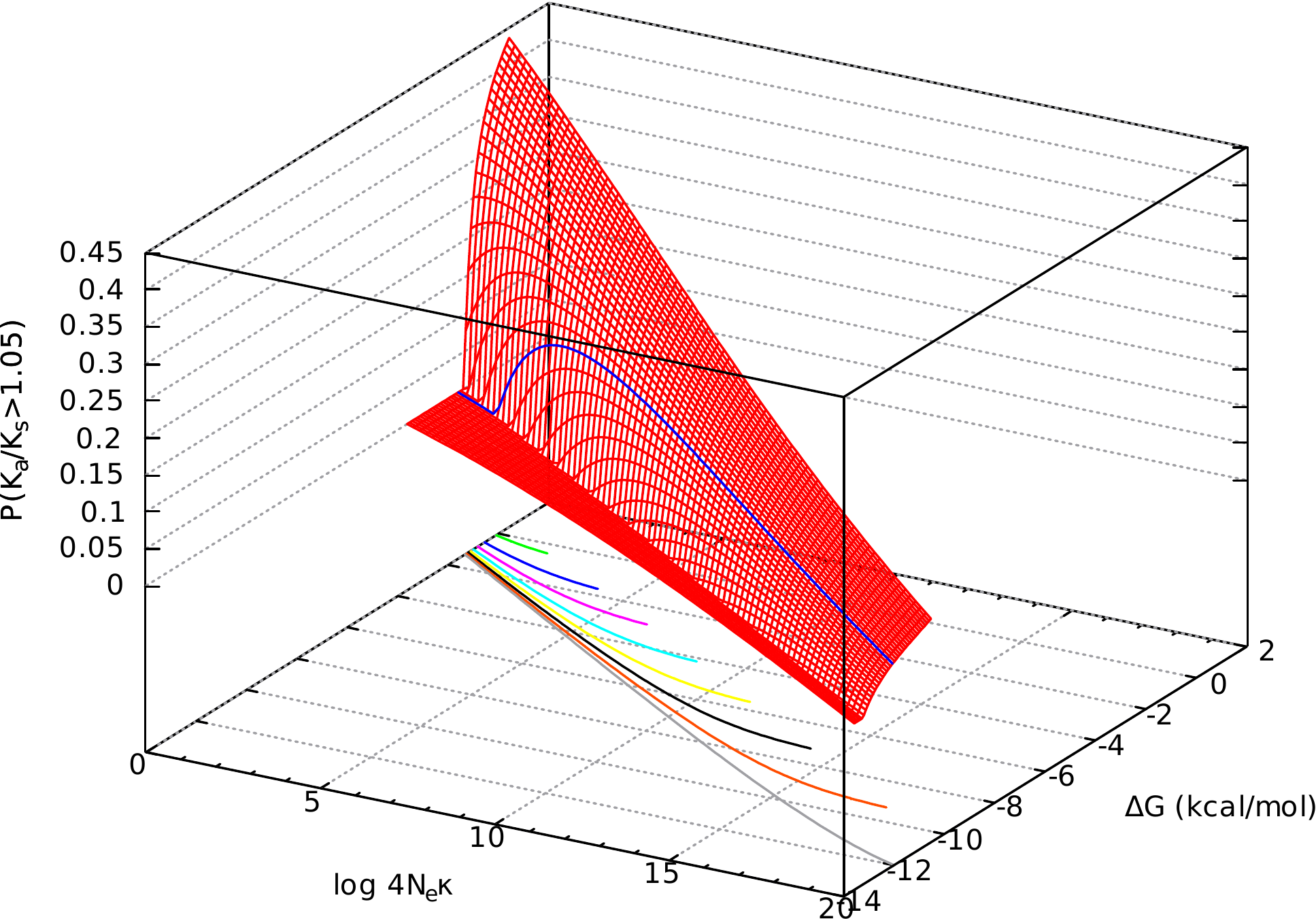}
}
\centerline{
\includegraphics*[width=85mm,angle=0]{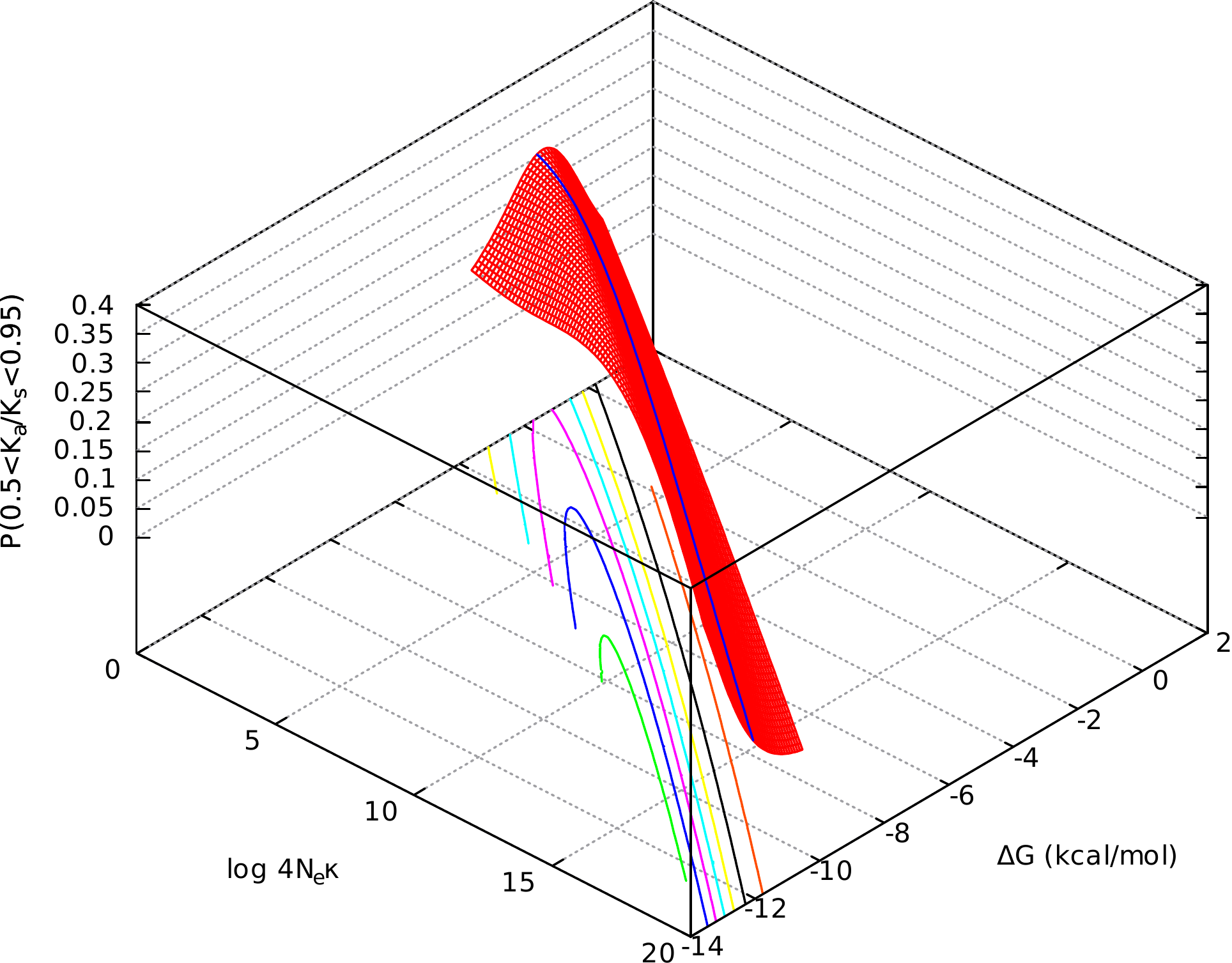}
\includegraphics*[width=90mm,angle=0]{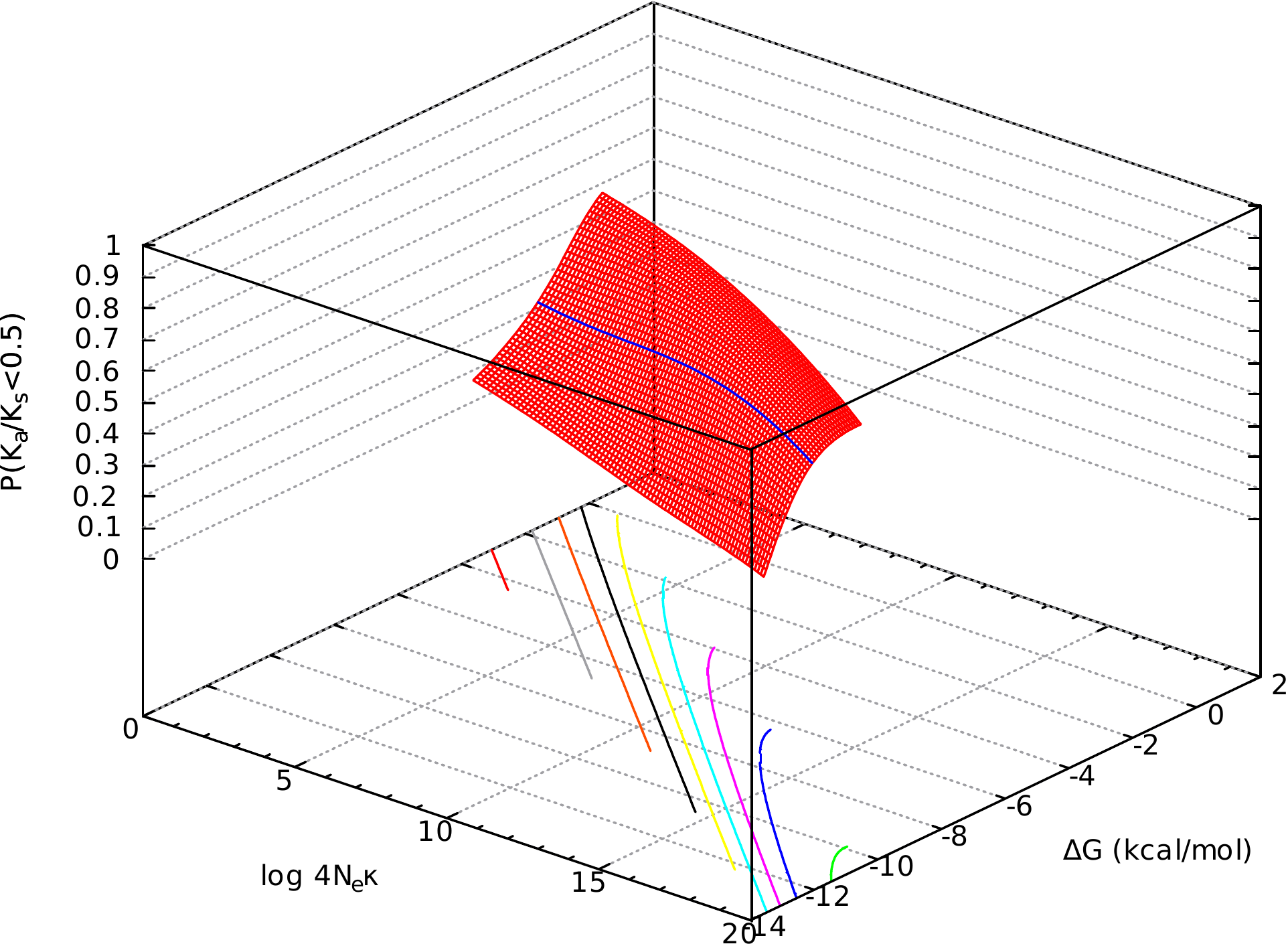}
}
} 
\vspace*{1em}
\caption{
\label{sfig: prob_of_each_selection_category_in_all_mutants;_dependence_on_kappa_and_dG}
\FigureLegends{
\BF{Dependence of the probability of each selection category in all mutants on $4N_e\kappa$ and $\Delta G$.}
A blue line on the surface grid shows $\Delta G=\Delta G_e$, which is the equilibrium value of $\Delta G$
in protein evolution.
The range of $\Delta G$ shown in the figures is
$| \Delta G - \Delta G_e | < 2 \cdot \Delta\Delta G^{\script{sd}}_{\script{fixed}}$, 
where $\Delta\Delta G^{\script{sd}}_{\script{fixed}}$ is the standard deviation of $\Delta\Delta G$ over fixed mutants at $ \Delta G = \Delta G_e$.
Arbitrarily, the value of $K_a/K_s$ is categorized into four classes;
negative, slightly negative, nearly neutral, and positive selection categories in which
$K_a/K_s$ is within the ranges of
$K_a/K_s \leq 0.5$, $0.5< K_a/K_s \leq 0.95$, $0.95< K_a/K_s \leq 1.05$, and $ 1.05 < K_a/K_s$,
respectively.
$\theta = 0.53$
is employed.
The kcal/mol unit is used for $\Delta G$.
} 
}
\end{figure*}

\clearpage
\FigureInLegends{\newpage}

\begin{figure*}[ht]
\FigureInLegends{
\centerline{
\includegraphics*[width=87mm,angle=0]{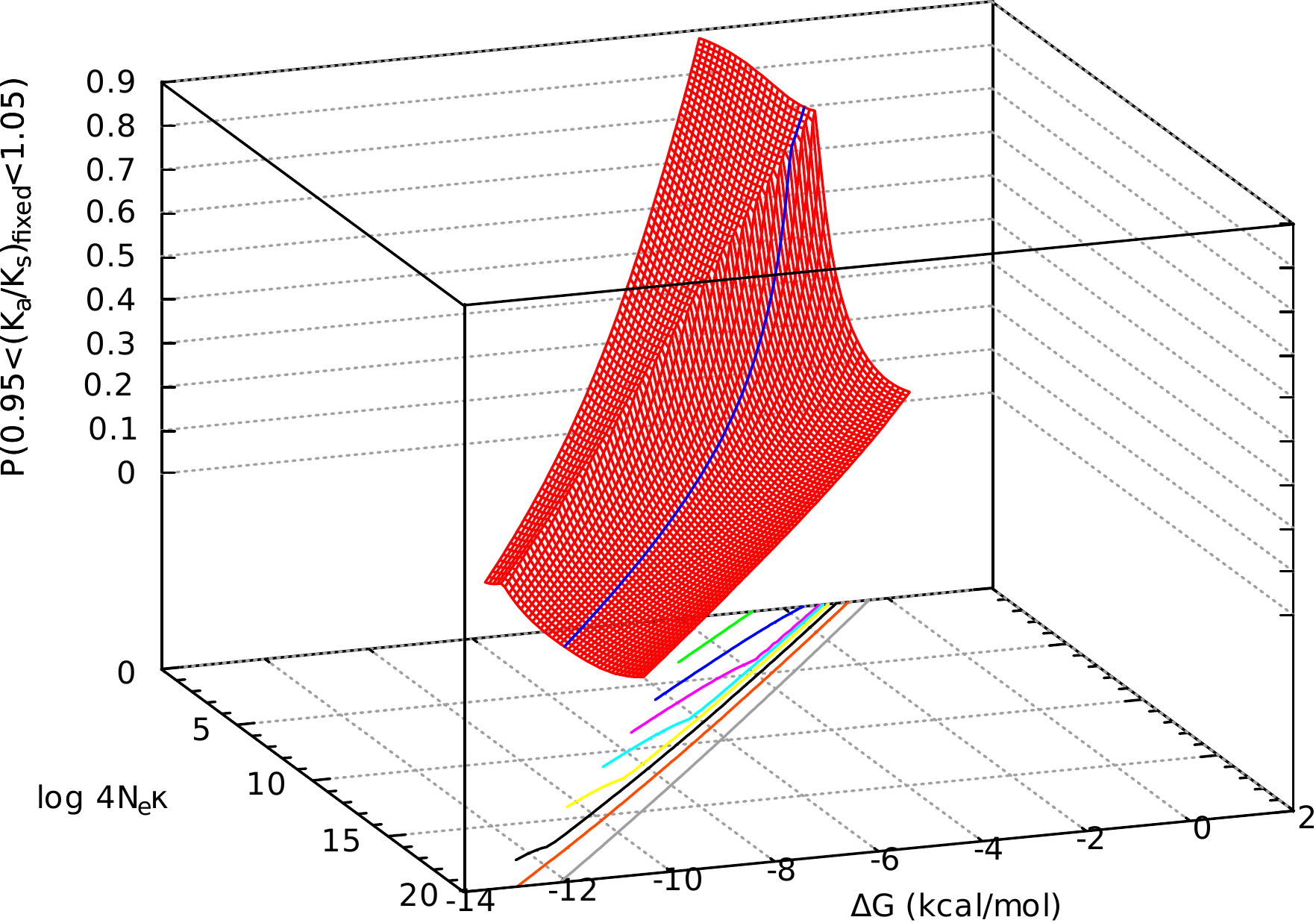}
\includegraphics*[width=90mm,angle=0]{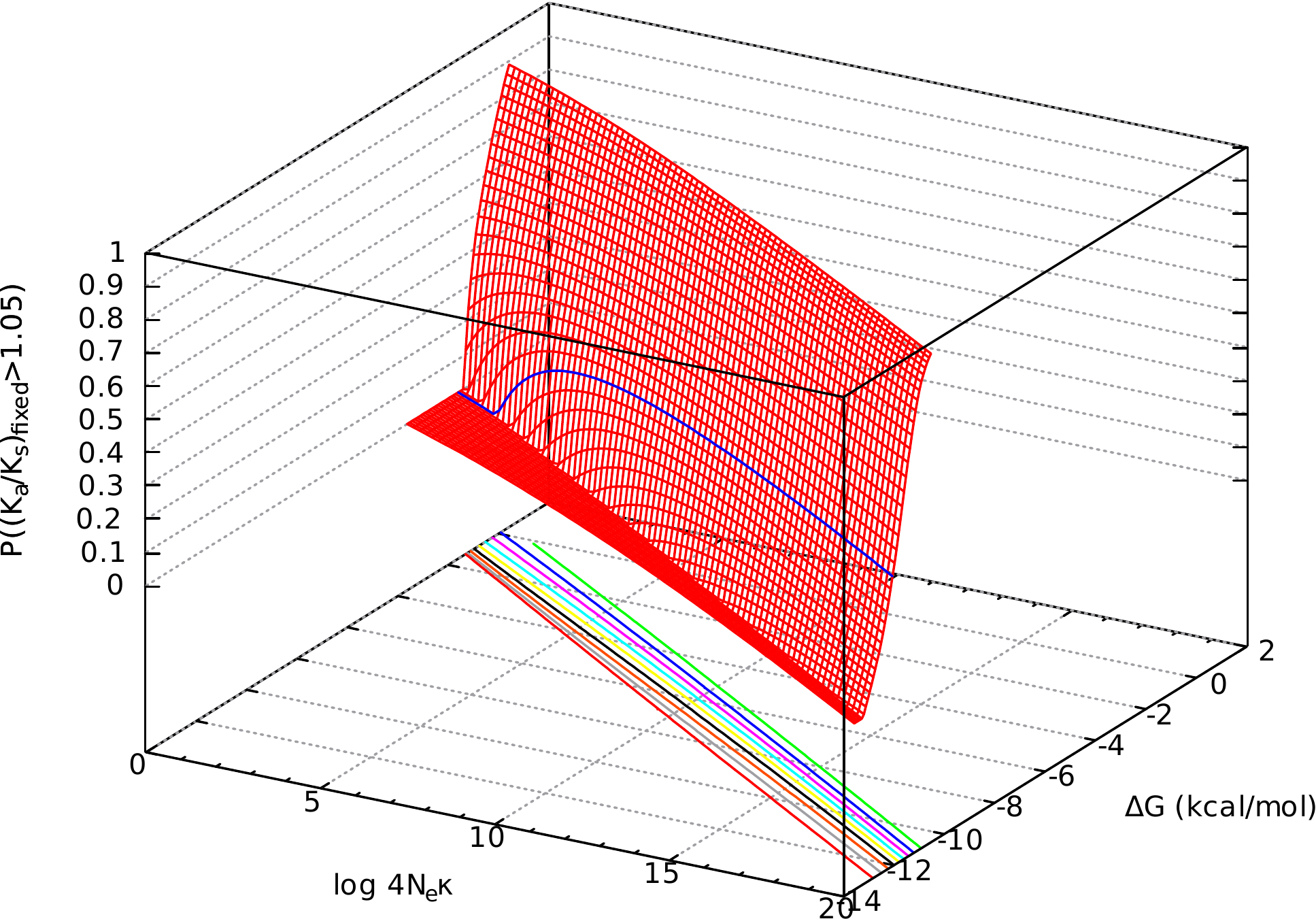}
}
\centerline{
\includegraphics*[width=90mm,angle=0]{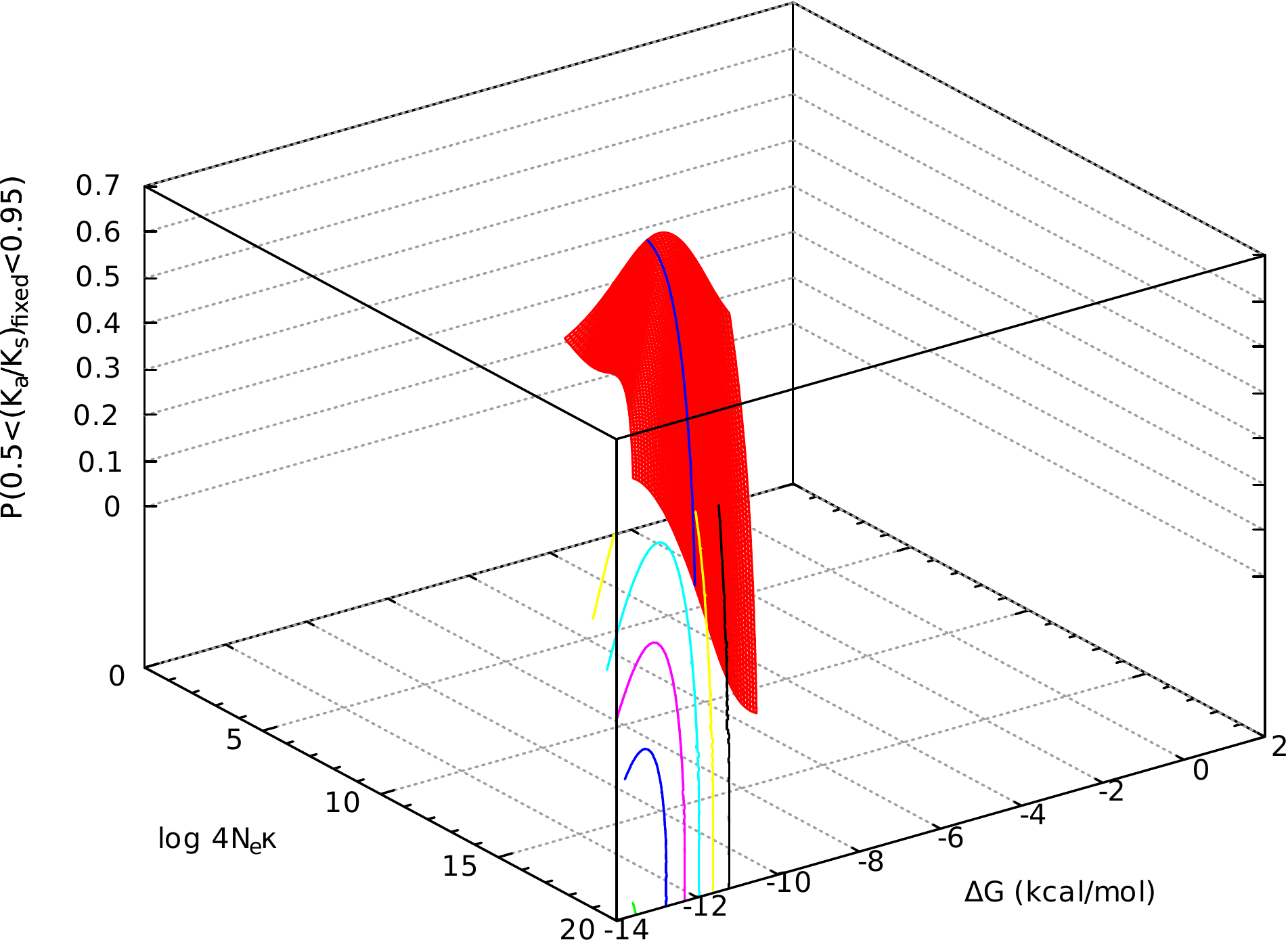}
\includegraphics*[width=90mm,angle=0]{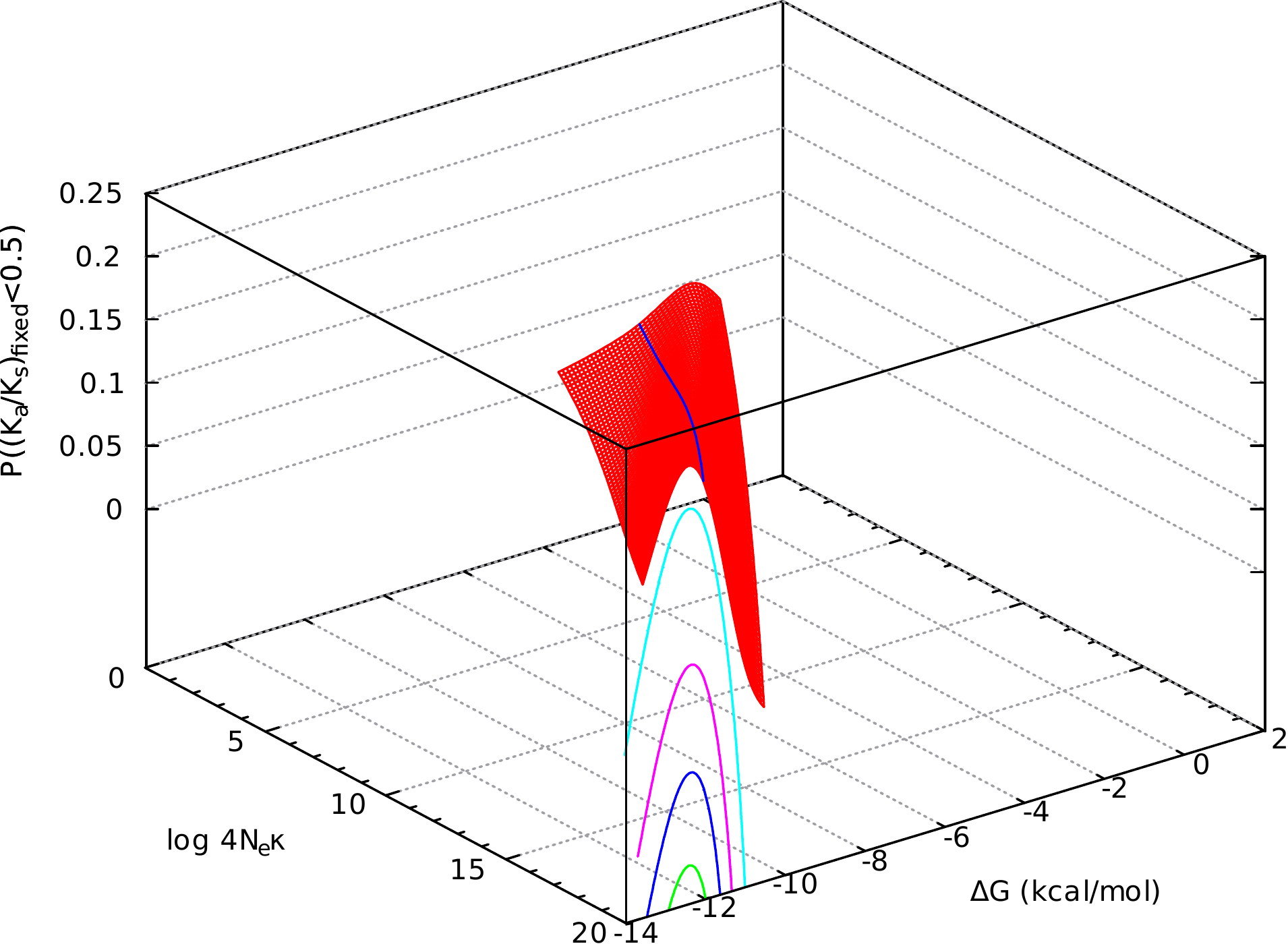}
}
} 
\vspace*{1em}
\caption{
\label{sfig: prob_of_each_selection_category_in_fixed_mutants;_dependence_on_kappa_and_dG}
\FigureLegends{
\BF{Dependence of the probability of each selection category in fixed mutants on $4N_e\kappa$ and $\Delta G$.}
A blue line on the surface grid shows $\Delta G=\Delta G_e$, which is the equilibrium value of $\Delta G$
in protein evolution.
The range of $\Delta G$ shown in the figures is
$| \Delta G - \Delta G_e | < 2 \cdot \Delta\Delta G^{\script{sd}}_{\script{fixed}}$,
where $\Delta\Delta G^{\script{sd}}_{\script{fixed}}$ is the standard deviation of $\Delta\Delta G$ over fixed mutants at $ \Delta G = \Delta G_e$.
Arbitrarily, the value of $K_a/K_s$ is categorized into four classes;
negative, slightly negative, nearly neutral, and positive selection categories in which
$K_a/K_s$ is within the ranges of
$K_a/K_s \leq 0.5$, $0.5< K_a/K_s \leq 0.95$, $0.95< K_a/K_s \leq 1.05$, and $ 1.05 < K_a/K_s$,
respectively.
$\theta = 0.53$
is employed.
The kcal/mol unit is used for $\Delta G$.
} 
}
\end{figure*}

\FigureInLegends{\newpage}

\begin{figure*}[ht]
\FigureInLegends{
\centerline{
\includegraphics*[width=80mm,angle=0]{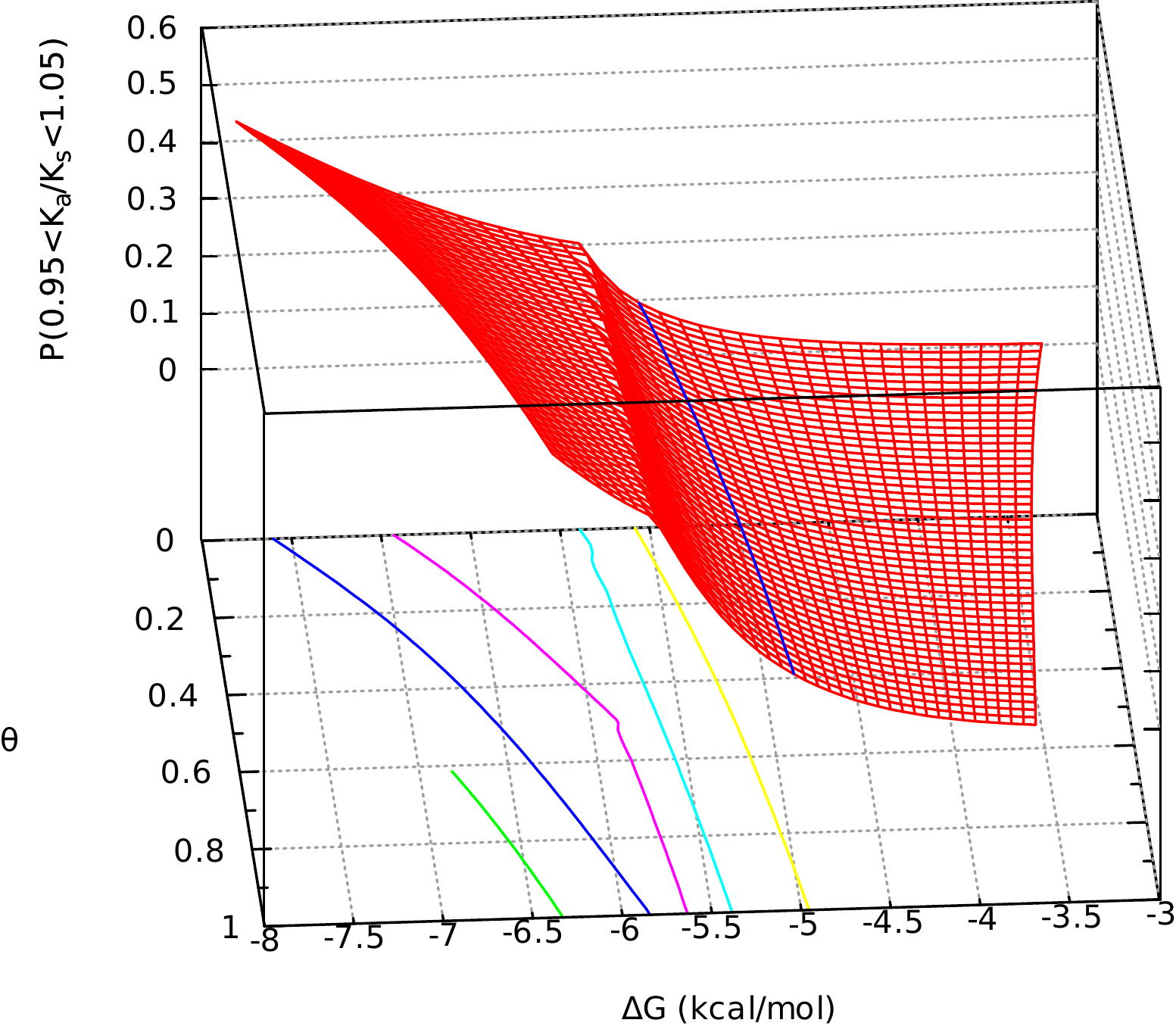}
\includegraphics*[width=90mm,angle=0]{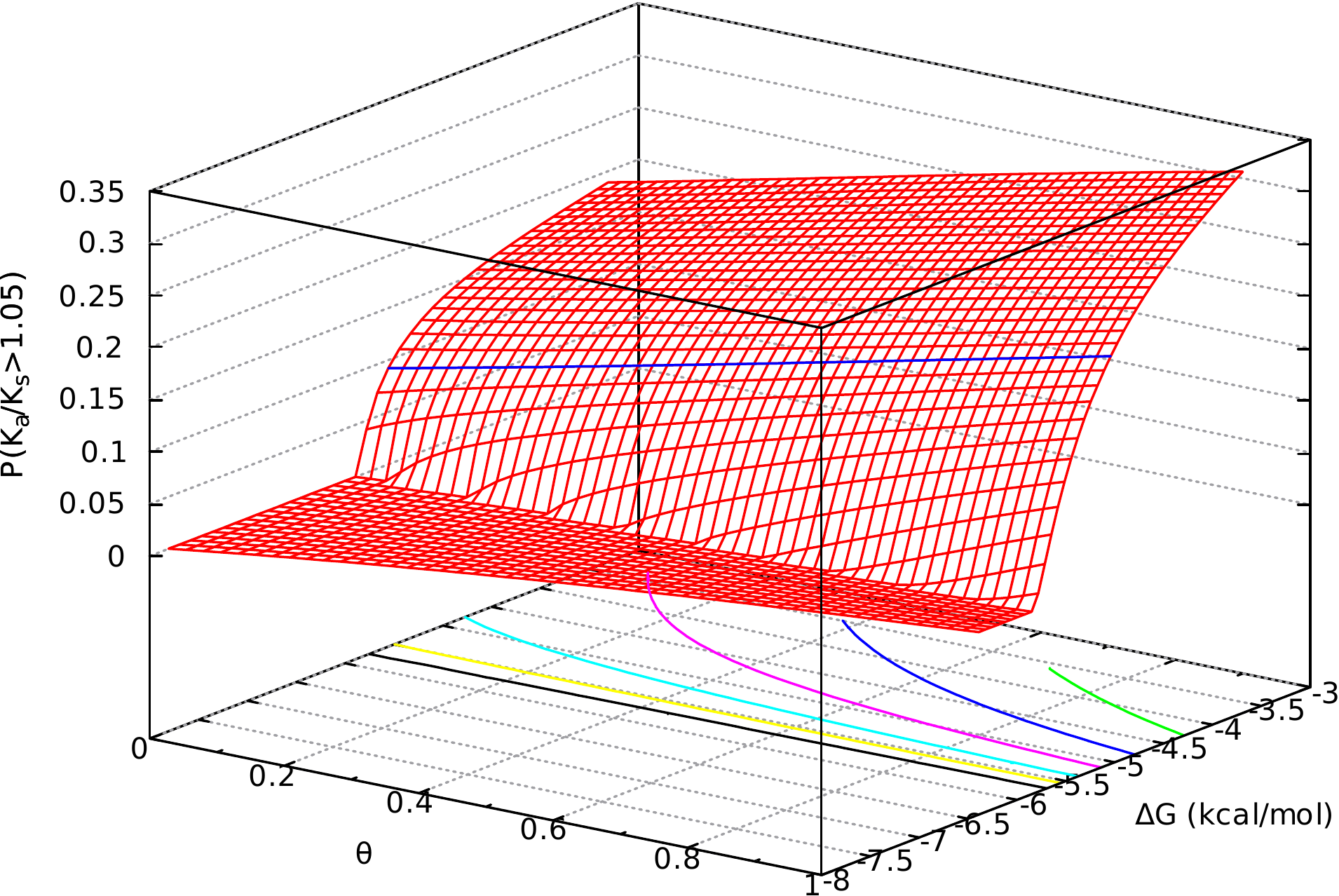}
}
\centerline{
\includegraphics*[width=83mm,angle=0]{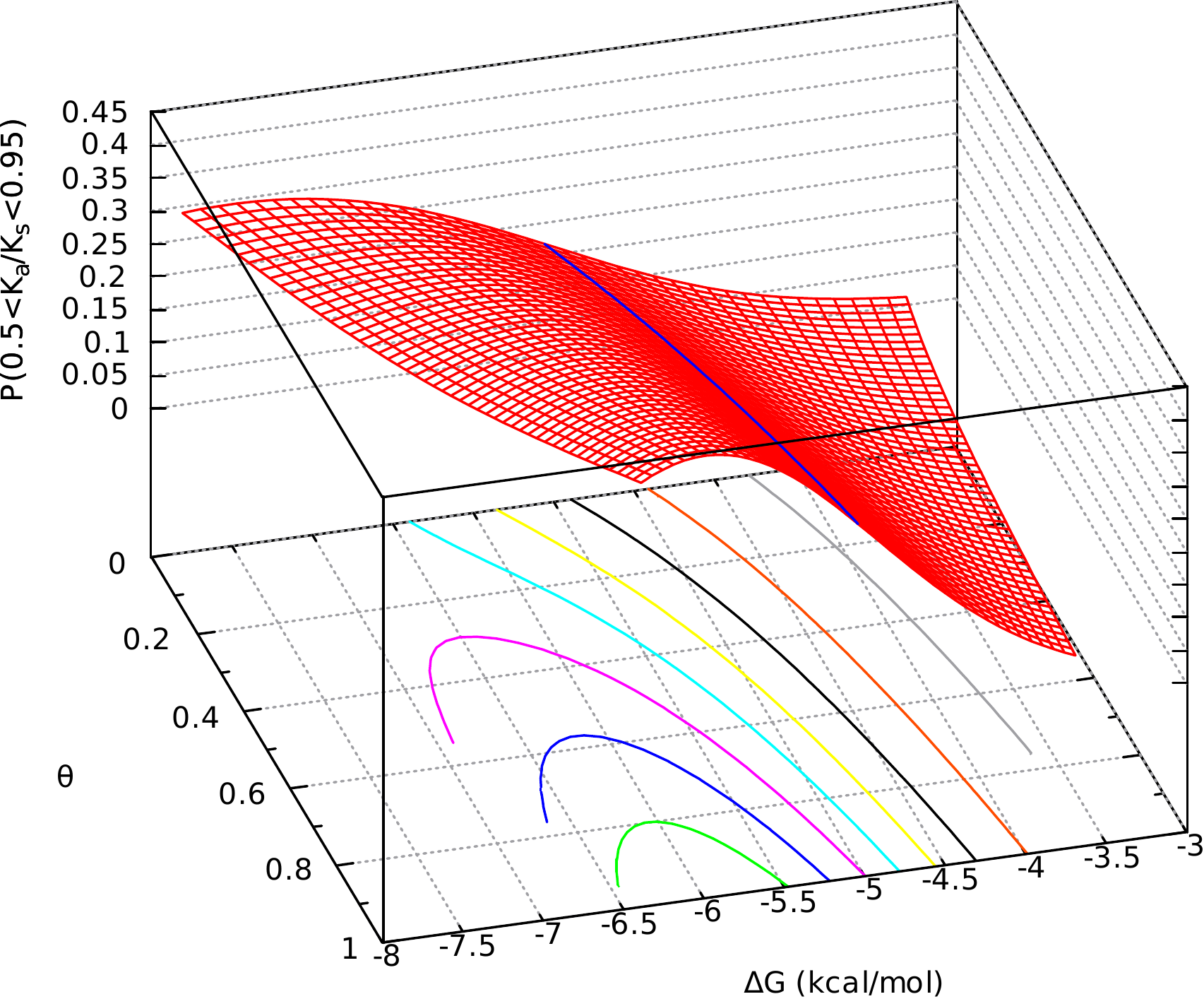}
\includegraphics*[width=90mm,angle=0]{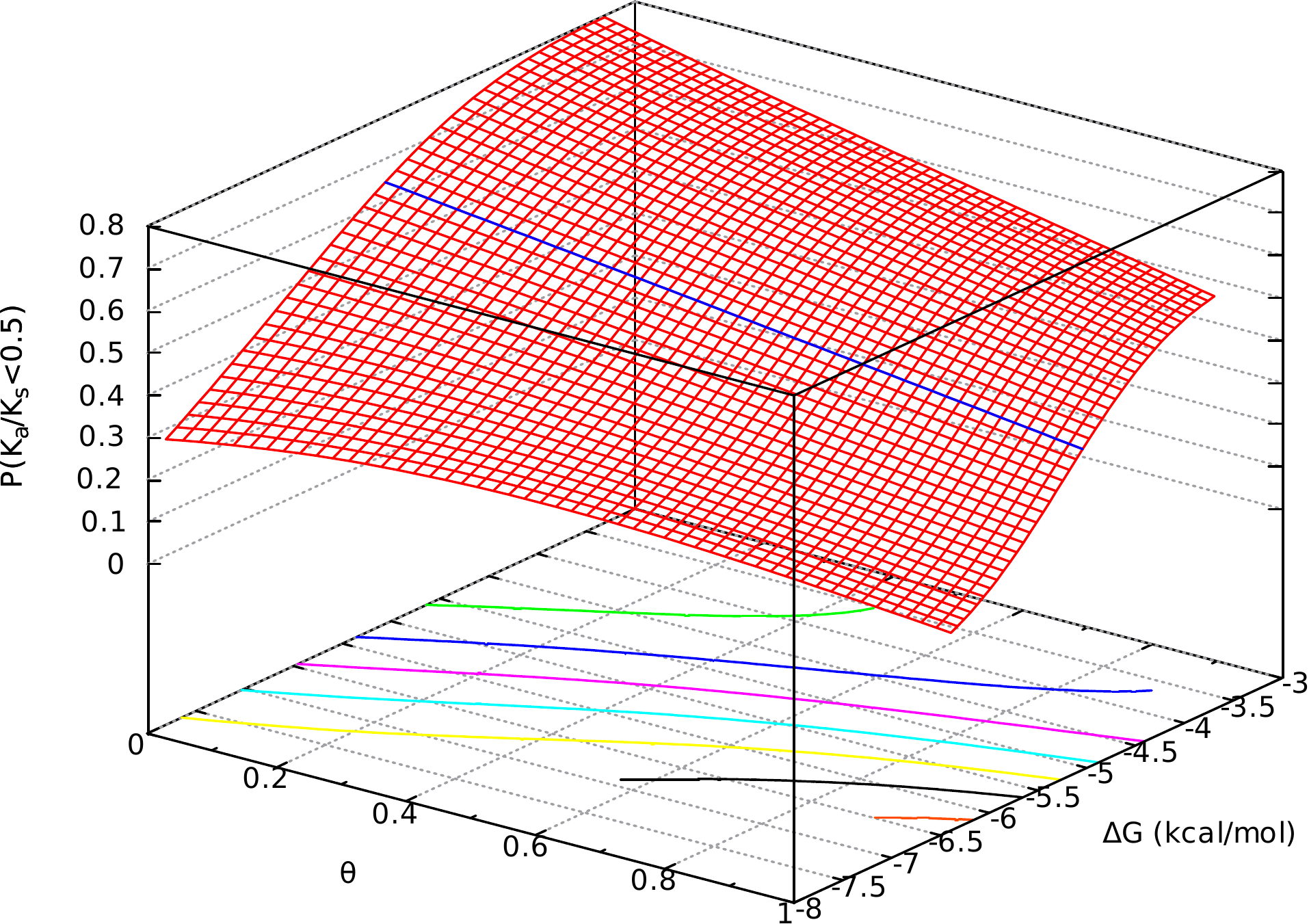}
}
} 
\vspace*{1em}
\caption{
\label{sfig: prob_of_each_selection_category_in_all_mutants;_dependence_on_theta_and_dG}
\FigureLegends{
\BF{Dependence of the probability of each selection category in all mutants on $\theta$ and $\Delta G$.}
A blue line on the surface grid shows $\Delta G=\Delta G_e$, which is the equilibrium value of $\Delta G$
in protein evolution.
The range of $\Delta G$ shown in the figures is
$| \Delta G - \Delta G_e | < 2 \cdot \Delta\Delta G^{\script{sd}}_{\script{fixed}}$,
where $\Delta\Delta G^{\script{sd}}_{\script{fixed}}$ is the standard deviation of $\Delta\Delta G$ over fixed mutants at $ \Delta G = \Delta G_e$.
Arbitrarily, the value of $K_a/K_s$ is categorized into four classes;
negative, slightly negative, nearly neutral, and positive selection categories in which
$K_a/K_s$ is within the ranges of
$K_a/K_s \leq 0.5$, $0.5< K_a/K_s \leq 0.95$, $0.95< K_a/K_s \leq 1.05$, and $ 1.05 < K_a/K_s$,
respectively.
$\log 4N_e\kappa = 7.550$ 
is employed.
The kcal/mol unit is used for $\Delta G$.
} 
}
\end{figure*}

\FigureInLegends{\newpage}

\begin{figure*}[ht]
\FigureInLegends{
\centerline{
\includegraphics*[width=83mm,angle=0]{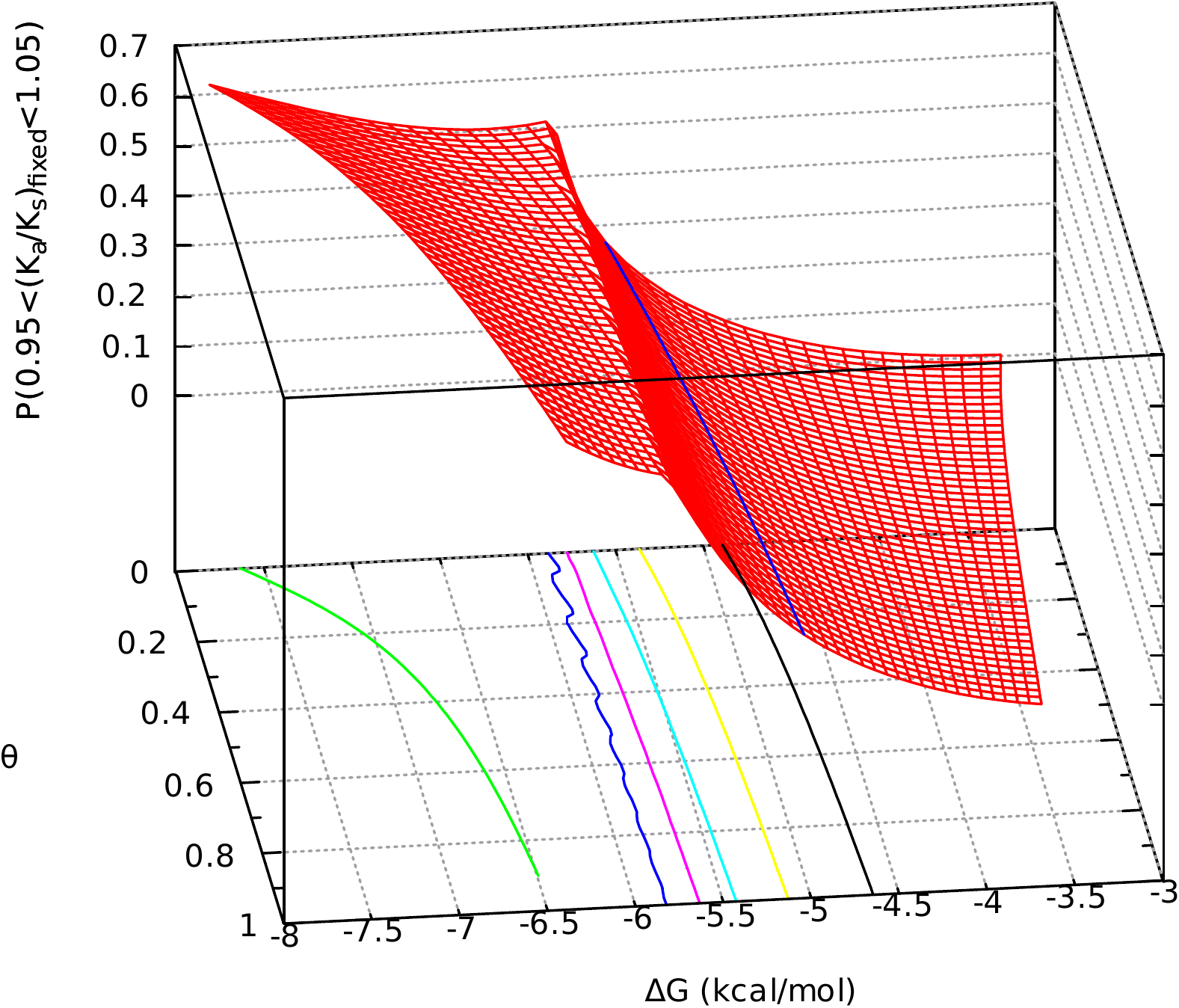}
\includegraphics*[width=90mm,angle=0]{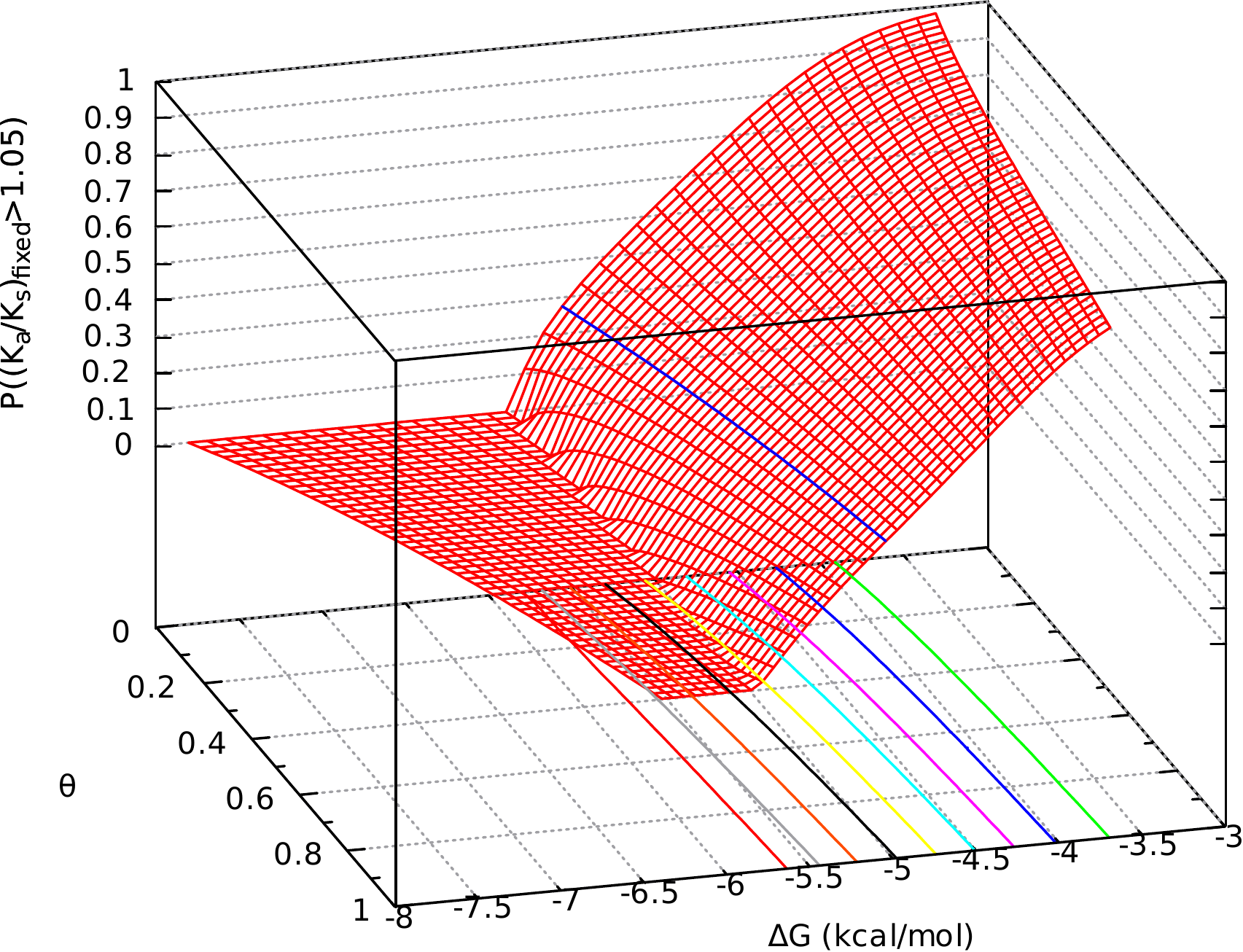}
}
\centerline{
\includegraphics*[width=89mm,angle=0]{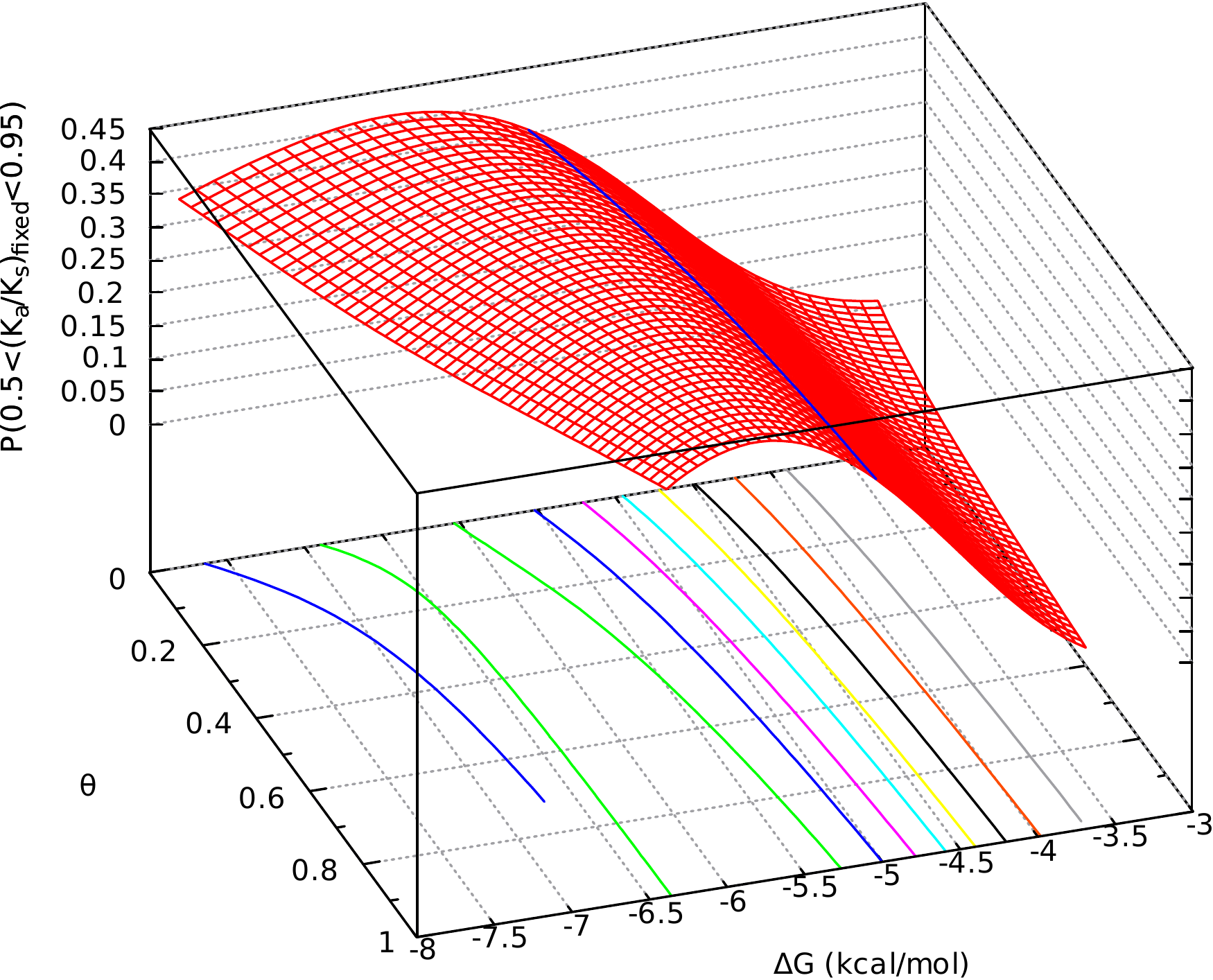}
\includegraphics*[width=90mm,angle=0]{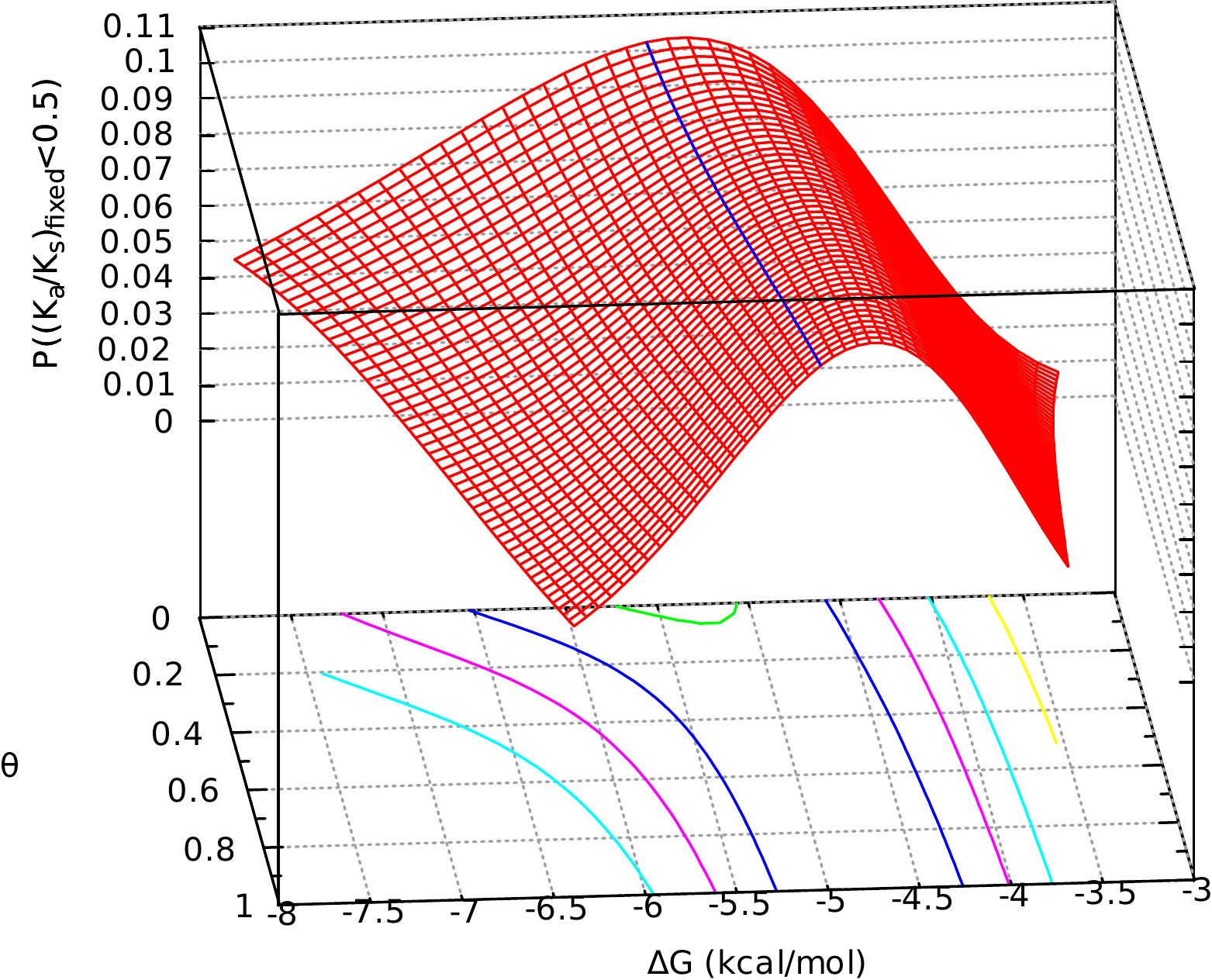}
}
} 
\vspace*{1em}
\caption{
\label{sfig: prob_of_each_selection_category_in_fixed_mutants;_dependence_on_theta_and_dG}
\FigureLegends{
\BF{Dependence of the probability of each selection category in fixed mutants on $\theta$ and $\Delta G$.}
The blue line on the surface grid shows $\Delta G=\Delta G_e$, which is the equilibrium value of $\Delta G$
in protein evolution.
The range of $\Delta G$ shown in the figures is
$| \Delta G - \Delta G_e | < 2 \cdot \Delta\Delta G^{\script{sd}}_{\script{fixed}}$,
where $\Delta\Delta G^{\script{sd}}_{\script{fixed}}$ is the standard deviation of $\Delta\Delta G$ over fixed mutants at $ \Delta G = \Delta G_e$.
Arbitrarily, the value of $K_a/K_s$ is categorized into four classes;
negative, slightly negative, nearly neutral, and positive selection categories in which
$K_a/K_s$ is within the ranges of
$K_a/K_s \leq 0.5$, $0.5< K_a/K_s \leq 0.95$, $0.95< K_a/K_s \leq 1.05$, and $ 1.05 < K_a/K_s$,
respectively.
$\log 4N_e\kappa = 7.550$
is employed.
The kcal/mol unit is used for $\Delta G$.
} 
}
\end{figure*}

\FigureInLegends{\clearpage\newpage}

\begin{figure*}[ht]
\FigureInLegends{
\centerline{
\includegraphics*[width=90mm,angle=0]{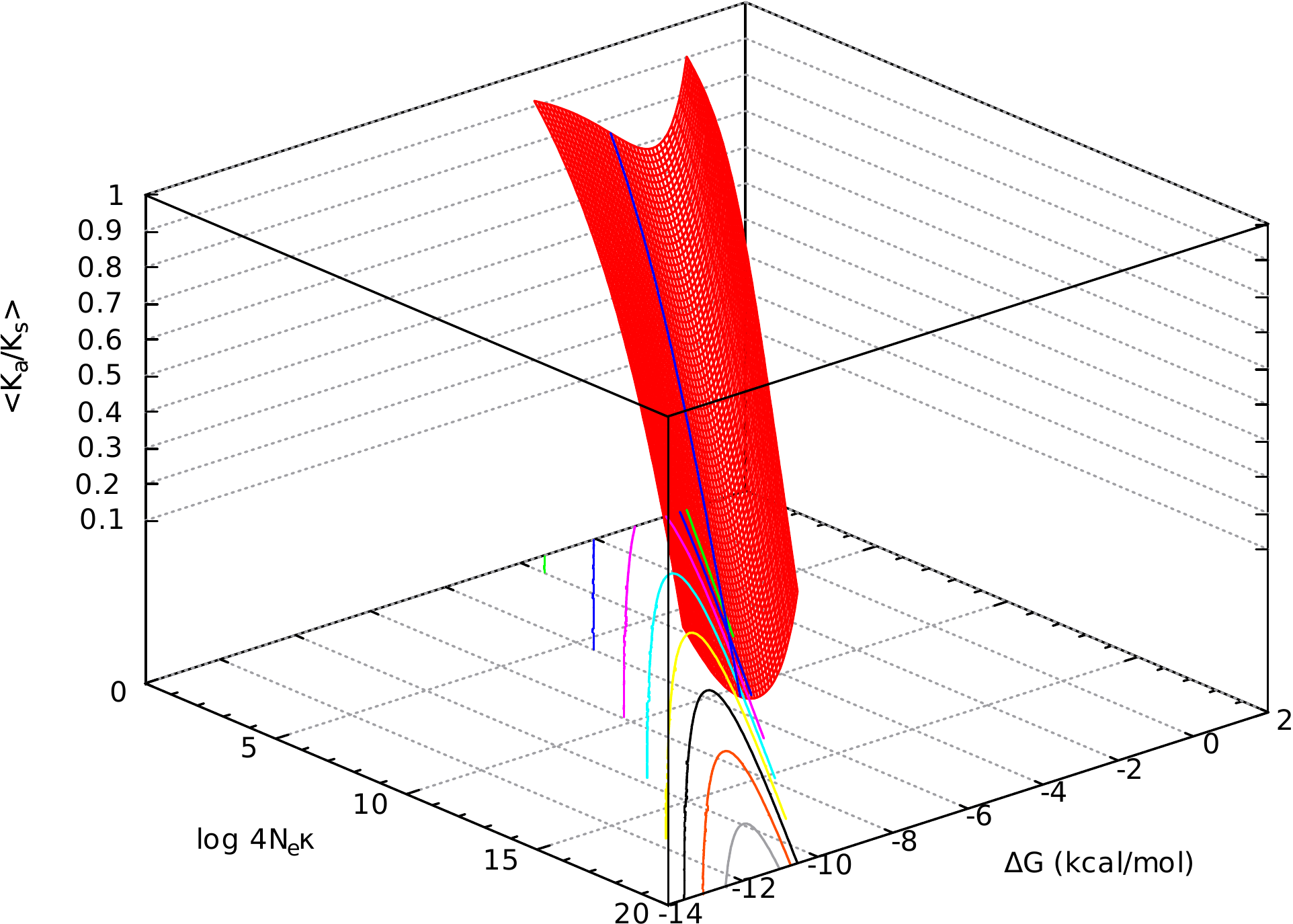}
\includegraphics*[width=85mm,angle=0]{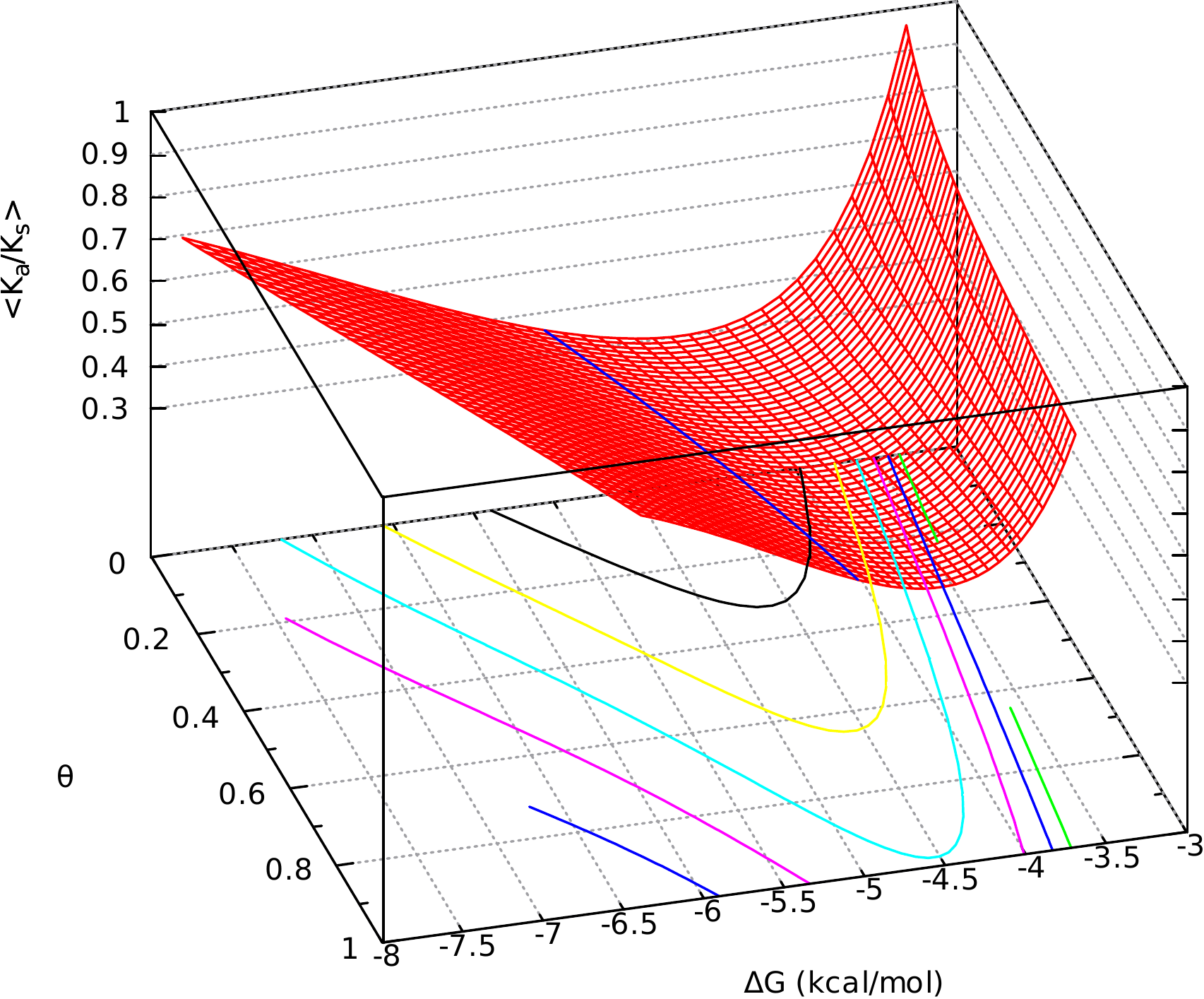}
}
\centerline{
\includegraphics*[width=88mm,angle=0]{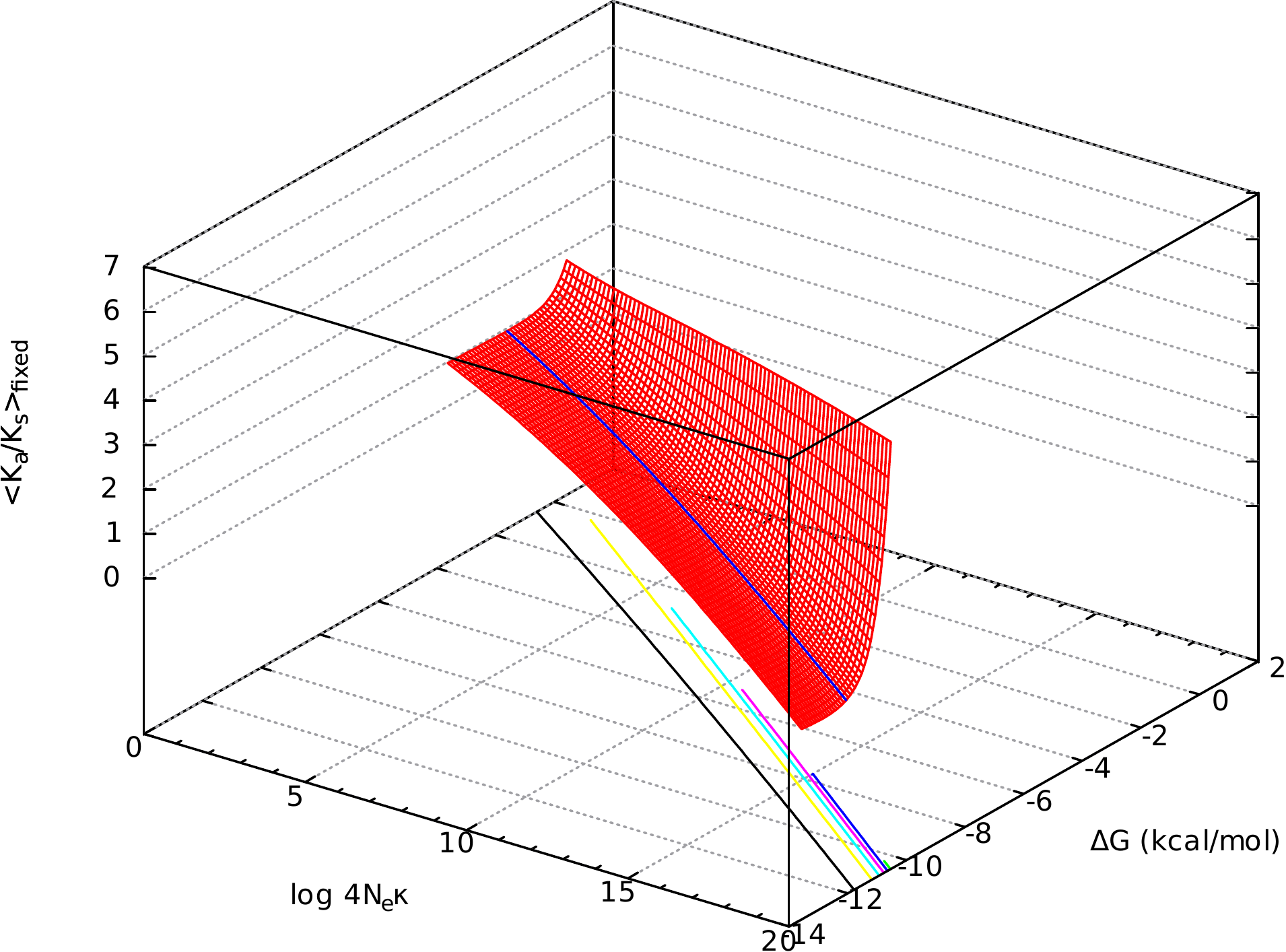}
\includegraphics*[width=90mm,angle=0]{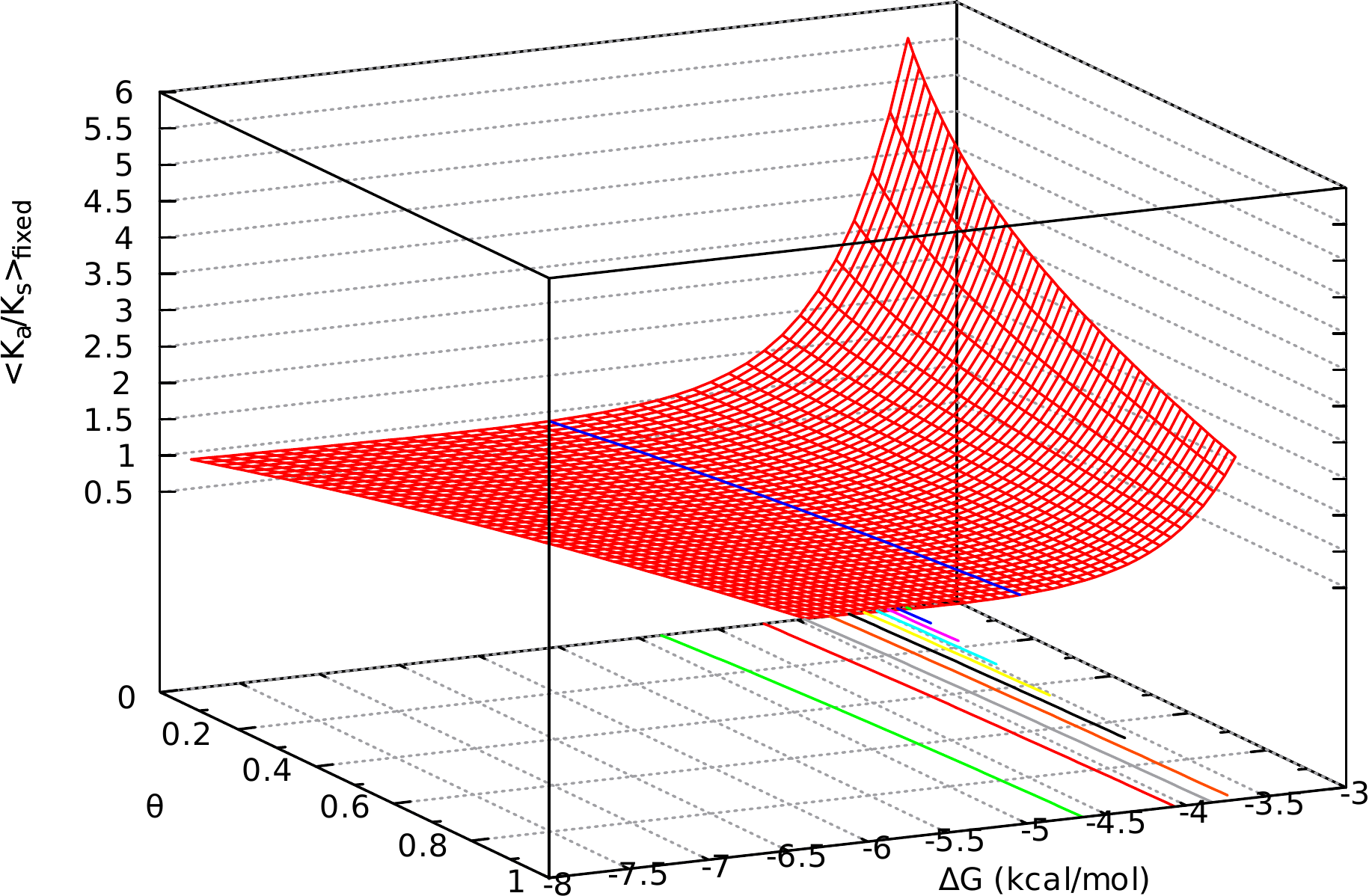}
}
} 
\vspace*{1em}
\caption{
\label{sfig: dependence_of_ave_Ka_over_Ks_on_dG}
\label{sfig: dependence_of_ave_Ka_over_Ks_fixed_on_dG}
\FigureLegends{
\BF{Dependence of the average of $K_a / K_s$ over all mutants or over fixed mutants only 
on protein stability, $\Delta G$, of the wild type.}
A blue line on the surface grid shows $\Delta G=\Delta G_e$, which is the equilibrium value of $\Delta G$
in protein evolution.
The range of $\Delta G$ shown in the figures is
$| \Delta G - \Delta G_e | < 2 \cdot \Delta\Delta G^{\script{sd}}_{\script{fixed}}$,
where $\Delta\Delta G^{\script{sd}}_{\script{fixed}}$ is the standard deviation of $\Delta\Delta G$ over fixed mutants at $\Delta G = \Delta G_e$.
Unless specified, 
$\log 4N_e\kappa = 7.550$ and $\theta = 0.53$
are employed.
The kcal/mol unit is used for $\Delta G$.
} 
}
\end{figure*}

\FigureInLegends{\newpage}

\begin{figure*}[ht]
\FigureInLegends{
\centerline{
\includegraphics*[width=90mm,angle=0]{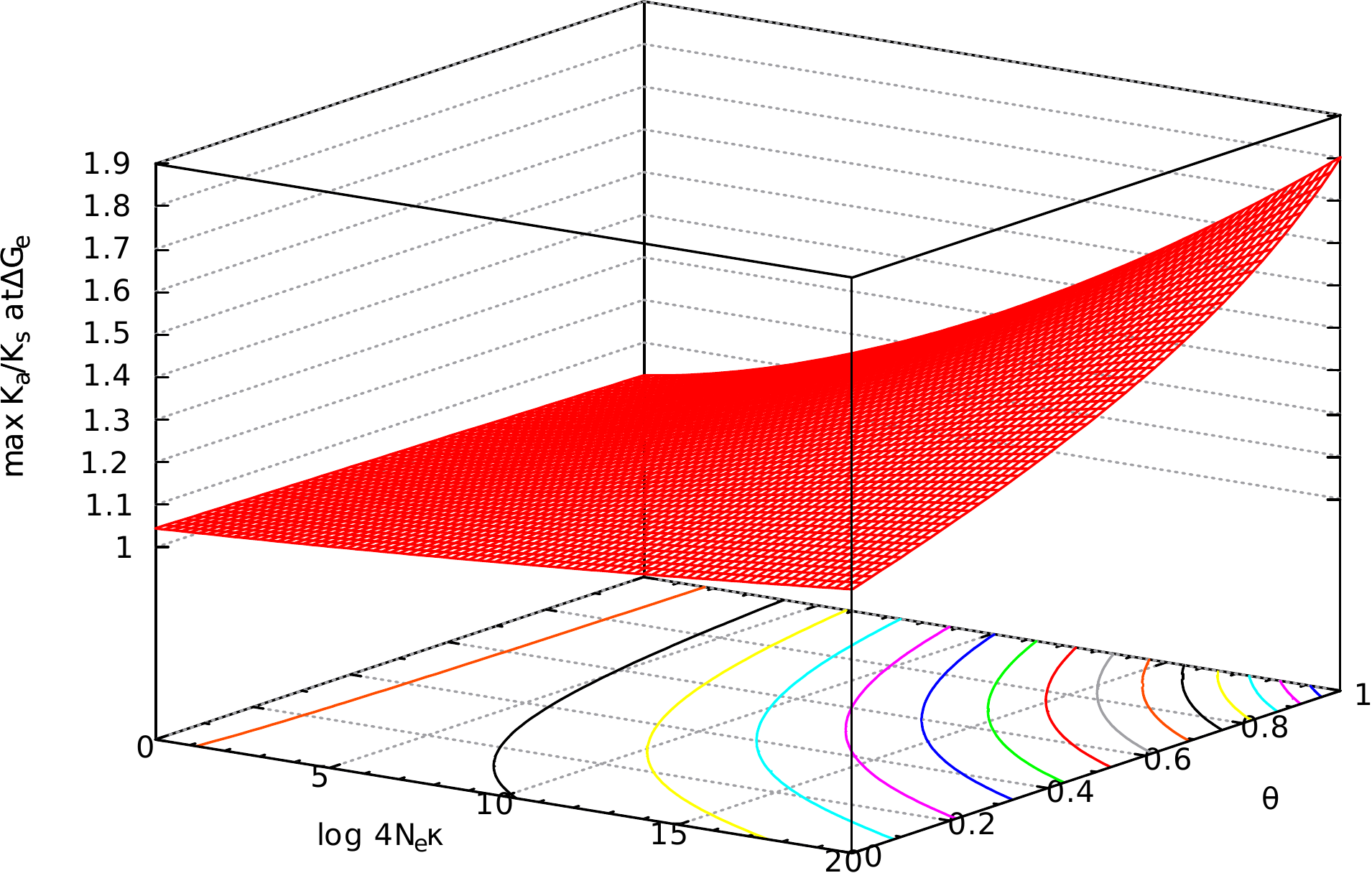}
\includegraphics*[width=90mm,angle=0]{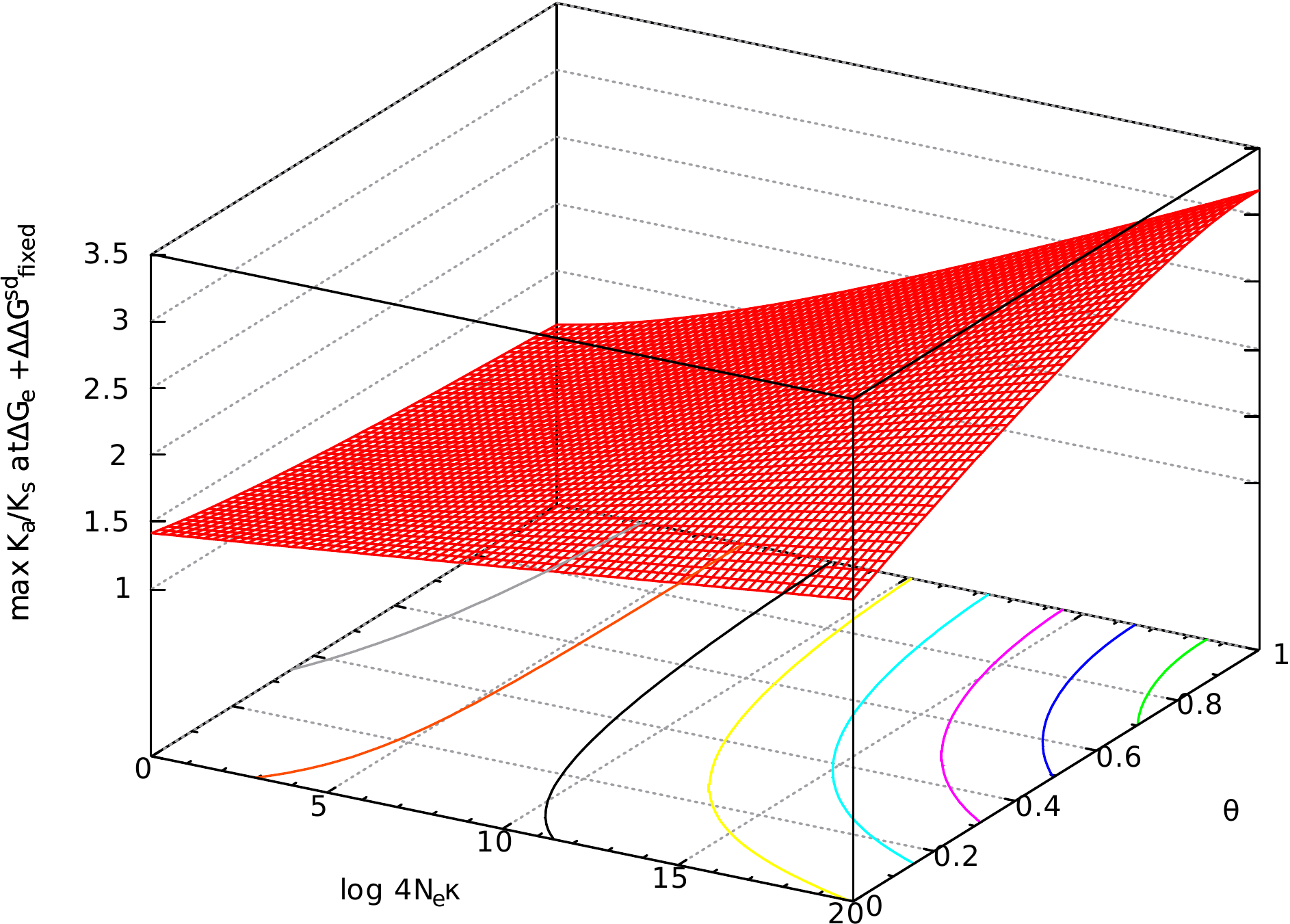}
}
} 
\vspace*{1em}
\caption{
\label{sfig: maximum_Ka_over_Ks_on_4Nekappa_and_theta}
\FigureLegends{
\BF{Dependence of $\max K_a / K_s$ on $4N_e\kappa$ and $\theta$.}
The maximum values of $K_a/K_s$, 
which correspond to the upper bound of selective advantage $s$ (\Eq{\ref{seq: s_max}}), 
at $\Delta G = \Delta G_e$ and at $\Delta G = \Delta G_e + \Delta\Delta G^{\script{sd}}_{\script{fixed}}$
are plotted as a function of $\log 4N_e\kappa$ and $\theta$; 
$\Delta\Delta G^{\script{sd}}_{\script{fixed}}$ is the standard deviation
of $\Delta\Delta G$ over fixed mutants at $\Delta G = \Delta G_e$.
} 
}
\end{figure*}

\FigureInLegends{\clearpage\newpage}

\begin{figure*}[ht]
\FigureInLegends{
\centerline{
\includegraphics*[width=86mm,angle=0]{FIGS_Corrected/max4Nes_on_kappa_and_T_type47}
\includegraphics*[width=90mm,angle=0]{FIGS/dGe_on_kappa_and_T_type47}
}
\centerline{
\includegraphics*[width=86mm,angle=0]{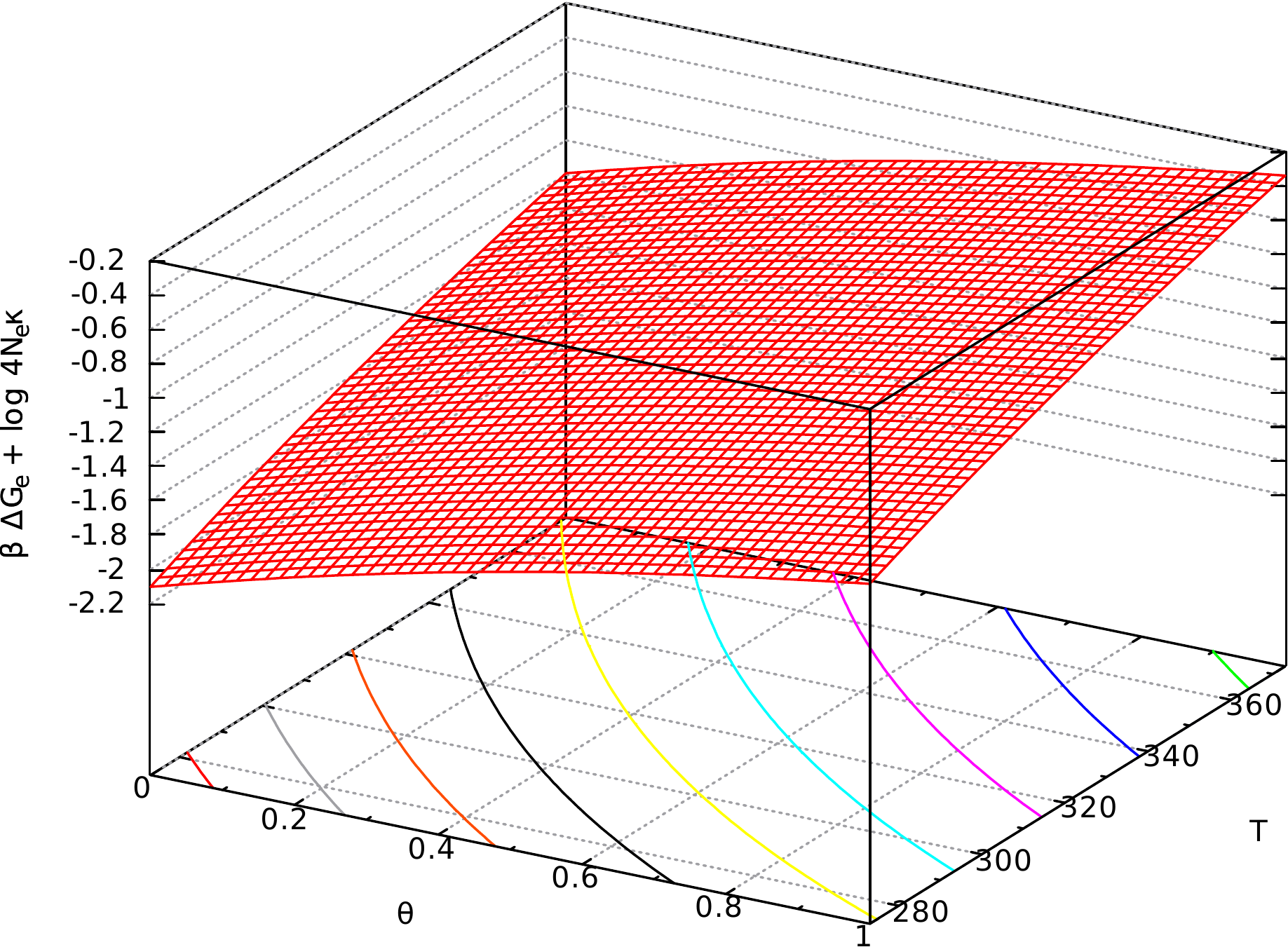}
\includegraphics*[width=88mm,angle=0]{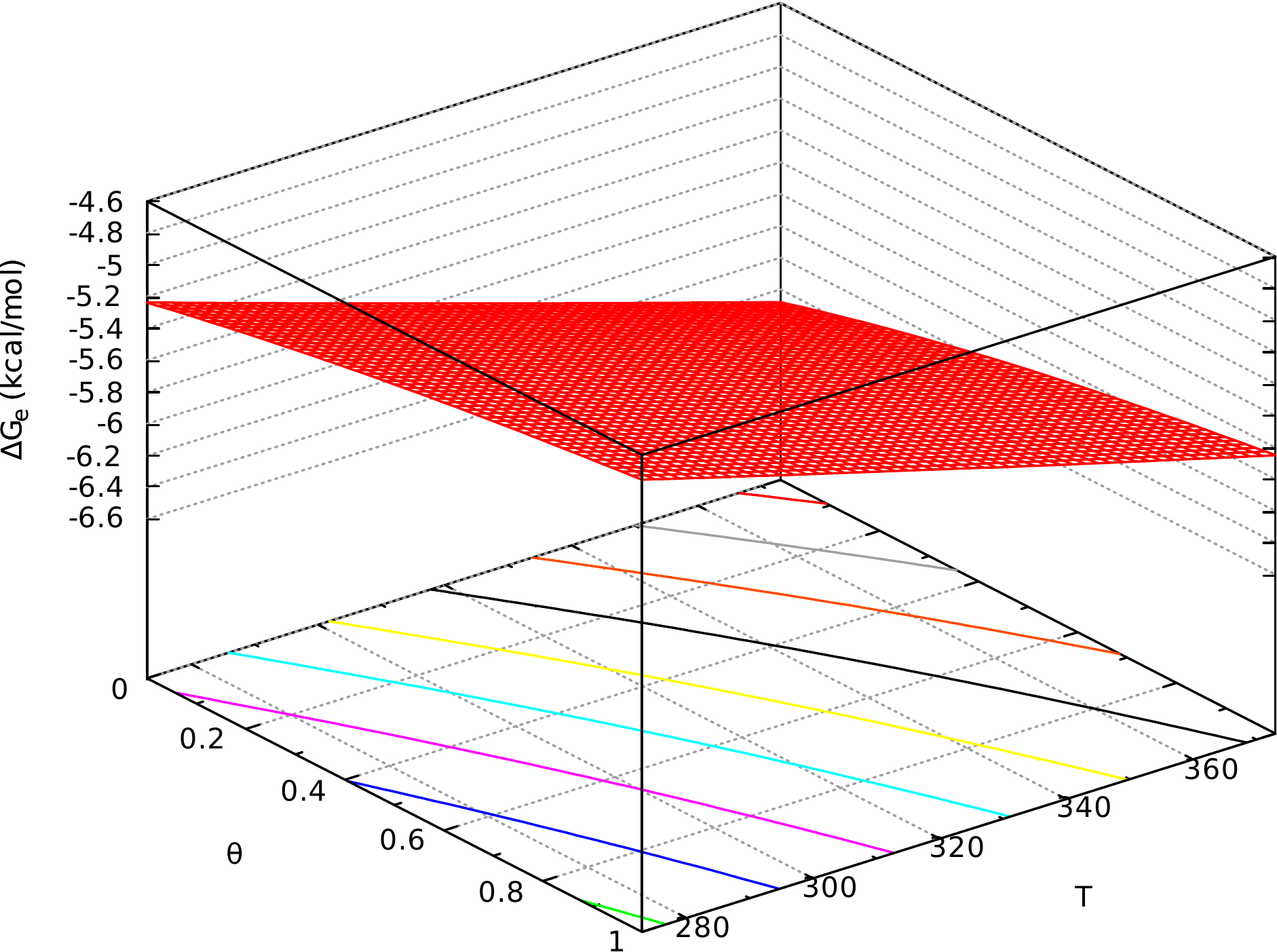}
}
} 
\vspace*{1em}
\caption{
\label{sfig: dependence_of_max4Nes_on_4Nekappa_and_T}
\label{sfig: dependence_of_dGe_on_4Nekappa_and_T}
\label{sfig: dependence_of_max4Nes_on_theta_and_T}
\label{sfig: dependence_of_dGe_on_theta_and_T}
\FigureLegends{
\Red{
\BF{Dependence of equilibrium stability, $\Delta G_e$, on parameters, $4N_e\kappa$, $\theta$ and $T$.}
} 
$\Delta G_e$ is the equilibrium value of folding free energy, $\Delta G$,
in protein evolution.
$T$ is absolute temperature; $\beta = 1 / kT$, where $k$ is the Boltzmann constant.
\EQUATIONS{\Ref{eq: def_bi-Gaussian}, \Ref{eq: def_ms} and \Ref{eq: def_mc}} are assumed for
the distribution of $\Delta\Delta G$ and its dependency on $\Delta G$; they are assumed to be independent of $T$.
Unless specified, 
$\log 4N_e\kappa = 7.550$ and $\theta = 0.53$
are employed.
The value of $\beta\Delta G_e + \log 4N_e\kappa$ is the upper bound of $\log 4N_e s$,
and would not depend on $\log 4N_e \kappa$ if the mean of $\Delta\Delta G$ in all arising mutants did not depend on $\Delta G$; 
see \Eq{\ref{seq: def_s}}.
The kcal/mol unit is used for $\Delta G_e$.
} 
}
\end{figure*}

\FigureInLegends{\clearpage\newpage}

\begin{figure*}[ht]
\FigureInLegends{
\centerline{
\includegraphics*[width=90mm,angle=0]{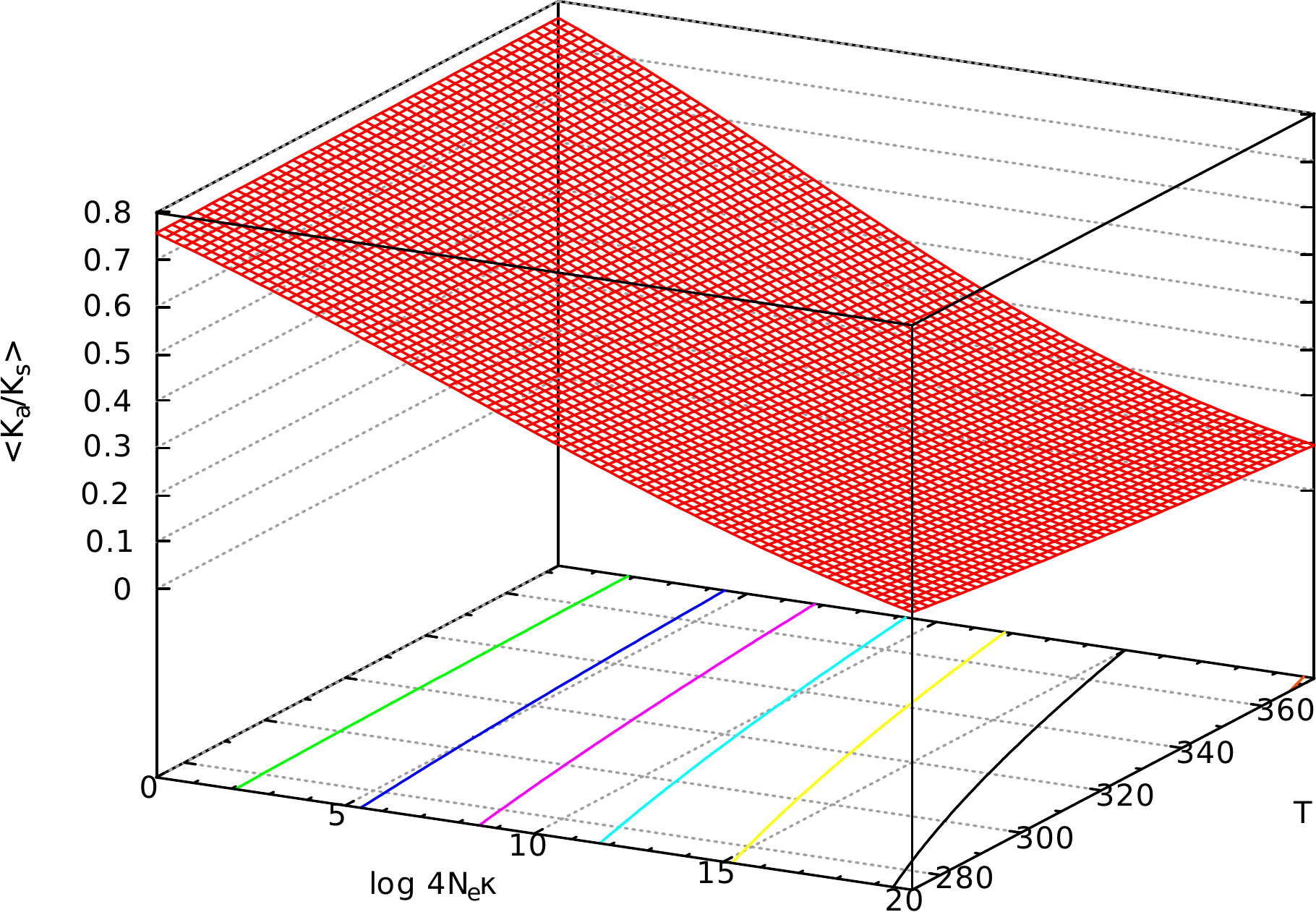}
\includegraphics*[width=86mm,angle=0]{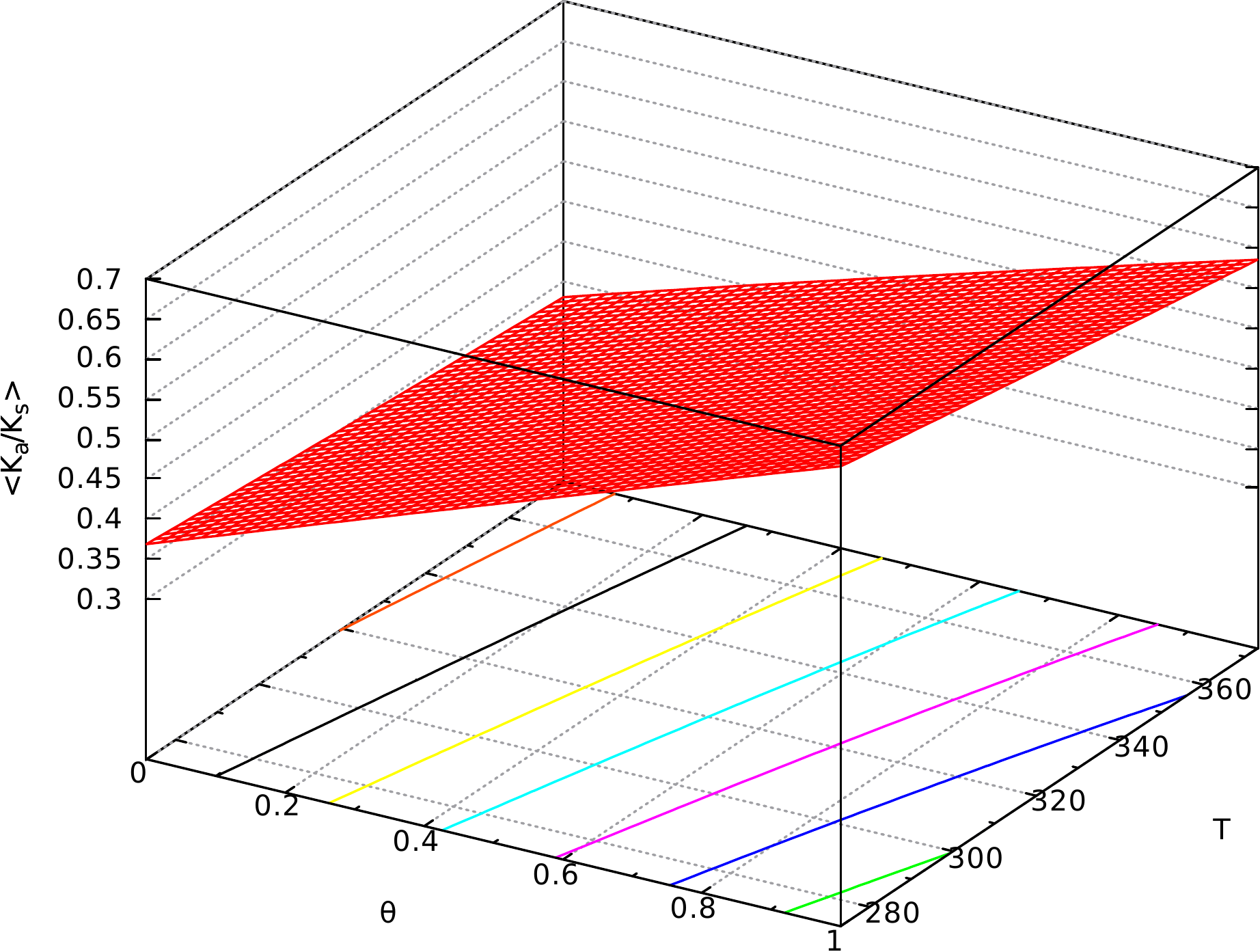}
}
\centerline{
\includegraphics*[width=90mm,angle=0]{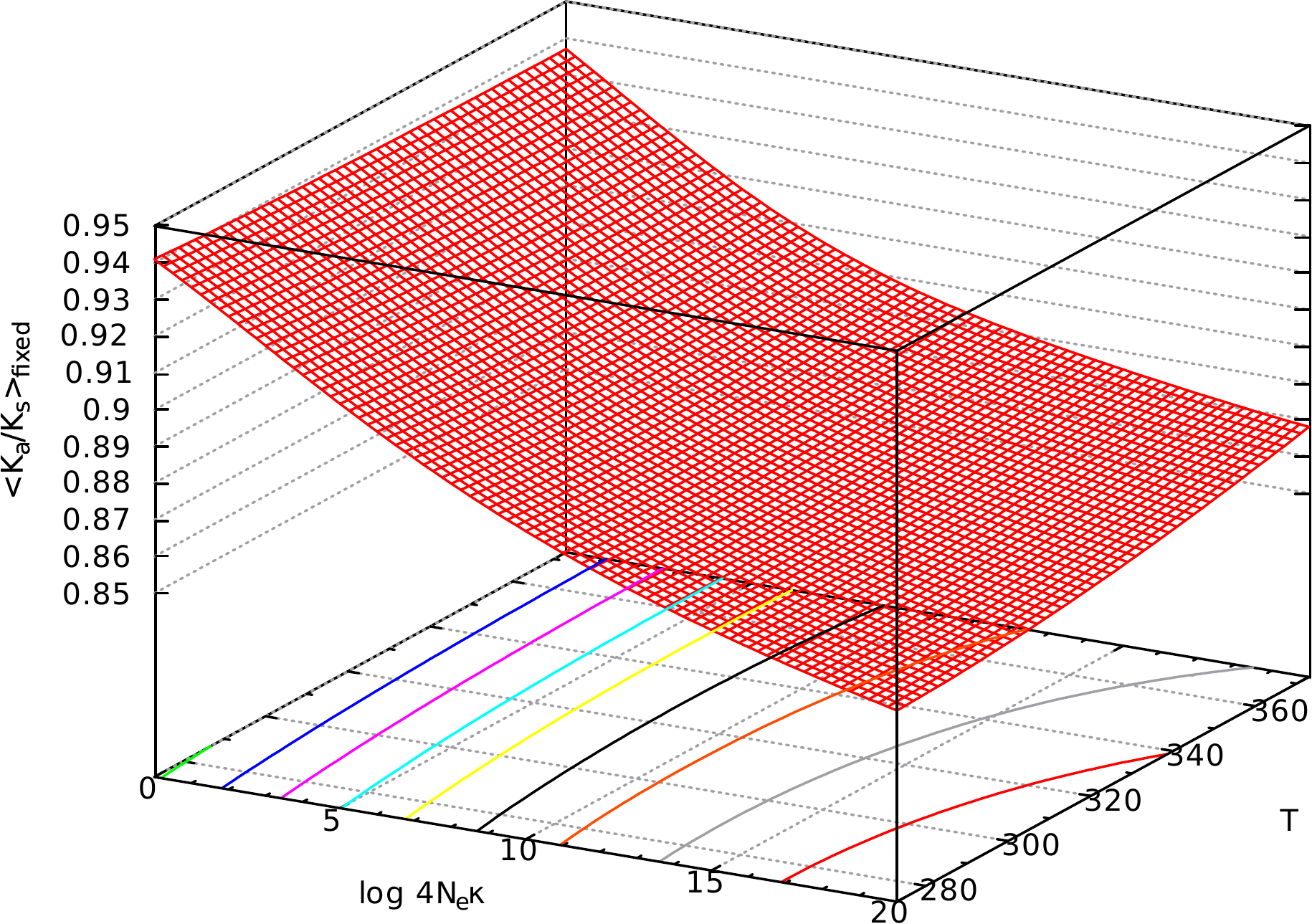}
\includegraphics*[width=86mm,angle=0]{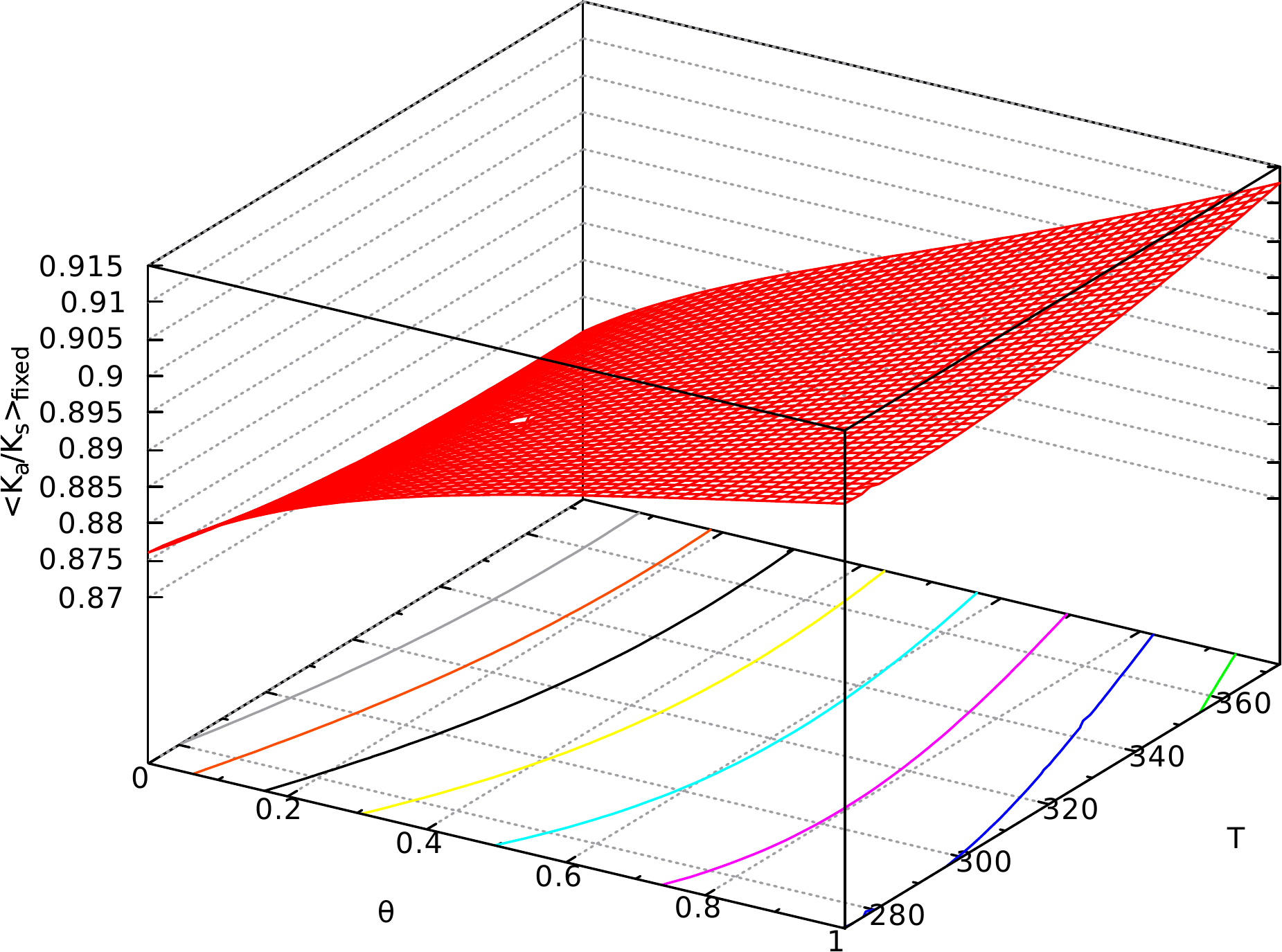}
}
} 
\vspace*{1em}
\caption{
\label{sfig: dependence_of_ave_Ka_over_Ks_on_4Nekappa_and_T}
\label{sfig: dependence_of_ave_Ka_over_Ks_on_theta_and_T}
\label{sfig: dependence_of_ave_Ka_over_Ks_fixed_at_4Nekappa_and_T}
\label{sfig: dependence_of_ave_Ka_over_Ks_fixed_at_theta_and_T}
\FigureLegends{
\Red{
\BF{The average of $K_a / K_s$ over all mutants or over fixed mutants only at equilibrium of protein stability, $\Delta G = \Delta G_e$: Dependence on temperature.}
} 
$T$ is absolute temperature.
\EQUATIONS{\Ref{eq: def_bi-Gaussian}, \Ref{eq: def_ms} and \Ref{eq: def_mc}} are assumed for
the distribution of $\Delta\Delta G$ and its dependency on $\Delta G$; they are assumed to be independent of $T$.
Unless specified, 
$\log 4N_e\kappa = 7.550$ and $\theta = 0.53$
are employed.
} 
}
\end{figure*}

\FigureInLegends{\clearpage\newpage}

\begin{figure*}[ht]
\FigureInLegends{
\centerline{
\includegraphics*[width=90mm,angle=0]{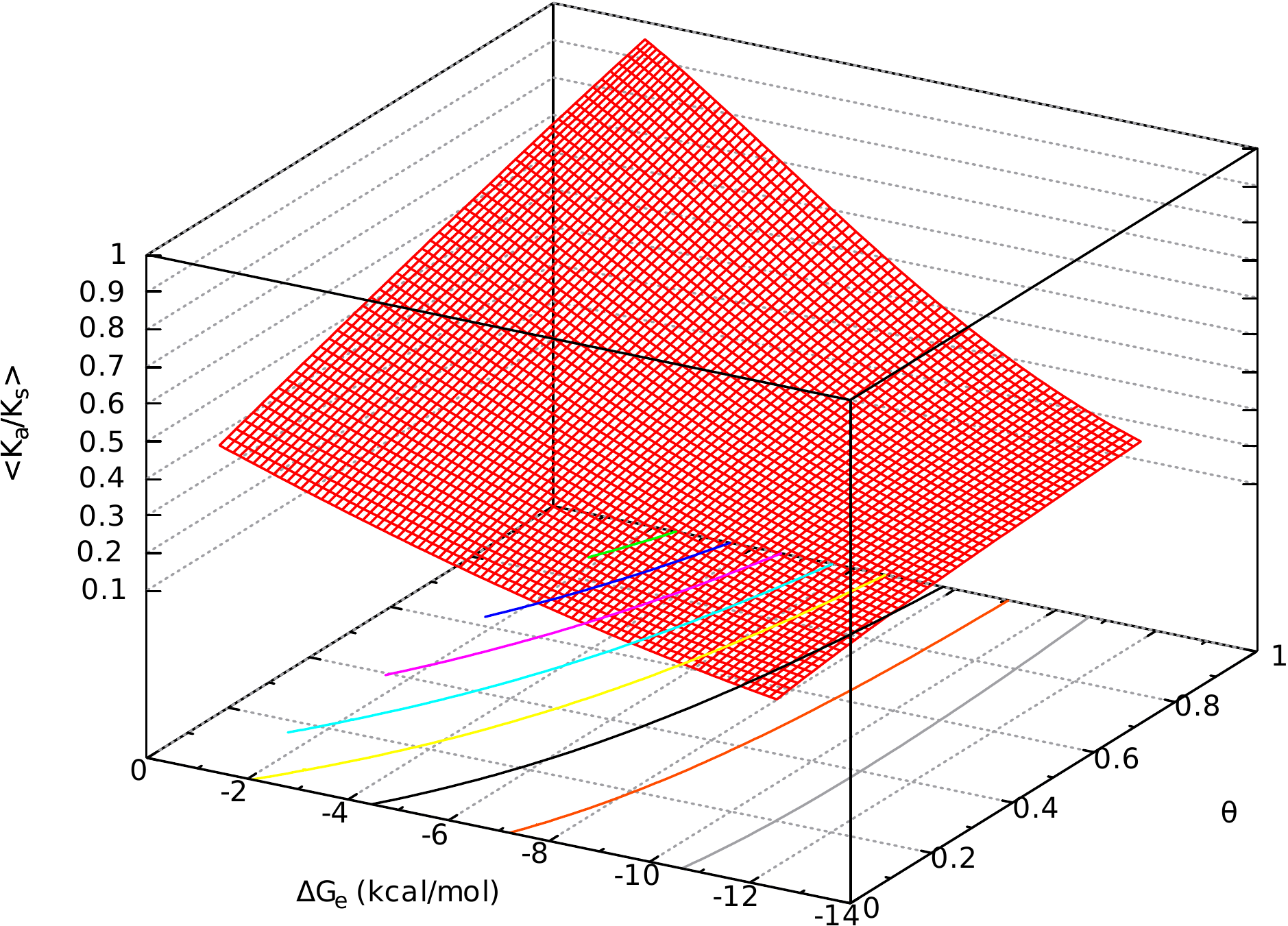}
}
} 
\vspace*{1em}
\caption{
\label{sfig: dependence_of_ave_Ka_over_Ks_on_dGe_and_theta}
\FigureLegends{
\BF{The average of $K_a / K_s$ over all mutants as a function of $\Delta G_e$ and $\theta$.}
} 
}
\end{figure*}

} 

\clearpage
\newpage
\pagestyle{empty}

\renewcommand{\FigureInLegends}[1]{#1}
\renewcommand{\FigureLegends}[1]{}
\setcounter{figure}{0}

\SupFig{

\FigureInLegends{\clearpage\newpage}

\begin{figure*}[ht]
\FigureInLegends{
\centerline{
\includegraphics*[width=90mm,angle=0]{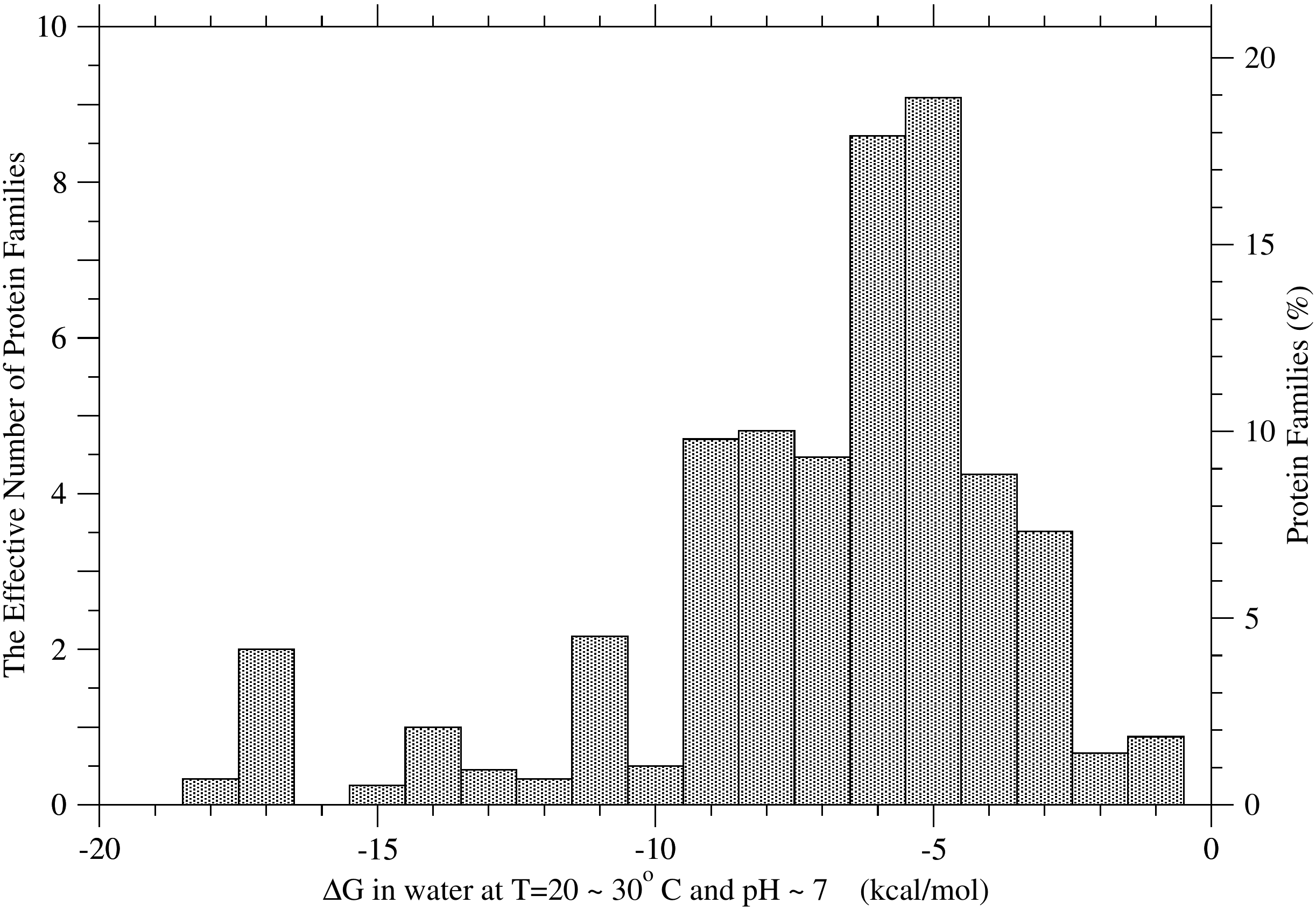}
}
} 
\vspace*{1em}
\caption{
\label{sfig: observed_distribution_of_protein_stabilities}
\FigureLegends{
\Red{
\BF{Distribution of folding free energies of monomeric protein families.}
} 
Stability data of monomeric proteins for which the item of dG\_H2O or dG
was obtained in the experimental condition of $6.7 \leq \textrm{pH} \leq 7.3$ 
and $20^{\circ}C \leq T \leq 30^{\circ}C$ and their folding-unfolding transition is two state and reversible
are extracted from the ProTherm\CITE{KBGPKUS:06};
in the case of dG only thermal transition data are used. 
Thermophilic proteins, and proteins observed with salts or additives are also removed.
An equal sampling weight is assigned to each species of homologous protein,
and the total sampling weight of each protein family is normalized to
one.  In the case in which multiple data exist for the same species of protein, 
its sampling weight is divided to each of the data. 
However, proteins whose stabilities are known 
may be samples biased from the protein universe.
The value, $\Delta G_e = -5.24$ kcal/mol, of equilibrium stability at the representative parameter values, $\log 4N_e\kappa = 7.550$
and $\theta = 0.53$, agrees with the most probable value of $\Delta G$ in the distribution above.
Also, the range of $\Delta G$ shown above is consistent with
that range, $-2$ to $-12.5$ kcal/mol, expected from the present model.
The kcal/mol unit is used for $\Delta G$.
A similar distribution was also compiled \CITE{ZCS:07}.
} 
}
\end{figure*}

\FigureInLegends{\clearpage\newpage}

\begin{figure*}[ht]
\FigureInLegends{
\centerline{
\includegraphics*[width=90mm,angle=0]{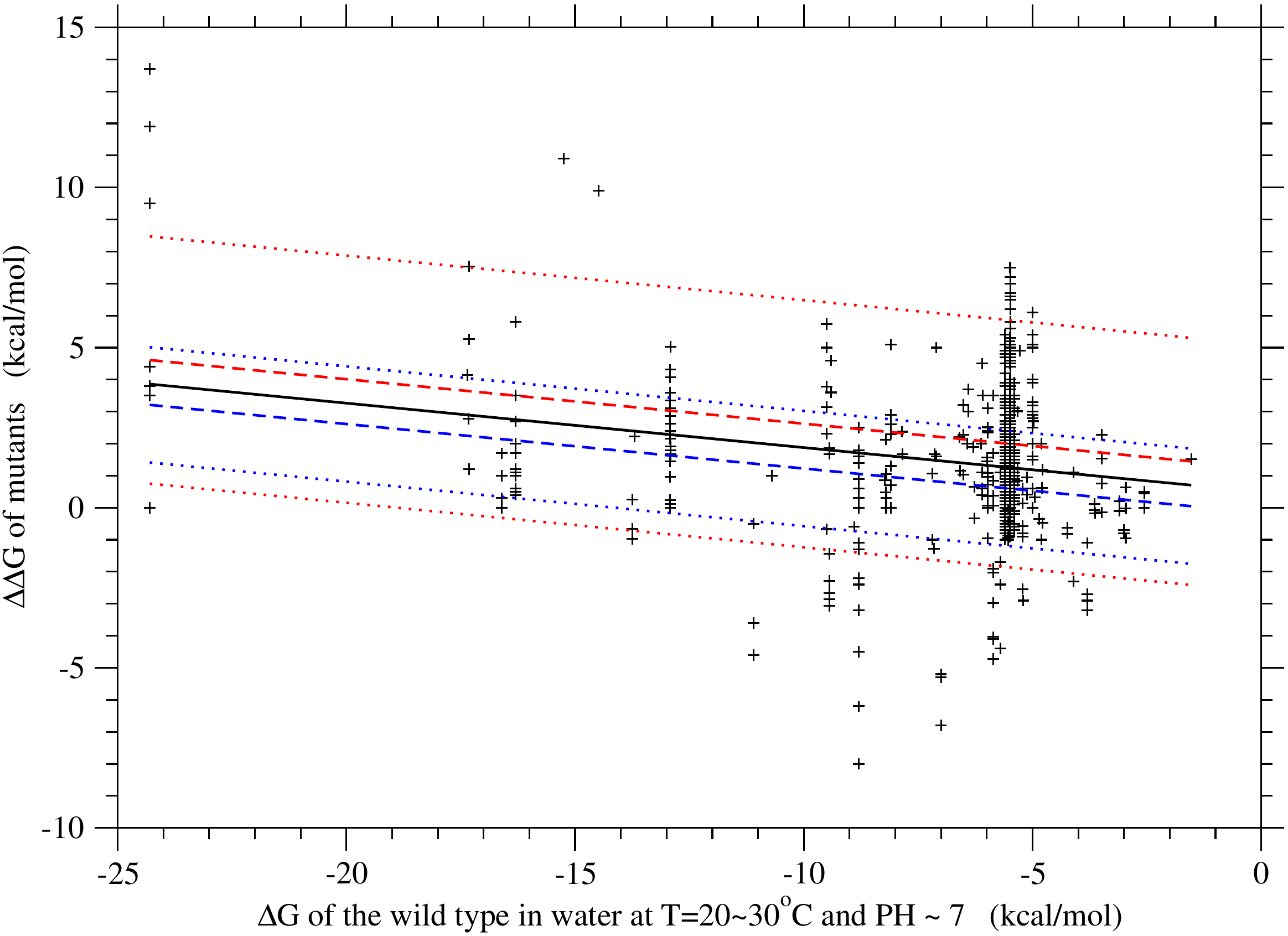}
}
} 
\vspace*{1em}
\caption{
\label{sfig: ddG_vs_dG}
\FigureLegends{
\Red{
\BF{Dependence of stability changes, $\Delta\Delta G$, due to single amino acid substitutions
on the protein stability, $\Delta G$, of the wild type.}
} 
A solid line shows the regression line, 
$\Delta\Delta G = -0.139 \Delta G + 0.490$; 
the correlation coefficient and p-value are equal to $-0.20$ and $< 10^{-7}$, respectively.
Broken lines show two means of bi-Gaussian distributions, $\mu_s$ in blue and $\mu_c$ in red.
Blue dotted lines show $\mu_s \pm 2 \sigma_s$ and red dotted lines $\mu_c \pm 2 \sigma_c$.
See \Eqs{\Ref{seq: def_bi-Gaussian}, \Ref{seq: def_ms} and \Ref{seq: def_mc}} for the bi-Gaussian distribution.
Stability data of single amino acid mutants for which the items dG\_H2O and ddG\_H2O or dG and ddG 
were obtained in the experimental condition of 
$6.7 \leq \textrm{pH} \leq 7.3$ and $20^{\circ}C \leq T \leq 30^{\circ}C$
and their folding-unfolding transitions are two state and reversible
are extracted from the ProTherm\CITE{KBGPKUS:06}.
In the case of dG only thermal transition data are used.
In the case in which multiple data exist for the same protein, only one of them
is used.
The kcal/mol unit is used for $\Delta\Delta G$ and $\Delta G$.
A similar distribution was also compiled \CITE{SRS:12}.
} 
}
\end{figure*}

\FigureInLegends{\newpage}

\begin{figure*}[ht]
\FigureInLegends{
\centerline{
\includegraphics*[width=90mm,angle=0]{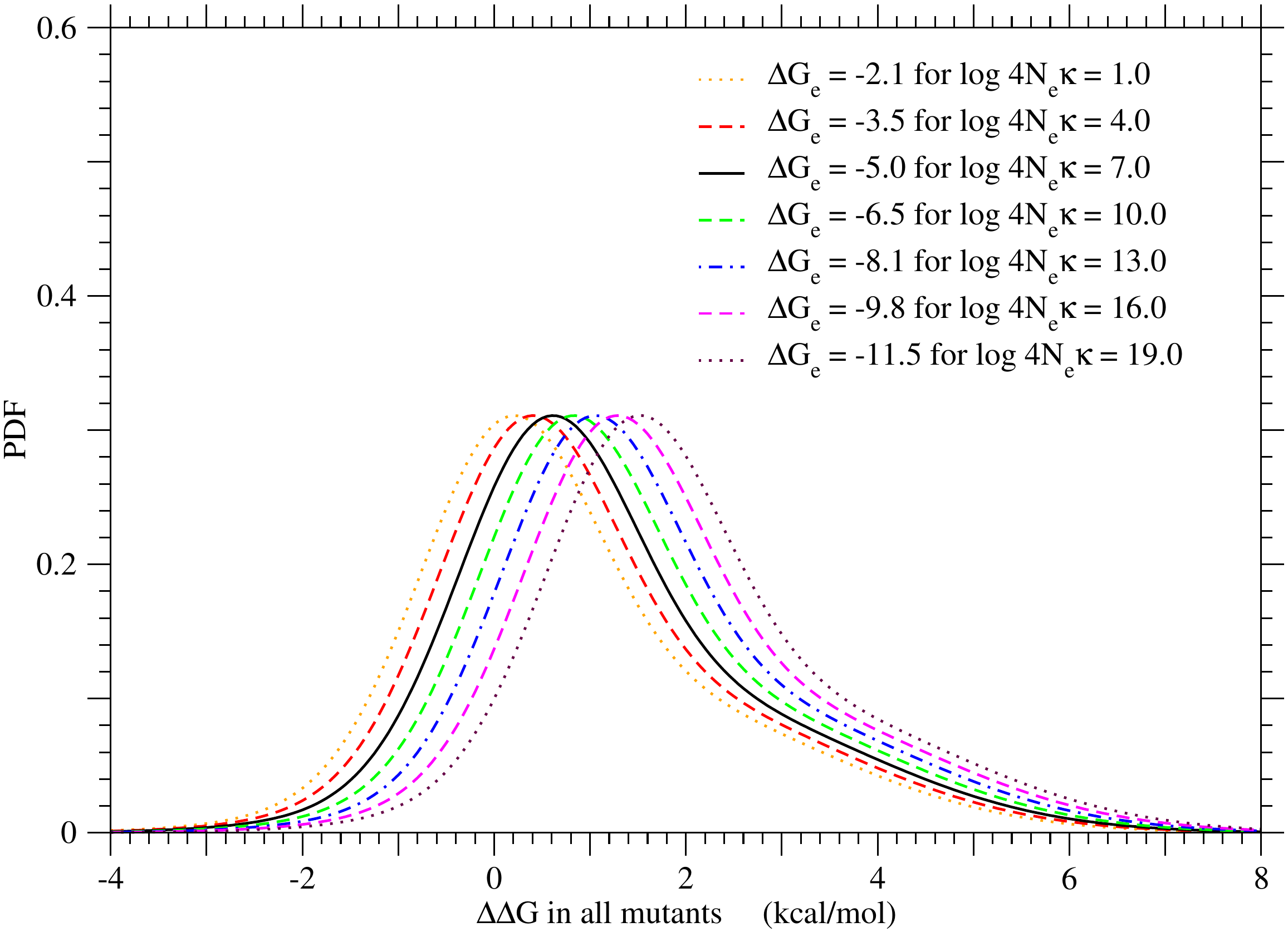}
\includegraphics*[width=90mm,angle=0]{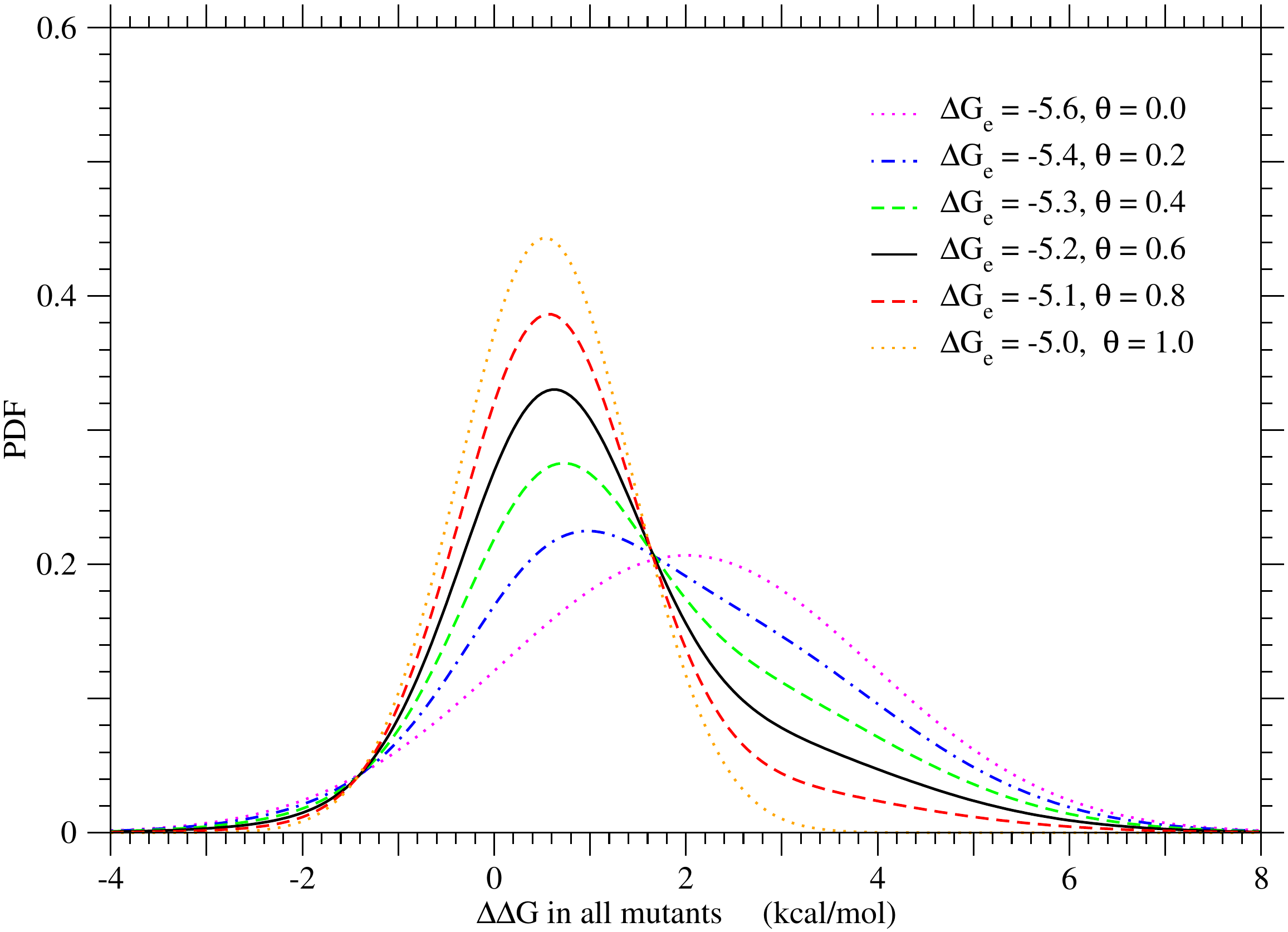}
}
\centerline{
\includegraphics*[width=90mm,angle=0]{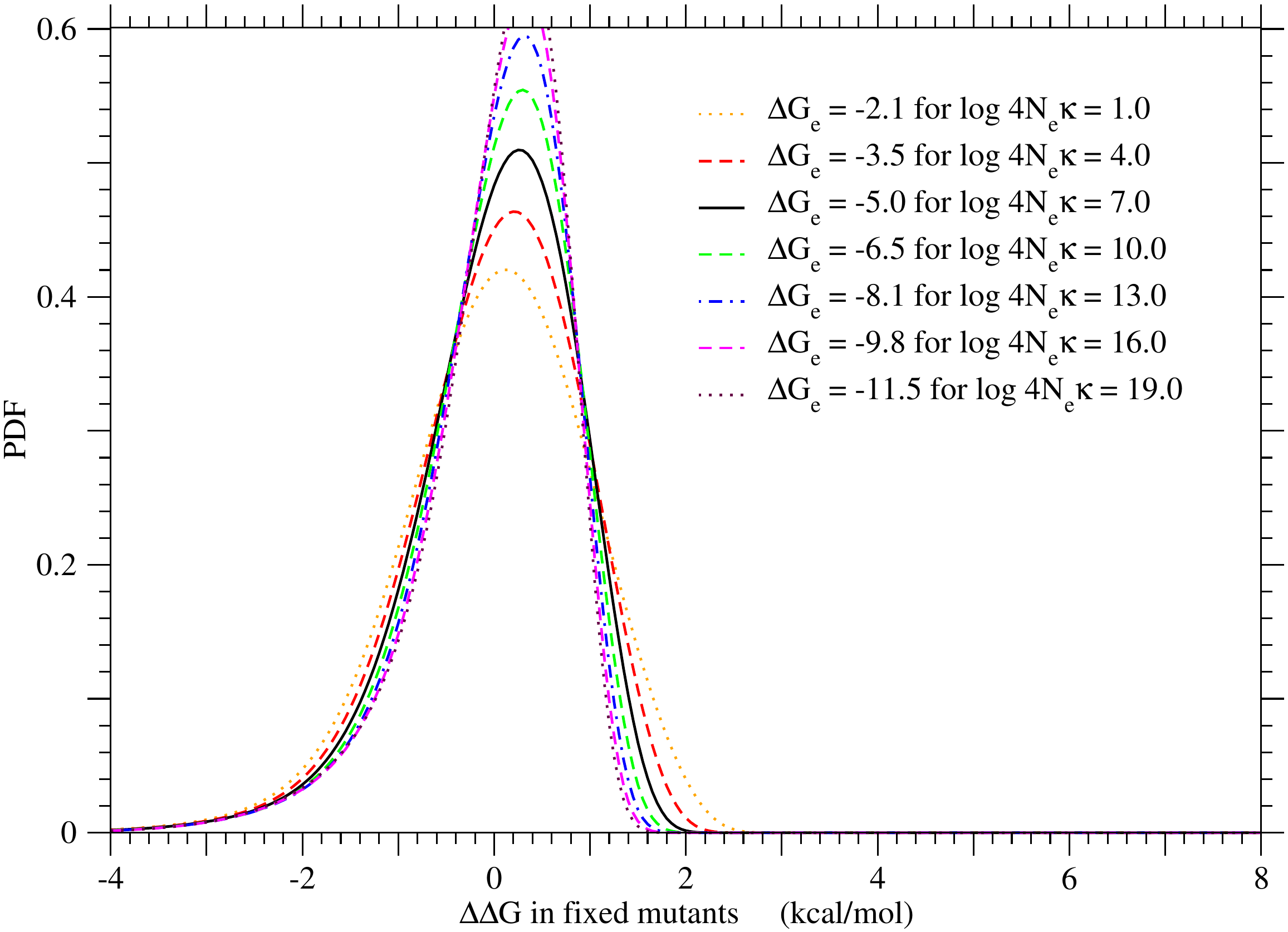}
\includegraphics*[width=90mm,angle=0]{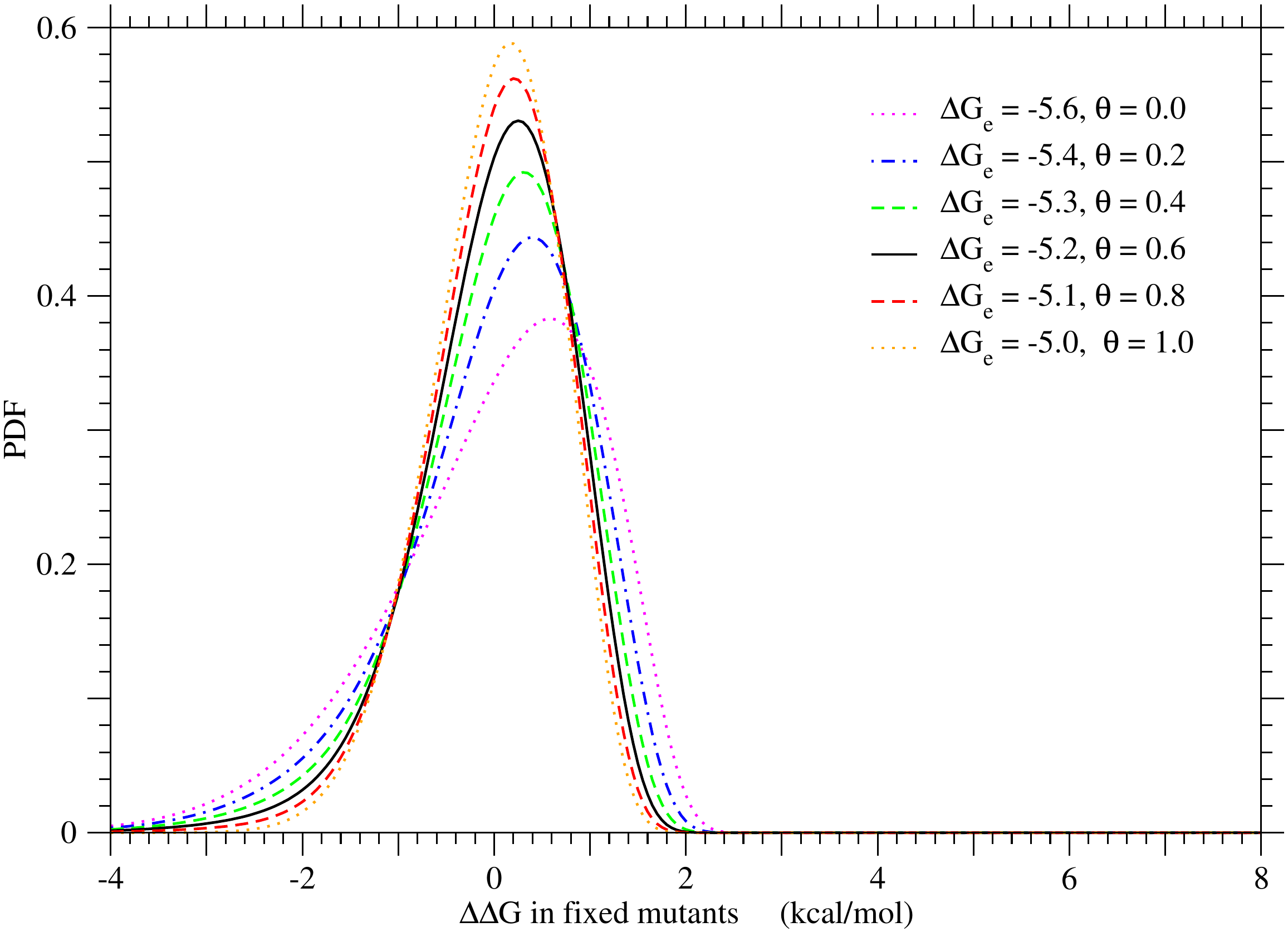}
}
} 
\vspace*{1em}
\caption{
\label{sfig: pdf_of_ddG_at_dGe}
\label{sfig: pdf_of_ddG_fixed_at_dGe}
\label{sfig: pdf_of_ddG}
\FigureLegends{
\BF{PDFs of stability changes, $\Delta\Delta G$, due to single amino acid substitutions
in all mutants and in fixed mutants
at equilibrium of protein stability, $\Delta G = \Delta G_e$.}
The PDF of $\Delta\Delta G$ due to single amino acid substitutions
in all arising mutants is assumed to be bi-Gaussian; see \Eq{\ref{seq: def_bi-Gaussian}}.
Unless specified, 
$\log 4N_e\kappa = 7.550$ and $\theta = 0.53$
are employed.
The kcal/mol unit is used for $\Delta\Delta G$ and $\Delta G_e$.
} 
}
\end{figure*}

\FigureInLegends{\newpage}

\begin{figure*}[ht]
\FigureInLegends{
\centerline{
\includegraphics*[width=90mm,angle=0]{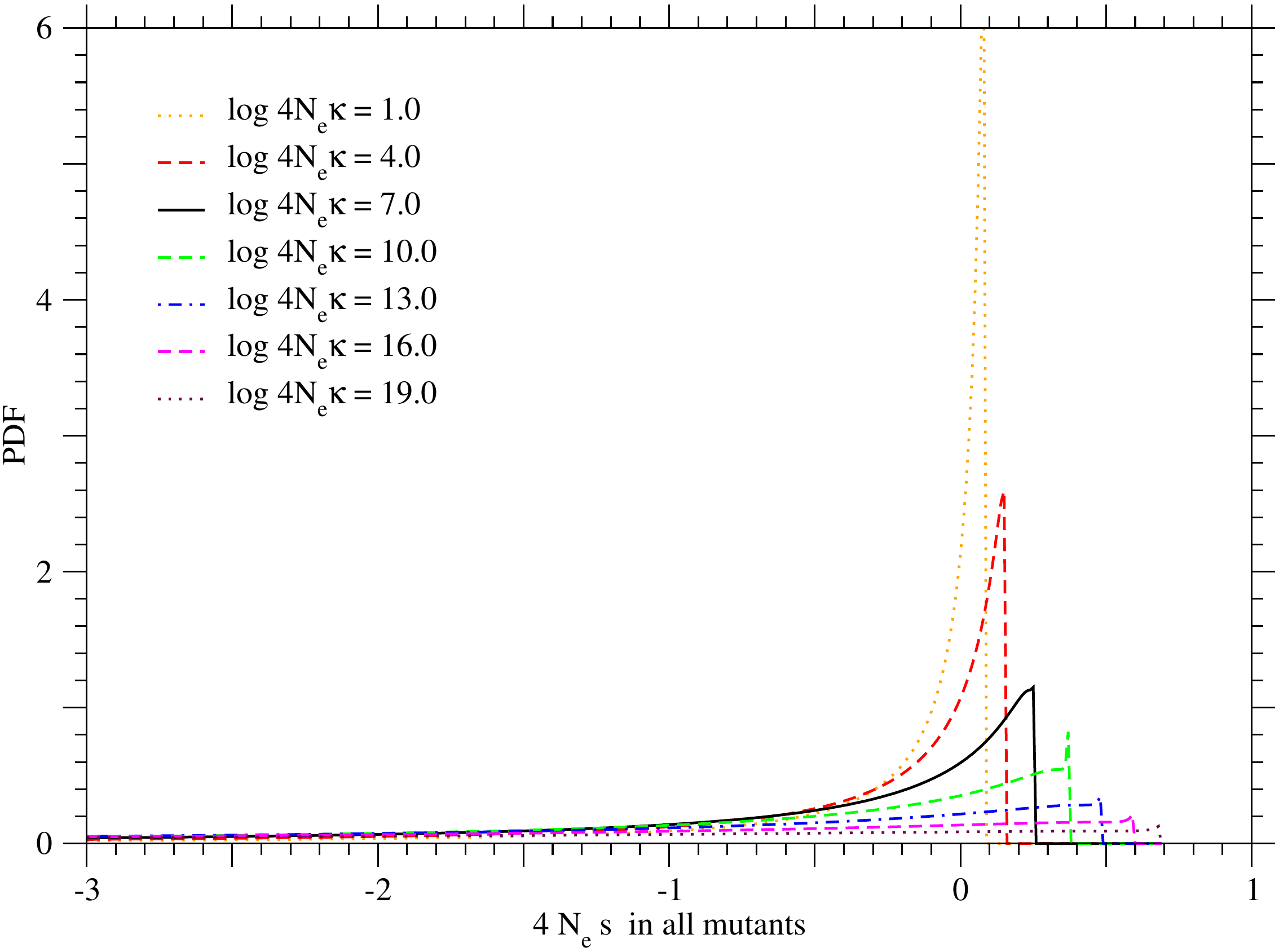}
\includegraphics*[width=90mm,angle=0]{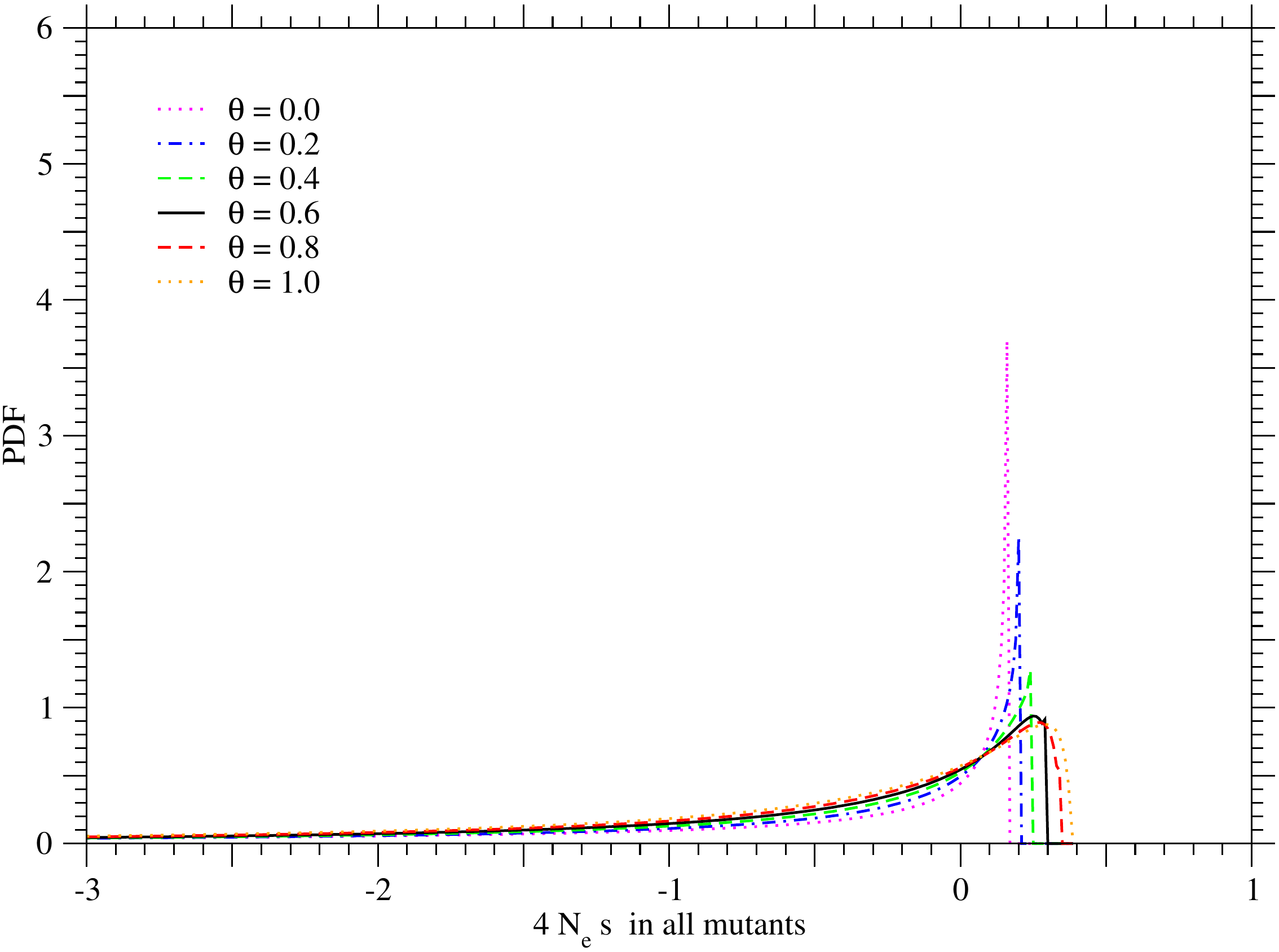}
}
\centerline{
\includegraphics*[width=90mm,angle=0]{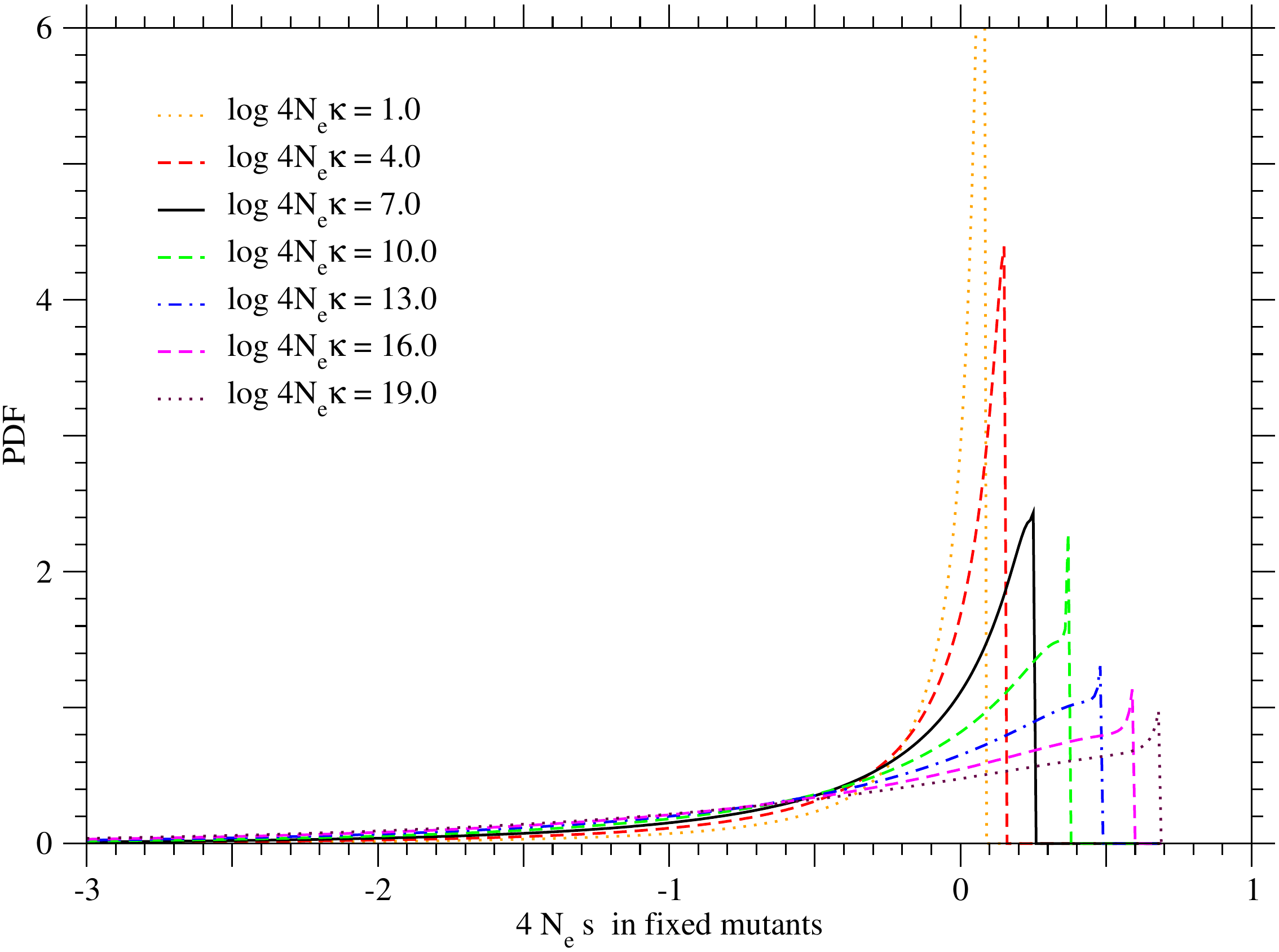}
\includegraphics*[width=90mm,angle=0]{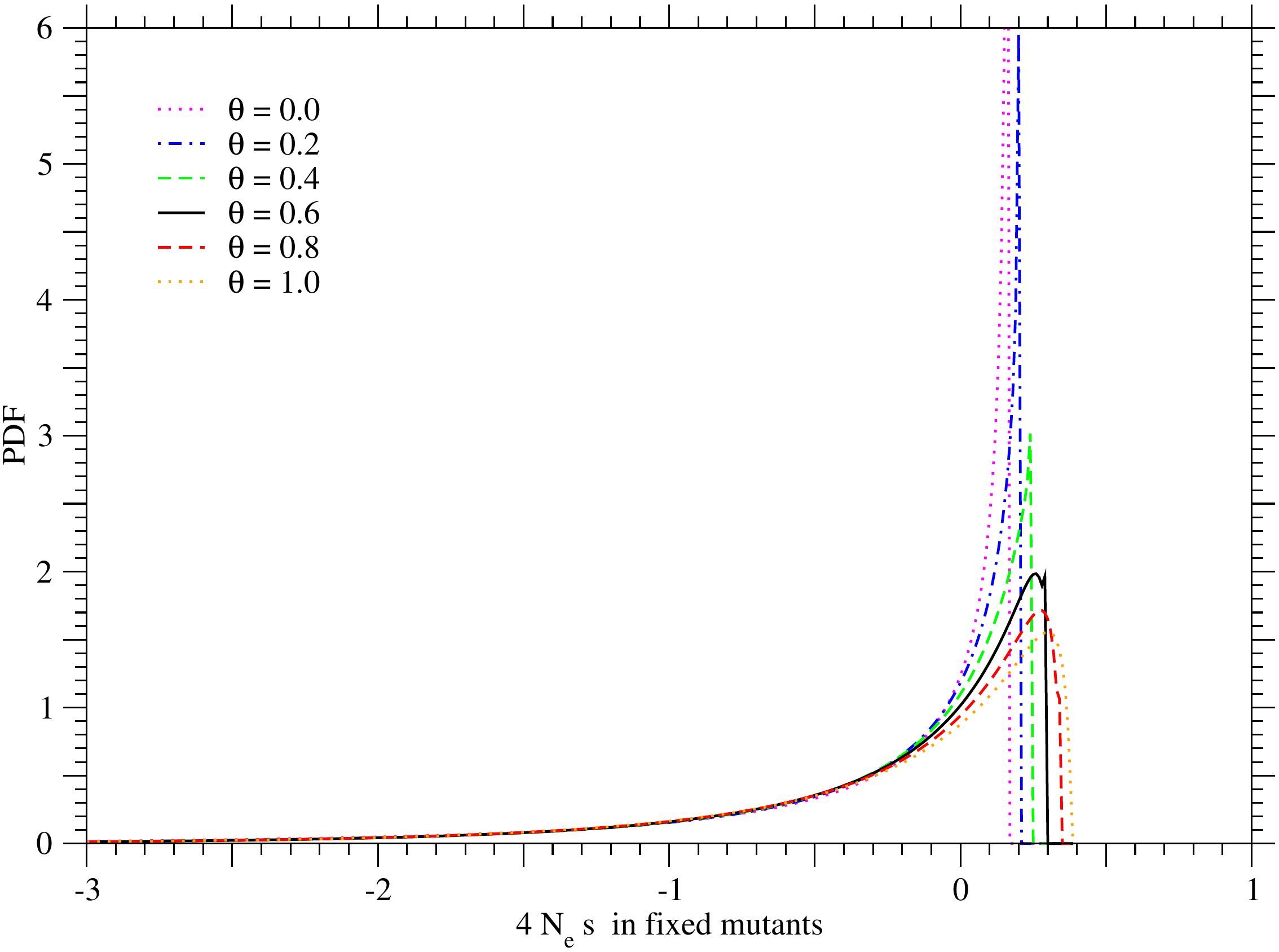}
}
} 
\vspace*{1em}
\caption{
\label{sfig: pdf_of_4sNe_at_dGe}
\label{sfig: pdf_of_4sNe}
\label{sfig: pdf_of_4sNe_fixed}
\label{sfig: pdf_of_4sNe_fixed_at_dGe}
\FigureLegends{
\BF{PDFs of $4N_e s$ in all mutants and in fixed mutants at equilibrium of protein stability, $\Delta G = \Delta G_e$.}
Unless specified, 
$\log 4N_e\kappa = 7.550$ and $\theta = 0.53$
are employed.
} 
}
\end{figure*}

\FigureInLegends{\newpage}

\begin{figure*}[ht]
\FigureInLegends{
\centerline{
\includegraphics*[width=90mm,angle=0]{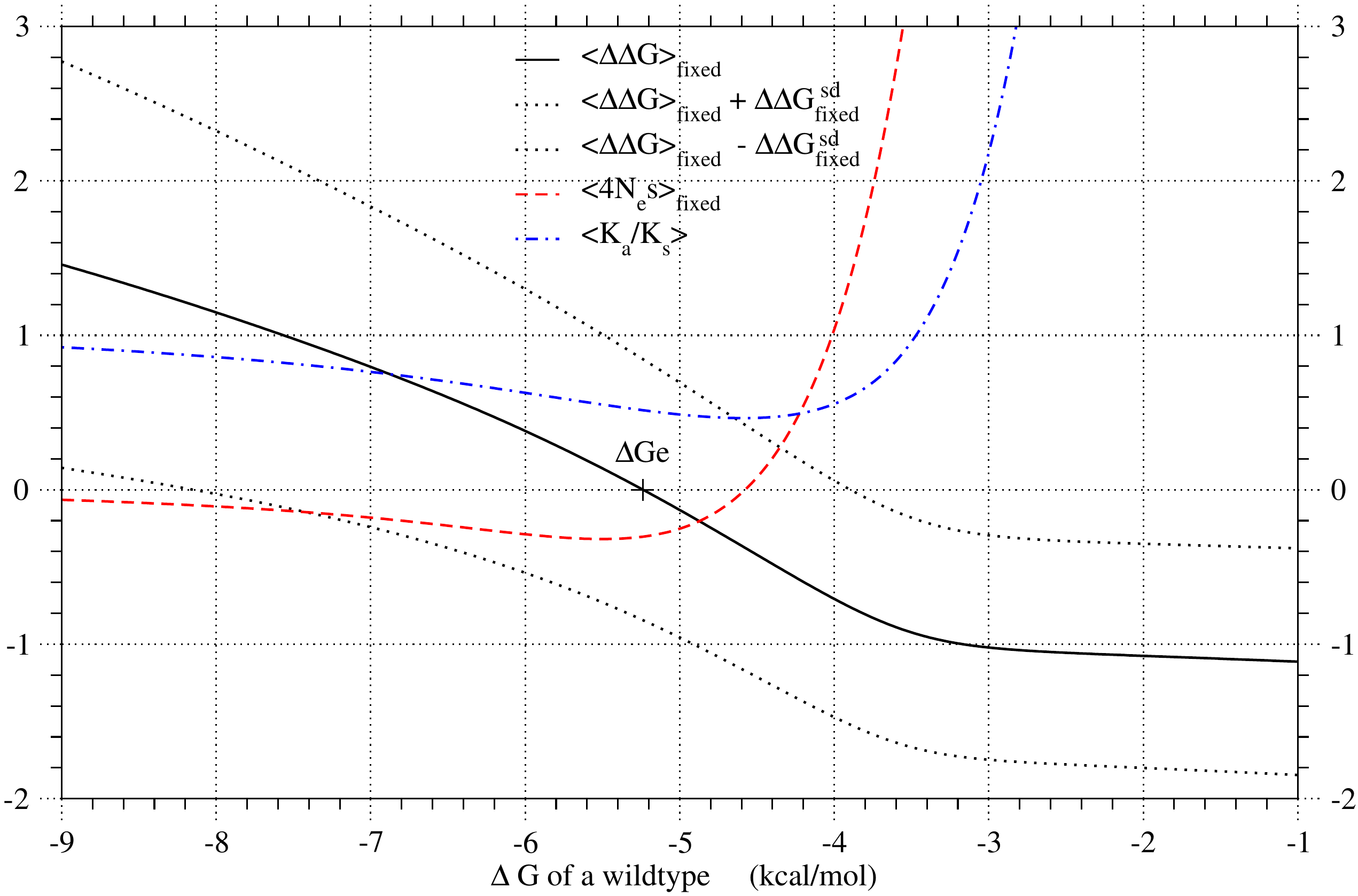}
}
} 
\vspace*{1em}
\caption{
\label{sfig: ave_ddG_vs_dG}
\label{sfig: ave_ddG_vs_dG_fixed}
\FigureLegends{
\BF{The average, $\langle \Delta\Delta G \rangle_{\script{fixed}}$, of stability changes over fixed mutants versus protein stability, $\Delta G$, of the wild type.}
$\Delta G_e$, where $\langle \Delta\Delta G \rangle = 0$, is the stable equilibrium value
of folding free energy, $\Delta G$, in protein evolution. 
The averages of $\Delta\Delta G$, $4N_e s$, and $K_a/Ks$ over fixed mutants are
plotted against protein stability, $\Delta G$, of the wild type by solid, broken, and dash-dot lines,
respectively.
Thick dotted lines show the values of $\langle \Delta\Delta G \rangle_{\script{fixed}} \pm \Delta\Delta G^{\script{sd}}_{\script{fixed}}$,
where $\Delta\Delta G^{\script{sd}}_{\script{fixed}}$ is the standard deviation of $\Delta\Delta G$ over fixed mutants.
$\log 4N_e\kappa = 7.550$ and $\theta = 0.53$
are employed.
The kcal/mol unit is used for $\Delta\Delta G$.
} 
}
\end{figure*}

\FigureInLegends{\newpage}

\begin{figure*}[ht]
\FigureInLegends{
\centerline{
\includegraphics*[width=90mm,angle=0]{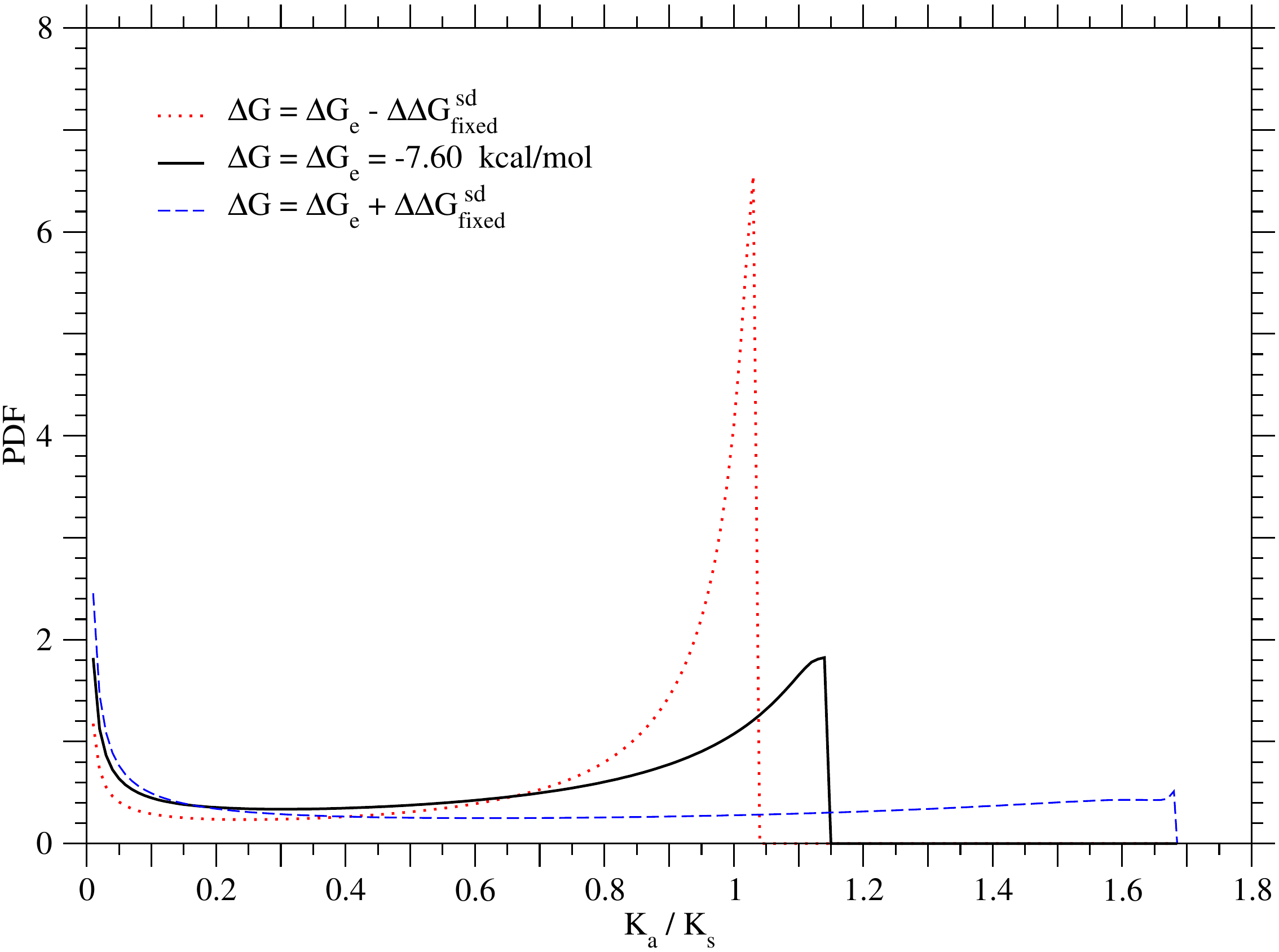}
}
\centerline{
\includegraphics*[width=90mm,angle=0]{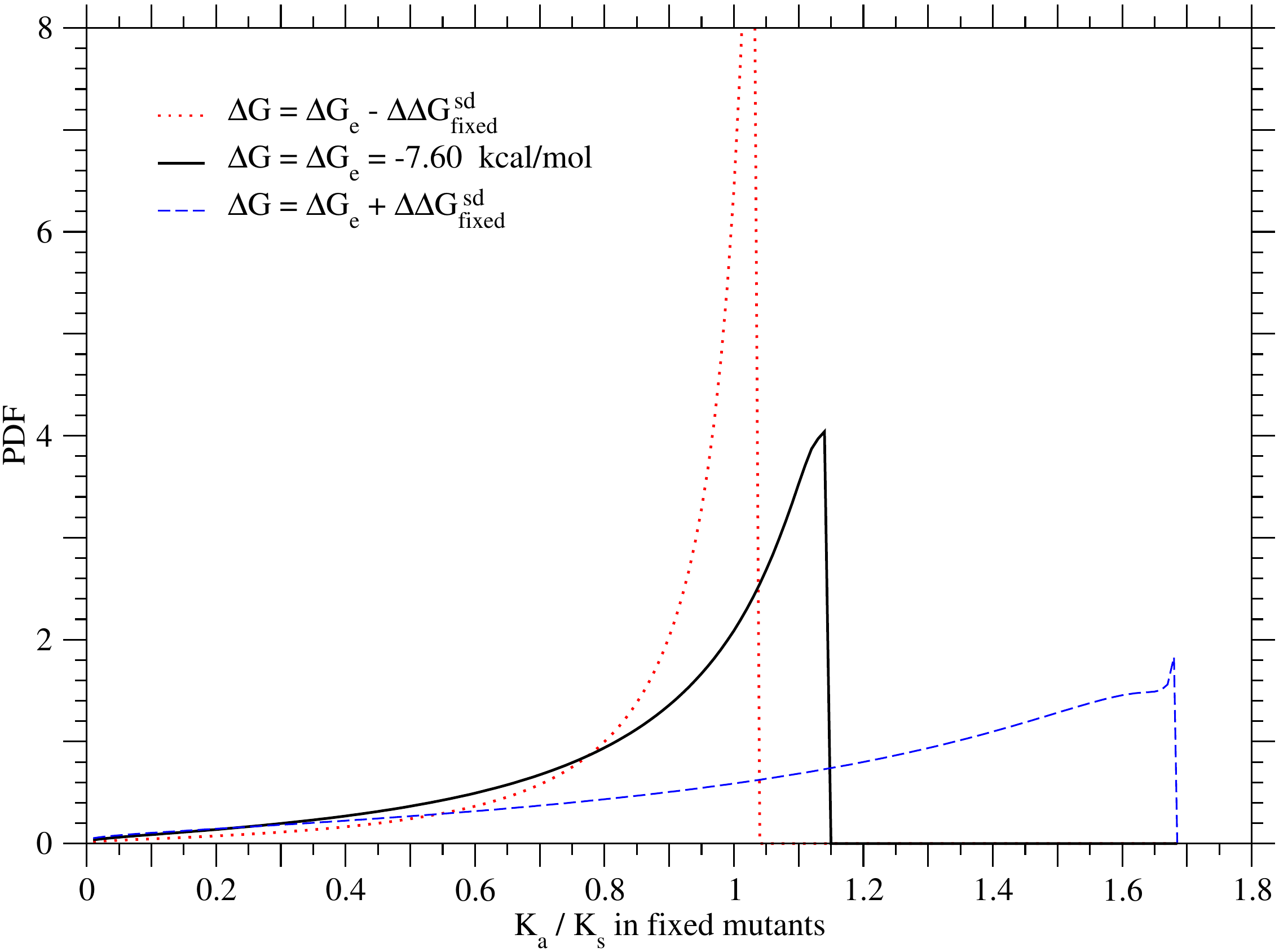}
}
} 
\vspace*{1em}
\caption{
\label{sfig: pdf_of_Ka_over_Ks_around_dGe}
\label{sfig: pdf_of_Ka_over_Ks_fixed_around_dGe}
\FigureLegends{
\BF{Dependence of the PDF of $K_a/K_s$ on protein stability, $\Delta G$, of the wild type in all mutants or in fixed mutants only.}
$\Delta\Delta G^{\script{sd}}_{\script{fixed}}$ is the standard deviation 
(0.84 kcal/mol)	
of $\Delta\Delta G$
over fixed mutants at $\Delta G = \Delta G_e$.
$\log 4N_e\kappa = 7.550$ and $\theta = 0.53$
are employed.
The kcal/mol unit is used for $\Delta G_e$.
} 
}
\end{figure*}

\FigureInLegends{\clearpage\newpage}

\begin{figure*}[ht]
\FigureInLegends{
\centerline{
\includegraphics*[width=90mm,angle=0]{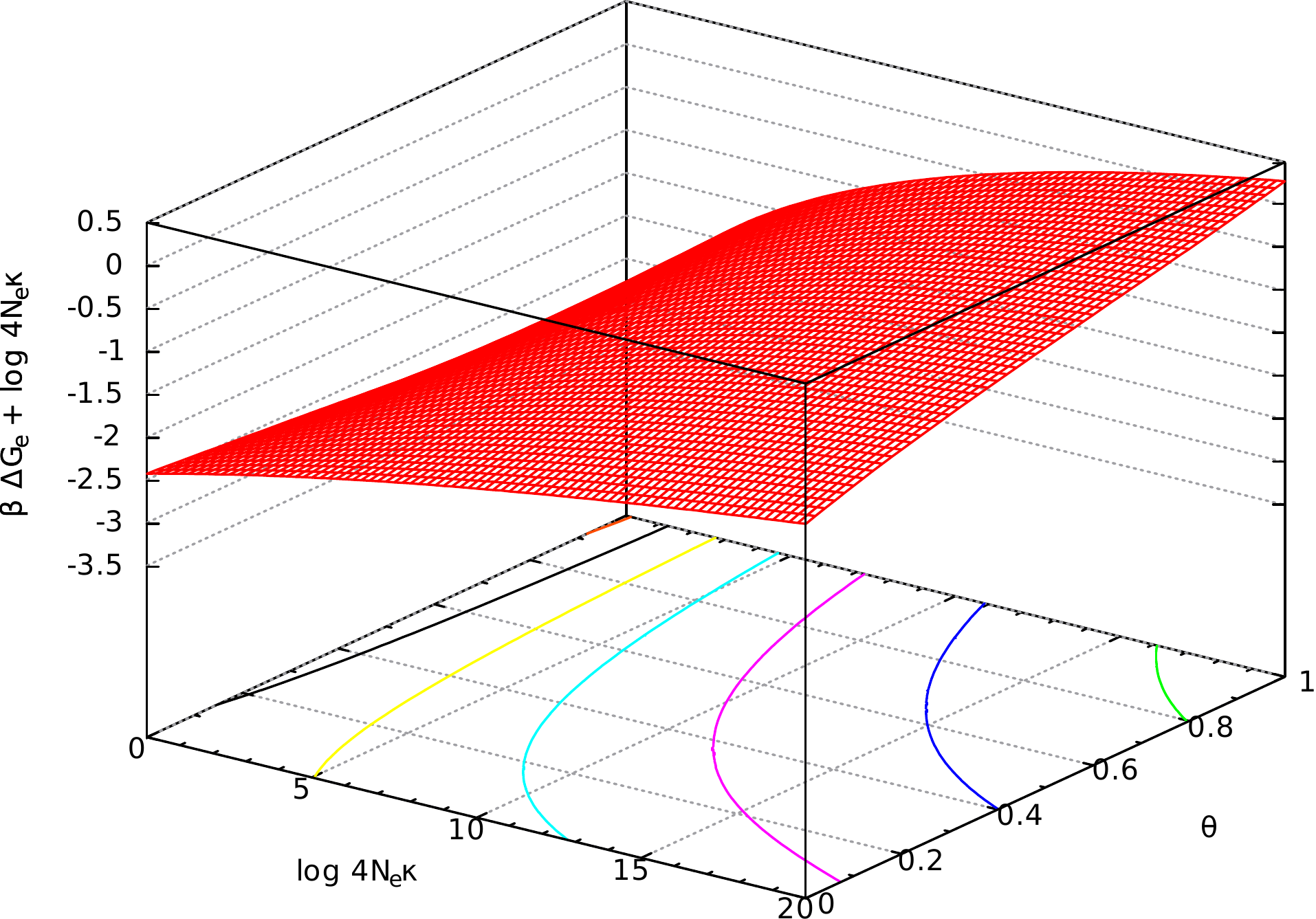}
\includegraphics*[width=90mm,angle=0]{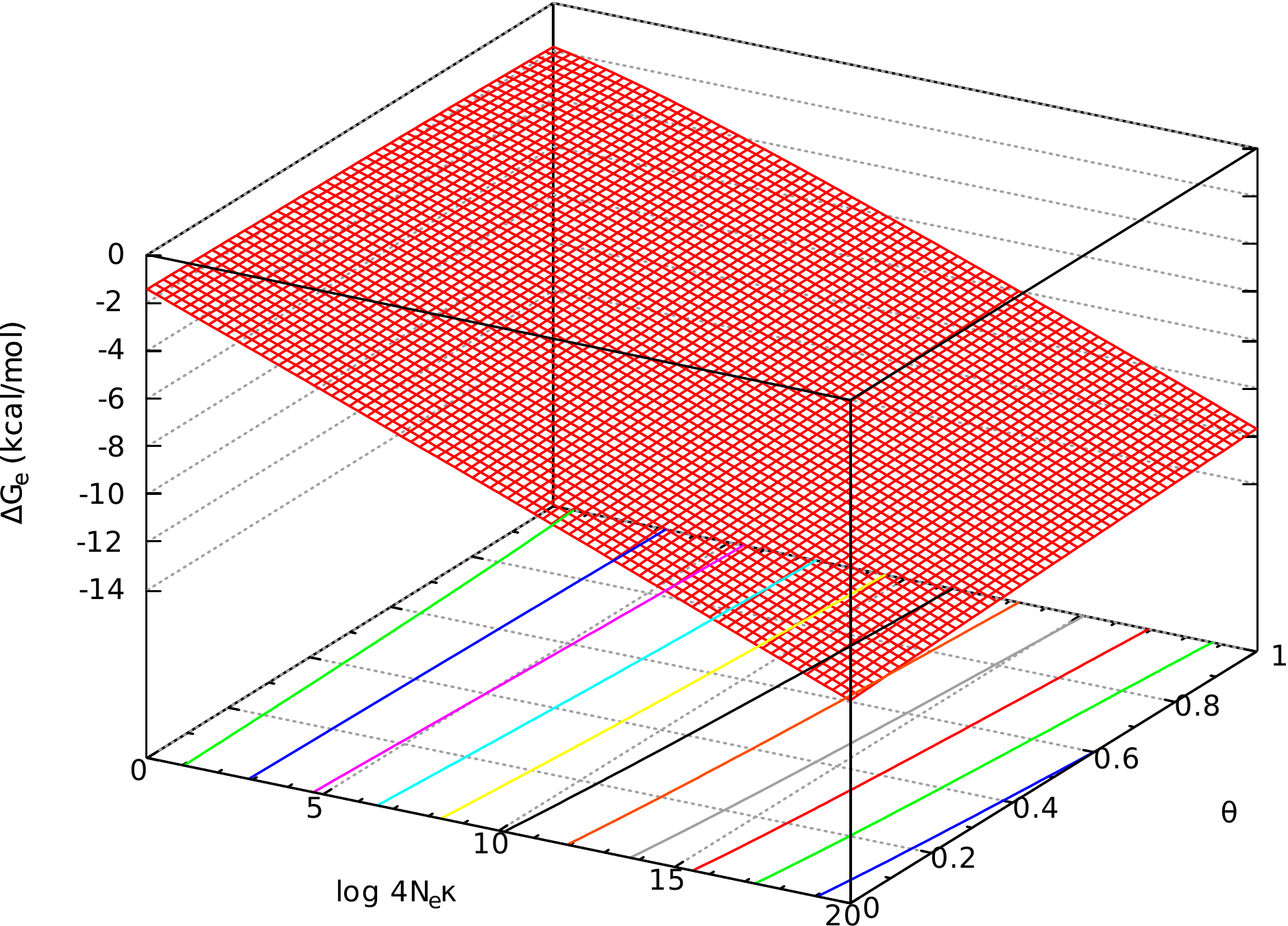}
}
} 
\vspace*{1em}
\caption{
\label{sfig: dependence_of_dGe_on_4Nekappa_and_theta}
\FigureLegends{
\BF{Dependence of equilibrium stability, $\Delta G_e$, on parameters, $4N_e\kappa$ and $\theta$.}
$\Delta G_e$ is the equilibrium value of folding free energy, $\Delta G$,
in protein evolution.
The value of $\beta\Delta G_e + \log 4N_e\kappa$ is the upper bound of $\log 4N_e s$,
and would be constant if the mean of $\Delta\Delta G$ in all arising mutants did not depend on $\Delta G$; 
see \Eq{\ref{seq: def_s}}.
The kcal/mol unit is used for $\Delta G_e$.
} 
}
\end{figure*}

\FigureInLegends{\clearpage\newpage}

\begin{figure*}[ht]
\FigureInLegends{
\centerline{
\includegraphics*[width=90mm,angle=0]{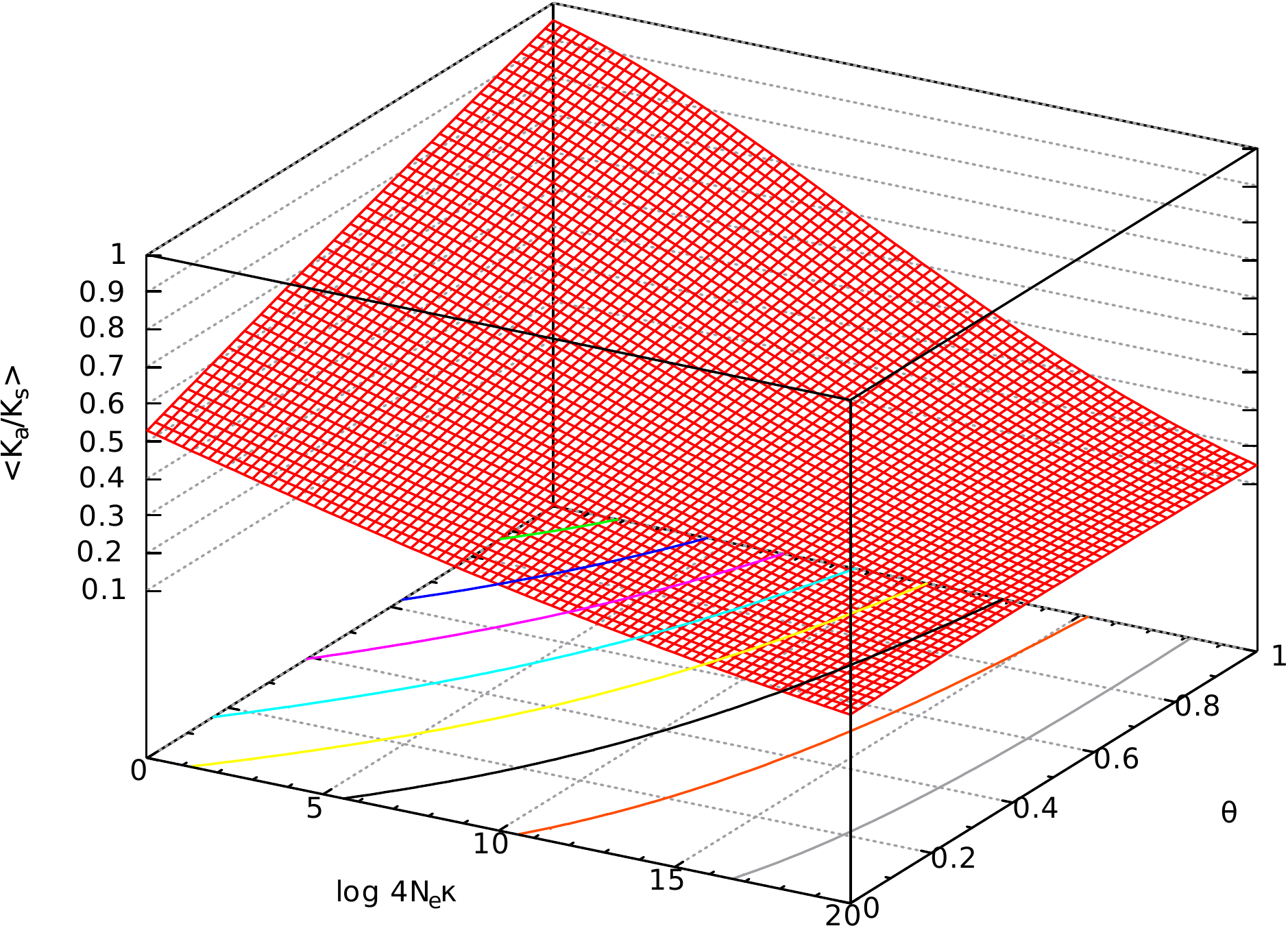}
\includegraphics*[width=90mm,angle=0]{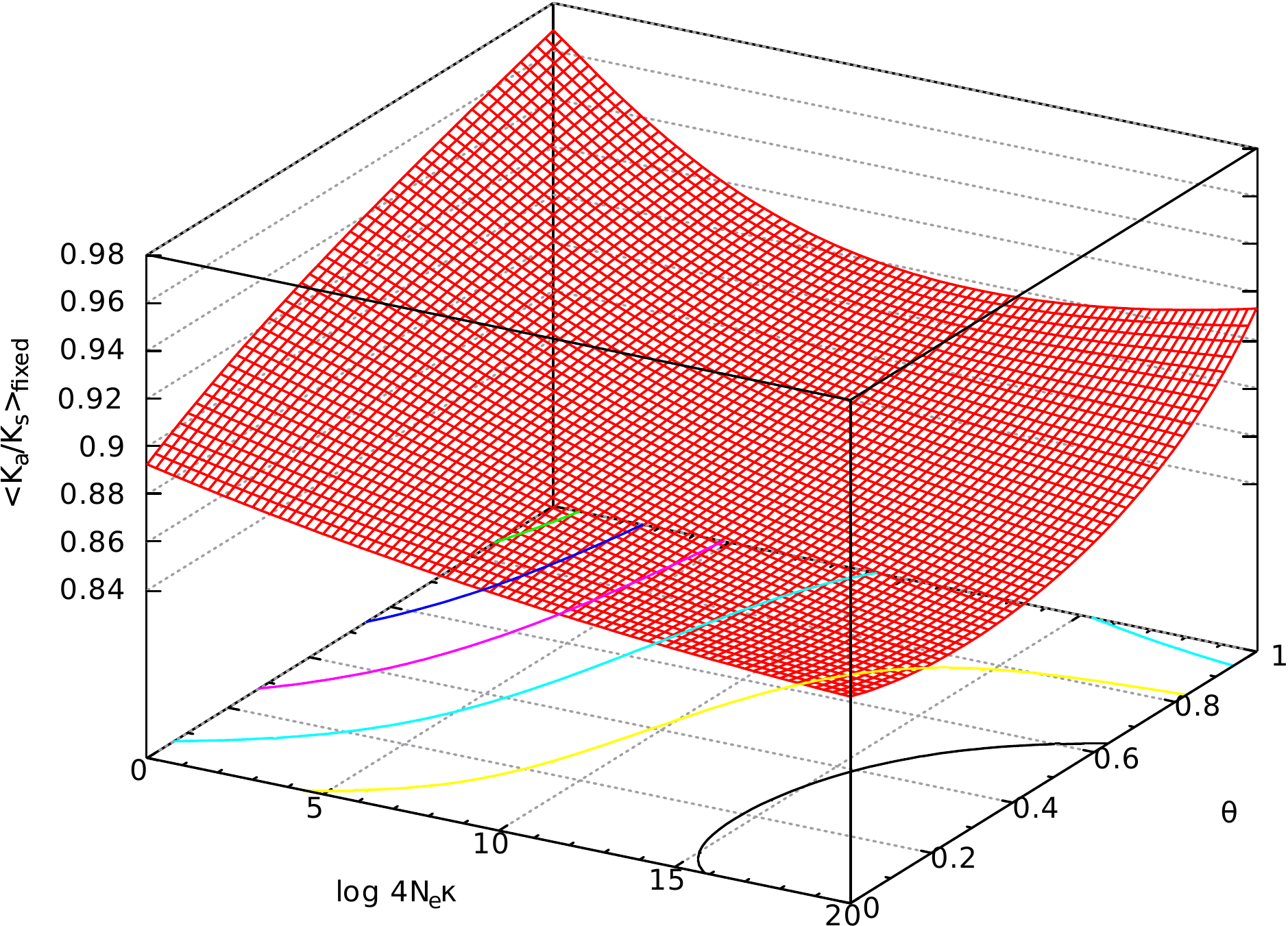}
}
} 
\vspace*{1em}
\caption{
\label{sfig: dependence_of_ave_Ka_over_Ks_on_4Nekappa_and_theta}
\label{sfig: dependence_of_ave_Ka_over_Ks_fixed_on_4Nekappa_and_theta}
\FigureLegends{
\BF{The average of $K_a / K_s$ over all mutants or over fixed mutants only at equilibrium of protein stability, $\Delta G = \Delta G_e$.}
} 
}
\end{figure*}

\FigureInLegends{\newpage}

\begin{figure*}[ht]
\FigureInLegends{
\centerline{
\includegraphics*[width=90mm,angle=0]{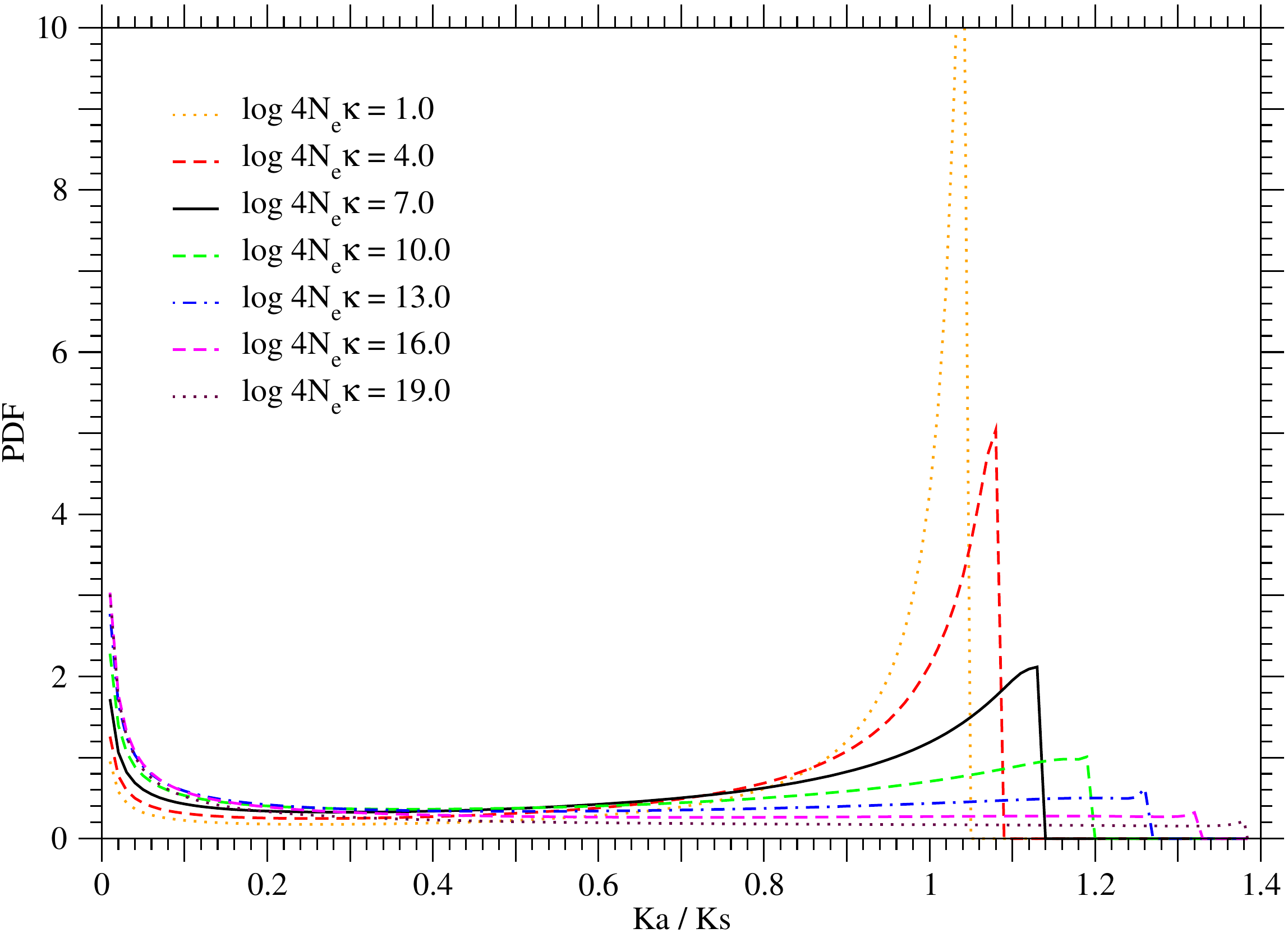}
\includegraphics*[width=90mm,angle=0]{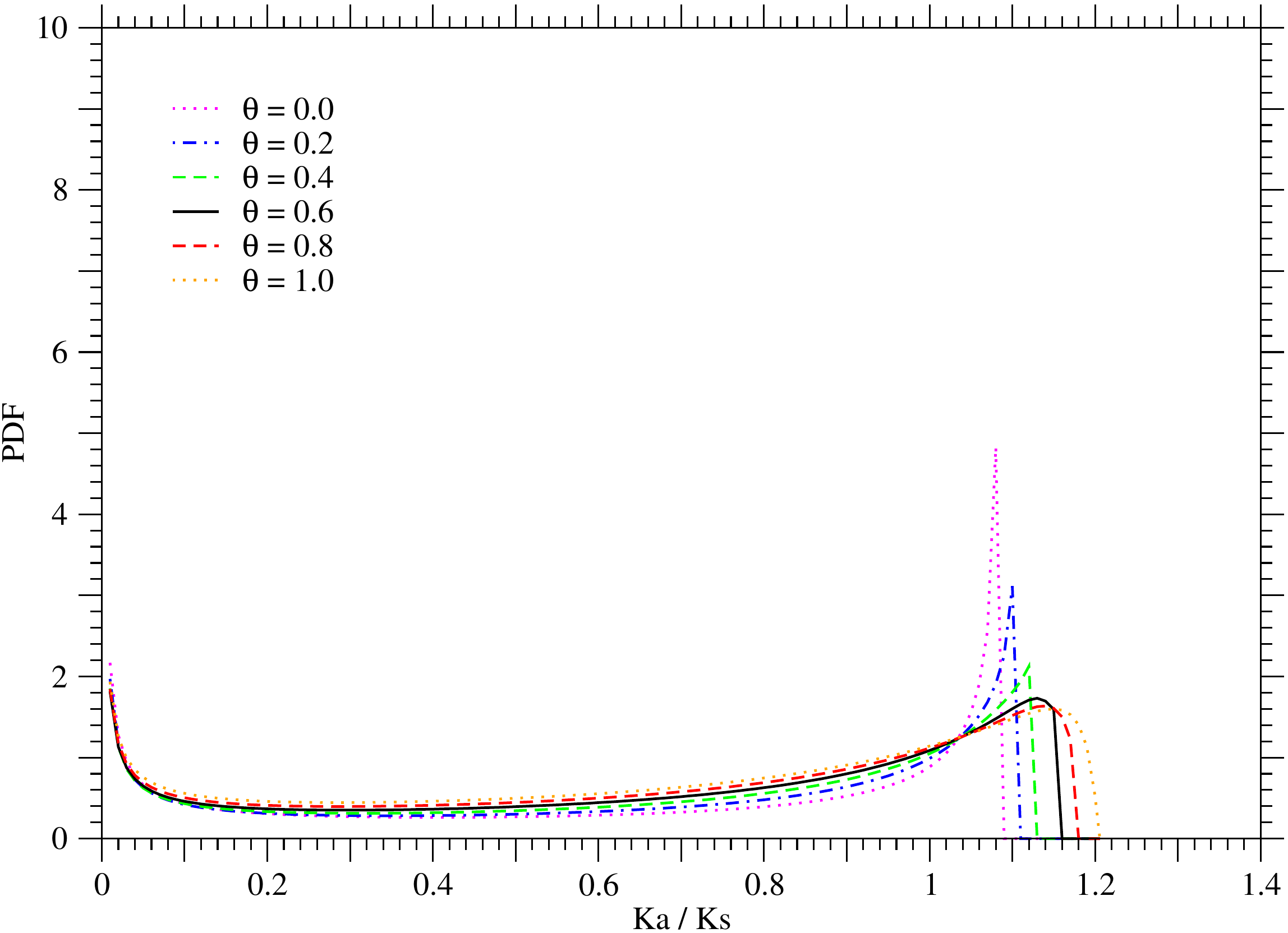}
}
\centerline{
\includegraphics*[width=90mm,angle=0]{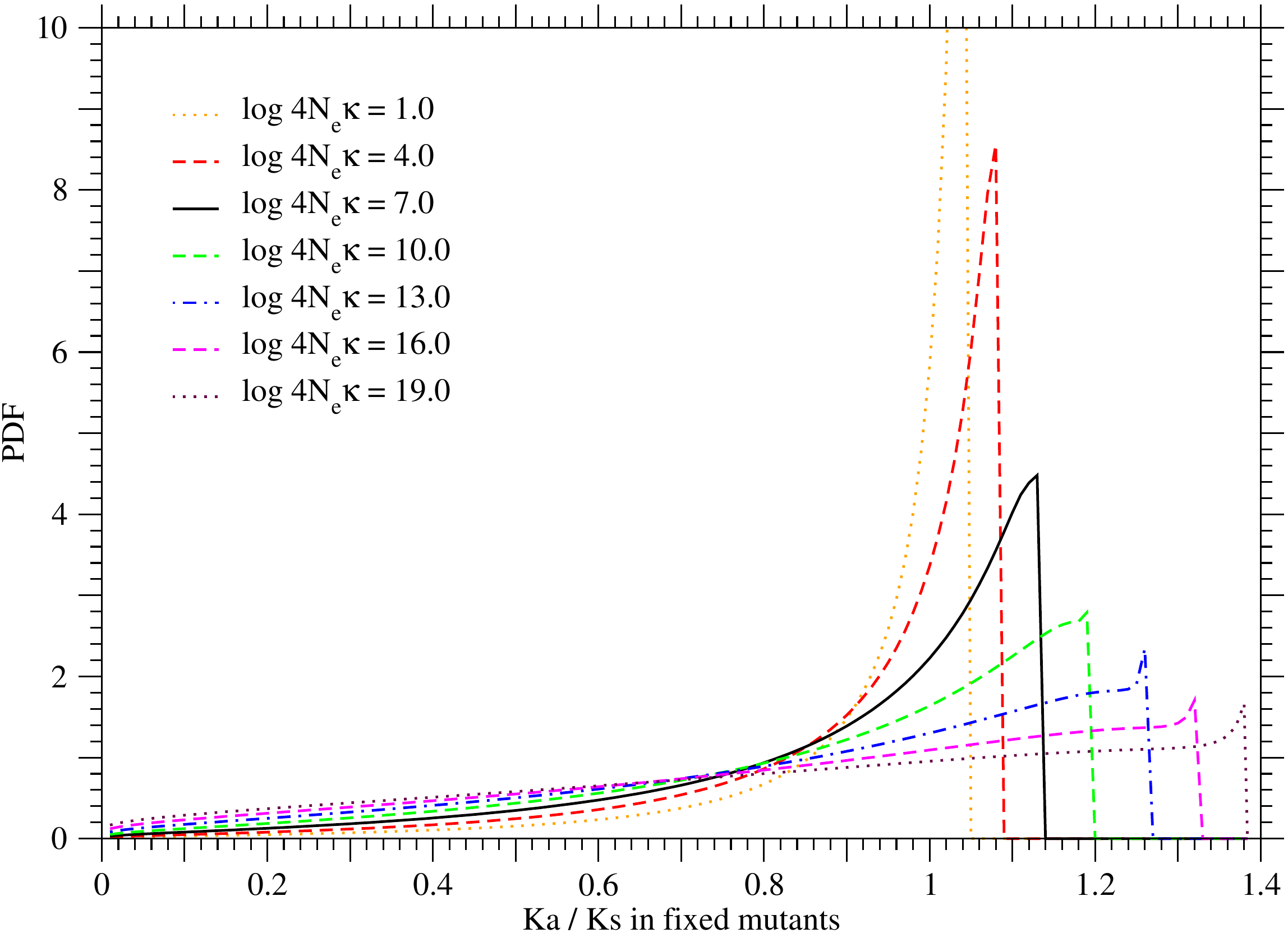}
\includegraphics*[width=90mm,angle=0]{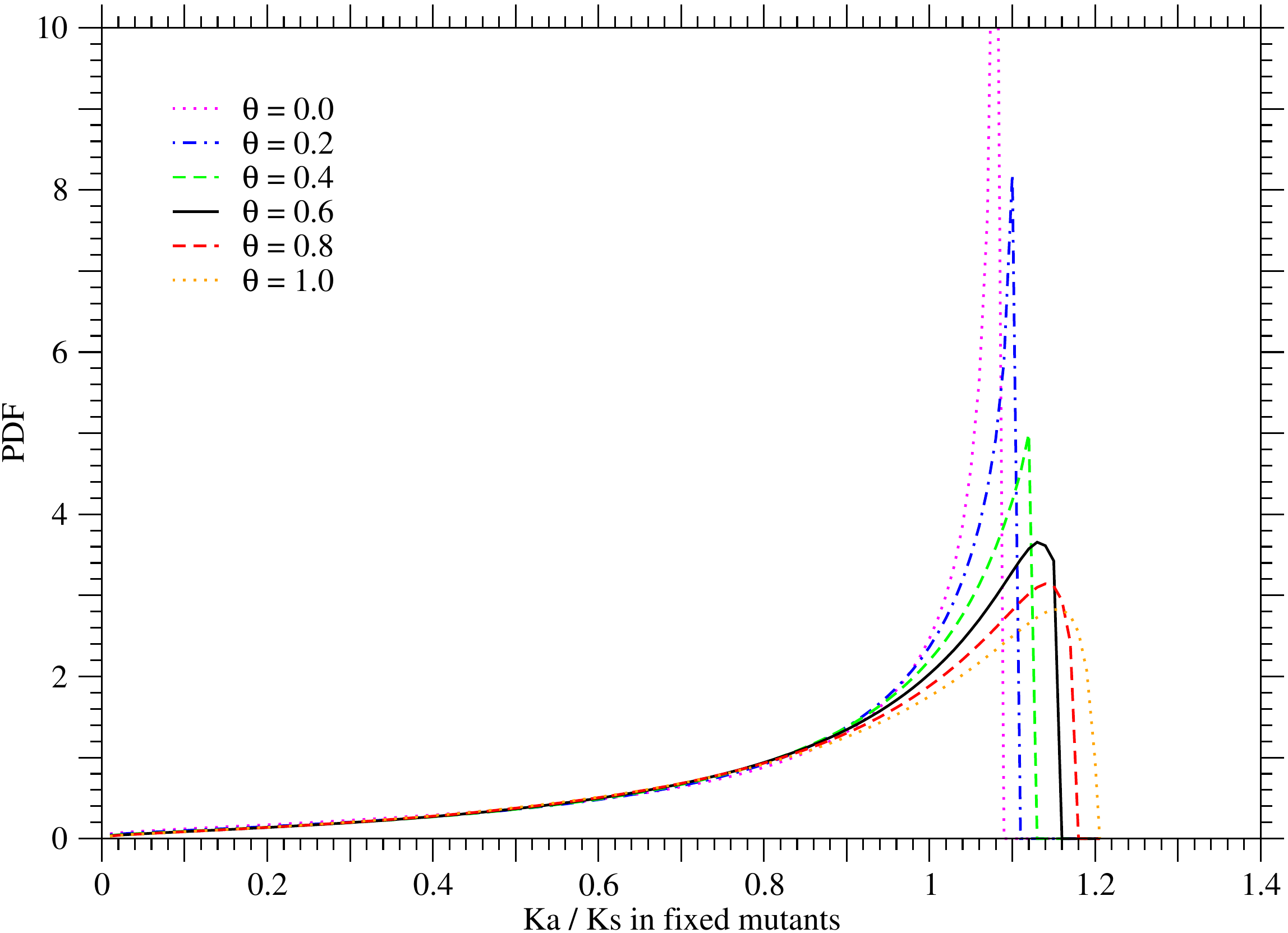}
}
} 
\vspace*{1em}
\caption{
\label{sfig: dependence_of_pdf_of_Ka_over_Ks_on_4Nekappa_and_theta}
\label{sfig: dependence_of_pdf_of_Ka_over_Ks_fixed_on_4Nekappa_and_theta}
\FigureLegends{
\BF{PDFs of $K_a / K_s$ in all mutants and in fixed mutants only
at equilibrium of protein stability, $\Delta G = \Delta G_e$.}
Unless specified, 
$\log 4N_e\kappa = 7.550$ and $\theta = 0.53$
are employed.
} 
}
\end{figure*}

\FigureInLegends{\newpage}

\begin{figure*}[ht]
\FigureInLegends{
\centerline{
\includegraphics*[width=87mm,angle=0]{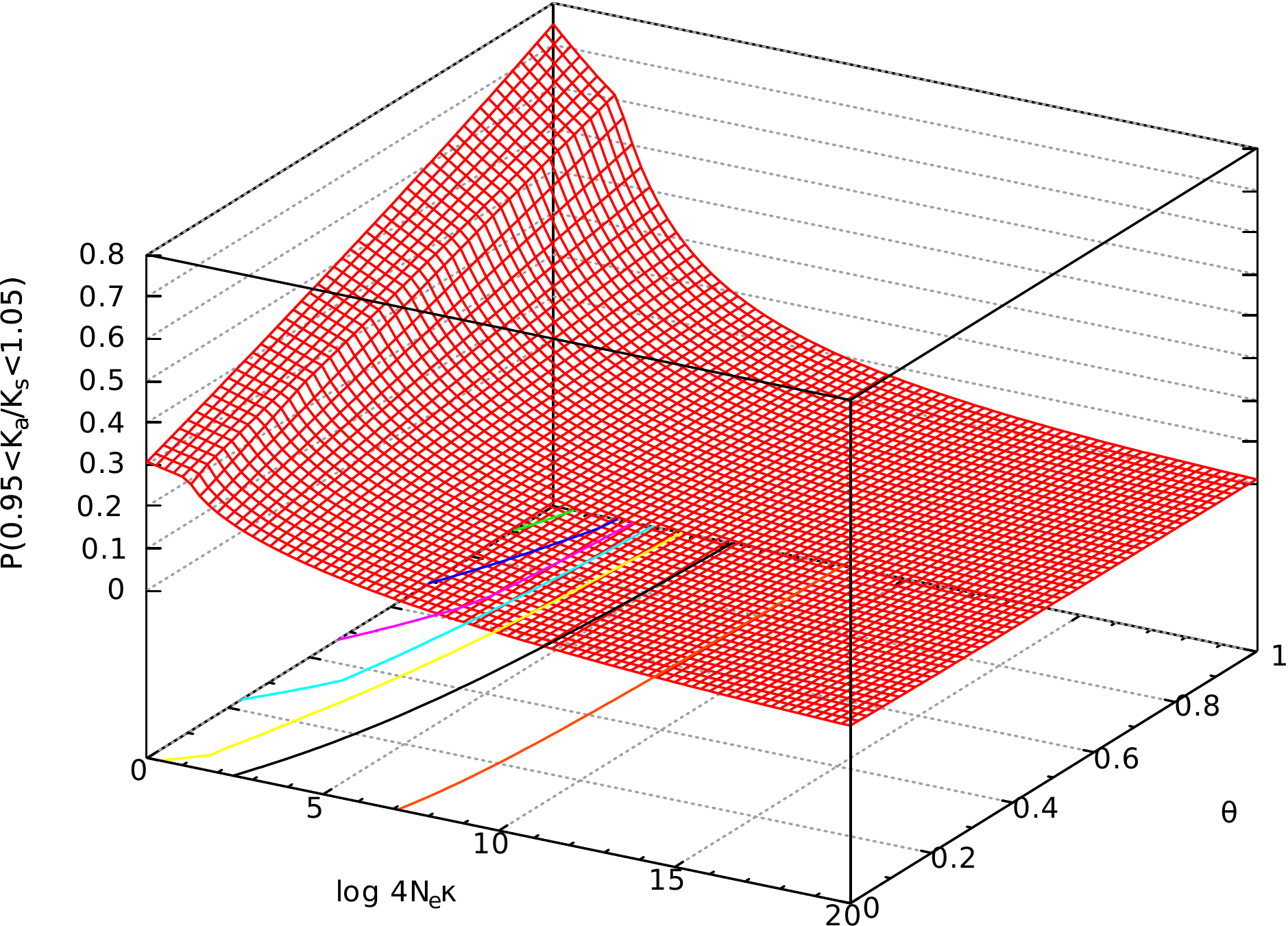}
\includegraphics*[width=90mm,angle=0]{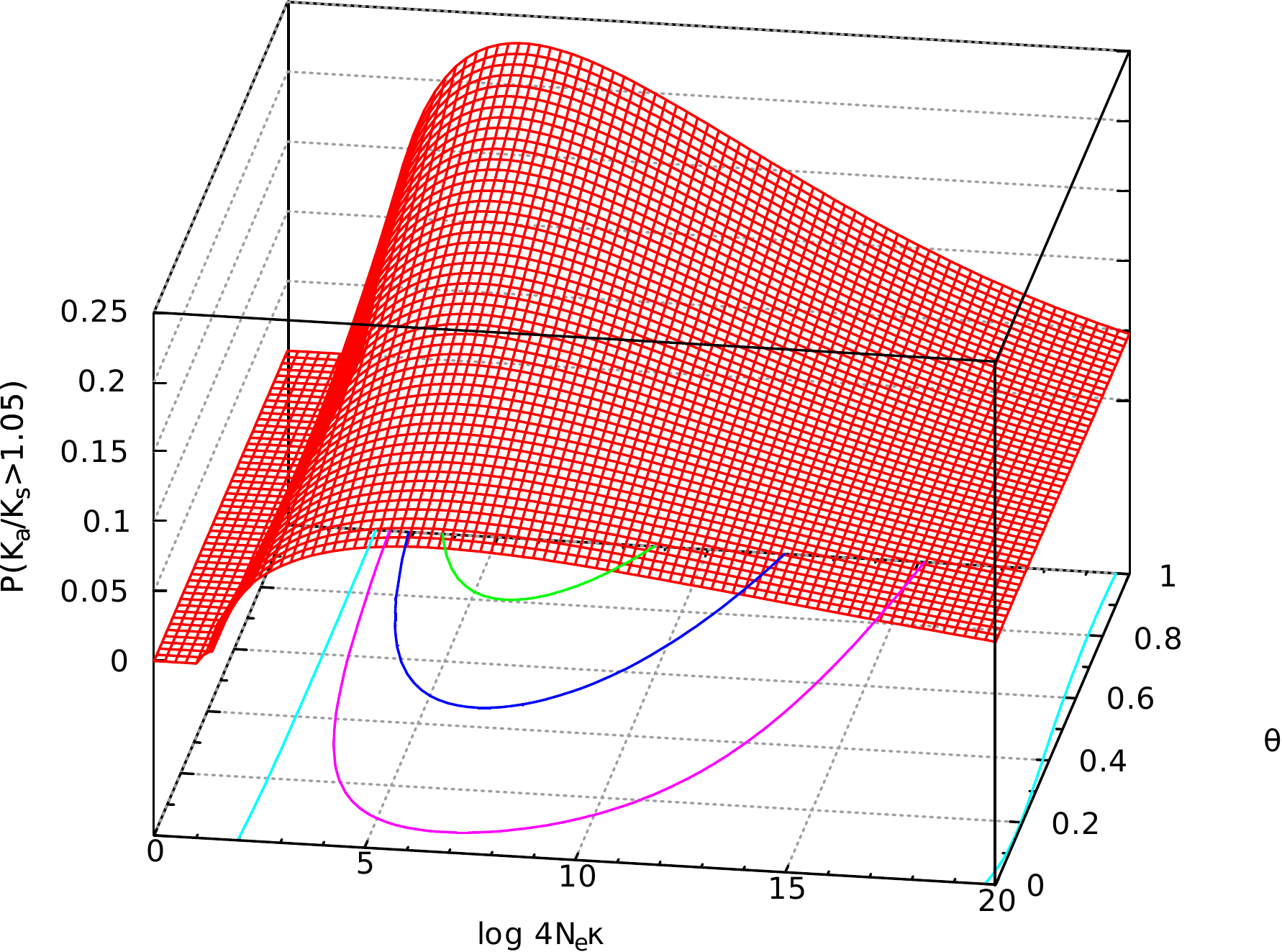}
}
\centerline{
\includegraphics*[width=88mm,angle=0]{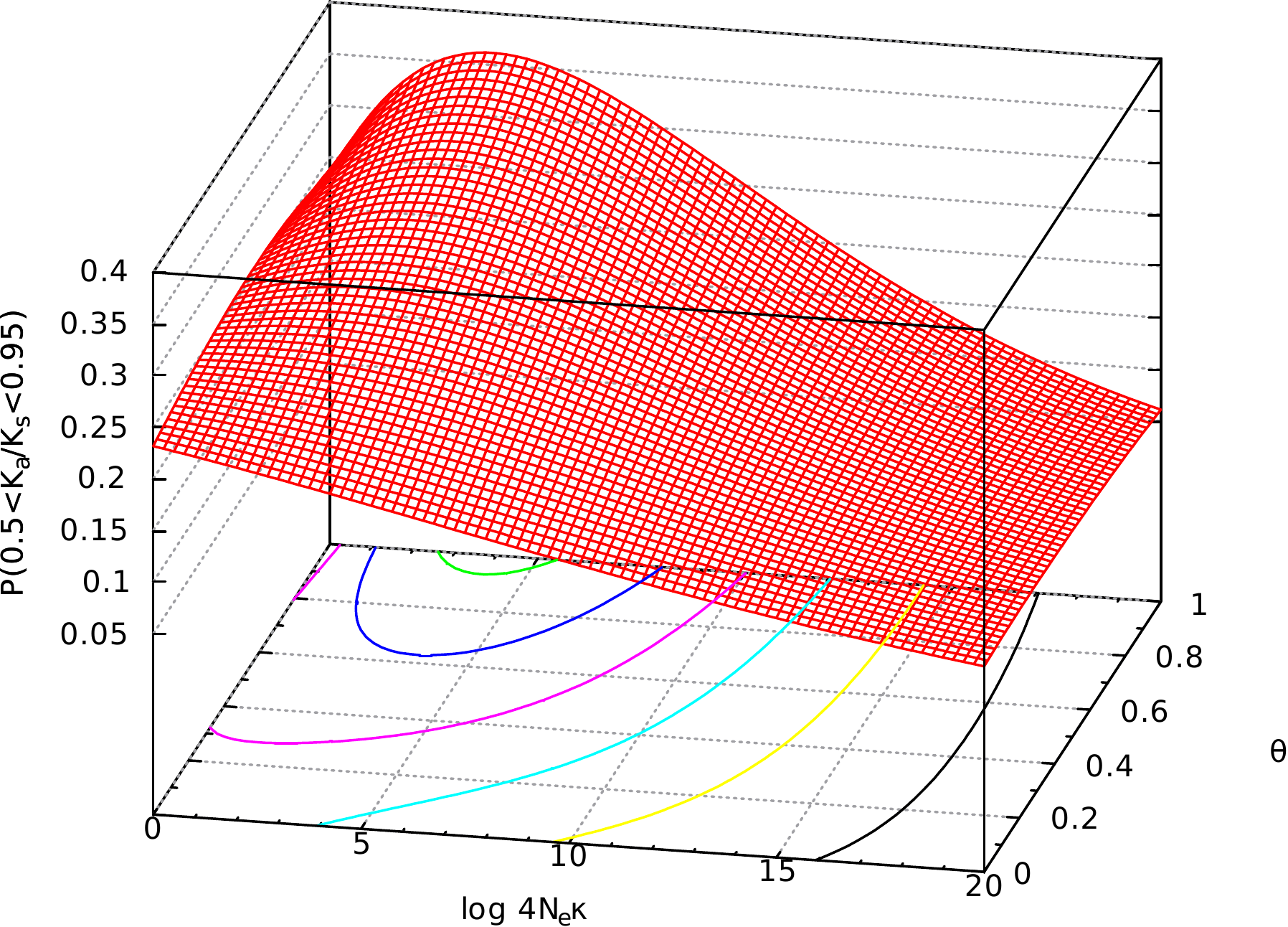}
\includegraphics*[width=90mm,angle=0]{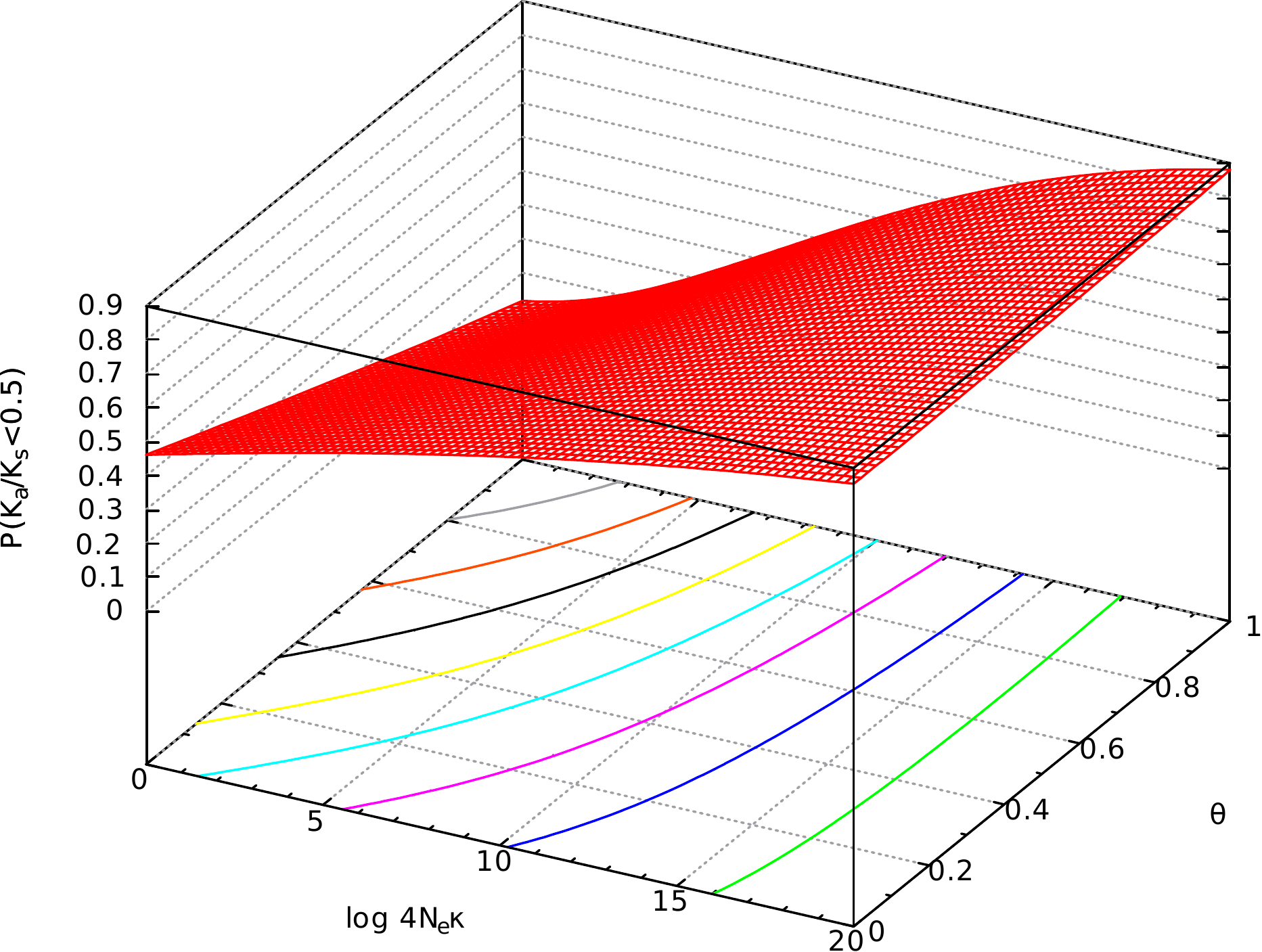}
}
} 
\vspace*{1em}
\caption{
\label{sfig: prob_of_each_selection_category_in_all_mutants}
\FigureLegends{
\BF{Probability of each selection category in all mutants at equilibrium of protein stability, $\Delta G = \Delta G_e$.}
Arbitrarily, the value of $K_a/K_s$ is categorized into four classes;
negative, slightly negative, nearly neutral, and positive selection categories in which
$K_a/K_s$ is within the ranges of
$K_a/K_s \leq 0.5$, $0.5< K_a/K_s \leq 0.95$, $0.95< K_a/K_s \leq 1.05$, and $ 1.05 < K_a/K_s$,
respectively.
} 
}
\end{figure*}

\FigureInLegends{\newpage}

\begin{figure*}[ht]
\FigureInLegends{
\centerline{
\includegraphics*[width=90mm,angle=0]{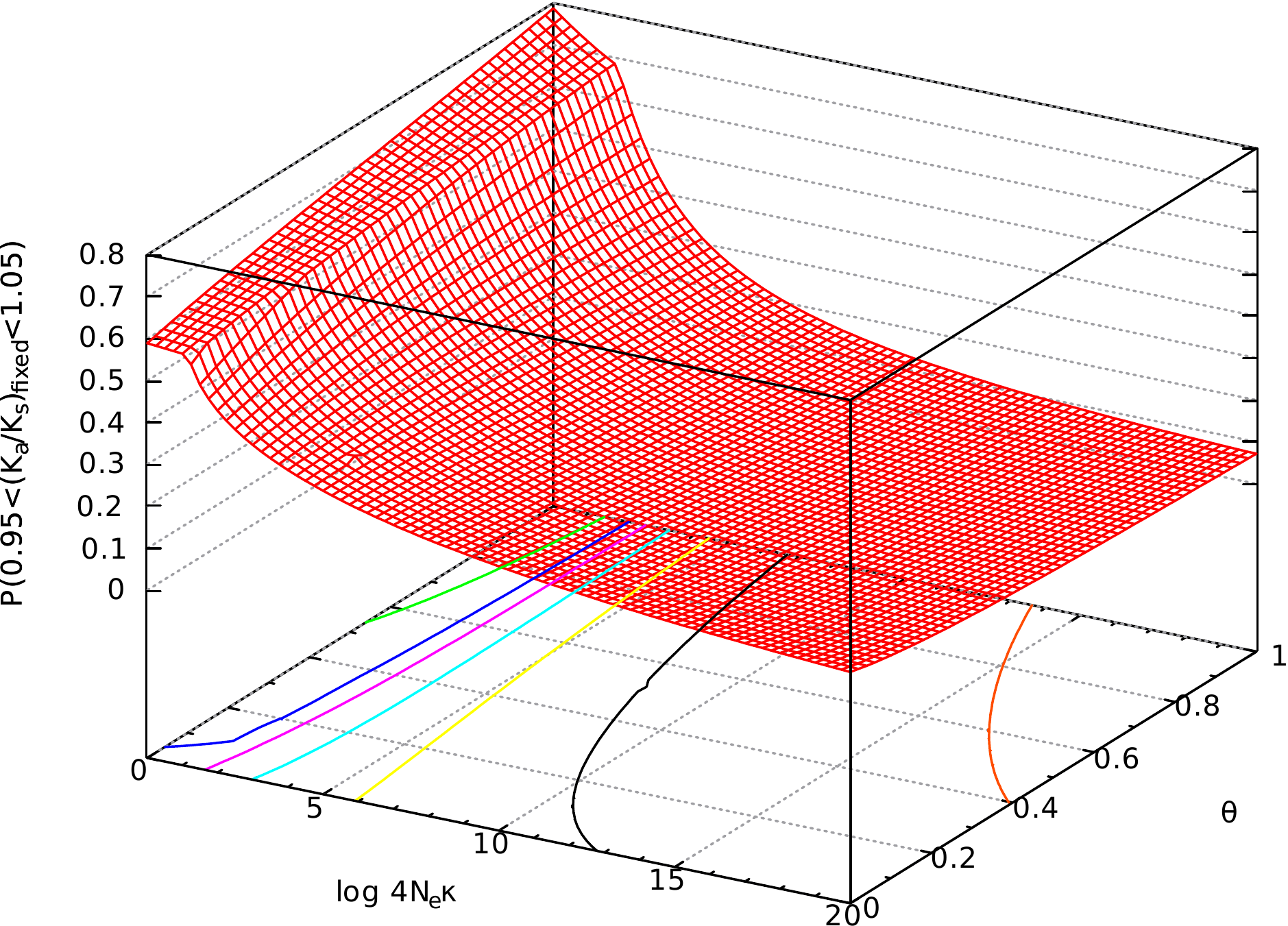}
\includegraphics*[width=90mm,angle=0]{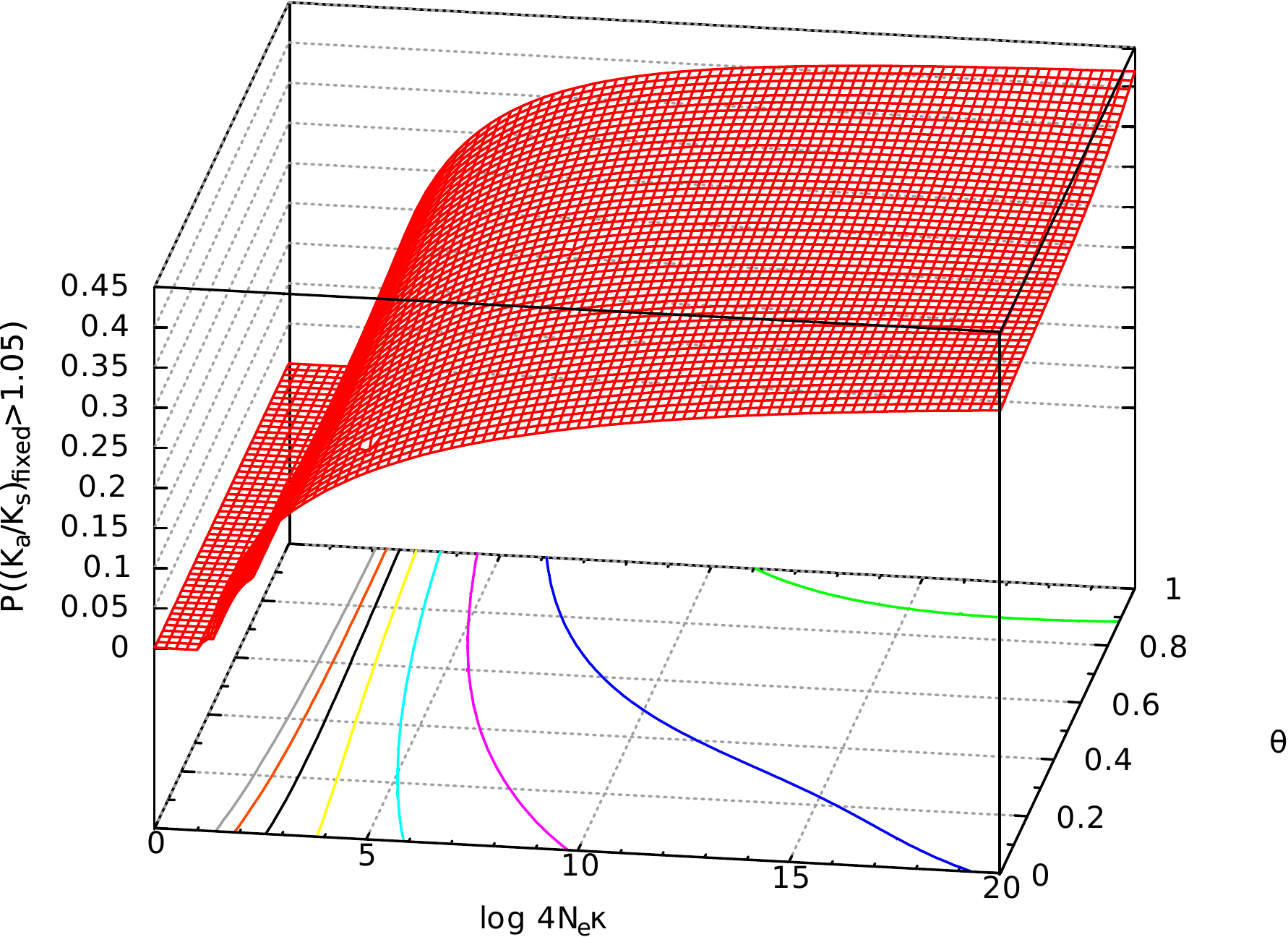}
}
\centerline{
\includegraphics*[width=90mm,angle=0]{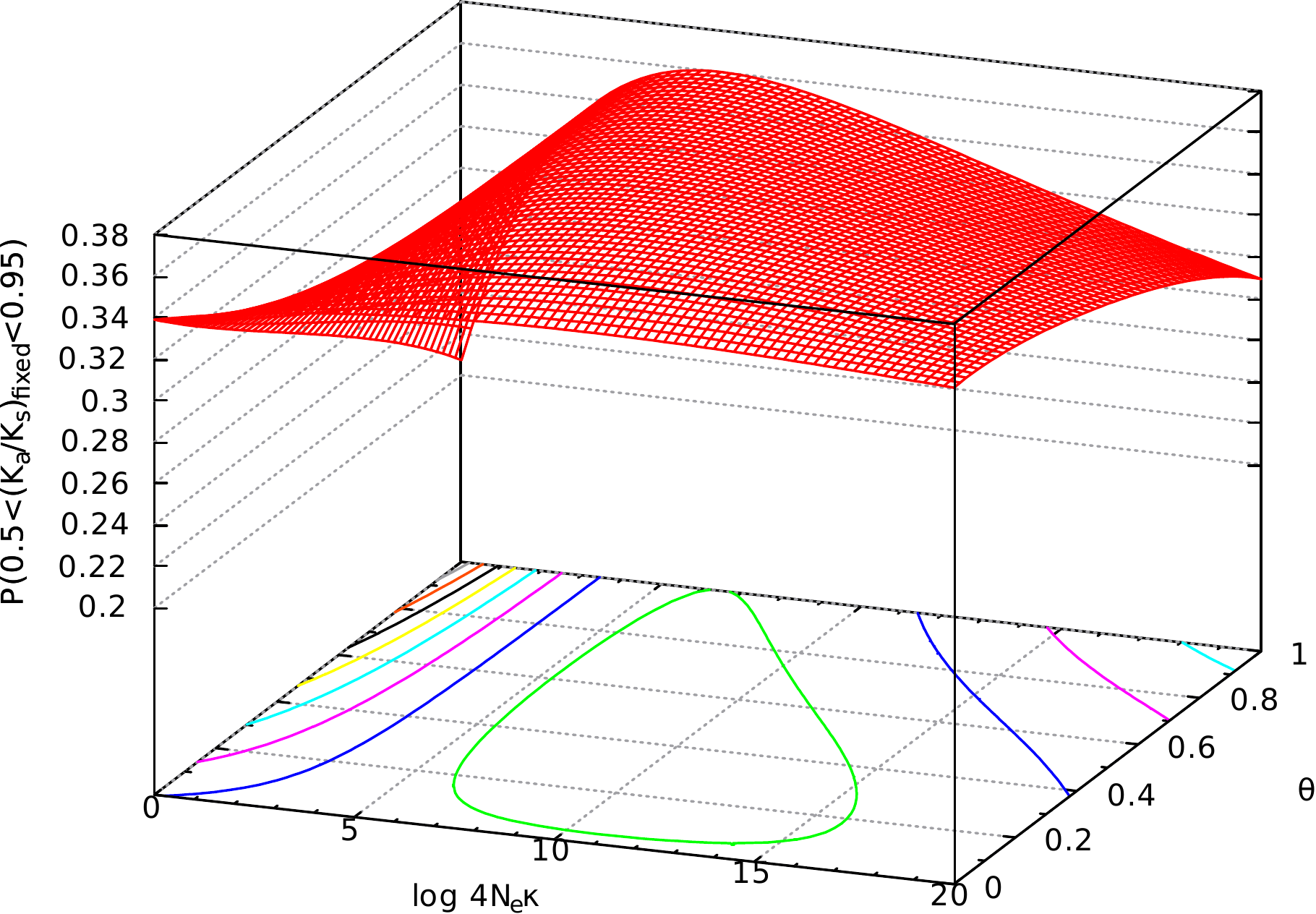}
\includegraphics*[width=90mm,angle=0]{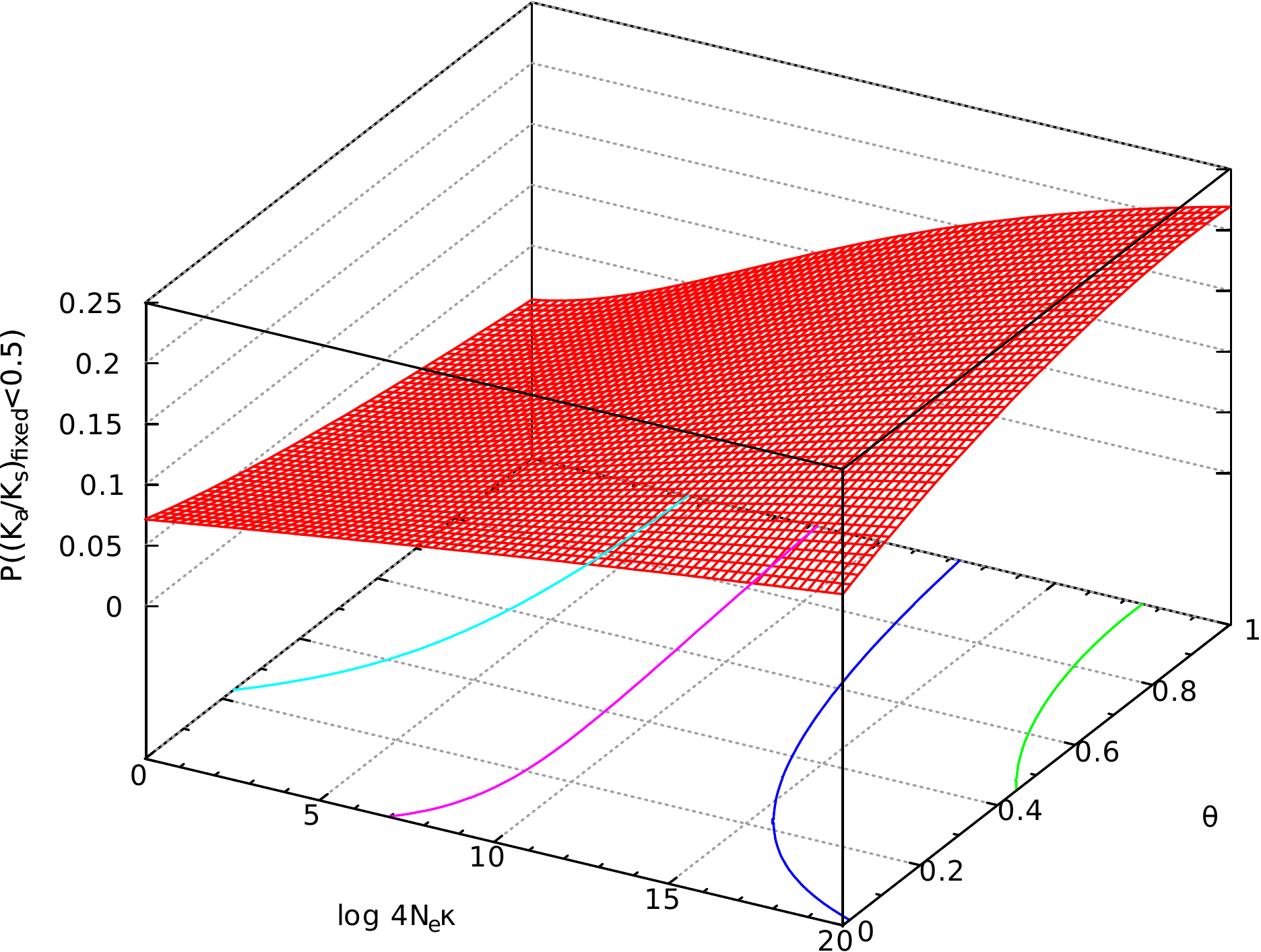}
}
} 
\vspace*{1em}
\caption{
\label{sfig: prob_of_each_selection_category_in_fixed_mutants}
\FigureLegends{
\BF{Probability of each selection category in fixed mutants at equilibrium of protein stability, $\Delta G = \Delta G_e$.}
Arbitrarily, the value of $K_a/K_s$ is categorized into four classes;
negative, slightly negative, nearly neutral, and positive selection categories in which
$K_a/K_s$ is within the ranges of
$K_a/K_s \leq 0.5$, $0.5< K_a/K_s \leq 0.95$, $0.95< K_a/K_s \leq 1.05$, and $ 1.05 < K_a/K_s$,
respectively.
} 
}
\end{figure*}

\FigureInLegends{\newpage}

\begin{figure*}[ht]
\FigureInLegends{
\centerline{
\includegraphics*[width=80mm,angle=0]{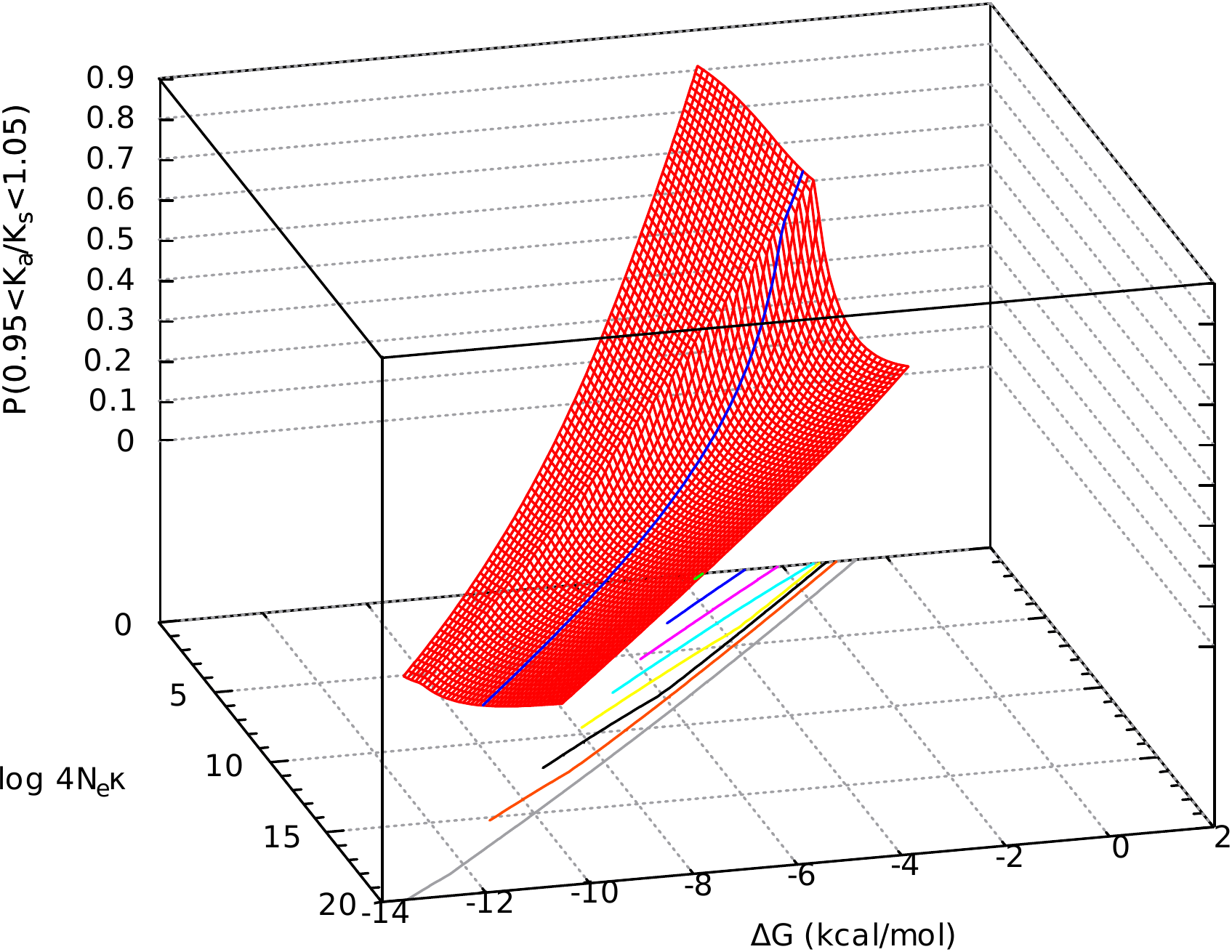}
\includegraphics*[width=90mm,angle=0]{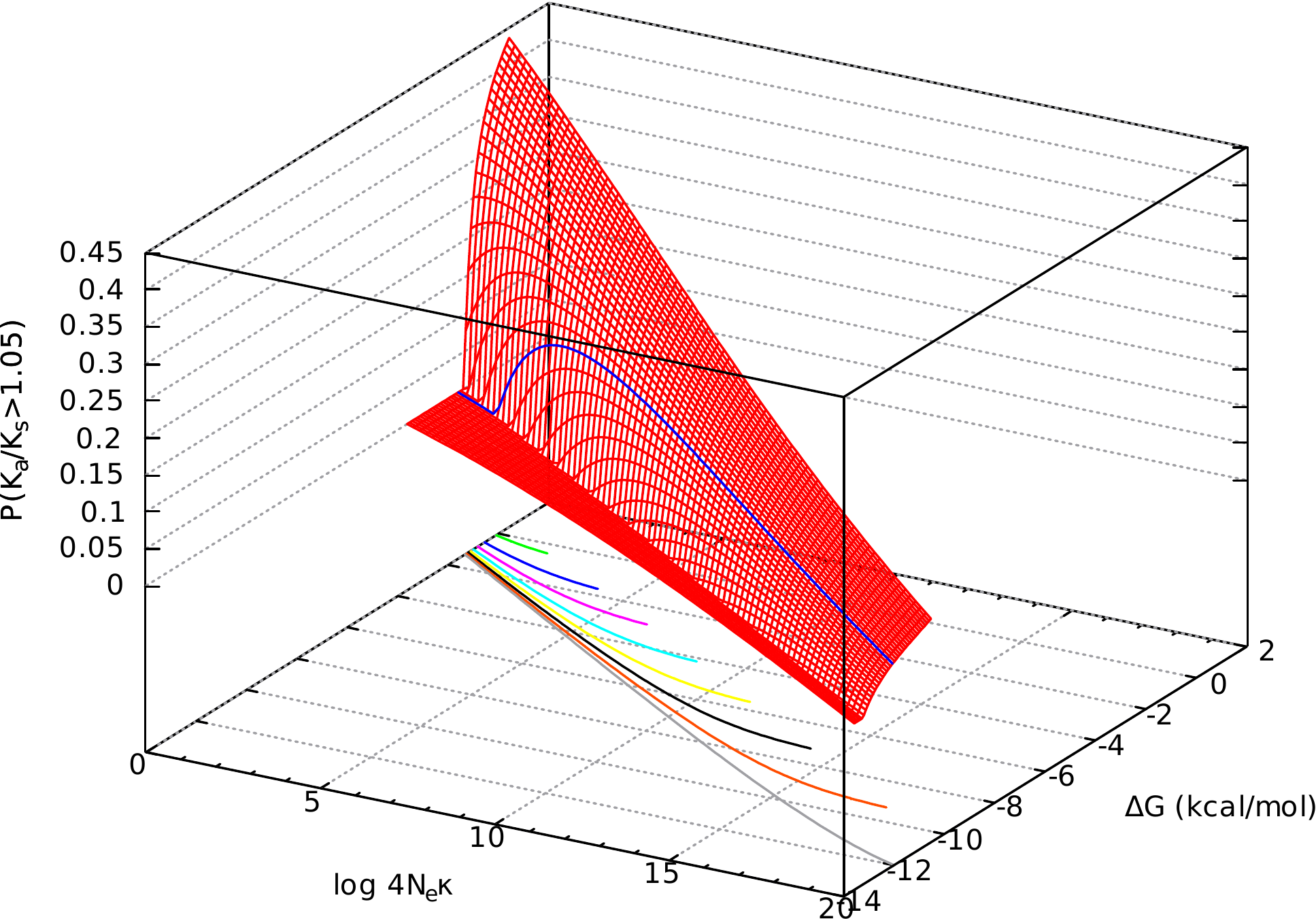}
}
\centerline{
\includegraphics*[width=85mm,angle=0]{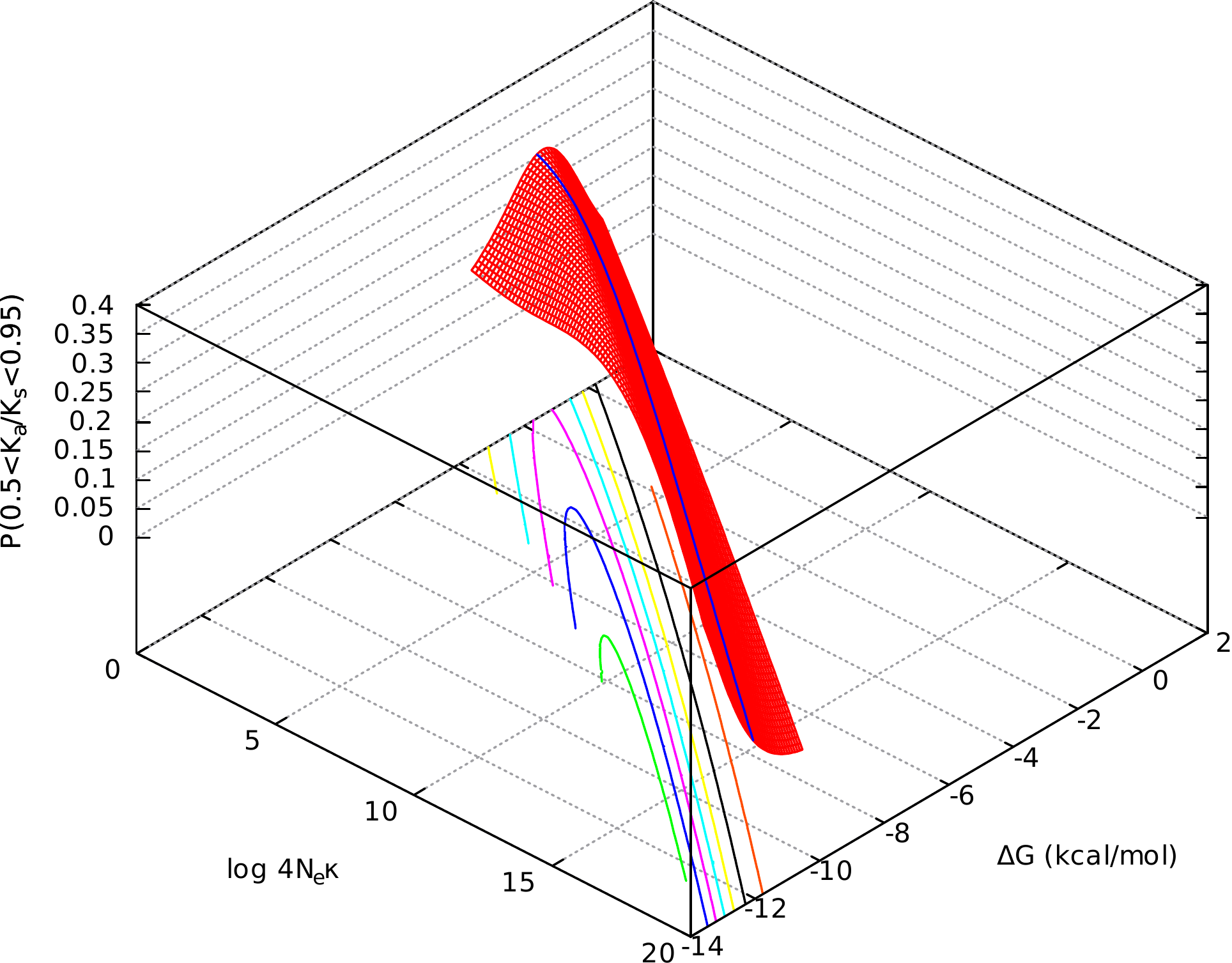}
\includegraphics*[width=90mm,angle=0]{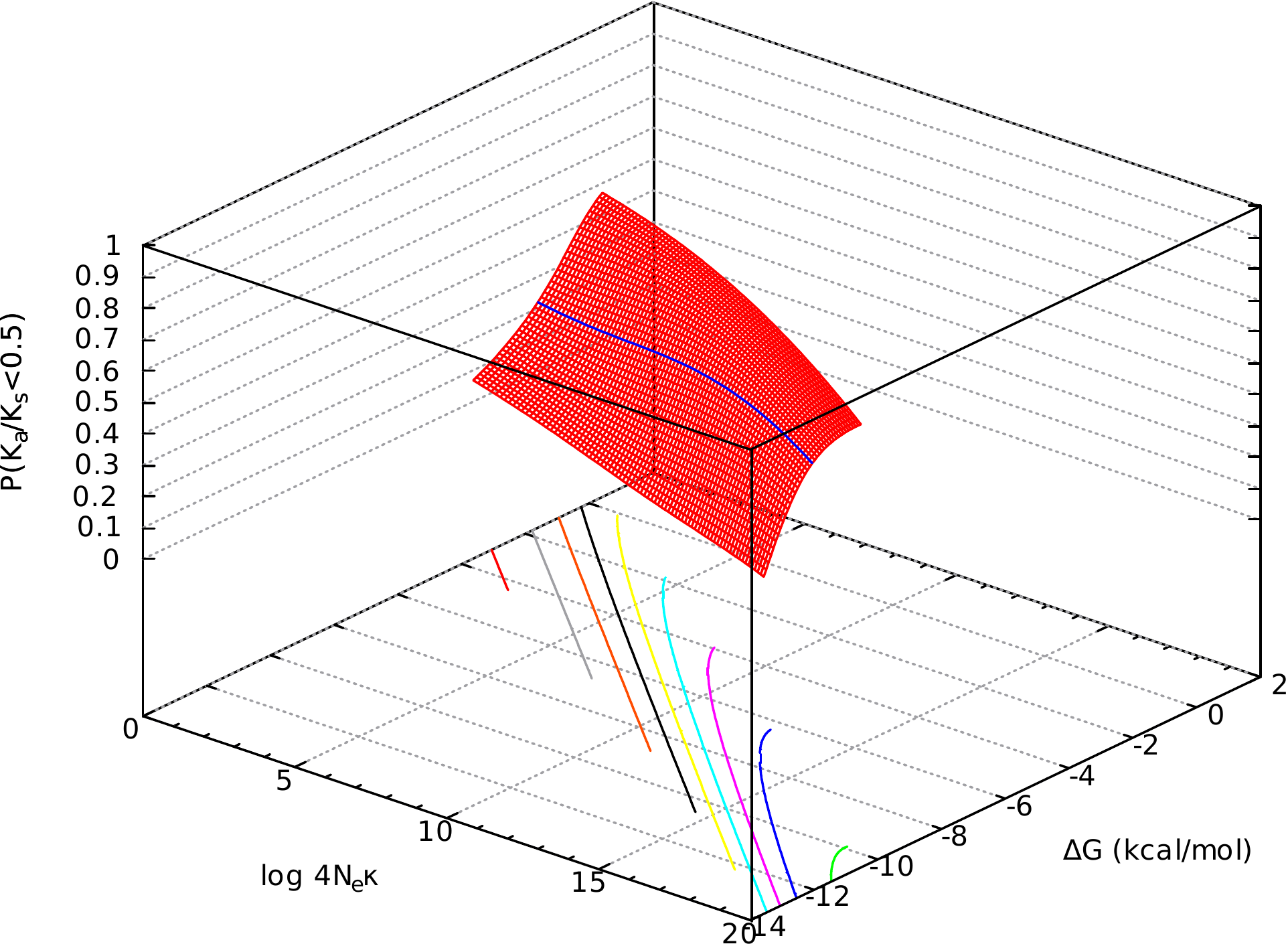}
}
} 
\vspace*{1em}
\caption{
\label{sfig: prob_of_each_selection_category_in_all_mutants;_dependence_on_kappa_and_dG}
\FigureLegends{
\BF{Dependence of the probability of each selection category in all mutants on $4N_e\kappa$ and $\Delta G$.}
A blue line on the surface grid shows $\Delta G=\Delta G_e$, which is the equilibrium value of $\Delta G$
in protein evolution.
The range of $\Delta G$ shown in the figures is
$| \Delta G - \Delta G_e | < 2 \cdot \Delta\Delta G^{\script{sd}}_{\script{fixed}}$, 
where $\Delta\Delta G^{\script{sd}}_{\script{fixed}}$ is the standard deviation of $\Delta\Delta G$ over fixed mutants at $ \Delta G = \Delta G_e$.
Arbitrarily, the value of $K_a/K_s$ is categorized into four classes;
negative, slightly negative, nearly neutral, and positive selection categories in which
$K_a/K_s$ is within the ranges of
$K_a/K_s \leq 0.5$, $0.5< K_a/K_s \leq 0.95$, $0.95< K_a/K_s \leq 1.05$, and $ 1.05 < K_a/K_s$,
respectively.
$\theta = 0.53$
is employed.
The kcal/mol unit is used for $\Delta G$.
} 
}
\end{figure*}

\clearpage
\FigureInLegends{\newpage}

\begin{figure*}[ht]
\FigureInLegends{
\centerline{
\includegraphics*[width=87mm,angle=0]{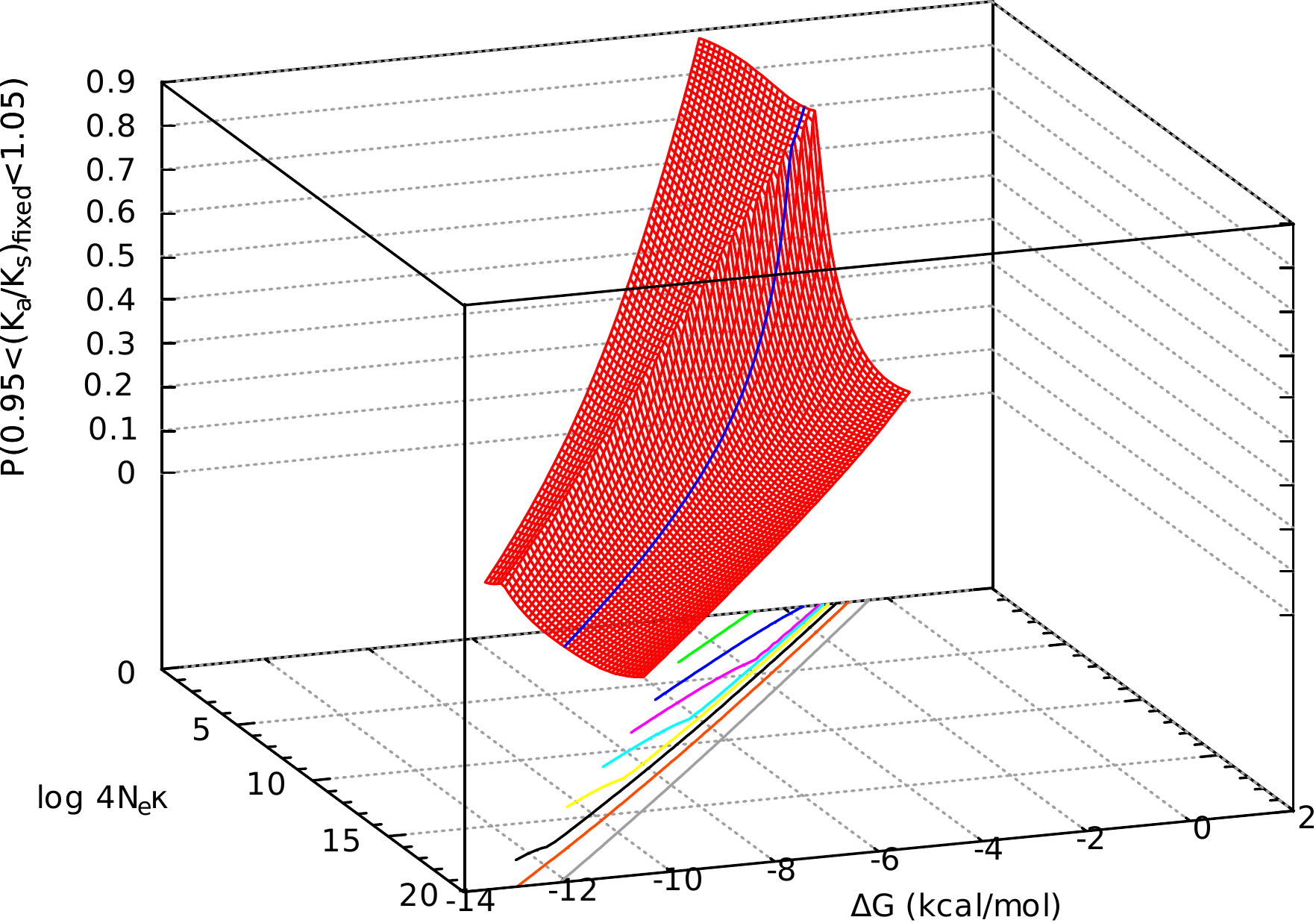}
\includegraphics*[width=90mm,angle=0]{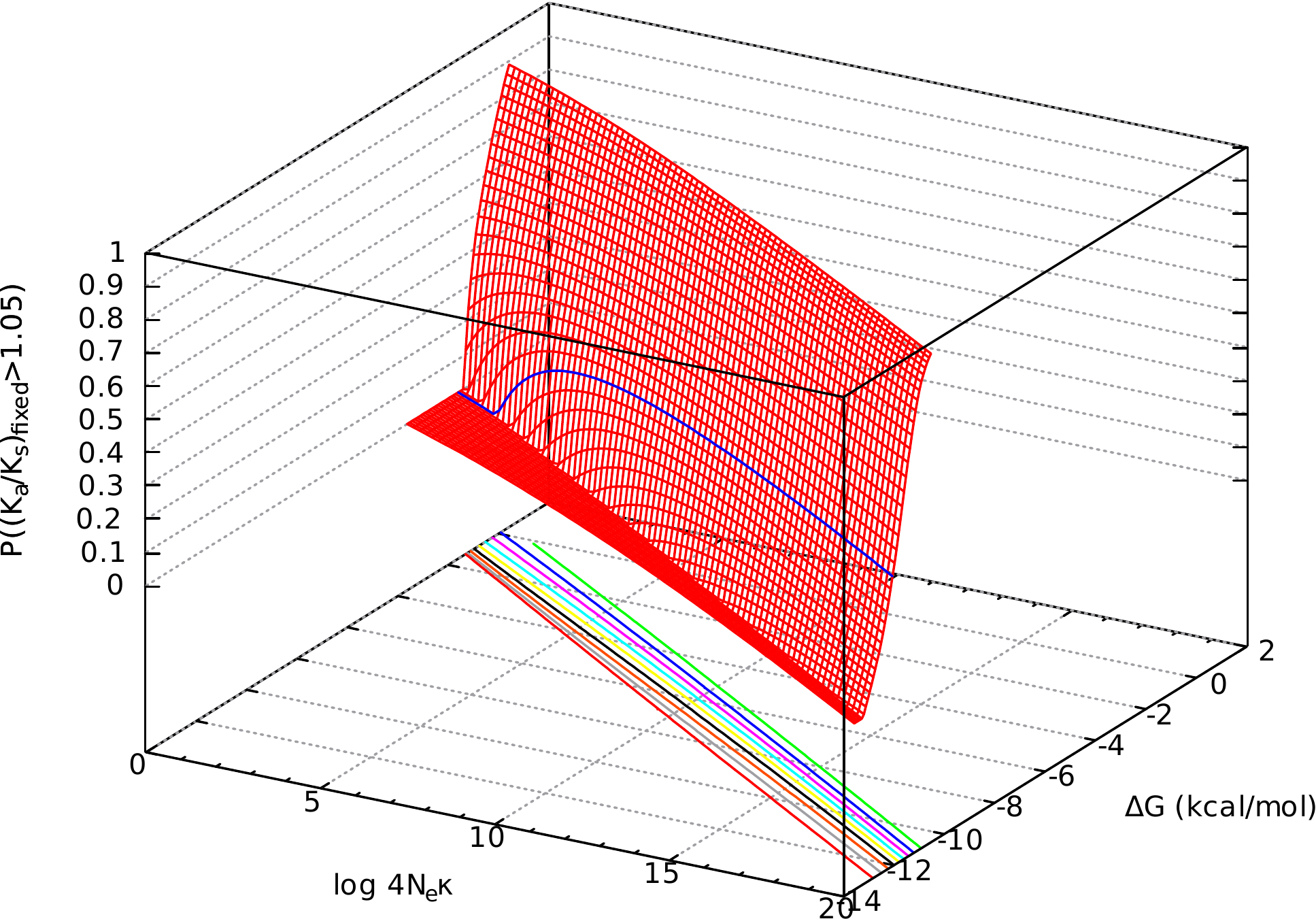}
}
\centerline{
\includegraphics*[width=90mm,angle=0]{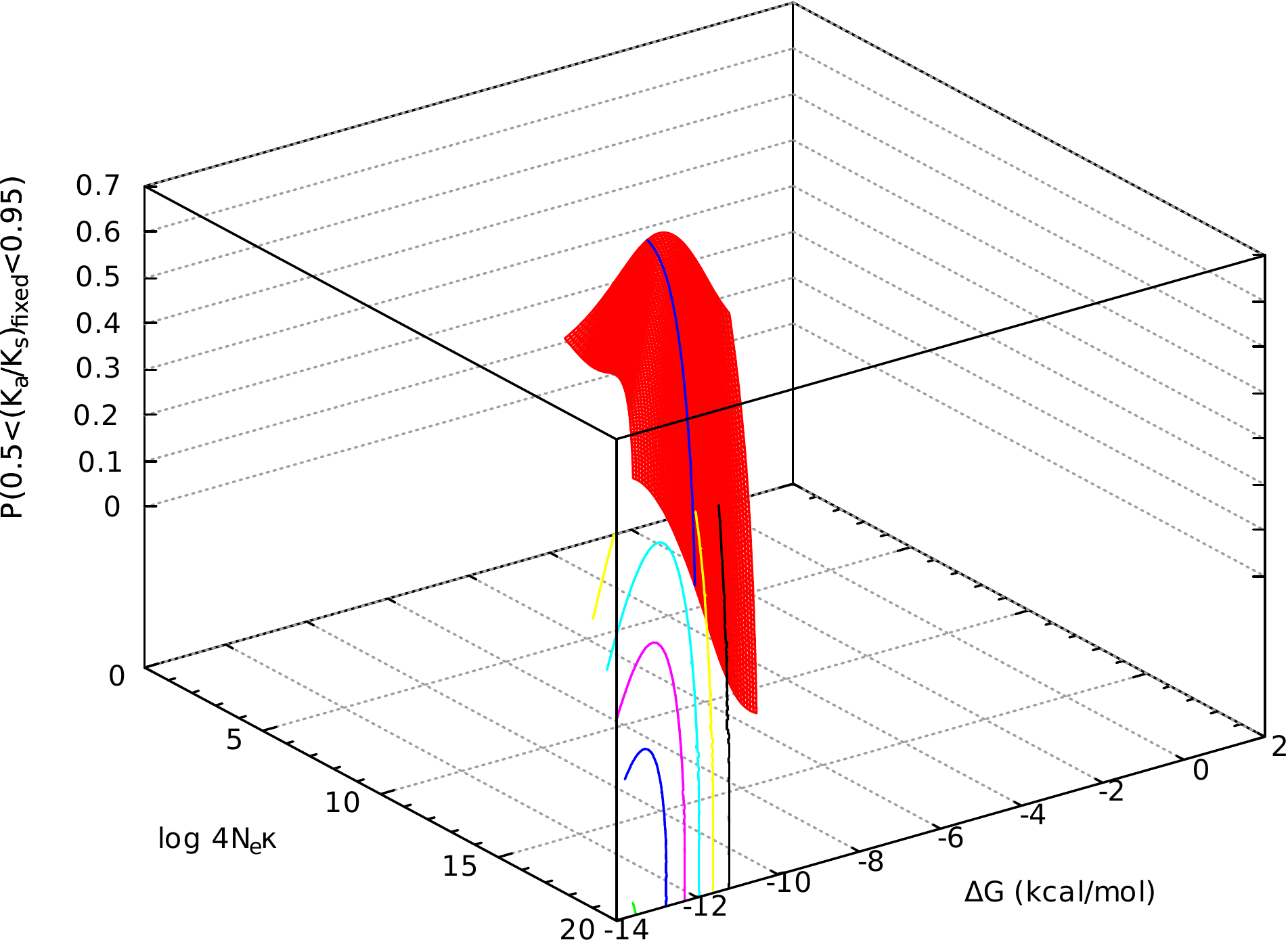}
\includegraphics*[width=90mm,angle=0]{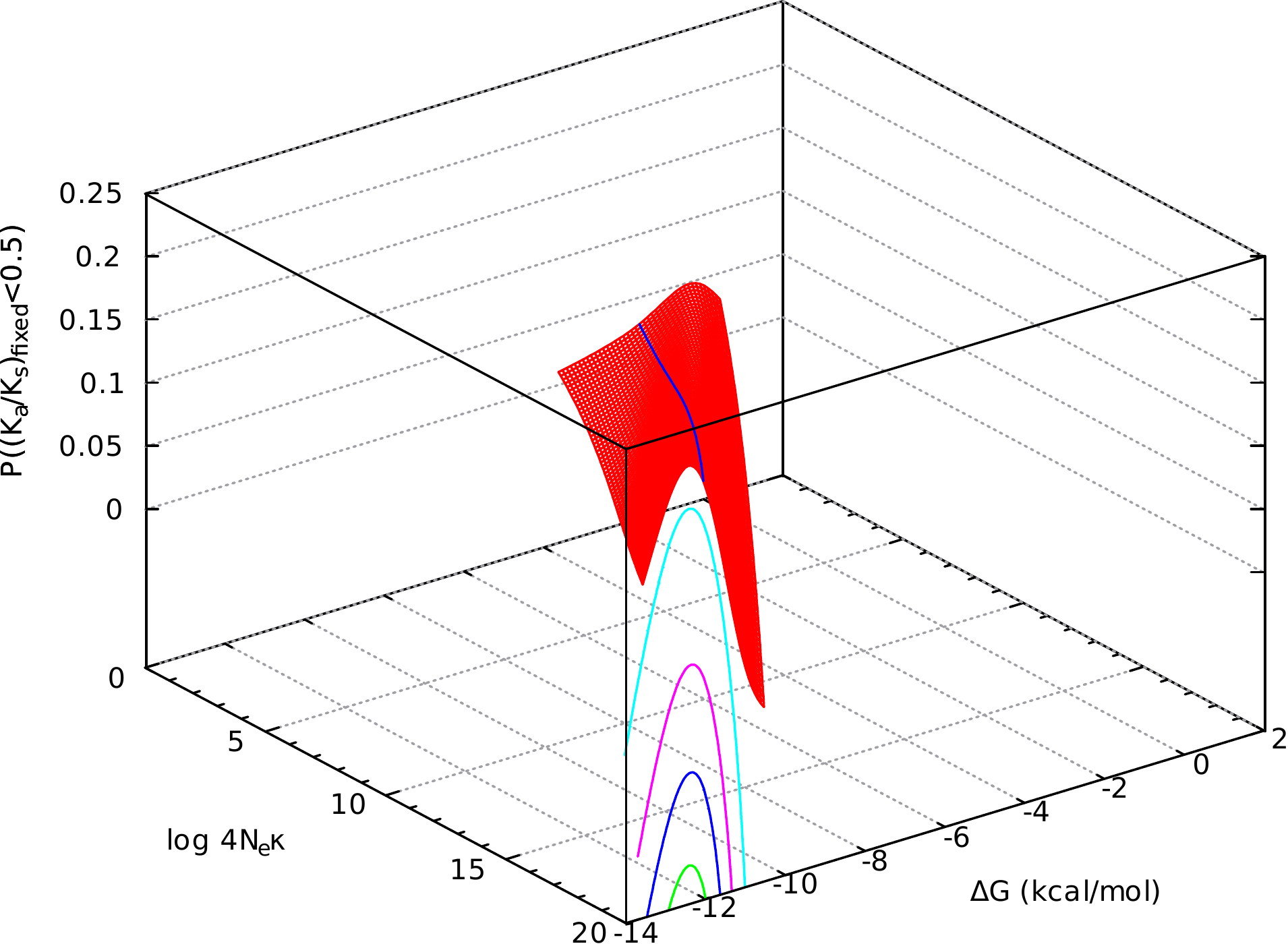}
}
} 
\vspace*{1em}
\caption{
\label{sfig: prob_of_each_selection_category_in_fixed_mutants;_dependence_on_kappa_and_dG}
\FigureLegends{
\BF{Dependence of the probability of each selection category in fixed mutants on $4N_e\kappa$ and $\Delta G$.}
A blue line on the surface grid shows $\Delta G=\Delta G_e$, which is the equilibrium value of $\Delta G$
in protein evolution.
The range of $\Delta G$ shown in the figures is
$| \Delta G - \Delta G_e | < 2 \cdot \Delta\Delta G^{\script{sd}}_{\script{fixed}}$,
where $\Delta\Delta G^{\script{sd}}_{\script{fixed}}$ is the standard deviation of $\Delta\Delta G$ over fixed mutants at $ \Delta G = \Delta G_e$.
Arbitrarily, the value of $K_a/K_s$ is categorized into four classes;
negative, slightly negative, nearly neutral, and positive selection categories in which
$K_a/K_s$ is within the ranges of
$K_a/K_s \leq 0.5$, $0.5< K_a/K_s \leq 0.95$, $0.95< K_a/K_s \leq 1.05$, and $ 1.05 < K_a/K_s$,
respectively.
$\theta = 0.53$
is employed.
The kcal/mol unit is used for $\Delta G$.
} 
}
\end{figure*}

\FigureInLegends{\newpage}

\begin{figure*}[ht]
\FigureInLegends{
\centerline{
\includegraphics*[width=80mm,angle=0]{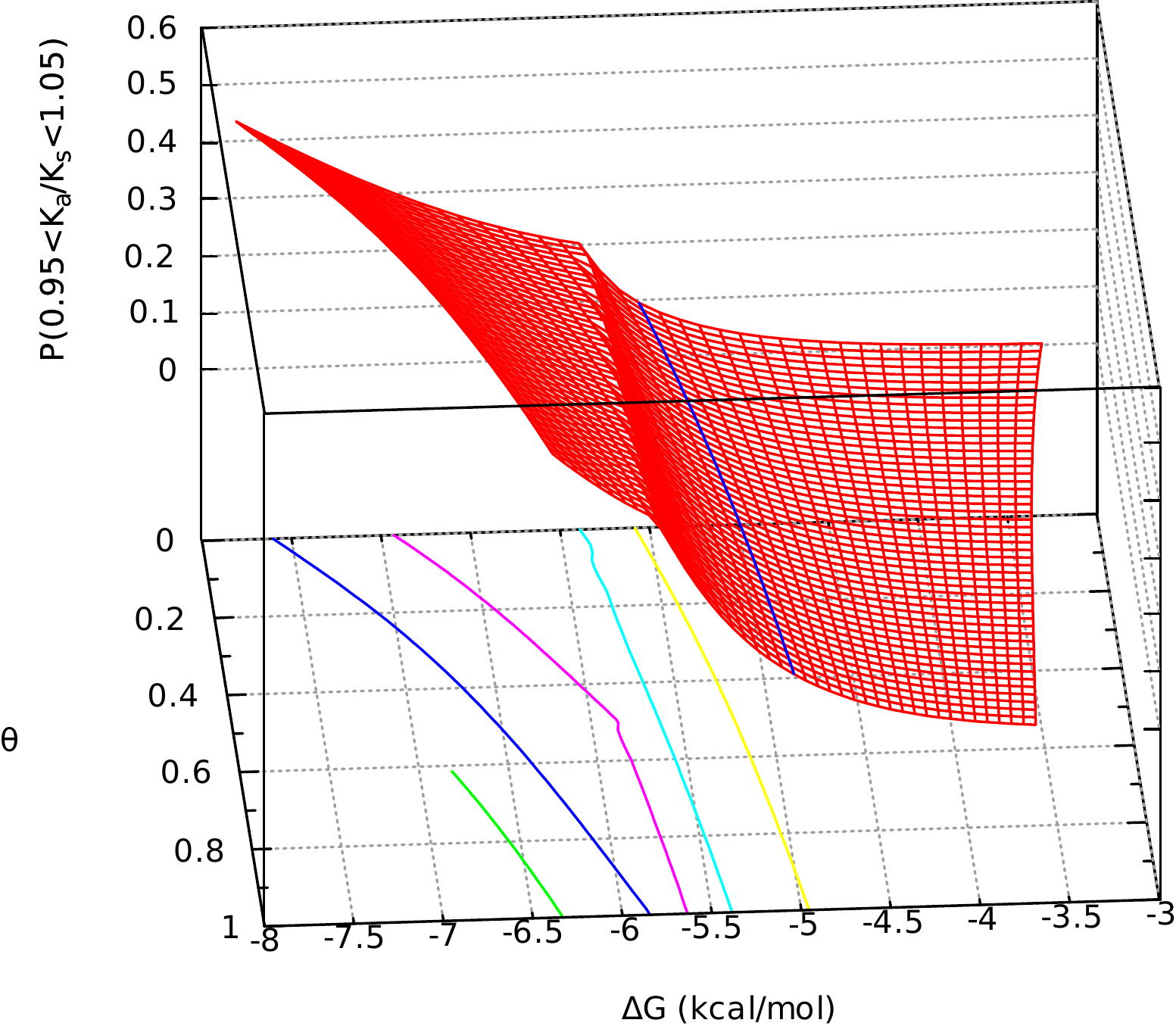}
\includegraphics*[width=90mm,angle=0]{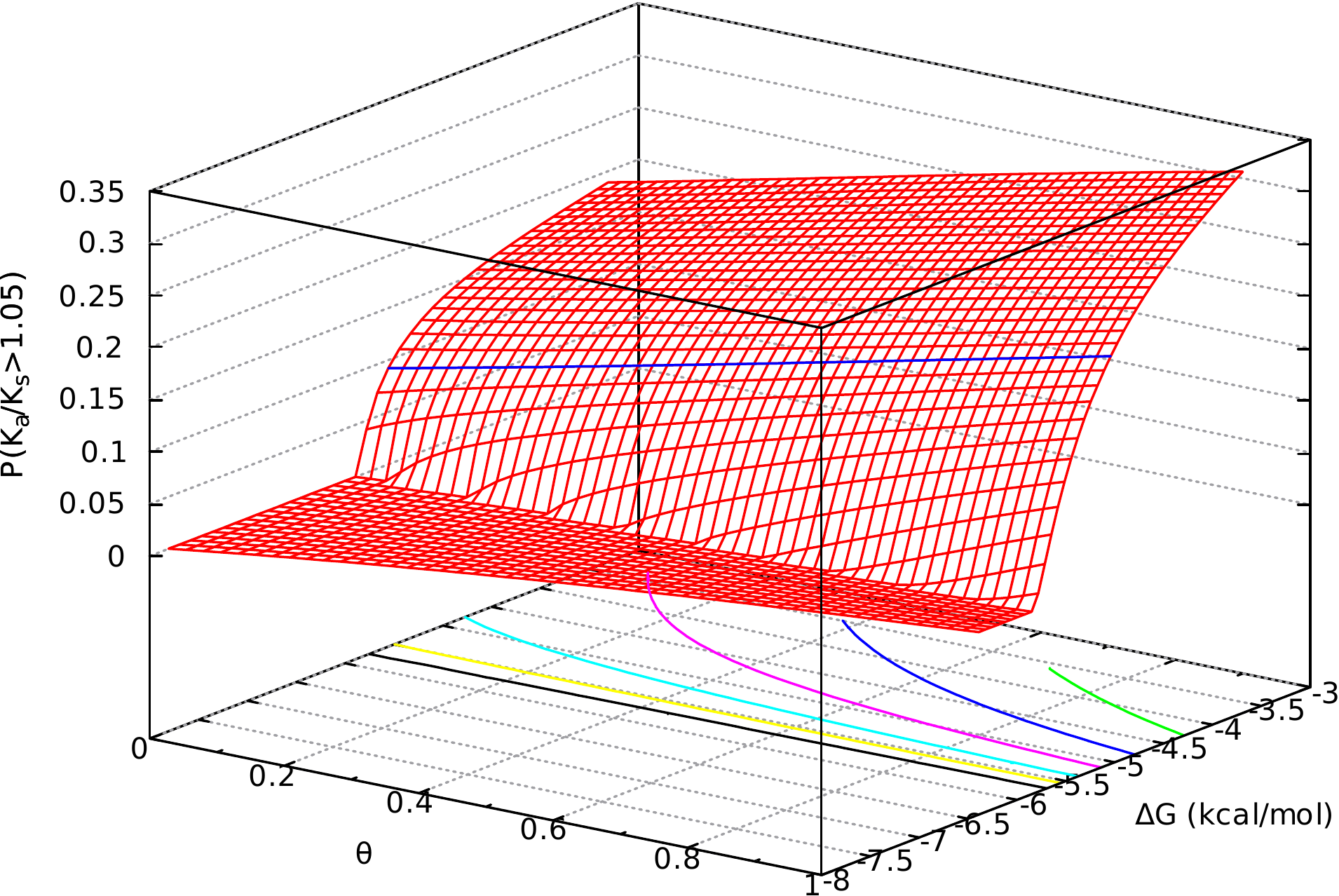}
}
\centerline{
\includegraphics*[width=83mm,angle=0]{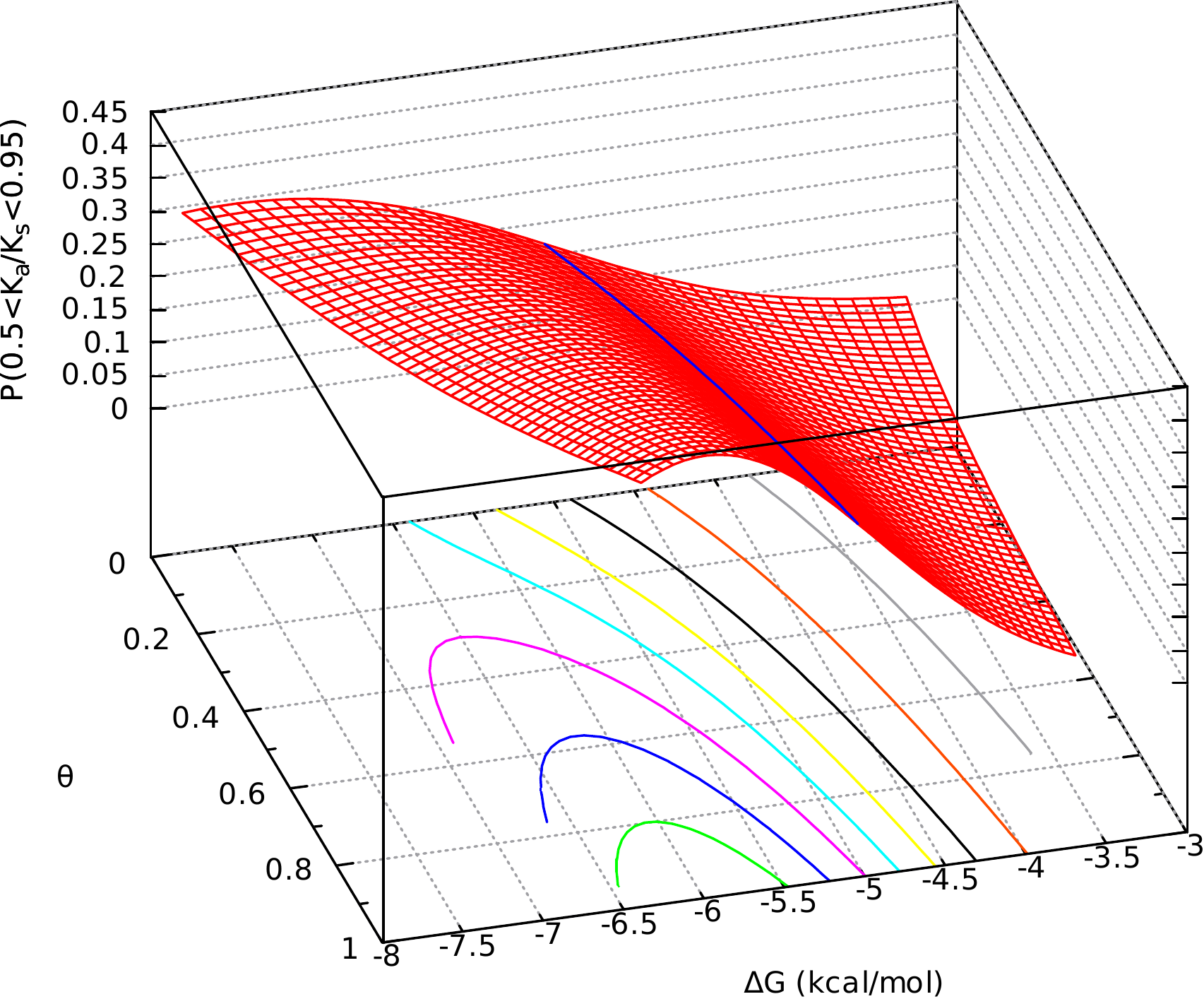}
\includegraphics*[width=90mm,angle=0]{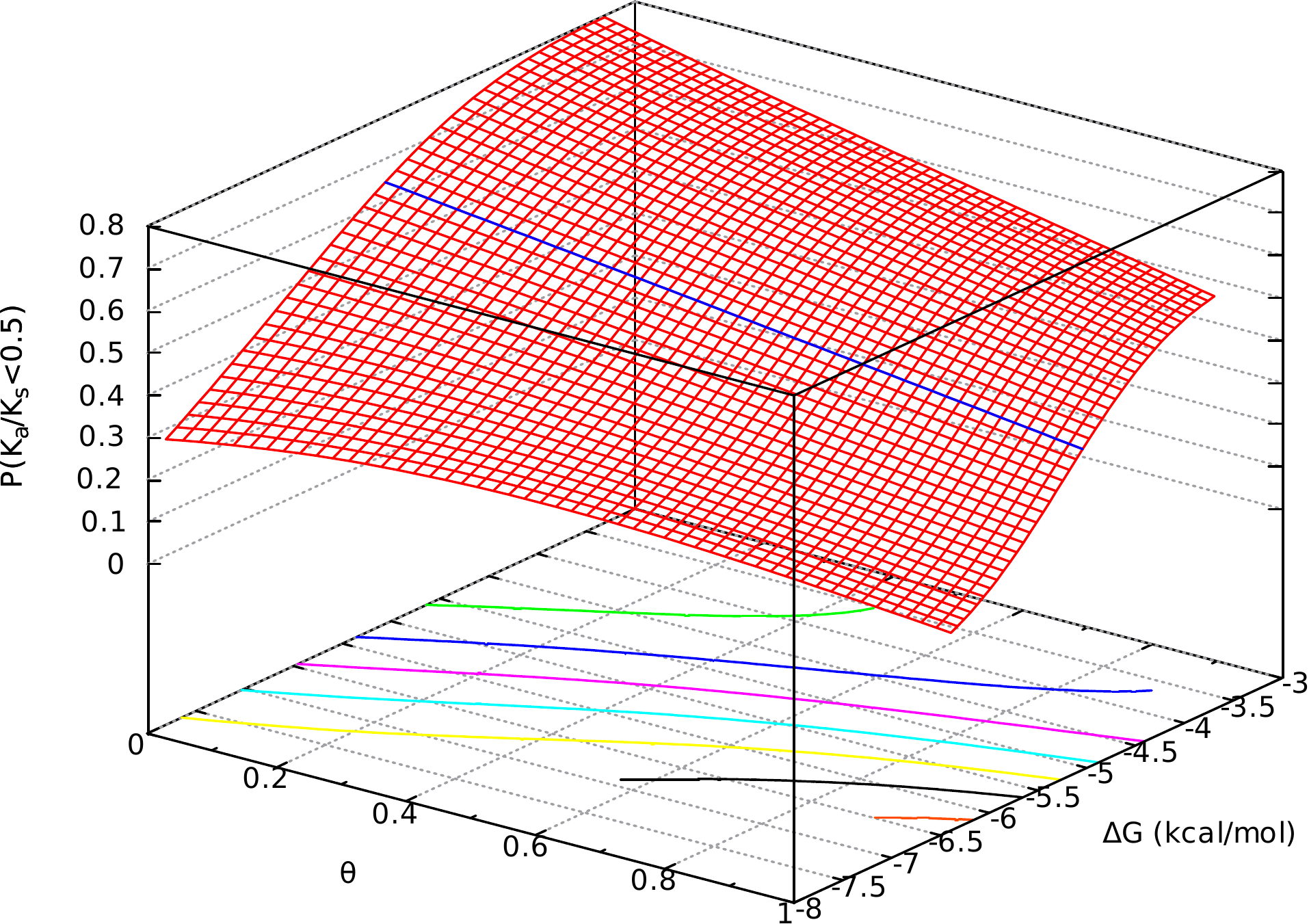}
}
} 
\vspace*{1em}
\caption{
\label{sfig: prob_of_each_selection_category_in_all_mutants;_dependence_on_theta_and_dG}
\FigureLegends{
\BF{Dependence of the probability of each selection category in all mutants on $\theta$ and $\Delta G$.}
A blue line on the surface grid shows $\Delta G=\Delta G_e$, which is the equilibrium value of $\Delta G$
in protein evolution.
The range of $\Delta G$ shown in the figures is
$| \Delta G - \Delta G_e | < 2 \cdot \Delta\Delta G^{\script{sd}}_{\script{fixed}}$,
where $\Delta\Delta G^{\script{sd}}_{\script{fixed}}$ is the standard deviation of $\Delta\Delta G$ over fixed mutants at $ \Delta G = \Delta G_e$.
Arbitrarily, the value of $K_a/K_s$ is categorized into four classes;
negative, slightly negative, nearly neutral, and positive selection categories in which
$K_a/K_s$ is within the ranges of
$K_a/K_s \leq 0.5$, $0.5< K_a/K_s \leq 0.95$, $0.95< K_a/K_s \leq 1.05$, and $ 1.05 < K_a/K_s$,
respectively.
$\log 4N_e\kappa = 7.550$ 
is employed.
The kcal/mol unit is used for $\Delta G$.
} 
}
\end{figure*}

\FigureInLegends{\newpage}

\begin{figure*}[ht]
\FigureInLegends{
\centerline{
\includegraphics*[width=83mm,angle=0]{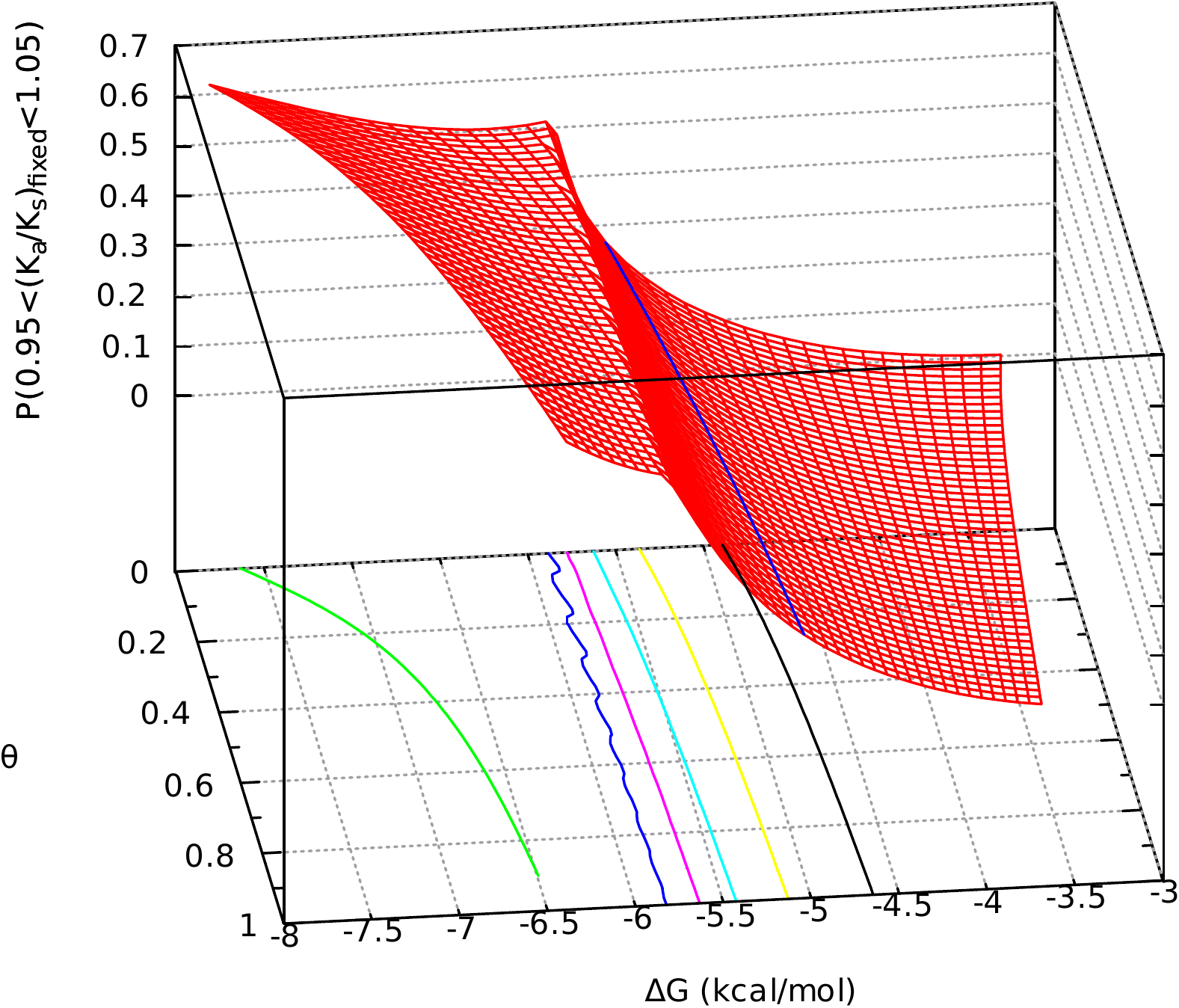}
\includegraphics*[width=90mm,angle=0]{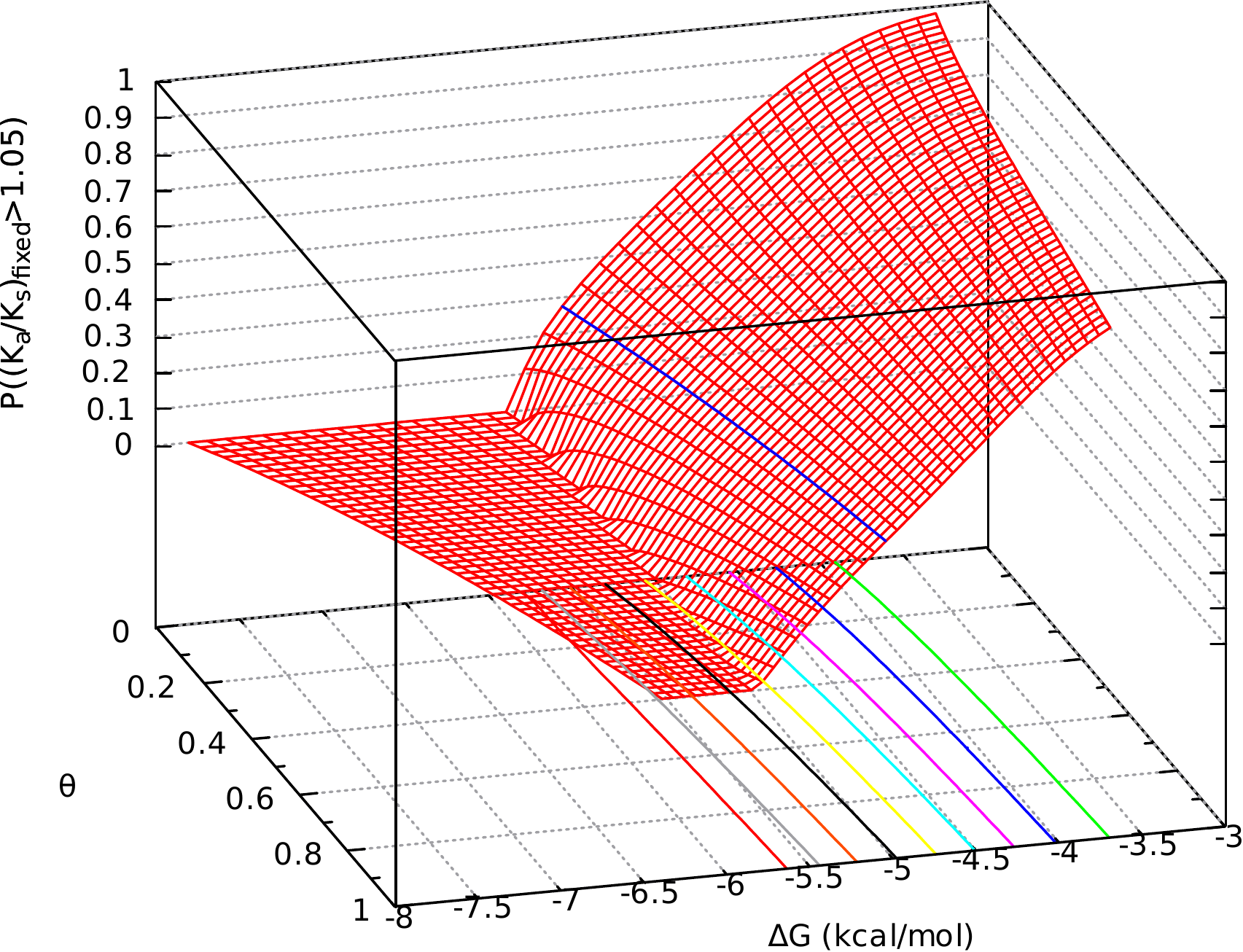}
}
\centerline{
\includegraphics*[width=89mm,angle=0]{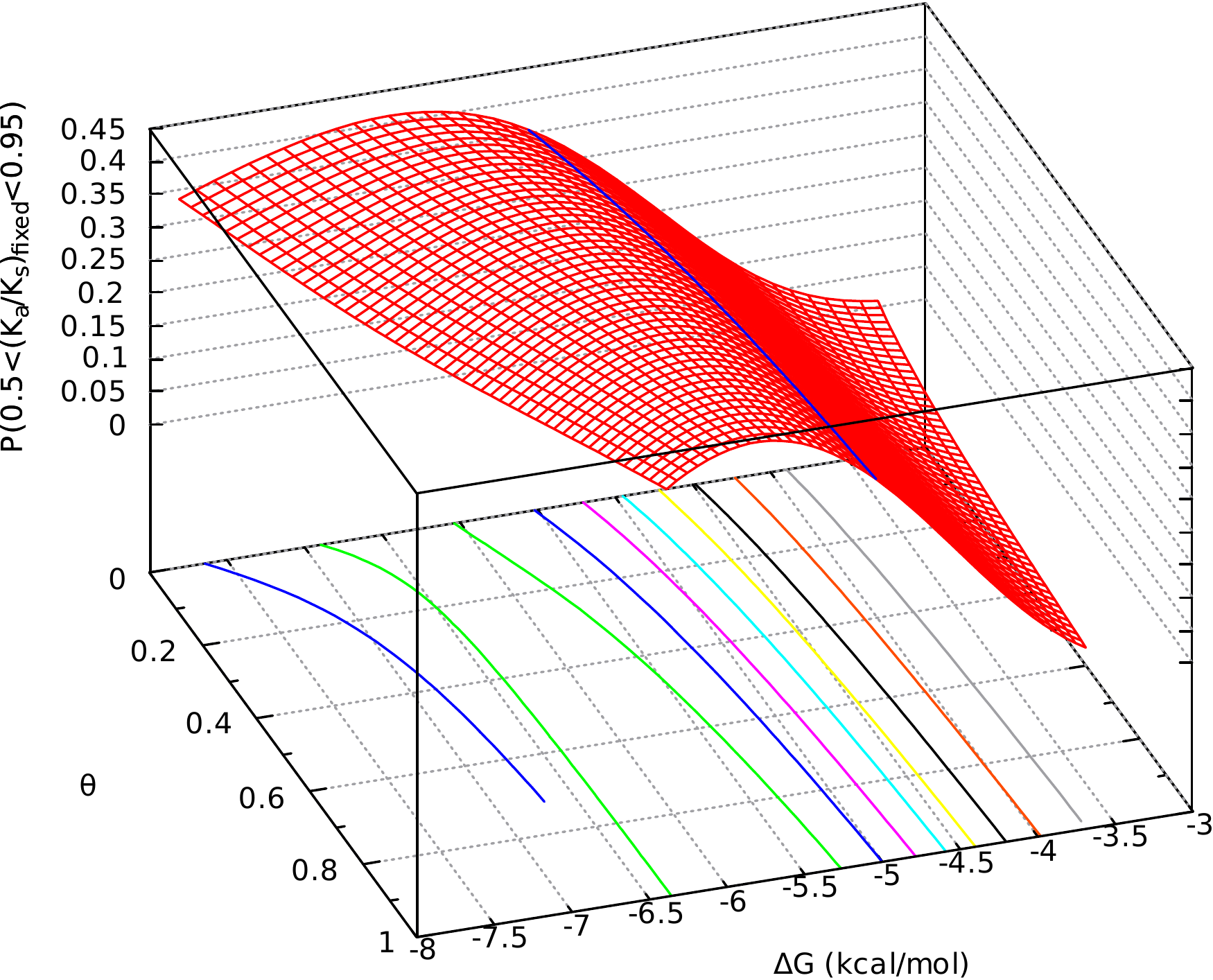}
\includegraphics*[width=90mm,angle=0]{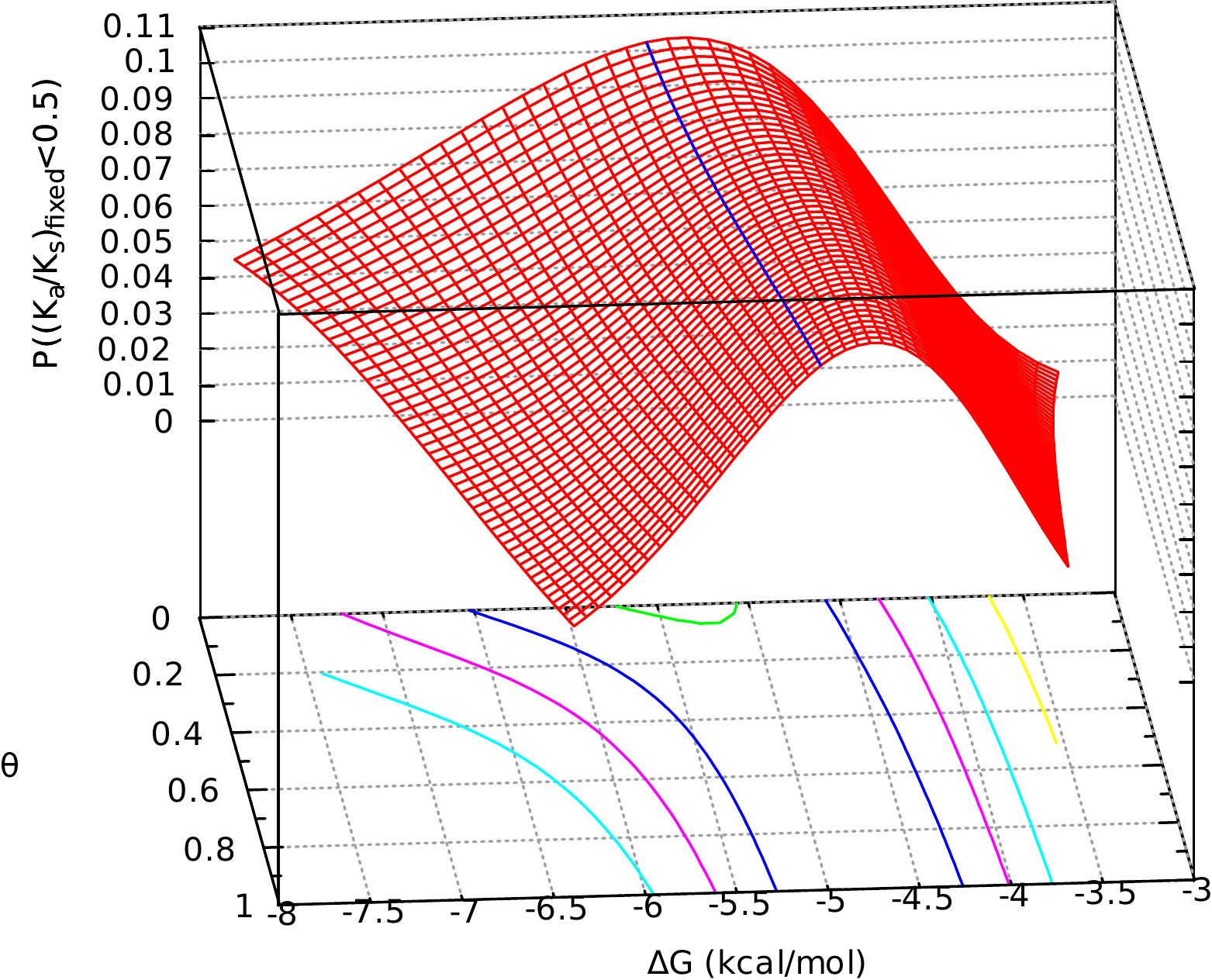}
}
} 
\vspace*{1em}
\caption{
\label{sfig: prob_of_each_selection_category_in_fixed_mutants;_dependence_on_theta_and_dG}
\FigureLegends{
\BF{Dependence of the probability of each selection category in fixed mutants on $\theta$ and $\Delta G$.}
The blue line on the surface grid shows $\Delta G=\Delta G_e$, which is the equilibrium value of $\Delta G$
in protein evolution.
The range of $\Delta G$ shown in the figures is
$| \Delta G - \Delta G_e | < 2 \cdot \Delta\Delta G^{\script{sd}}_{\script{fixed}}$,
where $\Delta\Delta G^{\script{sd}}_{\script{fixed}}$ is the standard deviation of $\Delta\Delta G$ over fixed mutants at $ \Delta G = \Delta G_e$.
Arbitrarily, the value of $K_a/K_s$ is categorized into four classes;
negative, slightly negative, nearly neutral, and positive selection categories in which
$K_a/K_s$ is within the ranges of
$K_a/K_s \leq 0.5$, $0.5< K_a/K_s \leq 0.95$, $0.95< K_a/K_s \leq 1.05$, and $ 1.05 < K_a/K_s$,
respectively.
$\log 4N_e\kappa = 7.550$
is employed.
The kcal/mol unit is used for $\Delta G$.
} 
}
\end{figure*}

\FigureInLegends{\clearpage\newpage}

\begin{figure*}[ht]
\FigureInLegends{
\centerline{
\includegraphics*[width=90mm,angle=0]{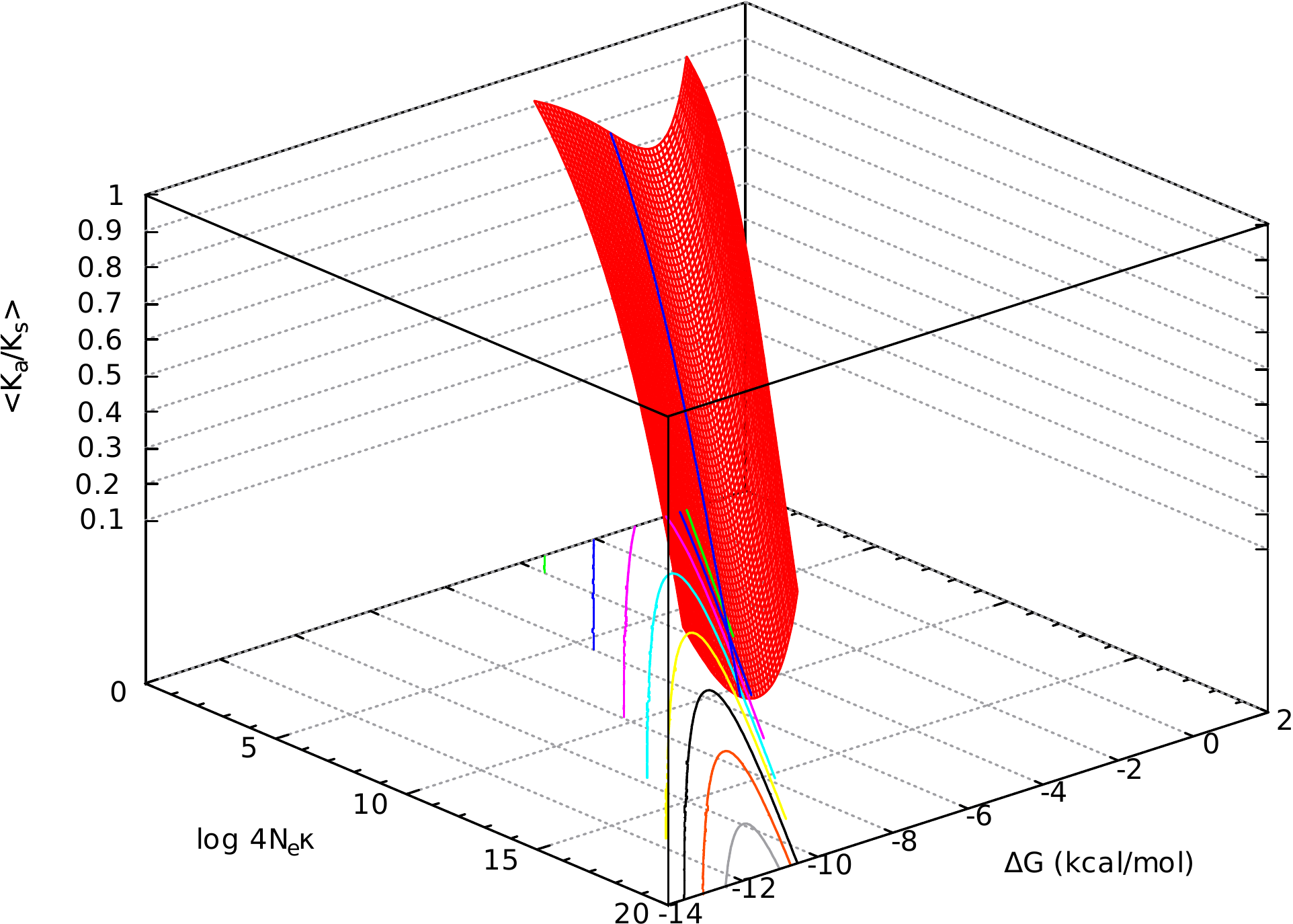}
\includegraphics*[width=85mm,angle=0]{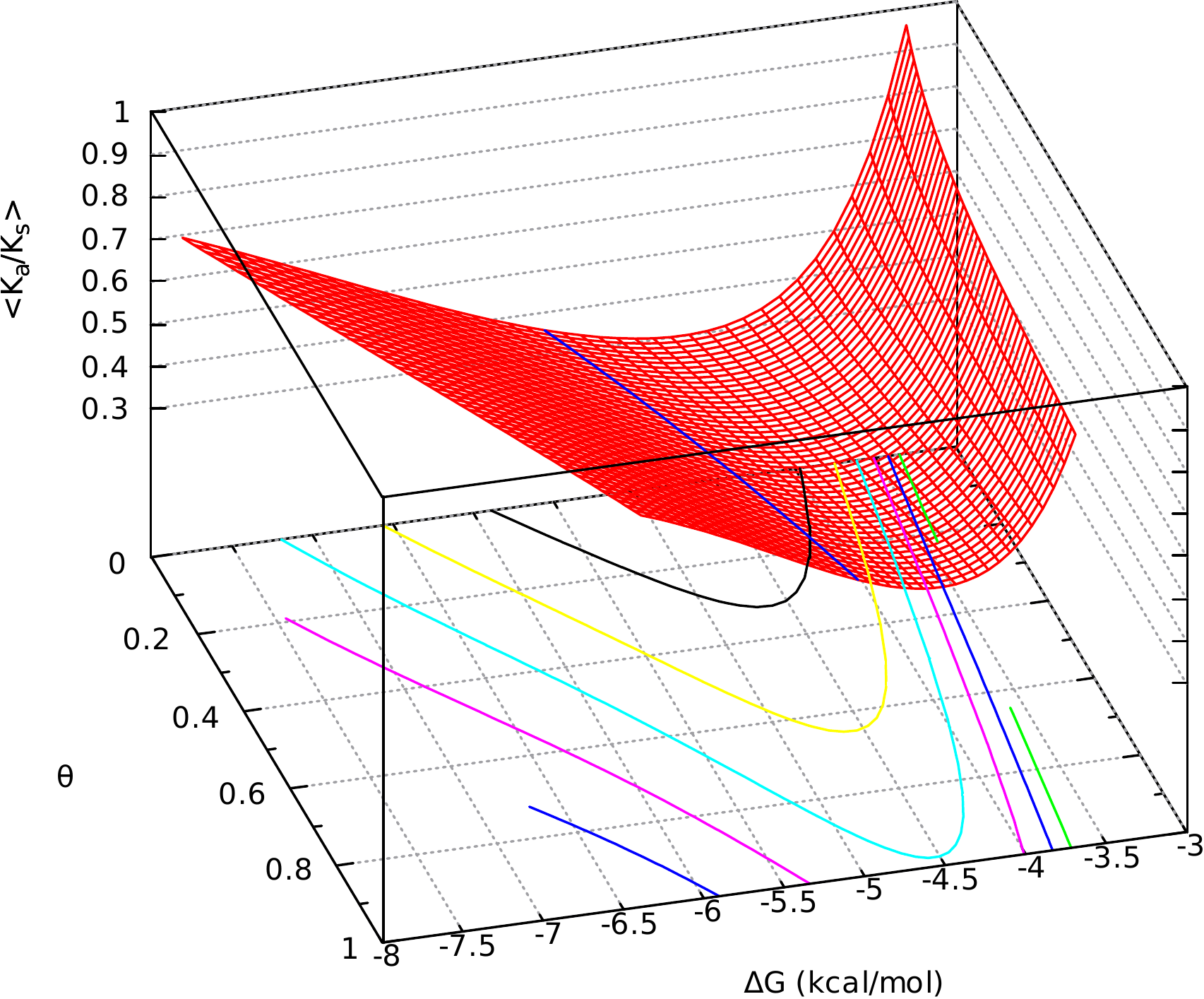}
}
\centerline{
\includegraphics*[width=88mm,angle=0]{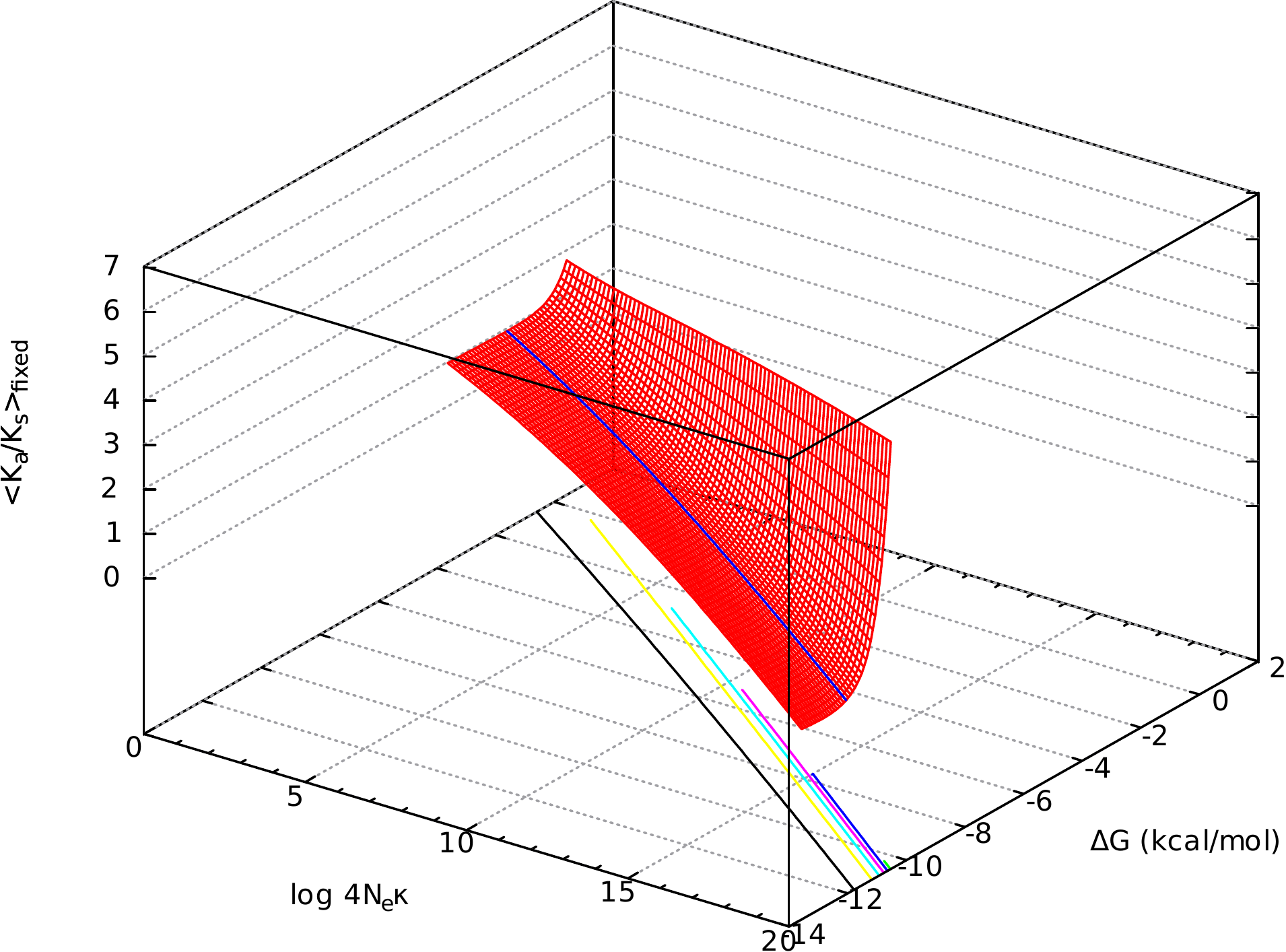}
\includegraphics*[width=90mm,angle=0]{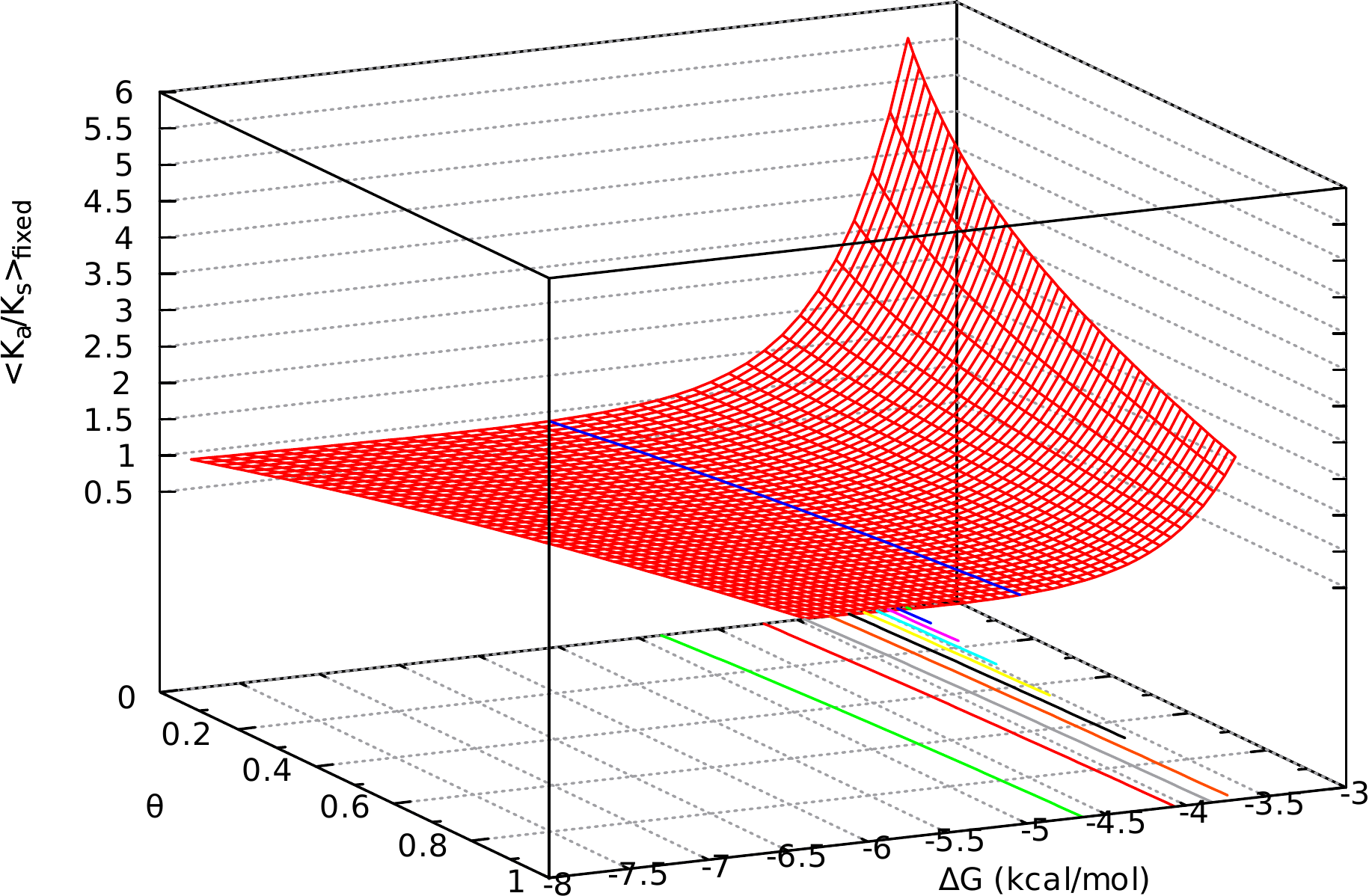}
}
} 
\vspace*{1em}
\caption{
\label{sfig: dependence_of_ave_Ka_over_Ks_on_dG}
\label{sfig: dependence_of_ave_Ka_over_Ks_fixed_on_dG}
\FigureLegends{
\BF{Dependence of the average of $K_a / K_s$ over all mutants or over fixed mutants only 
on protein stability, $\Delta G$, of the wild type.}
A blue line on the surface grid shows $\Delta G=\Delta G_e$, which is the equilibrium value of $\Delta G$
in protein evolution.
The range of $\Delta G$ shown in the figures is
$| \Delta G - \Delta G_e | < 2 \cdot \Delta\Delta G^{\script{sd}}_{\script{fixed}}$,
where $\Delta\Delta G^{\script{sd}}_{\script{fixed}}$ is the standard deviation of $\Delta\Delta G$ over fixed mutants at $\Delta G = \Delta G_e$.
Unless specified, 
$\log 4N_e\kappa = 7.550$ and $\theta = 0.53$
are employed.
The kcal/mol unit is used for $\Delta G$.
} 
}
\end{figure*}

\FigureInLegends{\newpage}

\begin{figure*}[ht]
\FigureInLegends{
\centerline{
\includegraphics*[width=90mm,angle=0]{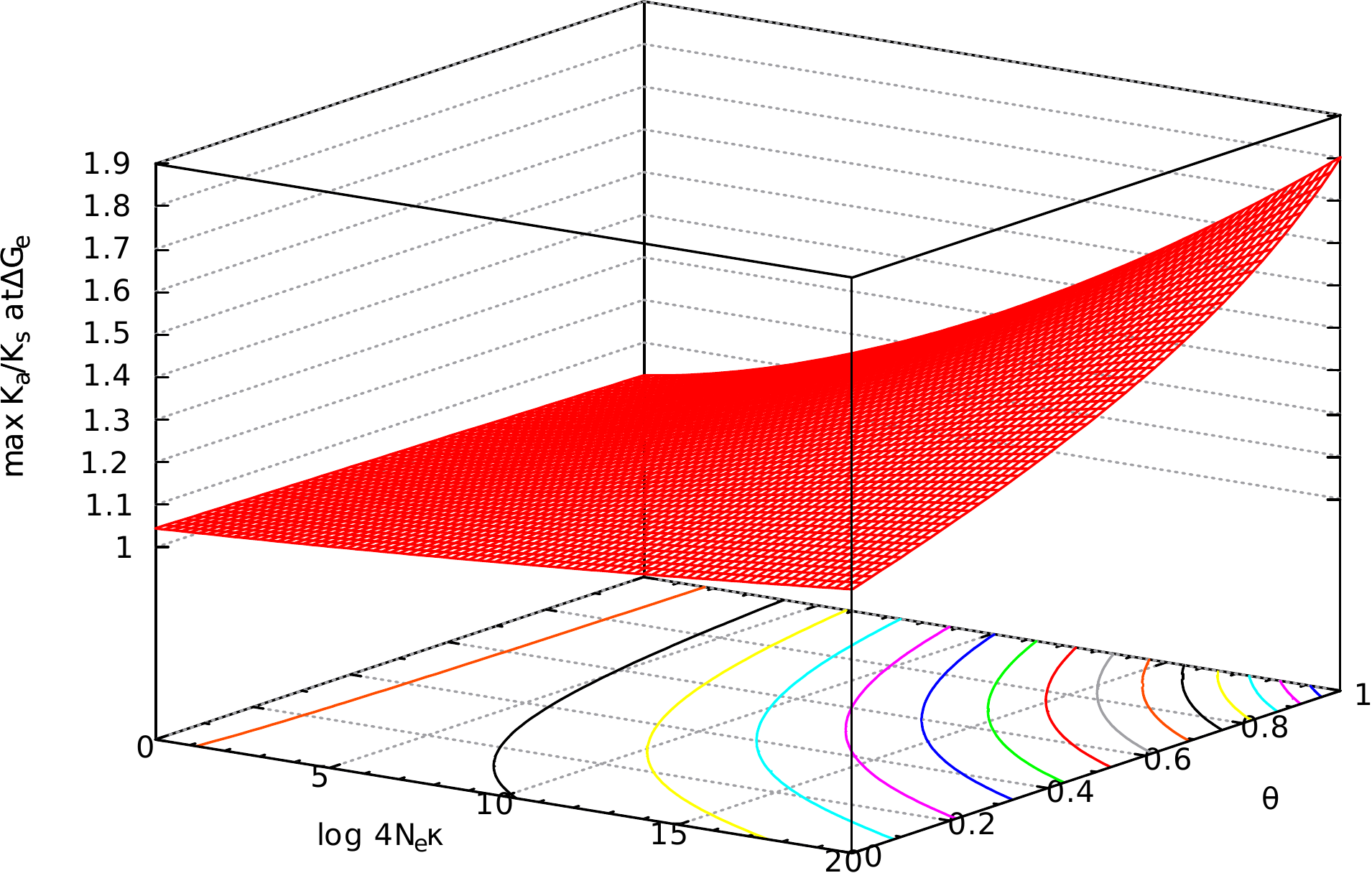}
\includegraphics*[width=90mm,angle=0]{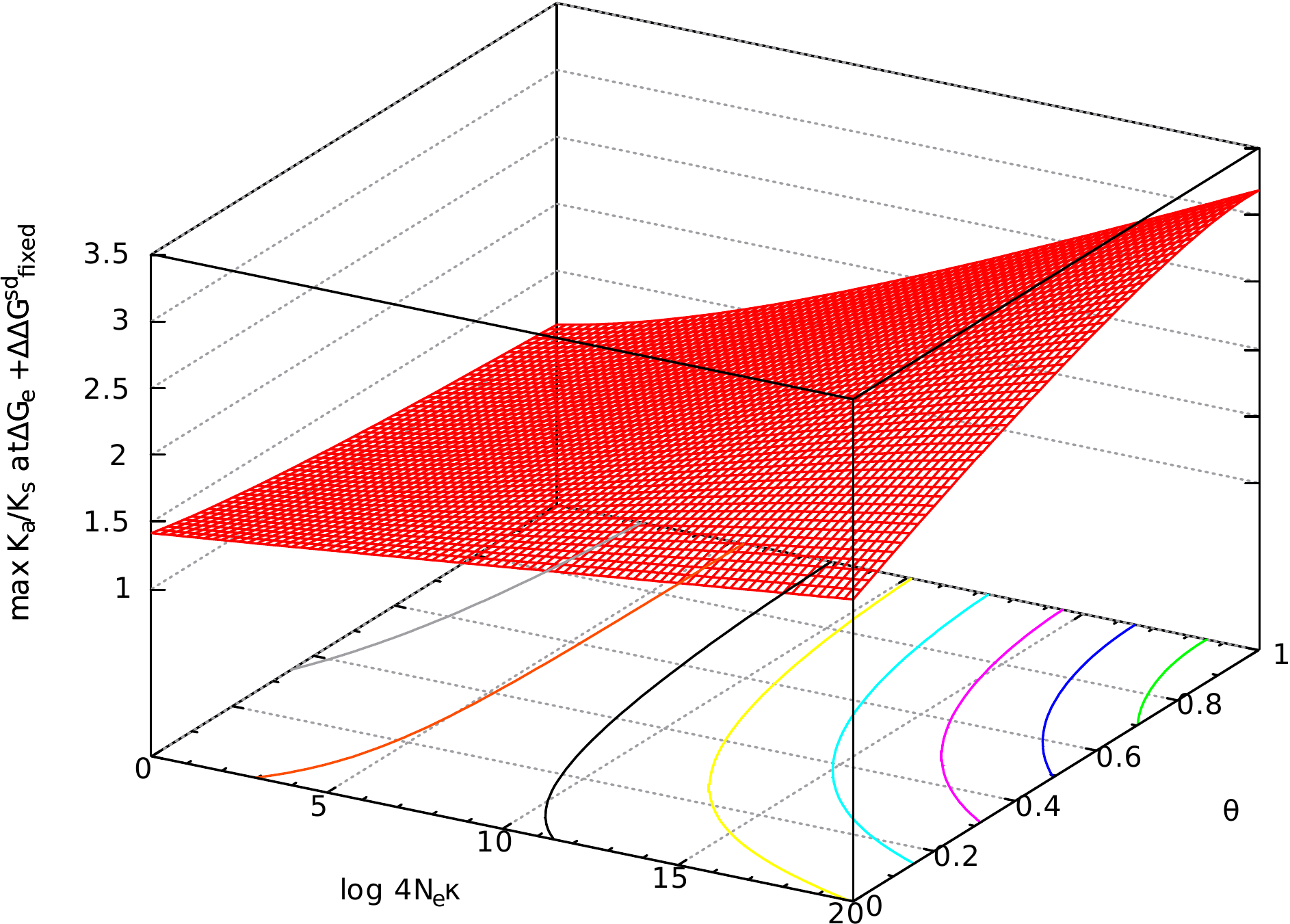}
}
} 
\vspace*{1em}
\caption{
\label{sfig: maximum_Ka_over_Ks_on_4Nekappa_and_theta}
\FigureLegends{
\BF{Dependence of $\max K_a / K_s$ on $4N_e\kappa$ and $\theta$.}
The maximum values of $K_a/K_s$, 
which correspond to the upper bound of selective advantage $s$ (\Eq{\ref{seq: s_max}}), 
at $\Delta G = \Delta G_e$ and at $\Delta G = \Delta G_e + \Delta\Delta G^{\script{sd}}_{\script{fixed}}$
are plotted as a function of $\log 4N_e\kappa$ and $\theta$; 
$\Delta\Delta G^{\script{sd}}_{\script{fixed}}$ is the standard deviation
of $\Delta\Delta G$ over fixed mutants at $\Delta G = \Delta G_e$.
} 
}
\end{figure*}

\FigureInLegends{\clearpage\newpage}

\begin{figure*}[ht]
\FigureInLegends{
\centerline{
\includegraphics*[width=86mm,angle=0]{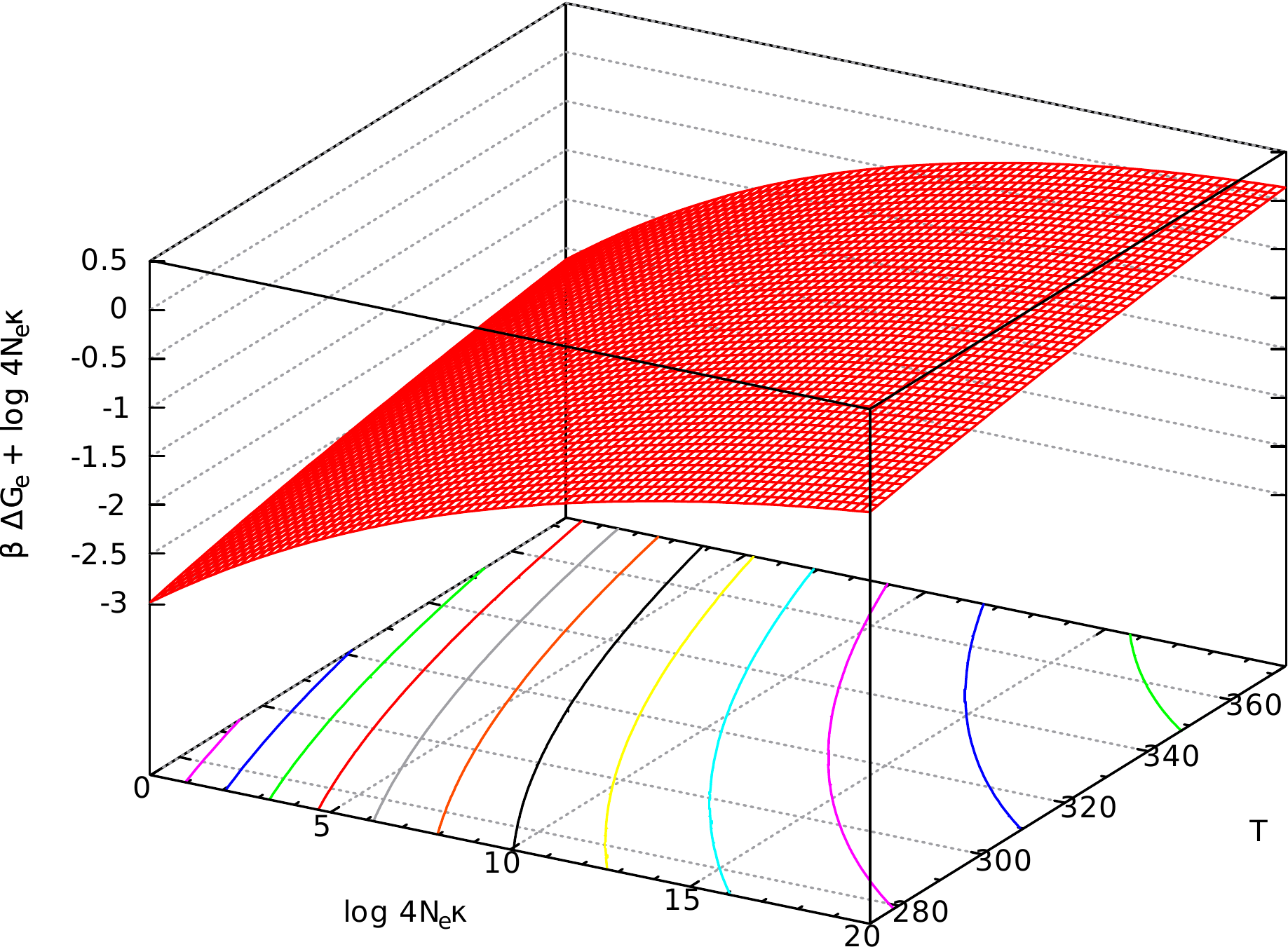}
\includegraphics*[width=90mm,angle=0]{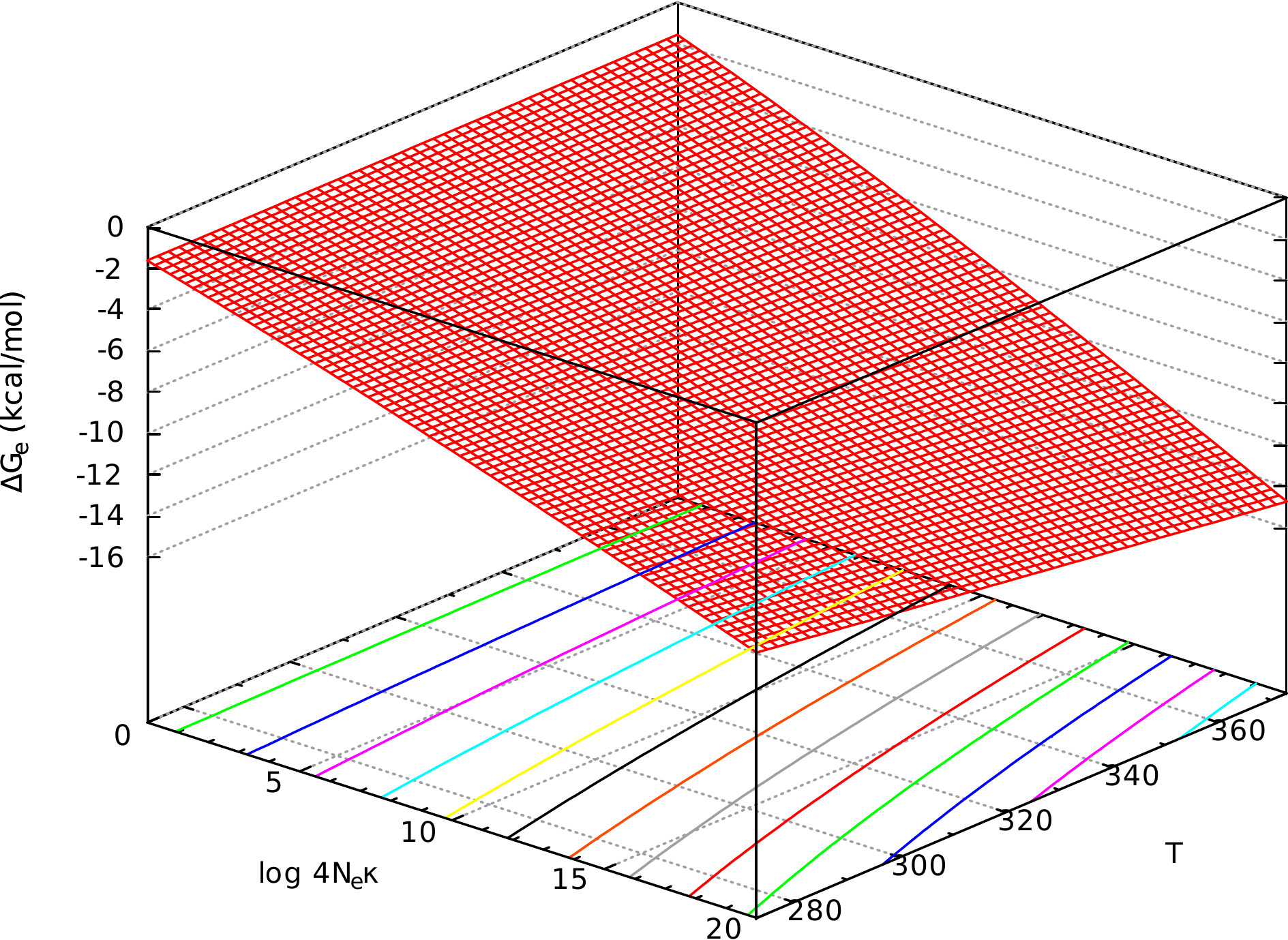}
}
\centerline{
\includegraphics*[width=86mm,angle=0]{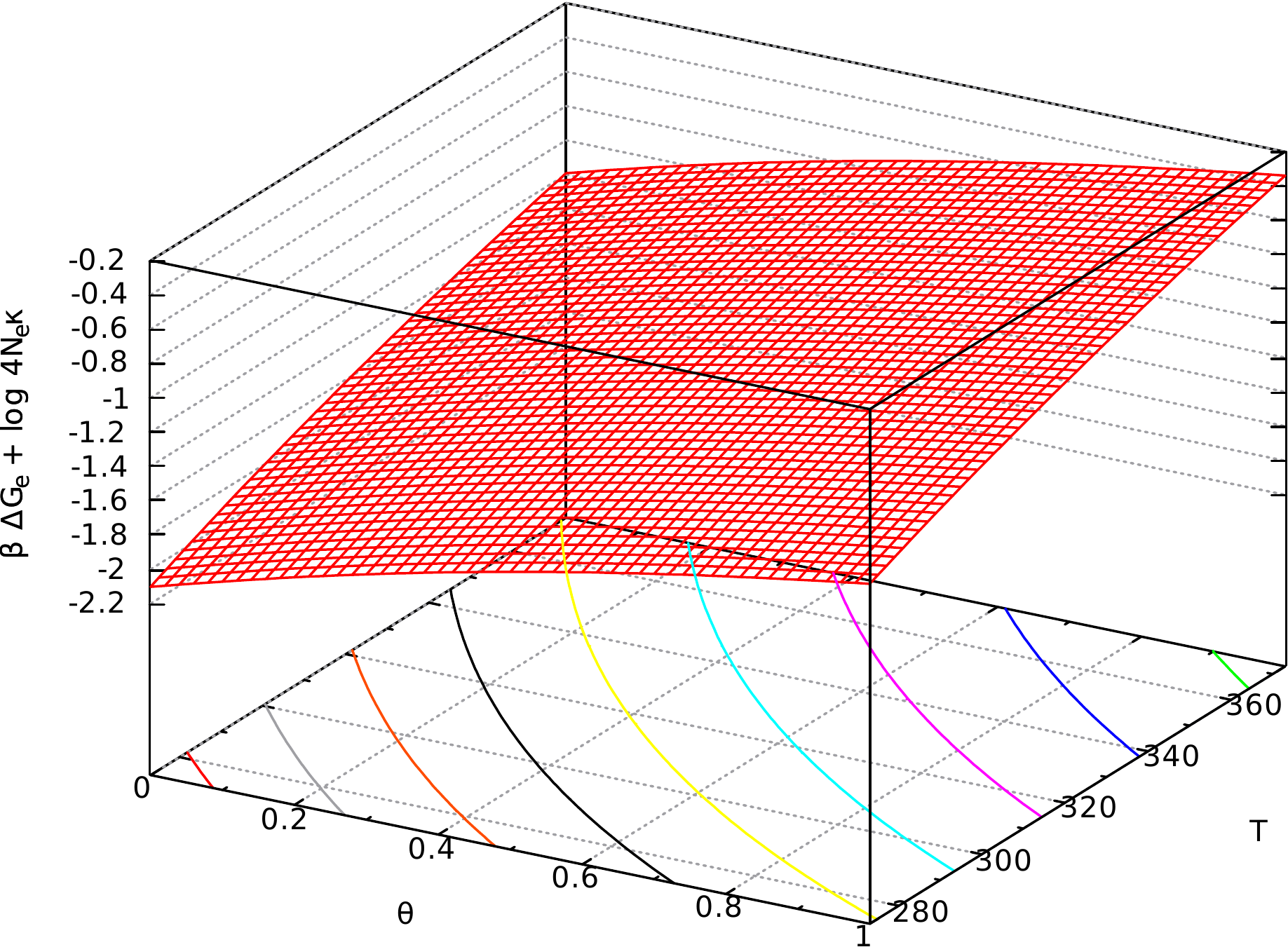}
\includegraphics*[width=88mm,angle=0]{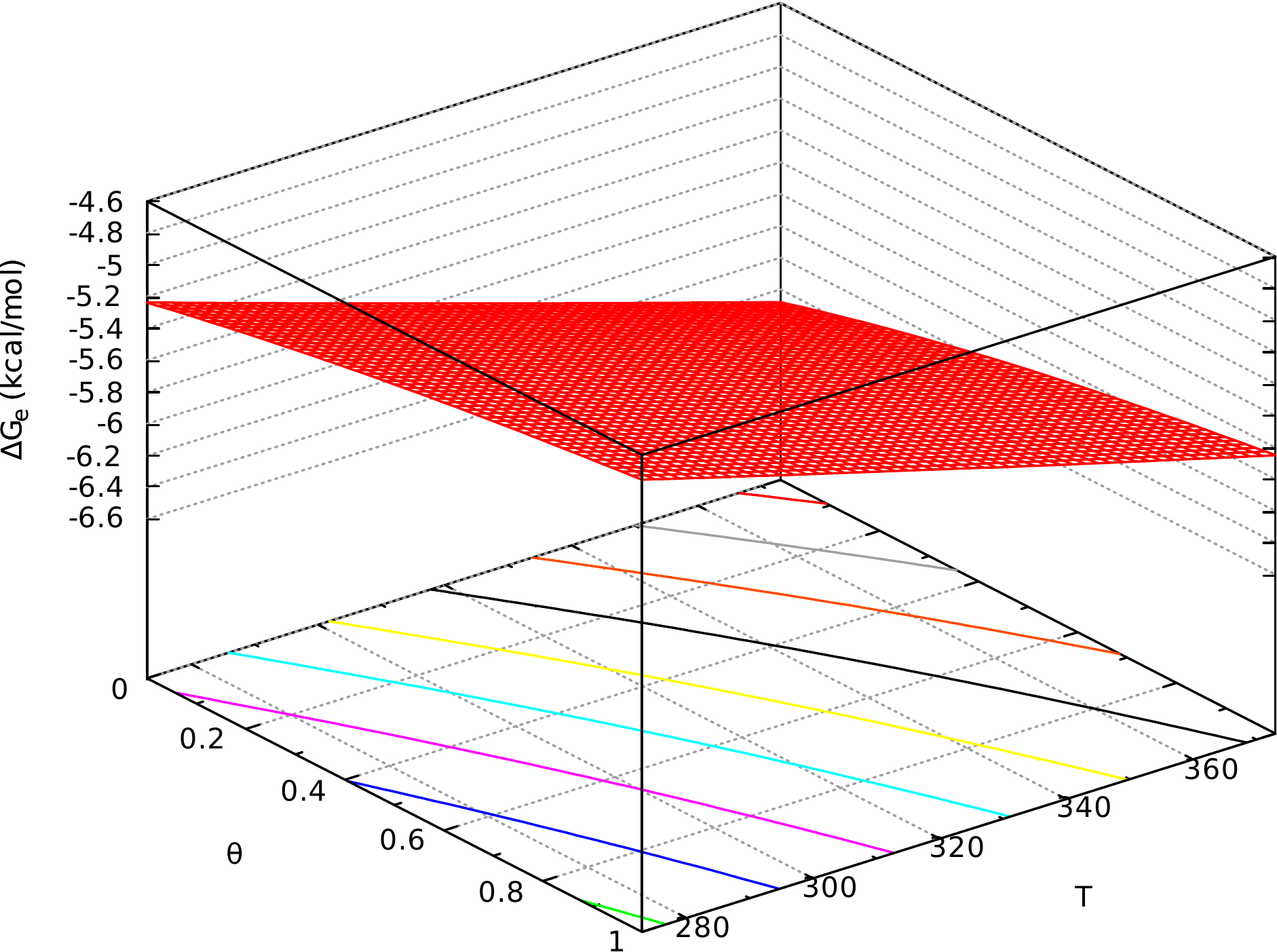}
}
} 
\vspace*{1em}
\caption{
\label{sfig: dependence_of_max4Nes_on_4Nekappa_and_T}
\label{sfig: dependence_of_dGe_on_4Nekappa_and_T}
\label{sfig: dependence_of_max4Nes_on_theta_and_T}
\label{sfig: dependence_of_dGe_on_theta_and_T}
\FigureLegends{
\Red{
\BF{Dependence of equilibrium stability, $\Delta G_e$, on parameters, $4N_e\kappa$, $\theta$ and $T$.}
} 
$\Delta G_e$ is the equilibrium value of folding free energy, $\Delta G$,
in protein evolution.
$T$ is absolute temperature; $\beta = 1 / kT$, where $k$ is the Boltzmann constant.
\EQUATIONS{\Ref{eq: def_bi-Gaussian}, \Ref{eq: def_ms} and \Ref{eq: def_mc}} are assumed for
the distribution of $\Delta\Delta G$ and its dependency on $\Delta G$; they are assumed to be independent of $T$.
Unless specified, 
$\log 4N_e\kappa = 7.550$ and $\theta = 0.53$
are employed.
The value of $\beta\Delta G_e + \log 4N_e\kappa$ is the upper bound of $\log 4N_e s$,
and would not depend on $\log 4N_e \kappa$ if the mean of $\Delta\Delta G$ in all arising mutants did not depend on $\Delta G$; 
see \Eq{\ref{seq: def_s}}.
The kcal/mol unit is used for $\Delta G_e$.
} 
}
\end{figure*}

\FigureInLegends{\clearpage\newpage}

\begin{figure*}[ht]
\FigureInLegends{
\centerline{
\includegraphics*[width=90mm,angle=0]{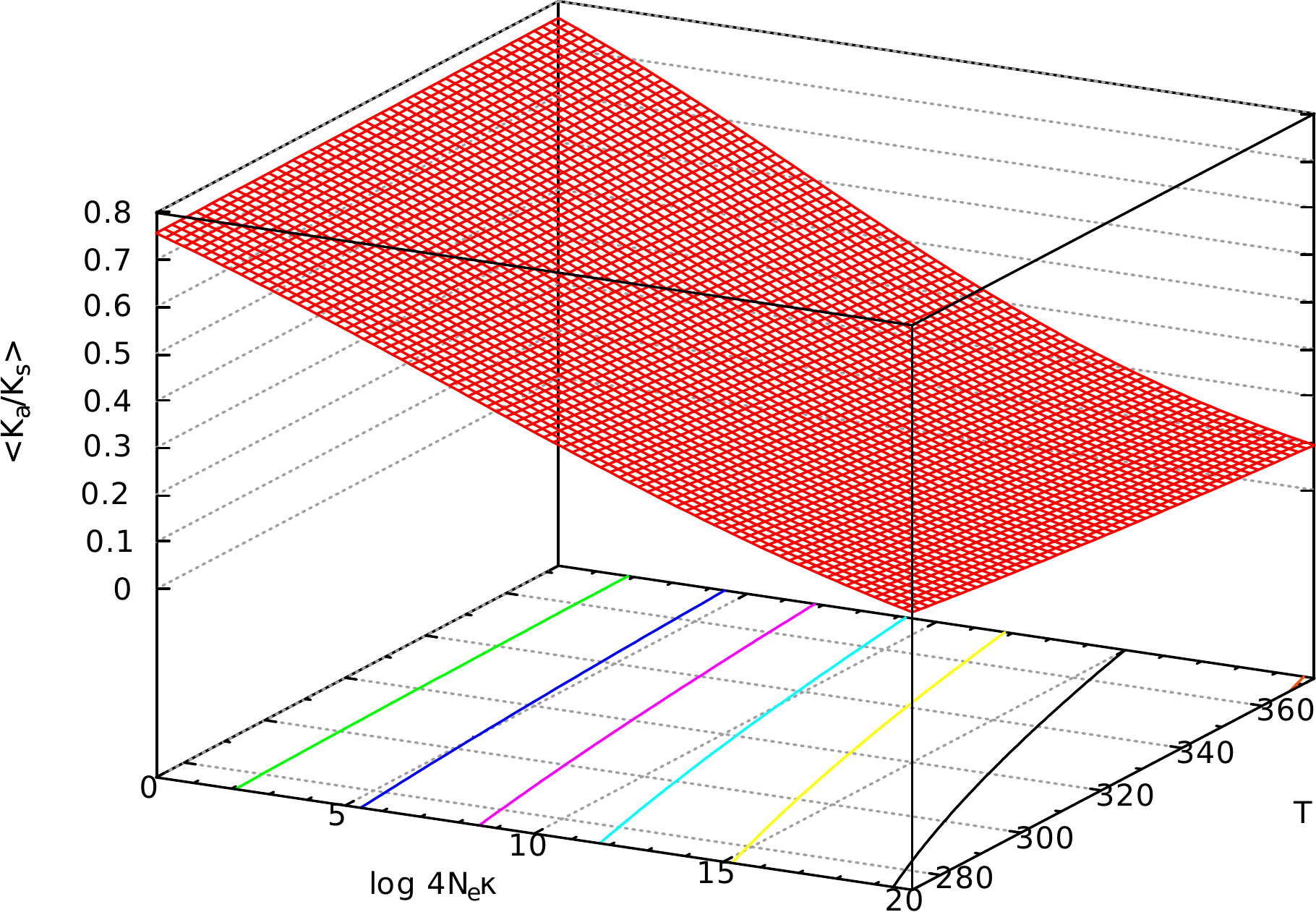}
\includegraphics*[width=86mm,angle=0]{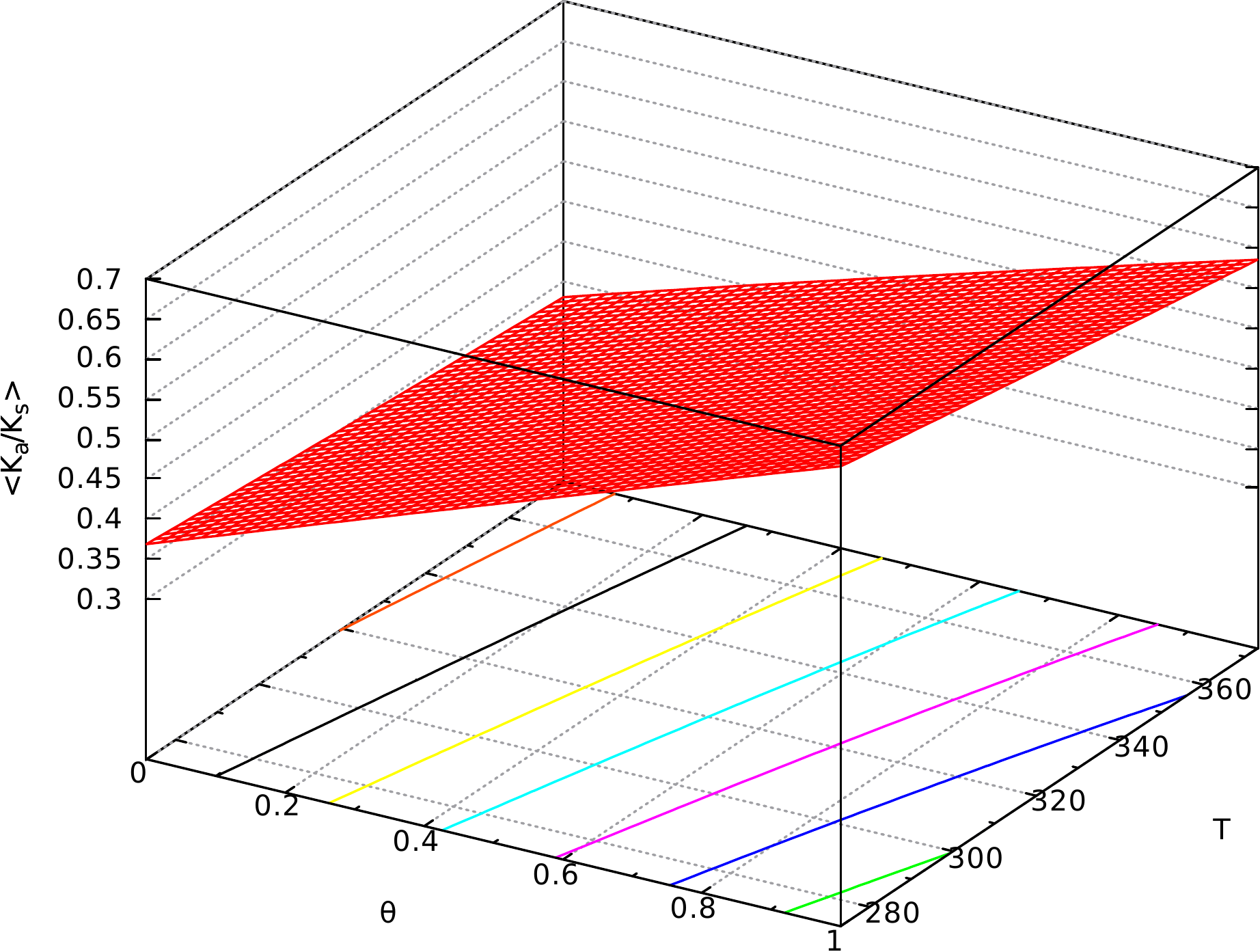}
}
\centerline{
\includegraphics*[width=90mm,angle=0]{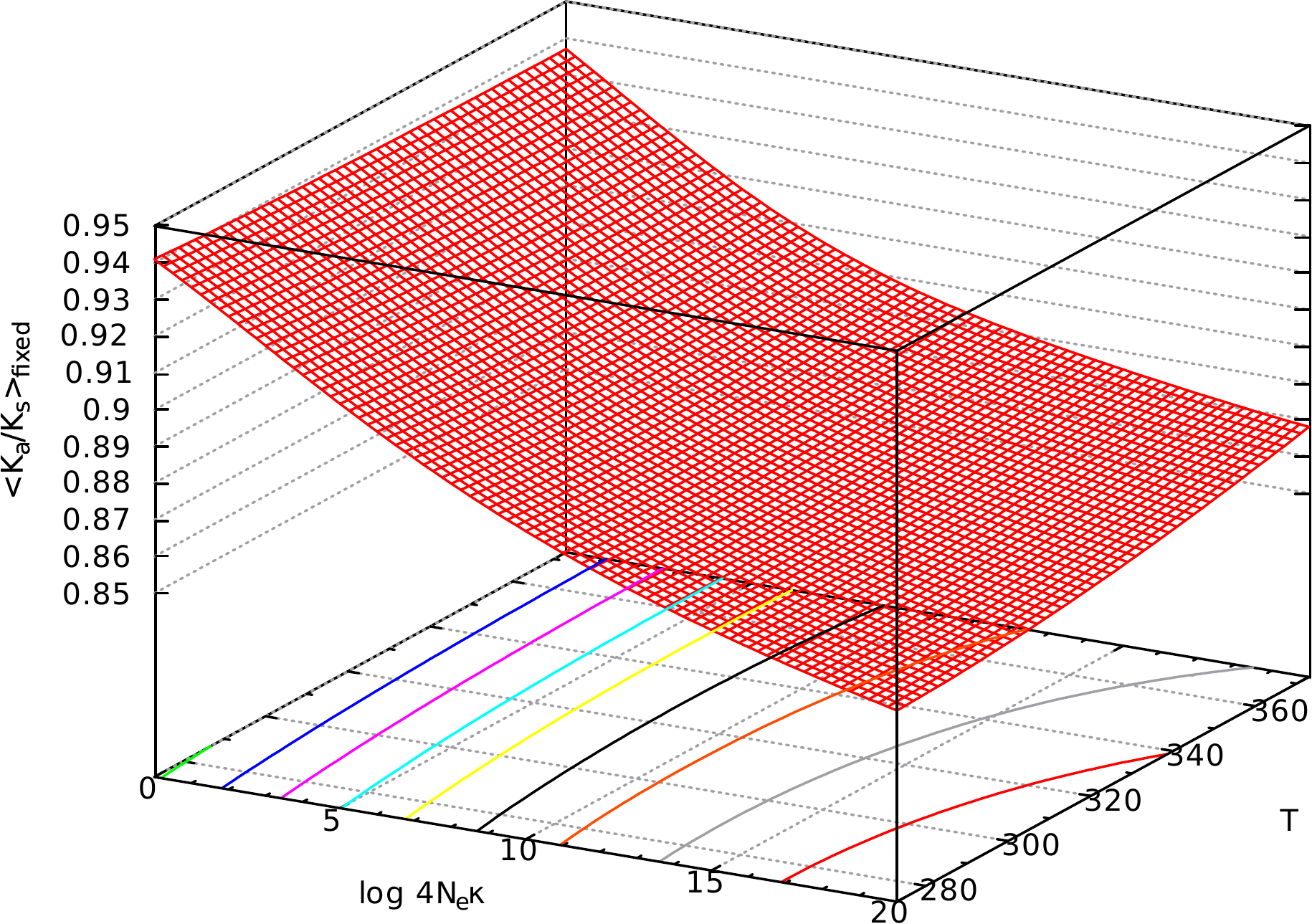}
\includegraphics*[width=86mm,angle=0]{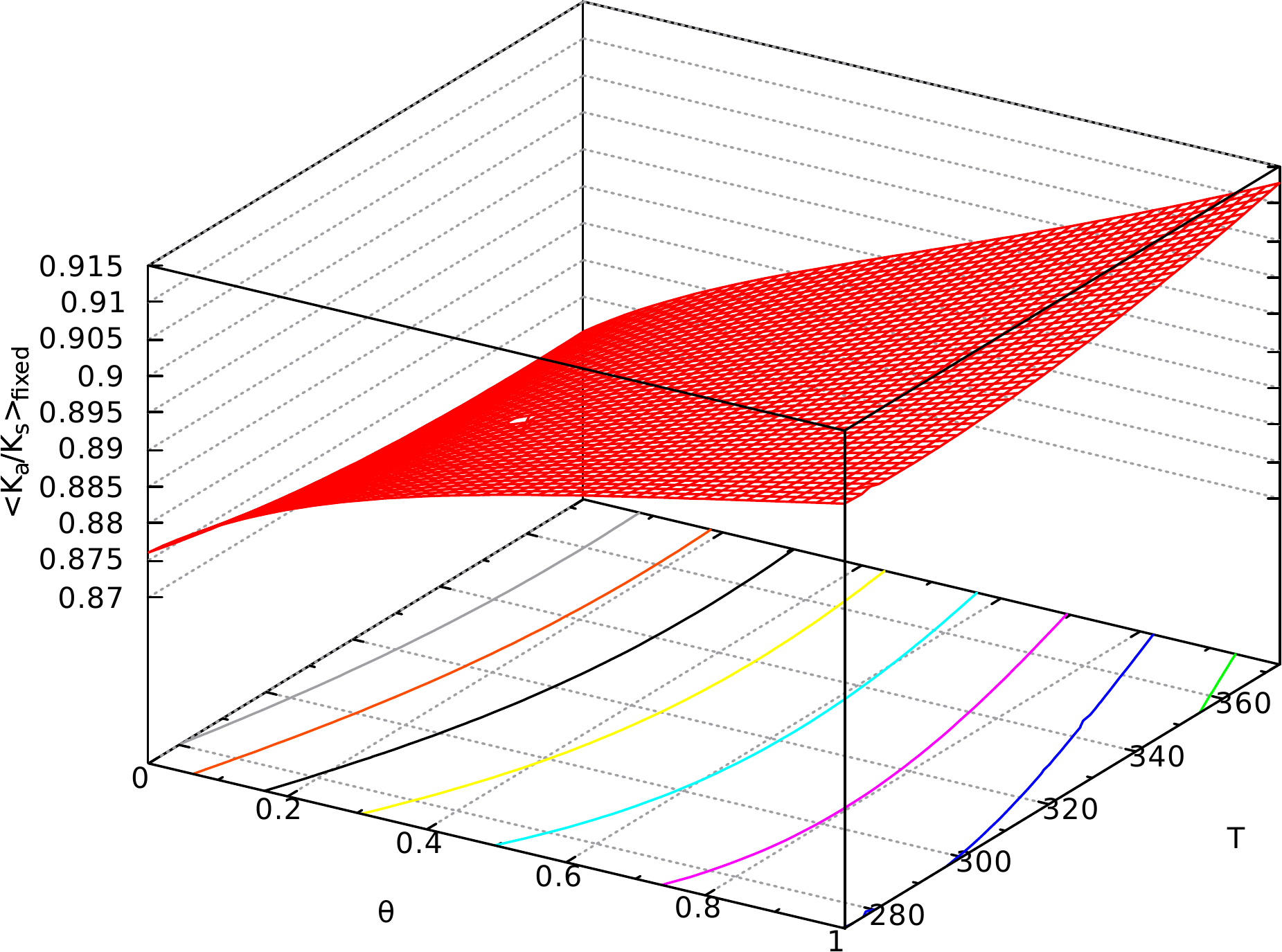}
}
} 
\vspace*{1em}
\caption{
\label{sfig: dependence_of_ave_Ka_over_Ks_on_4Nekappa_and_T}
\label{sfig: dependence_of_ave_Ka_over_Ks_on_theta_and_T}
\label{sfig: dependence_of_ave_Ka_over_Ks_fixed_at_4Nekappa_and_T}
\label{sfig: dependence_of_ave_Ka_over_Ks_fixed_at_theta_and_T}
\FigureLegends{
\Red{
\BF{The average of $K_a / K_s$ over all mutants or over fixed mutants only at equilibrium of protein stability, $\Delta G = \Delta G_e$: Dependence on temperature.}
} 
$T$ is absolute temperature.
\EQUATIONS{\Ref{eq: def_bi-Gaussian}, \Ref{eq: def_ms} and \Ref{eq: def_mc}} are assumed for
the distribution of $\Delta\Delta G$ and its dependency on $\Delta G$; they are assumed to be independent of $T$.
Unless specified, 
$\log 4N_e\kappa = 7.550$ and $\theta = 0.53$
are employed.
} 
}
\end{figure*}

\FigureInLegends{\clearpage\newpage}

\begin{figure*}[ht]
\FigureInLegends{
\centerline{
\includegraphics*[width=90mm,angle=0]{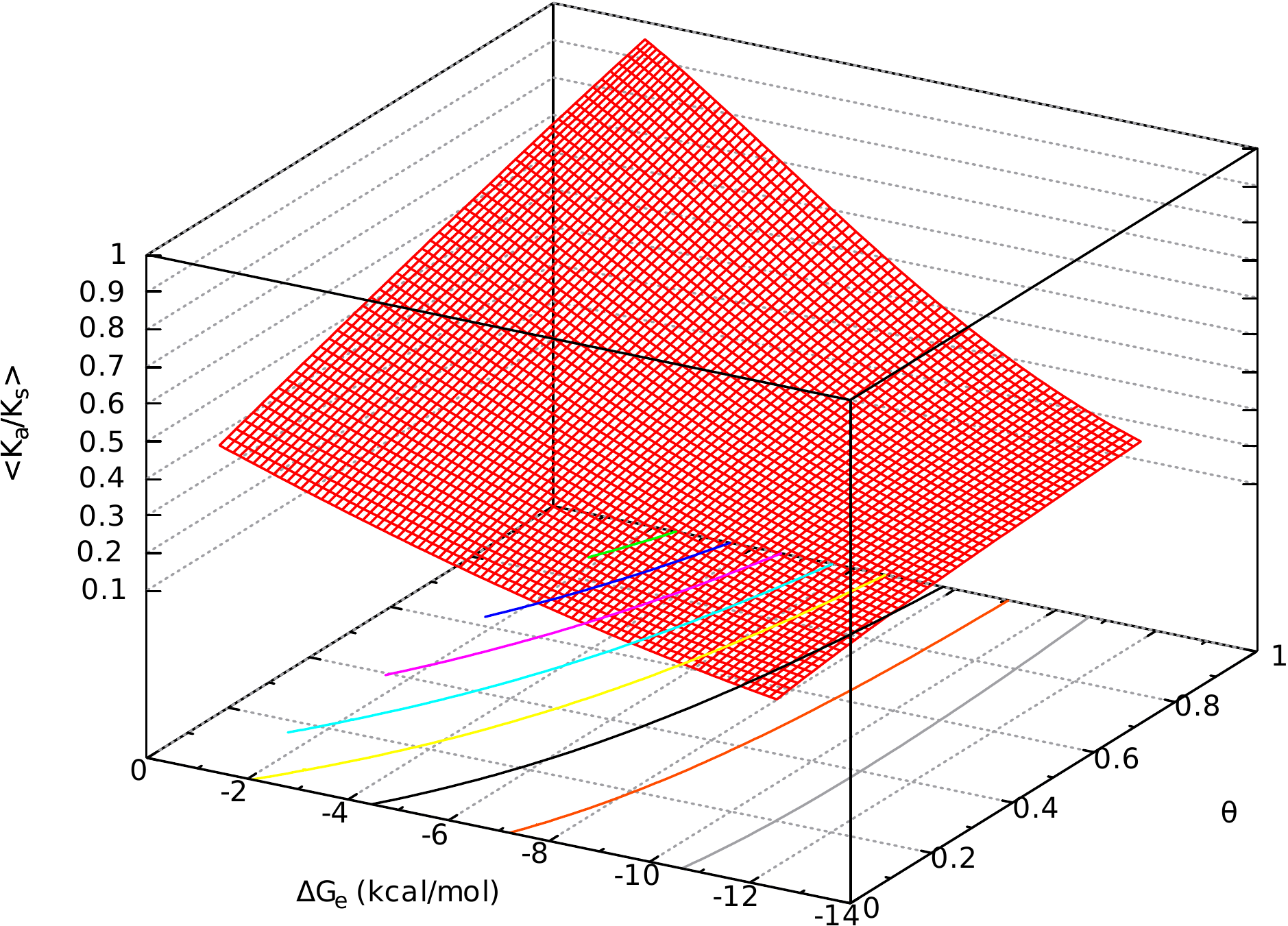}
}
} 
\vspace*{1em}
\caption{
\label{sfig: dependence_of_ave_Ka_over_Ks_on_dGe_and_theta}
\FigureLegends{
\BF{The average of $K_a / K_s$ over all mutants as a function of $\Delta G_e$ and $\theta$.}
} 
}
\end{figure*}

} 

\clearpage

} 

\pagestyle{empty}


\SupplementaryMaterial{

\newpage
\pagestyle{plain}
\setcounter{page}{1}
\renewcommand{\thepage}{S.\arabic{page}}

} 

\setcounter{equation}{0}
\renewcommand{\theequation}{S.\arabic{equation}}

\SupplementaryMaterial{
\begin{center}
\textbf{Supplementary material}	\\
for "Selection maintaining protein stability at equilibrium",	\\
	Journal Theoretical Biology 	\\
	(doi:10.1016/j.jtbi.2015.12.001)
\end{center}

\vspace*{1em}
\begin{center}
Sanzo Miyazawa	\\
sanzo.miyazawa@gmail.com
\end{center}

\vspace*{2em}

\subsection*{Malthusian fitness originating in protein stability in the present model}
 
For typical proteins whose folding free energy $\Delta G$ satisfies $\exp(\beta\Delta G) \ll 1$,
without loss of generality 
we can assume the Malthusian fitness of the single protein-coding genes to be equal to
\begin{eqnarray}
	m &=& - \kappa e^{\beta \Delta G}
	\hspace*{1em}
	\textrm{ with } \kappa \geq 0
	\label{seq: def_m}
\end{eqnarray}
where $\beta = 1 / kT$, $k$ is the Boltzmann constant, $T$ is absolute temperature,
and $\kappa$ is a parameter
whose meaning may depend on the situation; refer to Method for details.
If the fitness costs of
functional loss and toxicity due to misfolded proteins are both taken into account and assumed to be
additive in the Malthusian fitness scale,
$\kappa$ will be defined as
\begin{eqnarray}
        \kappa &=& c A + \gamma
        \label{seq: def_kappa}
\end{eqnarray}
where $c$ is fitness cost per misfolded protein\CITE{GDBWHD:11},
$A$ is the cellular abundance of a protein\CITE{GDBWHD:11},
and $\gamma$ is indispensability\CITE{DW:08} and
defined as $\gamma \equiv - \log ($deletion-strain growth rate / max growth rate$)$.
The parameter $\kappa$ is assumed in the present analysis to take
values in the range of
$0 \leq \log 4 N_e \kappa \leq 20$ with effective population size $N_e$,
taking account of the values of the parameters, $c \sim 10^{-4}$\CITE{DW:08}, $10<A<10^{6}$\CITE{GHBHBDOW:03},
$\gamma = 10$ for essential genes\CITE{DW:08}, and
$N_e \sim 10^4$ to $10^5$ for vertebrates, $\sim 10^5$ to $10^6$ for invertebrates,
$\sim 10^7$ to $10^8$ for unicellular eukaryotes, and $> 10^8$ for prokaryotes \CITE{LC:03}.

Then, the selective advantage of a mutant protein is given as follows 
in Malthusian parameters as a function of 
the folding free energy ($\Delta G$) of the wild-type protein,
the stability change ($\Delta\Delta G$) of a mutant protein,
and the parameter $\kappa$;
\begin{eqnarray}
	s
	&\equiv& m^{\script{mutant}} - m^{\script{wildtype}}
	= \kappa e^{\beta \Delta G} (1 - e^{\beta \Delta \Delta G} )
	\label{seq: def_s}
\end{eqnarray}
\EQUATION{\ref{seq: def_s}} indicates that $s$ is upper-bounded.  
\begin{eqnarray}
	s &\leq& \kappa e^{\beta \Delta G} 
		\label{seq: s_max}
\end{eqnarray}

\subsection*{Distribution of stability changes ($\Delta\Delta G$) due to single amino acid substitutions}

Here, according to \CITE{TSSST:07},
the distribution of folding free energy changes, $\Delta\Delta G$, of mutant proteins is 
assumed to be a bi-Gaussian function with
mean depending on $\Delta G$, 
in order to take into account the effects of structural constraint on evolutionary rate.
The probability density function (PDF), $p(\Delta \Delta G)$, 
of $\Delta \Delta G$ for nonsynonymous 
substitutions is defined as
\begin{eqnarray}
p(\Delta \Delta G) &=& \theta \mathcal{N}(\mu_s, \sigma_s)
		+ (1 - \theta) \mathcal{N}(\mu_c, \sigma_c)
	\label{seq: def_bi-Gaussian}
\end{eqnarray}
where $0 \leq \theta \leq 1$, and $\mathcal{N}(\mu, \sigma)$
is a normal distribution with mean $\mu$ and standard deviation $\sigma$.
One of the two Gaussian distributions above,
$\mathcal{N}(\mu_s, \sigma_s)$,
results from substitutions on protein surfaces and
is a narrow distribution with a mildly destabilizing mean $\Delta \Delta G$,
whereas the other, $\mathcal{N}(\mu_c, \sigma_c)$, due to substitutions in protein cores is a wider
distribution with a stronger destabilizing mean\CITE{TSSST:07}.
Since the majority of substitutions appear to be single nucleotide substitutions,
the values of standard deviations 
($\sigma_s$ and $\sigma_c$)
estimated in \CITE{TSSST:07} for single nucleotide substitutions
are employed here; in kcal/mol units,
\begin{eqnarray}
\mu_s &=& -0.139 \, \Delta G - 0.168	
\hspace*{1em}, \hspace*{1em}
\sigma_s = 0.90 
	\label{seq: def_ms}
	\\
\mu_c &=& -0.139 \, \Delta G + 1.232	
\hspace*{1em}, \hspace*{1em}
\sigma_c = 1.93 
	\label{seq: def_mc}
\end{eqnarray}
\Red{
To analyze the dependences of the means, $\mu_s$ and $\mu_c$, on $\Delta G$,
we plotted the observed values of $\Delta\Delta G$ of single amino acid mutants
against $\Delta G$ of the wild type, which are collected in the ProTherm
database\CITE{KBGPKUS:06}; the same analysis was done in \CITE{SRS:12}.
\Fig{\ref{sfig: ddG_vs_dG}} shows a significant dependence of $\Delta\Delta G$ on $\Delta G$;
the regression line is $\mu = -0.139 \Delta G + 0.490$.		
} 
The linear slopes
of $\mu_s$ and $\mu_c$ are taken to be equal to
the slope
($-0.139$)	
of the regression line.
The intercepts have been estimated
to satisfy the following two conditions.
\begin{enumerate}
\item \EQUATIONS{\Ref{seq: def_ms} and \Ref{seq: def_mc}} satisfy
$\mu_s(\Delta G_0) = 0.56$ and $\mu_c(\Delta G_0) = 1.96$,
which were estimated for single nucleotide substitutions in \CITE{TSSST:07},
at a certain value ($\Delta G_0$) of $\Delta G$.
\item The total mean of the two Gaussian functions agrees with
the regression line, $\mu = -0.139 \Delta G + 0.490$.	
The value of $\theta$ is taken to be $0.53$, which is
equal to
the average of $\theta$ over proteins used in \CITE{TSSST:07}.
\end{enumerate}
A representative value,
$7.550$,
of $\log 4N_e \kappa$ is determined
in such way that the equilibrium value of $\Delta G$
is equal to $\Delta G_{0} = -5.24$ introduced above.
\Red{
It is interesting that this value $\Delta G_e = -5.24$ kcal/mol
agrees with the most probable value of $\Delta G$ in the observed distribution of protein stabilities
shown in \Fig{\ref{fig: observed_distribution_of_protein_stabilities}}.
} 
The fraction $\theta$ of less-constrained residues such as most residues on protein surface 
is correlated with protein length for globular, monomeric proteins \CITE{TSSST:07}; 
$
\theta = 1.27 - 0.33 \cdot \log_{10}(\textrm{protein length}) 
			\textrm{ for }50 \leq \textrm{length} \leq 330 
$.
\CITE{TSSST:07}.  
However, residues taking part in protein--protein interactions
may be regarded as core residues rather than surface residues.

\subsection*{Probability distributions of selective advantage, fixation rate and $K_a/K_s$}

Now, we can consider the probability distributions of
characteristic quantities that describe the evolution of genes. 
First of all, the PDF of selective advantage $s$, $p(s)$, of mutant genes can be represented by
\begin{eqnarray}
p(s) &=& - p(\Delta \Delta G) \frac{d \Delta \Delta G}{d s}
	\nonumber
	\\
	&=& p(\Delta \Delta G) \frac{1}{\beta (\kappa e^{\beta\Delta G} - s)} 
	\label{seq: pdf_of_s}
\end{eqnarray}
where $\Delta \Delta G$
must be regarded as a function of $s$, that is,
$
\Delta \Delta G = \beta^{-1} \log (1 - s (\kappa \exp(\beta \Delta G))^{-1})
$.
The PDF of $4N_e s$, $p(4N_e s)$, may be more useful than $p(s)$.
\begin{eqnarray}
p(4N_e s) &=& p(s) \frac{1}{4N_e}
\end{eqnarray}

\Detail{
The landscape of the average selective advantage 
$\langle s(\kappa, \Delta G) \rangle$ on the $\kappa$ and
$\beta \Delta G$ was studied in detail by \CITE{SLS:13}. 
\begin{eqnarray}
	\langle s(\kappa, \Delta G) \rangle 
	&=& \kappa e^{\beta \Delta G} (1 - \langle e^{\beta \Delta \Delta G} \rangle )
\end{eqnarray}
If the PDF of $\Delta\Delta G$ is approximately Gaussian, 
the average $\langle \exp(\beta \Delta\Delta G) \rangle$ can be analytically calculated.
\begin{eqnarray}
\langle e^{\beta \Delta\Delta G} \rangle_{\mathcal{N}(m, \sigma)}	
	&=& e^{(\beta m + \beta^2 \sigma^2 / 2) }
\end{eqnarray}

} 

The fixation probability $u$ of a mutant gene with selective advantage $s$ 
and gene frequency $q$
in a duploid system is equal to\CITE{CK:70} 
\begin{eqnarray}
	u(4N_e s) &=& \frac{1 - e^{-4N_e s q} }{1 - e^{-4N_e s} } 
		\label{seq: fixation_prob}
\end{eqnarray}
where $q = 1/ (2N)$ for a single mutant gene in a population of size $N$.
Population size is taken to be $N = 10^6$.
Thus, the PDF of fixation probability $u$ can be represented by
\begin{eqnarray}
p(u)	&=& p(4N_e s) \frac{d 4N_e s}{d u} 
	\nonumber
	\\
	&=& p(4N_e s)
	 \frac{ ( e^{4N_e s} - 1)^2 e^{4N_e s (q - 1)} }
	{ q( e^{4N_e s} - 1) - (e^{4N_e s q} - 1) }
		\label{seq: pdf_of_fixation_prob}
\end{eqnarray}
where $s$ must be regarded as a function of $u$. 

The ratio of the substitution rate per nonsynonymous site ($K_a$) for nonsynonymous
substitutions with selective advantage s to the substitution rate per
synonymous site ($K_a$) for nonsynonymous substitutions with s = 0 is
\begin{eqnarray}
	\frac{K_a}{K_s} &=& \frac{u(4N_e s)}{u(0)} = \frac{u(4N_e s)}{q}
		\label{seq: def_ka_over_ks}
		\\
		&\simeq& \frac{4N_e s}{1 - e^{-4N_e s}} \hspace*{1em} \textrm{ for } \frac{| 4N_e s q| }{2} \ll 1 
\end{eqnarray}
assuming that synonymous substitutions are completely neutral 
and mutation rates at both types of sites are the same.
The PDF of $K_a/K_s$ is
\begin{eqnarray}
	p(K_a/K_s) &=& 
		p(u) \frac{d u}{d (K_a/K_s)} 
	= p(u) q  
		\label{seq: pdf_of_Ka_over_Ks}
	\\
	&\simeq& p(4N_e s) \frac{(e^{4N_e s} - 1)^2}{e^{4N_e s}(e^{4N_e s} - 1 - 4N_e s)} 
			\textrm{ for } \frac{| 4N_e s q |}{2} \ll 1
		\label{seq: approx_pdf_of_Ka_over_Ks}
\end{eqnarray}
In the range of $| 4N_e s q | / 2 \ll 1$, both $K_a/K_s$ and $p(K_a/K_s)$ do not depend on $q = 1/(2N)$.

\subsection*{Probability distributions of $\Delta\Delta G$, $4N_e s$, $u$, and $K_a/K_s$ in fixed mutant genes}

Now, let us think about fixed mutant genes. 
The PDF of the $\Delta\Delta G$ of fixed mutant genes is 
\begin{eqnarray}
p(\Delta\Delta G_{\script{fixed}})
	&=& p(\Delta \Delta G) 
	\frac{u(4N_e s(\Delta \Delta G))}{\langle u(4N_e s(\Delta \Delta G)) \rangle}
	\\
 \langle u \rangle &\equiv& \int_{-\infty}^{\infty} u(4N_e s) p(\Delta\Delta G) d\Delta\Delta G
        \\
		&=& \int_{-\infty}^{4N_e\kappa\exp(\beta\Delta G)} u(4N_e s) p(4N_es) d4N_es
\end{eqnarray}
Likewise, the PDF of the selective advantage 
of fixed mutant genes is 
\begin{eqnarray}
p(4N_e s_{\script{fixed}}) &=&
	p(4N_e s)
	\frac{u(4N_e s)}{\langle u(4N_e s) \rangle}
\end{eqnarray}
and those of the $u$ and $K_a/K_s$ of fixed mutant genes are
\begin{eqnarray}
	p( u_{\script{fixed}} ) &=&
		p(u) \frac{u}{\langle u \rangle}
		\\
	p( (\frac{K_a}{K_s})_{\script{fixed}}) &=&
		p(\frac{K_a}{K_s}) \frac{u}{\langle u \rangle}
		=
		p(\frac{K_a}{K_s}) \frac{\frac{K_a}{K_s}}{\langle \frac{K_a}{K_s} \rangle}
\end{eqnarray}

Then, the probabilities of $a< K_a/K_s <b$ and  the averages of $K_a/K_s$
over all mutants and also in fixed mutants can be calculated.
The average of $K_a/K_s$ in fixed mutants is equal to the ratio of the second moment to the first moment of $K_a/K_s$
in all arising mutants.
\begin{eqnarray}
\langle \frac{K_a}{K_s} \rangle_{\script{fixed}} &=& \langle (\frac{K_a}{K_s})^2 \rangle / \langle \frac{K_a}{K_s} \rangle
\end{eqnarray}

} 

\renewcommand{\FigureInLegends}[1]{}
\renewcommand{\FigureLegends}[1]{}

\SupplementaryMaterial{
\renewcommand{\FigureInLegends}[1]{#1}
\renewcommand{\FigureLegends}[1]{#1}
} 

\renewcommand{\TextFig}[1]{}
\renewcommand{\SupFig}[1]{#1}
\setcounter{figure}{0}

\renewcommand{\thefigure}{S.\arabic{figure}}

\SupFig{

\FigureInLegends{\clearpage\newpage}

\begin{figure*}[ht]
\FigureInLegends{
\centerline{
\includegraphics*[width=90mm,angle=0]{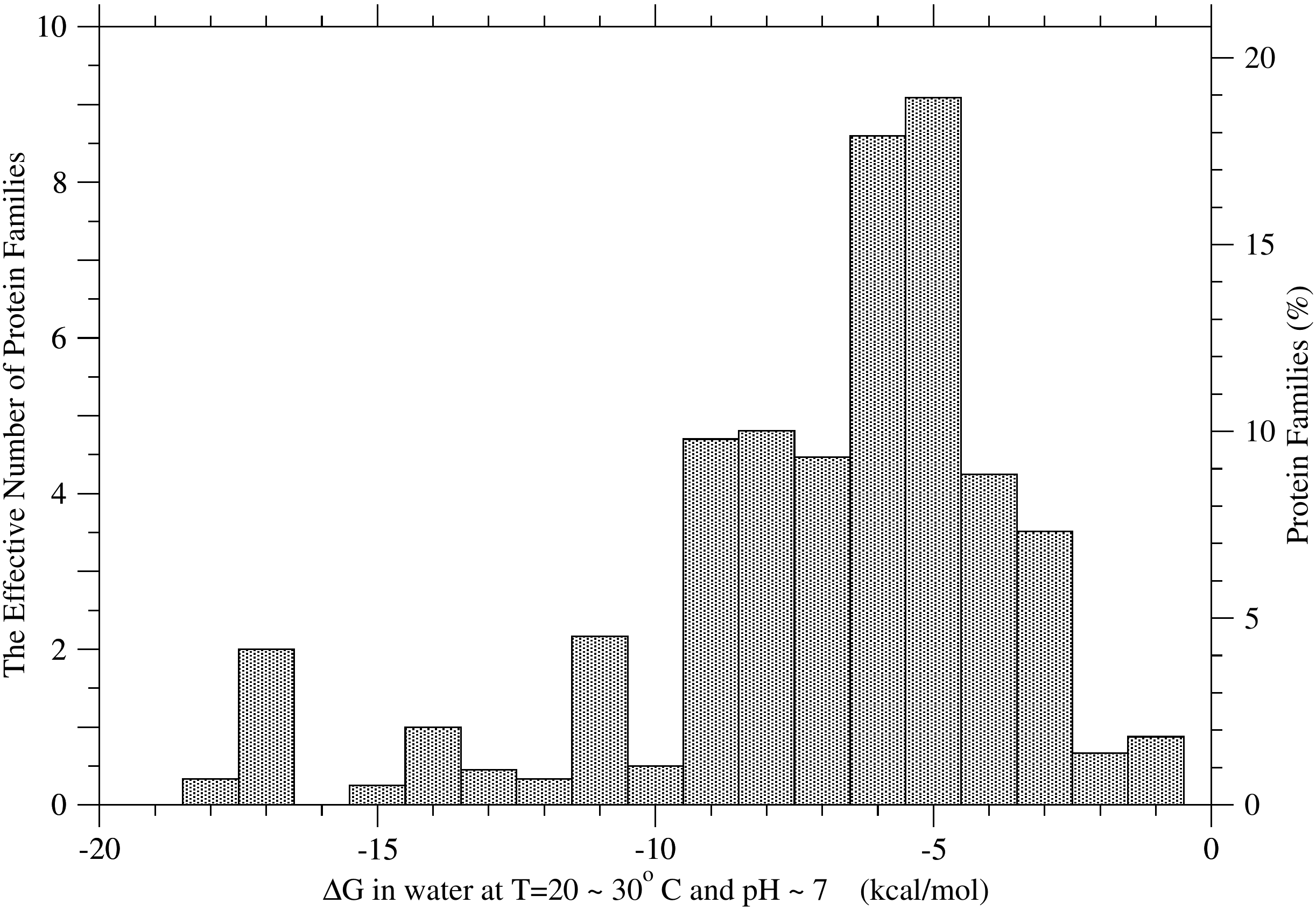}
}
} 
\vspace*{1em}
\caption{
\label{sfig: observed_distribution_of_protein_stabilities}
\FigureLegends{
\Red{
\BF{Distribution of folding free energies of monomeric protein families.}
} 
Stability data of monomeric proteins for which the item of dG\_H2O or dG
was obtained in the experimental condition of $6.7 \leq \textrm{pH} \leq 7.3$ 
and $20^{\circ}C \leq T \leq 30^{\circ}C$ and their folding-unfolding transition is two state and reversible
are extracted from the ProTherm\CITE{KBGPKUS:06};
in the case of dG only thermal transition data are used. 
Thermophilic proteins, and proteins observed with salts or additives are also removed.
An equal sampling weight is assigned to each species of homologous protein,
and the total sampling weight of each protein family is normalized to
one.  In the case in which multiple data exist for the same species of protein, 
its sampling weight is divided to each of the data. 
However, proteins whose stabilities are known 
may be samples biased from the protein universe.
The value, $\Delta G_e = -5.24$ kcal/mol, of equilibrium stability at the representative parameter values, $\log 4N_e\kappa = 7.550$
and $\theta = 0.53$, agrees with the most probable value of $\Delta G$ in the distribution above.
Also, the range of $\Delta G$ shown above is consistent with
that range, $-2$ to $-12.5$ kcal/mol, expected from the present model.
The kcal/mol unit is used for $\Delta G$.
A similar distribution was also compiled \CITE{ZCS:07}.
} 
}
\end{figure*}

\FigureInLegends{\clearpage\newpage}

\begin{figure*}[ht]
\FigureInLegends{
\centerline{
\includegraphics*[width=90mm,angle=0]{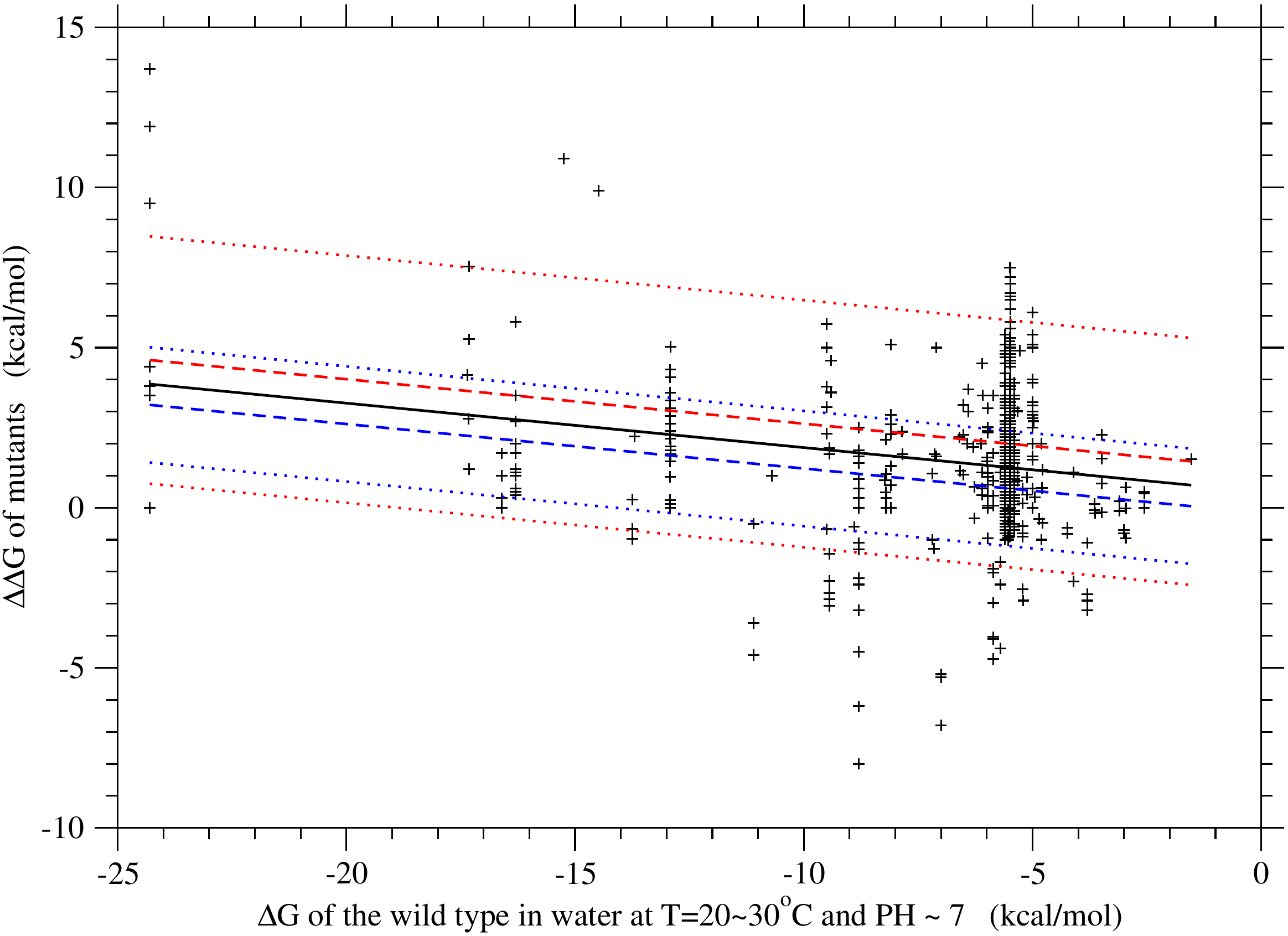}
}
} 
\vspace*{1em}
\caption{
\label{sfig: ddG_vs_dG}
\FigureLegends{
\Red{
\BF{Dependence of stability changes, $\Delta\Delta G$, due to single amino acid substitutions
on the protein stability, $\Delta G$, of the wild type.}
} 
A solid line shows the regression line, 
$\Delta\Delta G = -0.139 \Delta G + 0.490$; 
the correlation coefficient and p-value are equal to $-0.20$ and $< 10^{-7}$, respectively.
Broken lines show two means of bi-Gaussian distributions, $\mu_s$ in blue and $\mu_c$ in red.
Blue dotted lines show $\mu_s \pm 2 \sigma_s$ and red dotted lines $\mu_c \pm 2 \sigma_c$.
See \Eqs{\Ref{seq: def_bi-Gaussian}, \Ref{seq: def_ms} and \Ref{seq: def_mc}} for the bi-Gaussian distribution.
Stability data of single amino acid mutants for which the items dG\_H2O and ddG\_H2O or dG and ddG 
were obtained in the experimental condition of 
$6.7 \leq \textrm{pH} \leq 7.3$ and $20^{\circ}C \leq T \leq 30^{\circ}C$
and their folding-unfolding transitions are two state and reversible
are extracted from the ProTherm\CITE{KBGPKUS:06}.
In the case of dG only thermal transition data are used.
In the case in which multiple data exist for the same protein, only one of them
is used.
The kcal/mol unit is used for $\Delta\Delta G$ and $\Delta G$.
A similar distribution was also compiled \CITE{SRS:12}.
} 
}
\end{figure*}

\FigureInLegends{\newpage}

\begin{figure*}[ht]
\FigureInLegends{
\centerline{
\includegraphics*[width=90mm,angle=0]{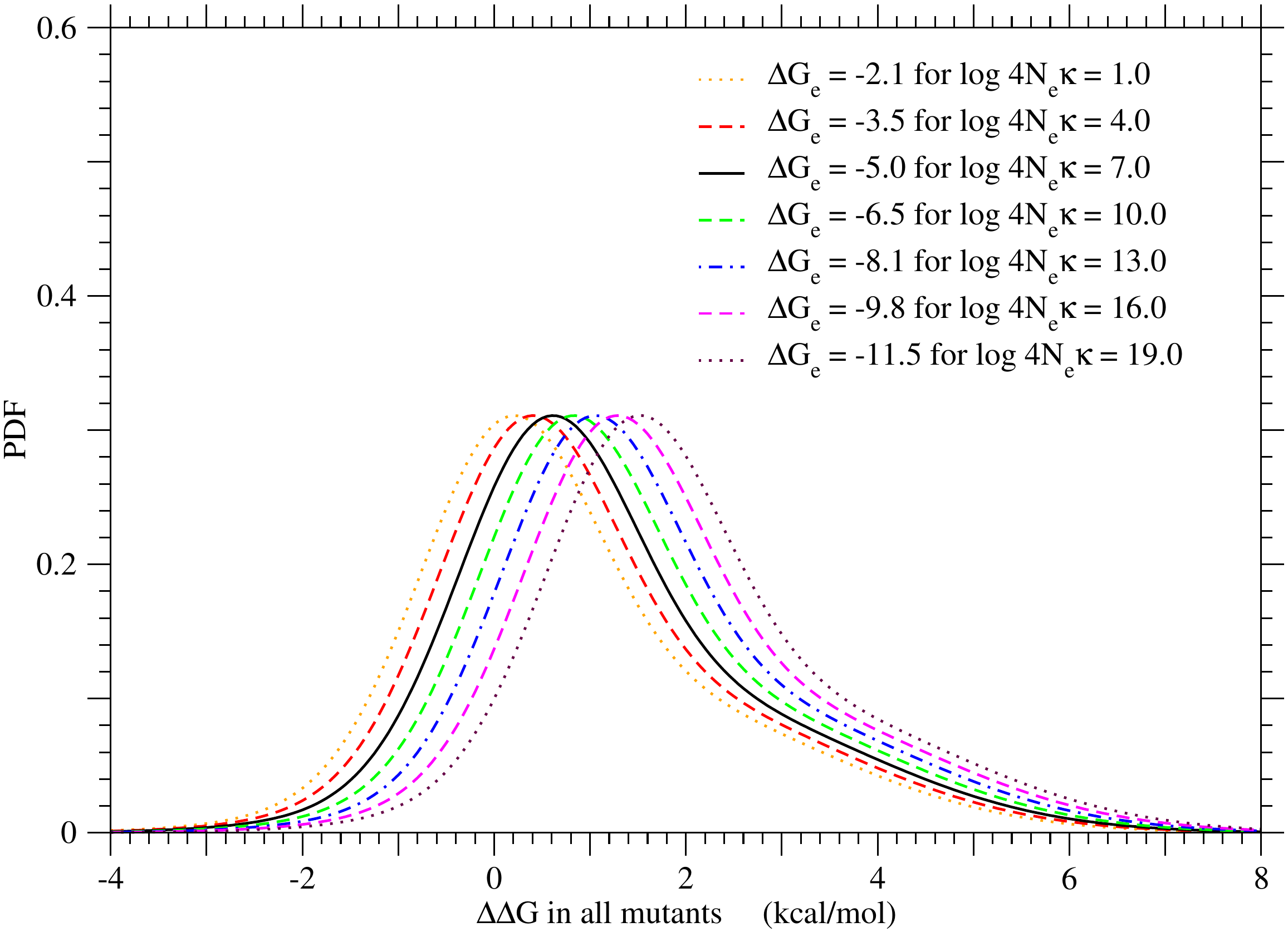}
\includegraphics*[width=90mm,angle=0]{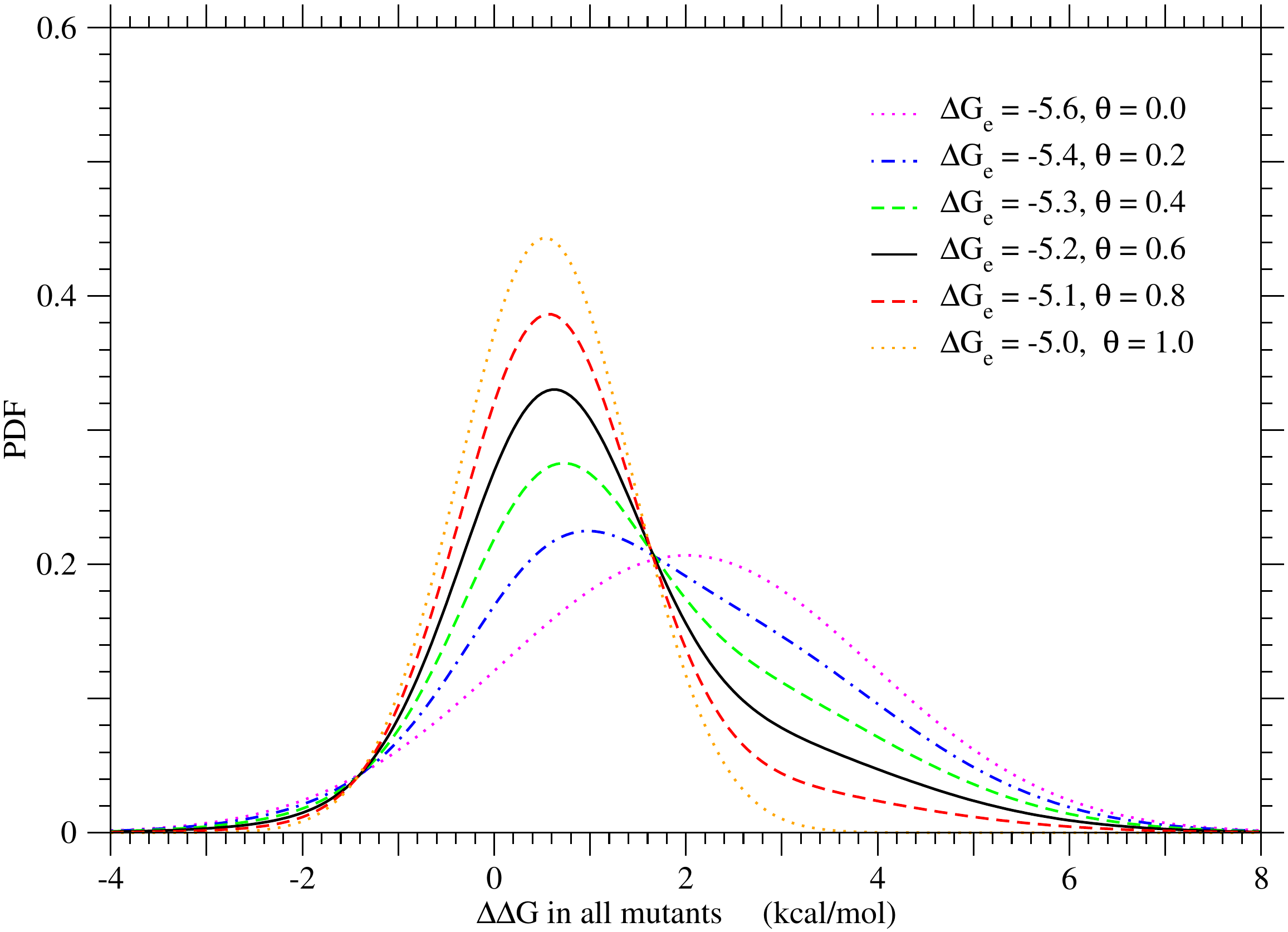}
}
\centerline{
\includegraphics*[width=90mm,angle=0]{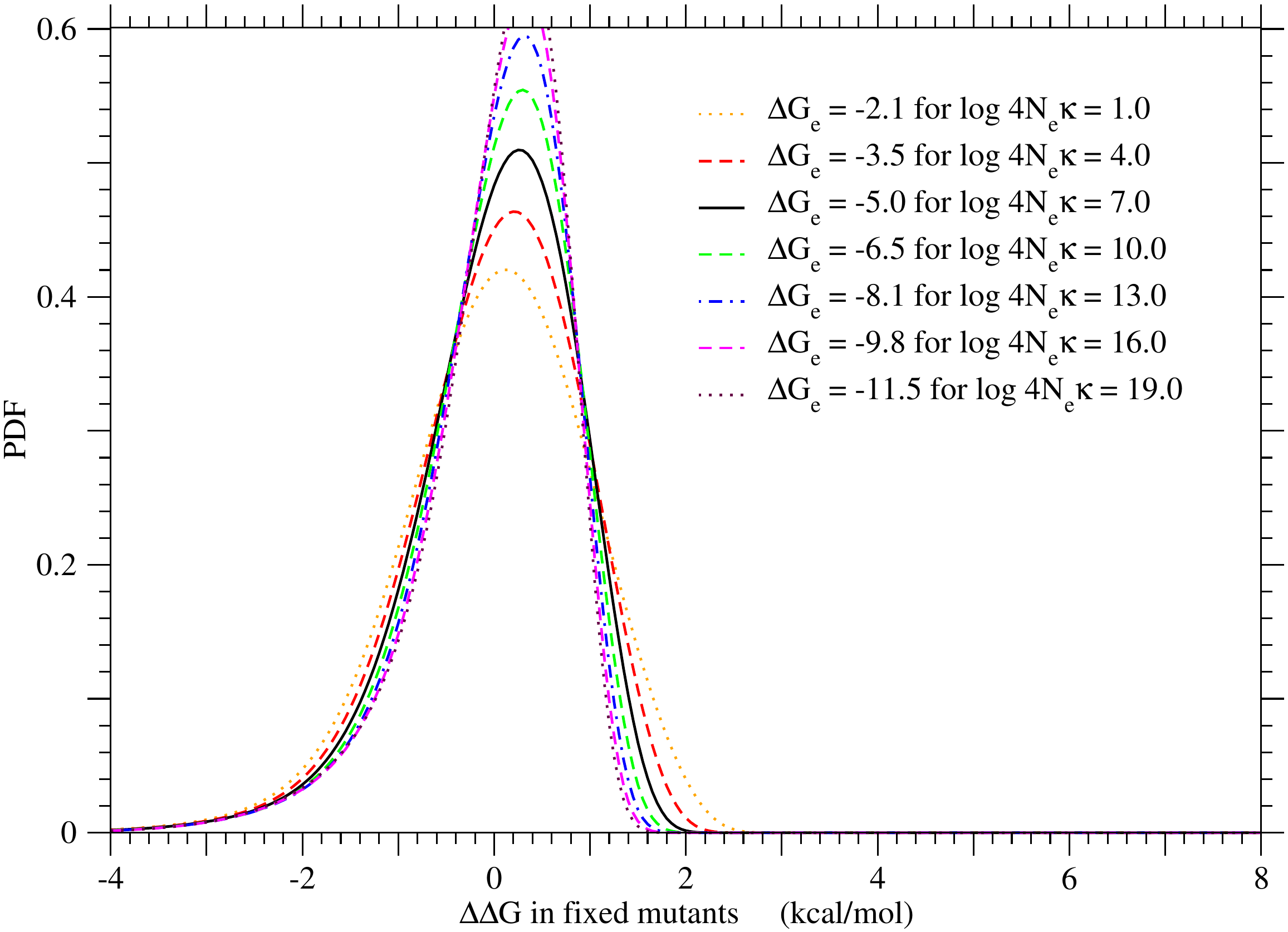}
\includegraphics*[width=90mm,angle=0]{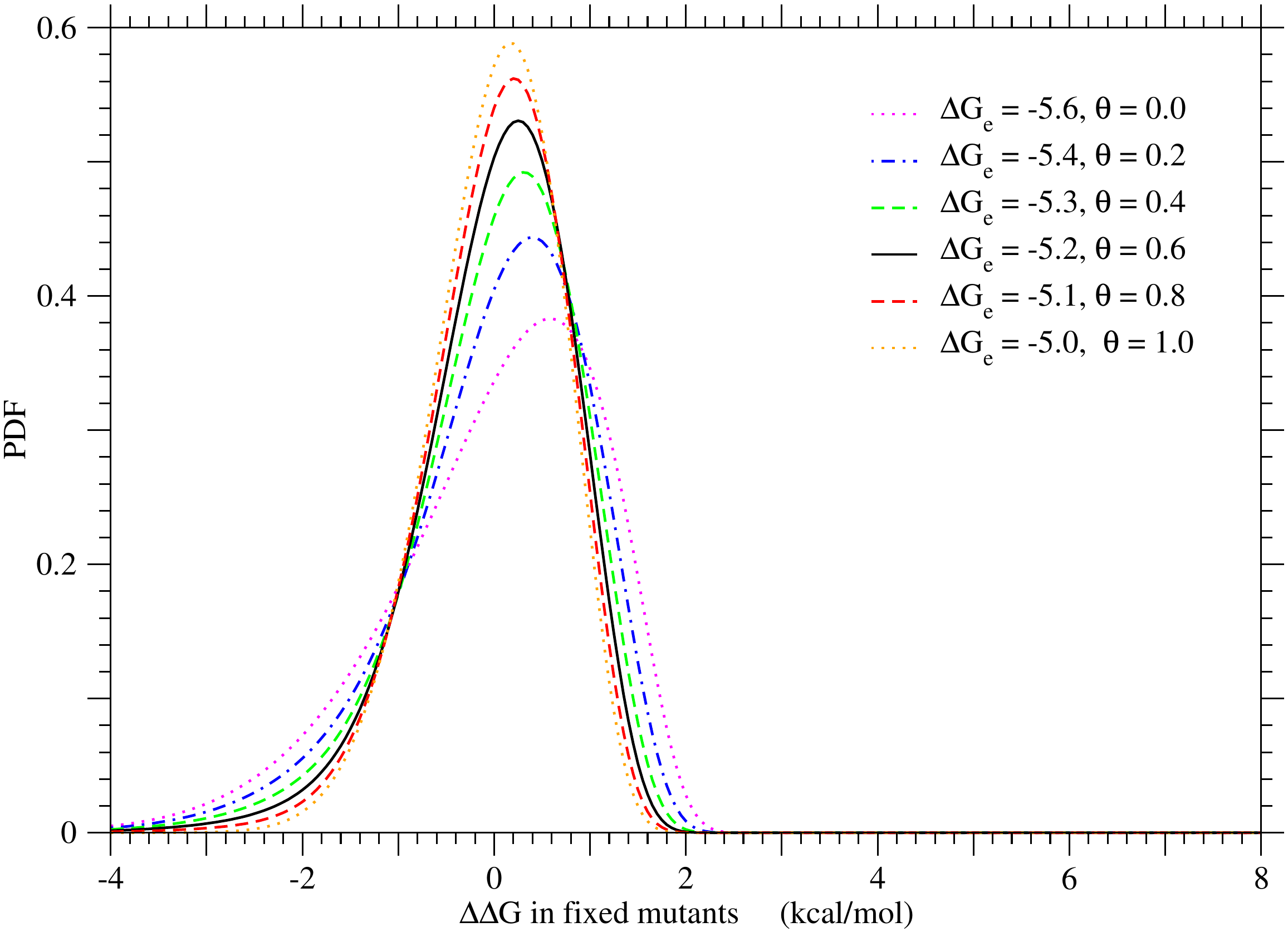}
}
} 
\vspace*{1em}
\caption{
\label{sfig: pdf_of_ddG_at_dGe}
\label{sfig: pdf_of_ddG_fixed_at_dGe}
\label{sfig: pdf_of_ddG}
\FigureLegends{
\BF{PDFs of stability changes, $\Delta\Delta G$, due to single amino acid substitutions
in all mutants and in fixed mutants
at equilibrium of protein stability, $\Delta G = \Delta G_e$.}
The PDF of $\Delta\Delta G$ due to single amino acid substitutions
in all arising mutants is assumed to be bi-Gaussian; see \Eq{\ref{seq: def_bi-Gaussian}}.
Unless specified, 
$\log 4N_e\kappa = 7.550$ and $\theta = 0.53$
are employed.
The kcal/mol unit is used for $\Delta\Delta G$ and $\Delta G_e$.
} 
}
\end{figure*}

\FigureInLegends{\newpage}

\begin{figure*}[ht]
\FigureInLegends{
\centerline{
\includegraphics*[width=90mm,angle=0]{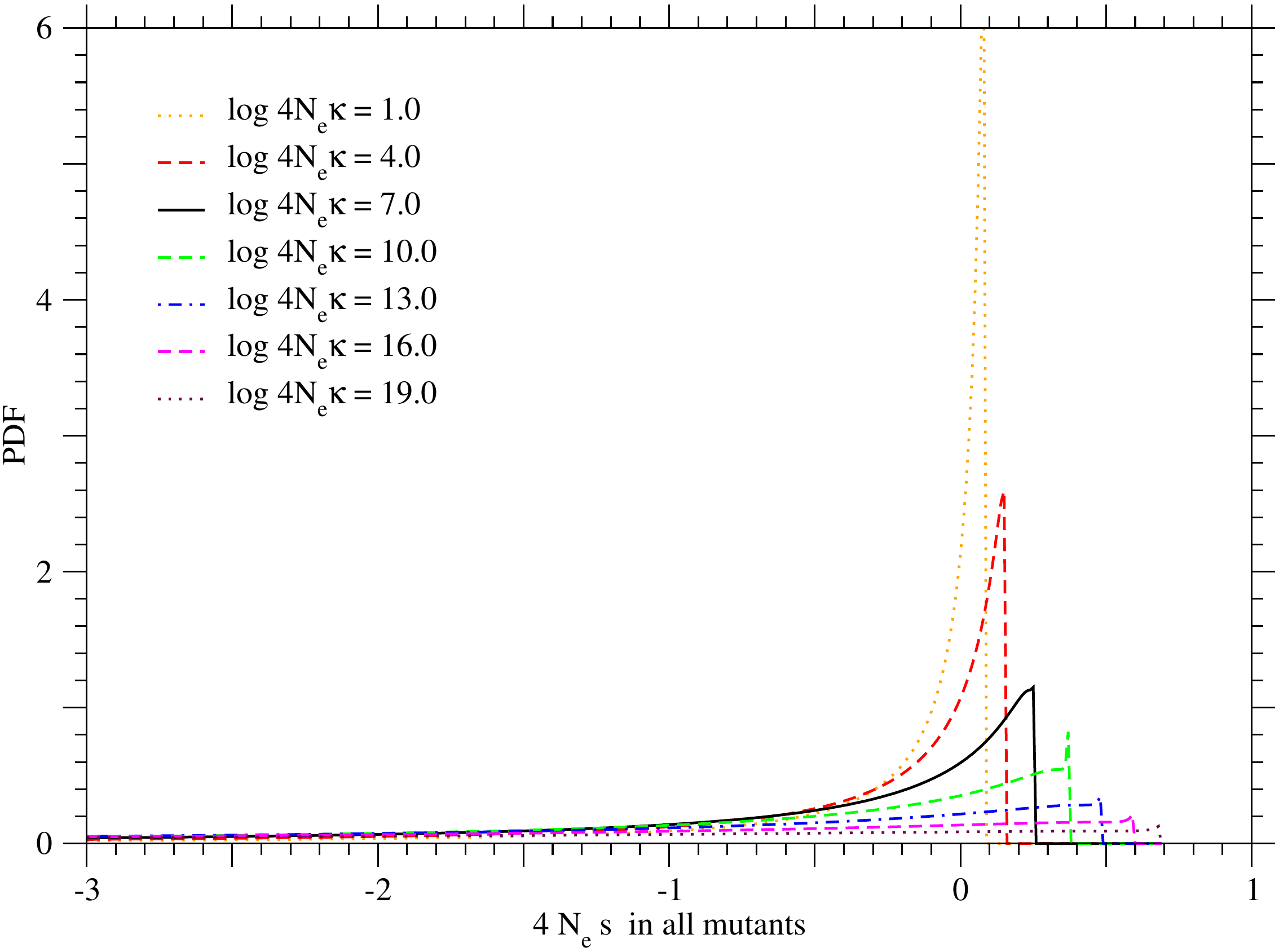}
\includegraphics*[width=90mm,angle=0]{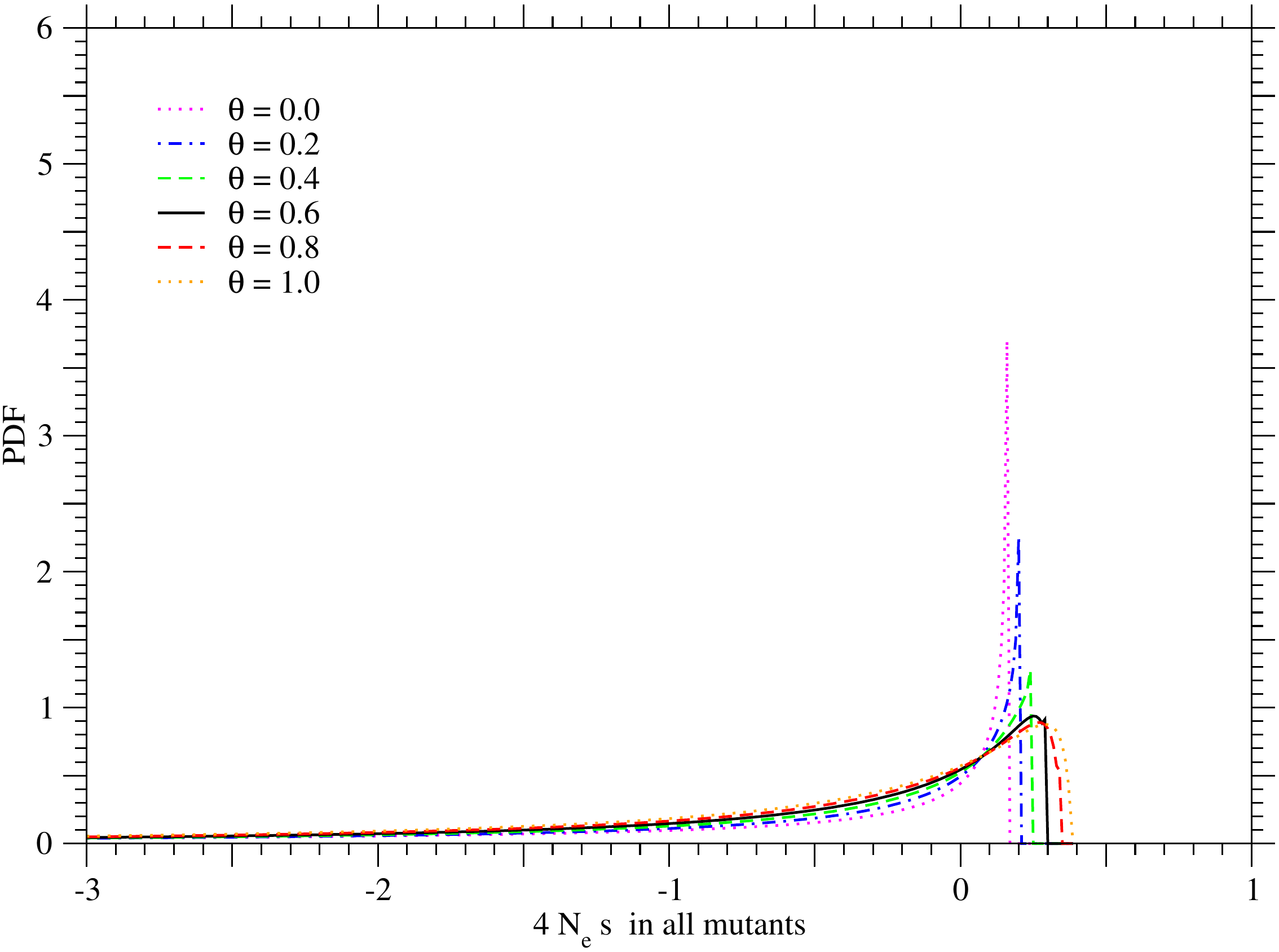}
}
\centerline{
\includegraphics*[width=90mm,angle=0]{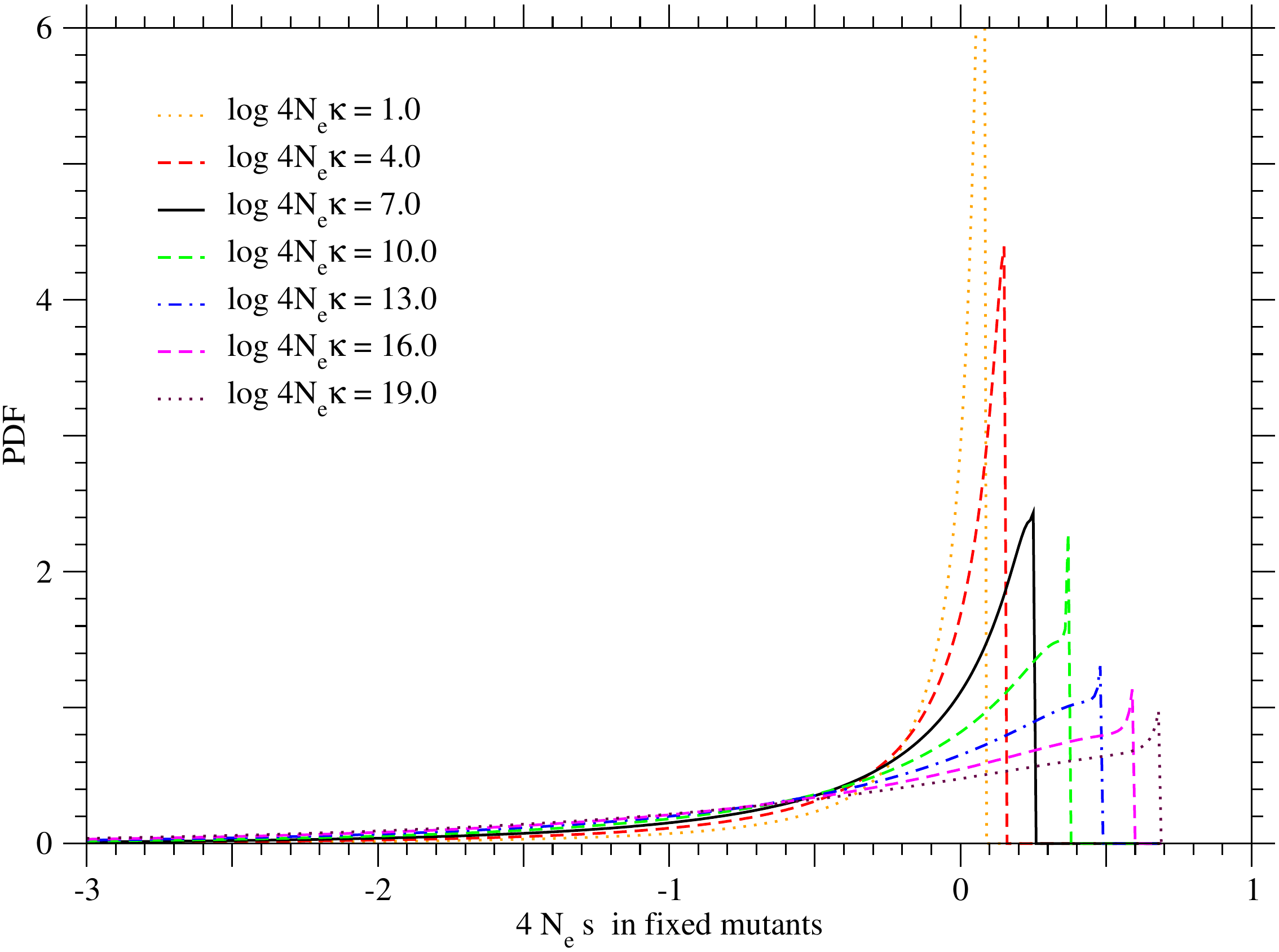}
\includegraphics*[width=90mm,angle=0]{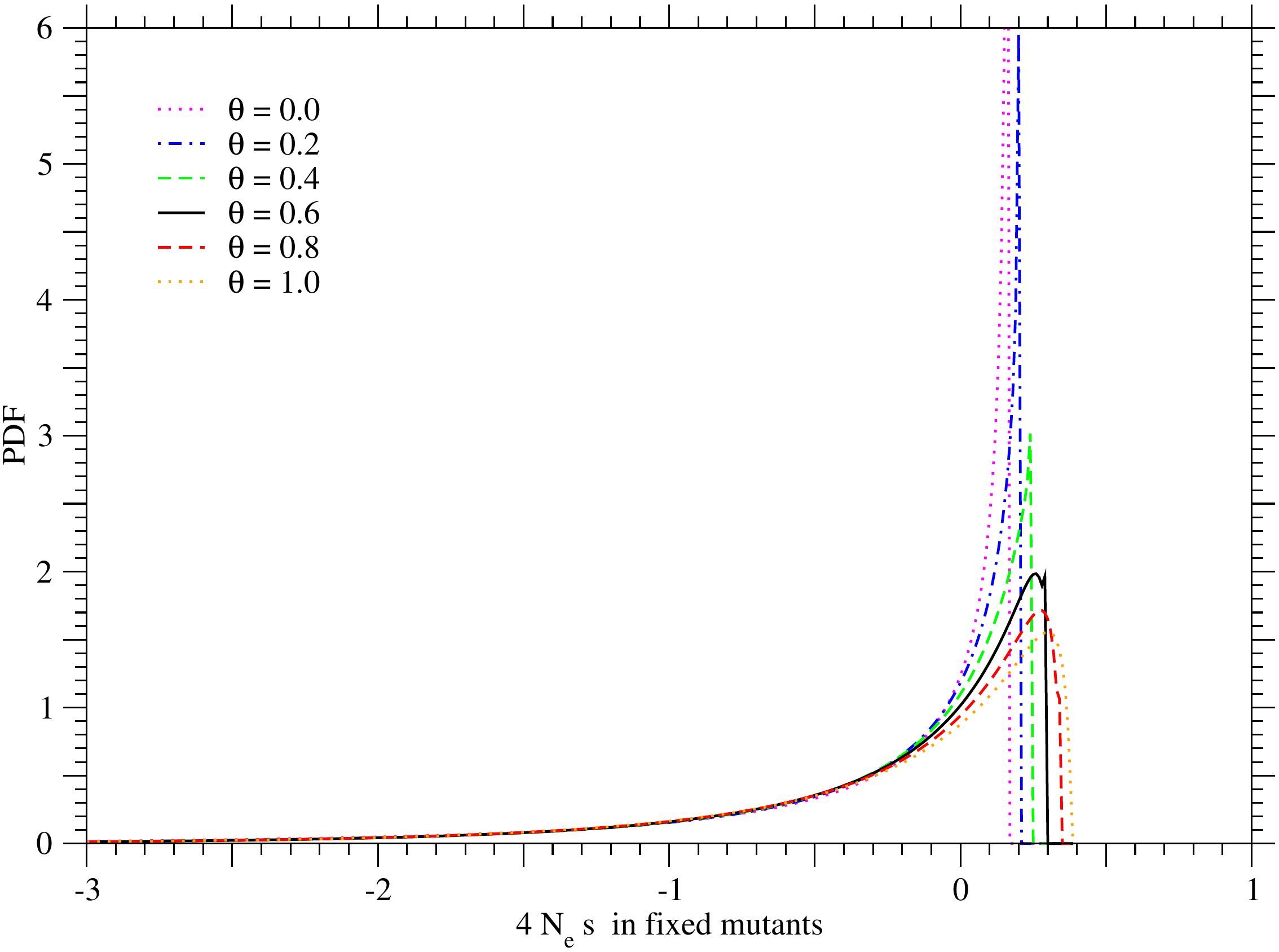}
}
} 
\vspace*{1em}
\caption{
\label{sfig: pdf_of_4sNe_at_dGe}
\label{sfig: pdf_of_4sNe}
\label{sfig: pdf_of_4sNe_fixed}
\label{sfig: pdf_of_4sNe_fixed_at_dGe}
\FigureLegends{
\BF{PDFs of $4N_e s$ in all mutants and in fixed mutants at equilibrium of protein stability, $\Delta G = \Delta G_e$.}
Unless specified, 
$\log 4N_e\kappa = 7.550$ and $\theta = 0.53$
are employed.
} 
}
\end{figure*}

\FigureInLegends{\newpage}

\begin{figure*}[ht]
\FigureInLegends{
\centerline{
\includegraphics*[width=90mm,angle=0]{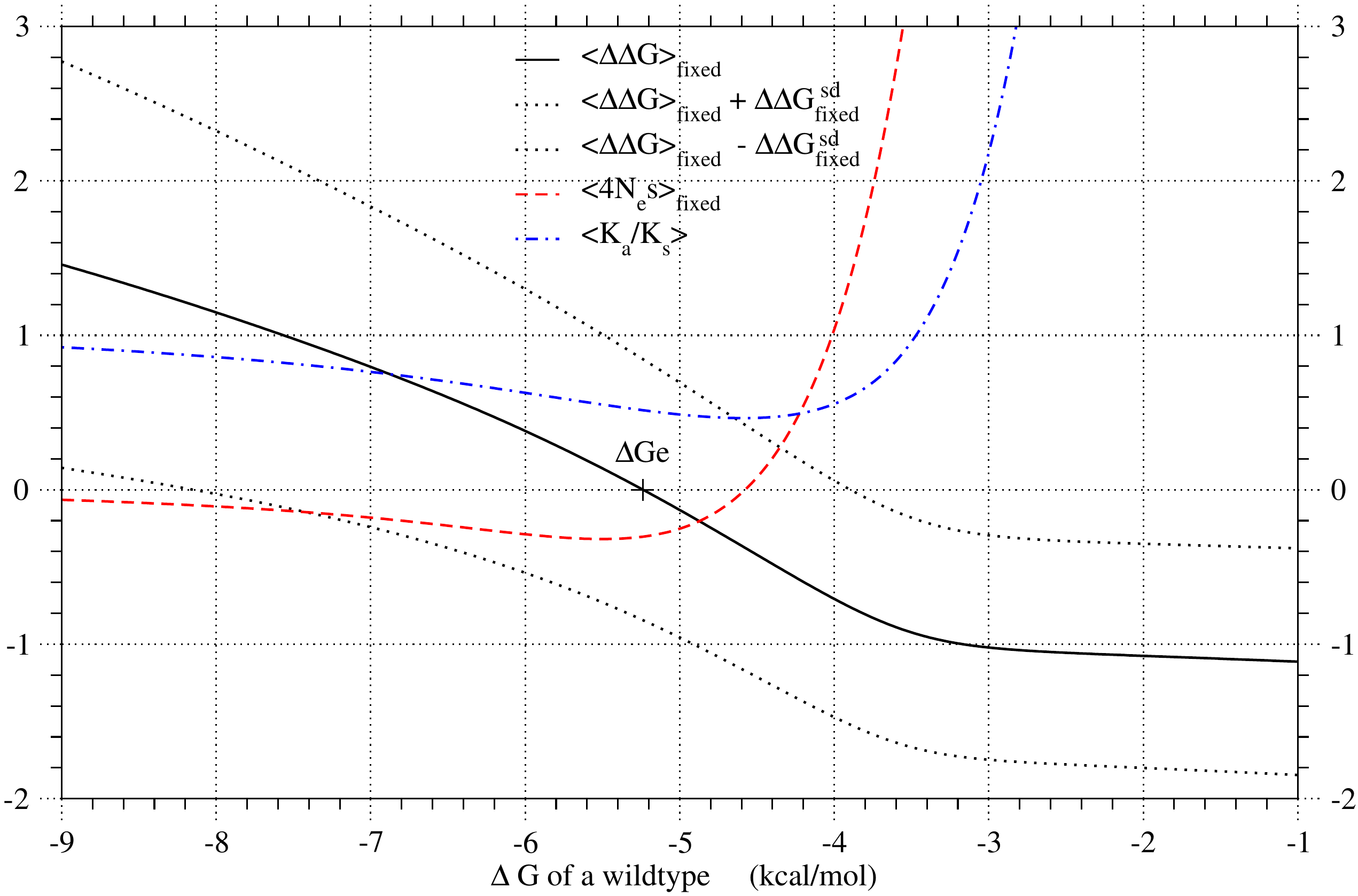}
}
} 
\vspace*{1em}
\caption{
\label{sfig: ave_ddG_vs_dG}
\label{sfig: ave_ddG_vs_dG_fixed}
\FigureLegends{
\BF{The average, $\langle \Delta\Delta G \rangle_{\script{fixed}}$, of stability changes over fixed mutants versus protein stability, $\Delta G$, of the wild type.}
$\Delta G_e$, where $\langle \Delta\Delta G \rangle = 0$, is the stable equilibrium value
of folding free energy, $\Delta G$, in protein evolution. 
The averages of $\Delta\Delta G$, $4N_e s$, and $K_a/Ks$ over fixed mutants are
plotted against protein stability, $\Delta G$, of the wild type by solid, broken, and dash-dot lines,
respectively.
Thick dotted lines show the values of $\langle \Delta\Delta G \rangle_{\script{fixed}} \pm \Delta\Delta G^{\script{sd}}_{\script{fixed}}$,
where $\Delta\Delta G^{\script{sd}}_{\script{fixed}}$ is the standard deviation of $\Delta\Delta G$ over fixed mutants.
$\log 4N_e\kappa = 7.550$ and $\theta = 0.53$
are employed.
The kcal/mol unit is used for $\Delta\Delta G$.
} 
}
\end{figure*}

\FigureInLegends{\newpage}

\begin{figure*}[ht]
\FigureInLegends{
\centerline{
\includegraphics*[width=90mm,angle=0]{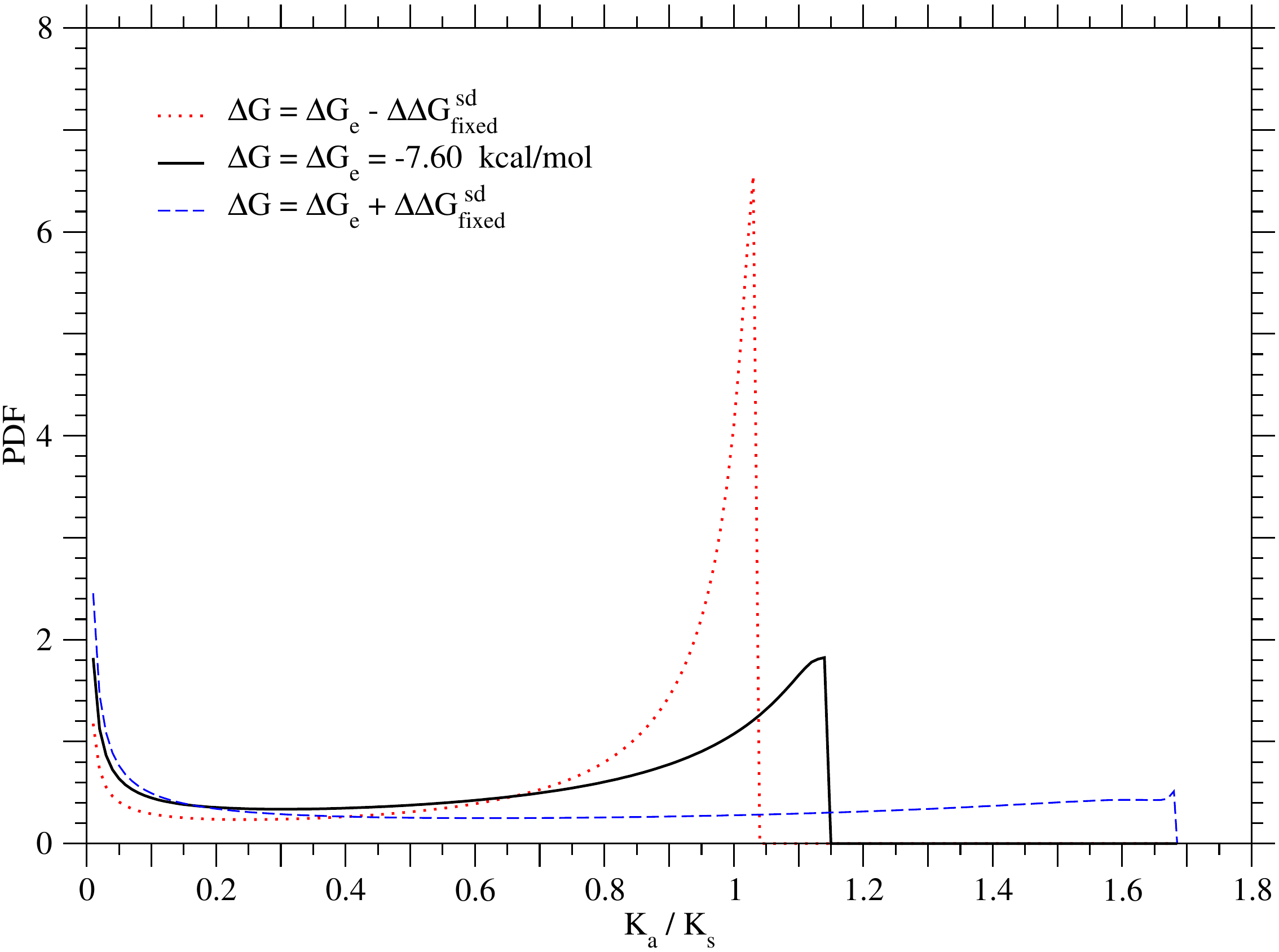}
}
\centerline{
\includegraphics*[width=90mm,angle=0]{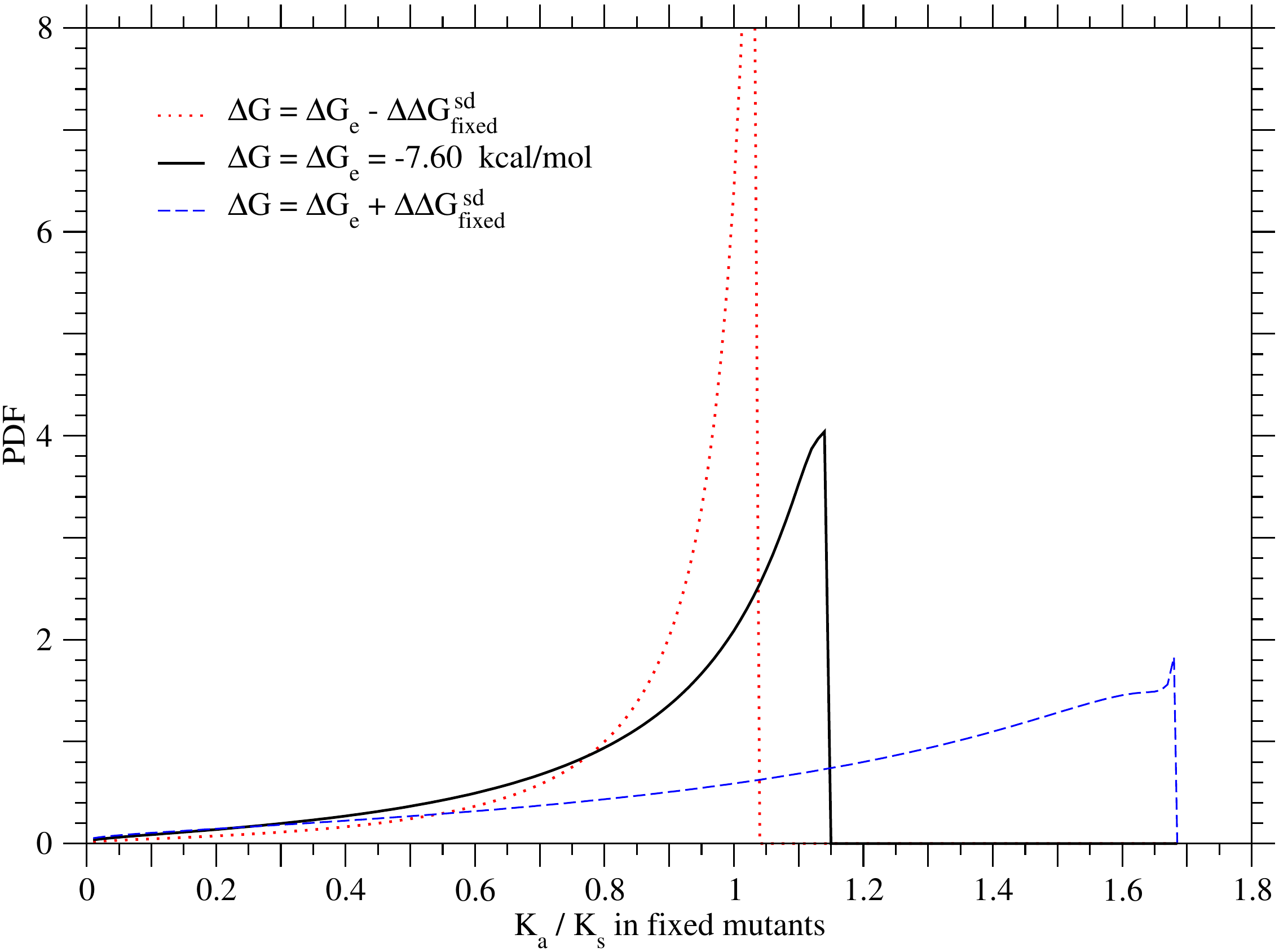}
}
} 
\vspace*{1em}
\caption{
\label{sfig: pdf_of_Ka_over_Ks_around_dGe}
\label{sfig: pdf_of_Ka_over_Ks_fixed_around_dGe}
\FigureLegends{
\BF{Dependence of the PDF of $K_a/K_s$ on protein stability, $\Delta G$, of the wild type in all mutants or in fixed mutants only.}
$\Delta\Delta G^{\script{sd}}_{\script{fixed}}$ is the standard deviation 
(0.84 kcal/mol)	
of $\Delta\Delta G$
over fixed mutants at $\Delta G = \Delta G_e$.
$\log 4N_e\kappa = 7.550$ and $\theta = 0.53$
are employed.
The kcal/mol unit is used for $\Delta G_e$.
} 
}
\end{figure*}

\FigureInLegends{\clearpage\newpage}

\begin{figure*}[ht]
\FigureInLegends{
\centerline{
\includegraphics*[width=90mm,angle=0]{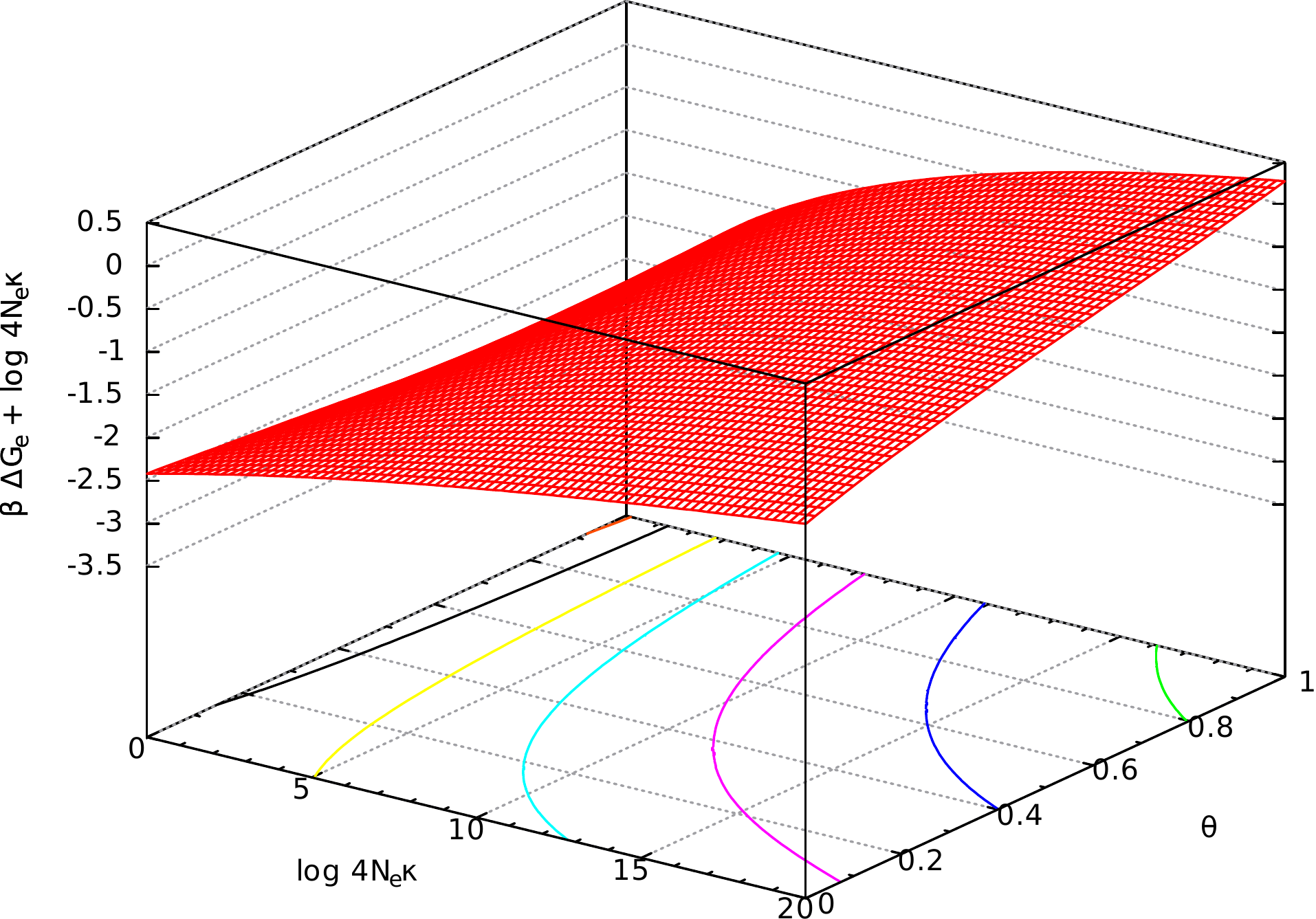}
\includegraphics*[width=90mm,angle=0]{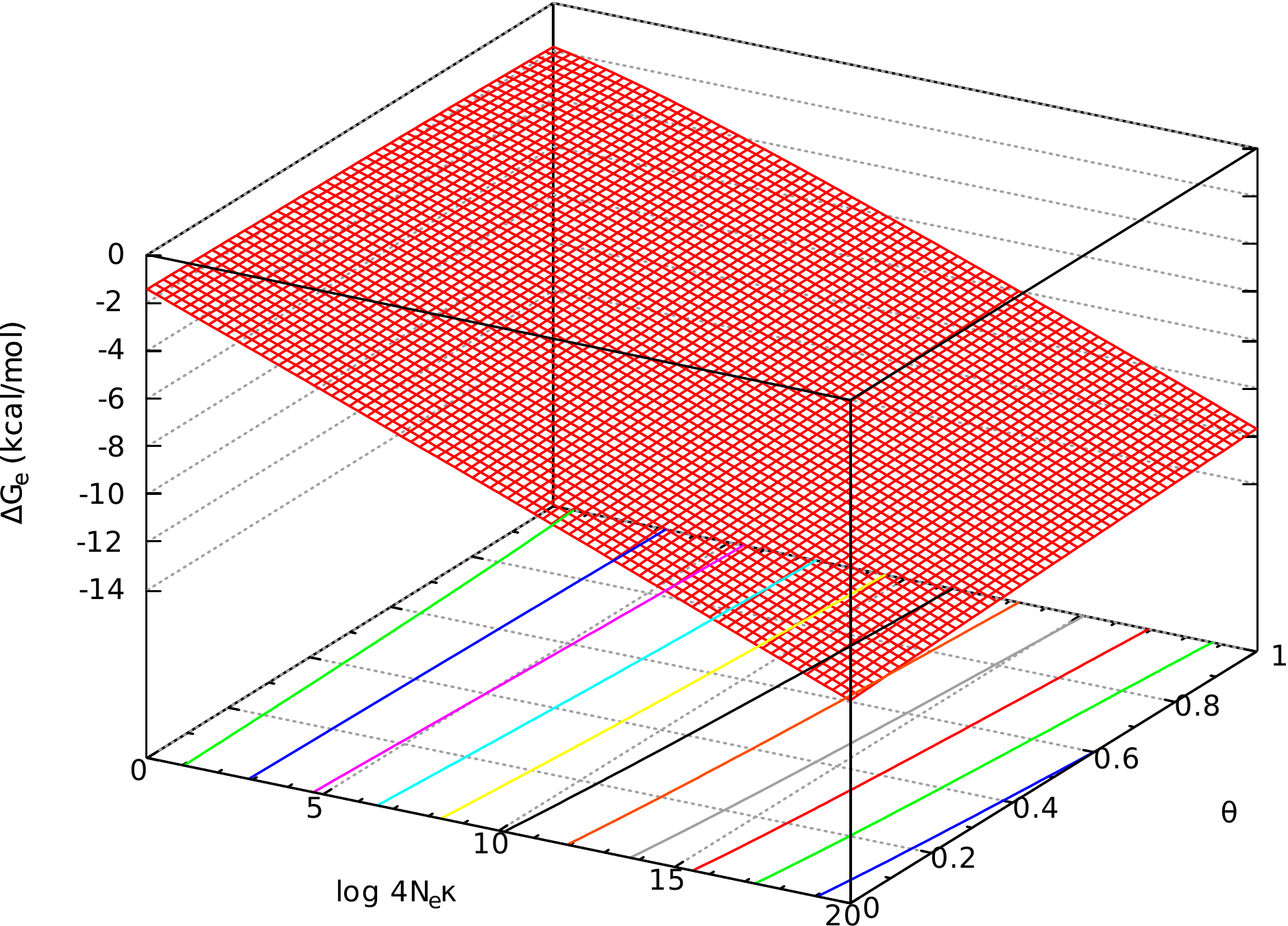}
}
} 
\vspace*{1em}
\caption{
\label{sfig: dependence_of_dGe_on_4Nekappa_and_theta}
\FigureLegends{
\BF{Dependence of equilibrium stability, $\Delta G_e$, on parameters, $4N_e\kappa$ and $\theta$.}
$\Delta G_e$ is the equilibrium value of folding free energy, $\Delta G$,
in protein evolution.
The value of $\beta\Delta G_e + \log 4N_e\kappa$ is the upper bound of $\log 4N_e s$,
and would be constant if the mean of $\Delta\Delta G$ in all arising mutants did not depend on $\Delta G$; 
see \Eq{\ref{seq: def_s}}.
The kcal/mol unit is used for $\Delta G_e$.
} 
}
\end{figure*}

\FigureInLegends{\clearpage\newpage}

\begin{figure*}[ht]
\FigureInLegends{
\centerline{
\includegraphics*[width=90mm,angle=0]{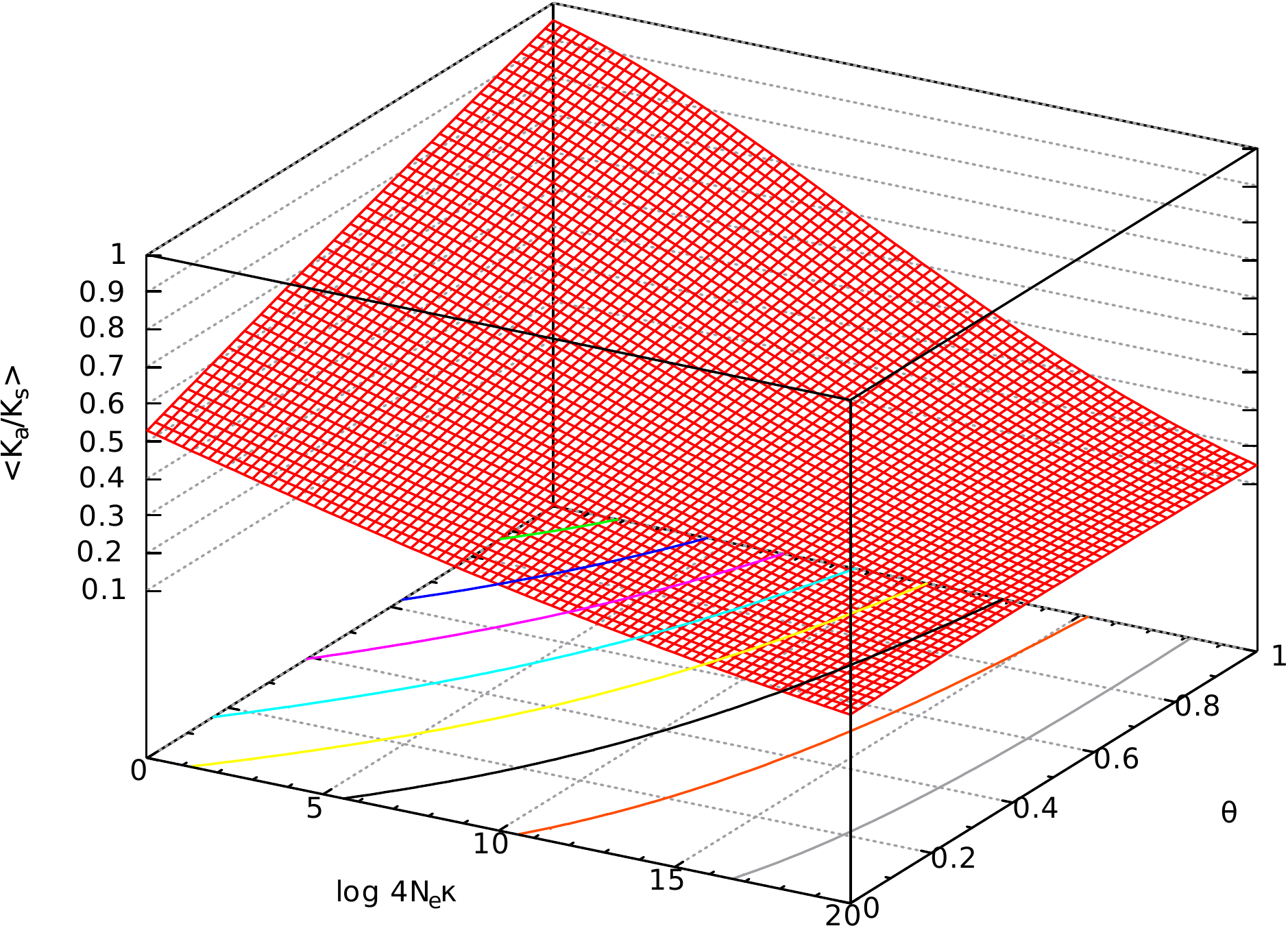}
\includegraphics*[width=90mm,angle=0]{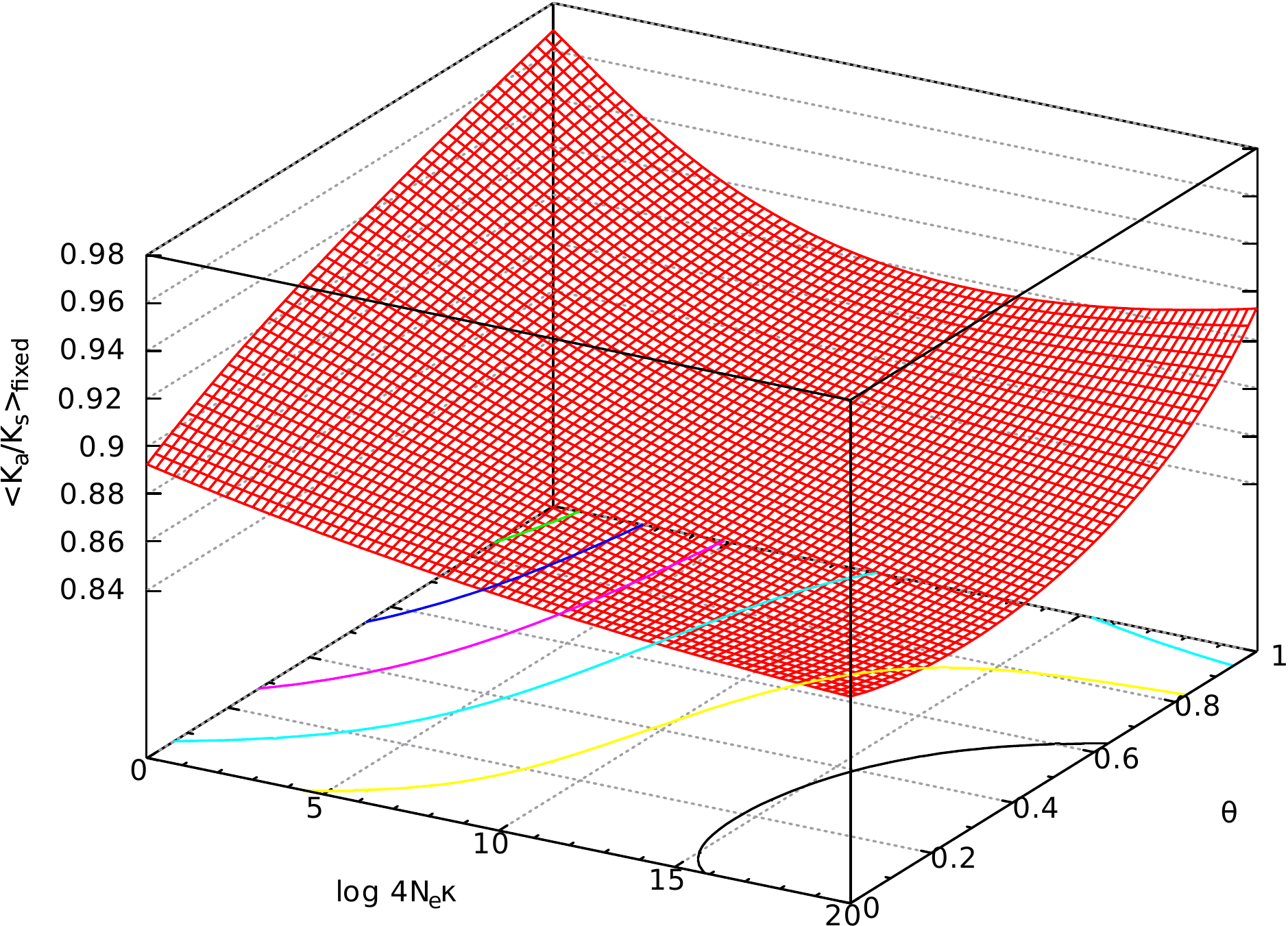}
}
} 
\vspace*{1em}
\caption{
\label{sfig: dependence_of_ave_Ka_over_Ks_on_4Nekappa_and_theta}
\label{sfig: dependence_of_ave_Ka_over_Ks_fixed_on_4Nekappa_and_theta}
\FigureLegends{
\BF{The average of $K_a / K_s$ over all mutants or over fixed mutants only at equilibrium of protein stability, $\Delta G = \Delta G_e$.}
} 
}
\end{figure*}

\FigureInLegends{\newpage}

\begin{figure*}[ht]
\FigureInLegends{
\centerline{
\includegraphics*[width=90mm,angle=0]{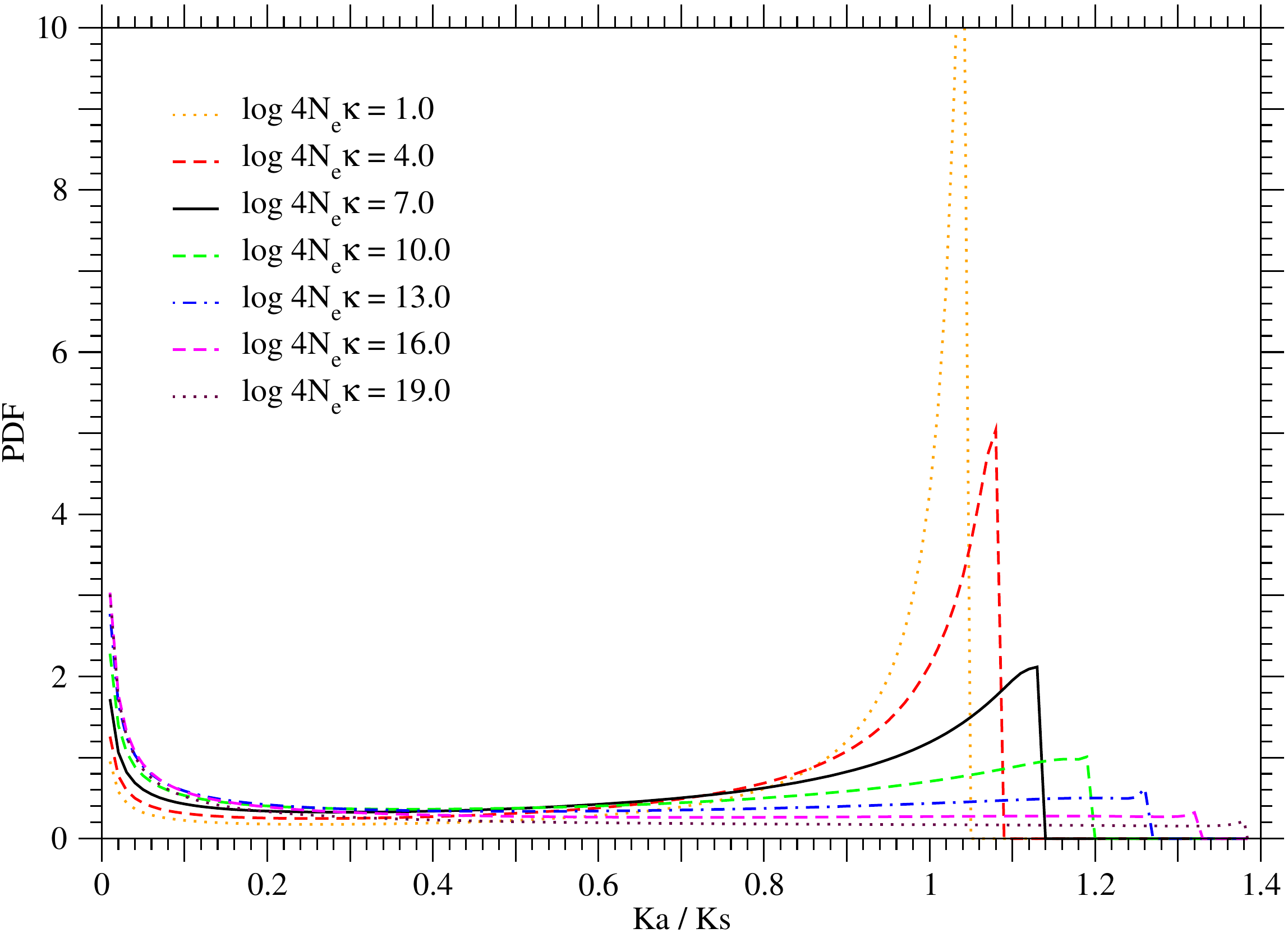}
\includegraphics*[width=90mm,angle=0]{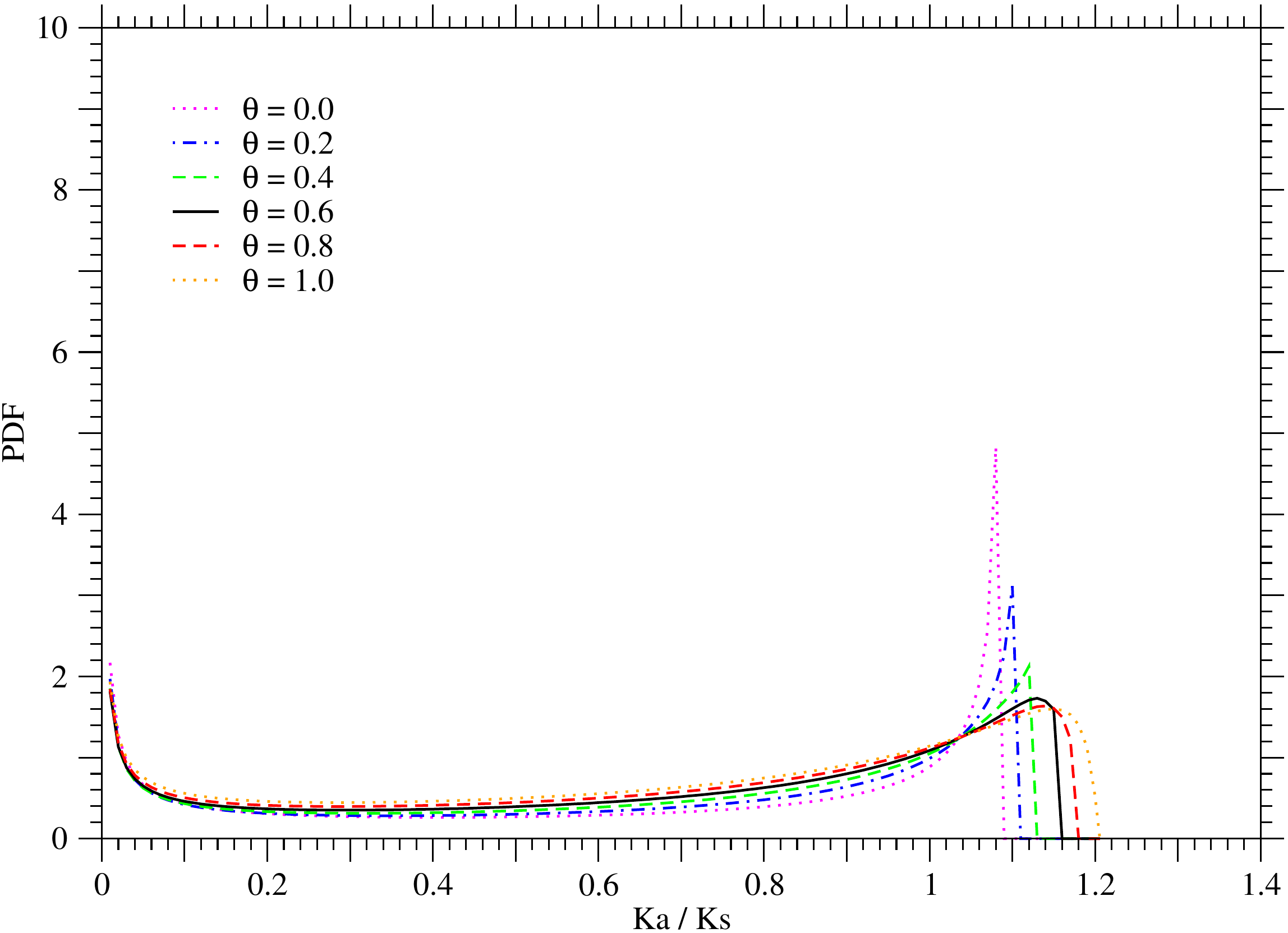}
}
\centerline{
\includegraphics*[width=90mm,angle=0]{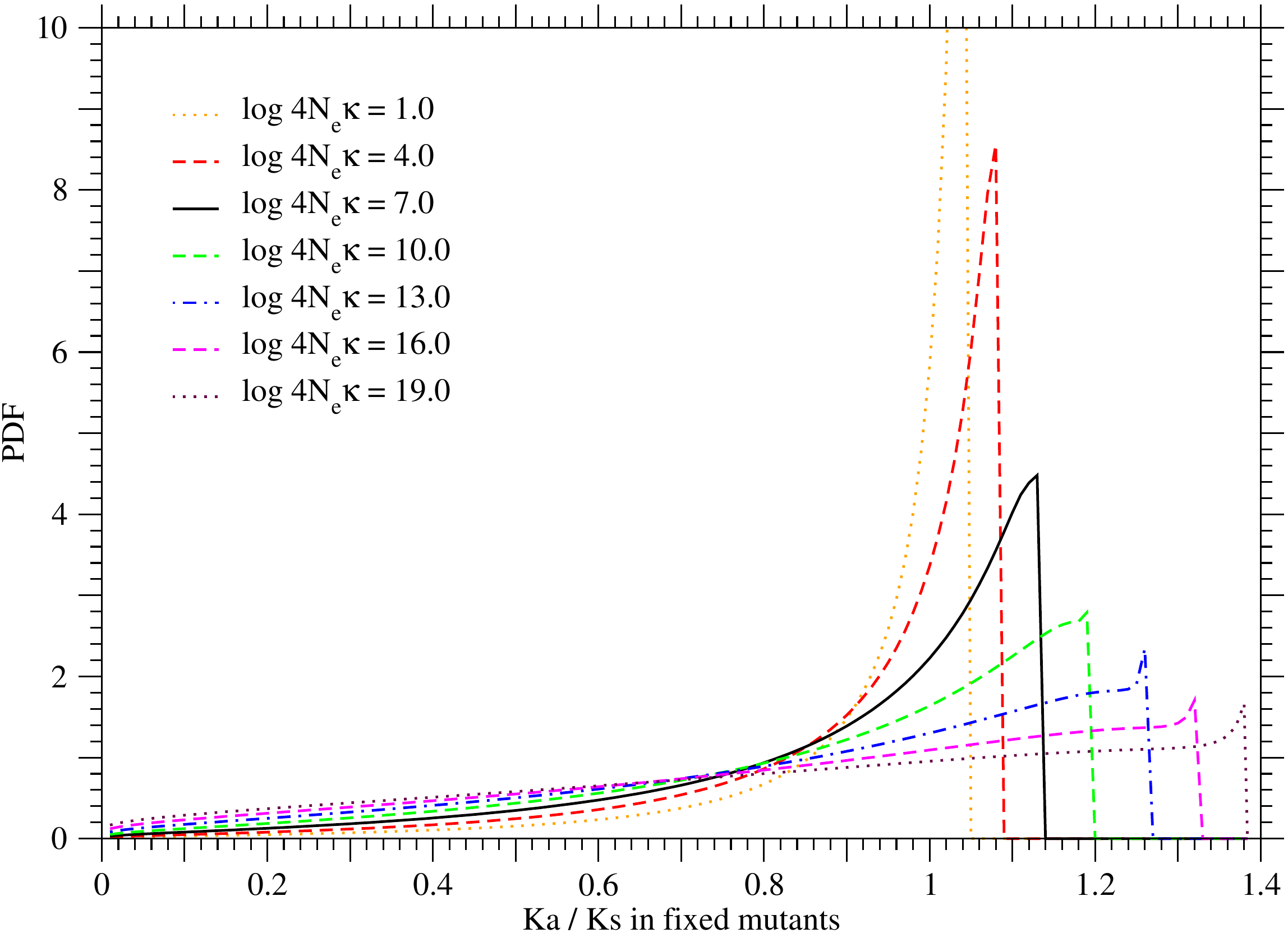}
\includegraphics*[width=90mm,angle=0]{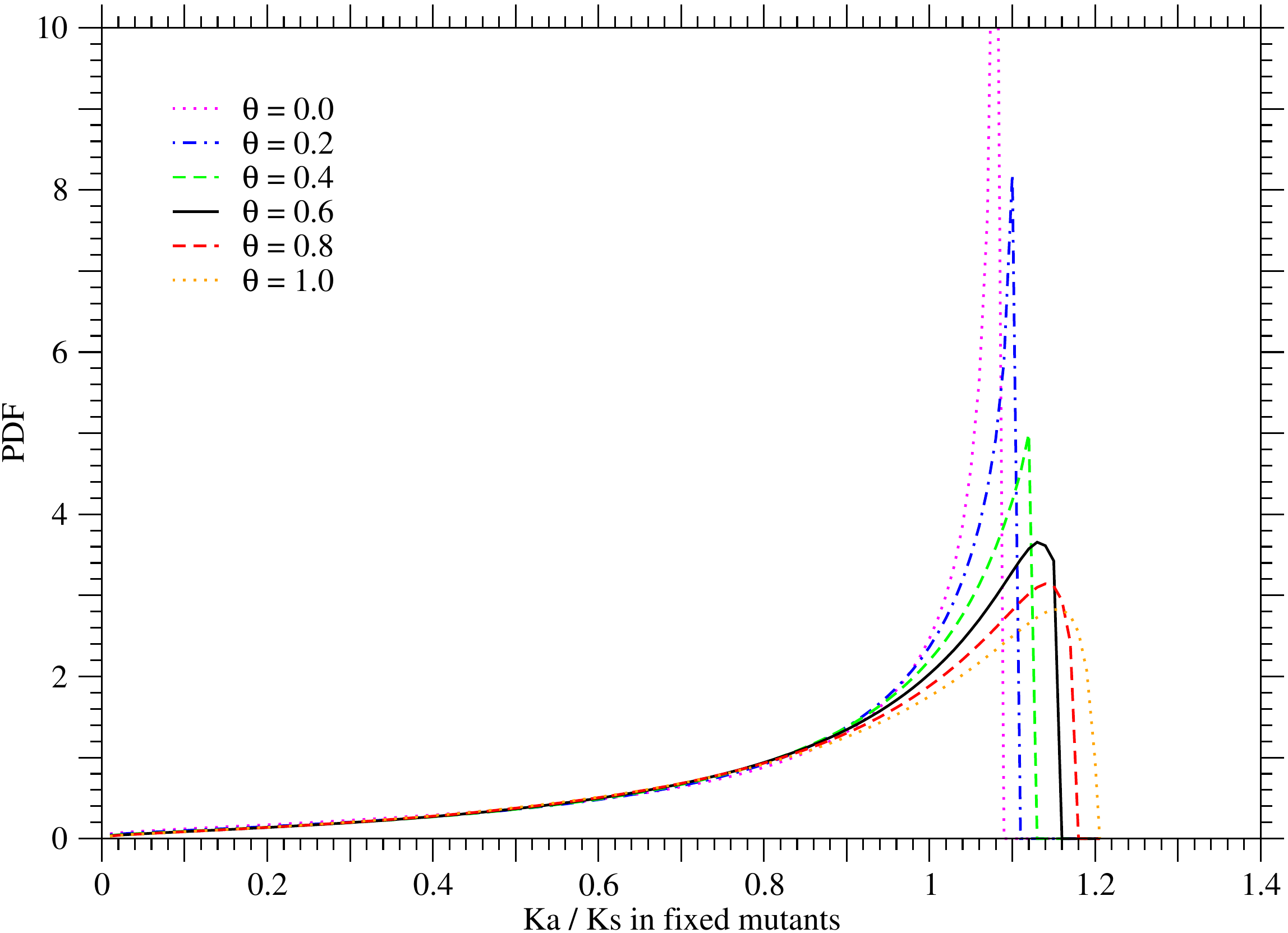}
}
} 
\vspace*{1em}
\caption{
\label{sfig: dependence_of_pdf_of_Ka_over_Ks_on_4Nekappa_and_theta}
\label{sfig: dependence_of_pdf_of_Ka_over_Ks_fixed_on_4Nekappa_and_theta}
\FigureLegends{
\BF{PDFs of $K_a / K_s$ in all mutants and in fixed mutants only
at equilibrium of protein stability, $\Delta G = \Delta G_e$.}
Unless specified, 
$\log 4N_e\kappa = 7.550$ and $\theta = 0.53$
are employed.
} 
}
\end{figure*}

\FigureInLegends{\newpage}

\begin{figure*}[ht]
\FigureInLegends{
\centerline{
\includegraphics*[width=87mm,angle=0]{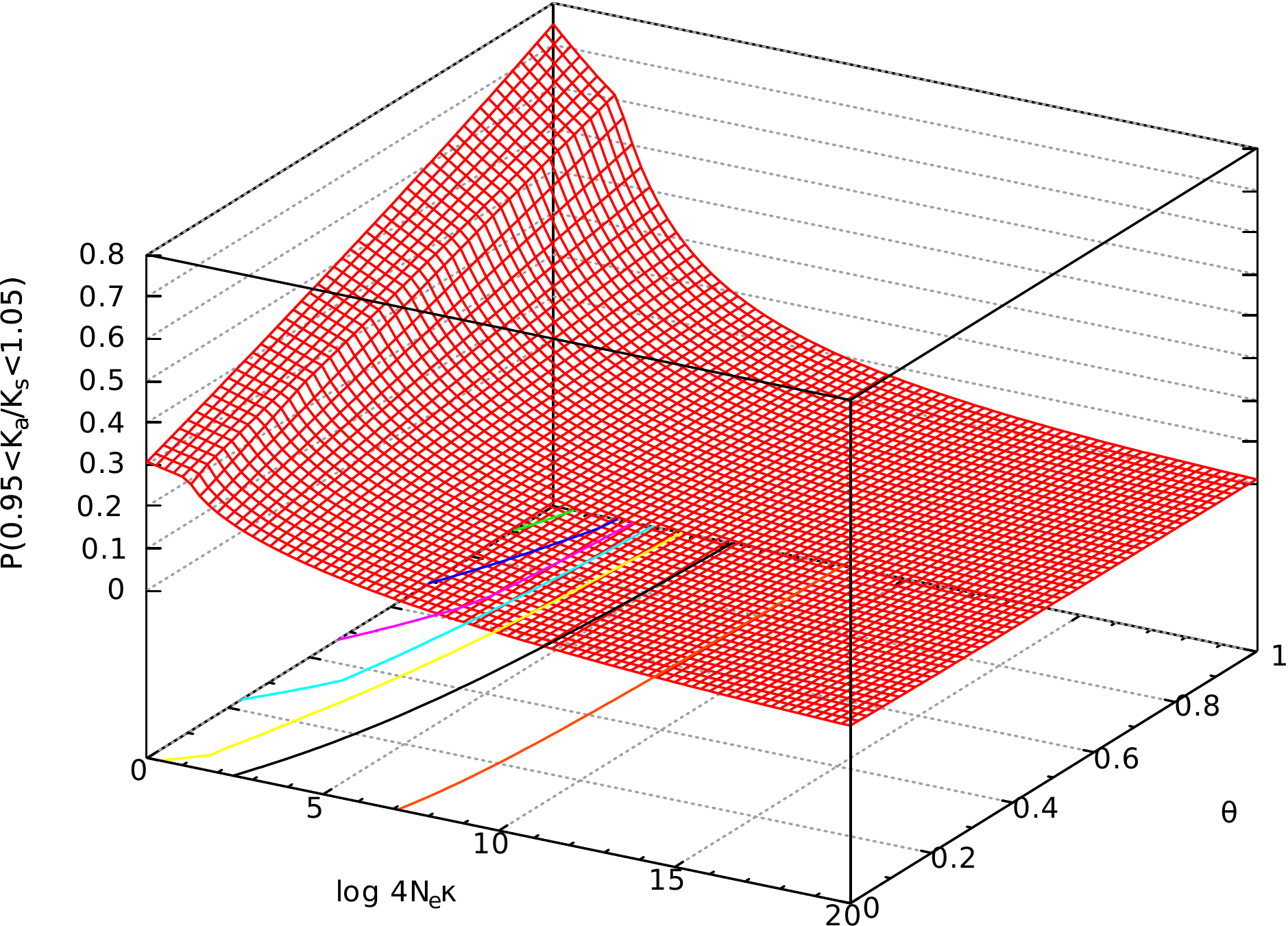}
\includegraphics*[width=90mm,angle=0]{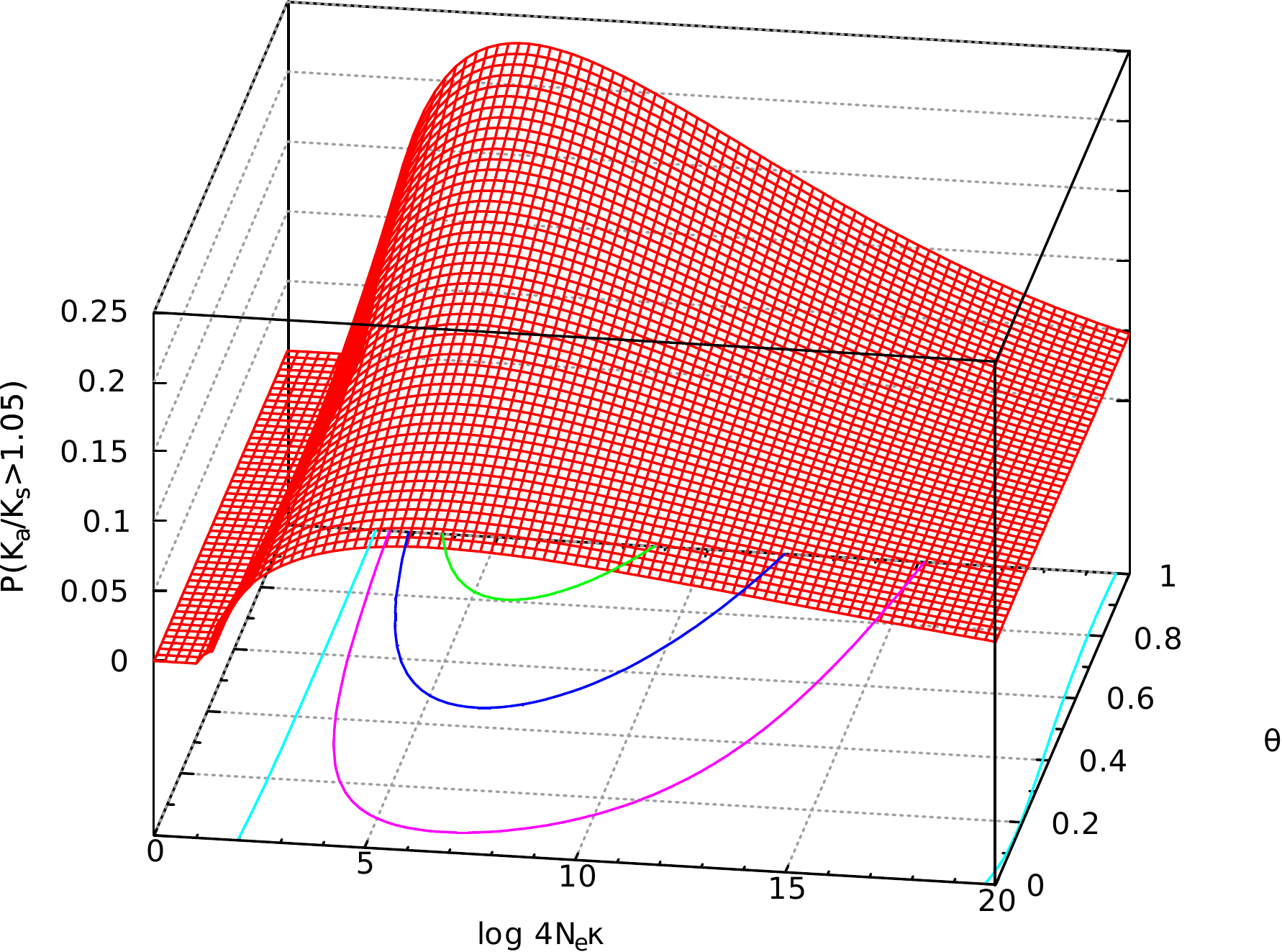}
}
\centerline{
\includegraphics*[width=88mm,angle=0]{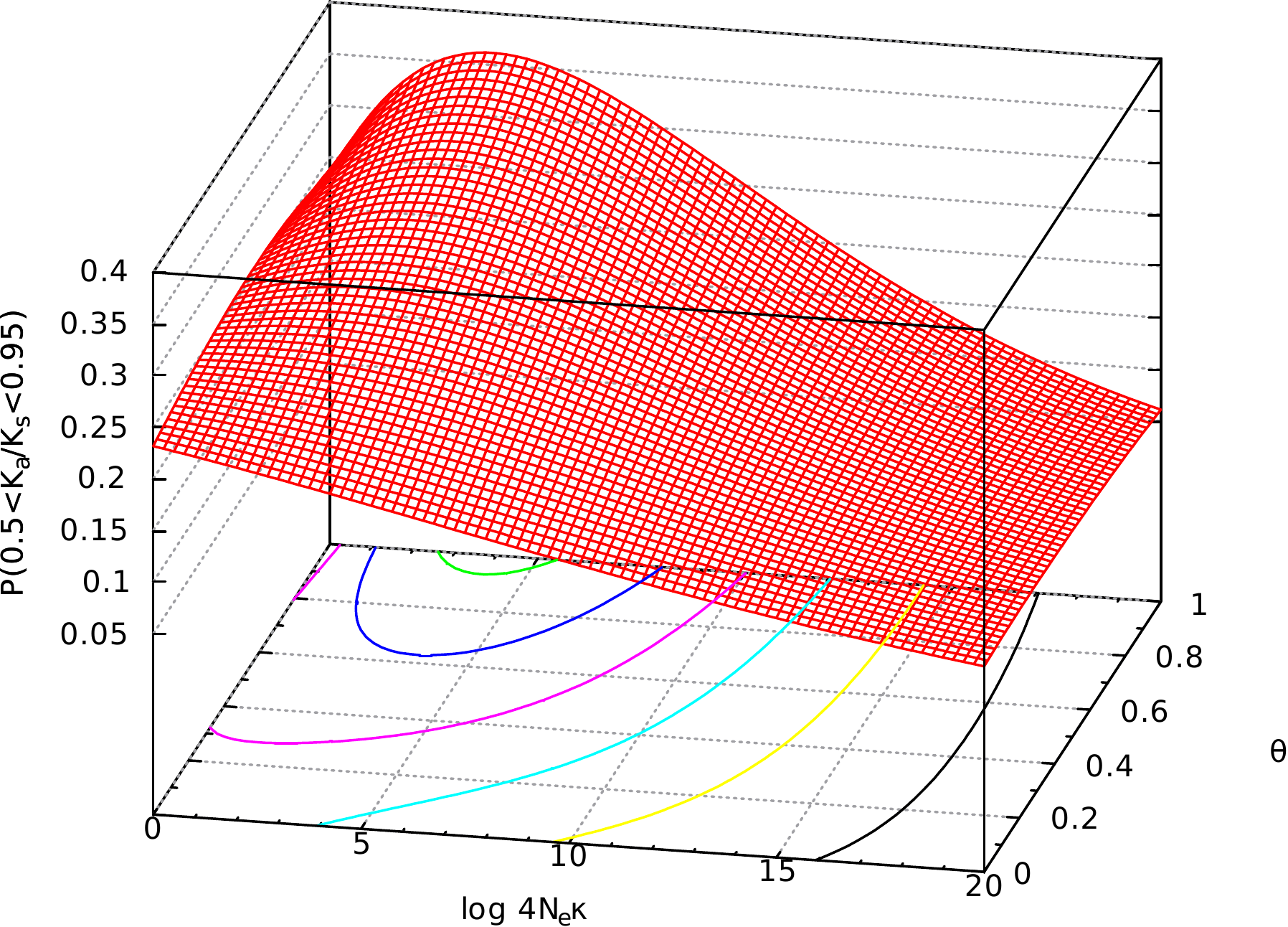}
\includegraphics*[width=90mm,angle=0]{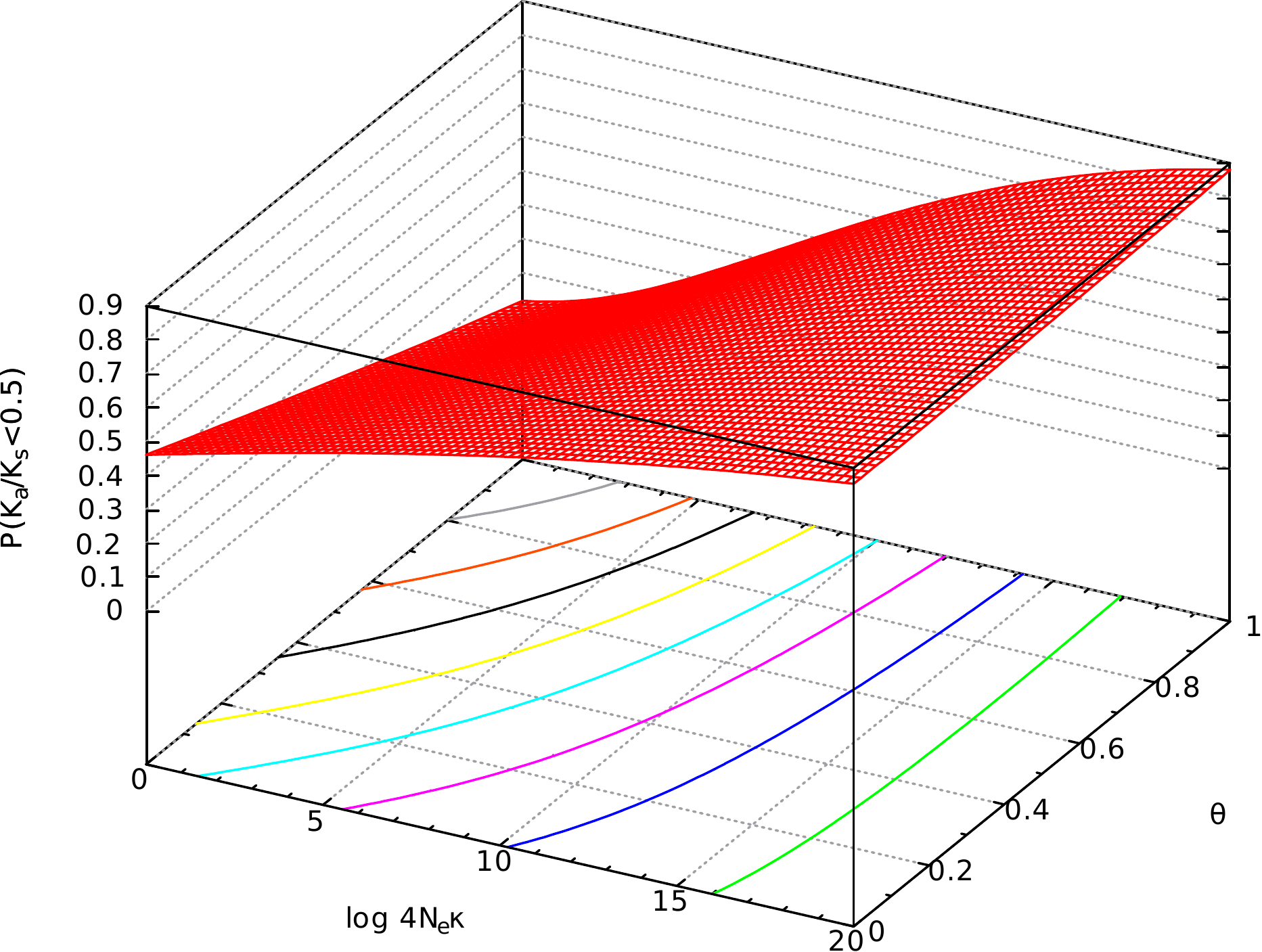}
}
} 
\vspace*{1em}
\caption{
\label{sfig: prob_of_each_selection_category_in_all_mutants}
\FigureLegends{
\BF{Probability of each selection category in all mutants at equilibrium of protein stability, $\Delta G = \Delta G_e$.}
Arbitrarily, the value of $K_a/K_s$ is categorized into four classes;
negative, slightly negative, nearly neutral, and positive selection categories in which
$K_a/K_s$ is within the ranges of
$K_a/K_s \leq 0.5$, $0.5< K_a/K_s \leq 0.95$, $0.95< K_a/K_s \leq 1.05$, and $ 1.05 < K_a/K_s$,
respectively.
} 
}
\end{figure*}

\FigureInLegends{\newpage}

\begin{figure*}[ht]
\FigureInLegends{
\centerline{
\includegraphics*[width=90mm,angle=0]{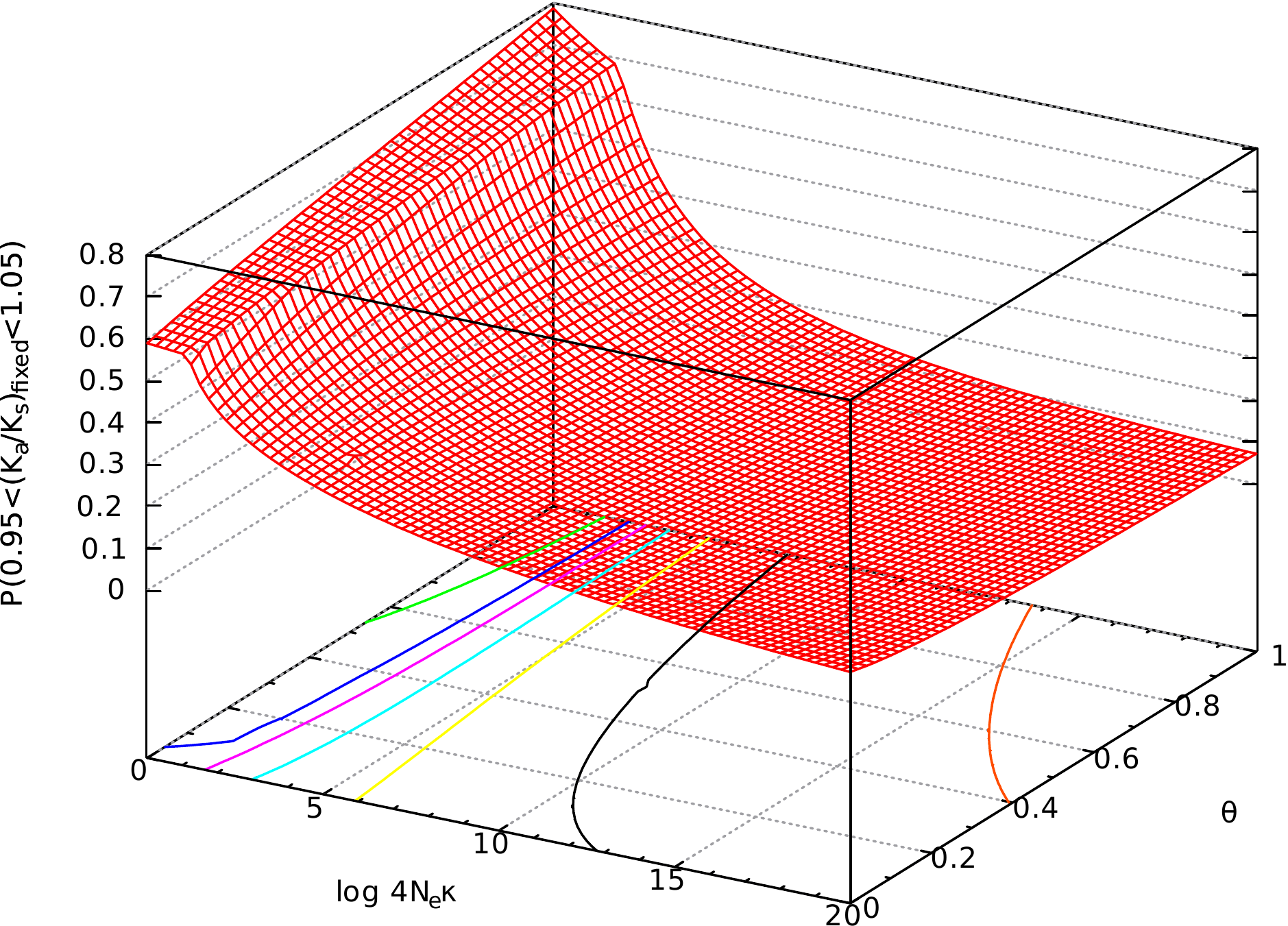}
\includegraphics*[width=90mm,angle=0]{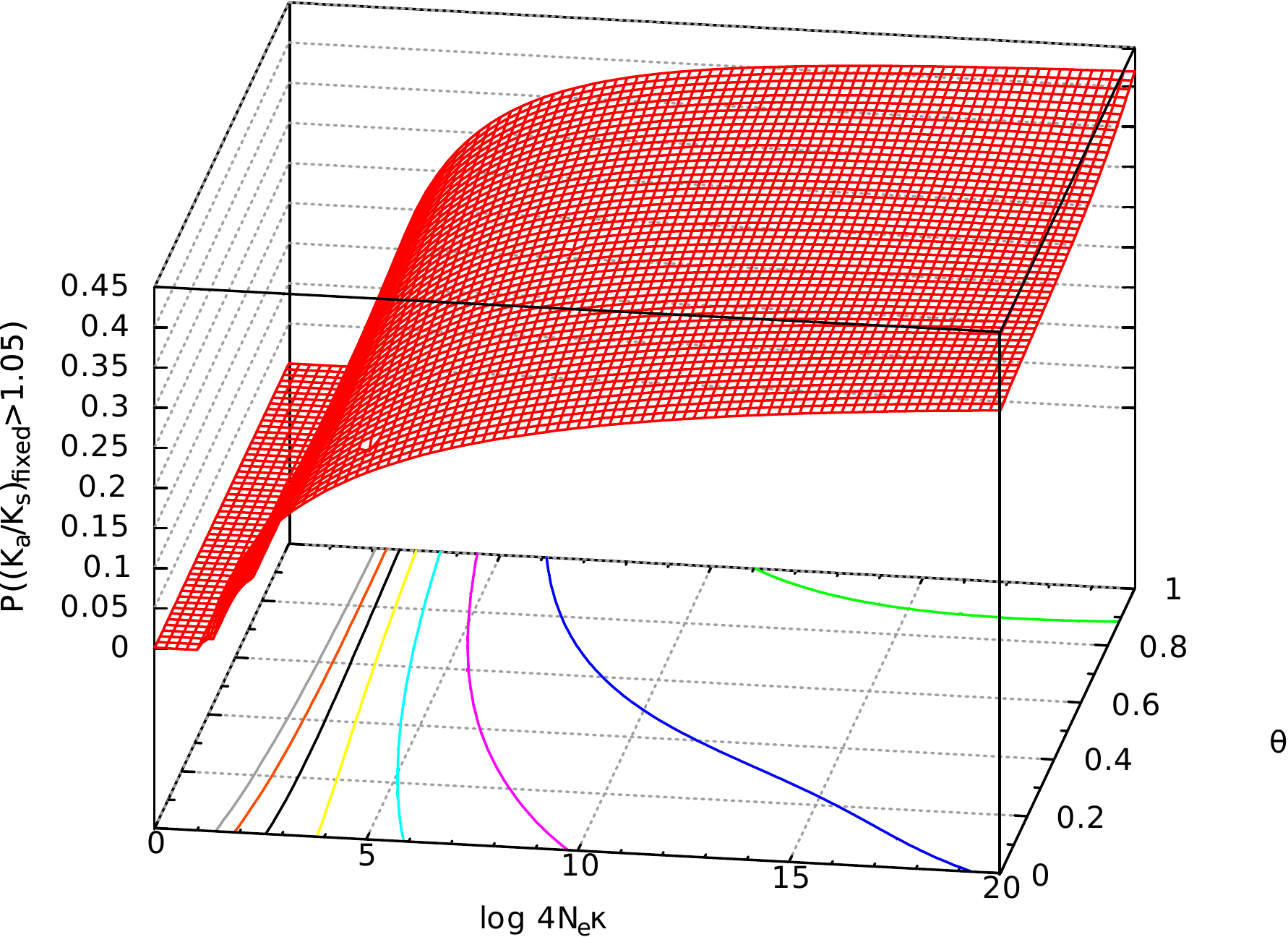}
}
\centerline{
\includegraphics*[width=90mm,angle=0]{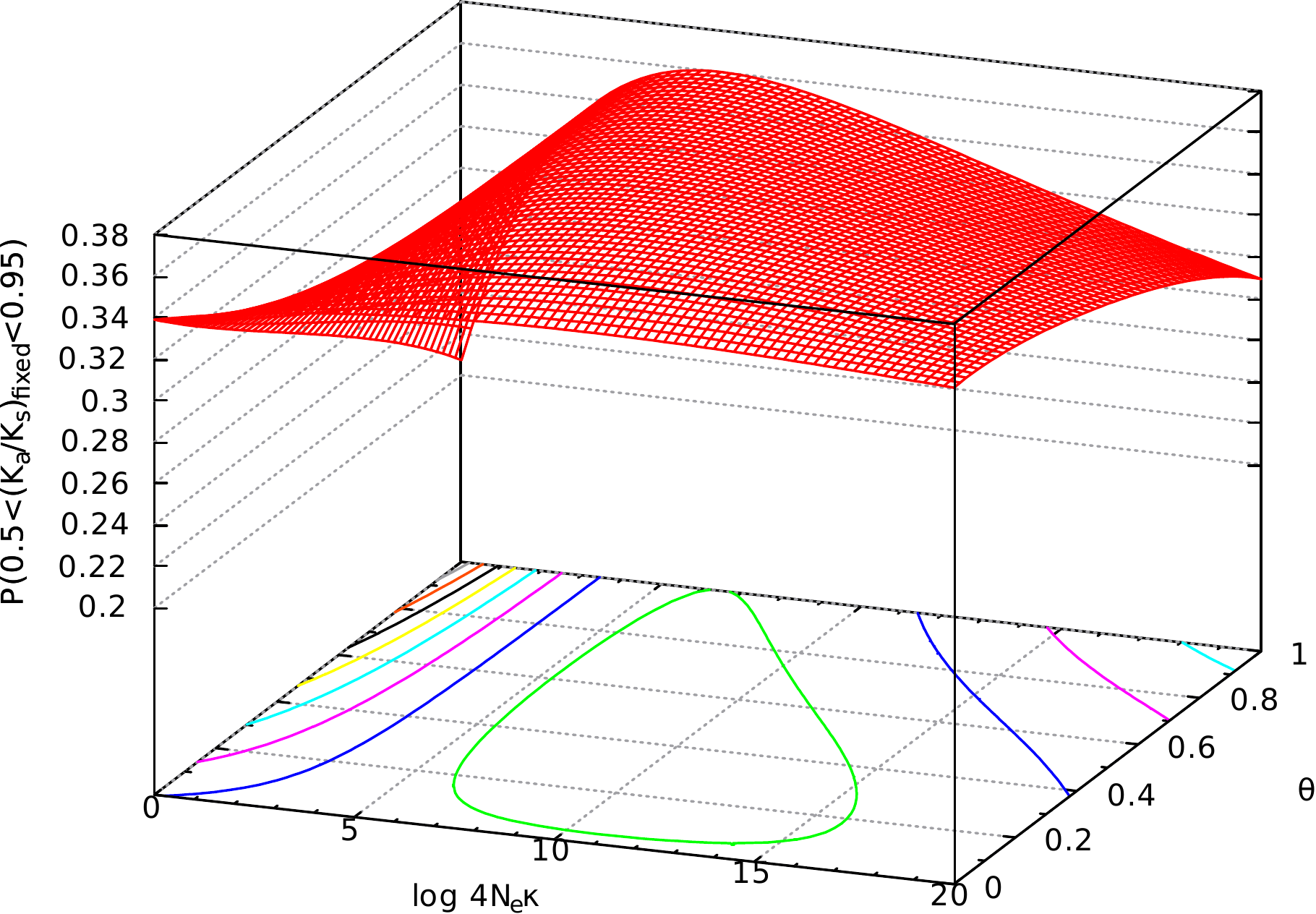}
\includegraphics*[width=90mm,angle=0]{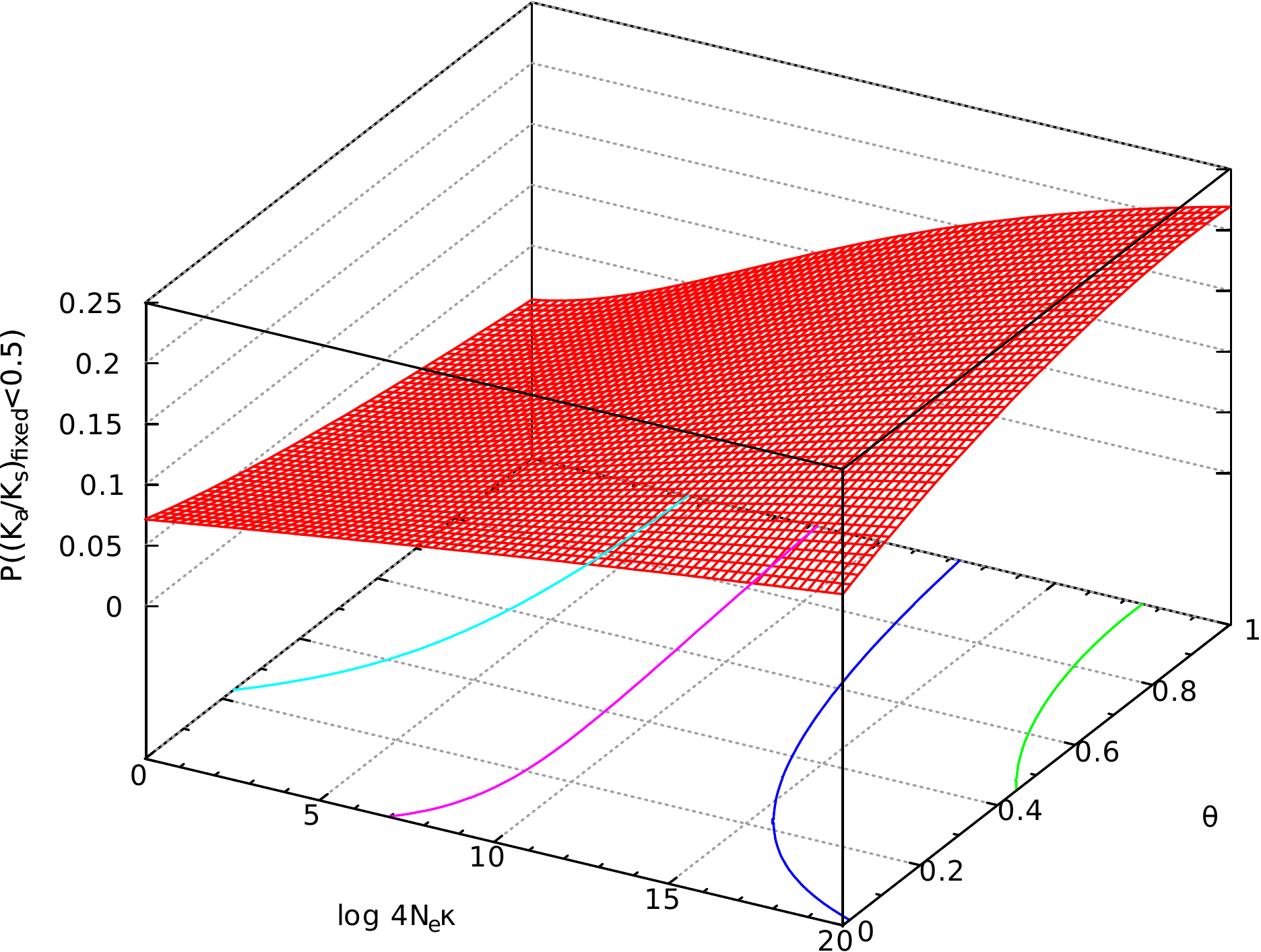}
}
} 
\vspace*{1em}
\caption{
\label{sfig: prob_of_each_selection_category_in_fixed_mutants}
\FigureLegends{
\BF{Probability of each selection category in fixed mutants at equilibrium of protein stability, $\Delta G = \Delta G_e$.}
Arbitrarily, the value of $K_a/K_s$ is categorized into four classes;
negative, slightly negative, nearly neutral, and positive selection categories in which
$K_a/K_s$ is within the ranges of
$K_a/K_s \leq 0.5$, $0.5< K_a/K_s \leq 0.95$, $0.95< K_a/K_s \leq 1.05$, and $ 1.05 < K_a/K_s$,
respectively.
} 
}
\end{figure*}

\FigureInLegends{\newpage}

\begin{figure*}[ht]
\FigureInLegends{
\centerline{
\includegraphics*[width=80mm,angle=0]{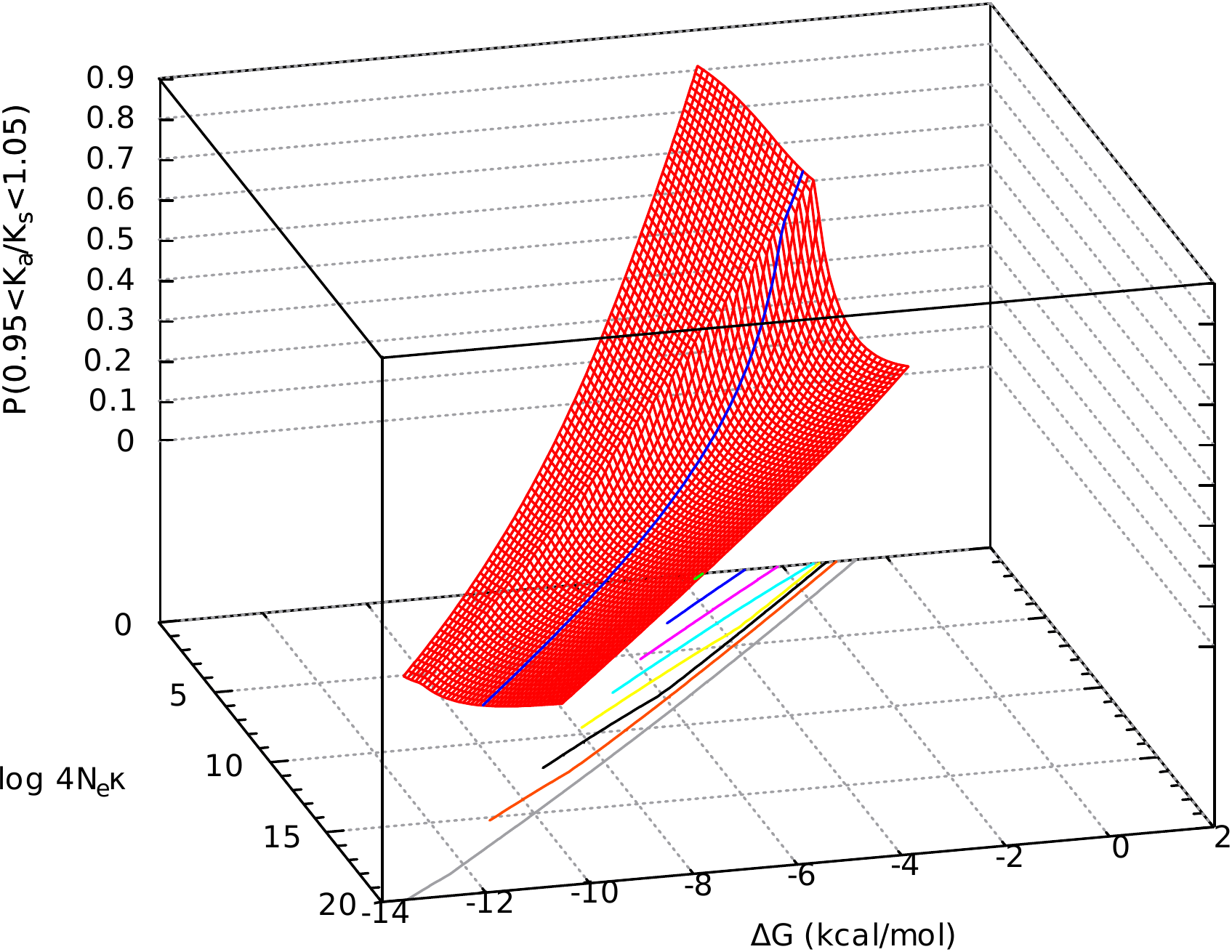}
\includegraphics*[width=90mm,angle=0]{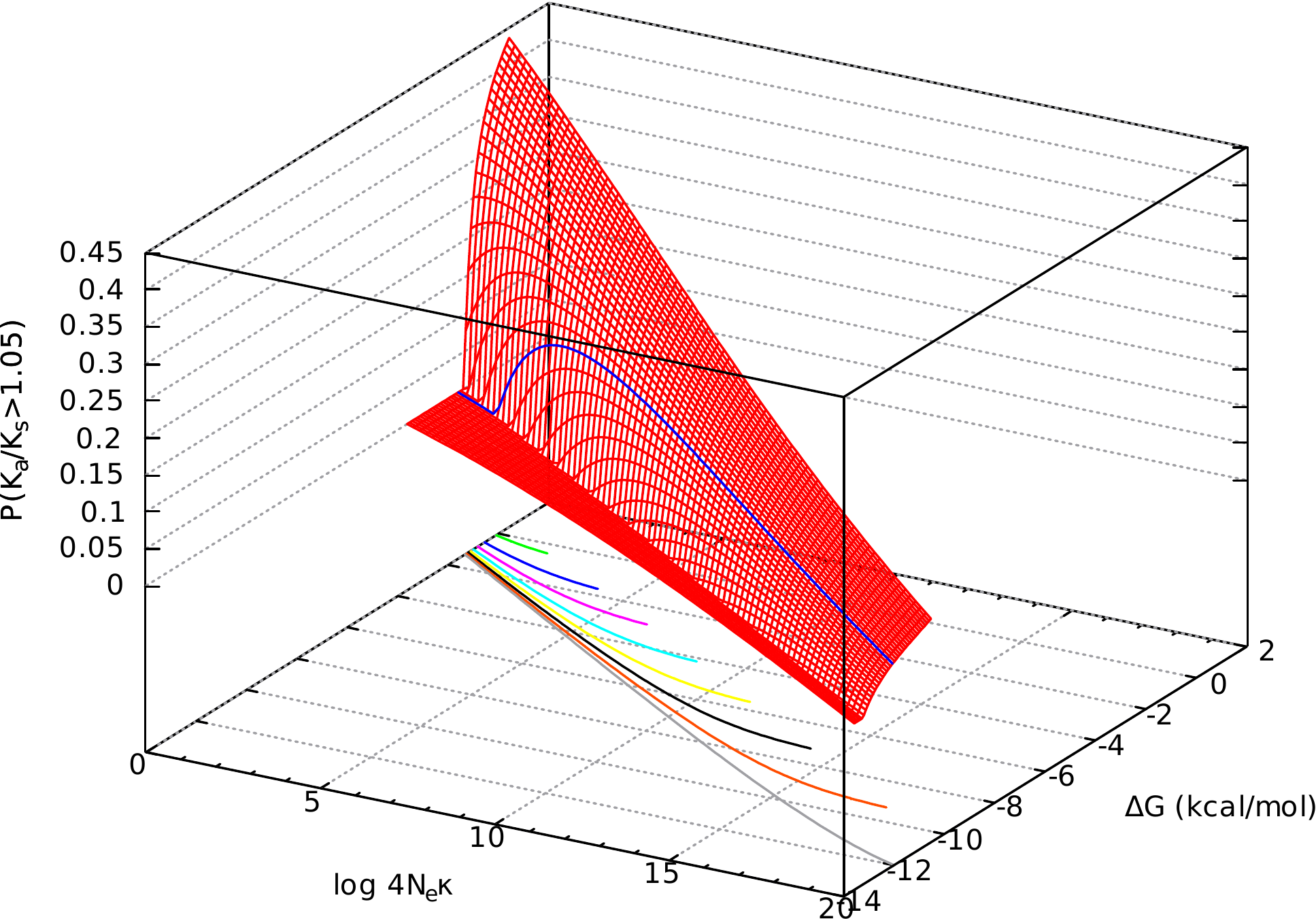}
}
\centerline{
\includegraphics*[width=85mm,angle=0]{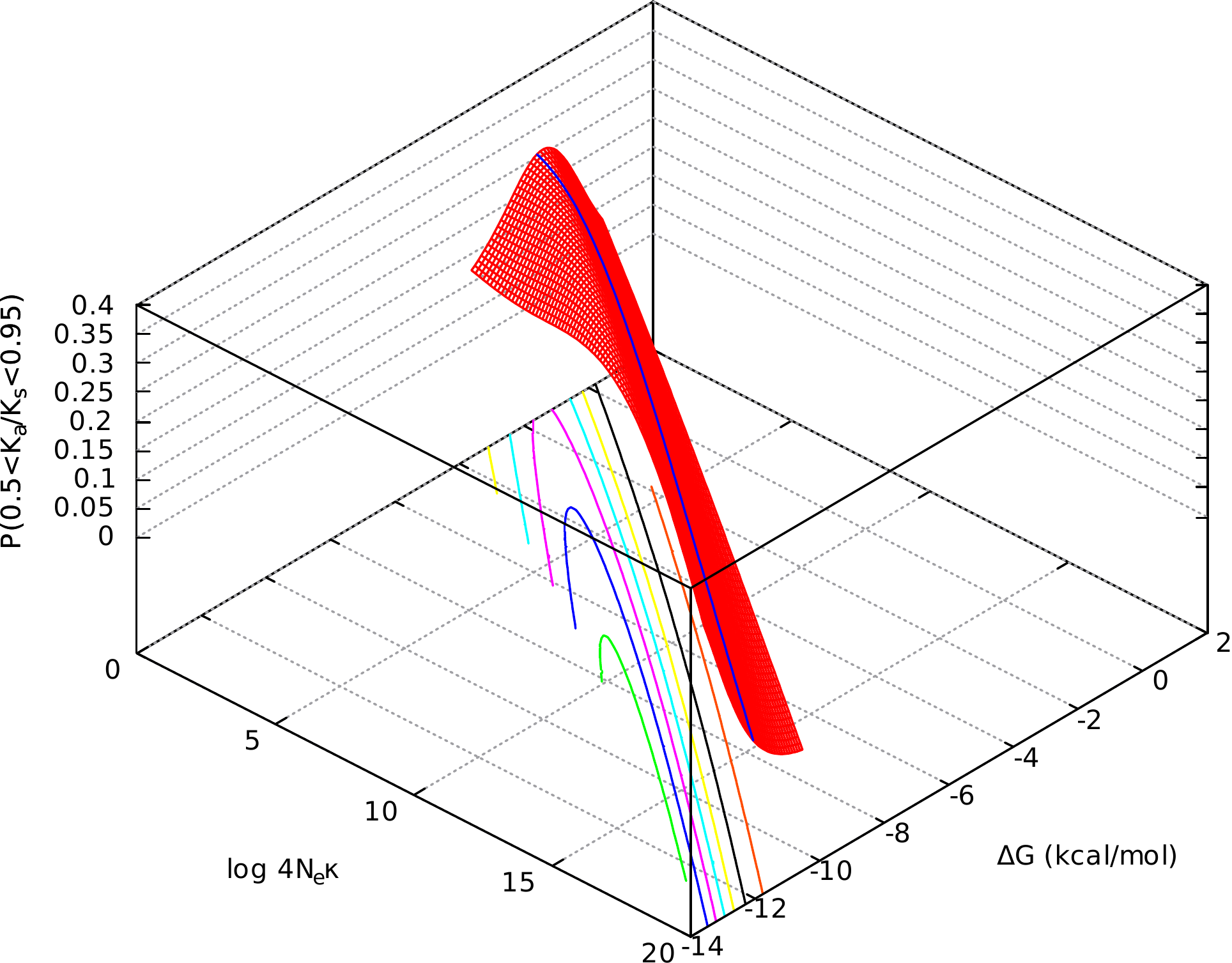}
\includegraphics*[width=90mm,angle=0]{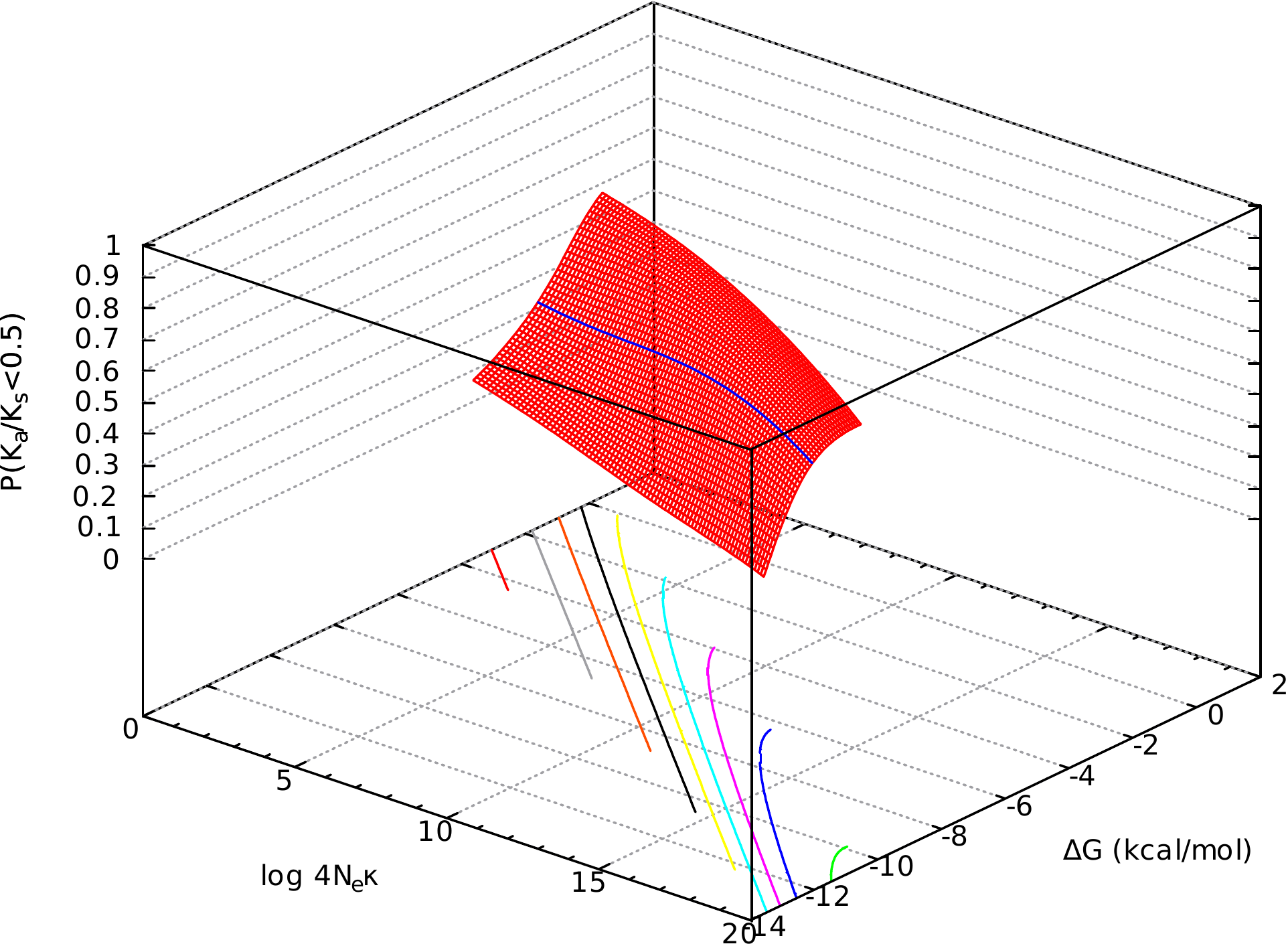}
}
} 
\vspace*{1em}
\caption{
\label{sfig: prob_of_each_selection_category_in_all_mutants;_dependence_on_kappa_and_dG}
\FigureLegends{
\BF{Dependence of the probability of each selection category in all mutants on $4N_e\kappa$ and $\Delta G$.}
A blue line on the surface grid shows $\Delta G=\Delta G_e$, which is the equilibrium value of $\Delta G$
in protein evolution.
The range of $\Delta G$ shown in the figures is
$| \Delta G - \Delta G_e | < 2 \cdot \Delta\Delta G^{\script{sd}}_{\script{fixed}}$, 
where $\Delta\Delta G^{\script{sd}}_{\script{fixed}}$ is the standard deviation of $\Delta\Delta G$ over fixed mutants at $ \Delta G = \Delta G_e$.
Arbitrarily, the value of $K_a/K_s$ is categorized into four classes;
negative, slightly negative, nearly neutral, and positive selection categories in which
$K_a/K_s$ is within the ranges of
$K_a/K_s \leq 0.5$, $0.5< K_a/K_s \leq 0.95$, $0.95< K_a/K_s \leq 1.05$, and $ 1.05 < K_a/K_s$,
respectively.
$\theta = 0.53$
is employed.
The kcal/mol unit is used for $\Delta G$.
} 
}
\end{figure*}

\clearpage
\FigureInLegends{\newpage}

\begin{figure*}[ht]
\FigureInLegends{
\centerline{
\includegraphics*[width=87mm,angle=0]{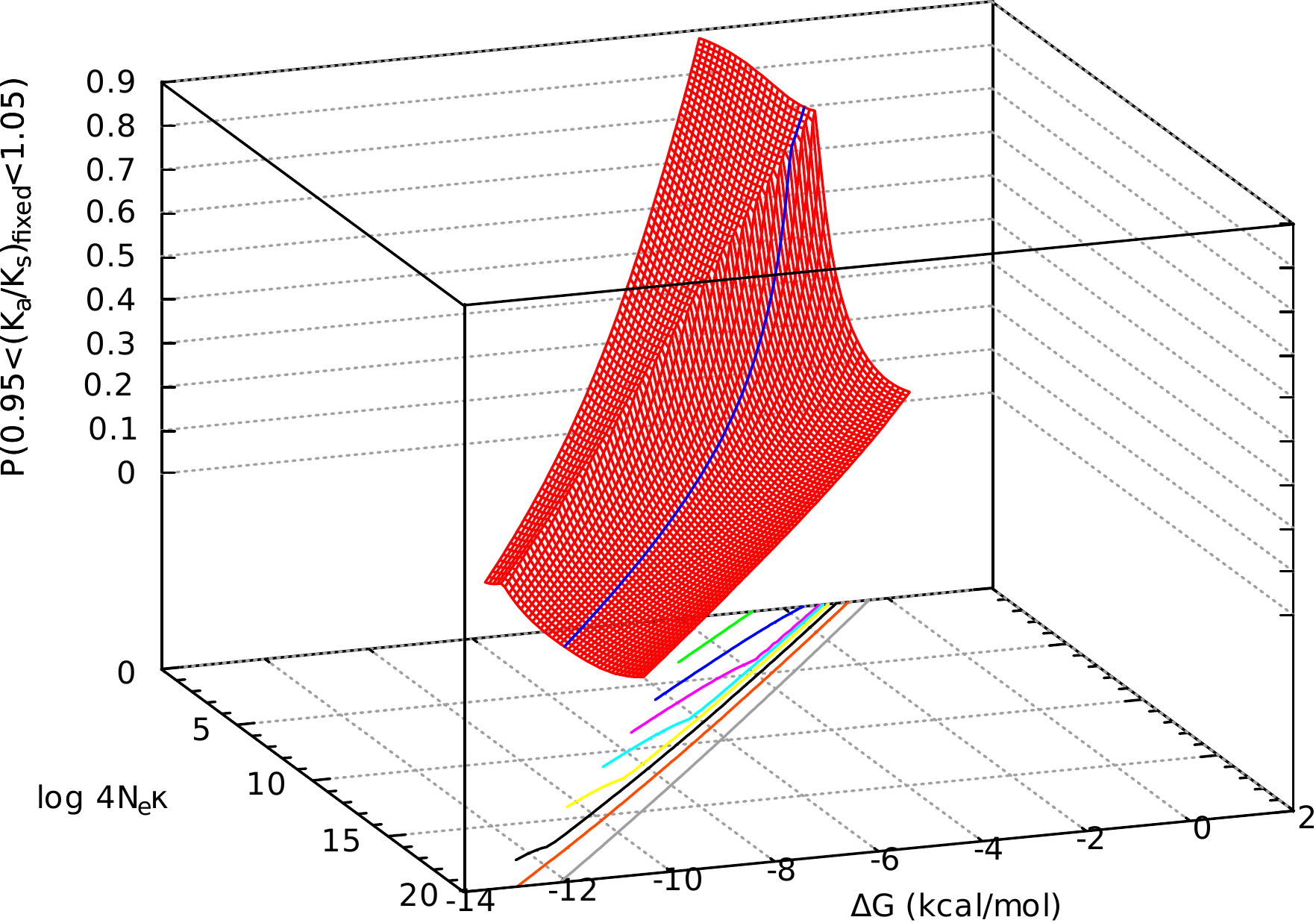}
\includegraphics*[width=90mm,angle=0]{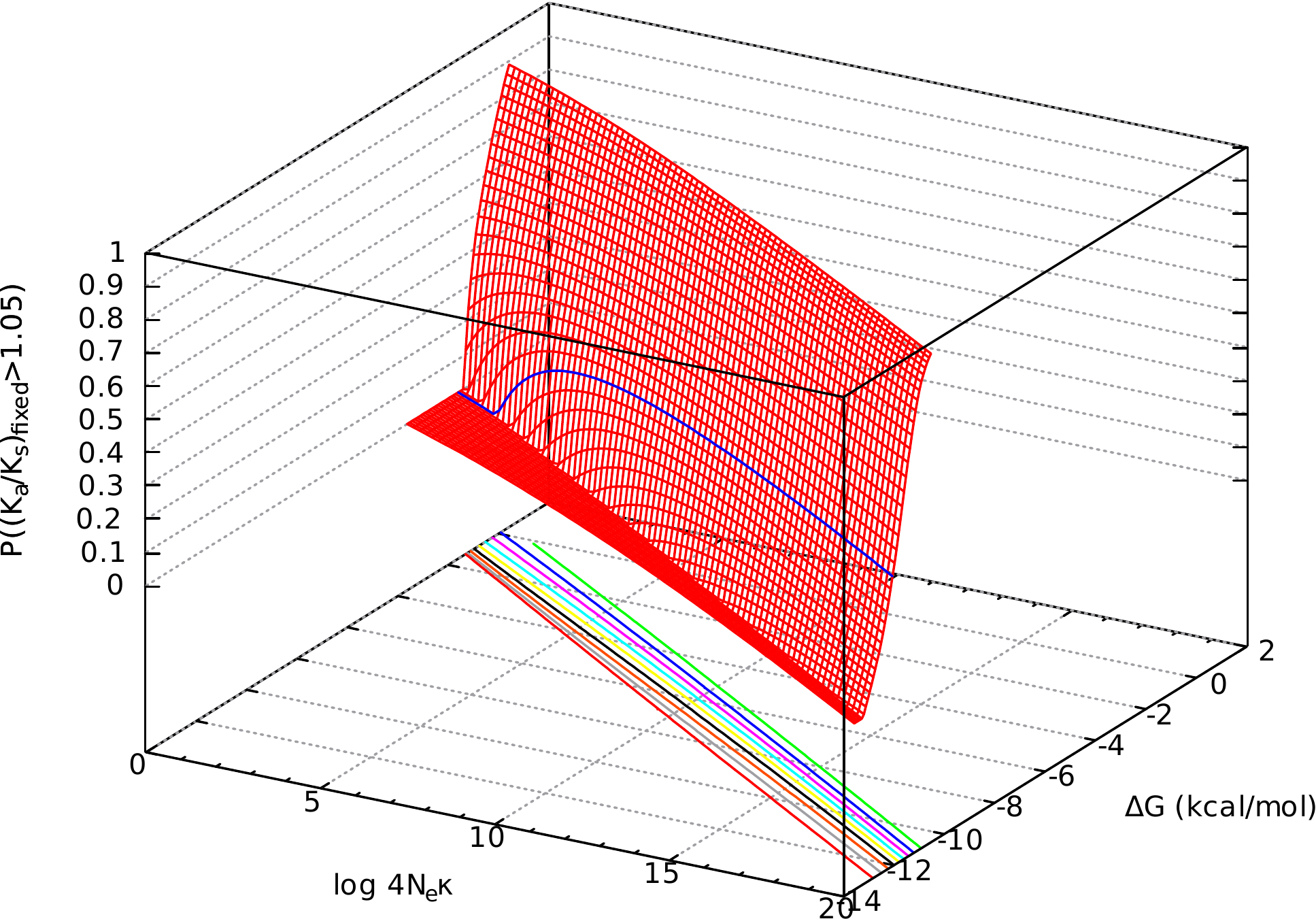}
}
\centerline{
\includegraphics*[width=90mm,angle=0]{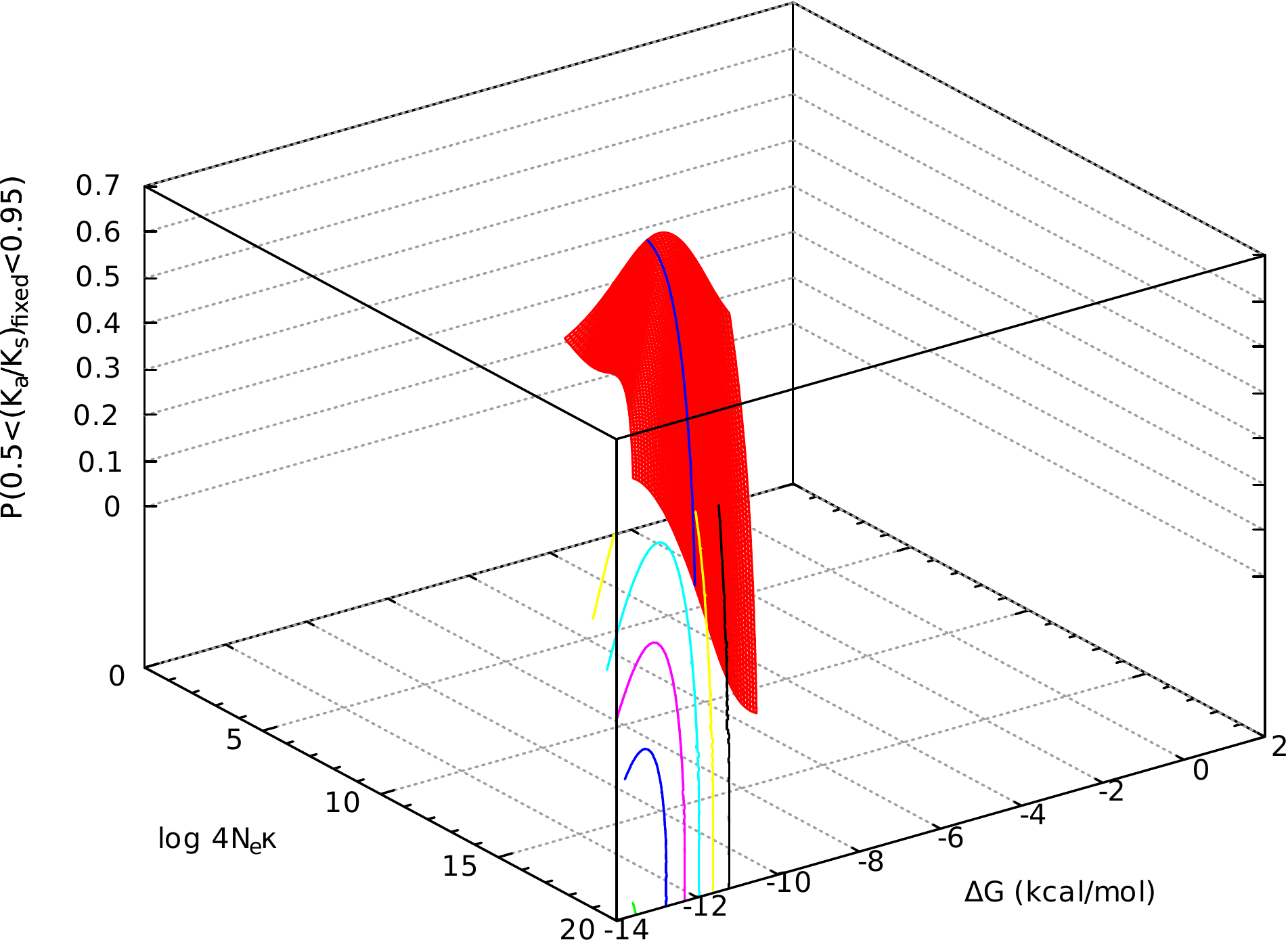}
\includegraphics*[width=90mm,angle=0]{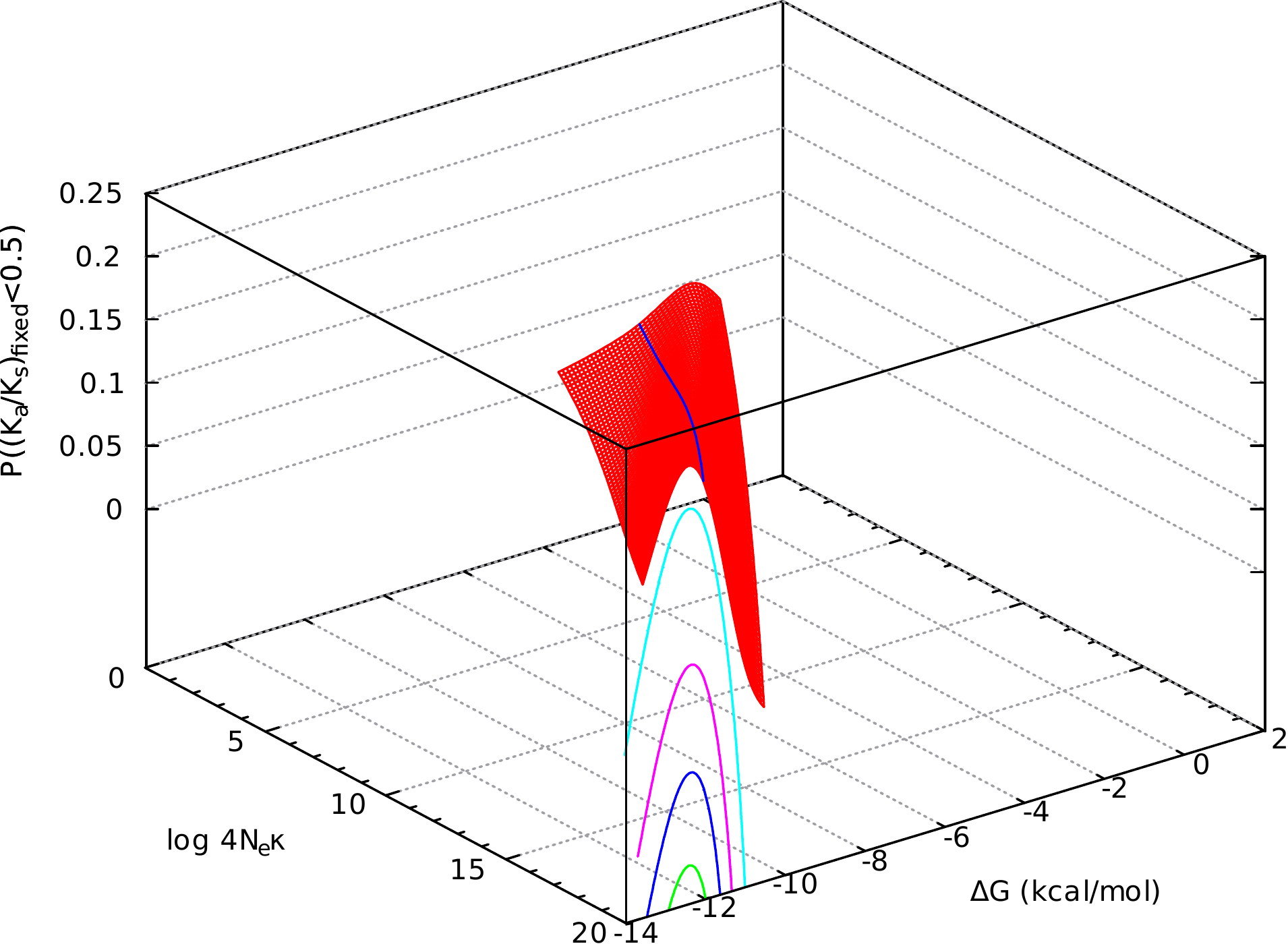}
}
} 
\vspace*{1em}
\caption{
\label{sfig: prob_of_each_selection_category_in_fixed_mutants;_dependence_on_kappa_and_dG}
\FigureLegends{
\BF{Dependence of the probability of each selection category in fixed mutants on $4N_e\kappa$ and $\Delta G$.}
A blue line on the surface grid shows $\Delta G=\Delta G_e$, which is the equilibrium value of $\Delta G$
in protein evolution.
The range of $\Delta G$ shown in the figures is
$| \Delta G - \Delta G_e | < 2 \cdot \Delta\Delta G^{\script{sd}}_{\script{fixed}}$,
where $\Delta\Delta G^{\script{sd}}_{\script{fixed}}$ is the standard deviation of $\Delta\Delta G$ over fixed mutants at $ \Delta G = \Delta G_e$.
Arbitrarily, the value of $K_a/K_s$ is categorized into four classes;
negative, slightly negative, nearly neutral, and positive selection categories in which
$K_a/K_s$ is within the ranges of
$K_a/K_s \leq 0.5$, $0.5< K_a/K_s \leq 0.95$, $0.95< K_a/K_s \leq 1.05$, and $ 1.05 < K_a/K_s$,
respectively.
$\theta = 0.53$
is employed.
The kcal/mol unit is used for $\Delta G$.
} 
}
\end{figure*}

\FigureInLegends{\newpage}

\begin{figure*}[ht]
\FigureInLegends{
\centerline{
\includegraphics*[width=80mm,angle=0]{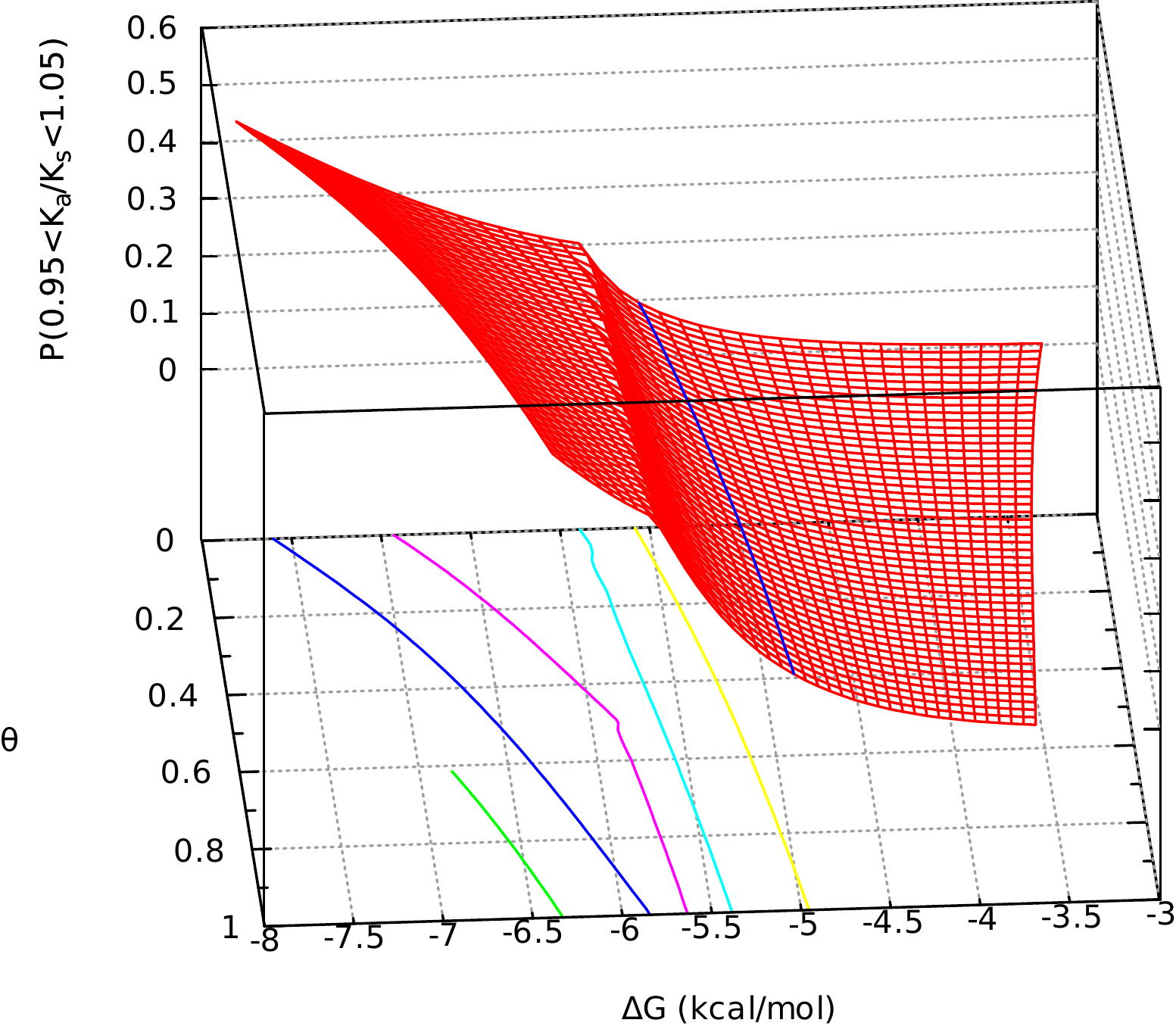}
\includegraphics*[width=90mm,angle=0]{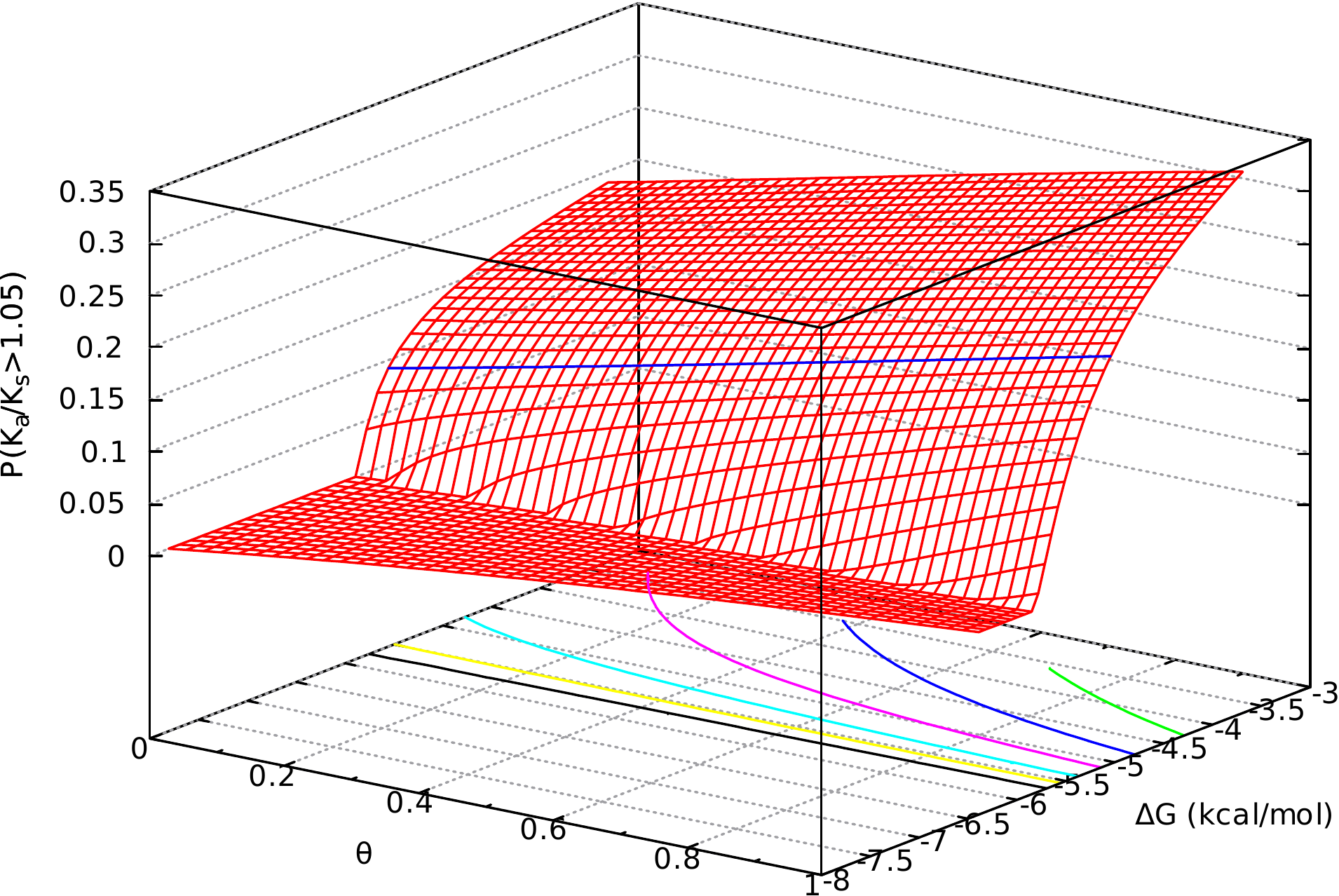}
}
\centerline{
\includegraphics*[width=83mm,angle=0]{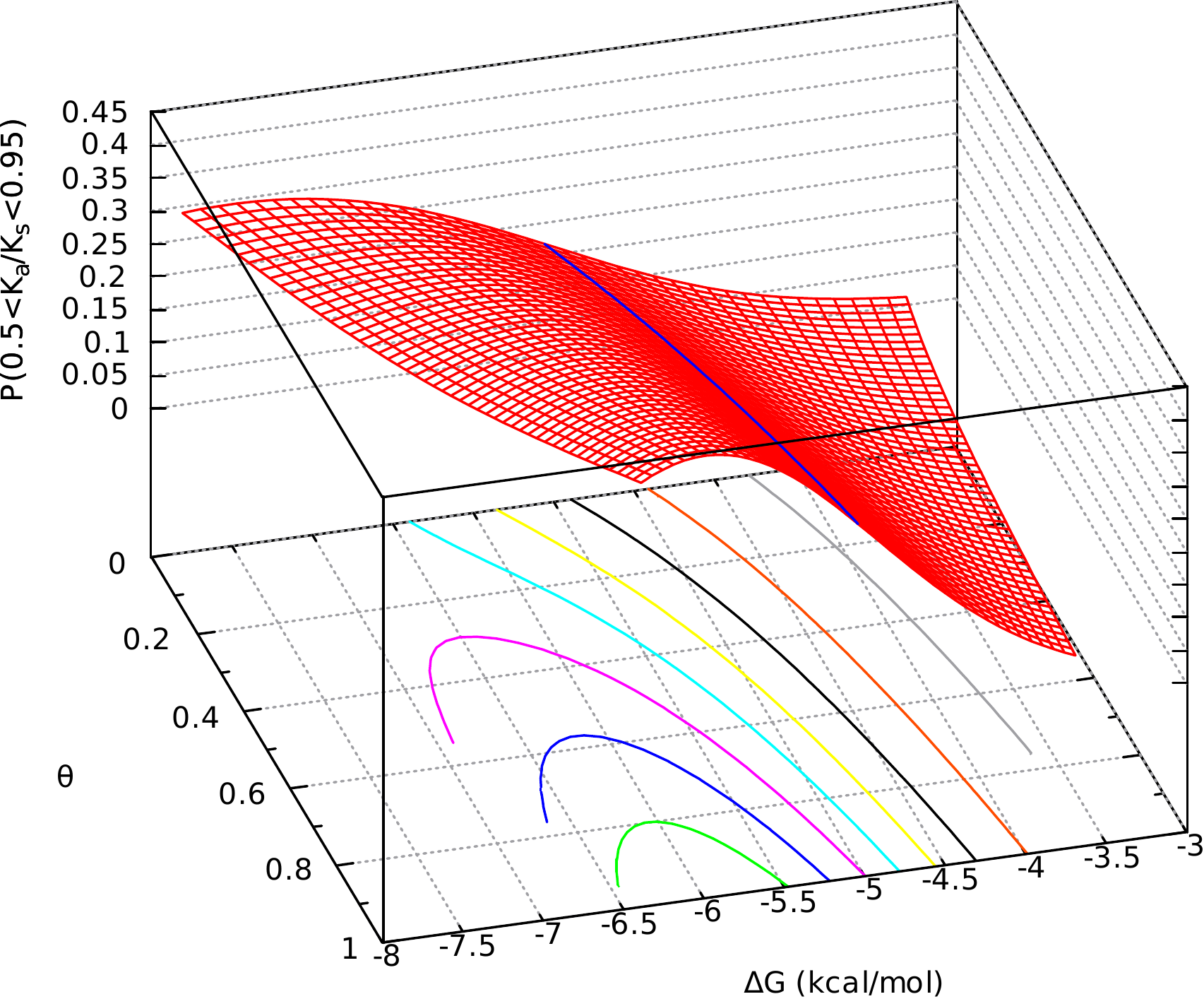}
\includegraphics*[width=90mm,angle=0]{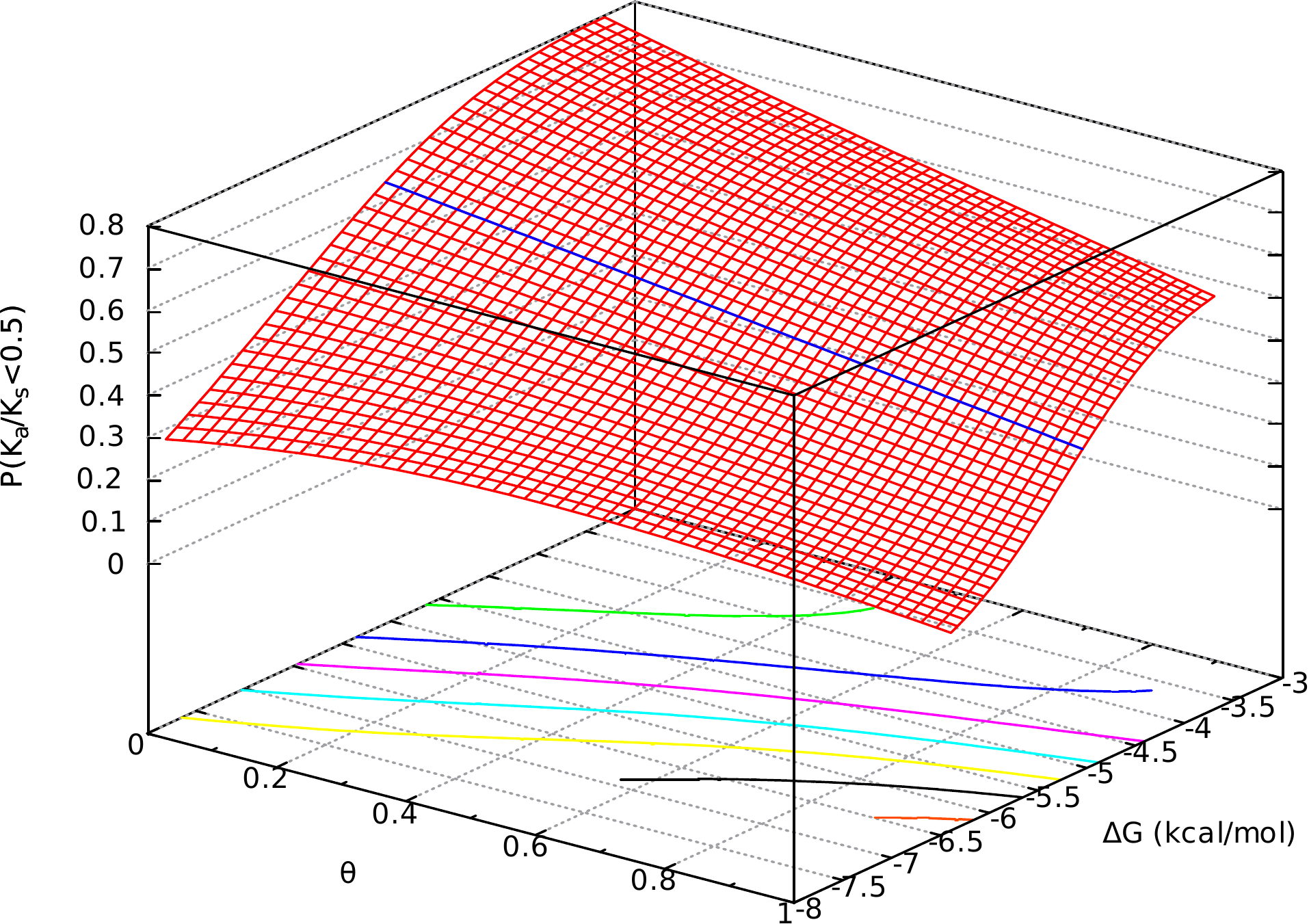}
}
} 
\vspace*{1em}
\caption{
\label{sfig: prob_of_each_selection_category_in_all_mutants;_dependence_on_theta_and_dG}
\FigureLegends{
\BF{Dependence of the probability of each selection category in all mutants on $\theta$ and $\Delta G$.}
A blue line on the surface grid shows $\Delta G=\Delta G_e$, which is the equilibrium value of $\Delta G$
in protein evolution.
The range of $\Delta G$ shown in the figures is
$| \Delta G - \Delta G_e | < 2 \cdot \Delta\Delta G^{\script{sd}}_{\script{fixed}}$,
where $\Delta\Delta G^{\script{sd}}_{\script{fixed}}$ is the standard deviation of $\Delta\Delta G$ over fixed mutants at $ \Delta G = \Delta G_e$.
Arbitrarily, the value of $K_a/K_s$ is categorized into four classes;
negative, slightly negative, nearly neutral, and positive selection categories in which
$K_a/K_s$ is within the ranges of
$K_a/K_s \leq 0.5$, $0.5< K_a/K_s \leq 0.95$, $0.95< K_a/K_s \leq 1.05$, and $ 1.05 < K_a/K_s$,
respectively.
$\log 4N_e\kappa = 7.550$ 
is employed.
The kcal/mol unit is used for $\Delta G$.
} 
}
\end{figure*}

\FigureInLegends{\newpage}

\begin{figure*}[ht]
\FigureInLegends{
\centerline{
\includegraphics*[width=83mm,angle=0]{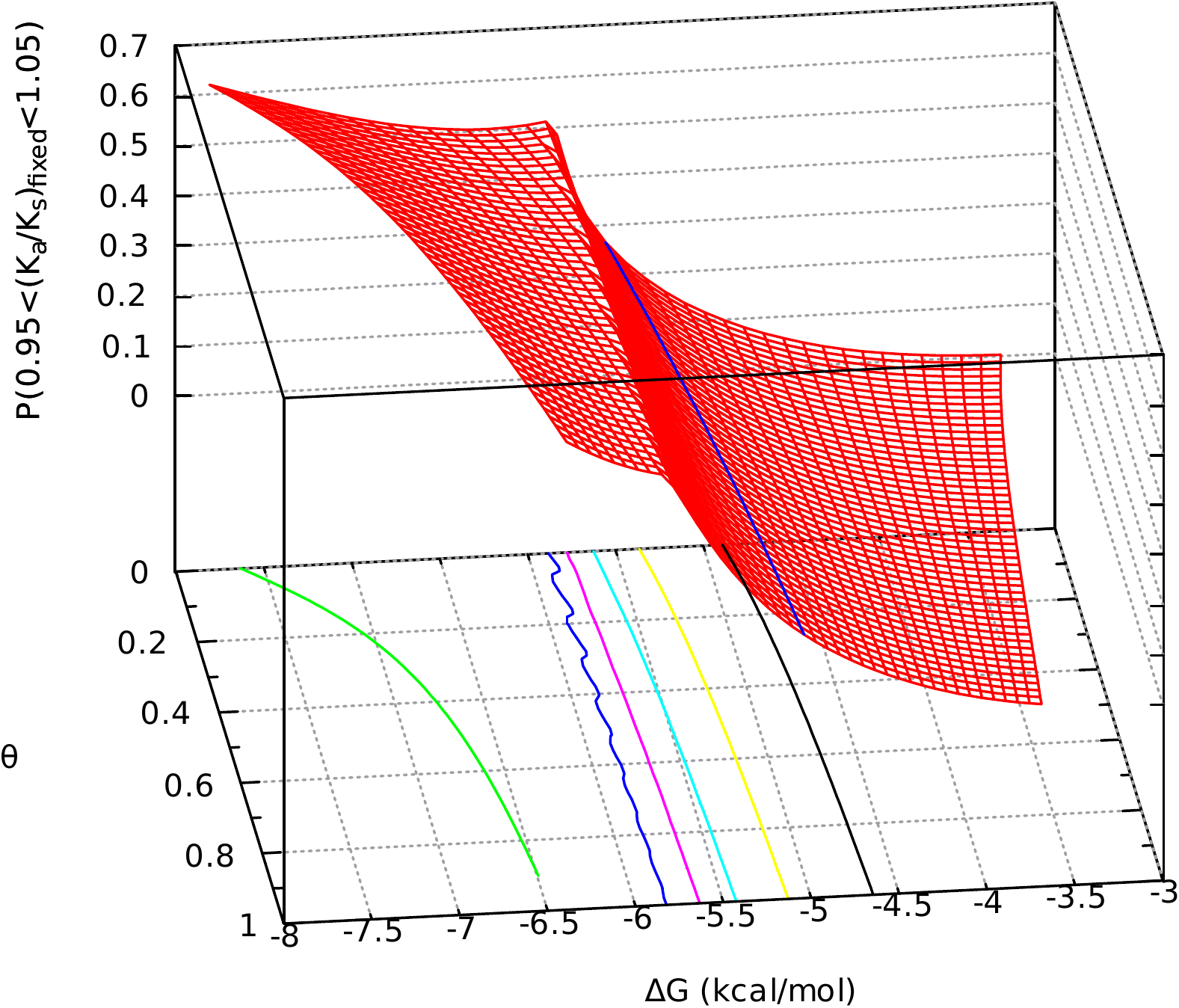}
\includegraphics*[width=90mm,angle=0]{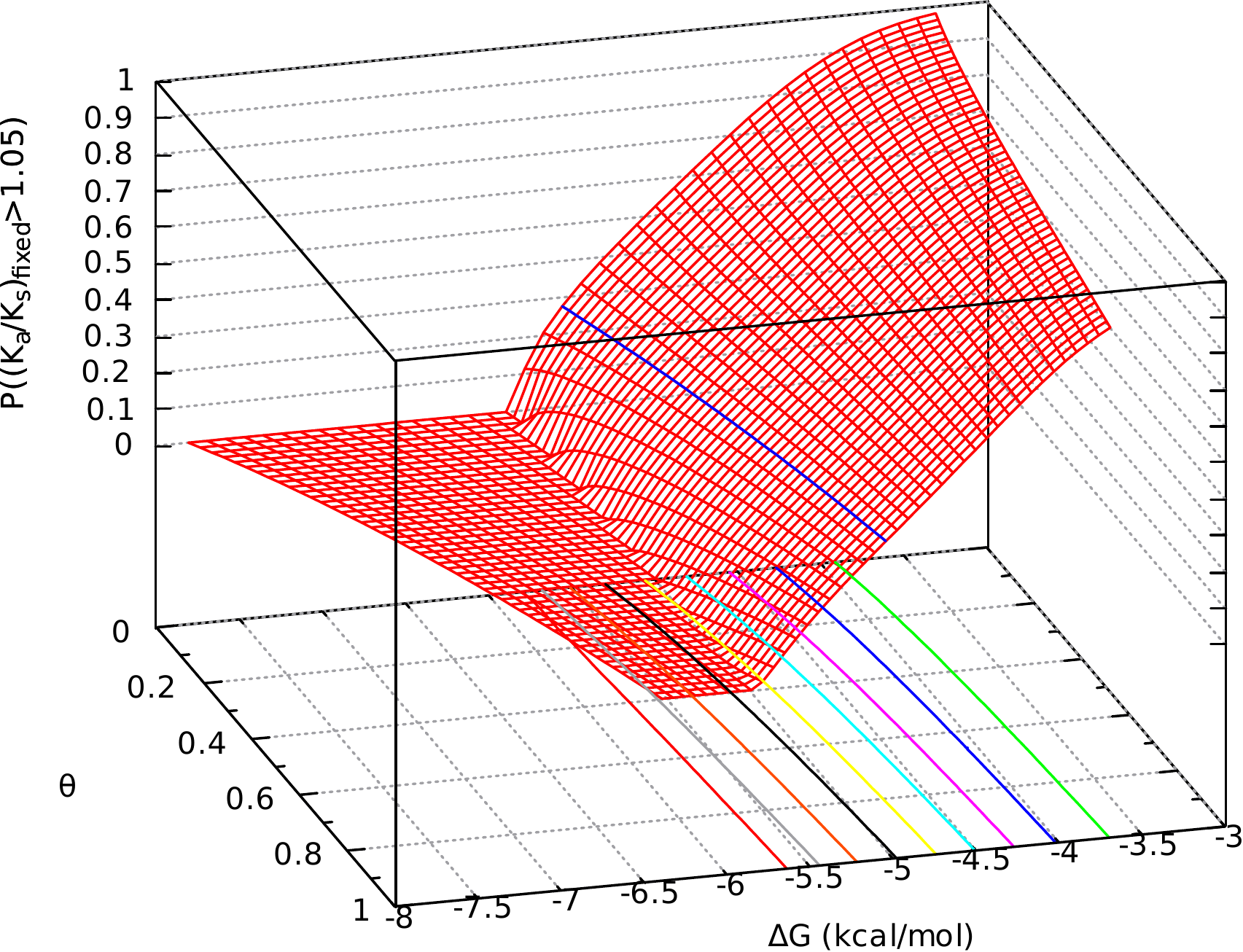}
}
\centerline{
\includegraphics*[width=89mm,angle=0]{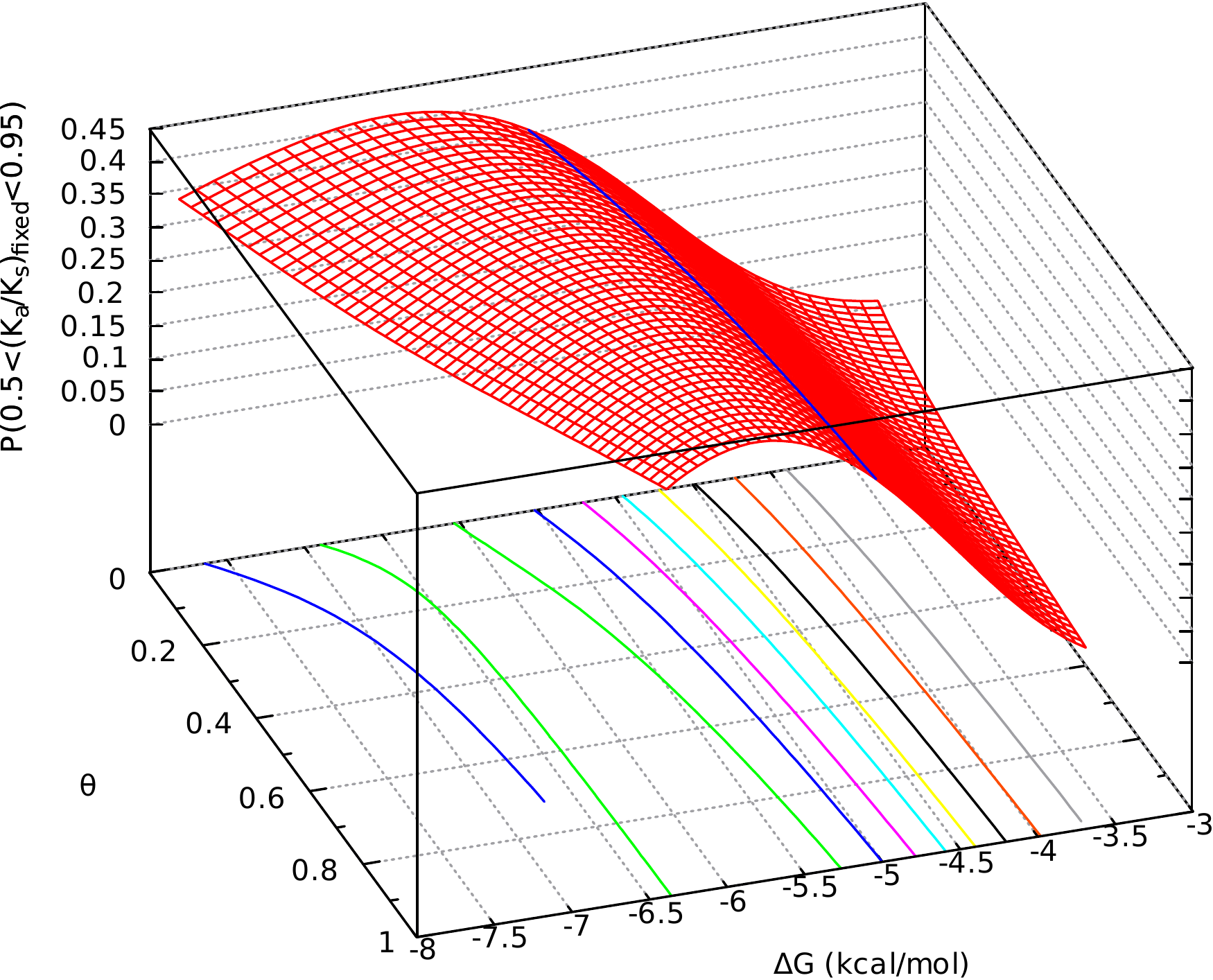}
\includegraphics*[width=90mm,angle=0]{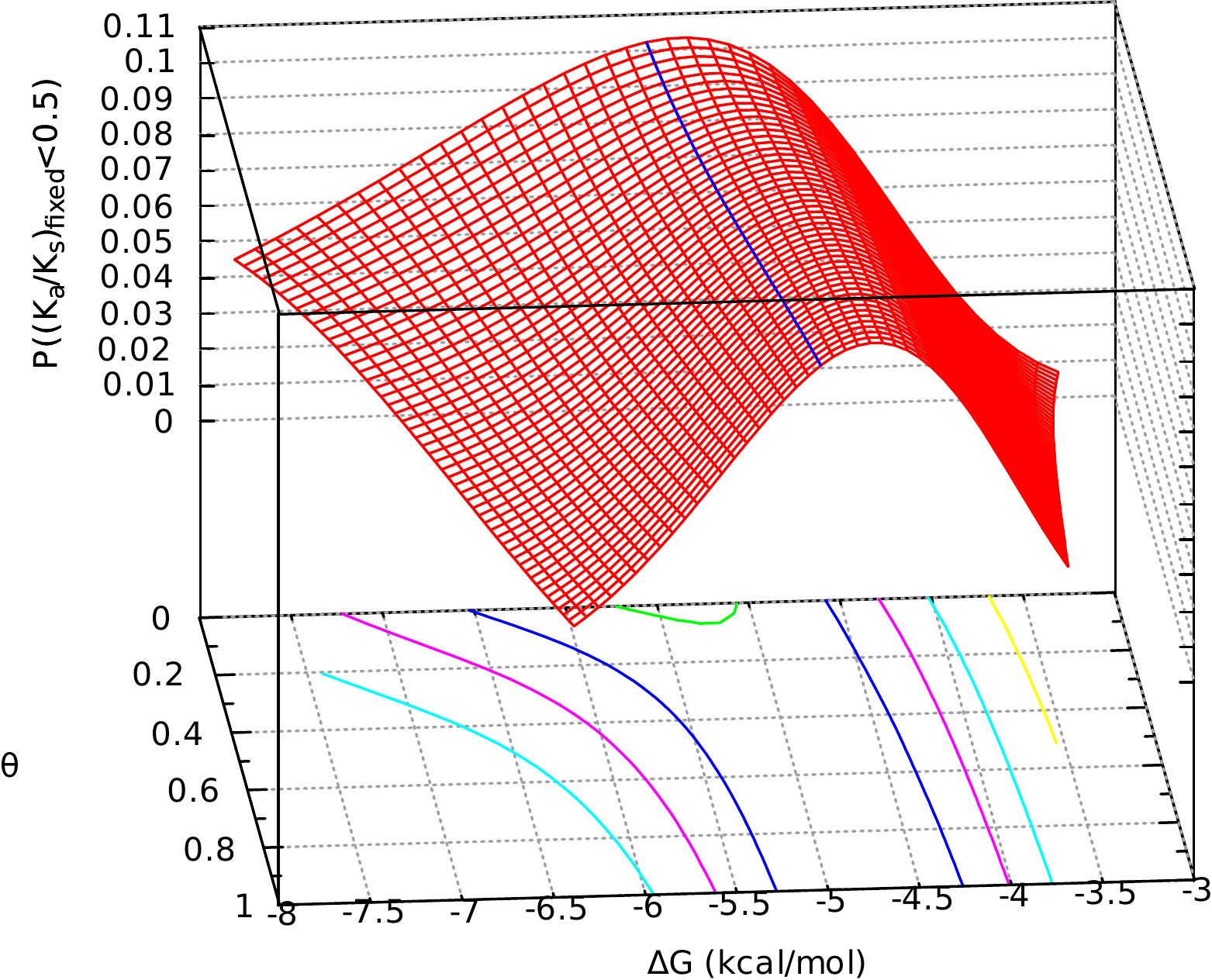}
}
} 
\vspace*{1em}
\caption{
\label{sfig: prob_of_each_selection_category_in_fixed_mutants;_dependence_on_theta_and_dG}
\FigureLegends{
\BF{Dependence of the probability of each selection category in fixed mutants on $\theta$ and $\Delta G$.}
The blue line on the surface grid shows $\Delta G=\Delta G_e$, which is the equilibrium value of $\Delta G$
in protein evolution.
The range of $\Delta G$ shown in the figures is
$| \Delta G - \Delta G_e | < 2 \cdot \Delta\Delta G^{\script{sd}}_{\script{fixed}}$,
where $\Delta\Delta G^{\script{sd}}_{\script{fixed}}$ is the standard deviation of $\Delta\Delta G$ over fixed mutants at $ \Delta G = \Delta G_e$.
Arbitrarily, the value of $K_a/K_s$ is categorized into four classes;
negative, slightly negative, nearly neutral, and positive selection categories in which
$K_a/K_s$ is within the ranges of
$K_a/K_s \leq 0.5$, $0.5< K_a/K_s \leq 0.95$, $0.95< K_a/K_s \leq 1.05$, and $ 1.05 < K_a/K_s$,
respectively.
$\log 4N_e\kappa = 7.550$
is employed.
The kcal/mol unit is used for $\Delta G$.
} 
}
\end{figure*}

\FigureInLegends{\clearpage\newpage}

\begin{figure*}[ht]
\FigureInLegends{
\centerline{
\includegraphics*[width=90mm,angle=0]{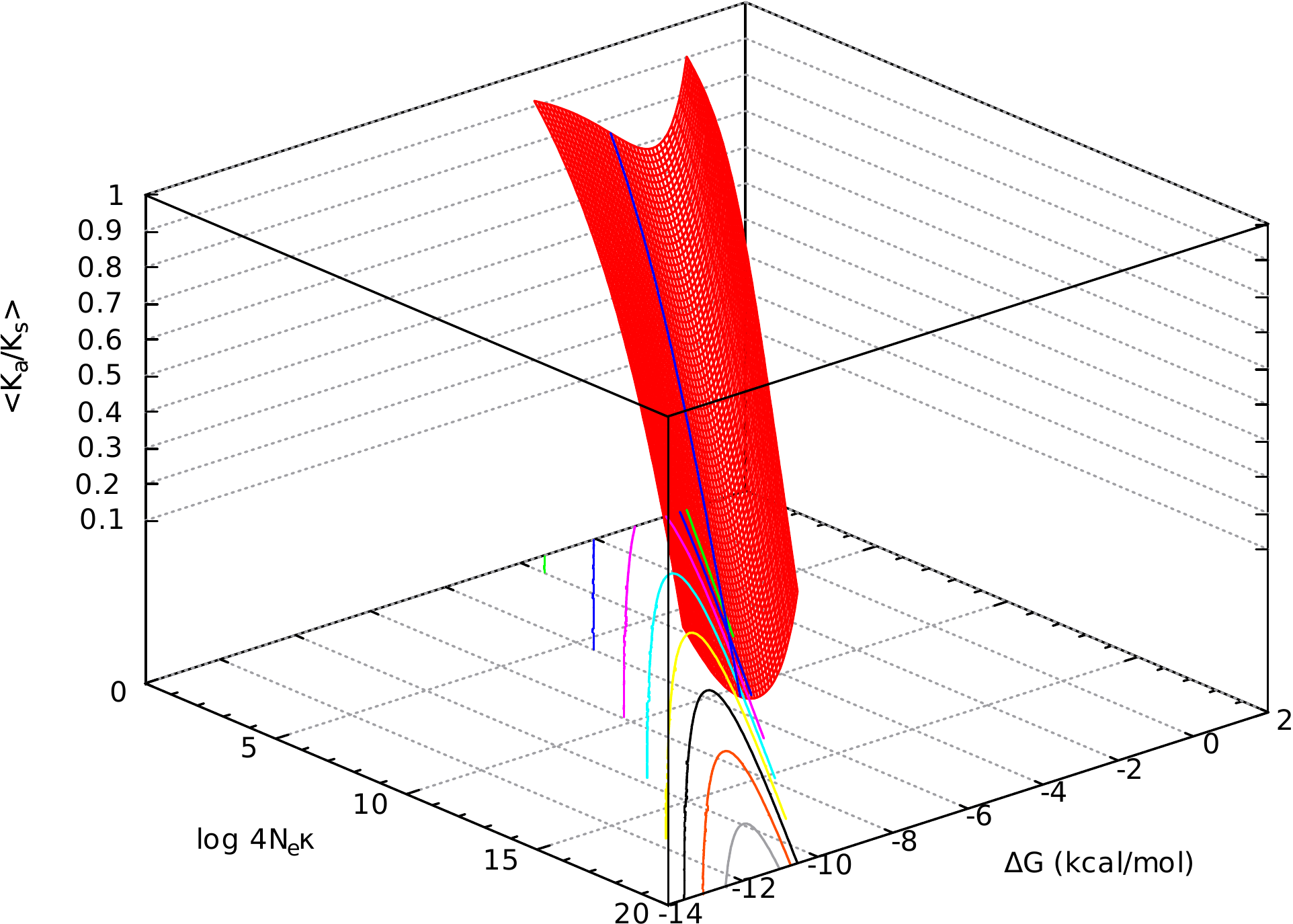}
\includegraphics*[width=85mm,angle=0]{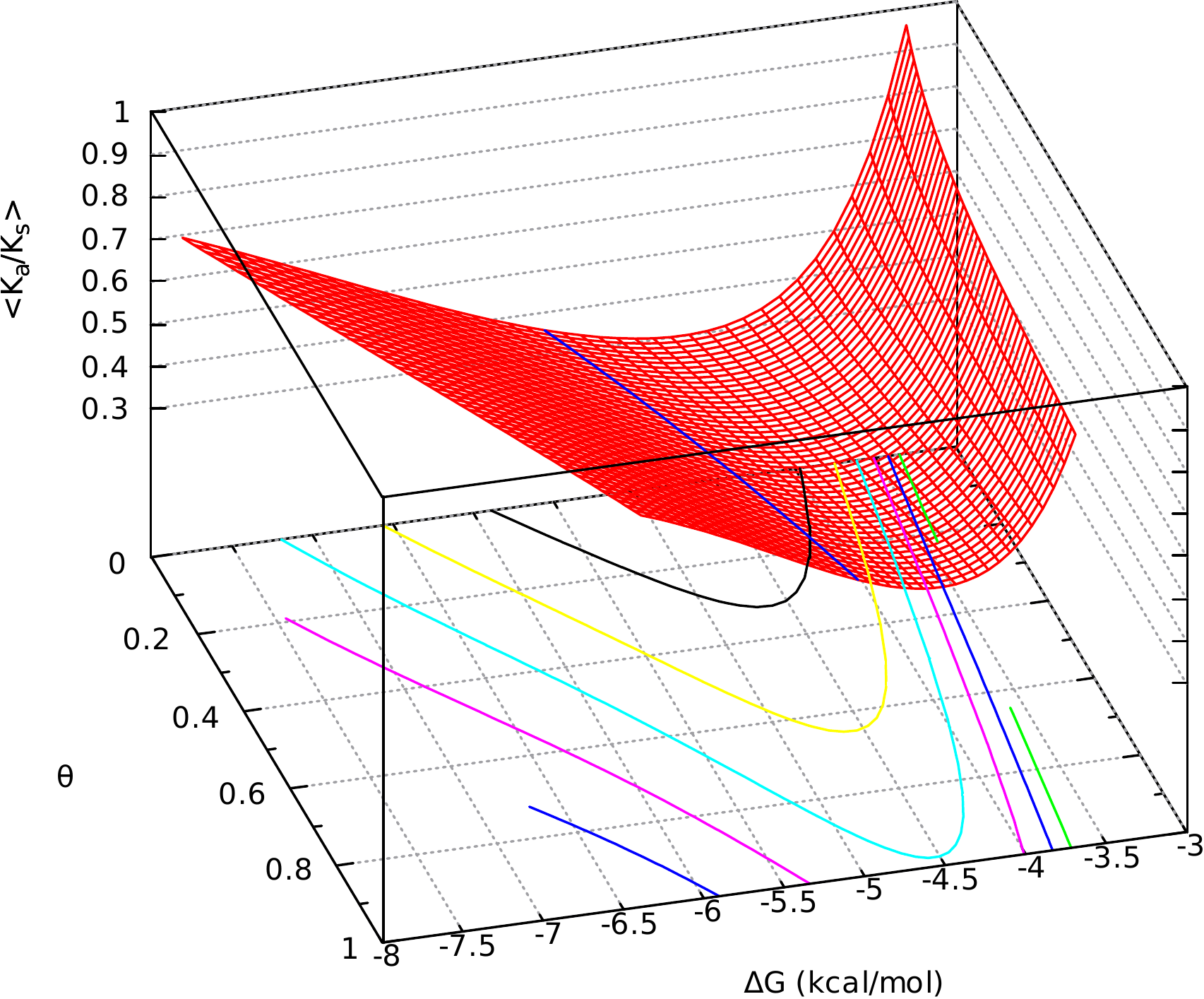}
}
\centerline{
\includegraphics*[width=88mm,angle=0]{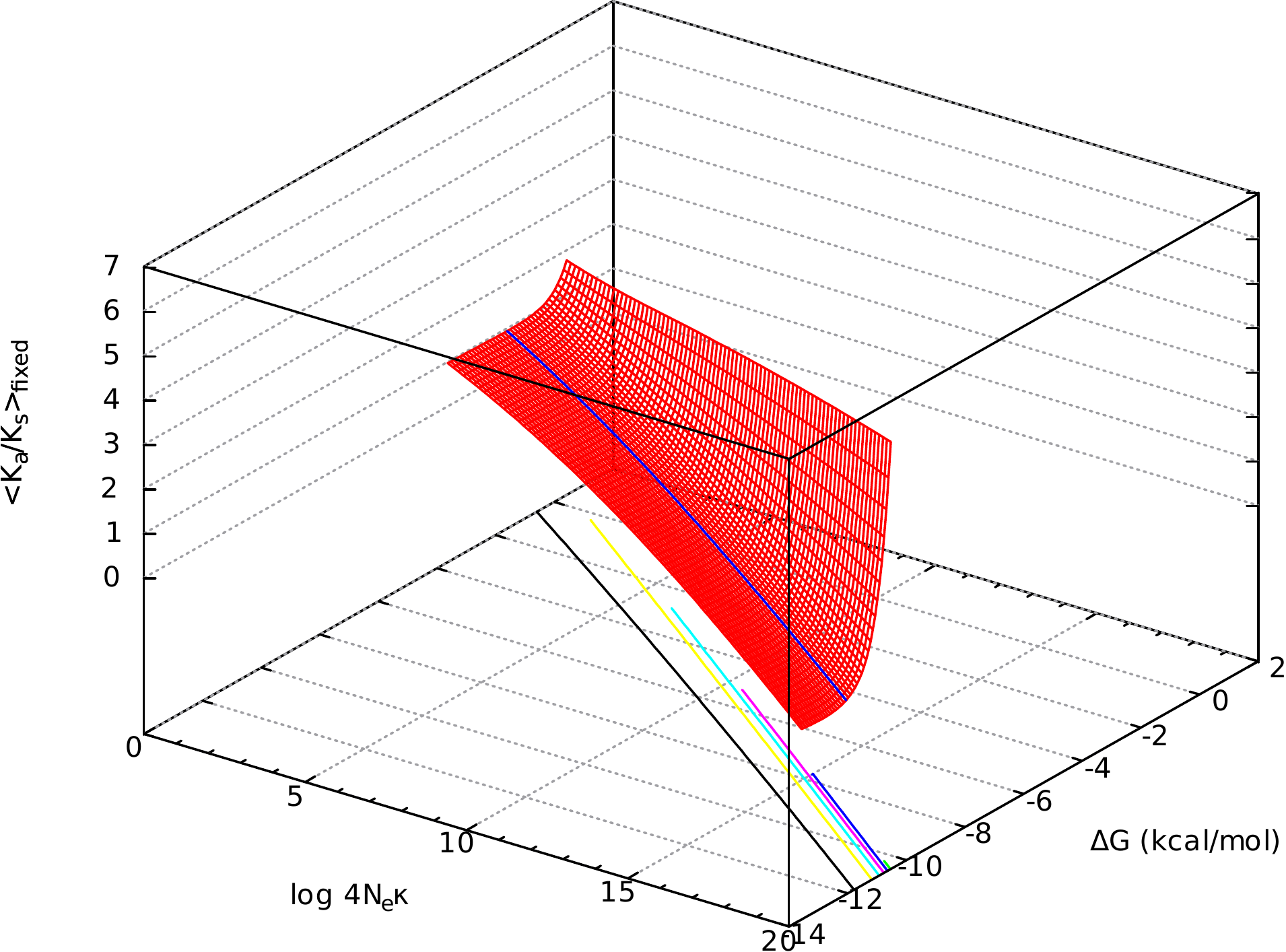}
\includegraphics*[width=90mm,angle=0]{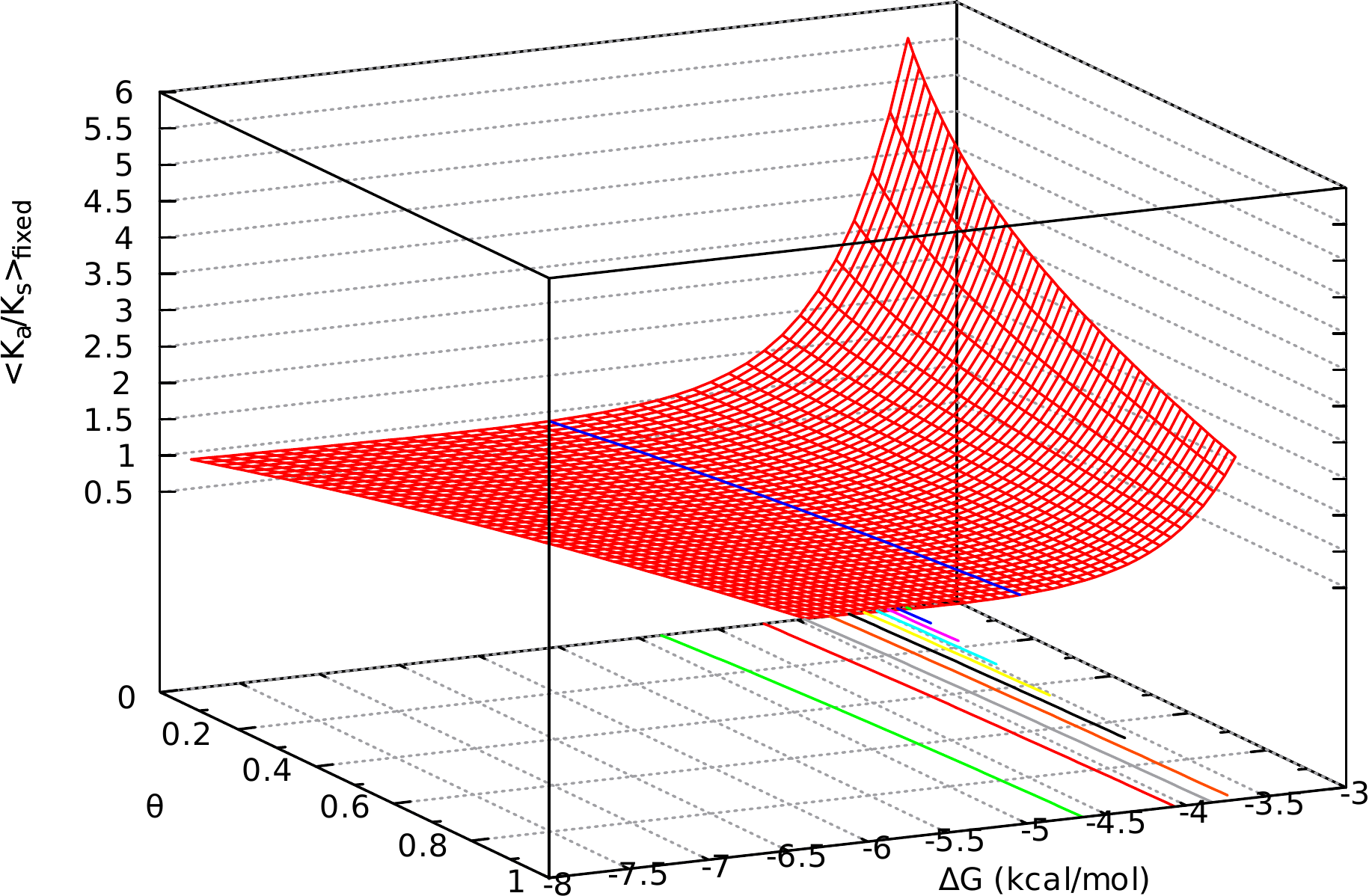}
}
} 
\vspace*{1em}
\caption{
\label{sfig: dependence_of_ave_Ka_over_Ks_on_dG}
\label{sfig: dependence_of_ave_Ka_over_Ks_fixed_on_dG}
\FigureLegends{
\BF{Dependence of the average of $K_a / K_s$ over all mutants or over fixed mutants only 
on protein stability, $\Delta G$, of the wild type.}
A blue line on the surface grid shows $\Delta G=\Delta G_e$, which is the equilibrium value of $\Delta G$
in protein evolution.
The range of $\Delta G$ shown in the figures is
$| \Delta G - \Delta G_e | < 2 \cdot \Delta\Delta G^{\script{sd}}_{\script{fixed}}$,
where $\Delta\Delta G^{\script{sd}}_{\script{fixed}}$ is the standard deviation of $\Delta\Delta G$ over fixed mutants at $\Delta G = \Delta G_e$.
Unless specified, 
$\log 4N_e\kappa = 7.550$ and $\theta = 0.53$
are employed.
The kcal/mol unit is used for $\Delta G$.
} 
}
\end{figure*}

\FigureInLegends{\newpage}

\begin{figure*}[ht]
\FigureInLegends{
\centerline{
\includegraphics*[width=90mm,angle=0]{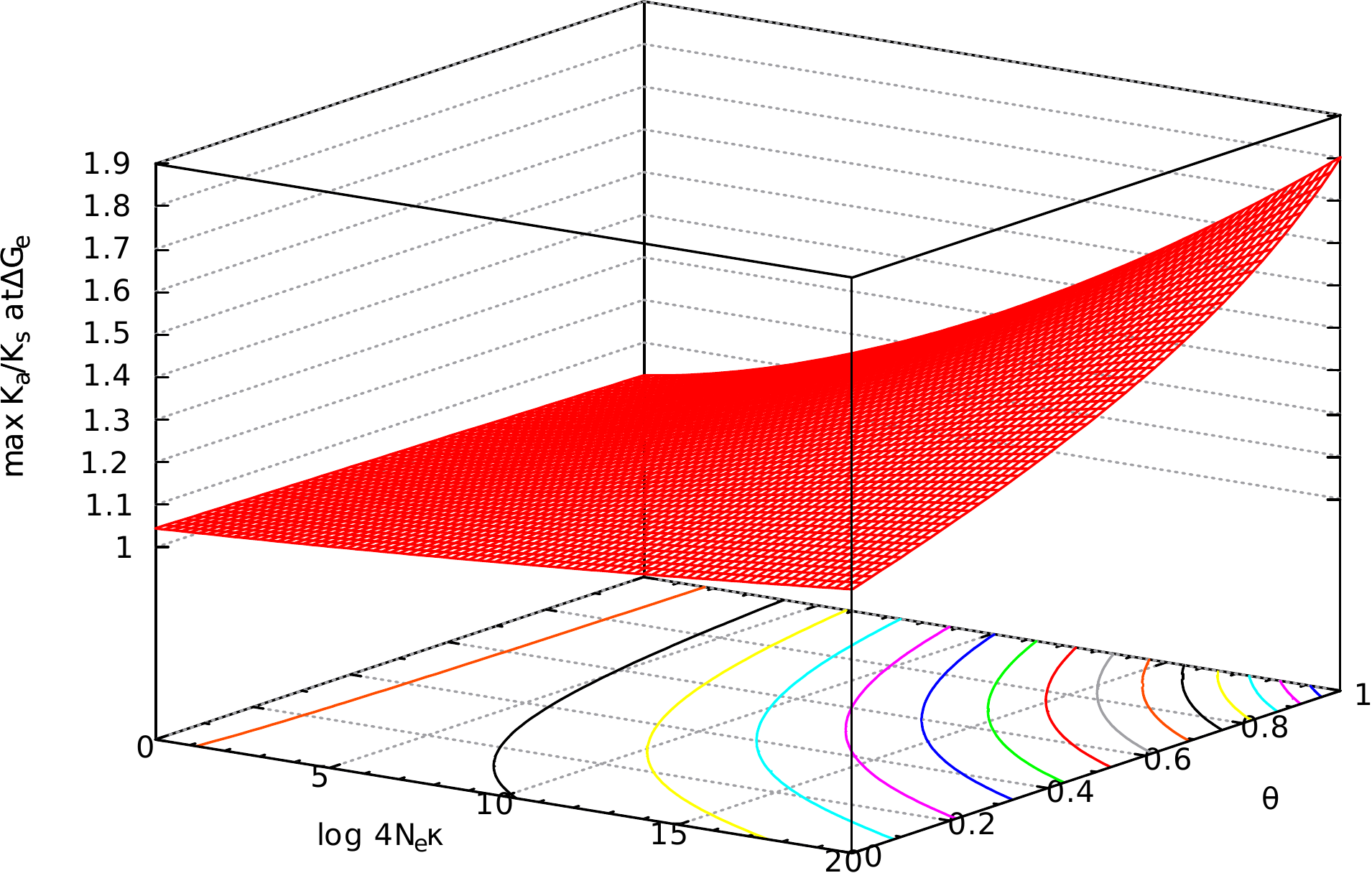}
\includegraphics*[width=90mm,angle=0]{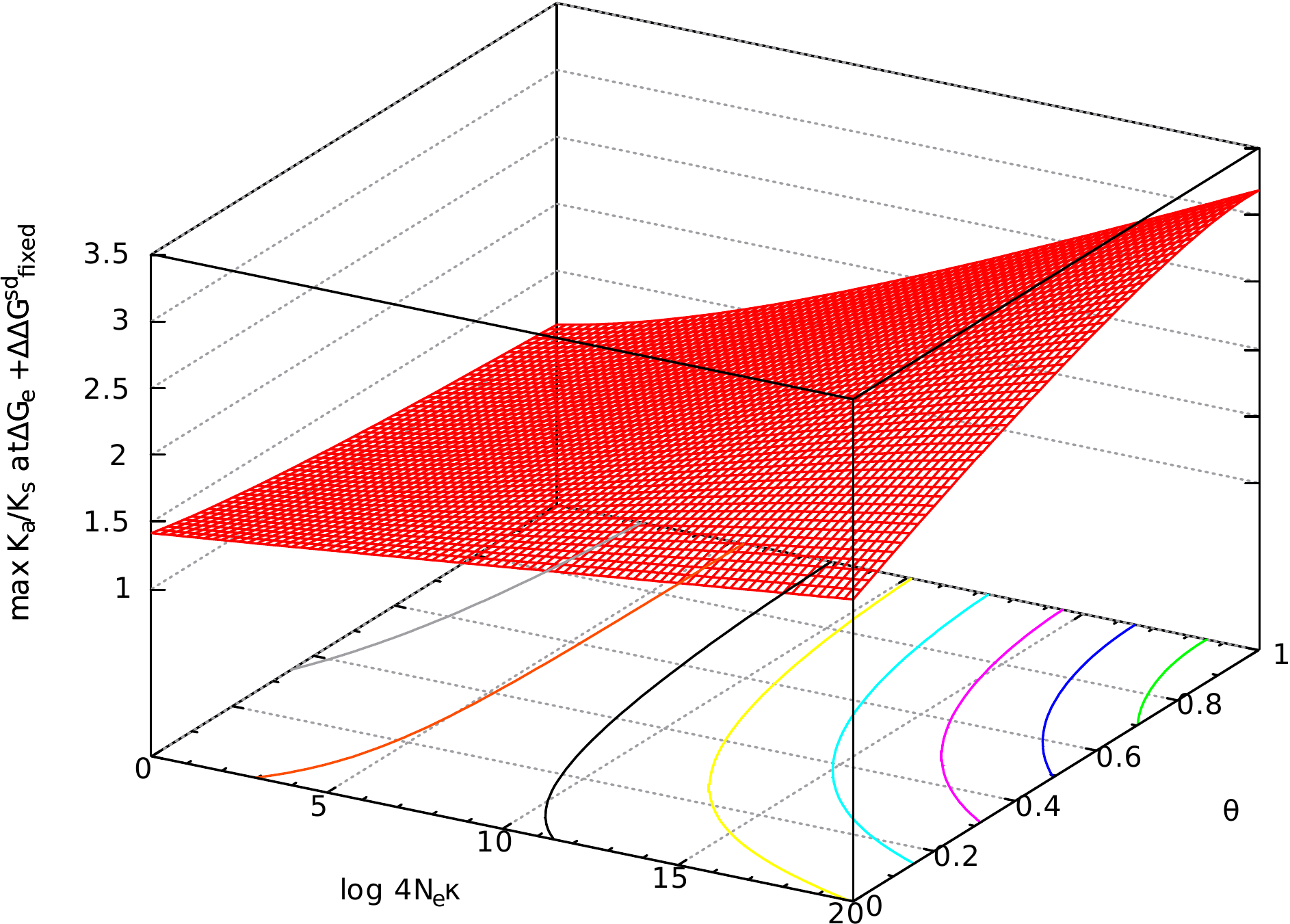}
}
} 
\vspace*{1em}
\caption{
\label{sfig: maximum_Ka_over_Ks_on_4Nekappa_and_theta}
\FigureLegends{
\BF{Dependence of $\max K_a / K_s$ on $4N_e\kappa$ and $\theta$.}
The maximum values of $K_a/K_s$, 
which correspond to the upper bound of selective advantage $s$ (\Eq{\ref{seq: s_max}}), 
at $\Delta G = \Delta G_e$ and at $\Delta G = \Delta G_e + \Delta\Delta G^{\script{sd}}_{\script{fixed}}$
are plotted as a function of $\log 4N_e\kappa$ and $\theta$; 
$\Delta\Delta G^{\script{sd}}_{\script{fixed}}$ is the standard deviation
of $\Delta\Delta G$ over fixed mutants at $\Delta G = \Delta G_e$.
} 
}
\end{figure*}

\FigureInLegends{\clearpage\newpage}

\begin{figure*}[ht]
\FigureInLegends{
\centerline{
\includegraphics*[width=86mm,angle=0]{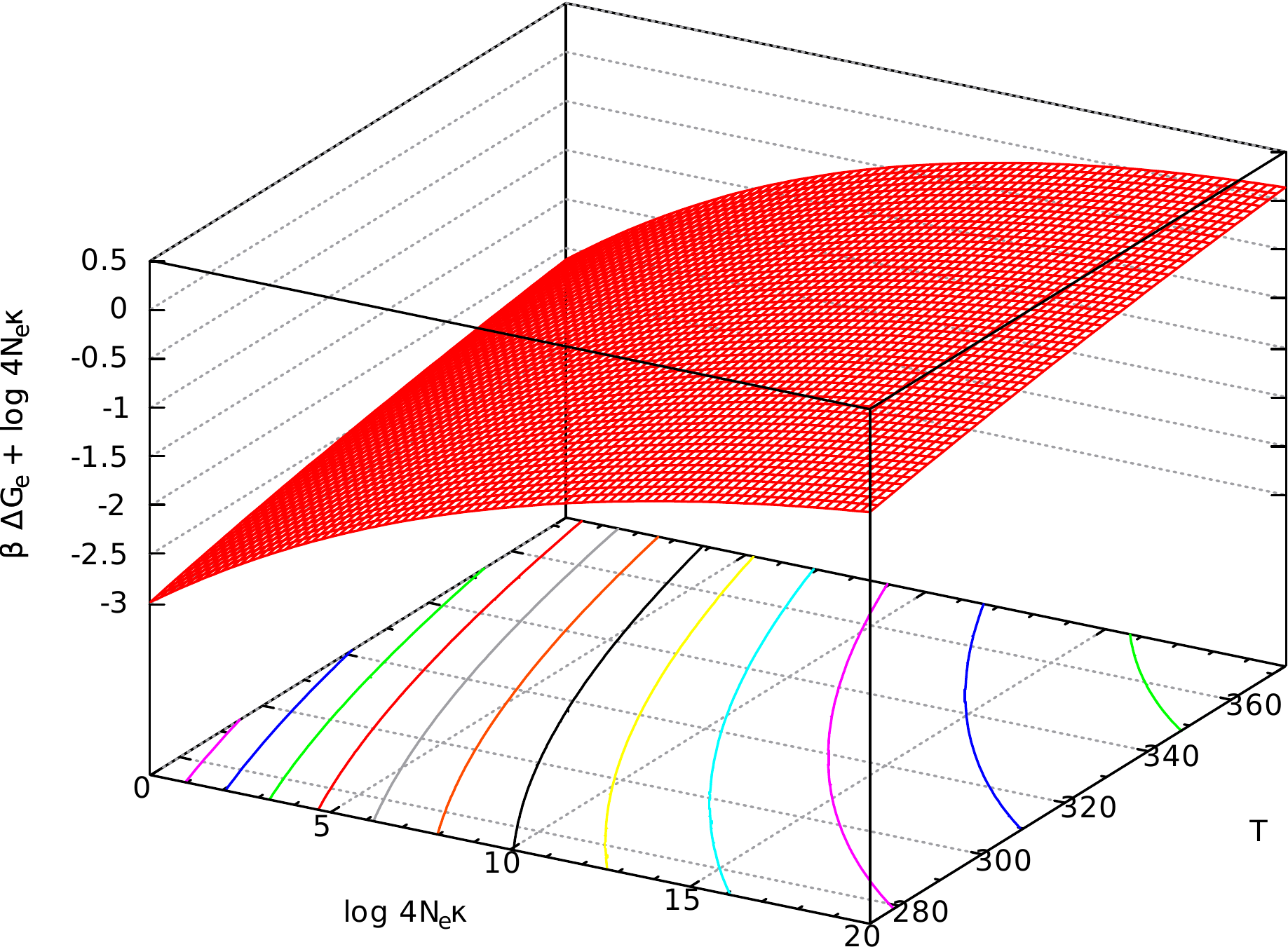}
\includegraphics*[width=90mm,angle=0]{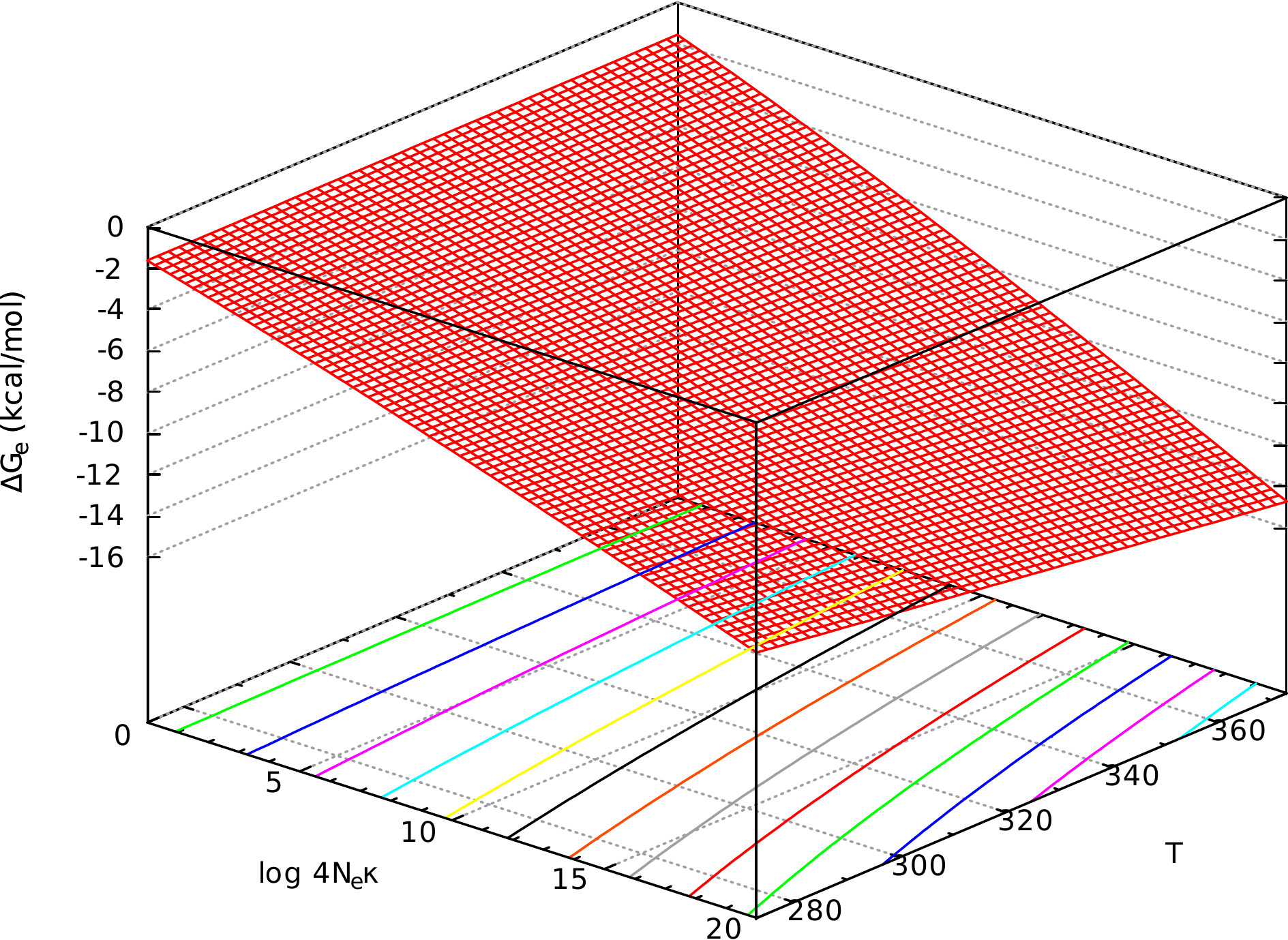}
}
\centerline{
\includegraphics*[width=86mm,angle=0]{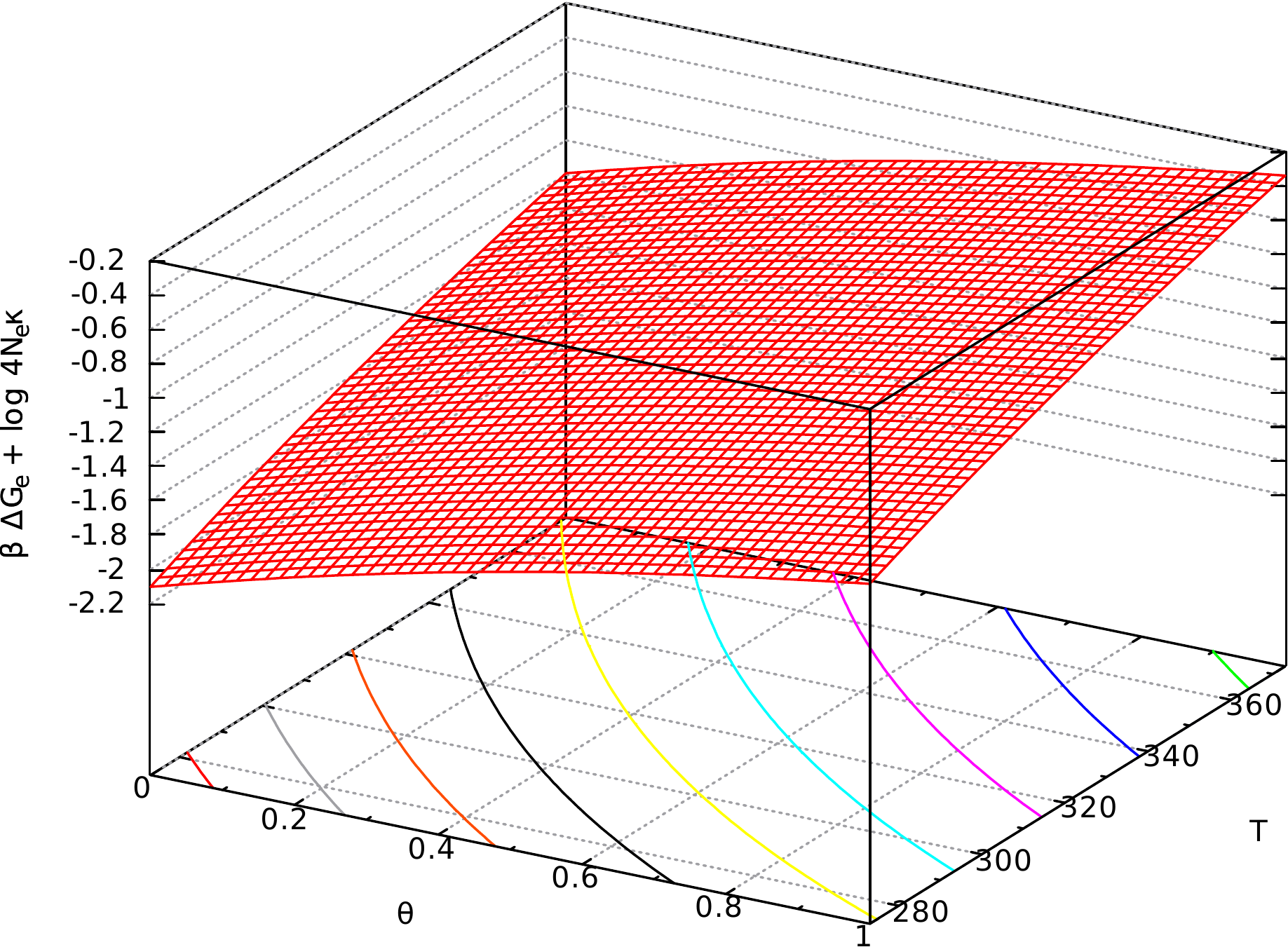}
\includegraphics*[width=88mm,angle=0]{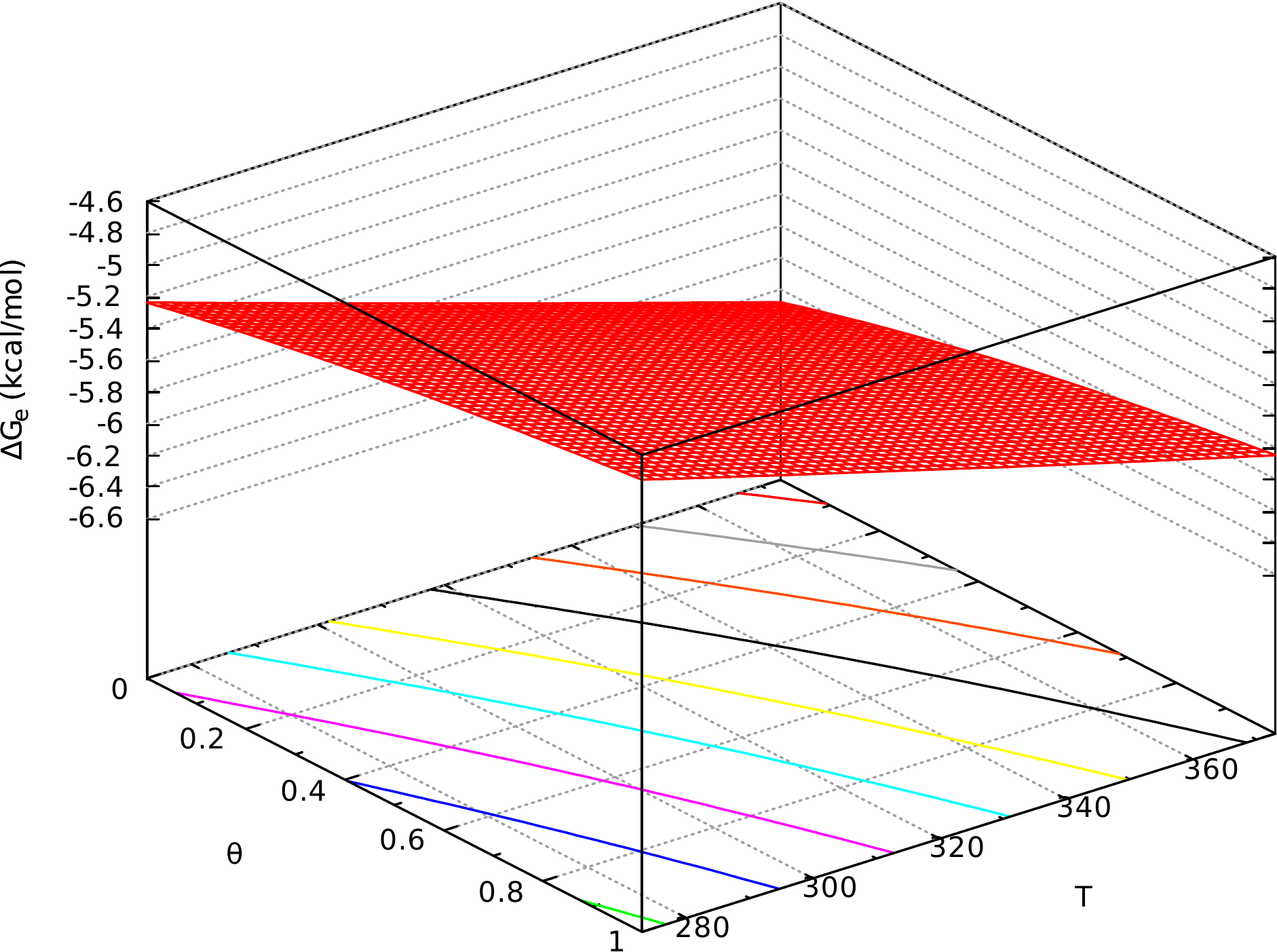}
}
} 
\vspace*{1em}
\caption{
\label{sfig: dependence_of_max4Nes_on_4Nekappa_and_T}
\label{sfig: dependence_of_dGe_on_4Nekappa_and_T}
\label{sfig: dependence_of_max4Nes_on_theta_and_T}
\label{sfig: dependence_of_dGe_on_theta_and_T}
\FigureLegends{
\Red{
\BF{Dependence of equilibrium stability, $\Delta G_e$, on parameters, $4N_e\kappa$, $\theta$ and $T$.}
} 
$\Delta G_e$ is the equilibrium value of folding free energy, $\Delta G$,
in protein evolution.
$T$ is absolute temperature; $\beta = 1 / kT$, where $k$ is the Boltzmann constant.
\EQUATIONS{\Ref{eq: def_bi-Gaussian}, \Ref{eq: def_ms} and \Ref{eq: def_mc}} are assumed for
the distribution of $\Delta\Delta G$ and its dependency on $\Delta G$; they are assumed to be independent of $T$.
Unless specified, 
$\log 4N_e\kappa = 7.550$ and $\theta = 0.53$
are employed.
The value of $\beta\Delta G_e + \log 4N_e\kappa$ is the upper bound of $\log 4N_e s$,
and would not depend on $\log 4N_e \kappa$ if the mean of $\Delta\Delta G$ in all arising mutants did not depend on $\Delta G$; 
see \Eq{\ref{seq: def_s}}.
The kcal/mol unit is used for $\Delta G_e$.
} 
}
\end{figure*}

\FigureInLegends{\clearpage\newpage}

\begin{figure*}[ht]
\FigureInLegends{
\centerline{
\includegraphics*[width=90mm,angle=0]{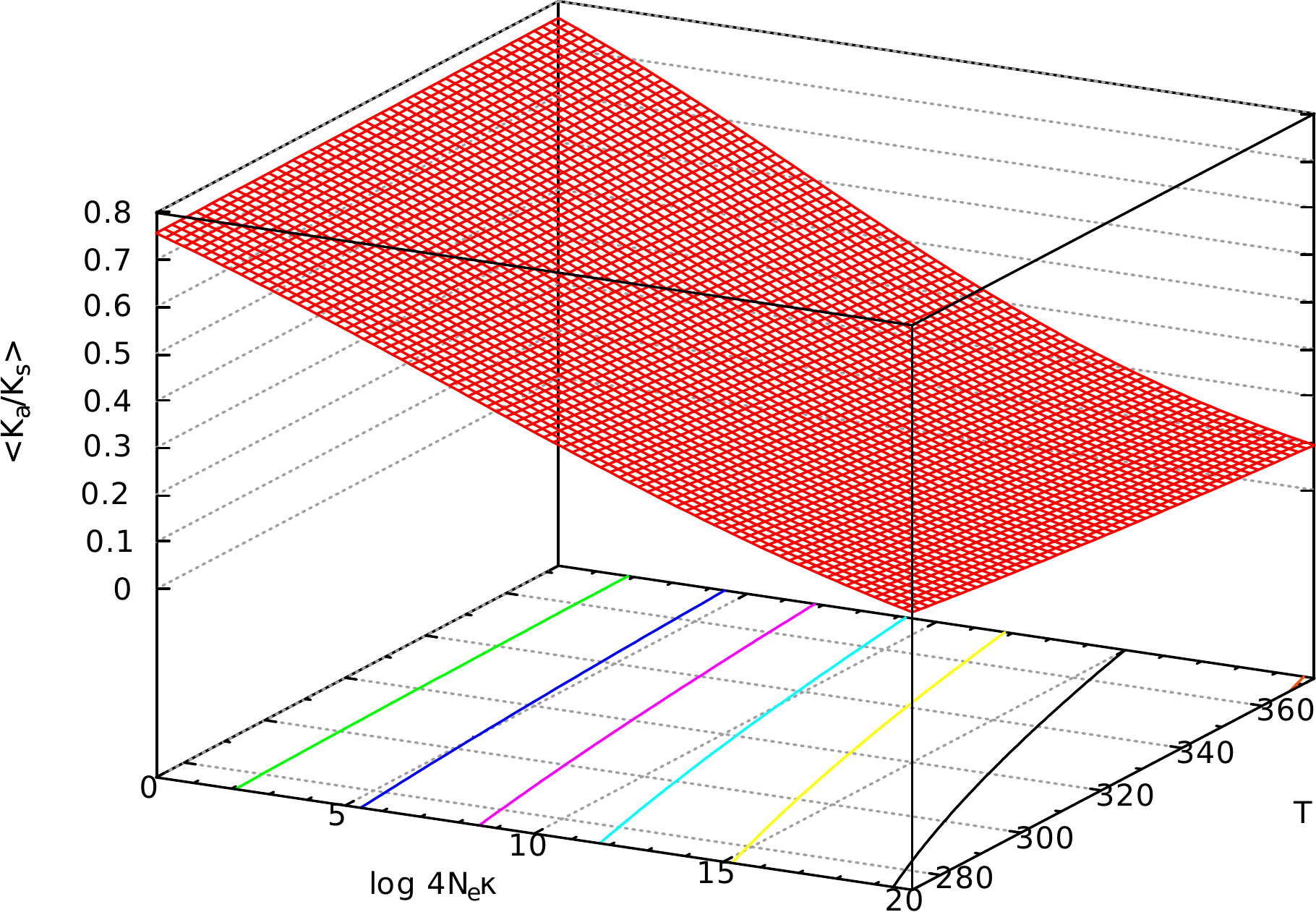}
\includegraphics*[width=86mm,angle=0]{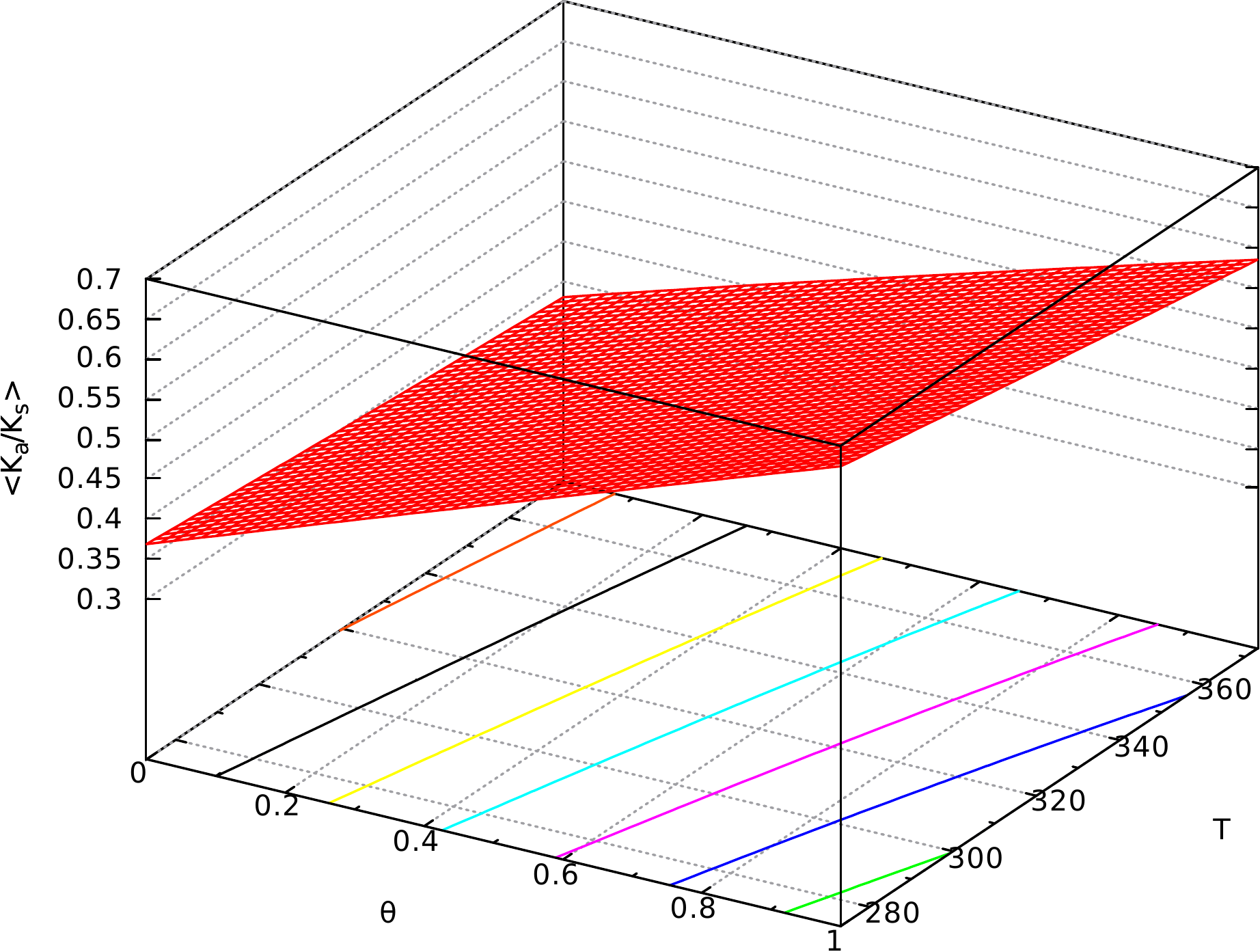}
}
\centerline{
\includegraphics*[width=90mm,angle=0]{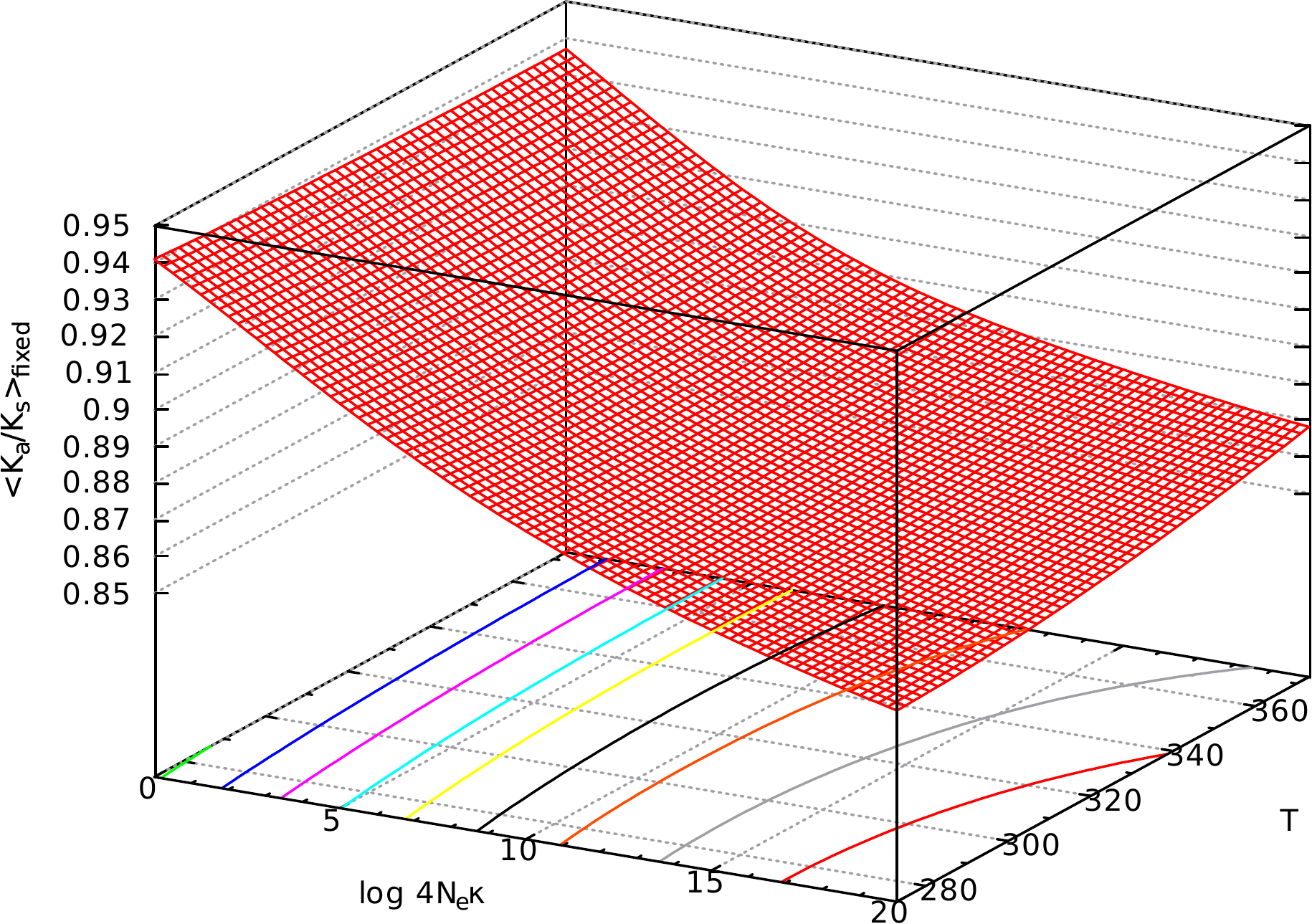}
\includegraphics*[width=86mm,angle=0]{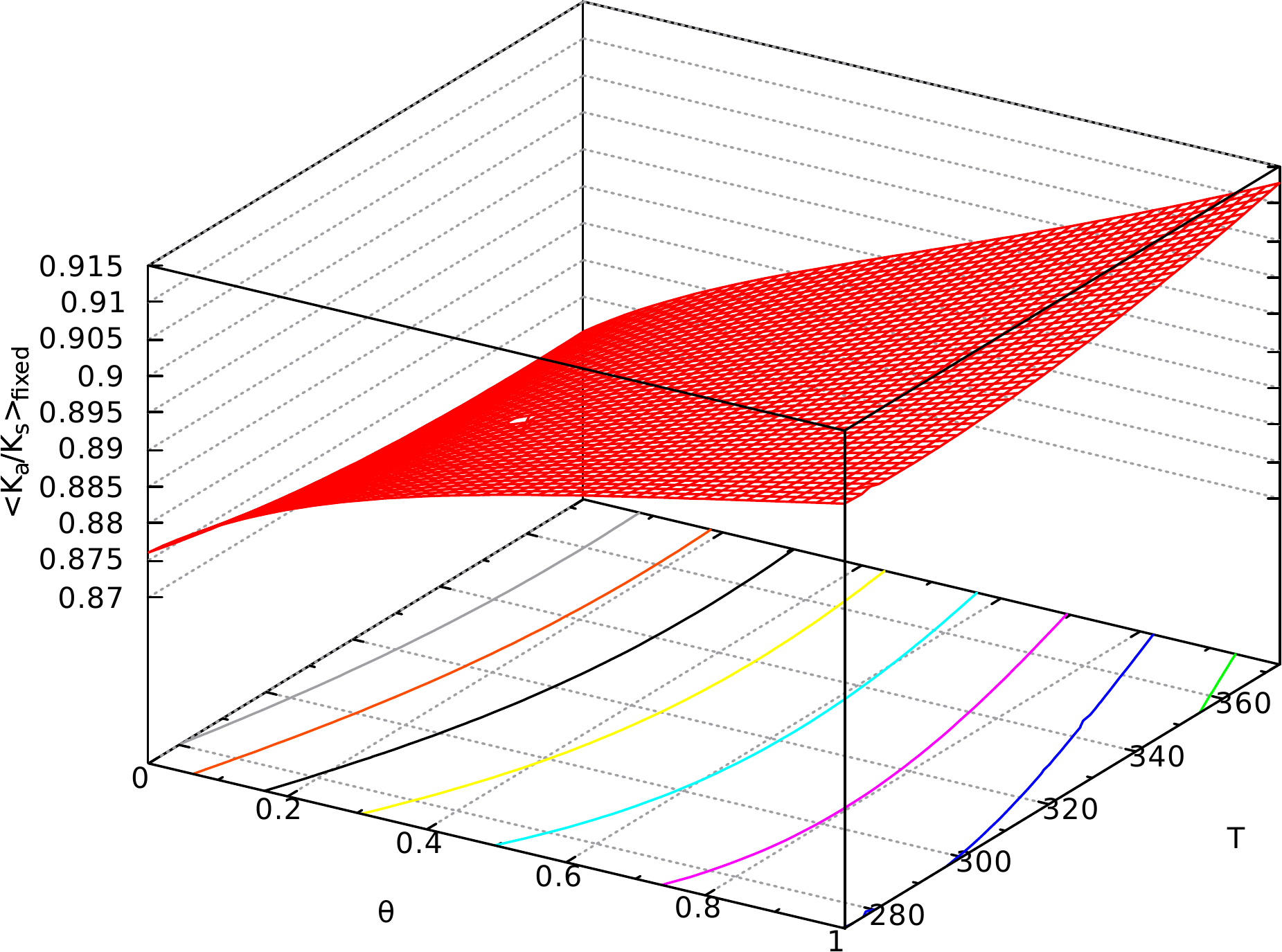}
}
} 
\vspace*{1em}
\caption{
\label{sfig: dependence_of_ave_Ka_over_Ks_on_4Nekappa_and_T}
\label{sfig: dependence_of_ave_Ka_over_Ks_on_theta_and_T}
\label{sfig: dependence_of_ave_Ka_over_Ks_fixed_at_4Nekappa_and_T}
\label{sfig: dependence_of_ave_Ka_over_Ks_fixed_at_theta_and_T}
\FigureLegends{
\Red{
\BF{The average of $K_a / K_s$ over all mutants or over fixed mutants only at equilibrium of protein stability, $\Delta G = \Delta G_e$: Dependence on temperature.}
} 
$T$ is absolute temperature.
\EQUATIONS{\Ref{eq: def_bi-Gaussian}, \Ref{eq: def_ms} and \Ref{eq: def_mc}} are assumed for
the distribution of $\Delta\Delta G$ and its dependency on $\Delta G$; they are assumed to be independent of $T$.
Unless specified, 
$\log 4N_e\kappa = 7.550$ and $\theta = 0.53$
are employed.
} 
}
\end{figure*}

\FigureInLegends{\clearpage\newpage}

\begin{figure*}[ht]
\FigureInLegends{
\centerline{
\includegraphics*[width=90mm,angle=0]{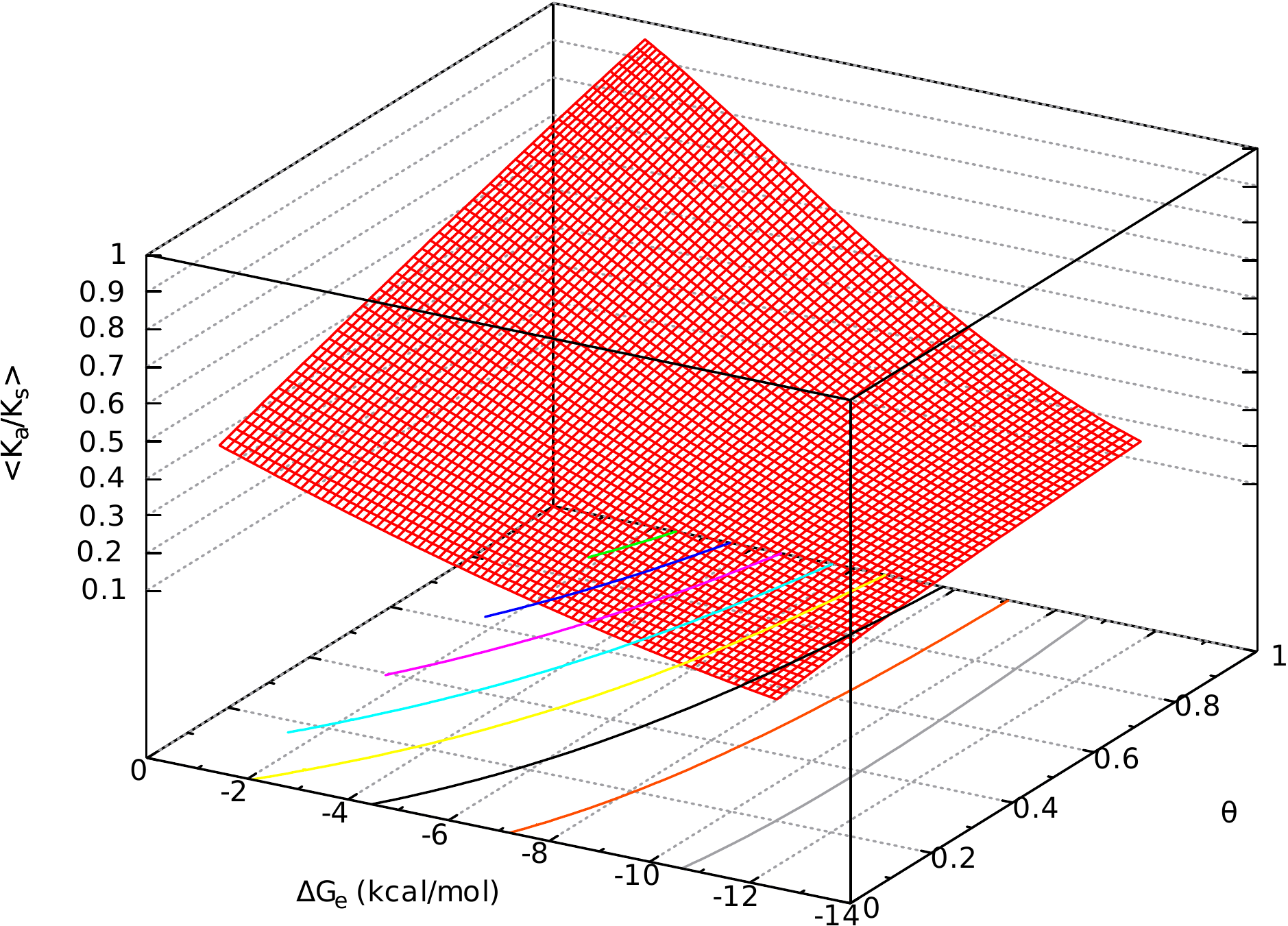}
}
} 
\vspace*{1em}
\caption{
\label{sfig: dependence_of_ave_Ka_over_Ks_on_dGe_and_theta}
\FigureLegends{
\BF{The average of $K_a / K_s$ over all mutants as a function of $\Delta G_e$ and $\theta$.}
} 
}
\end{figure*}

} 


\end{document}